\documentclass[twocolumn,epjc3-upd]{svjour3}  
\pdfoutput=1
\smartqed  

\RequirePackage[T1]{fontenc}
\RequirePackage[utf8]{inputenc}
\RequirePackage{fix-cm}

\RequirePackage{graphicx,amsmath,amssymb,wrapfig,axodraw,color}
\RequirePackage{epsfig,cite}
\RequirePackage{mathptmx}      

\newcommand{\bi}{\begin{itemize}}
 \newcommand{\ei}{\end{itemize}}
\newcommand{\te}{\tilde e}

\def\c1p{C1^\prime}
\def\msq3{\overline{m}_{\tilde{q}}(3)}
\newcommand{\tG}{\tilde G}

\newcommand{\ttau}{\tilde \tau}
\newcommand{\tmu}{\tilde \mu}
\newcommand{\tg}{\tilde g}
\newcommand{\tnu}{\tilde\nu}

\newcommand{\tq}{\tilde q}

\newcommand{\alst}{\stackrel{<}{\sim}}
\newcommand{\agt}{\stackrel{>}{\sim}}
\newcommand{\be}{\begin{equation}}  
\newcommand{\ee}{\end{equation}}  
\newcommand{\bea}{\begin{eqnarray}}  
\newcommand{\eea}{\end{eqnarray}}  
\newcommand{\BE}{\begin{equation}}
\newcommand{\BC}{\begin{center}}
\newcommand{\EC}{\end{center}}
\newcommand{\BEA}{\begin{eqnarray}}
\newcommand{\EEA}{\end{eqnarray}}
\newcommand{\VL}{\left( \begin{array}{c}}
\newcommand{\VR}{\end{array} \right)}
\newcommand{\KKL}{\left[}
\newcommand{\KKR}{\right]}
\newcommand{\KL}{\left(}
\newcommand{\KR}{\right)}

\newcommand{\tz}{\tilde\chi^0}
\newcommand{\aeff}{\al_{\rm eff}}
\newcommand{\Ga}{\Gamma}
\newcommand{\ga}{\gamma}
\providecommand{\ee}{e^+e^-}
\providecommand{\mW}{\MW}
\newcommand{\sweff}{\sin^2{\theta^\ell}_{\mathrm{eff}}}
\newcommand{\MW}{M_W}
\newcommand{\wz}{\sqrt{2}}
\newcommand{\MZ}{M_Z}
\newcommand{\MA}{M_A}
\newcommand{\sw}{s_W}
\newcommand{\cw}{c_W}
\newcommand{\MH}{M_H}
\newcommand{\mh}{m_h}
\newcommand{\mHp}{m_{H^\pm}}
\newcommand{\MHp}{M_{H^\pm}}
\newcommand{\Xt}{X_t}
\newcommand{\mb}{m_{b}}
\newcommand{\tab}{\tan \beta}
\newcommand{\CTb}{\cot \beta}
\newcommand{\Sbe}{\sin \beta}
\newcommand{\Sa}{\sin \alpha}
\newcommand{\Ca}{\cos \alpha}
\newcommand{\Cb}{\cos \beta}
\newcommand{\gf}{G_F}
\newcommand{\SQb}{\sin^2\!\beta}
\newcommand{\CQb}{\cos^2\!\beta}
\newcommand{\CQZb}{\cos^2 2\beta}
\newcommand{\mste}{m_{\tilde{t}_1}}
\newcommand{\mstz}{m_{\tilde{t}_2}}
\newcommand{\mhmax}{$\mh^{\rm max}$}
\newcommand{\br}{{\rm BR}}
\newcommand{\als}{\alpha_s}
\newcommand{\alt}{\alpha_t}
\newcommand{\alb}{\alpha_b}
\def\refeq#1{\mbox{Eq.~(\ref{#1})}}
\def\refta#1{\mbox{Tab.~\ref{#1}}}
\def\reffi#1{\mbox{Fig.~\ref{#1}}}
\def\refeqs#1{\mbox{Eqs.~(\ref{#1})}}
\def\refse#1{\mbox{Sect.~\ref{#1}}}
\def\citere#1{\mbox{Ref.~\cite{#1}}}
\def\citeres#1{\mbox{Refs.~\cite{#1}}}

\newcommand{\cp}{{\cal CP}}
\newcommand{\Mh}{M_h}

\newcommand{\MHexp}{125}
\newcommand{\mt}{m_t}
\newcommand{\At}{A_t}
\newcommand{\mcha}[1]{m_{\tilde \chi^\pm_{#1}}}
\newcommand{\msqd}{m_{\tilde{q}_3}}
\newcommand{\gev}{\,\, \mathrm{GeV}}
\newcommand{\mev}{\,\, \mathrm{MeV}}
\newcommand{\tev}{\,\, \mathrm{TeV}}
\newcommand{\si}{\sigma}
\newcommand{\de}{\delta}
\newcommand{\De}{\Delta}
\newcommand{\db}{\De_b}
\newcommand{\cL}{{\cal L}}
\newcommand{\mbms}{\overline{m}_b}
\newcommand{\lsim}
{\;\raisebox{-.3em}{$\stackrel{\displaystyle <}{\sim}$}\;}
\newcommand{\gsim}
{\;\raisebox{-.3em}{$\stackrel{\displaystyle >}{\sim}$}\;}
\newcommand{\al}{\alpha}
\newcommand{\non}{\nonumber}
\newcommand{\Stop}{\tilde{t}}
\newcommand{\Stope}{\tilde{t}_1}
\newcommand{\Stopz}{\tilde{t}_2}
\newcommand{\Sbot}{\tilde{b}}
\newcommand{\mgl}{m_{\tilde{g}}}
\newcommand{\cH}{{\cal H}}
\newcommand{\cHe}{\cH_1}
\newcommand{\cHz}{\cH_2}
\newcommand{\msbar}{$\overline{\rm{MS}}$}

\newcommand{\seffsf}[1]{\sin\!^2\theta^{#1}_{{\rm eff}}}
\newcommand{\sinleff}{\seffsf{\ell}}
\newcommand{\xspace}{\,}
\renewcommand{\gev}{\ensuremath{\mathrm{Ge\kern -0.1em V}}\xspace}
\renewcommand{\mev}{\ensuremath{\mathrm{Me\kern -0.1em V}}\xspace}
\newcommand{\mc}{\ensuremath{\overline{m}_c}\xspace}
\renewcommand{\mb}{\ensuremath{\overline{m}_b}\xspace}
\renewcommand{\mt}{\ensuremath{m_{t}}\xspace}
\newcommand{\dalphaHadMZ}{\ensuremath{\Delta\alpha_{\rm
had}^{(5)}(M_Z^2)}\xspace}

\newcommand{\deltatheo}{\ensuremath{\delta_{\rm th}}\xspace}
\newcommand{\Rfit}{{\em R}fit\xspace}

\newcommand{\ft}{\footnotesize}

\def\order#1{\ensuremath{{\cal O}(#1)}}
\journalname{Eur. Phys. J. C}
\begin{document}

\title{Physics at the $e^+ e^-$ Linear Collider \thanksref{}}

\titlerunning{Linear Collider Physics}        

\author{Editors: G.~Moortgat-Pick\thanksref{e1,addr1,addr2}
\and H.~Baer\thanksref{addr3} 
\and M.~Battaglia\thanksref{addr4} 
\and G.~Belanger\thanksref{addr5} 
\and K.~Fujii\thanksref{addr6} 
\and J.~Kalinowski\thanksref{addr7}
\and S.~Heinemeyer\thanksref{addr8} 
\and Y.~Kiyo\thanksref{addr9} 
\and K.~Olive\thanksref{addr10}
\and F.~Simon\thanksref{addr11} 
\and P.~Uwer\thanksref{addr12} 
\and D.~Wackeroth\thanksref{addr13} 
\and P.M.~Zerwas\thanksref{addr2}
\\
Specific Contributions: 
A.~Arbey\thanksref{addr44a,addr44b,addr44c}
\and M.~Asano\thanksref{addr15}
\and J.~Bagger\thanksref{addr50,addr62}
\and P.~Bechtle\thanksref{addr16}
\and A.~Bharucha\thanksref{addr17,addr60}
\and J.~Brau\thanksref{addr51}
\and F.~Br{\"u}mmer\thanksref{addr18}
\and S.Y.~Choi\thanksref{addr19}        
\and A.~Denner\thanksref{addr20}
\and K.~Desch\thanksref{addr16}
\and S.~Dittmaier\thanksref{addr21}
\and U.~Ellwanger\thanksref{addr22} 
\and C.~Englert\thanksref{addr23}
\and A.~Freitas\thanksref{addr24}
\and I.~Ginzburg\thanksref{addr25}
\and S.~Godfrey\thanksref{addr26}
\and N.~Greiner\thanksref{addr11,addr2}
\and C.~Grojean\thanksref{addr2,addr28}
\and M.~Gr{\"u}newald\thanksref{addr29}
\and J.~Heisig\thanksref{addr30}
\and A.~H{\"o}cker\thanksref{addr31}
\and S.~Kanemura\thanksref{addr32}
\and K.~Kawagoe\thanksref{addr53}
\and R.~Kogler\thanksref{addr33}
\and M.~Krawczyk\thanksref{addr7}
\and A.S.~Kronfeld\thanksref{addr55,addr61}
\and J.~Kroseberg\thanksref{addr16}
\and S.~Liebler\thanksref{addr1,addr2}
\and J.~List\thanksref{addr2}
\and F.~Mahmoudi\thanksref{addr44a,addr44b,addr44c}
\and Y.~Mambrini\thanksref{addr22}
\and S.~Matsumoto\thanksref{addr37}
\and J.~Mnich\thanksref{addr2}
\and K.~M{\"o}nig\thanksref{addr2}
\and M.M.~M{\"u}hlleitner\thanksref{addr39}
\and R.~P{\"o}schl\thanksref{addr47}
\and W.~Porod\thanksref{addr20}
\and S.~Porto\thanksref{addr1}
\and K.~Rolbiecki\thanksref{addr7,addr40}
\and M.~Schmitt\thanksref{addr41}
\and P.~Serpico\thanksref{addr5}
\and M.~Stanitzki\thanksref{addr2}
\and O.~St\aa{}l\thanksref{addr59}
\and T.~Stefaniak\thanksref{addr4}
\and D.~St{\"o}ckinger\thanksref{addr43}
\and G.~Weiglein\thanksref{addr2}
\and G.W.~Wilson\thanksref{addr45}
\and L.~Zeune\thanksref{addr46}\\
LHC contacts:
F. Moortgat\thanksref{addr31}
\and
S. Xella\thanksref{addr49}\\
Advisory Board:
J.~Bagger\thanksref{addr50,addr62}
\and J.~Brau\thanksref{addr51}
\and J.~Ellis\thanksref{addr52,addr31}
\and K.~Kawagoe\thanksref{addr53}
\and S.~Komamiya\thanksref{addr54}
\and A.S.~Kronfeld\thanksref{addr55,addr61}
\and J.~Mnich\thanksref{addr2}
\and M.~Peskin\thanksref{addr56}
\and D.~Schlatter\thanksref{addr31}
\and A.~Wagner\thanksref{addr2,addr33}
\and H.~Yamamoto\thanksref{addr58}}

\thankstext{e1}{e-mail: gudrid.moortgat-pick@desy.de}

\institute{II. Institute of Theo. Physics, University
  of Hamburg, D-22761 Hamburg,
  Germany \label{addr1}
           \and
           Deutsches Elektronen Synchrotron (DESY), 
Hamburg und Zeuthen, D-22603 Hamburg, Germany
 \label{addr2}
           \and
           Dept. of Physics and Astronomy, University of Oklahoma, Norman, 
OK73019, USA\label{addr3}
\and
Santa Cruz Institute for Particle Physics, University of California Santa 
Cruz, Santa Cruz CA, USA\label{addr4}
\and
Laboratoire de Physique Theorique 
(LAPTh), Universit\'e Savoie Mont Blanc, 
CNRS, B.P.110, Annecy-le-Vieux F-74941, France\label{addr5}
\and
High Energy Accelerator Research Organization (KEK), Tsukuba, Japan\label{addr6}
\and
Faculty of Physics, University of Warsaw, 02093 Warsaw, Poland\label{addr7}
\and
Instituto de F\'isica de Cantabria (CSIC-UC), E-39005 Santander, 
Spain\label{addr8}
\and
Department of Physics, Juntendo University,
Inzai, Chiba 270-1695, Japan\label{addr9}
\and
William I. Fine Theoretical Physics Institute, School of Physics and Astronomy,
University of Minnesota, Minneapolis, MN 55455, USA\label{addr10}
\and
Max-Planck-Institut f\"ur Physik, 80805 M\"unchen, Germany\label{addr11}
\and
Humboldt-Universit\"at zu Berlin, Institut f\"ur Physik, D-12489 Berlin, 
Germany\label{addr12}
\and
Department of Physics, SUNY at Buffalo, Buffalo, NY 14260-1500, USA\label{addr13}
\and
Physikalisches Institut and Bethe Center for Theoretical Physics
Universit\"at Bonn, D-53115 Bonn, Germany\label{addr15}
\and
Physikalisches Institut, University of Bonn, Bonn, Germany\label{addr16}
\and
Physik Department T31, Technische Universit\"at M\"unchen,
D-85748 Garching, Germany\label{addr17}
\and
LUPM, UMR 5299, Universit\'{e} de Montpellier II et CNRS, 34095 Montpellier, 
France
\label{addr18}
\and
Department of Physics, Chonbuk National University, Jeonju 
561-756, Republic of Korea\label{addr19}
\and
Universit\"at W\"urzburg, Institut f\"ur Theoretische Physik 
und Astrophysik, D-97074 W\"urzburg, Germany\label{addr20}
\and
Physikalisches Institut, Albert–Ludwigs–Universit\"at Freiburg, 
D–79104 Freiburg, Germany\label{addr21}
\and
Laboratoire de Physique, 
UMR 8627, CNRS, Universite de Paris-Sud, 91405 Orsay, France\label{addr22}
\and
SUPA, School of Physics and Astronomy,
University of Glasgow, Glasgow G12 8QQ, United Kingdom\label{addr23}
\and
PITT PACC,
Department of Physics \& Astronomy, Univ. of Pittsburgh, Pittsburgh, 
PA 15260, USA\label{addr24}
\and
Sobolev Institute of Mathematics and Novosibirsk State University, 
Novosibirsk, 630090, Russia\label{addr25}
\and
Ottawa-Carleton Institute for Physics, Department of Physics,
Carleton University, Ottawa, Canada K1S 5B6\label{addr26}
\and
ICREA at IFAE, Universitat Autonoma de Barcelona, E-08193 Bellaterra,
Spain\label{addr28}
\and
University College Dublin, Dublin, Ireland\label{addr29}
\and
Institute for Theoretical Particle Physics and Cosmology, RWTH 
Aachen University, 52056 Aachen, Germany\label{addr30}
\and
CERN, Geneva, Switzerland\label{addr31}
\and
Department of Physics, University of Toyama, 3190 Gofuku, Toyama, 
930-8555, Japan\label{addr32}
\and
University of Hamburg, Hamburg, Germany\label{addr33}
\and
Kavli IPMU (WPI), The University of Tokyo, Kashiwa, Chiba 277-8583, 
Japan\label{addr37}
\and
Institute for Theoretical Physics, Karlsruhe Institute of Technology,
76128 Karlsruhe, Germany\label{addr39}
\and
Instituto de Fisica Teorica, IFT-UAM/CSIC, Universidad
Autonoma de Madrid, Cantoblanco, 28049 Madrid, Spain\label{addr40}
\and
Department of Physics and Astronomy, Northwestern University, Evanston, 
IL 60091,USA\label{addr41}
\and
The Oskar Klein Centre, Department of Physics, Stockholm University,
SE-106 91 Stockholm, Sweden\label{addr59}
\and
Institut f\"ur Kern- und Teilchenphysik, TU Dresden, 01069 Dresden, 
Germany\label{addr43}
\and
Universit{\' e} de Lyon, Universit{\' e} Lyon 1, 69622 Villeurbonne Cedex, 
France\label{addr44a}
\and
Centre de Recherche Astrophysique de Lyon, CNRS, UMR 5574, 
69561 Saint-Genis Laval Cedex, France\label{addr44b}
\and
Ecole Normale Sup{\' e}rieure de Lyon, France
\label{addr44c}
\and
Department of Physics and Astronomy, university of Kansas, 
Lawrence, KS 66045, USA\label{addr45}
\and
ITFA, University of Amsterdam, Science Park 904, 1018 XE, Amsterdam, 
The Netherlands\label{addr46}
\and
Laboratoire de L'accelerateur Lineaire (LAL), CNRS/IN2P3,
Orsay, France\label{addr47}
\and
Niels Bohr Institute, University of Copenhagen, Kobenhavn, Denmark\label{addr49}
\and
Department of Physics and Astronomy, Johns Hopkins University
Baltimore, MD 21218, USA\label{addr50}
\and
Department of Physics, University of Oregon, Eugene, OR 97403, USA\label{addr51}
\and
Theoretical Particle Physics and Cosmology Group, Department of Physics,
King’s College London, Strand, London WC2R 2LS, U.K\label{addr52}
\and
Department of Physics, Kyushu University,
6-10-1 Hakozaki, Higashi-ku, 812-8581 Fukuoka, Japan\label{addr53}
\and
Department of Physics, Graduate School of Science, and 
International Center for Elementary Particle Physics, The University of Tokyo,
Tokyo 113-0033, Japan\label{addr54}
\and
Theoretical Physics Department, Fermi National Accelerator
Laboratory, Batavia, Illinois, USA
\label{addr55}
\and
SLAC, Stanford University, Menlo Park, California 94025 USA\label{addr56}
\and
Department of Physics, Tohoku University, Sendai, Miyagi, Japan\label{addr58}
\and
CNRS, Aix Marseille U., U. de Toulon, CPT, F-13288,
Marseille, France\label{addr60}
\and
Institute for Advanced Study, Technische Universit\"at M\"unchen,
D-85748 Garching, Germany\label{addr61}
\and
TRIUMF, Vancouver, BC V6T 2A3, Canada\label{addr62}
}

\date{Received: 25 March 2015 / Accepted: 9 May 2015}

\maketitle

\begin{abstract}
A comprehensive review of physics at an $e^+e^-$ Linear Collider in the
 energy range of $\sqrt{s}=92$~GeV--3~TeV is presented in view of
 recent and expected LHC results, experiments from low energy as well as
astroparticle physics.
 The report focuses in particular on  
Higgs boson, top quark and electroweak precision physics, but also 
discusses several models of beyond the Standard Model physics such
as Supersymmetry, 
little Higgs models and extra gauge bosons. The connection to cosmology 
has been analyzed as well.

\end{abstract}
 
\tableofcontents
 \section{Executive Summary}

\subsection{Introduction}
\label{intro}

With the discovery of a Higgs boson with a mass
of about $m_H= 125$~GeV based on data runs
at the Large Hadron Collider in its first stage  at $\sqrt{s}=7$ and 
$8$~TeV,  
the striking concept of explaining `mass' as consequence of a 
spontaneously broken symmetry received a decisive push forward.
The significance of this discovery was acknowledged by the award of the
Nobel prize for physics to P.~Higgs and F.~Englert in 2013
\cite{Englert:1964et,Higgs:1964ia,Higgs:1964pj,Guralnik:1964eu}.
The underlying idea of the Brout-Englert-Higgs (BEH) mechanism is 
the existence of a self-interacting Higgs field with a specific potential. 
The peculiar property of this Higgs field is that it is non-zero  
in the vacuum. In other words the Higgs field provides the vacuum a 
structure. The relevance of such a field 
not only for our understanding of matter but also 
for the history of the Universe is immanent.

The discovery of a Higgs boson as the materialization of the Higgs field
was the first important step in accomplishing our present level of
understanding of the fundamental interactions of nature and the
structure of matter that is adequately described by the Standard Model
(SM). In the SM the constituents of matter are fermions, leptons and quarks,
classified in three families with identical quantum properties. The 
electroweak and strong interactions are transmitted via the gauge bosons  
described by gauge field theories with the fundamental 
symmetry group $SU(3)_C\times SU(2)_L\times U(1)_Y$.

However, the next immediate steps are to answer the 
following questions:
\begin{itemize}
\item Is there  just one Higgs?
\item Does the Higgs field associated to the discovered particle
 really cause the corresponding couplings with all particles?
  Does it provide the right structure of the vacuum?
\item Is it a SM Higgs (width, couplings, spin)?
Is it a pure CP-even Higgs boson 
as predicted in the SM, or is it a Higgs boson from an extended Higgs sector, 
possibly with some admixture of a CP-odd component?
To which model beyond the Standard Model (BSM) does it point?
\end{itemize}
\begin{sloppypar}
In order to definitively establish the mechanism of electroweak symmetry
breaking, all Higgs boson properties (mass, width, couplings, quantum numbers)
 have to be precisely measured
and 
compared with the mass of the 
corresponding particles.
\end{sloppypar}

The LHC has excellent prospects 
for the future runs\footnote{As one example for a recent and comprehensive
 review of the LHC run-1 results, see ~\cite{lhcbook} and 
references therein.} 
2 and 3 where proton-proton beams collide
with an energy of $\sqrt{s}=$13~TeV starting in spring 2015, continued 
by runs with a foreseen high luminosity upgrade in the following
decade~\cite{Lamont-eps-proc}.  High--energy $e^+e^-$-colliders have
already been essential instruments in the past to search for the
fundamental constituents of matter and establish their interactions.
The most advanced design for a future lepton collider is the
International Linear Collider (ILC) that is laid out for the energy
range of $\sqrt{s}=90$~GeV--1~TeV~\cite{AguilarSaavedra:2001rg,DBD}.  In
case a drive beam accelerator technology can be applied, 
an energy frontier of about $3$~TeV might be accessible with the Compact 
Linear Collider (CLIC)\cite{Linssen:2012hp}.

At an $e^+e^-$ linear collider (LC) one expects rather clean
experimental conditions compared to the conditions at the LHC where
one has many overlapping events due to the QCD background from
concurring events. A direct consequence is that one does not need any
trigger at an LC but can use all data for physics analyses.  Due to the
collision of point-like particles the physics processes take place at
the precisely and well-defined initial energy $\sqrt{s}$, both stable
and measureable up to the per mille level. The energy at the LC
is tunable which
offers to perform precise energy scans and to optimize the kinematic conditions
for the different
physics processes, respectively. In addition, the beams can be
polarized: the electron beam up to about 90\%, the positron beam up to
about 60\%. With such a high degree of polarization, the initial state
is precisely fixed and well-known. Due to all these circumstances the final
states are generally fully reconstructable so that numerous observables as 
masses, total cross sections but also 
differential energy and angular distributions are available for data analyses.

The quintessence of LC physics at the precision frontier is
high luminosity and beam
polarization, tunable energy, precisely defined initial state and
clear separation of events via excellent detectors.
The experimental conditions
that are necessary to fulfill the physics requirements have been defined
in the LC scope documents\cite{scope}.

Such clean experimental conditions for high precision measurements at
a LC are the {\it `sine qua non'} for resolving the current puzzles
and open questions. They allow to analyze the physics data in a
particularly model-independent approach.
The compelling physics case for a LC has been described in numerous
publications as, for instance
\cite{AguilarSaavedra:2001rg,Abe:2001wn,Abe:2001gc,DBD}, a short and
compact overview is given in~\cite{Diberder}.

Although the SM has been tremendously successful and its predictions
experimentally been tested with accuracies at the quantum level, 
i.e. significantly
below the 1-percent level, the SM cannot be regarded as the final 
theory describing all aspects of Nature. 
Astro-physical measurements~\cite{Xenon100,subaru} are 
consistent with a universe that
contains only 4\% of the total energy composed of
ordinary mass but hypothesize the existence
of dark matter (DM) 
accounting for 22\% of the total energy that is responsible for
gravitational effects although no visible mass can be seen.  
Models accounting for dark matter  
can easily be embedded within BSM theories as, for instance,  
supergravity\cite{Kallosh:2014xwa}.  
The
strong belief in BSM physics is further supported by the absence of
gauge coupling unification in the SM as well as its failure to explain
the observed existing imbalance between baryonic and antibaryonic
matter in our universe.
Such facets together with the experimental data strongly
support the interpretation that the SM picture is not
complete but constitutes only a low-energy limit of an all-encompassing
`theory of everything', embedding gravity and quantum theory to describe all 
physical aspects of the universe.
Therefore experimental hints for BSM physics 
 are expected to manifest themselves at future colliders
and model-independent strategies are crucial to determine the 
underlying structure of the model. 

{\it A priori} there are only two approaches 
to reveal signals of new physics and to manifest the model of 
BSM at future experiments. 
Since the properties of the matter and gauge particles in the SM may 
be affected by the new energy scales, a 'bottom-up' approach consists in 
performing high precision studies of the top, Higgs and electroweak 
gauge bosons. Deviations from those measurements 
to SM predictions reveal hints to BSM physics.
Under the assumption that future experiments can be performed at 
energies high enough to cross new thresholds, a 'top-down' approach 
becomes also feasible where the new particles or interactions can be 
produced and studied directly.

\begin{sloppypar}
Obviously, the complementary search strategies at lepton and hadron 
colliders are predestinated for such  successful dual approaches. 
A successful high-energy
LC was already realized
in the 90s with
the construction and running of the SLAC Linear Collider (SLC) that
delivered up to $5\times 10^{10}$ particles per
pulse.
Applying in addition highly polarized electrons enabled the
SLC to provide the best single measurement of the electroweak mixing
angle with $\delta \sin^2\theta_W$ $\sim 0.00027$.
\end{sloppypar}

However, such a high precision manifests
a still-existing inconsistency, 
namely  the well--known discrepancy 
between the left-right polarization asymmetry at the Z-pole
measured at
SLC and the forward-backward asymmetry measured at LEP\cite{LEPEWWG}.  
Both values lead
to measured values of the electroweak mixing angle $\sin^2\theta_{\rm eff}$
that differ by more than 3-$\sigma$ and point to different predictions for
the Higgs mass, see Sect.\ref{sec:quantum} for more details. 
Clarifying the central value as well as improving the
precision is essential for testing the consistence of the SM as
well as BSM models. 

Another example for the relevance of highest precision 
measurements and their interplay
with most accurate theoretical predictions at the quantum level is 
impressively demonstrated in the interpretation of the muon anomalous
moment $g_{\mu}-2$~\cite{Bennett:2006}. 
The foreseen run of the $g_{\mu}-2$ experiment at
Fermilab, starting in 2017 \cite{FNALProposal,Mibe}, 
will further improve the current experimental precision by about a factor four
 and will set substantial bounds to many new
physics models via their high sensitivity to virtual effects of new
particles.

The LC concept has been proposed already
in 1965~\cite{Tigner:1965wf} for providing electron 
beams with high enough quality
for collision experiments. In \cite{Amaldi:1975hi} this concept has been 
proposed for collision experiments at high energies
in order to avoid the energy loss via synchrotron radiation: this energy loss
per turn scales with $E^4/r$, where $E$ denotes the beam energy and $r$ the 
bending radius. 
The challenging problems at the LC compared to 
circular colliders, however, 
are the luminosity and the energy transfer to the beams.
The luminosity is given by
\begin{equation}
{\cal L}\sim \frac{\eta P N_e}{\sigma_{xy} E_{c.m.}}\label{intro-eq-lumi},
\end{equation}
where $P$ denotes the required power with efficiency $\eta$, $N_e$ the
charge per bunch, $E_{c.m.}$ the center-of-mass energy and
$\sigma_{xy}$ the transverse geometry of the beam size.  From
eq.~(\ref{intro-eq-lumi}), it is obvious that flat beams and high
bunch charge allow  high luminosity with lower required beam
power $P_b=\epsilon P$.  The current designs for a high luminosity
$e^+e^-$ collider, ILC or CLIC, 
is perfectly aligned with such arguments.  One
expects an efficiency factor of about $\eta\sim20\%$ 
for the discussed designs.


The detectors are designed to improve the momentum resolution
from tracking by a factor 10 and the jet energy resolution by a factor
3 (in comparison with the CMS detector) and excellent $\tau^{\pm}$-, 
$b$, $\bar{b}$- and 
$c$, $\bar{c}$-tagging capabilities~\cite{DBD}, are
expected. 

As mentioned before, another novelty is the availability of 
the polarization of both beams, which can
precisely project out the interaction vertices and can analyze its
chirality directly.  

The experimental conditions to achieve such an
unprecedented precision frontier at high energy are high luminosity
(even about three orders of magnitude more particles per pulse,
$5\times 10^{13}$ than at the SLC), polarized electron/positron beams, 
tunable energy,
luminosity and beam energy stability below $0.1\%$ level~\cite{scope}.
Assuming a finite total overall running time it is a critical
 issue to divide up the available time between the different energies, 
polarizations and running options in order to maximize the physical results.
Several running scenarios are thoroughly studied~\cite{parametergroup}.

\begin{sloppypar}
In the remainder of this chapter we
summarize the physics highlights of this report. The corresponding 
 details can be found in the following chapters.
Starting with the three safe pillars of LC physics ---Higgs-, Top- and 
electroweak high precision physics---
chapter~\ref{hewsb} provides a comprehensive overview about 
the physics of electroweak symmetry breaking.
Recent developments in LHC analyses as well as on the theory side are
included, alternatives to the 
Higgs models are discussed. 
Chapter~\ref{Top-QCD} 
covers QCD and in particular
top quark physics. 
The LC will also 
set a new frontier in experimental precision physics and has a 
striking potential for 
discoveries in indirect searches. In chapter~\ref{sec:quantum} 
the impact of electroweak precision observables and their
interpretation within BSM physics are discussed. 
Supersymmetry (SUSY) is a well-defined example for physics beyond the
SM with high predictive power. Therefore in chapter~\ref{susy}
the potential of a LC for unravelling and determining the underlying
structure in different SUSY models is discussed.  Since many aspects of new
physics have strong impact on astroparticle physics and cosmology,
chapter \ref{astro} provides an overview in this regard.
\end{sloppypar}

The above mentioned safe physics topics can be realized at best
at different energy stages at the linear collider. The possible staged energy
approach for a LC is therefore
ideally suited to address all the different physics
topics. For some specific physics questions
very high luminosity is required 
and in
this context also a high-luminosity option at the LC is discussed, see 
\cite{parametergroup} for technical details.
The expected physics results of the high-luminosity LC
was studied in different
working group reports~\cite{jimbrau,ref:snowmass_higgs}, cf. Sect.\ref{sec:ewsb3}.

Such an optimization of the different running options of a LC
depends on the still awaited physics demands.
The possible physics outcome of different running
scenarios at the LC are currently under study\cite{parametergroup}, but
fixing the final running strategy is not yet advisable. 

One should note, however, that such a  large machine flexibility 
is one of the striking features of a LC.


\subsection{Physics highlights}
\label{exec}

Many of the examples shown in this review are based on results of
\cite{DBD,Linssen:2012hp,Moortgat-Pick:2013awa,lcnotes}
and references therein.\\

\noindent {\bf Higgs physics}\\[.5em]
The need for precision studies of the new boson, 
compatible with a SM-like Higgs, illuminates already
the clear path for taking data at different energy stages at the LC.

\begin{sloppypar}
For a Higgs boson with a mass of
$~125$~GeV, the
first envisaged energy stage is at about $\sqrt{s}=250$~GeV: the
dominant Higgsstrahlung process peaks at $\sqrt{s}=240$~GeV.
This energy stage allows the model-independent measurement of the
cross section $\sigma(HZ)$ with an accuracy of about 2.6\%,
cf.\ Sect.\ref{sec:ewsb3}. This quantity is the crucial ingredient for all
further Higgs analyses, in particular for deriving the total width via
measuring
the ratio of the partial width and the corresponding branching ratio. 
Already at this stage many couplings 
can be determined with  high accuracy in a model-independent way: 
a striking example is the precision 
of $1.3\%$ that can be expected for the 
coupling $g_{HZZ}$, see Sect.\ref{sec:ewsb3} 
for more details.
\end{sloppypar}

The precise determination of the mass is of interest in its own
right. However, it has also high impact for probing the Higgs physics,
since $m_H$ is a crucial input parameter. For instance, the branching
ratios $H\to ZZ^*$, $WW^*$ are very sensitive to $m_H$: a change in
$m_H$ by 200 MeV shifts $BR(H\to ZZ^*$ by 2.5\%.  Performing accurate
threshold scans enables the most precise mass measurements of $\delta
m_H=40$~MeV.  Furthermore and --of more fundamental relevance-- such
threshold scans in combination with measuring different angular
distributions allow a model-independent and unique determination of
the spin.
%

Another crucial quantity in the Higgs sector is 
the total width $\Gamma_H$ of the Higgs boson. 
The prediction in the SM is
$\Gamma_H=$4.07~MeV for $m_H=125$~GeV~\cite{Denner:2011mq}.  
The direct measurement of
such a small width is neither possible at the LHC nor at the LC
since it is much smaller than any detector resolution.
Nevertheless, at the LC a model-in\-de\-pen\-dent determination of 
$\Gamma_H$ can be achieved using the absolute measurement of Higgs 
branching ratios together with measurements of the corresponding 
partial widths. An essential
input quantity in this context is again the precisely measured total cross
section of the Higgsstrahlung process. 
At $\sqrt{s}=500$~GeV, one can derive the total width $\Gamma_H$ with 
a precision of $5\%$ based on a combination of the  
$H\to ZZ^*$ and $WW^*$ channels. Besides this 
model-in\-de\-pen\-dent determination, which is unique to the LC, 
constraints on the total width can also be obtained at the LC 
from a combination of on-- and off--shell Higgs 
contributions~\cite{Liebler:2015aka} in a 
similar way as at the LHC~\cite{Caola:2013yja}. The latter method, 
however, relies on certain theoretical assumptions, and also 
in terms of the achievable accuracy 
it is not competitive 
with the model-in\-de\-pen\-dent measurement based on the
production cross section $\sigma(ZH)$\cite{Liebler:2015aka}.

At higher energy such off-shell decays of the Higgs boson to
pairs of $W$ and $Z$ bosons 
offer access to the kinematic dependence of higher dimensional operators 
involving the Higgs boson. This dependence allows for example the test of
unitarity in BSM models~\cite{Gainer:2014hha,Ghezzi:2014qpa}.

In order to really establish the mechanism of electroweak symmetry breaking 
it is not only important to measure all couplings but also to
measure the Higgs potential:
$$ 
V=\frac{1}{2} m_H^2 \Phi^2_H + \lambda v \Phi_H^3+\frac{1}{4} \kappa \Phi_H^4, 
\label{intro-eq-higgspot}
$$ where $v=246$~GeV.  It is essential
to measure the trilinear coupling rather accurate
in order to test whether the
observed Higgs boson originates from a field that is in concordance
with the
observed particle masses and the predicted electroweak symmetry breaking 
mechanism\footnote{The quartic coupling will not be accessible
either at the LHC or at an LC. Even at the high-luminosity
 Large Hadron Collider (HL-LHC), 
i.e. the LHC at $\sqrt=14$~TeV but with a tenfold increase in luminosity, 
there does not exist an analysis how to 
 get access to this coupling.}
Since
the cross section for double Higgsstrahlung is small but has a
maximum of about 0.2~fb at $\sqrt{s}=500$~GeV
 for 
$m_H=125$~GeV, this energy stage is required to enable a first
measurement of this coupling. The uncertainty scales with $\Delta
\lambda / \lambda=1.8 \Delta \sigma/ \sigma$.  
New 
involved analyses methods in full simulations aim at a
precision of 20\% at $\sqrt{s}= 500$~GeV. Better accuracy one 
could get applying the full
LC programme and going also to higher energy, $\sqrt{s}=1$~TeV.

\begin{sloppypar}
Another very crucial quantity is accessible at $\sqrt{s}=500$~GeV: 
the $t\bar{t}H$--coupling. Measuring the top-Yukawa coupling
is a challenging endeavour since it is overwhelmed from
$t\bar{t}$-background. At the LHC one expects an
accuracy of 25\% on basis of 300 fb$^{-1}$ and under optimal assumptions
and neglecting the error from theory uncertainties.  At the LC already 
at the energy stage of $\sqrt{s}=500$~GeV, it is expected to achieve an
accuracy of $\Delta g_{ttH}/g_{ttH}\sim 10$\%, see Sect.\ref{hewsb}.  
This energy stage is close to the threshold of
$ttH$ production, therefore
the cross section for this process should 
be small. But thanks to QCD-induced threshold effects the cross section 
gets enhanced and
such an accuracy should be achievable with 1 ab$^{-1}$ at the LC. It is of great
importance to measure this Yukawa coupling with high precision in
order to test the Higgs mechanism and verify the measured top mass
$m_t=y_{ttH} v/\sqrt{2}$.  The precise determination of the top Yukawa coupling 
opens a sensitive window to new physics and
admixtures of non-SM contributions. For instance, in the general
Two-Higgs-Doublet model the deviations with respect to the SM value 
of this coupling
can typically be as large as $\sim 20\%$.
\end{sloppypar}

\begin{figure}[htb!]
\hspace*{-.5cm}
\begin{center}
\includegraphics[width=7cm,height=5.5cm]{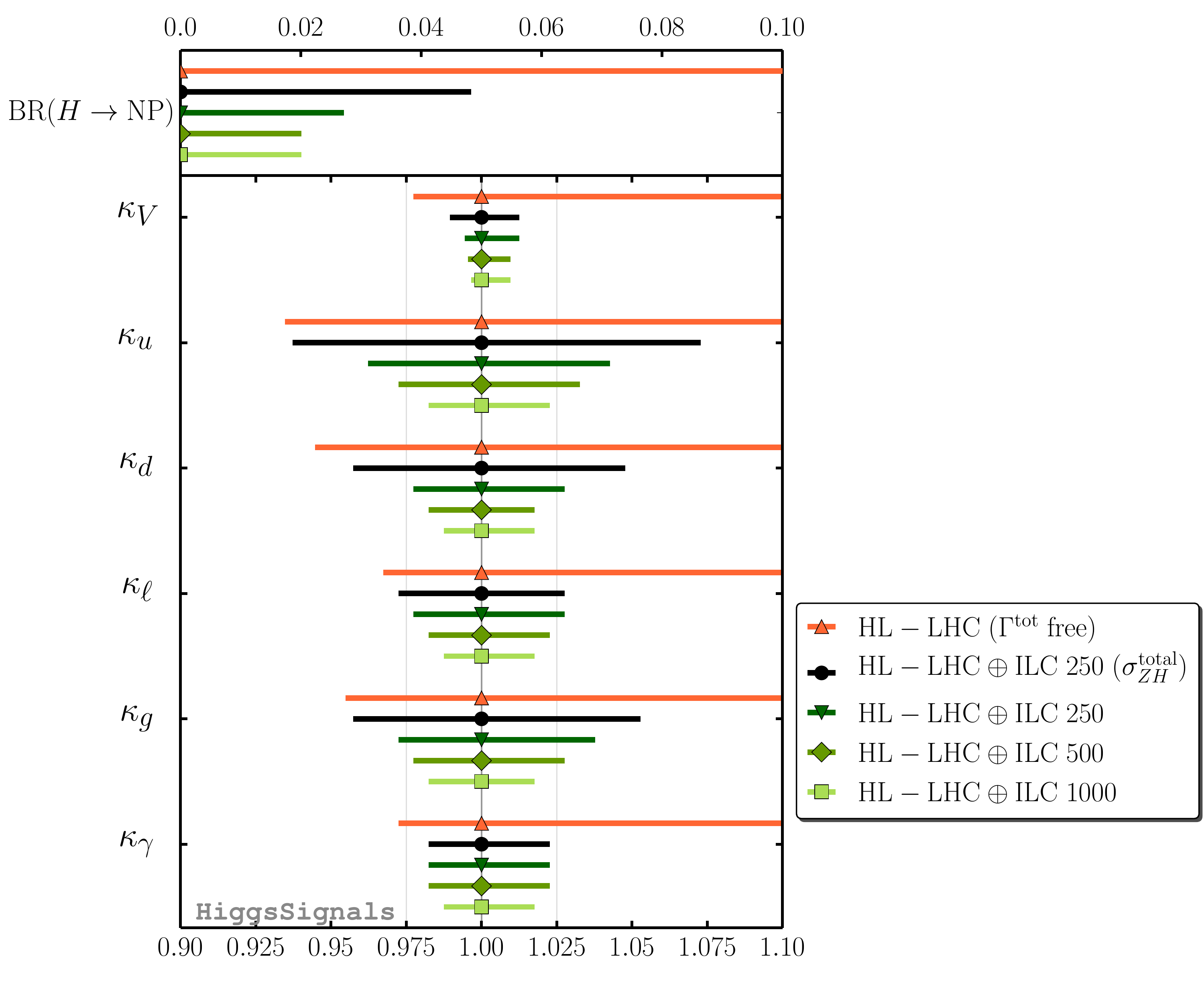}
\vspace*{-.5em}
\caption{The achievable precision in the different Higgs couplings at
  the LHC on bases of 3 ab$^{-1}$ and 50\% improvement in the
theoretical uncertainties in comparison with the
 the different energy stages at the ILC.
In the final LC stage all couplings can be
obtained in the 1--2\% range, some even better.~\cite{Bechtle:2014ewa}.}
\label{higgscoup}
\end{center}
\end{figure}

Since for a fixed $m_H$ all Higgs couplings are specified in the SM,
it is not possible to perform a fit within this model. In order to test
the compatibility of the SM Higgs predictions with the experimental
data, the LHC Higgs Cross Section Group proposed 'coupling scale
factors'~\cite{LHCHiggsCrossSectionWorkingGroup:2012nn,Heinemeyer:2013tqa}.  
These scale factors $\kappa_i$ ($\kappa_i=1$ corresponds to
the SM) dress the predicted Higgs cross section and partial widths.
Applying such a $\kappa$--framework, the following assumptions have been made:
there is only one 125~GeV state responsible for the signal with a 
coupling structure identical to the SM Higgs, i.e. a pure CP-even state,
and the zero width approximation can be applied. Usually, in addition the 
theory assumption $\kappa_{W,Z}<1$ (corresponds to an assumption on the total 
width) has to be made. Using, however, LC data and exploiting the precise
 measurement of $\sigma(HZ)$, this theory assumption can be dropped and all 
couplings can be obtained with an unprecedented precision of at least
1--2\%, see
Fig.\ref{higgscoup}\cite{Bechtle:2014ewa} and Sect.~\ref{hewsb} for further
details. 

\begin{sloppypar}
Another important property of the Higgs boson that has to be
determined is the CP quantum number.  In the SM the Higgs should be a pure
CP-even state.  In BSM models, however, the observed boson state {\it
  a priori} can be any admixture of CP-even and CP-odd
states, it is of high interest to determine limits on this admixture.
The $HVV$ couplings project out only the CP-even
components, therefore the degree of CP admixture cannot be tackled via
analysing these couplings. The measurements of CP-odd observables are
mandatory to reveal the Higgs CP-properties: for instance, the decays
of the Higgs boson into $\tau$ leptons provides the
possibility to construct unique CP--odd observables via the polarization
vector of the $\tau$'s, see further details in Sect.\ref{hewsb}.
\end{sloppypar}

\begin{figure}[htb!]
\begin{center}
\includegraphics[width=6.7cm,height=5.7cm]{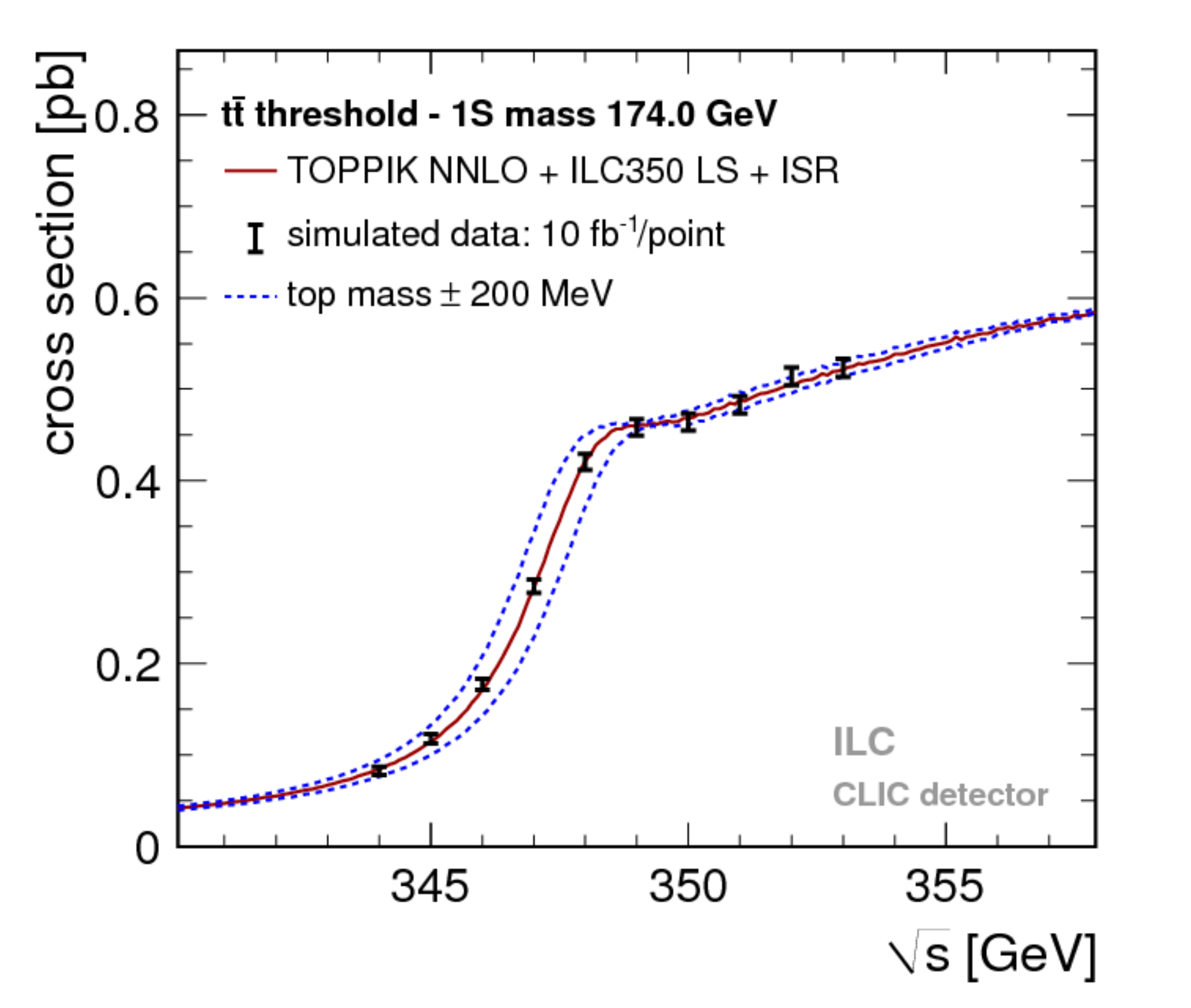}
\hspace{.3em}
\caption{Simulated measurement of the background-subtracted 
$t\bar{t}$ cross section with 10 fb$^{-1}$ per data point, 
assuming a top-quark mass of 174 GeV in the 1s scheme with the 
ILC luminosity spectrum for the CLIC-ILD detector~\cite{Seidel:2013sqa}.
}
\label{topthres} 
\end{center}
\end{figure}

\vspace{.5cm}
\noindent{\bf Top quark physics}\\[.5em] Top quark physics is another
rich field of phenomenology. It opens at $\sqrt{s}=350$~GeV. The mass
of the top quark itself has high impact on the physics analysis.  In
BSM physics $m_t$ is often the crucial parameter in loop corrections
to the Higgs mass. In each model where the Higgs boson mass is not a
free parameter but predicted in terms of the other model parameters,
the top quark mass enters the respective loop diagrams to the fourth
power, see Sect.\ref{sec:quantum} for details. Therefore the
interpretation of consistency tests of the electroweak precision
observables $m_W$, $m_Z$, $\sin^2\theta_{\rm eff}$ and $m_H$ require
the most precise knowledge on the top quark mass. The
top quark is not an asymptotic state and $m_t$ depends on the
renormalization scheme. Therefore a clear definition of the used top quark
mass is needed.  Measuring the mass via a threshold scan allows to
relate the measured mass uniquely to the well-defined
$m_t^{\overline{MS}}$ mass, see Fig.\ref{topthres}. 
 Therefore, this procedure is advantageous
compared to measurements via continuum observables.  It is expected to
achieve an unprecedented accuracy of $\Delta
m_{t}^{\overline{MS}}=100$~MeV via threshold scans. This uncertainty
contains already theoretical as well as experimental uncertainties.
Only such a high accuracy enables sensitivity to loop corrections for
electroweak precision observables.
Furthermore the accurate determination is also decisive for tests of
the vacuum stability within the SM.

\begin{figure}[htb]
\includegraphics[width=9cm,height=9cm]{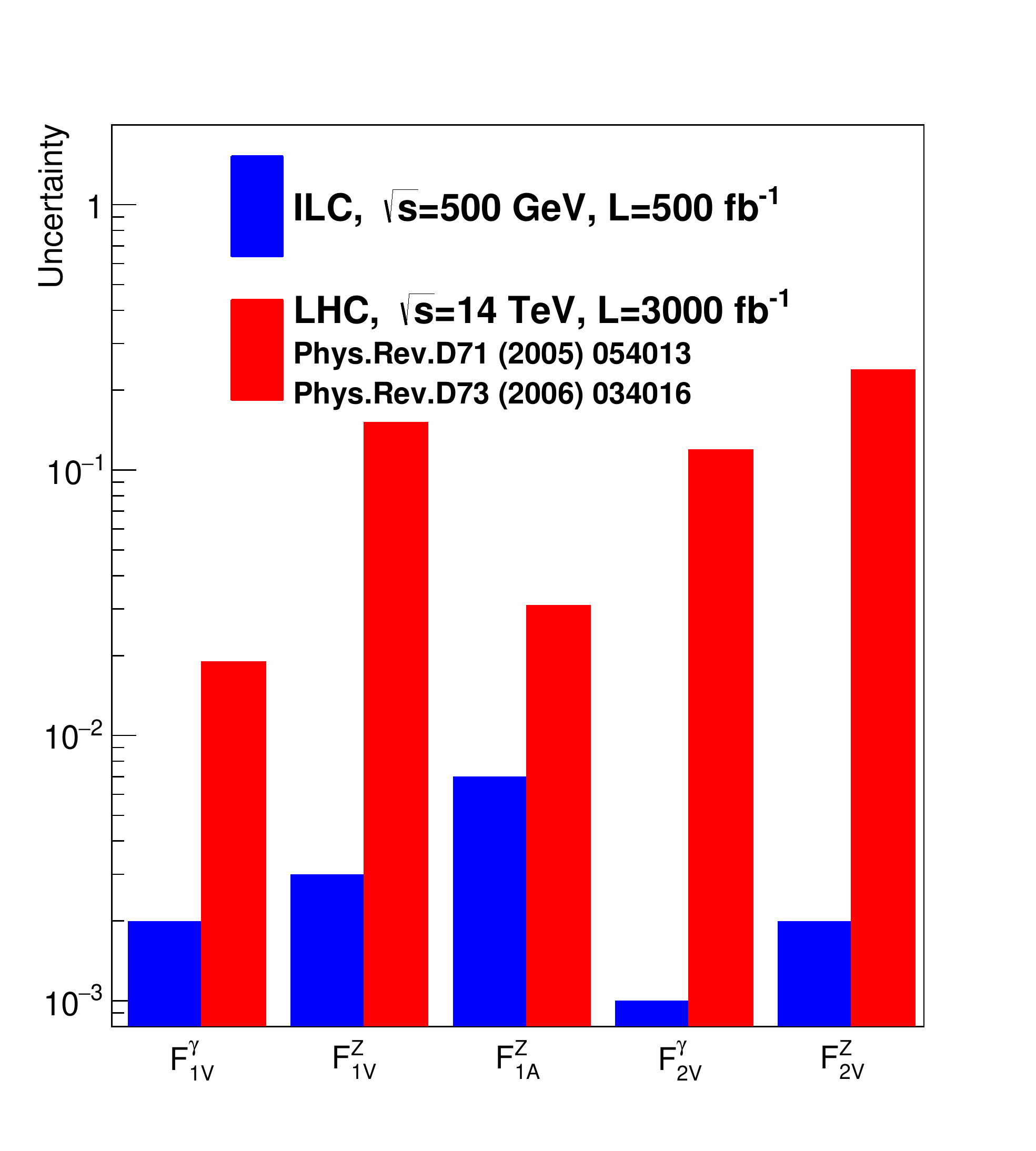}
\caption{Statistical precision on CP--conserving form factors expected
  at the LHC~\cite{ref:juste2005} and at the ILC~\cite{Amjad:2014fwa}.  The LHC
  results assume an integrated luminosity of ${\cal
    L}=300$~fb$^{-1}$. The results for the ILC are based on an
  integrated luminosity of ${\cal L}=500$~fb$^{-1}$ at
  $\sqrt{s}=500$~GeV and a beam polarization of $P_{e^-}=\pm 80\%$,
  $P_{e^+}=\mp 30\%$~\cite{Amjad:2014fwa}.}
\label{topcoup} 
\end{figure}

A sensitive window to BSM physics is
opened by the analysis of the top quark couplings. Therefore a precise determination
of all 
SM top-quark couplings together with the search for anomalous couplings is
crucial and can be performed very accurately at $\sqrt{s}=500$~GeV. 
Using the form-factor decomposition of the electroweak 
top quark couplings, it has been shown that one 
can improve the accuracy for the determination of the
couplings~\cite{Amjad:2014fwa}
by about one order of magitude at the LC
compared to studies at the LHC, see Fig.~\ref{topcoup} and Sect.~\ref{Top-QCD}.
%

\vspace{.5cm}
\noindent{\bf Beyond Standard Model Physics -- ``top-down'' }\\
\noindent Supersymmetry\\[.5em]
The SUSY concept is one of the most popular extensions of
the SM since it can close several open questions of the SM:
achieving gauge unification, providing dark matter candidates,
stabilizing the Higgs mass, embedding new sources for $CP$-violation
and also potentially
neutrino mixing.  However, the symmetry has to be
broken and the mechanism for symmetry breaking is completely unknown.
Therefore the most general parametrization allows around 100 
new parameters. In order to enable phenomenological
interpretations, for instance, at the LHC, strong restrictive
assumptions on the SUSY mass spectrum are set. However, as long as it
is not possible to describe the SUSY breaking mechanism within a full
theory, data interpretations based on these assumptions should be
regarded as a pragmatic approach. Therefore
the rather high limits obtained at the
LHC for some coloured particles exclude neither the concept of
SUSY as such, nor do they 
exclude light electroweak particles, nor relatively light
scalar quarks of the third generation.

Already the energy stage at $\sqrt{s}=350$~GeV provides a 
representative open window for the direct production of light
SUSY particles, for instance, light higgsino-like scenarios,
leading to signatures with only soft photons. 
The resolution of such signatures will be extremely challenging 
at the LHC but is feasible at the LC via  
the ISR method, as discussed in Sect.\ref{susy}.

\begin{figure}[htb]
\vspace{-3cm}
\hspace{-3cm}
\includegraphics[height=.5\textheight,width=.6\textheight]{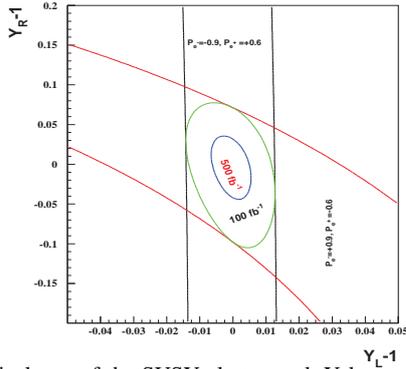}
\vspace*{-4cm}
\caption{Equivalence of the SUSY electroweak 
Yukawa couplings $g_{\tilde{W}}$, $g_{\tilde{B}}$
with the SU(2), U(1) gauge couplings $g$, $g'$. 
Shown are the contours of the polarized
cross sections $\sigma_L(e^+e^-\to \tilde{\chi}^0_1\tilde{\chi}^0_2)$ and $\sigma_R(e^+e^-\to \tilde{\chi}^0_1\tilde{\chi}^0_2)$
in the plane of the SUSY electroweak 
Yukawa couplings normalized to the gauge couplings,
$Y_L=g_{\tilde{W}}/g$, $Y_R=g_{\tilde{B}}/g'$\cite{Choi:2001ww} for a scenario 
with the electroweak spectrum similar 
to the reference point SPS1a.}
\label{fig:susy_yukawa}
\end{figure}

\begin{figure}[hb]
\vspace{3cm}
\hspace{2cm}
\begin{picture}(6,10)
\setlength{\unitlength}{1cm}
\put(-.1,0){\mbox{\includegraphics[height=.14\textheight]{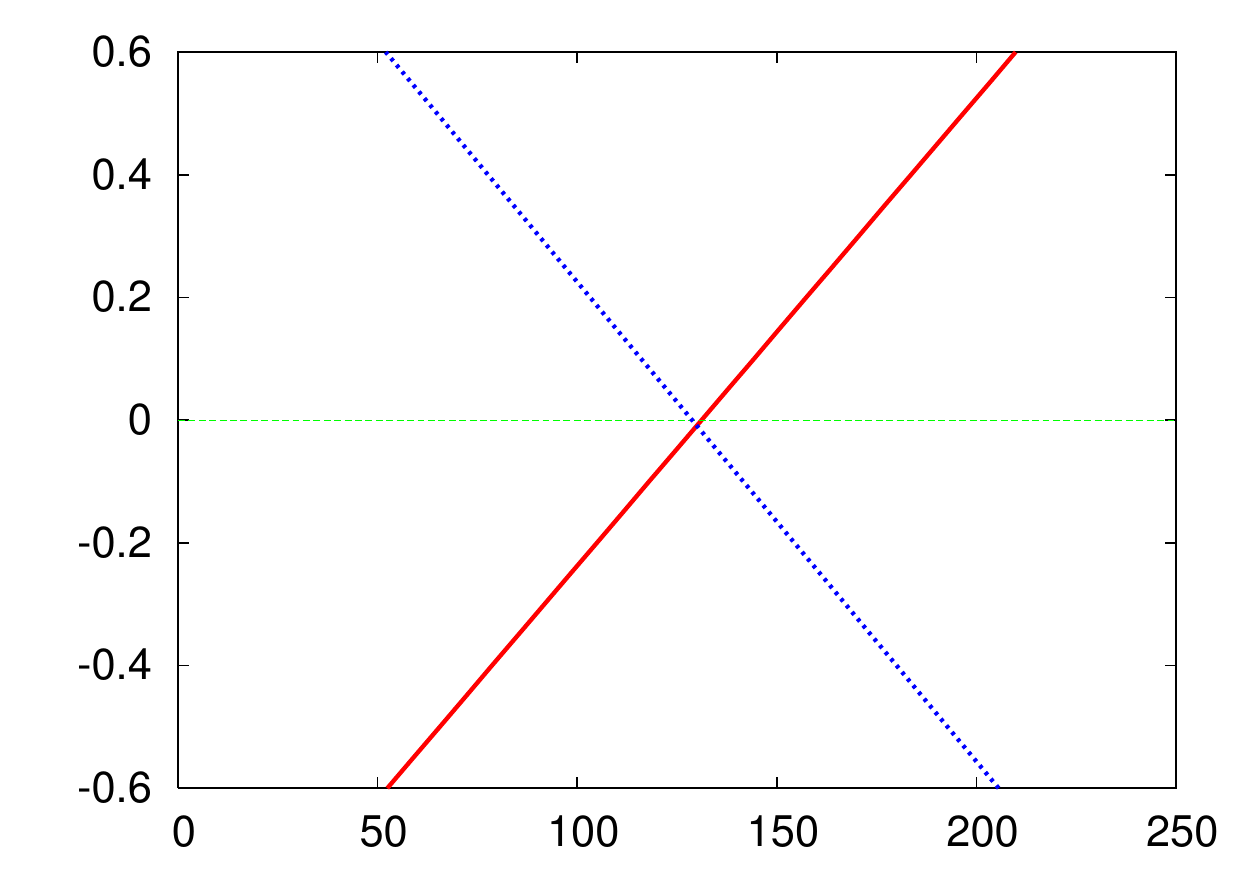}}}
\put(0.4,.0){\mbox{\includegraphics[height=.015\textheight,width=.22\textheight]{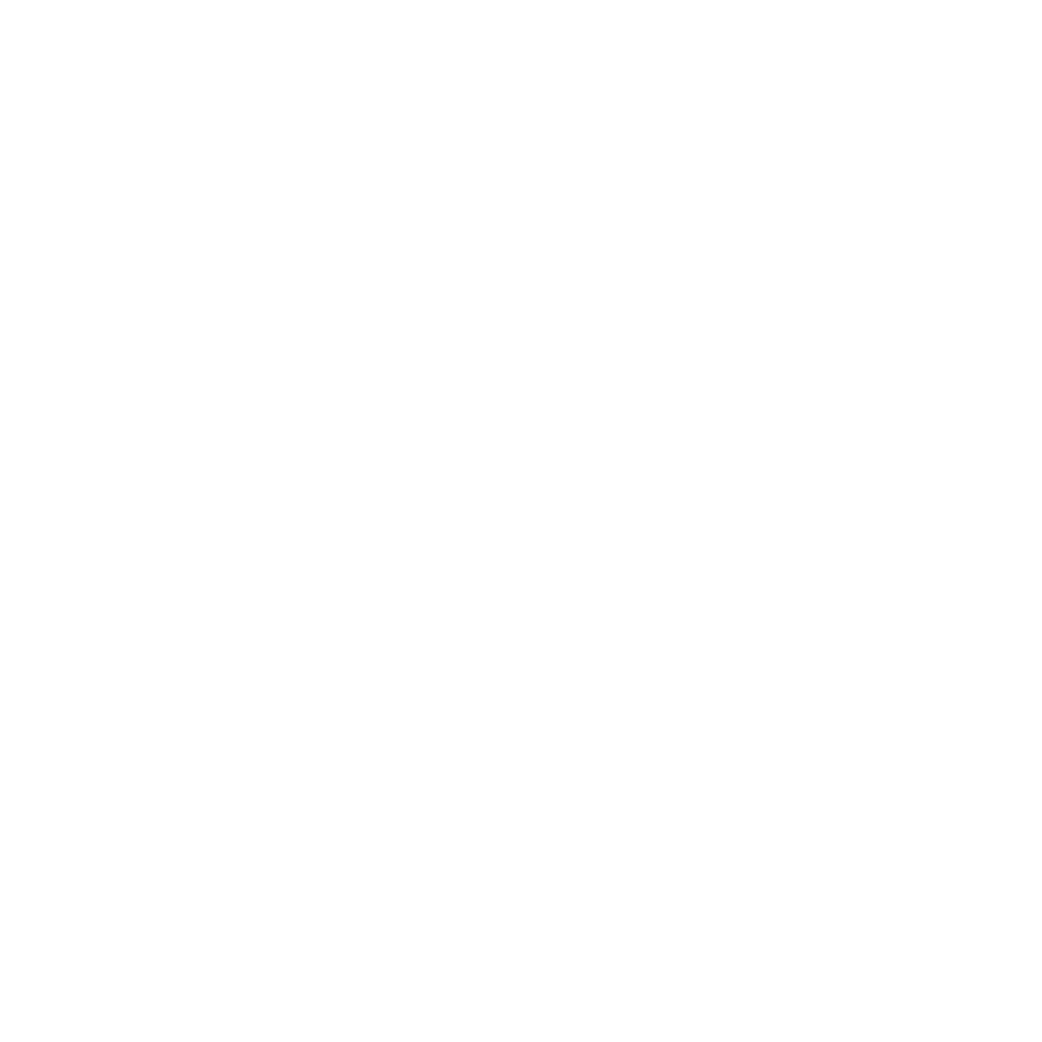}}}
\put(-.1,-3){\mbox{\includegraphics[height=.14\textheight]{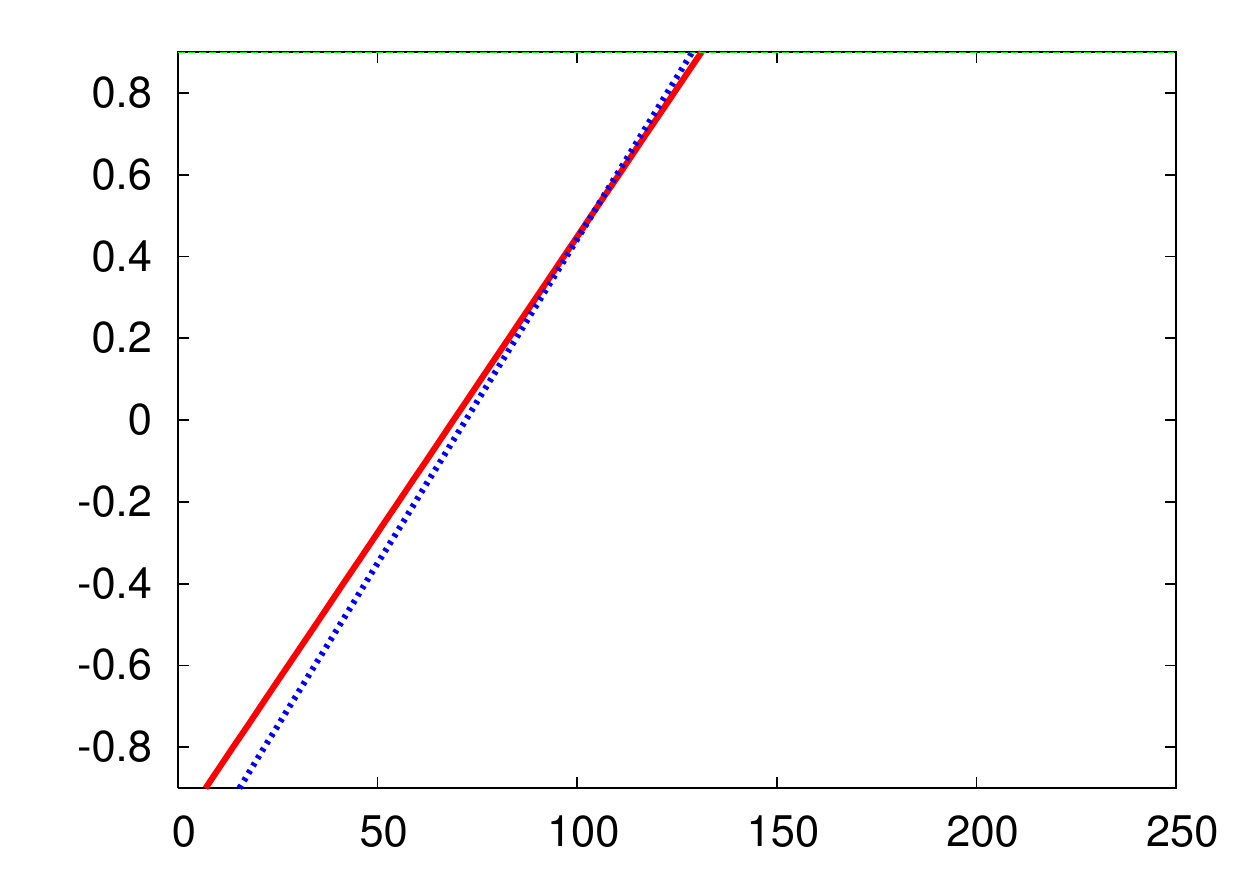}}}
\put(-.25,1.7){\footnotesize $P_{e^+}$}
\put(-.25,-1.3){\footnotesize $P_{e^-}$}
\put(-.4,0.1){\tiny $0.9\to$}
\put(4.5,1.7){\scriptsize $P_{e^-}=+0.9$}
\put(3.4,2.4){\scriptsize\color{red} $\tilde{e}^+_L\tilde{e}^-_R$}
\put(3.4,1){\scriptsize\color{blue} $\tilde{e}^+_R\tilde{e}^-_R$}
\put(2.7,-.3){\scriptsize\color{red}$\tilde{e}^+_L\tilde{e}^-_R$}
\put(1.3,-.3){\scriptsize\color{blue} $\tilde{e}^+_R\tilde{e}^-_R$}
\put(2.43,.7){\Large$\uparrow$}
\put(2.4,-2.5){\scriptsize $\sqrt{s}=500$~GeV}
\put(4.5,-1.3){\scriptsize $P_{e^+}=0$}
\put(0.6,3.4){\scriptsize $e^+e^-\to\tilde{e}^+_{L,R}\tilde{e}^-_{R}\to e^+e^- \tilde{\chi}^0_1\tilde{\chi}^0_1$}
\put(1.5,-3.2){\scriptsize cross section [fb]}
\end{picture}
\vspace{3.5cm}
\caption{Polarized cross sections versus $P_{e^-}$ (bottom panel) and
$P_{e^+}$ (top panel) for $e^+e^-\to \tilde{e}\tilde{e}$-production with direct
decays in $\tilde{\chi}^0_1 e$ in a scenario where the non-coloured spectrum is
similar to a SPS1a-modified scenario but with  $m_{\tilde{e}_L}=200$~GeV,
$m_{\tilde{e}_R}=195$~GeV. The associated chiral quantum numbers of 
the scalar SUSY partners $\tilde{e}_{L,R}$} can be tested via polarized 
$e^{\pm}$-beams.
\label{fig:susy_selectrons}
\end{figure}
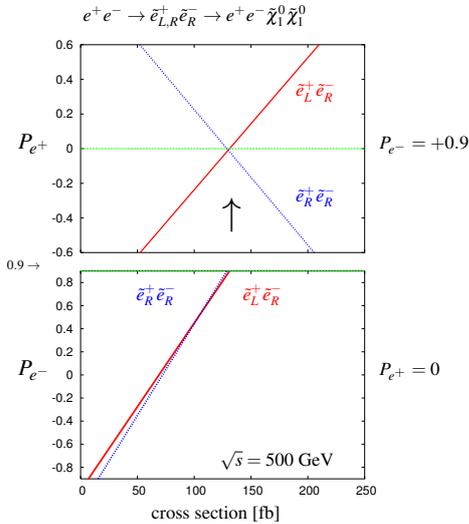

\begin{sloppypar}
Another striking feature of the LC physics potential
is the capability to
test predicted properties of new physics candidates. For instance, in
SUSY models one essential paradigm is that the coupling structure of
the SUSY particle is identical to its SM partner particle. That means, 
for instance, that the
SU(3), SU(2) and U(1) gauge couplings $g_S$, $g$ and $g'$ have to be
identical to the corresponding SUSY Yukawa couplings $g_{\tilde{g}}$,
  $g_{\tilde{W}}$ and $g_{\tilde{B}}$.  These tests are of fundamental
    importance to establish the theory. Testing, in particular, 
the SUSY electroweak Yukawa coupling is a unique feature of LC physics.
Under the assumption that the $SU(2)$
and $U(1)$ parameters have been determined in the gaugino/higgsino
sector (see Sect.\ref{sec:susy5}), the identity of
the Yukawa and the gauge couplings via measuring polarized cross 
sections can be successfully performed:
depending on the electron 
(and positron) beam polarization and on the luminosity, a 
per-cent-level precision
can be achieved: see Fig.~\ref{fig:susy_yukawa}.
\end{sloppypar}
Another important and unique feature of the LC potential is to test
experimentally the quantum numbers of new physics candidates.  For
instance, a particularly challenging measurement is the determination
of the chiral quantum numbers of the SUSY partners of the
fermions. These partners are predicted to be scalar particles and to
carry the chiral quantum numbers of their Standard Model partners.
In $e^+e^-$ collisions,
the associated production reactions $e^+e^-\rightarrow
\tilde{e}_L^+\tilde{e}_R^-$,\ $\tilde{e}_R^+\tilde{e}_L^-$ occur only
via $t$-channel exchange, where the $e^\pm$ are directly
coupled to their SUSY partners $\tilde{e}^{\pm}$. Separating the associated
 pairs, 
the chiral quantum numbers can be tested via the polarization of 
$e^\pm$ since chirality corresponds to
helicity in the high energy limit. 
As can be seen in Fig.~\ref{fig:susy_selectrons},
the polarization of both beams is absolutely essential to separate the
pair $\tilde{e}_L\tilde{e}_R$\cite{MoortgatPick:2005cw} and to test the 
associated quantum numbers.

\newpage
\begin{sloppypar}
\noindent{ Dark Matter Physics}\\[.5em]
Weakly interacting massive particles (WIMPs) are the favourite
candidates as components of the cold dark matter.  Neutral particles
that interact only weakly provide roughly the correct relic density in
a natural way.  Since there are no candidates for dark matter in
the SM, the strong observational evidence for dark matter
clearly points to physics beyond the SM. Due to precise results from
cosmological observations, for instance~\cite{Hinshaw:2012aka}, bounds
on the respective cross section and the mass of the dark matter
candidates can be set in the different models. Therefore, in many
models only rather light candidates are predicted, i.e.\ with a mass
around the scale of electroweak symmetry breaking or even lighter.
That means, for instance for SUSY models with R-parity conservation,
that the lightest SUSY particle, should be within the kinematical
reach of the ILC. The lowest threshold for such processes is pair
production of the WIMP particle.  Since such a final state, however,
escapes detection, the process is only visible if accompanied by a
radiative photons at the LC, that recoil against the WIMPs, for
instance, the process $e^+e^-\to \gamma\chi\chi$\cite{Birkedal:2004xn}, 
where $\chi$ denotes the 
WIMP particle in general with a spin $S_{\chi}=0,\frac{1}{2},1$. 
Such a process can be realized in SUSY models, 
in universal extra dimensions, little Higgs theories etc.
The
dominant SM background is radiative neutrino production, which can,
efficiently be suppressed via the use of beam polarization.
\end{sloppypar}

\begin{figure}[tbh!]
\begin{center}
\hspace{2cm}
\includegraphics[width=7.0cm]{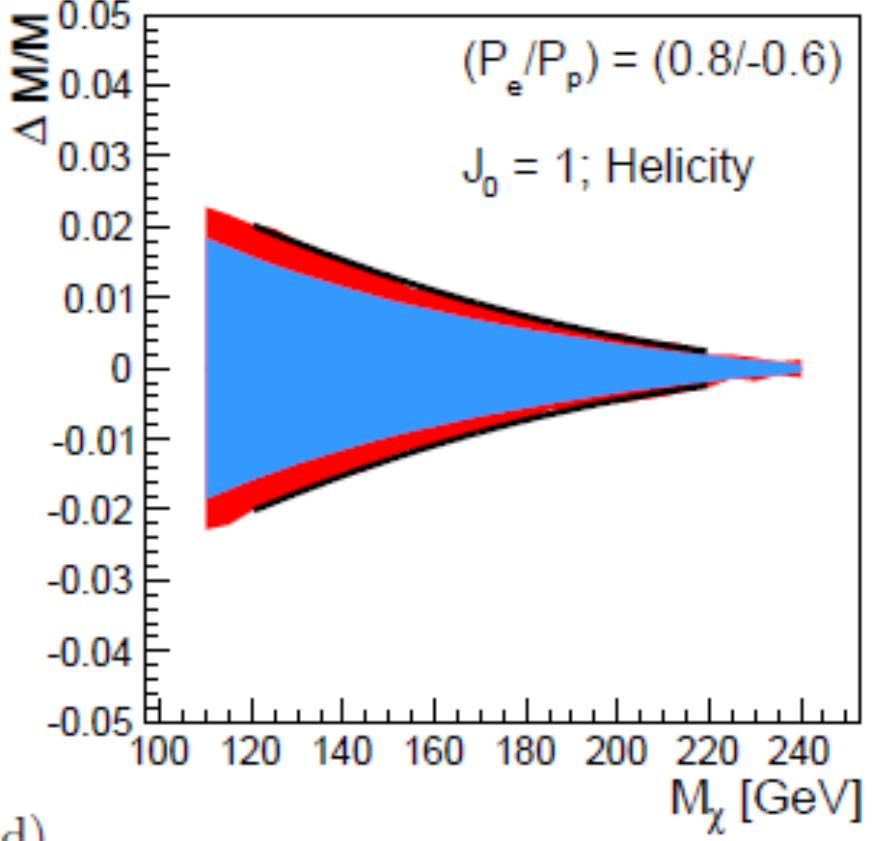}
\end{center}
\caption{WIMP mass as a function of the 
mass for p-wave ($J_0=1$) annihilation and under the assumption that 
WIMP couplings are helicity- and parity-conserving in the
process $e^+e^-\to \gamma \chi\chi$\cite{Birkedal:2004xn}. 
With an integrated luminosity of ${\cal L}=500$~fb$^{-1}$ and 
polarized beams with 
$P_{e^-}=+80\%$, $P_{e^+}=-60\%$ with $\Delta P/P=0.25\%$ the reconstructed
WIMP mass 
can be determined with an relative accuracy of the order 
of 1\%~\cite{Bartels:2012ex}.
The blue area shows the systematic uncertainty and the red bands the 
additional statistical contribution. The dominant sources of 
systematic uncertainties are $\Delta P/P$ and the shape of the beam 
energy spectrum.
}
\label{fig:dm-ilc}
\end{figure}

The present dark matter density depends strongly on the cross section for 
WIMP annihilation into SM particles
(assuming that there exist only one single
WIMP particle $\chi$ and ignoring coannihilation processes between the WIMP and 
other exotic particles)
in the limit when the colliding $\chi$'s are 
non-relativistic~\cite{Birkedal:2004xn},
depending on s- or p-wave contributions and on the WIMP mass.
Due to the excellent resolution at the LC the WIMP mass can be determined with
relative accuracy of the order of 1\%, see Fig.\ref{fig:dm-ilc}.

\begin{figure}[tbh!]
\begin{center}
\includegraphics[width=5.0cm]{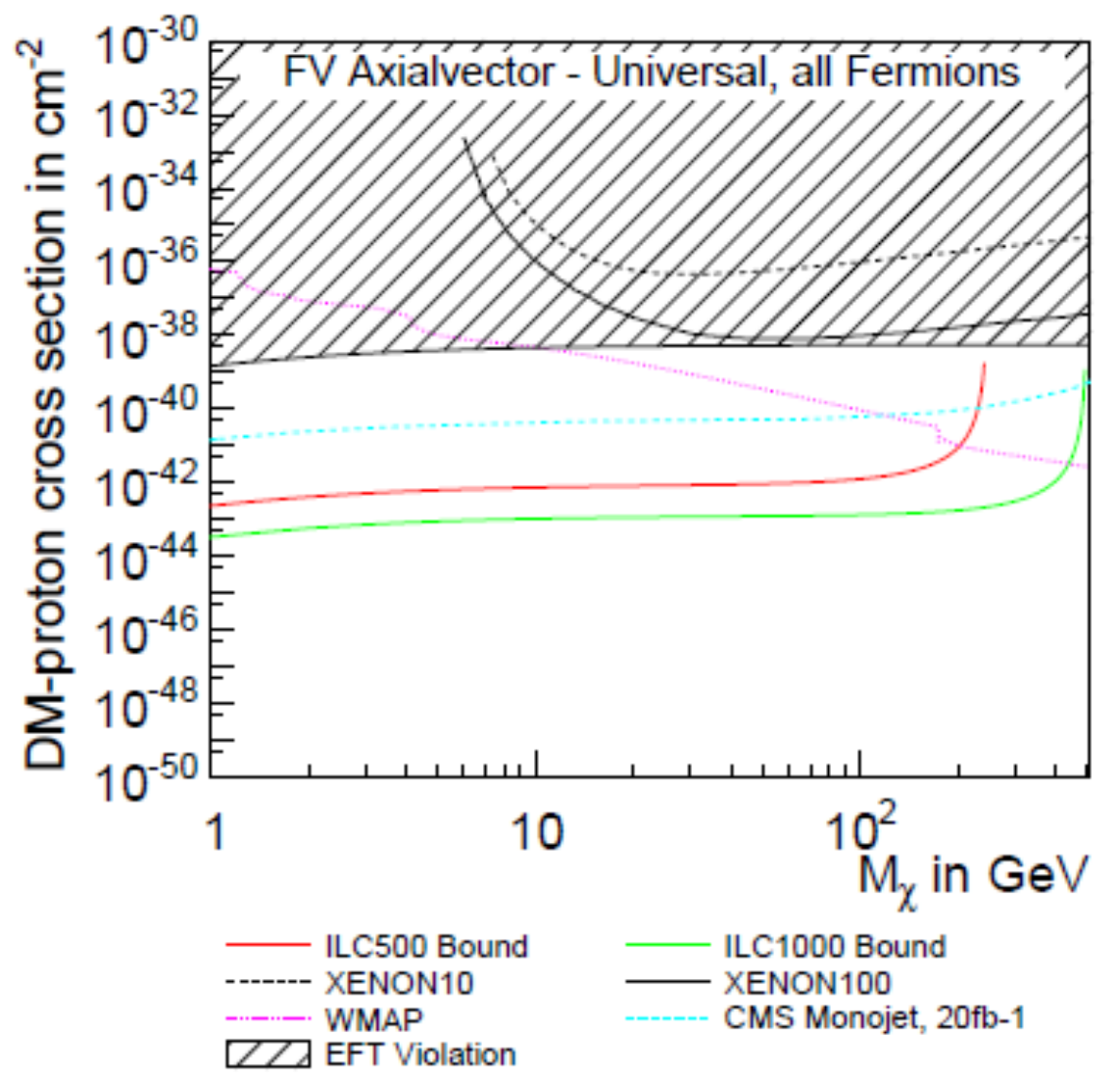}
\includegraphics[width=5.0cm]{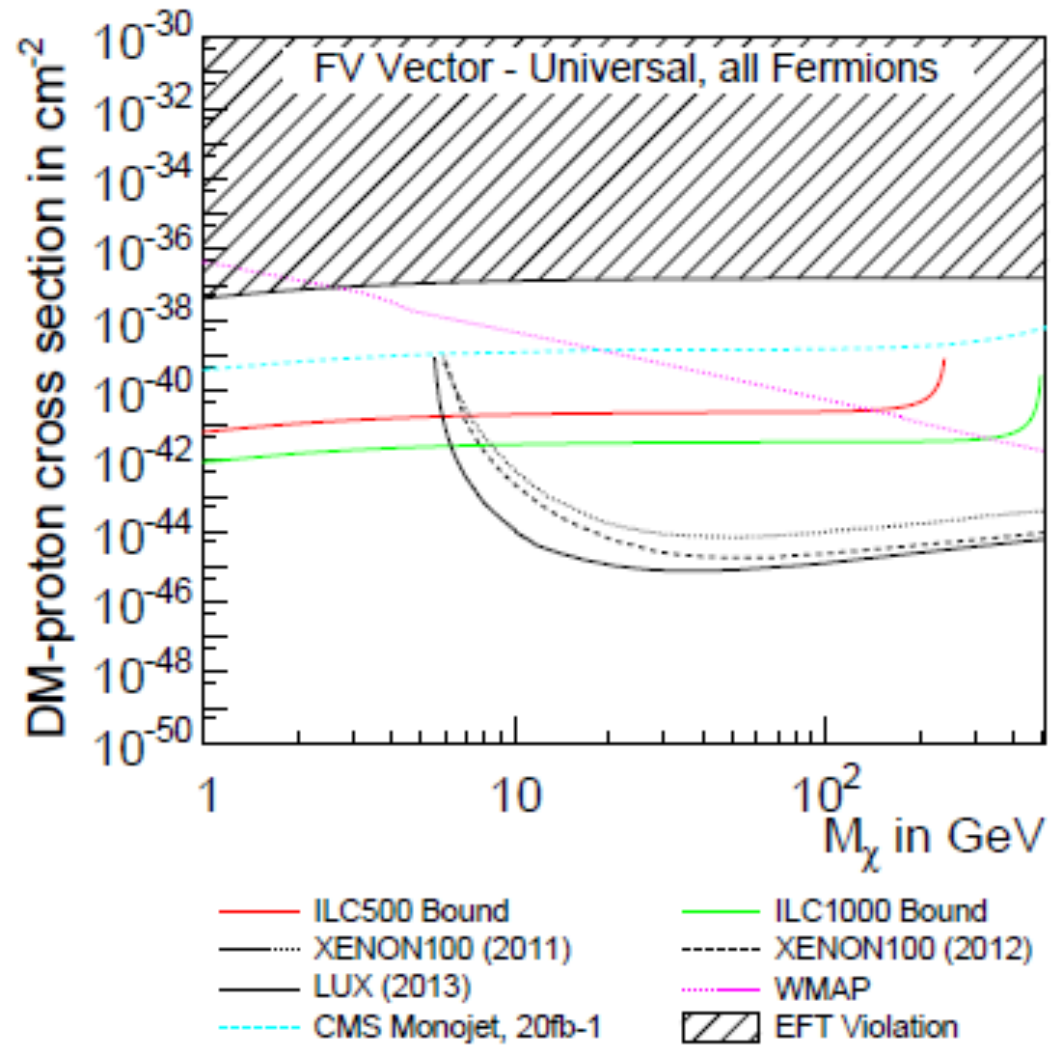}
\end{center}
\caption{
Combined limits for
fermionic dark matter models. The process $e^+e^-\to\chi\chi\gamma$
is assumed to be detected only by the hard photon. The analysis has been
modeled
correspondingly to~\cite{Bartels:2012ex} and is based on
${\cal L}=500$~fb$^{-1}$ at $\sqrt{s}=500$~GeV and $\sqrt{s}=1$TeV and
different polarizations~\cite{Dreiner:2012xm}.}
\label{fig:dm-spin}
\end{figure}

\begin{sloppypar}
Following another approach and parametrizing dark matter interactions
in the form of effective operators, a non-relativistic approximation is
not required and the derived bounds can be compared with experimental
bounds from direct detection. 
Assuming  that the dark matter particles only interact with SM
fields via heavy mediators that are kinematically not accessible at the ILC,
 it was 
shown in~\cite{Dreiner:2012xm} that the ILC could nevertheless
probe effective 
WIMP couplings $G^{\rm ILC}_{\rm max}=g_ig_j/M^2 = 10^{-7}$~GeV$^{-2}$ 
(vector or scalar mediator case), 
or $G^{\rm ILC}_{\rm max}=g_ig_j/M = 10^{-4}$~GeV$^{-1}$ (fermionic mediator case).
The direct detection searches give much stronger
bounds on spin independent ('vector') than on spin dependent
('axialvector') interactions under the 
simplifying assumption that all SM
particles couple with the same strength to the dark matter 
candidate ('universal coupling').
If the WIMP particle is rather light ($<10$~GeV) the ILC offers a
unique opportunity to search for dark matter candidates beyond any
other experiment, even for spin independent interactions, cf.\
Fig.~\ref{fig:dm-spin} (upper panel). 
In view of spin dependent interactions the 
ILC searches are also superior for heavy WIMP particles, 
see Fig.\ref{fig:dm-spin} (lower panel).
\end{sloppypar}

\begin{figure}[tbh!]
\begin{center}
\hspace{2cm}
\includegraphics[width=7.0cm]{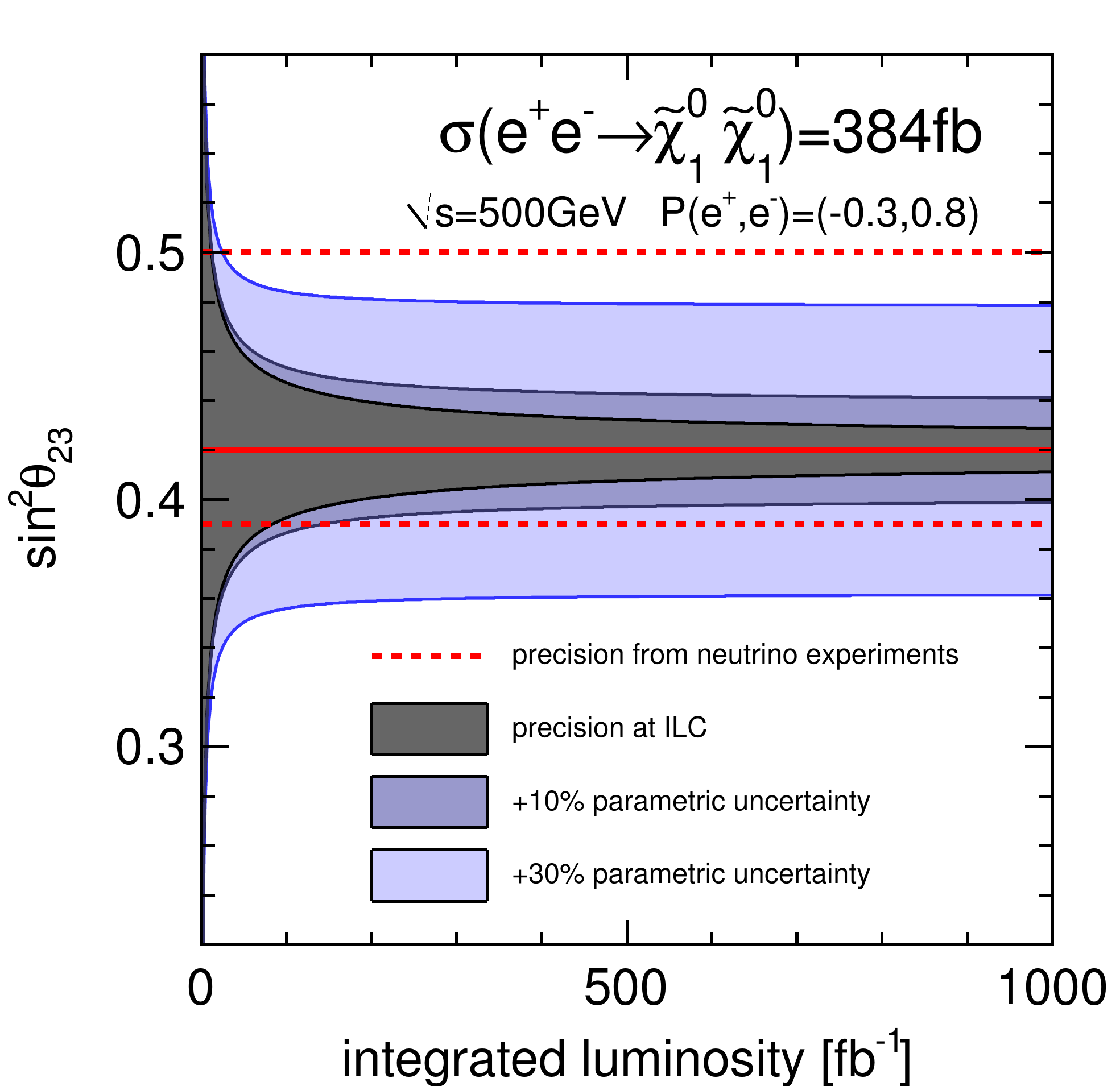}
\end{center}
\caption{Achievable precision on $\sin^2\theta_{23}$ from bilinear R-parity
violating decays
  of the $\tilde{\chi}^0_1$ as a function of the produced number of
  neutralino pairs compared to the current precision from neutrino
  oscillation measurements~\cite{List:2013dga}.}
\label{fig:susy-BRpV}
\end{figure}

\noindent Neutrino mixing angle\\[.5em]
Another interesting question is how to explain the observed neutrino mixing 
and mass patterns in a more complete theory. 
SUSY with broken R-parity allows to embed
and to predict such an hierarchical pattern. The mixing between 
neutralinos and neutrinos puts strong relations between 
the LSP branching ratios and neutrino mixing angles.
For instance, the solar neutrino mixing angle $\sin^2\theta_{23}$
is accessible via measuring the ratio of the branching fractions for
$\tilde{\chi}^0_1 \to W^\pm\mu^\mp$ and $W^\pm\tau^\mp$.
Performing an experimental analysis at $\sqrt{s}=500$~GeV
allows to determine the neutrino
mixing angle $\sin^2 \theta_{23}$ up to a percent-level
precision, as illustrated in Fig.~\ref{fig:susy-BRpV}\cite{List:2013dga}.
This direct relation 
between neutrino physics and high-energy physics is striking. 
It allows to directly test whether 
the measured neutrino mixing angles can be embedded within a 
theoretical model of high predictive power, namely 
a bilinear R-parity violation model in SUSY, based on
precise measurements of neutralino branching 
ratios~\cite{Porod:2000hv,Hirsch:2003fe}
 at a future $e^+e^-$
linear collider.

\begin{figure}[htb!]
\begin{center}
\includegraphics[width=6.7cm,height=5.7cm]{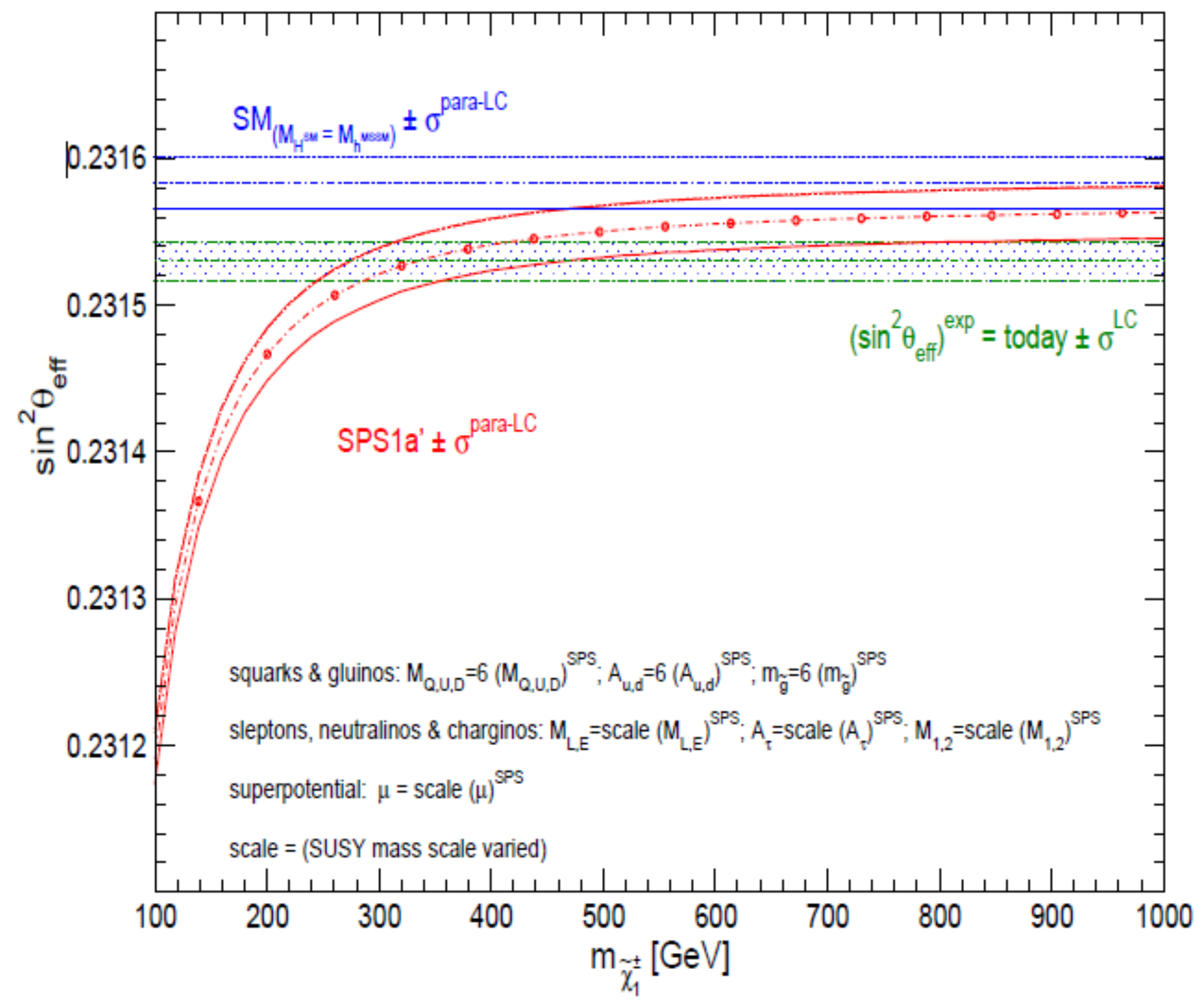}
\hspace{.3em}
\caption{
Theoretical prediction for $\sin^2\theta_{\rm eff}$ in the SM and the 
MSSM (including
prospective parametric theoretical uncertainties) compared to
the experimental precision at the LC with GigaZ option.  
A SUSY inspired scenario SPS~1a$'$ has been used, 
where the coloured SUSY particles masses
are fixed to 6~times their SPS~1a$'$ values. The other mass 
parameters are varied with a common scale factor.}
\label{fig:arne} 
\end{center}
\end{figure}

\vspace{.5cm}
\begin{sloppypar}
\noindent{\bf Beyond Standard Model Physics -- ``bottom-up''}\\
 \noindent Electroweak Precision Observables\\[.5em]
 Another compelling
physics case for the LC can be made for the measurement of Electroweak
Precision Observables (EWPO) at $\sqrt{s}\approx 92$~GeV (Z-pole) and
$\sqrt{s}\approx 160$~GeV ($WW$ threshold), where a new level of
precision can be reached.  Detecting with highest precision any
deviations from the SM predictions provides traces of new physics
which could lead to groundbreaking discoveries.  Therefore,
particularly in case no further discovery is made from
the LHC data, it will be beneficial to perform
such high precision measurements at these low energies.  Many new
physics models, including those of extra large dimensions, of extra
gauge bosons, of new leptons, of SUSY, etc., can lead to measurable
contributions to the electroweak mixing angle even if the scale of the
respective new physics particles are in the multi-TeV range,
i.e. out of range of the high luminosity LHC. Therefore the potential
of the LC to measure this quantity with an unprecedented precision,
i.e. of about one order of magnitude better than at LEP/SLC offers to
enter a new precision frontier.  With such a high precision
--mandatory are high luminosity and both beams polarized-- one gets
sensitivity to even virtual effects from BSM where the particles are
beyond the kinematical reach of the $\sqrt{s}=500$~GeV LC and the
LHC. In Fig.~\ref{fig:arne} the prediction for 
$\sin^2\theta_{\rm eff}$ as a function of the lighter chargino mass
$m_{\tilde{\chi}^{+}_1}$ is shown. The MSSM prediction is compared
with the prediction in the SM assuming the experimental resolution
expected at GigaZ. In this scenario no coloured SUSY particles would
be observed at the LHC but the LC could resolve indirect effects 
of
SUSY up to $m_{\tilde{\chi}^{+}_1}\le 500$~GeV via the measurement of 
 $\sin^2\theta_{\rm  eff}$ with unprecedented precision at the low energy option 
GigaZ, see Sect.~\ref{sec:quantum} for details.
\end{sloppypar}
The
possibility to run with high luminosity and both beam polarized
on these low energies is essential in these regards.\\

\begin{figure}[htb!]
\begin{center}
\includegraphics[width=6.7cm,height=5.7cm]{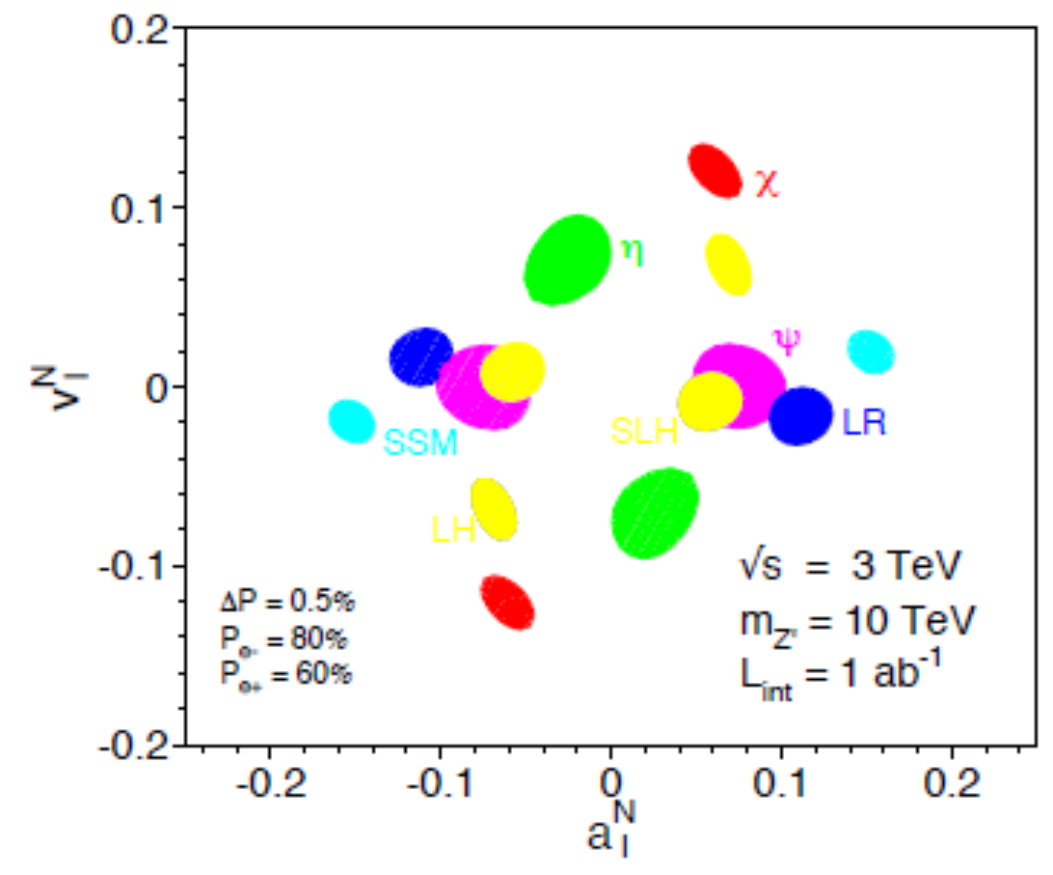}
\hspace{.3em}
\caption{
New gauge bosons in the $\mu^+\mu^-$ channel. The plot shows the expected resolution at CLIC with $\sqrt{s}=3$~TeV and ${\cal L}=1$~ab$^{-1}$ on the 
`normalized' vector $v_f^n=v'_f\sqrt{s/(m_Z'^2-s)}$ and axial-vector 
$a_f^n=a'_f\sqrt{s/(m_Z'^2-s)}$ couplings to a 10~TeV $Z'$ 
in terms of the SM couplings $v'_f$, $a'_f$. 
The mass of $Z'$ is assumed to be unknown, nevertheless the couplings can 
be determined up to a two-fold ambiguity.
The colours denote different $Z'$ models~\cite{Linssen:2012hp}.}
\label{zp-coupling} 
\end{center}
\end{figure}

\noindent Extra Gauge Bosons\\[.5em] One should stress that not only
SUSY theories can be tested via indirect searches, but also other
models, for instance, models with large extra dimensions or models
with extra $Z'$, see Fig.\ref{zp-coupling}, where the mass of the $Z'$
boson is far beyond the direct kinematical reach of the LHC and the
LC and therefore is assumed to be unknown.  
Because of the clean LC environment, one even can determine the
vector and axial vector coupling of such a $Z'$ model.\\

\noindent {\bf Synopsis}\\[.5em] 
The full Higgs and top quark physics programme as well as the
promising programme on 
dark matter and 
BSM
physics should be accomplished with the higher energy LC
set-up at 1~TeV. Model-independent parameter determination is 
essential for the
crucial identification of the underlying model. Accessing a large part of the
particle spectrum of a new physics model would nail down the
structure of the underlying physics. But measuring
already only the light part of 
the spectrum with high precision and model-independently
can provide substantial information. 
Table~\ref{tab:sum1} gives an overview of the different physics topics and
the required energy stages.
The possibility of an tunable
energy in combination with polarized beams,
 is particularly beneficial to successfully accomplish the
comprehensive physics programme at high-energy physics collider and to 
fully exploit the complete physics potential of the future Linear Collider.

{\small
\begin{table*}
\begin{tabular}{|l|c|c|c|c|c|c|c|}
\hline
$\sqrt{s}/$GeV: & 92,160 & 240 & 350 & 500 & 1000 & 3000 & 
threshold scans required\\ \hline
\multicolumn{1}{|l}{\bf Higgs} & \multicolumn{7}{l|}{}\\ \hline
$m_H$          & -- & x & x & x & x  &x & x \\
$\Gamma_{tot}$ & -- & --& x & x &  & &  \\
$g_{c,b}$    & -- & x & x & x & & x & \\
$g_{ttH}$      & -- & --&-- & x & x & & \\ 
$g_{HHH}$      & -- & --&-- & x & x & x & \\
$m_{H,A}^{SUSY}$   & -- & --&-- & x & x & x & x \\  
\hline
\multicolumn{1}{|l}{\bf Top} & \multicolumn{7}{l|}{}      \\ \hline
$m_{t}^{th}$   & -- & -- & x & & &  &x\\ 
$m_{t}^{cont}$ & -- & -- & --& x & (x) & (x) & \\
$A_{FB}^t$ & -- & -- & x & x & &&\\
$g_{Z,\gamma}$ & -- & -- & --& x &  & & \\
$g_{FCNC}$     & -- & -- & --& x & x & (?) & \\ \hline
\multicolumn{1}{|l}{\bf Electroweak Precision Observables} & \multicolumn{7}{l|}{} \\ \hline
 $\sin^2\theta_{\rm eff}$(Z-pole) & x &  &&&&(x)& \\
$m_W^{th}$          & x & &&&& & x \\ 
$m_W^{cont}$          &  & x&x&x&(x)&(x)&  \\ 
$\Gamma_Z$          & x &  &  &  &&&x \\
$A_{LR}$   & x & &  &  &  &  & \\ 
$A_{FB}$  & x &  &&&&& \\ \hline
\multicolumn{1}{|l}{\bf SUSY} & \multicolumn{7}{l|}{} \\ \hline 
indirect search   & x  & x & x &  &  &  &  \\ 
direct search     & -- & -- & x  & x & x & x & x \\ 
light higgsinos & -- & -- & x & x &&&x \\
parameter determination & -- & -- & x & x & x & & x \\
quantum numbers & -- & -- & x & x & x & & x\\
extrapolations & -- & -- & -- & x & x & x & x\\\hline
\multicolumn{1}{|l}{\bf $\nu$ mixing} & \multicolumn{7}{l|}{} \\ \hline
$\theta^2_{23}$    &--  & -- & x & x && &\\ \hline
\multicolumn{1}{|l}{\bf Dark Matter} & \multicolumn{7}{l|}{}  \\ \hline
effective-field-theory    & -- & -- & -- & x & x &x& \\ 
non-relativistic & -- & -- & x & x & x &x& \\ \hline
\multicolumn{1}{|l}{\bf Extra gauge bosons} & \multicolumn{7}{l|}{}\\ \hline 
indirect search $m_{z'}$& x & -- & -- & x & x & x& \\
\phantom{indirect search}$v'_f$, $a'_f$ &-- & -- & -- & x & x & (x) & \\
\phantom{indirect search}$m_{W'}$ & x & -- & -- & x  & x  & x& \\
direct search & -- & -- & -- & -- & -- & x & x \\ \hline
\end{tabular}
\caption{Physics topics where the $e^+e^-$-Linear Collider provides substantial
results at the different energy stages that are complementary to the LHC. 
The examples are 
described in the following chapters as well as 
in~\cite{AguilarSaavedra:2001rg,Abe:2001wn,Abe:2001gc,BrauJames:2007aa,Djouadi:2007ik,Linssen:2012hp,DBD,Diberder,parametergroup,Moortgat-Pick:2013awa,lcnotes,jimbrau}.
\label{tab:sum1}
}
\end{table*}
}


\section[Higgs and Electroweak Symmetry Breaking]{Higgs and Electroweak Symmetry Breaking\protect\footnotemark}
\footnotetext{Editors: K. Fujii, S. Heinemeyer, P.M. Zerwas\footnotemark \\
Contributing authors: M.~Asano, K.~Desch, U.~Ellwanger,
C.~Englert, I.~Ginzburg, C.~Grojean, 
S.~Kanemura, M.~Krawczyk, J.~Kroseberg, S.~Matsumoto,
M.M.~M\"uhlleitner, M.~Stanitzki}\footnotetext{Cooperation, 
including R\'{e}sum\'{e}, in early phase of the report.}
\label{hewsb}

\noindent
After a brief description of the physical
basis of the Higgs mechanism, we summarize the crucial results for Higgs
properties in the Standard Model as expected from measurements at LHC
and ILC/CLIC, based on the respective reports. Extensions of the SM
Higgs sector are sketched thereafter, discussed thoroughly in the
detailed reports which follow: portal models requiring analyses of
invisible Higgs decays, supersymmetry scenarios as generic
representatives of weakly coupled Higgs sectors, and finally strong
interaction elements as suggested by Little Higgs models and composite
models motivated by extended space dimensions.

\subsection{R\'{e}sum\'{e}\protect\footnotemark}
\footnotetext{ Keisuke Fujii, Sven Heinemeyer, Peter M. Zerwas}
\label{sec:resumee}
\noindent 
The Brout-Englert-Higgs mechanism 
\cite{Englert:1964et,Higgs:1964ia,Higgs:1964pj,Higgs:1966ev,Guralnik:1964eu} 
is a central element of particle physics. Masses are introduced 
consistently in gauge theories for vector bosons, leptons and quarks, 
and the Higgs boson itself, by transformation of the interaction energy
between the initially massless fields and the vacuum expectation value of
the Higgs-field. The
non-zero value of the Higgs field in the vacuum, at the minimum of the 
potential breaking the electroweak symmetry, is generated by 
self-interactions of the Higgs field. The framework of the Standard Model 
(SM)\cite{Glashow:1961tr,Salam:1968rm,Weinberg:1967tq} demands the 
physical Higgs boson as a new scalar degree of freedom, supplementing 
the spectrum of vectorial gauge bosons and spinorial matter particles.  \\

This concept of mass generation has also been applied, {\it mutatis mutandis}, 
to extended theories into which the SM may be embedded. 
The new theory may remain weakly interacting up to the grand-unification 
scale, or even the Planck scale, as familiar in particular from supersymmetric theories, 
or novel strong interactions may become effective already close to the
TeV regime. In such theories the Higgs sector is enlarged compared 
with the SM. A spectrum of several Higgs particles is 
generally predicted, the lightest particle often with properties 
close to the SM Higgs boson, and others with masses typically 
in the TeV regime. \\

A breakthrough on the path to establishing the Higgs mechanism experimentally 
has been achieved by observing at LHC \cite{Aad:2012gk,Chatrchyan:2012ufa} a new particle 
with a mass of about 125 GeV and couplings to electroweak gauge bosons and 
matter particles compatible, {\it cum grano salis}, with expectations 
for the Higgs boson in the (SM)~\cite{ATLAS-CONF-2014-009,CMS:2014ega,Aad:2013xqa,Chatrchyan:2012jja}.\\

\subsubsection{Zeroing in on the Higgs particle of the SM}

\noindent
Within the SM the Higgs mechanism is realized by introducing 
a scalar weak-isospin-doublet. Three Goldstone degrees of freedom are absorbed 
for generating the longitudinal components of the massive electroweak $W^\pm,Z$ 
bosons, and one degree of freedom is realized as a scalar physical particle 
unitarizing the theory properly. After the candidate particle has been found, 
three steps are necessary to establish the relation with the Higgs mechanism: \\[-2mm]

-- {\it{The mass, the lifetime (width) and the spin/$CP$ quantum numbers must be 
              measured as general characteristics of the particle;}}  \\[-3mm]

-- {\it{The couplings of the Higgs particle to electroweak gauge bosons
              and to leptons/quarks must be proven to rise (linearly)
              with their masses;}} \\[-3mm] 
  
-- {\it{The self-coupling of the Higgs particle, responsible for the  
              potential which generates the non-zero vacuum value of the Higgs field, 
              must be established.}} \\

When the mass of the Higgs particle is fixed, all its properties are pre-determined.
The spin/$CP$ assignement $J^{CP} = 0^{++}$ is required for an isotropic and $C,P$-even 
vacuum. Gauge interactions of the vacuum Higgs-field with the electroweak bosons and 
Yukawa interactions with the leptons/quarks generate the masses which in turn
determine the couplings of the Higgs particle to all SM particles. Finally,
the self-interaction potential, which leads to the non-zero vacuum value  
$v$ of the Higgs field, being responsible for breaking the electroweak symmetries, 
is determined by the Higgs mass, and, as a result, the trilinear and quadrilinear 
Higgs self-interactions are fixed. \\  

Since the Higgs mechanism provides the closure of the SM,
the experimental investigation of the mechanism, connected with
precision measurements{\footnote{Experimental results and simulations
    quoted in this introduction, as well as the large {\it corpus} of
    original theoretical studies in this field, are referenced
    properly in the review articles included subsequently in this
    section.}} of the properties of the Higgs particle, is mandatory
for the understanding of the microscopic laws of nature as formulated
at the electroweak scale. However, even though the SM is
internally consistent, the large number of parameters, {\it notabene}
mass and mixing parameters induced in the Higgs sector, suggests the
embedding of the SM into a more comprehensive theory
(potentially passing on the way through even more complex structures).
Thus
observing specific patterns in the Higgs sector could hold
essential clues to this underlying theory. \\

\begin{figure} [h]
\begin{center}
\includegraphics[width=0.8\hsize]{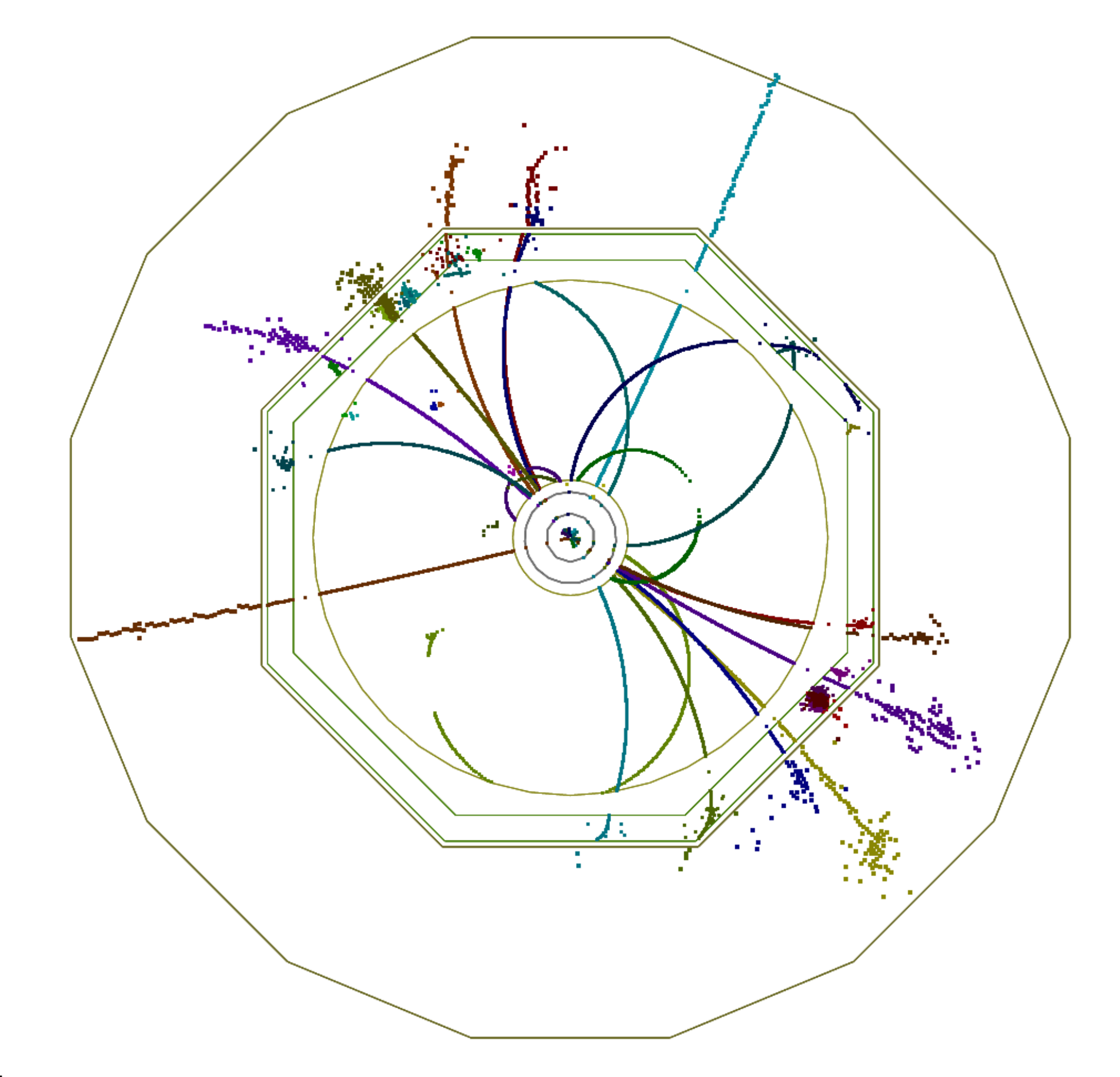}
\end{center}
\begin{center}
\includegraphics[width=7.5cm]{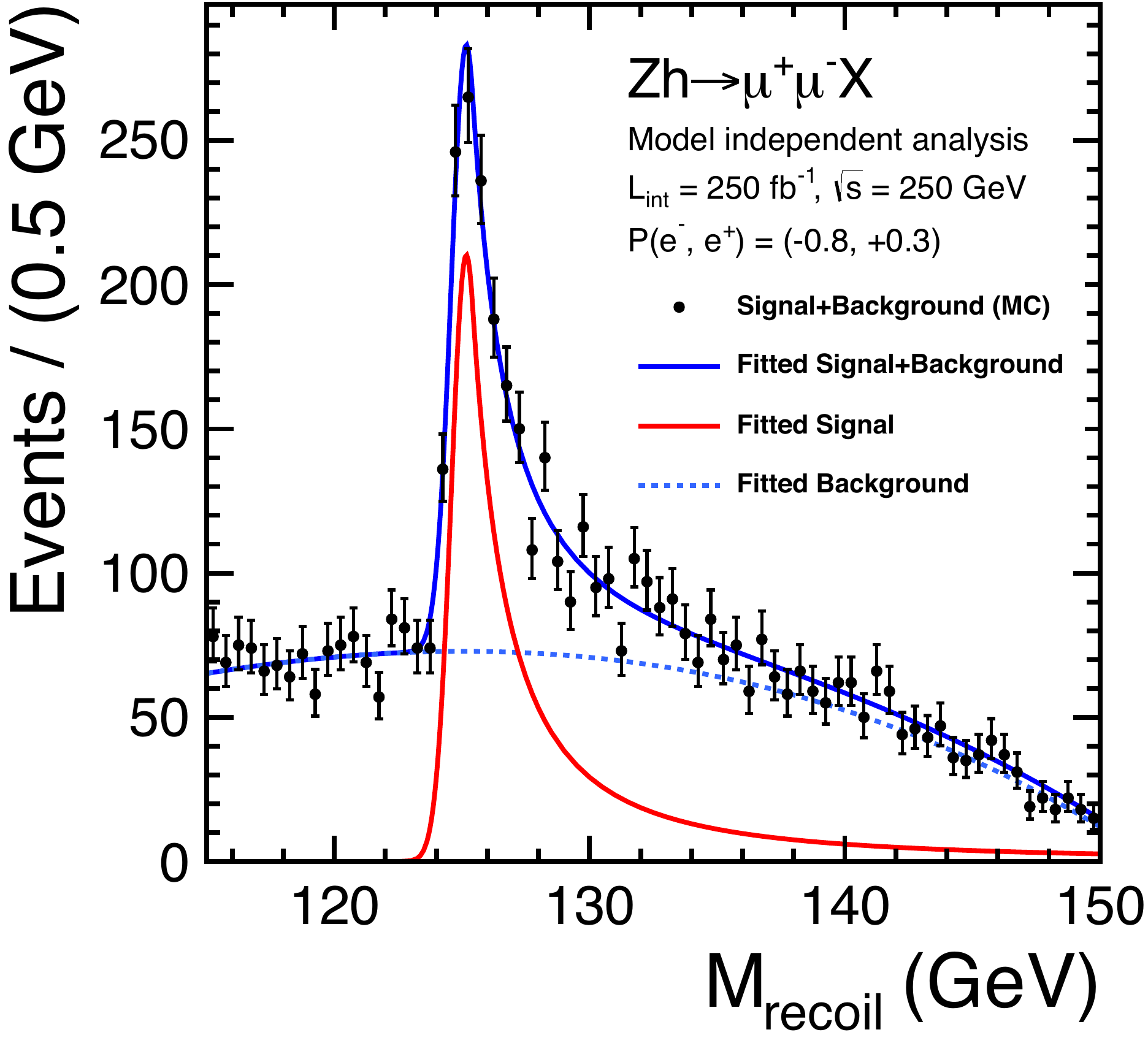}
\end{center}
\caption{{\it Upper plot: Event in Higgs-strahlung $e^+ e^- \to ZH \to (\mu^+ \mu^-)(jet\; jet)$
                  for a Higgs mass of 125 GeV at a collider energy of 500 GeV;
              lower plot: Distribution of the recoiling Higgs decay jets.}}
\label{fig:Higgsstrhlg}
\end{figure}

The SM Higgs boson can be produced through several channels in $pp$ collisions
at LHC, with gluon fusion providing by far the maximum rate for intermediate masses. 
In $e^+ e^-$ collisions the central channels \cite{Ellis:1975ap,Ioffe:1976sd,Jones:1979bq,
Cahn:1983ip,Kilian:1995tr} are
\begin{eqnarray}
{\text{Higgs-strahlung}}\;   &:& e^+ e^- \to Z + H                    \\
{\text{W-boson\, fusion}}\;  &:& e^+ e^- \to {\bar{\nu}}_e \nu_e + H  \,,
\end{eqnarray}
with cross sections for a Higgs mass $M_H = 125$ GeV as shown in Tab.{\ref{tab:cross}} for
the LC target energies of 250 GeV, 500 GeV, 1 TeV and 3 TeV.
\begin{table} [h]
\begin{center}
{\tiny
\begin{tabular}{|l||c|c|c|c|}
\hline         
                     & $\!$ 250 GeV $\!$    & $\!$ 500 GeV $\!$    & $\!$ 1 TeV $\!$   &  $\!$ 3 TeV  $\!$     \\     
\hline\hline
$\sigma[e^+e^- \to ZH]$        &   318         &   95.5          &  22.3           &  2.37           \\
$\sigma[e^+e^- \to \bar{\nu}_e \nu_e H]$   &       36.6      &   163          &   425       &      862        \\
\hline
\end{tabular}
\label{cross}
}
\end{center}
\caption{\it Cross sections in units of fb for Higgs-strahlung and $W$-boson fusion of Higgs bosons in the 
SM for a set of typical ILC/CLIC energies with beam polarizations: $P(e^-,e^+)=(-0.8,+0.3)$ for ILC at $250$ and $500\,$GeV, $(-0.8,+0.2)$ for ILC at $1\,$TeV, and $(-0.8,0)$ for CLIC at $3\,$TeV.}
\label{tab:cross}
\end{table}
By observing the $Z$-boson in Higgs-strahlung, {\it cf.}
Fig.{\ref{fig:Higgsstrhlg}}, the properties of the Higgs boson in
the recoil state can be studied experimentally
in a model-independent way. \\

\noindent
{\it a) Higgs particle: mass and $J^{CP}$}  \\[-2mm]

\noindent
Already for quite some time, precision analyses of the electroweak parameters, like the
$\rho$-parameter, suggested an SM Higgs mass of less than 161 GeV in the intermediate range
\cite{LEPEWWG}, above the lower LEP2 limit of 114.4 GeV \cite{ADLO} (for a review see
\cite{Schlatter:2011ia}). The mass of the new particle observed close to 125 GeV at LHC, 
agrees nicely with this expectation. \\  

The final accuracy for direct measurements of an SM Higgs mass of 125 GeV is predicted 
at LHC/HL-LHC and LC in the bands
\begin{eqnarray}
{\text{LHC / HL-LHC}}\, \;:\ M_H &=& 125\pm 0.1 / 0.05\; {\rm GeV} \\
            {\rm LC} \,\;:\ M_H &=& 125\pm 0.03\; {\rm GeV}       \,.  
\end{eqnarray}
\vskip5mm
Extrapolating the Higgs self-coupling associated with this mass value to the Planck scale, 
a value remarkably close to zero emerges \cite{Cabibbo:1979ay,Sher:1988mj,Degrassi:2012ry}. \\

\begin{figure}[h]                                                                                          
\begin{center}
\includegraphics[width=0.80\hsize]{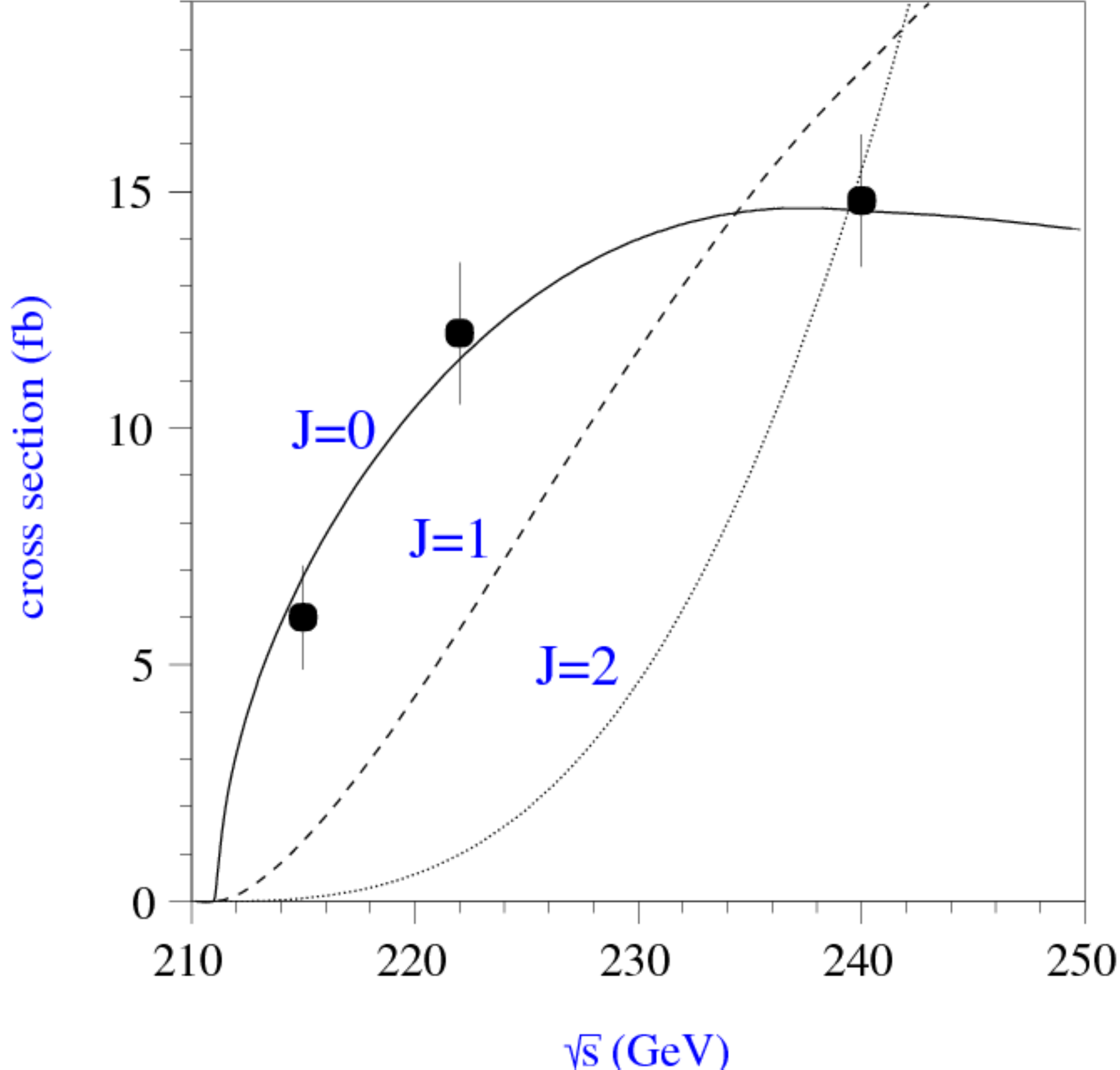}
\end{center}
\hspace{-2cm}
\includegraphics[width=1.15\hsize]{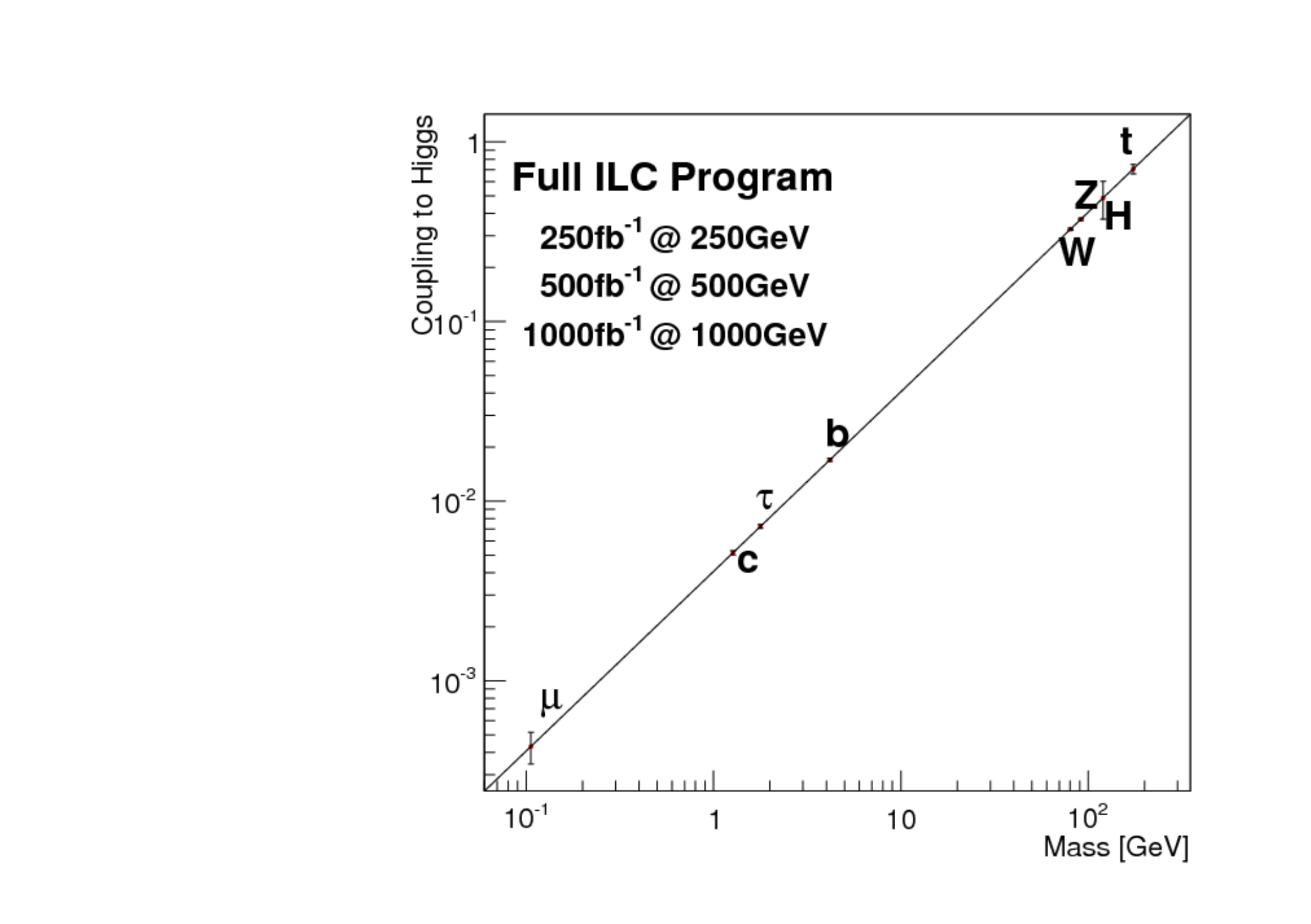}
\caption{{\it Upper plot: Threshold rise of the Cross section for
    Higgs-strahlung $e^+ e^- \to ZH$ corresponding to Higgs spin = 0,
    1, 2, complemented by the analysis of angular correlations; lower plot:
    Measurements of Higgs couplings as a function of particle
    masses.}}
\label{fig:spin}
\end{figure}

Various methods can be applied for confirming the $J^{CP} = 0^{++}$ quantum numbers 
of the Higgs boson. While $C = +$ follows trivially from the $H \to \gamma\gamma$
decay mode, correlations among the particles in decay final states and between 
initial and final states, as well as threshold effects in Higgs-strahlung \cite{Barger:1993wt},
{\it cf.} Fig.{\ref{fig:spin}} (upper plot), can be exploited for measuring these quantum numbers.\\

\noindent
{\it b) Higgs couplings to SM particles} \\[-2mm]

\noindent
Since the interaction between SM particles $x$
 and the vacuum Higgs-field generates 
the fundamental
SM masses, the coupling between SM particles and the physical Higgs particle, defined
dimensionless, is determined by their masses:
\begin{equation}
g_{Hxx} = [\sqrt{2} G_F]^\frac{1}{2}\, M_x   \,,
\label{Hcplg}
\end{equation}
%
the coefficient fixed in the SM by the vacuum
field $v = [\sqrt{2} G_F]^\frac{-1}{2}$. This fundamental relation 
is a cornerstone of the Higgs mechanism. 
It can be studied experimentally by measuring production cross sections 
and decay branching ratios. \\

\begin{sloppypar}
At hadron colliders the twin observable $\sigma \times BR$ is measured
for narrow states,
and ratios of Higgs couplings are accessible directly. Since in a
model-independent analysis $BR$ potentially includes invisible decays
in the total width, absolute values of the couplings can only be
obtained with rather large errors. This problem can be solved in $e^+
e^-$ colliders where the invisible Higgs decay branching ratio can be
measured directly in Higgs-strahlung. Expectations for measurements at
LHC (HL-LHC) and linear colliders are collected in
Tab.{\ref{tab:cplgs}}. The rise of the Higgs couplings
with the masses is demonstrated for LC measurements impressively in
Fig.{\ref{fig:spin}} (lower plot).  
\end{sloppypar}

\begin{table}[h]
\begin{tabular}{|l||c|c||c||c|}
\hline
%
Coupling            		&  LHC     		& HL-LHC      	& LC        &  HL-LHC + LC               \\
\hline\hline
$HWW$               		&  4-6\%         	& 2-5\%         	&  0.3\%	&  0.1\%                     \\
$HZZ$               		&  4-6\%          	& 2-4\%           	&  0.5\%	&  0.3\%                     \\
$Htt$               			&  14-15\%     	& 7-10\%   	&  1.3\% 	&  1.3\%                     \\
$Hbb$               		&  10-13\%      	& 4-7\%		&  0.6\% 	&  0.6\%                     \\
$H\tau\tau$         		&  6-8\%         	& 2-5\%    	&  1.3\%	&  1.2\%                     \\
\hline
$H\gamma\gamma$     	&  5-7\%     	&  2-5\%          	&  3.8\%  	&  3.0\%                      \\
$Hgg$               		&  6-8\%     	&  3-5\%      	&  1.2\% 	& 1.1\%                       \\
\hline
$H {\rm invis}$   		&  ---              	& ---              	&   0.9\%  & 0.9\%                            \\
\hline
\end{tabular}
\caption{\it Expected accuracy with which fundamental and derived Higgs couplings can be measured; the deviations
             are defind as $\kappa:=g/g_{SM}=1\pm\Delta$ compared to the SM at the LHC/HL-LHC, LC and 
             in combined analyses of the HL-LHC and LC\cite{ref:snowmass_higgs}.
             The fit assumes generation universality: $\kappa_u\equiv \kappa_c\equiv \kappa_t$, $\kappa_d\equiv\kappa_s\equiv \kappa_b$, and $\kappa_\mu\equiv \kappa_\tau$.
             The 95\% C.L. upper limit of potential couplings to invisible channels is also given.}
\label{tab:cplgs}
\end{table}

A special role is played by the loop-induced $\gamma\gamma$ width which can most accurately 
be measured by Higgs fusion-formation in a photon collider. \\  

From the cross section measured in $WW$-fusion the partial width $\Gamma[WW^\ast]$
can be derived and, at the same time, from the Higgs-strahlung process the decay 
branching ratio $BR[WW^\ast]$ can be determined so that the total width follows 
immediately from
\begin{equation}
\Gamma_{tot}[H] = \Gamma[WW^\ast] / BR[WW^\ast]   \,.  
\end{equation}
\begin{sloppypar}
Based on the expected values at LC, the total width of the SM Higgs particle
at 125 GeV is derived as $\Gamma_{tot}[H] = 4.1\,{\rm MeV}$$[1\pm 5\%]$. 
Measurements based on off-shell production of Higgs bosons provide only a
very rough upper bound on the total width. 
\\
\end{sloppypar}

Potential deviations of the couplings from the SM values can be
attributed to the impact of physics beyond the SM.
Parameterizing these effects, as naturally expected in dimensional
operator expansions, by $g_H = g_H^{SM} [1 + v^2 / \Lambda^2_\ast]$,
the BSM scale is estimated to $\Lambda_\ast >$ 550 GeV for an accuracy
of 20\% in the measurement of the coupling, and 2.5 TeV for 1\%, see also
\cite{Englert:2014uua}.  The
shift in the coupling can be induced either by mixing effects or by
loop corrections to the Higgs vertex. Such mixing effects
are well known in the supersymmetric Higgs sector where in the
decoupling limit the mixing parameters in the Yukawa vertices approach
unity as $\sim v^2/m^2_A$. Other mixing effects are induced in Higgs
portal models and strong interaction Higgs models with either
universal or non-universal shifts of the couplings at an amount $\xi =
(v/f)^2$, which is determined by the Goldstone scale $f$ of global
symmetry breaking in the strong-interaction sector; with $f \sim$ 1
TeV, vertices may be modified up to the level of 10\%. Less promising
is the second class comprising loop corrections of Higgs vertices.
Loops, generated for example by the exchange of new $Z'$-bosons, are
suppressed by the numerical coefficient $4\pi^2$ (reduced in addition
by potentially weak couplings). Thus the accessible mass range, $M <
\Lambda_\ast / 2\pi \sim$ 250 GeV,
can in general be covered easily by direct LHC searches. \\

\noindent
{\it c) Higgs self-couplings} \\[-2mm]

\noindent
The self-interaction of the Higgs field,
\begin{eqnarray}
V &=& \lambda\, [\,|\phi|^2 - v^2 /2\,]^2                                \,,
\end{eqnarray}
is responsible for electroweak symmetry breaking by shifting the vacuum state of
minimal energy from zero to $v/\sqrt{2} \simeq$ 174~GeV. The quartic form of the potential, 
required to render the theory renormalizable, generates trilinear and quadrilinear 
self-couplings when $\phi \to [v+H]/\sqrt{2}$ is shifted to the physical Higgs field $H$. 
The strength of the couplings are determined uniquely by the Higgs mass, with $M_H^2 = 2 \lambda v^2$:
\begin{equation}
\lambda_3 = M_H^2 / 2 v, \quad \lambda_4 = M_H^2 / 8 v^2 \quad {\text{and}} \quad 
                               \lambda_{n > 4} = 0                                   \,.
\end{equation}
The trilinear Higgs coupling can be measured in Higgs pair-production \cite{Djouadi:1999gv}. 
Concerning the LHC, 
the cross section is small and thus the high luminosity of HL-LHC is needed to achieve some 
sensitivity to the coupling. Prospects are brighter in Higgs pair-production 
in Higgs-strahlung and $W$-boson fusion of $e^+ e^-$ collisions, {\it i.e.} 
$e^+ e^- \to Z + H^\ast \to Z + HH$, {\it etc}. In total, 
a precision of
\begin{eqnarray}
{\text{LC}}\;            &:& \lambda_3 = 10-13\%                                                                         
\end{eqnarray}
may be expected. On the other hand, the cross section for triple Higgs production
is so small, $\mathcal{O}$(ab), that the measurement of $\lambda_4$ values 
near the SM prediction will not be feasible at either type of colliders.  \\[2mm]

\noindent
{\it d) Invisible Higgs decays} \\[-2mm]

\begin{sloppypar}
\noindent
The observation of cold dark matter suggests the existence of a hidden sector 
with {\it a priori} unknown, potentially high complexity. The Higgs field of the SM 
can be coupled to a corresponding Higgs field in the hidden sector, $\tilde{\mathcal{V}} = \eta
|\phi_{SM}|^2 |\phi_{hid}|^2$, in a form compatible with all standard symmetries. Thus a portal 
could be opened from the SM to the hidden sector 
\cite{Binoth:1996au,Patt:2006fw}. Analogous mixing 
with radions is predicted in theories incorporating extra-space dimensions. The mixing of the Higgs
fields in the two sectors induces potentially small universal changes in the observed Higgs 
couplings to the SM particles and, moreover, Higgs decays to invisible hidden states
(while this channel is opened in the canonical SM only indirectly by neutrino decays of $Z$ pairs).
Both signatures are a central target for experimentation at LC, potentially allowing
the first sighting of a new world of matter in the Higgs sector. \\[2mm] 
\end{sloppypar}   

\noindent
In summary, essential elements of the Higgs mechanism in the SM 
can be determined at $e^+ e^-$ linear colliders in the 250 to 500 GeV and 1 to 3 TeV modes 
at high precision. Improvements on the fundamental parameters by nearly an order of magnitude
can be achieved in such a faciliy. Thus a fine-grained picture of the Higgs
sector as third component 
of the SM can be drawn at a linear collider, completing the theory 
of matter and forces at the electroweak scale. First glimpses of a sector 
beyond the SM are possible by observing deviations from the SM picture
at scales far beyond those accessible at colliders directly. \\

\subsubsection{Supersymmetry scenarios}

\noindent
The hypothetical extension of the SM to a supersymmetric theory
\cite{Golfand:1971iw,Wess:1974tw} is intimately connected with the Higgs sector. 
If the SM is embedded in a
grand unified scenario, excessive fine-tuning in radiative corrections would be 
needed to keep the Higgs mass near the electroweak scale, {\it i.e.} fourteen
orders of magnitude below the grand unification scale. A stable bridge can be
constructed however in a natural way if matter and force fields are assigned 
to fermion-boson symmetric multiplets with masses not spread more than order TeV. 
In addition, by switching the mass (squared) of a scalar field from positive to
negative value when evolved from high to low scales, supersymmetry offers an
attractive physical explication of the Higgs mechanism. It should be 
noted that supersymmetrization of the SM is not the only solution 
of the hierarchy problem, however it joins in nicely with arguments of highly
precise unification of couplings, the approach to gravity in local supersymmetry,
and the realization of cold dark matter. Even though not yet backed at present by the direct
experimental observation of supersymmetric particles, supersymmetry remains an 
attractive extension of the SM, offering solutions to a variety of
fundamental physical problems. \\

To describe the Higgs interaction with matter fields by a superpotential, 
and to keep the theory anomaly-free, at least two independent Higgs iso-doublets 
must be introduced, coupling separately to up- and down-type matter fields.
They are extended eventually by additional scalar superfields, {\it etc}. \\[2mm]

\noindent
{\it a) Minimal supersymmetric model MSSM} \\[-2mm]

\noindent
Extending the SM fields to super-fields and adding a second Higgs doublet
defines the
minimal supersymmetric standard model (MSSM). After gauge symmetry breaking, three 
Goldstone components out of the eight scalar fields are aborbed to provide masses 
to the electroweak gauge bosons while five degrees of freedom are realized 
as new physical fields, corresponding to two neutral $CP$-even scalar particles 
$h^0,H^0$; one neutral $CP$-odd scalar particle $A^0$; and a pair of charged 
$H^\pm$ scalar particles \cite{Gunion:1984yn,Gunion:1986nh,Gunion:1989we,
Djouadi:2005gj}. \\

\begin{sloppypar}
Since the quadrilinear Higgs couplings are predetermined by the (small) gauge 
couplings, the mass of the lightest Higgs particle is small. The bound, $M_{h^0}
< M_Z | \cos 2\beta |$ at lowest order, with $\tan\beta$ accounting for Goldstone - 
Higgs mixing, is significantly increased however to $\sim$ 130 GeV by radiative 
corrections, adding a contribution of order $3 M^4_t/2 \pi^2 v^2\, \log M^2_{\tilde{t}}/M^2_t
+ mix$ for large top and stop masses. To reach a value of 125 GeV, large stop masses
and/or large trilinear couplings are required in mixings.\\
\end{sloppypar}

Predictions for production and decay amplitudes deviate, in general, from the 
SM not only because of modified tree couplings but also due to
additional loop contributions, as $\tilde{\tau}$ loops in the $\gamma\gamma$
decay mode of the lightest Higgs boson. \\ 

To accomodate a 125 GeV Higgs boson in minimal supergravity the quartet of heavy 
Higgs particles $H^0,A^0,H^\pm$ is shifted to the decoupling regime 
with order TeV masses. The properties of the lightest Higgs boson $h^0$ are very close 
in this regime to the properties of the SM Higgs boson.  \\

\begin{figure} [h]
%
\begin{center}
\includegraphics[width=0.95\hsize]{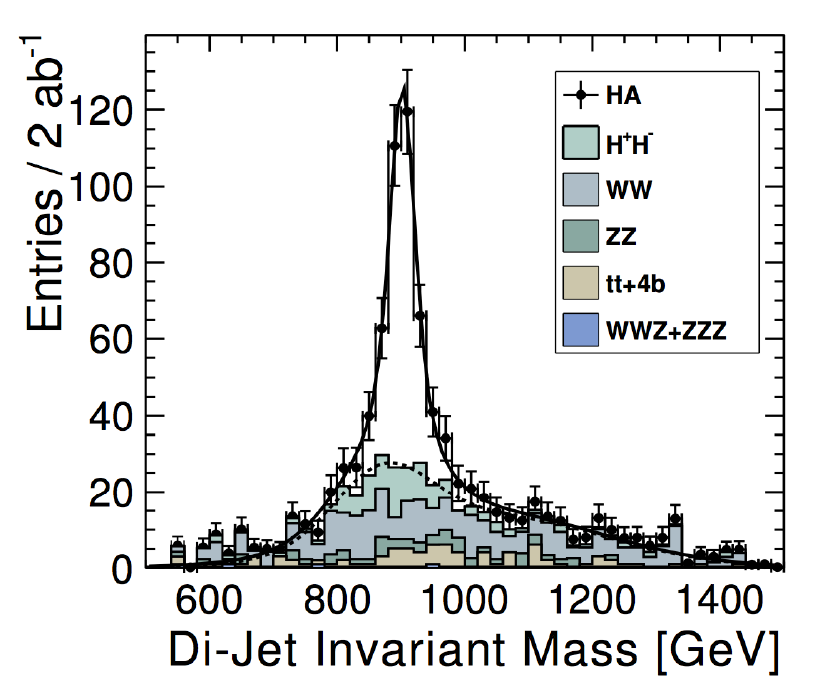}
\end{center}
\begin{center}
\includegraphics[width=0.95\hsize]{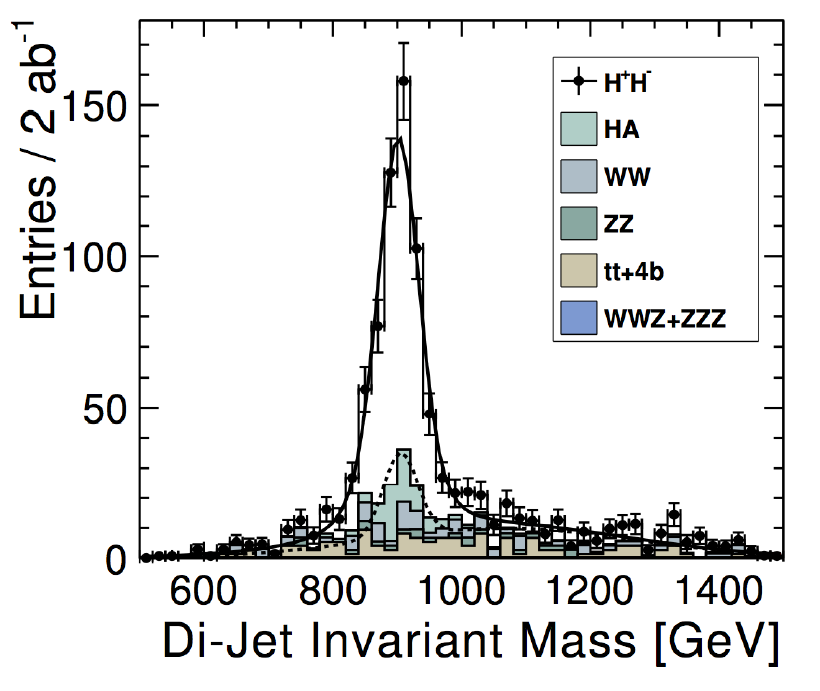}
\end{center}
\caption{{\it Upper plot: Reconstructed 2-jet invariant mass for associated production: $e^+ e^- \to AH \to b\bar{b}b\bar{b}$                                                                                          
                  for a Higgs mass of 900 GeV at a collider energy of 3 TeV; 
              lower plot: Similar plot for $e^+ e^- \to H^+H^- \to t\bar{b}\bar{t}b$.}}
\label{fig:AH}
\end{figure}

\begin{sloppypar}
The heavy Higgs boson quartet is difficult to search for at LHC. In fact, these
particles cannot be detected in a blind wedge which opens at 200 GeV for intermediate
values of the mixing parameter $\tan\beta$ and which covers the parameter space for
masses beyond 500~GeV. At the LC, Higgs-strahlung $e^+ e^- \to Z\,h^0$ is supplemented by
Higgs pair-production:
\begin{eqnarray}
~~~~~~~~~~ && e^+ e^- \to A^0\,H^0 \;\, {\text{and}} \;\, H^+ \, H^-   \,  
\end{eqnarray}
providing a rich source of heavy Higgs particles in $e^+ e^-$ collisions for masses 
$M < \sqrt{s}/2$, {\it cf.} Fig.{\ref{fig:AH}}. Heavy Higgs masses come with 
$ZAH$ couplings of the order of gauge couplings so that the cross sections 
are large enough for copious production of heavy neutral $CP$ even/odd and charged
Higgs boson pairs. \\ 
\end{sloppypar}

Additional channels open in single Higgs production $\gamma\gamma \to A^0,H^0$, 
completely exhausting the multi-TeV energy potential ${\sqrt{s}}_{\gamma\gamma}$ 
of a photon collider.   \\[2mm]

\noindent
{\it b) Extended supersymmetry scenarios} \\[-2mm]

\noindent
The minimal supersymmetry model is quite restrictive by connecting the quadrilinear
couplings with the gauge couplings, leading naturally to a small Higgs mass, and 
grouping the heavy Higgs masses close to each other. The simplest extension 
of the system introduces an additional iso-scalar Higgs field 
\cite{Fayet:1976cr,Ellwanger:2009dp}, the next-to-minimal model (NMSSM). 
This extension augments the Higgs spectrum by two additional physical states, 
$CP$-even and $CP$-odd, which mix with the corresponding MSSM-type states. \\

The bound on the mass of the lightest MSSM Higgs particle is alleviated 
by contributions from the trilinear Higgs couplings in the superpotential
(reducing the amount of 'little fine-tuning' in this theory).  Loop contributions
to accomodate a 125 GeV Higgs boson are reduced so that the bound 
on stop masses is lowered to about 100 GeV as a result. \\

The additional parameters in the NMSSM render the predictions for production
cross sections and decay branching ratios more flexible, so that an increased
rate of $pp \to Higgs \to \gamma\gamma$, for instance, can be accomodated 
more easily than within the MSSM. \\[2mm]

\noindent
Motivations for many other extensions of the Higgs sector have been presented 
in the literature. Supersymmetry provides an attractive general framework in 
this context. The new structures could be so rich that the clear
experimental environment of $e^+ e^-$ collisions is needed to map out this 
Higgs sector and to unravel its underlying physical basis. \\

\subsubsection{Composite Higgs bosons}   

\noindent
Not long after pointlike Higgs theories had been introduced to generate the breaking
of the electroweak symmetries, alternatives have been developed based on novel strong
interactions \cite{Weinberg:1975gm,Susskind:1978ms}. The breaking of global symmetries 
in such theories gives rise to massless Goldstone bosons which can be absorbed 
by gauge bosons to generate their masses. This concept had been expanded later 
to incorporate also light Higgs bosons with mass in the intermediate range. Generic 
examples for such theories are Little Higgs Models and theories formulated 
in higher dimensions, which should be addressed briefly as generic examples. \\

\noindent
{\it a) Little Higgs models} \\[-2mm]

\noindent
If new strong interactions are introduced at a scale of a few 10 TeV, the breaking 
of global symmetries generates a Goldstone scale $f$ typically reduced by one order of
magnitude, {\it i.e.} at a few TeV. The spontaneous breaking of large global groups
leads to an extended scalar sector with Higgs masses generated radiatively 
at the Goldstone scale. The lightest Higgs mass is delayed, by contrast, acquiring mass 
at the electroweak scale only through collective symmetry breaking at higher oder.  \\

Such a scenario \cite{Han:2005ru} can be realized, for instance, in minimal form as a non-linear  
sigma model with a global SU(5) symmetry broken down to SO(5). After separating the Goldstone modes 
which provide masses to gauge bosons, ten Higgs bosons emerge in this scenario which split 
into an isotriplet $\Phi$, including a pair of doubly charged $\Phi^{\pm\pm}$ states with TeV scale masses, 
and the light standard doublet $h$. The properties of $h$ are affected at the few per-cent level 
by the extended spectrum of the fermion and gauge sectors. The new TeV triplet Higgs bosons 
with doubly charged scalars can be searched for very effectively in pair production 
at LC in the TeV energy range. \\[2mm]

\noindent
{\it b) Relating to higher dimensions} \\[-2mm]

\noindent
An alternative approach emerges out of gauge theories formulated in 5-dimensional
Anti-de-Sitter space. The AdS/CFT correspondence relates this theory to a
4-dimensional strongly coupled theory, the 5-th components of the gauge fields
interpreted as Goldstone modes in the strongly coupled 4-dimensional sector. 
In this picture the light Higgs boson appears as a composite state with properties 
deviating to order $(v/f)^2$ from the standard values \cite{Espinosa:2012qj},
either universally or non-universally with alternating signs for vector bosons and
fermions.  

%

\subsection{The SM Higgs at the LHC: Status and Prospects\protect\footnotemark}
\footnotetext{J\"urgen Kroseberg}
\label{sec:ewsb2}
%
\noindent
In July 2012 the ATLAS and CMS experiments at the LHC announced the 
discovery of a new particle with a mass of about 
125 GeV that provided a compelling candidate for the Higgs boson in the framework of the Standard Model of particle
physics (SM). Both experiments found consistent evidence from a combination of searches for three decay modes, 
$H\to\gamma\gamma$,  \mbox{$H\to ZZ\to 4l$}  and 
$H\to WW\to 2 l2\nu$ ($l=e,\mu$), with event rates and properties 
in agreement with SM predictions for Higgs boson production and decay.
These findings, which were based on proton-proton collision data
recorded at centre-of-mass energies of 7 and 8 TeV and corresponding
to an integrated luminosity of about 10~fb$^{-1}$ per experiment,
received a lot of attention both within and outside the particle
physics community and were eventually published
in~\cite{Aad:2012tfa,Chatrchyan:2012ufa,atlSCI,Chatrchyan:2013lba}.

Since then, the LHC experiments have concluded their first phase of
data taking ("Run1") and significantly larger data sets corresponding
to about 25~fb$^{-1}$ per experiment have been used to perform further
improved analyses enhancing the signals in previously observed decay
channels, establishing evidence of other decays and specific
production modes as well as providing more precise measurements of the
mass and studies of other properties of the new
particle. Corresponding results, some of them still preliminary, form
the basis of the first part of this section, which summarises the
status of the ATLAS and CMS analyses of the Higgs boson candidate
within the SM.

The second part gives an outlook on Higgs boson studies during the second phase 
("Run2") of the LHC operation scheduled to start later this year and the long-term potential 
for an upgraded high-luminosity LHC.

While the following discussion is restricted to analyses within the
framework of the SM, the consistency of the observed Higgs boson
candidate with SM expectations (as evaluated
in~\cite{Dittmaier:2011ti,Dittmaier:2012vm,Heinemeyer:2013tqa} and
references therein) does not exclude that extensions of the SM with a
richer Higgs sector are realised in nature and might show up
experimentally at the LHC. Thus, both the ATLAS and CMS collaborations
have been pursuing a rich program of analyses that search for
deviations from the SM predictions and for additional Higgs bosons in
the context of models beyond the SM. A review of this work is,
however, beyond the scope of this section.

\begin{figure}
\centering
\includegraphics[width=0.8\columnwidth]{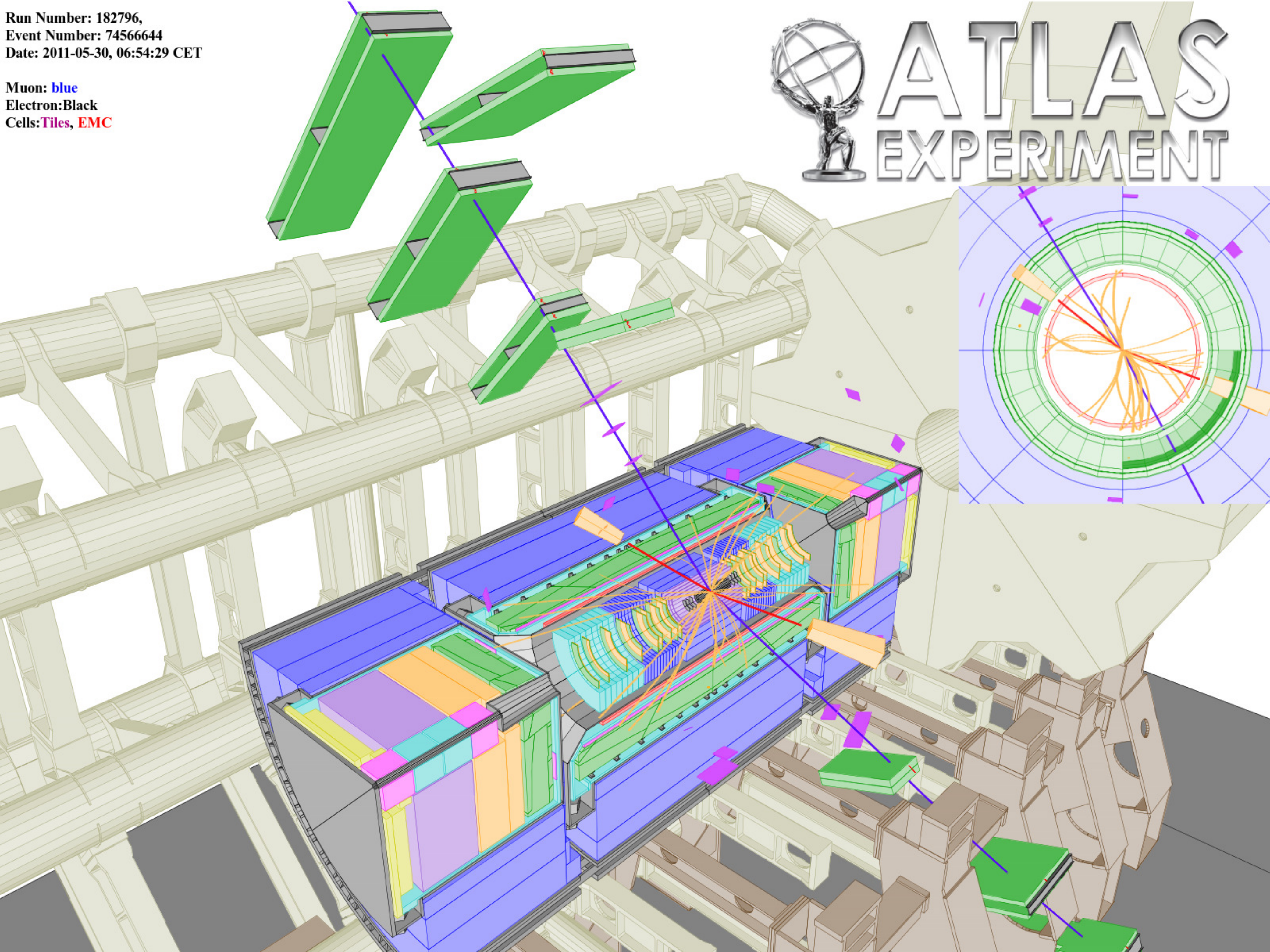}
\includegraphics[width=0.8\columnwidth]{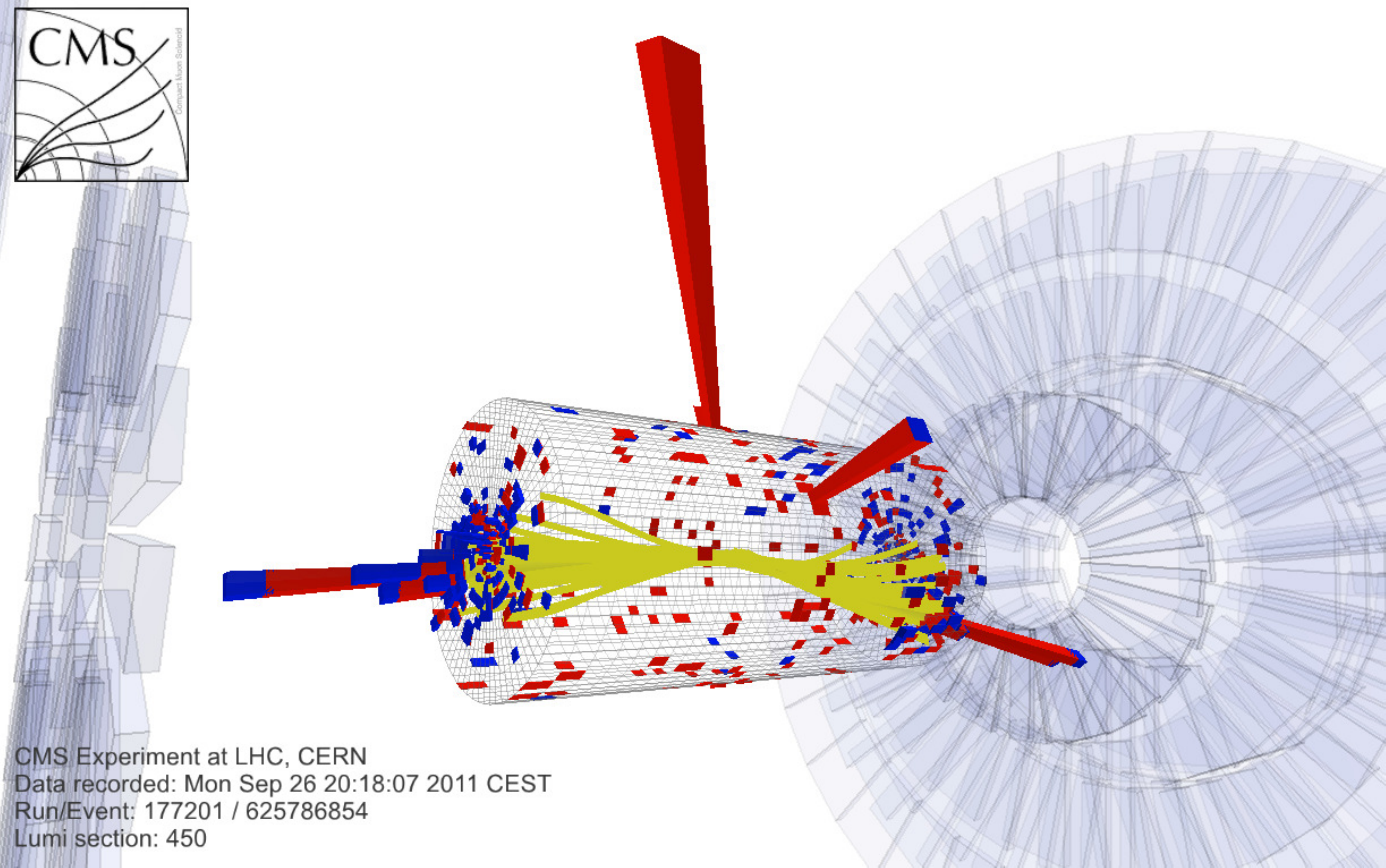}
\caption{Displays of example Higgs boson candidate events. {Top:} $H\to ZZ\to 2\mu2e$ candidate in the ATLAS 
detector; {bottom:} VBF $H\to\gamma\gamma$ candidate in the CMS detector.
}
\label{fig:lhc1}        
\end{figure}


\subsubsection{Current Status}
\label{subsubsec:lhcsm}
The initial SM Higgs boson searches at the LHC were designed for  a fairly large Higgs mass window between 100 
and 600 GeV, most of which was excluded by the ATLAS and CMS results based on the data sets recorded in 
2011~\cite{ATLAS:2012ae,Chatrchyan:2012tx}. In the following we focus on the analyses including the full 2012 
data and restrict the discussion to decay channels relevant to the discovery and subsequent study of the 125~GeV 
Higgs boson.


\subsubsection*{Relevant decay channels}
For all decay channels described below, the analysis strategies have evolved over time in similar ways. Early searches 
were based on inclusive analyses of the Higgs boson decay products.  With larger data sets, these were replaced by 
analyses in separate categories corresponding to different event characteristics and background composition. Such 
categorisation significantly increases the signal sensitivity and can also be used to separate different production processes, 
which is relevant for the current and future studies of the Higgs boson couplings discussed below. Also, with larger data 
sets and higher complexity of the analyses, it became increasingly important to model the background contributions from 
data control regions instead of relying purely on simulated events. Another common element is the application of
multivariate techniques in more recent analyses.
Still, the branching ratios, detailed signatures and relevant background processes for different decays differ substantially; 
two example Higgs boson production and decay candidate event displays are shown in Fig.~\ref{fig:lhc1}. Therefore, the 
experimental approaches 
and resulting information on the 125 GeV Higgs boson 
vary as well: 
\begin{figure}
\centering
\includegraphics[width=0.77\columnwidth]{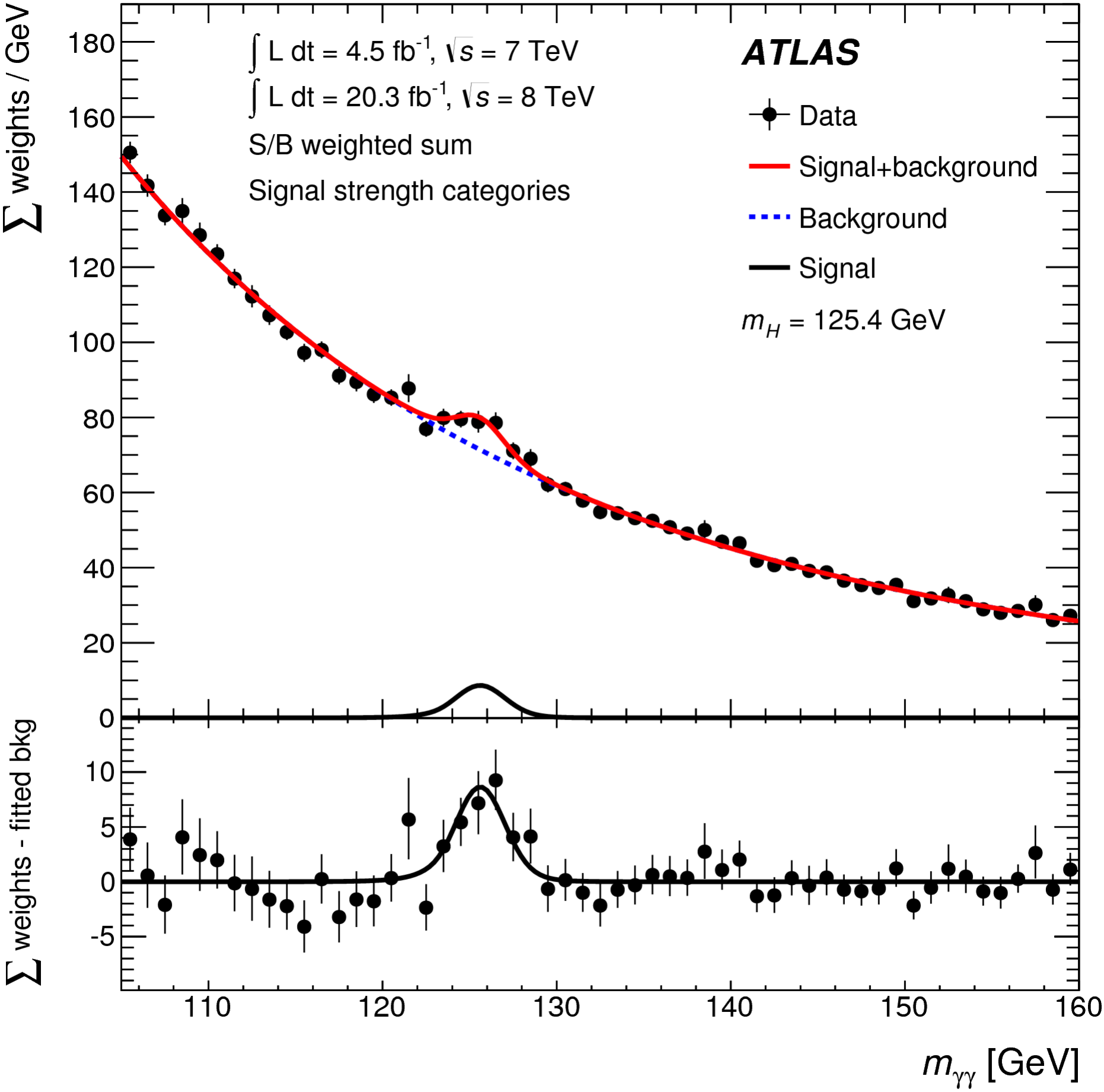}
\includegraphics[width=0.77\columnwidth]{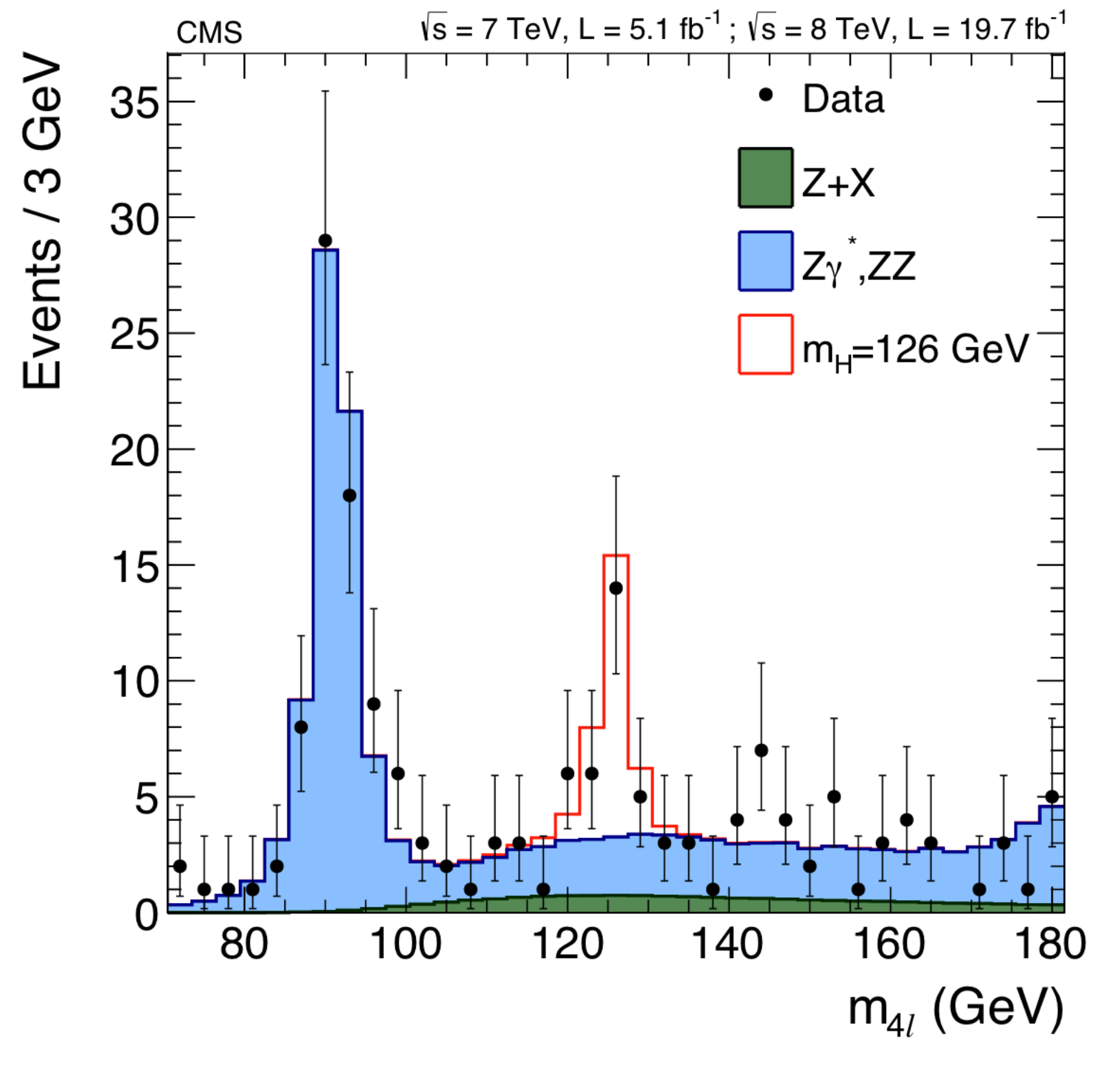} 
\includegraphics[width=0.77\columnwidth]{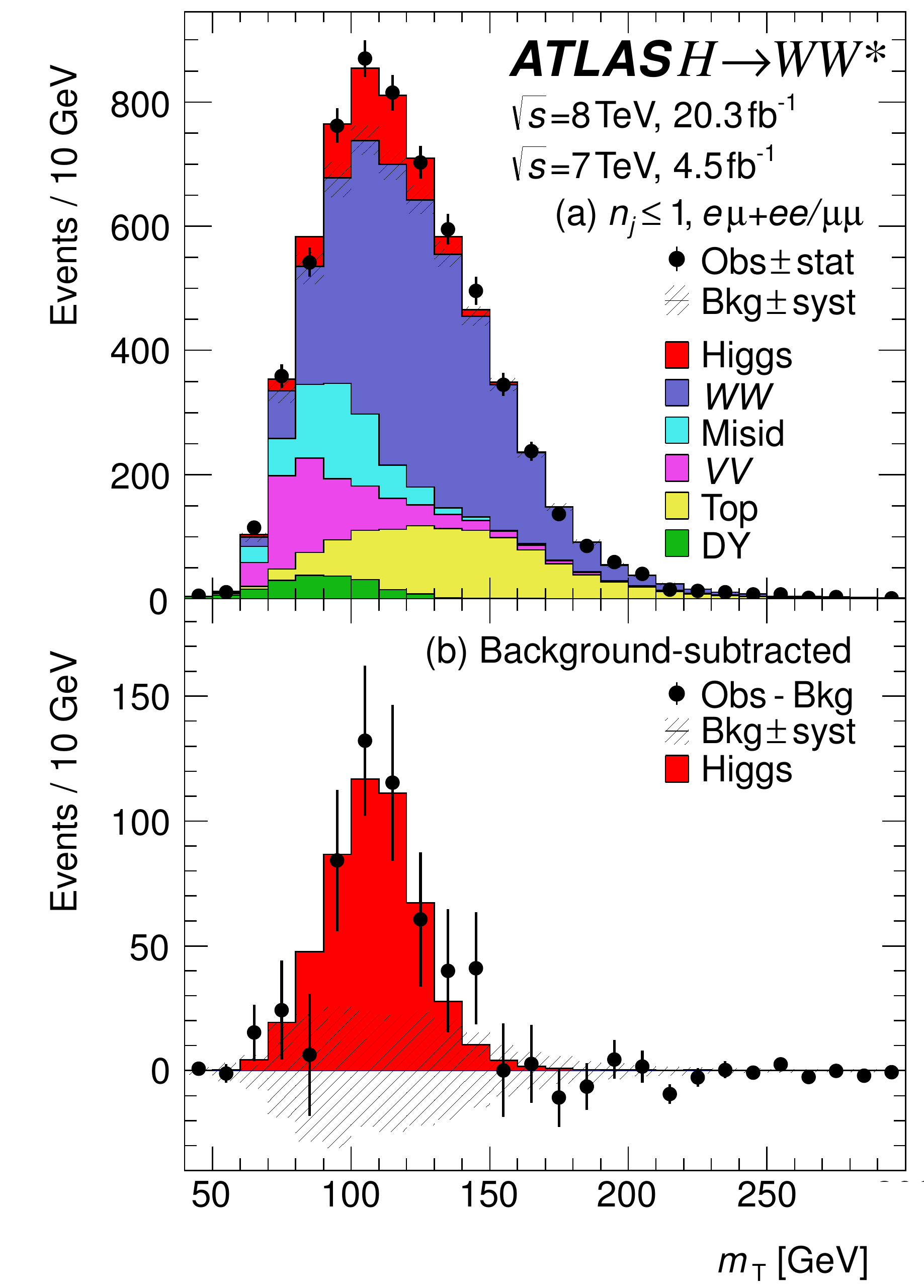}
\caption{Reconstructed distributions of the Higgs boson candidate decay products for the complete 2011/12 data, 
expected backgrounds, and simulated signal from {top:} the \mbox{ATLAS} $H\to\gamma\gamma$~\cite{Aad:2014eha}, 
{centre:} the CMS $H\to ZZ\to 4\ell$~\cite{Chatrchyan:2013mxa}, and {bottom:} the 
ATLAS $H\to WW\to 2\ell2\nu$~\cite{ATLAS:2014aga} analyses. 
}
\label{fig:lhc2}        
\end{figure}
%

\begin{itemize}
\item $H\to\gamma\gamma$: 
The branching fraction is very small but the two high-energy photons provide a clear experimental signature
and a good mass resolution. 
Relevant background processes are diphoton continuum production as well as photon-jet and dijet events. 
The most recent ATLAS~\cite{Aad:2014eha} and CMS~\cite{Khachatryan:2014ira1} analyses yield signals 
with significances of $5.2\sigma$ and $5.7\sigma$, respectively, where $4.6\sigma$ and $5.2\sigma$ are 
expected. 

\item $H\to ZZ\to 4\ell$: Also this decay combines a small branching fraction with a clear experimental signature and a 
good mass resolution. 
The selection of events with two pairs of isolated, same-flavour, opposite-charge electrons or muons  
results in the largest signal-to-background ratio of all currently considered Higgs boson decay channels. The remaining background 
originates mainly from  conti\-nuum $ZZ$, $Z$+jets and $t\bar{t}$ production processes.
ATLAS~\cite{Aad:2014eva} and CMS~\cite{Chatrchyan:2013mxa} report observed (expected) signal significances of 
$8.1\sigma$ ($6.2\sigma$) and $6.8\sigma$ ($6.7\sigma$).

\item $H\to WW\to2\ell2\nu$: The main advantage of this decay is its large rate, and the two oppositely charged 
leptons from the $W$ decays provide a good experimental handle. However, due to the two undetectable final-state 
neutrinos it is not possible to reconstruct a narrow mass 
peak. The dominant background processes are $WW$, $Wt$, and $t\bar{t}$ production. The observed (expected) 
ATLAS~\cite{ATLAS:2014aga} and CMS~\cite{Chatrchyan:2013iaa} signals have significances of 
$6.1\sigma$ ($5.8\sigma$) and $4.3\sigma$ ($5.8\sigma$).

\end{itemize}
 \noindent
Fig.~\ref{fig:lhc2} shows reconstructed Higgs candidate mass distributions from ATLAS and CMS searches for 
$H\to\gamma\gamma$ and $H\to ZZ\to 4\ell$, respectively, as well as the \mbox{ATLAS}  $H\to WW\to2\ell2\nu$ 
transverse mass distribution. 
%
Other bosonic decay modes
are searched for as well but these analyses are not yet sensitive to a SM Higgs boson observation. 

\begin{figure}[h!]
\centering
\hspace*{0cm}\includegraphics[width=0.7\columnwidth]{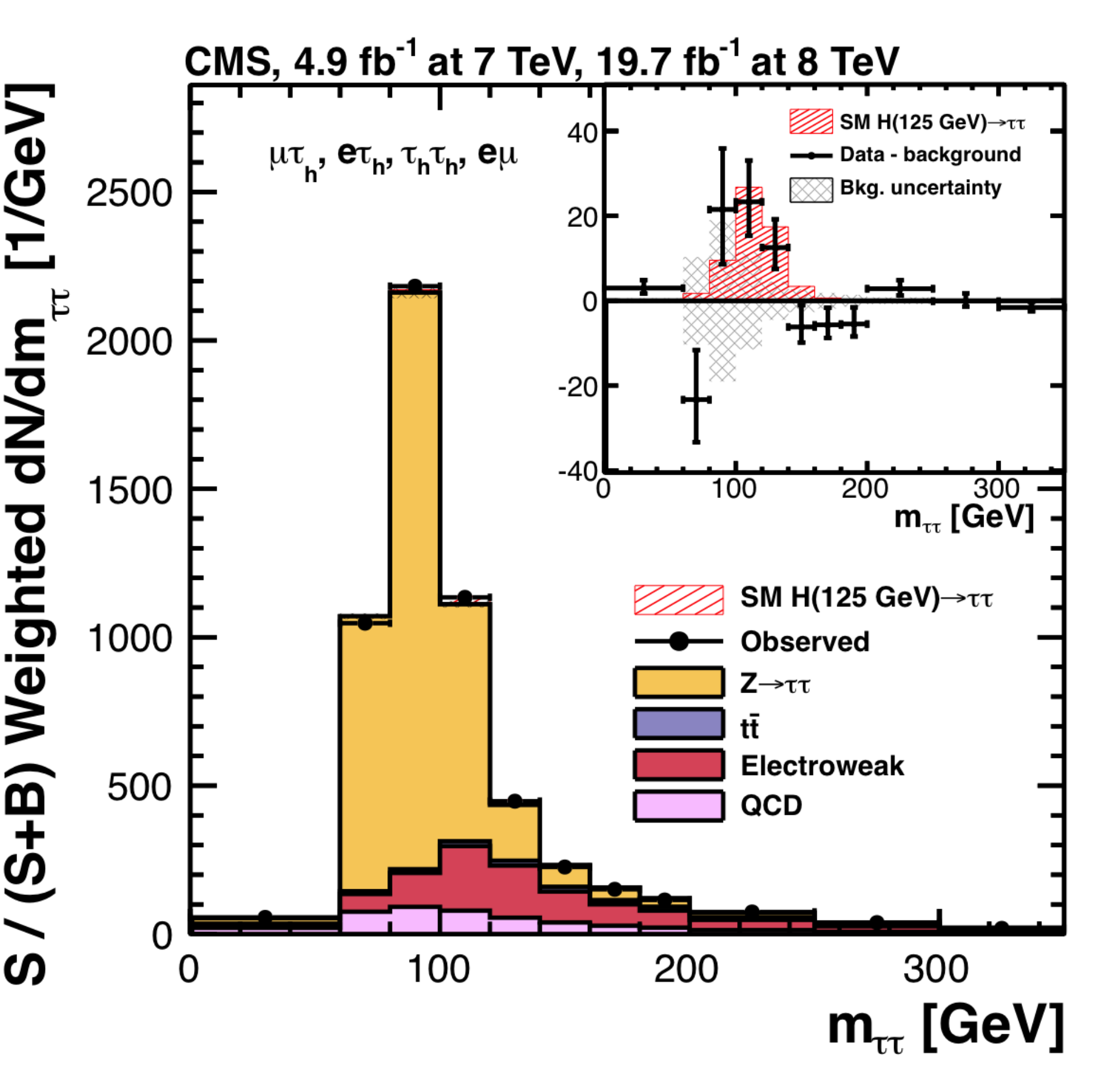}
\includegraphics[width=0.7\columnwidth]{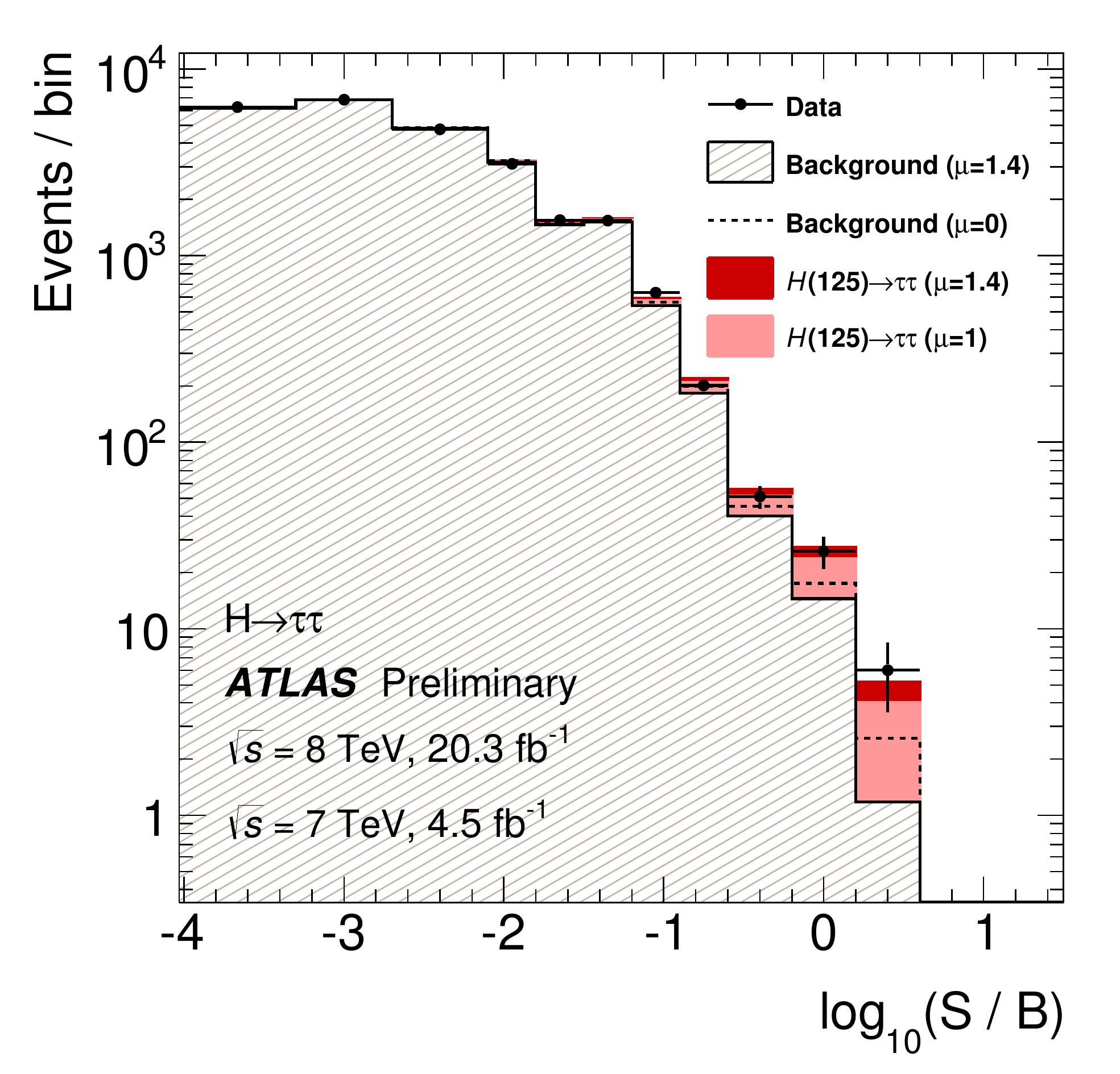}
\caption{
Evidence for the decay $H\to\tau\tau$. {Top}: 
CMS observed and predicted $m_{\tau\tau}$ distributions~\cite{Chatrchyan:2014nva}. The distributions obtained in 
each category of each channel are weighted by the ratio between the expected signal and signal-plus-background 
yields in the category. The inset shows the corresponding difference between the observed data and expected 
background distributions, together with the signal distribution for a SM Higgs boson at $m_H$=125 GeV; 
{bottom}: ATLAS event yields as a function of log(S/B), where S (signal yield) and B (background yield) are
taken from the corresponding bin in the distribution of the relevant BDT output discriminant~\cite{Aad:2015vsa}.
}
\label{fig:lhc3}       	
\end{figure}

\begin{itemize}
\item $H\to bb$: For a Higgs boson mass of 125~GeV this is the dominant Higgs boson decay mode. The experimental signature of
$b$ quark jets alone 
is difficult to exploit at the LHC, though, so that current analyses focus on the Higgs production associated with a vector 
boson $Z$ or $W$. Here, diboson, vector boson+jets and top production processes constitute the relevant backgrounds. 

\item $H\to\tau\tau$: All combinations of hadronic and leptonic $\tau$-lepton decays are used to search for a broad 
excess in the $\tau\tau$ invariant mass spectrum. The dominant and irreducible background is coming from 
\mbox{$Z\to\tau\tau$} decays; further background contributions arise from processes with a vector boson and jets, top 
and diboson production.
\end{itemize}
While searches for $H\to bb$ decays~\cite{Aad:2014xzb,Chatrchyan:2013zna} have not yet resulted in significant signals, first evidence for direct Higgs 
boson decays to fermions has been reported by both ATLAS and CMS following analyses of $\tau\tau$ final 
states.
The CMS results~\cite{Chatrchyan:2014nva} are predominantly based
on fits to the reconstructed $\tau\tau$ invariant mass distributions, whereas the  ATLAS 
analysis~\cite{Aad:2015vsa} uses the output of boosted decision trees (BDTs) 
throughout for the statistical analysis of the selected 
data. ATLAS (CMS) find signals with a significance of 4.5$\sigma$ (3.5$\sigma$), 
where 3.4$\sigma$ (3.7$\sigma$)  are expected, cf.~Fig.~\ref{fig:lhc3}. 
In~\cite{Chatrchyan:2014vua} CMS present the combination of their $H\to\tau\tau$ and $H\to bb$ 
analyses yielding an observed (expected) signal significance of $3.8\sigma$ ($4.4\sigma$).
Searches for other fermionic decays 
are performed as well but are not yet sensitive to the observation of 
the SM Higgs boson.

%

In the following, we summarise the status of SM Higgs boson analyses of the full 2011/12 data sets with ATLAS and 
CMS. The discussion is based on preliminary combinations of ATLAS and published CMS results collected in \cite{ATLAS-CONF-2015-007} 
and \cite{Khachatryan:2014jba}, respectively;  an ATLAS publication of Higgs boson mass measurements~\cite{Aad:2014aba}; 
\mbox{ATLAS}~\cite{Aad:2015xua} and 
CMS~\cite{Khachatryan:2014iha} constraints on the Higgs boson
width; studies of the Higgs boson spin and parity by CMS~\cite{Khachatryan:2014kca} and 
ATLAS~\cite{Aad:2013xqa,Aad:2015rwa,ATLAS-CONF-2015-008}; and other results on specific aspects or 
channels referenced later in this section. 


\subsubsection*{Signal strength}
For a given Higgs boson mass, the parameter $\mu$ is defined as the observed Higgs boson production strength normalised to the SM
expectation. 
\begin{figure}[b!]
\centering
\includegraphics[width=1.0\columnwidth]{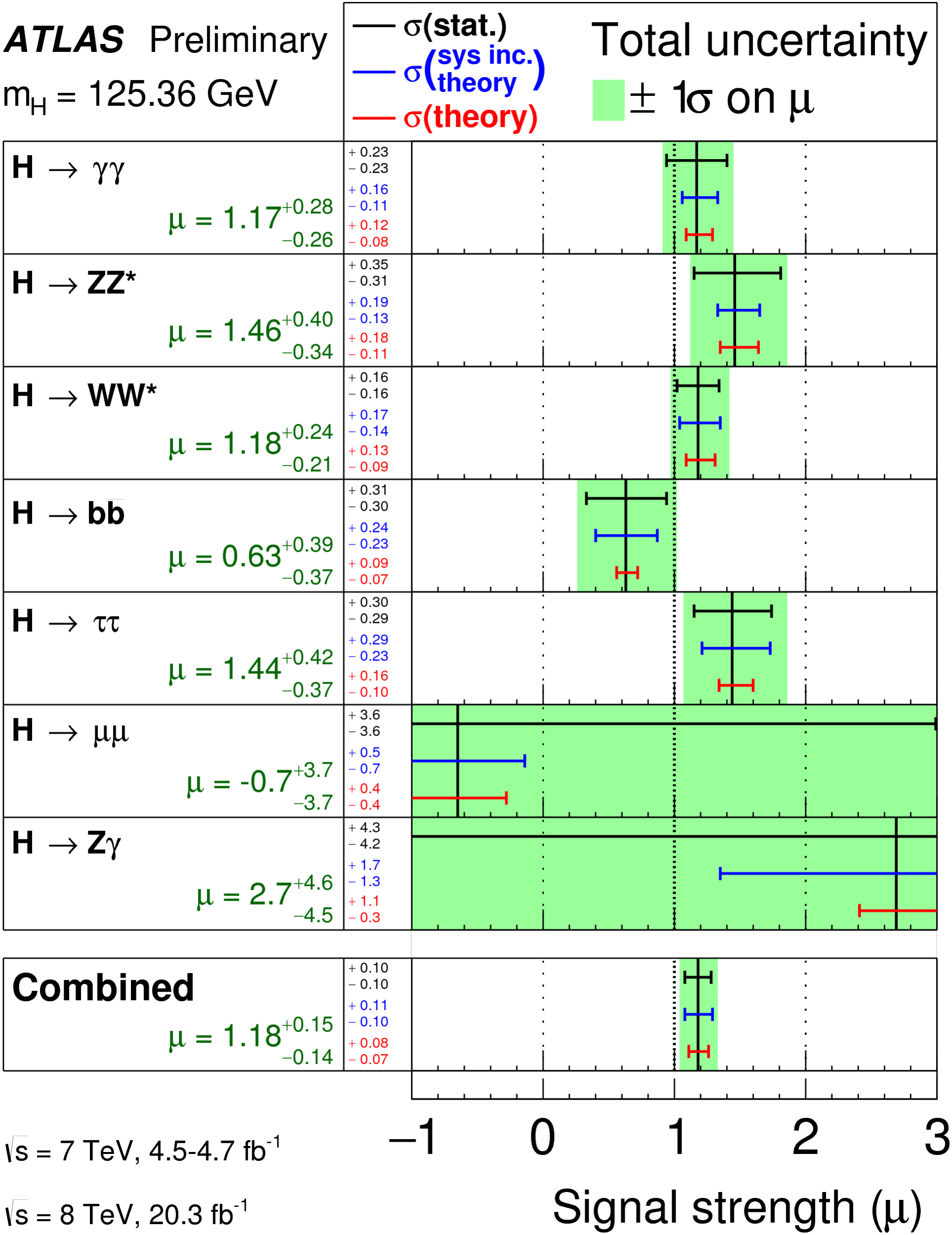}
\caption{Higgs boson signal strength as measured by ATLAS for different decay channels~\cite{ATLAS-CONF-2015-007}. 
}
\label{fig:lhc4}       
\end{figure}
\begin{figure}[t!]
\hspace*{-0.5cm}
\includegraphics[width=1.\columnwidth]{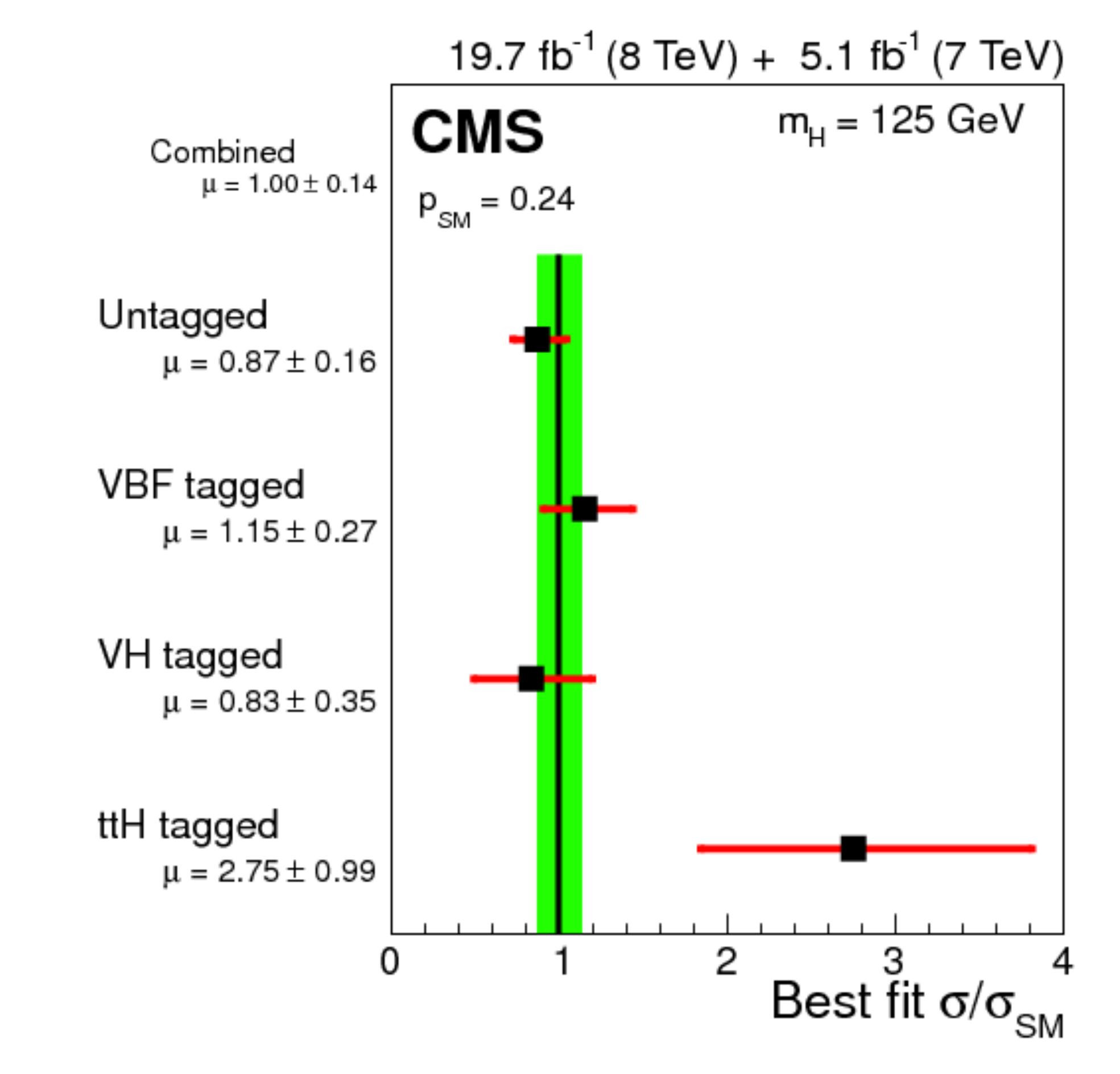}
\caption{Higgs boson production strength, normalised to the SM expectation, based on CMS 
analyses~\cite{Khachatryan:2014jba}, for a combination of analysis categories related to different production modes.
}
\label{fig:lhc5}       
\end{figure}
Thus, $\mu=1$ reflects the SM expectation and $\mu=0$ corresponds to the background-only hypothesis.

Fixing the Higgs boson mass to the measured value and considering the decays $H\to\gamma\gamma$, 
$H\to ZZ\to 4\ell$,  $H\to WW \to2\ell2\nu$, $H\to bb$, and $H\to\tau\tau$, ATLAS report~\cite{ATLAS-CONF-2015-007} 
a preliminary overall production strength of $$\mu=1.18^{+0.15}_{-0.14};$$ the separate combination of the bosonic 
and fermionic decay modes yields $\mu=1.35^{+0.21}_{-0.20}$ and $\mu=1.09^{+0.36}_{-0.32}$, respectively. The 
corresponding CMS result~\cite{Khachatryan:2014jba} is $$\mu=1.00\pm0.13.$$ 
Good consistency is found, for both experiments, across different decay modes and analyses categories related to 
different production modes, see Figs.~\ref{fig:lhc4} and \ref{fig:lhc5}.  
%

\begin{figure}[b!]
\centering
\includegraphics[width=1.0\columnwidth]{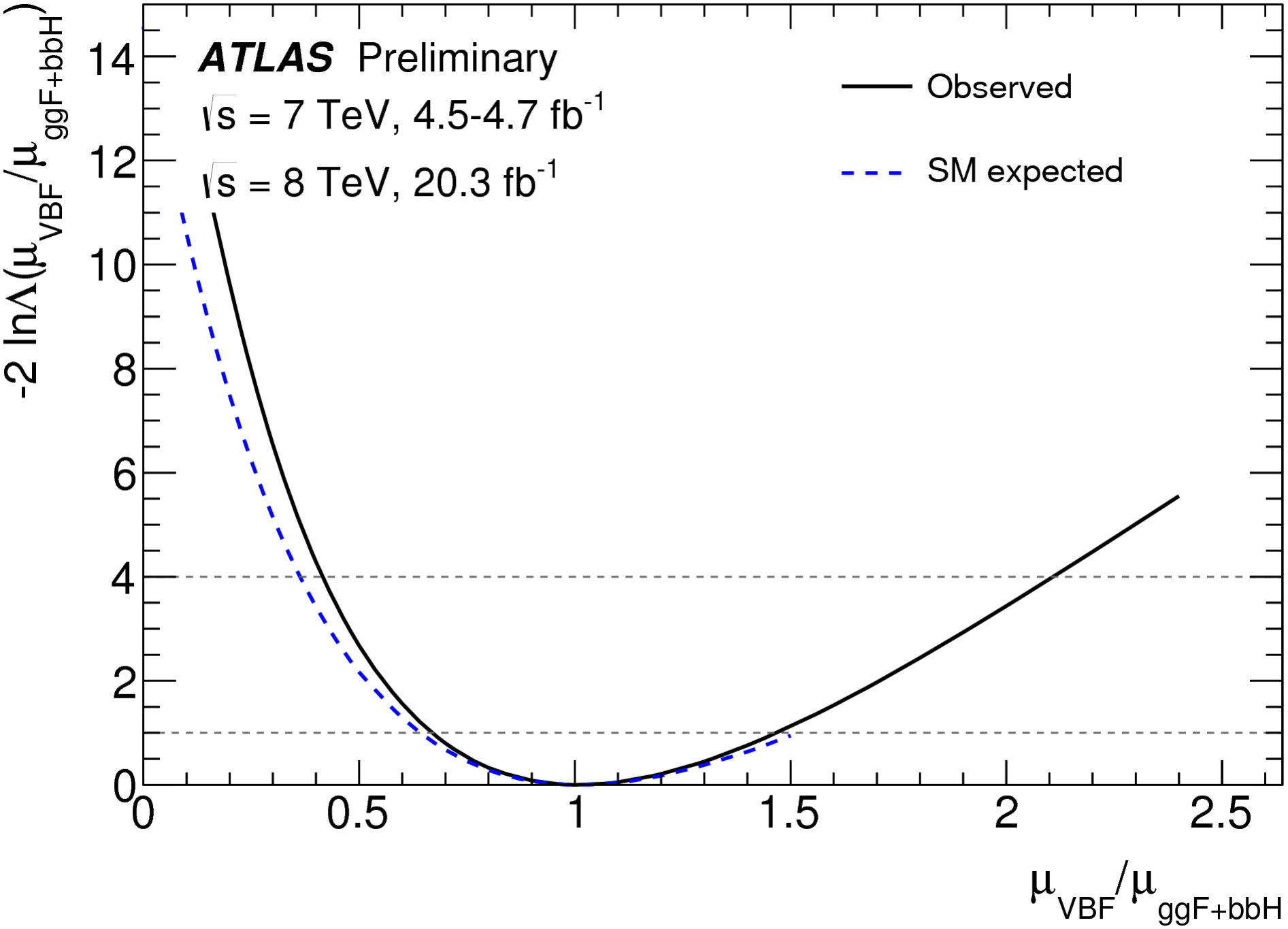}
\caption{
Likelihood for the ratio $\mu_{\mbox{VBF}}/\mu_{\mbox{ggF+ttH}}$ obtained by ATLAS for the combination of the 
$H\to\gamma\gamma$,  
$ZZ\to4\ell$ and $WW\to2\nu2\ell$ channels and $m_H$ = 125.5 GeV~\cite{ATLAS-CONF-2015-007}. 
}
\label{fig:lhc6}       
\end{figure} 
\mbox{ATLAS} and CMS have  also studied the relative contributions  from production mechanisms mediated by vector 
bosons (VBF and VH processes) and gluons (ggF and $ttH$ processes),
respectively. 
For example, Fig.~\ref{fig:lhc6} shows ATLAS results
constituting a $4.3\sigma$ evidence that part of the Higgs boson
production proceeds via VBF processes~\cite{ATLAS-CONF-2015-007}.


\subsubsection*{Couplings to other particles}
The Higgs boson couplings to other particles enter the observed signal strengths via both the Higgs production and 
decay.  Leaving other SM characteristics unchanged, in particular assuming the observed Higgs boson candidate to 
be a single, narrow, CP-even scalar state, its couplings are tested by introducing free parameters  $\kappa_X$ for 
each particle $X$, such that the SM predictions for production cross sections and decay widths are modified by 
a multiplicative factor  $\kappa^2_X$. This includes effective coupling modifiers $\kappa_{g}$, $\kappa_\gamma$ for 
the  loop-mediated interaction with gluons and photons. An additional scale factor modifies the total Higgs boson width 
by $\kappa^2_H$.
  
\begin{figure}[h!]
\centering
\includegraphics[width=1.00\columnwidth]{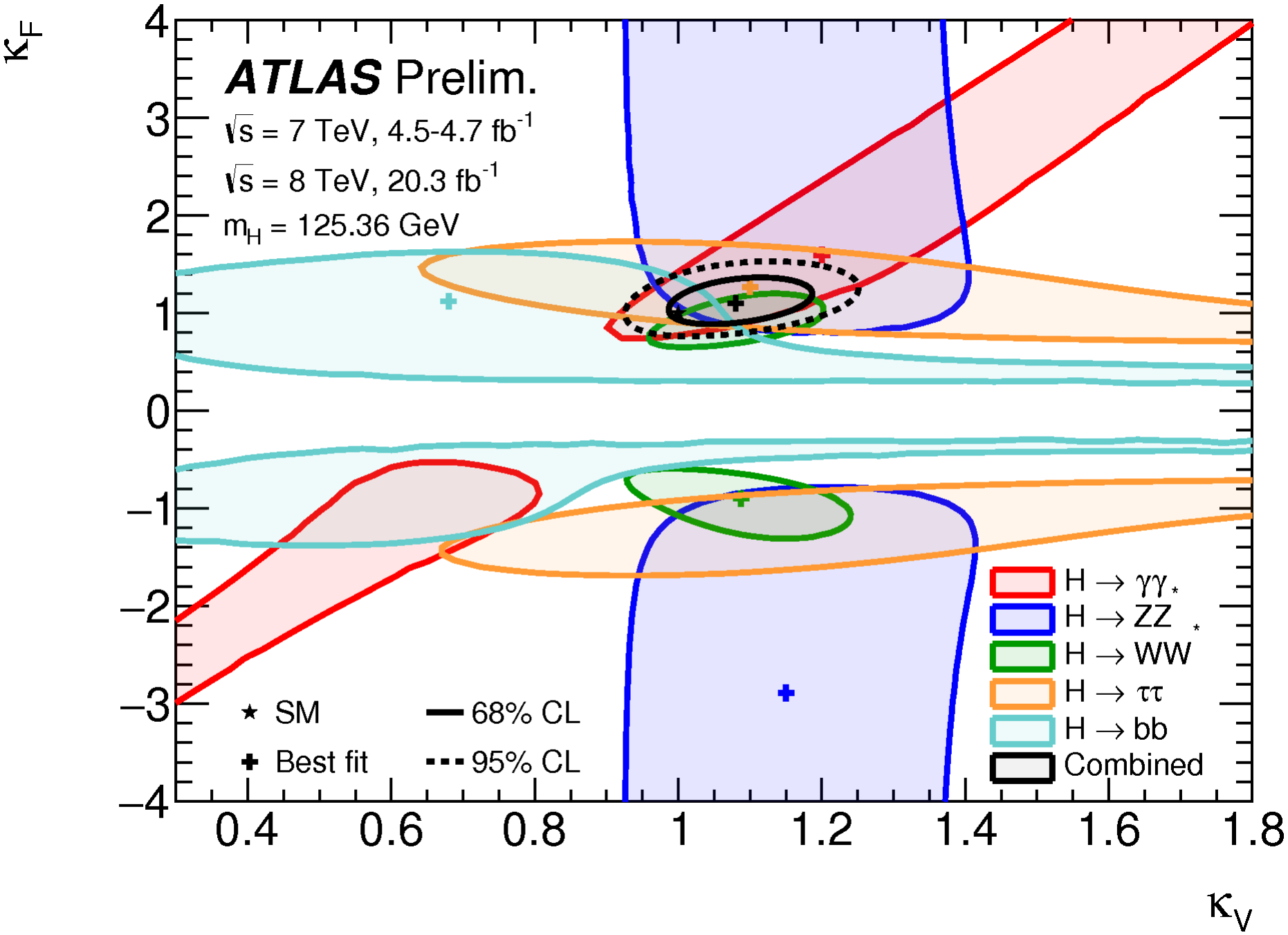}
\caption{
Preliminary ATLAS results of fits for a two-parameter benchmark model that probes different coupling strength scale factors common for fermions 
 ($\kappa_F$) and vector bosons ( $\kappa_V$), respectively, assuming only SM contributions to the total width.
 Shown  are 68\% and 95\% C.L .contours of the two-dimensional fit; overlaying the 68\% CL contours derived from the individual channels and their combination.
The best-fit result (x) and the SM expectation (+) are also indicated~\cite{ATLAS-CONF-2015-007}.
}
\label{fig:lhc7}        
\end{figure}

\begin{sloppypar}
Several different set of assumptions, detailed in~\cite{LHCHiggsCrossSectionWorkingGroup:2012nn} 
and~\cite{Heinemeyer:2013tqa}, 
form the basis of such coupling analyses. For example, a fit 
to the ATLAS data~\cite{ATLAS-CONF-2015-007} assuming common scale factors $\kappa_F$ and $\kappa_V$ for all fermions 
and bosons, respectively, yields the results depicted in Fig.~\ref{fig:lhc7}. 
\end{sloppypar}

%

\begin{figure}[t!]
\centering
\includegraphics[width=0.85\columnwidth]{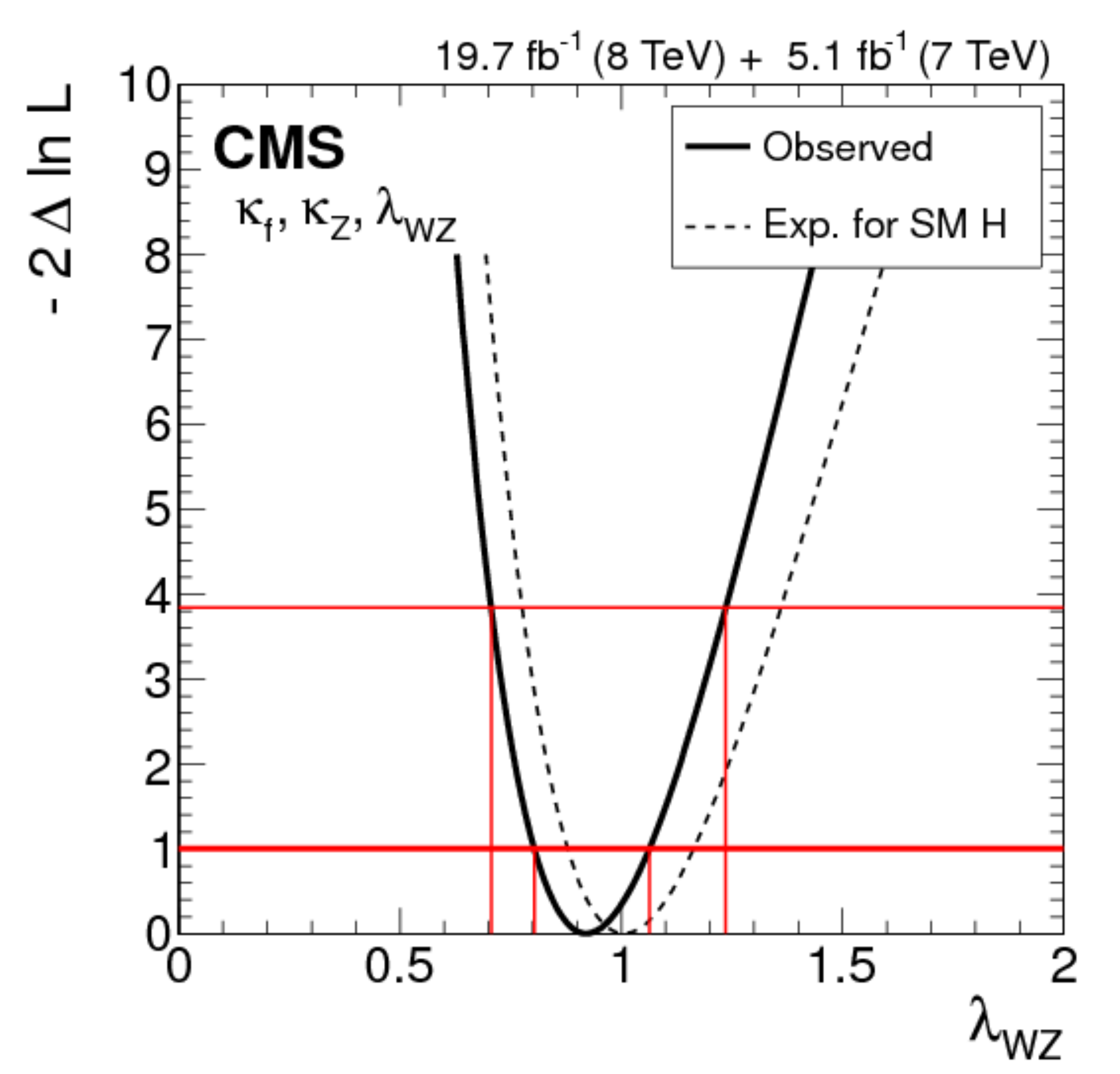}
\caption{
Test of custodial symmetry: CMS likelihood scan of the ratio $\lambda_{WZ}$, where SM coupling of the Higgs bosons to fermions 
are assumed~\cite{Khachatryan:2014jba}.
}
\label{fig:lhc8}        
\end{figure}

Within the SM, 
$\lambda_{WZ}=\kappa_W/\kappa_Z=1$
is implied by custodial symmetry. Agreement with this prediction is 
found by both CMS, see Fig.~\ref{fig:lhc8}, and ATLAS. Similar ratio analyses are performed for the couplings  to 
leptons and quarks ($\lambda_{lq}$) as well as to down and up-type fermions ($\lambda_{du}$).

\begin{figure}[h!]
\centering
\includegraphics[width=0.85\columnwidth]{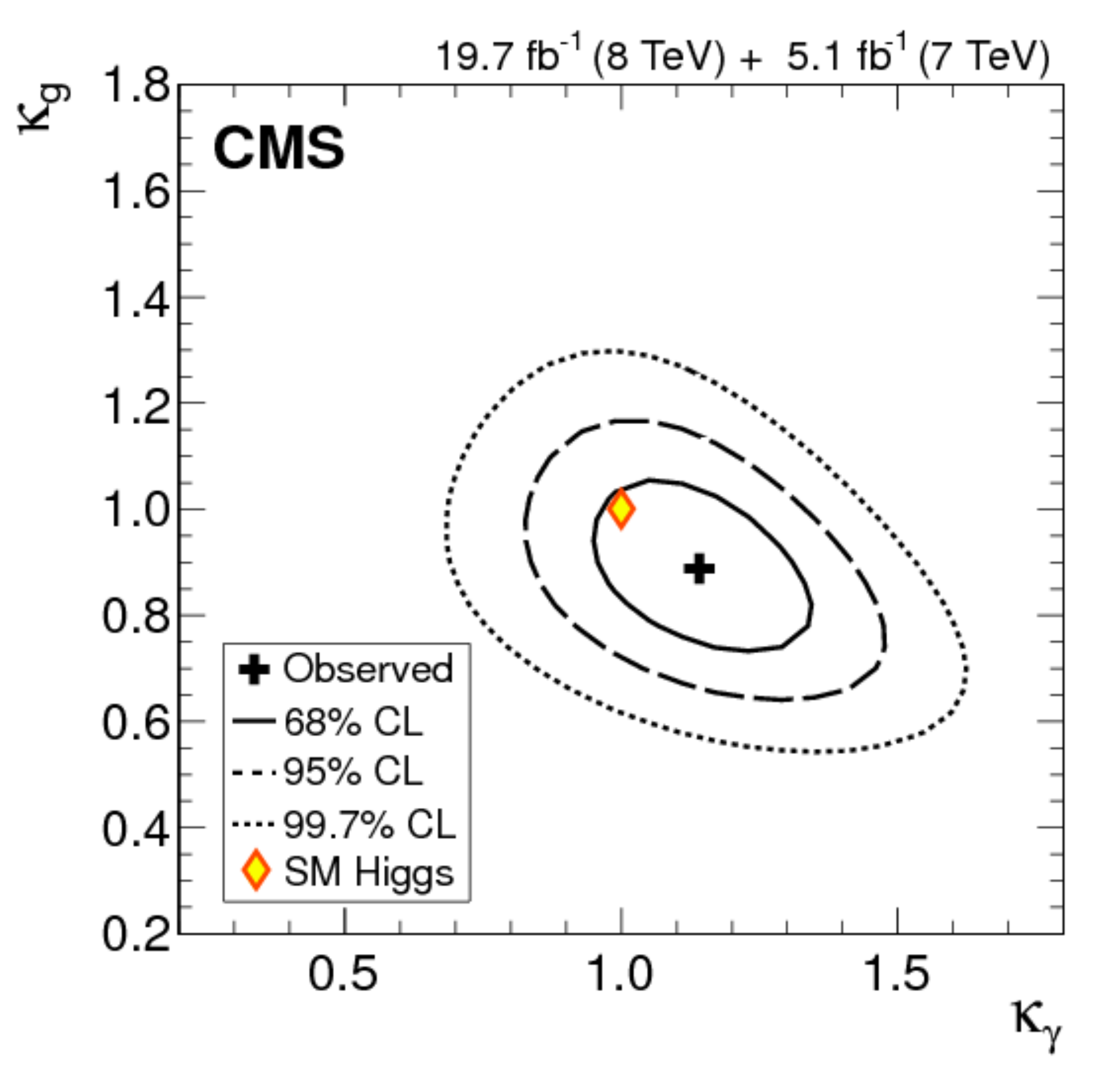}
\caption{
Constraining BSM contributions to particle loops: CMS 2d likelihood scan of gluon and photon coupling modifiers $\kappa_{g}$, 
$\kappa_{\gamma}$~\cite{Khachatryan:2014jba}.}
\label{fig:lhc9}        
\end{figure}

Within a scenario where all modifiers $\kappa$ except for $\kappa_{g}$ and $\kappa_{\gamma}$ are fixed to 1, 
contributions from beyond-SM particles to the loops that mediate the $ggH$ and $H\gamma\gamma$ interactions 
can be constrained; a corresponding CMS result~\cite{Khachatryan:2014jba} is shown in Fig.~\ref{fig:lhc9}.

\begin{figure}[h!]
\centering
\includegraphics[width=0.95\columnwidth]{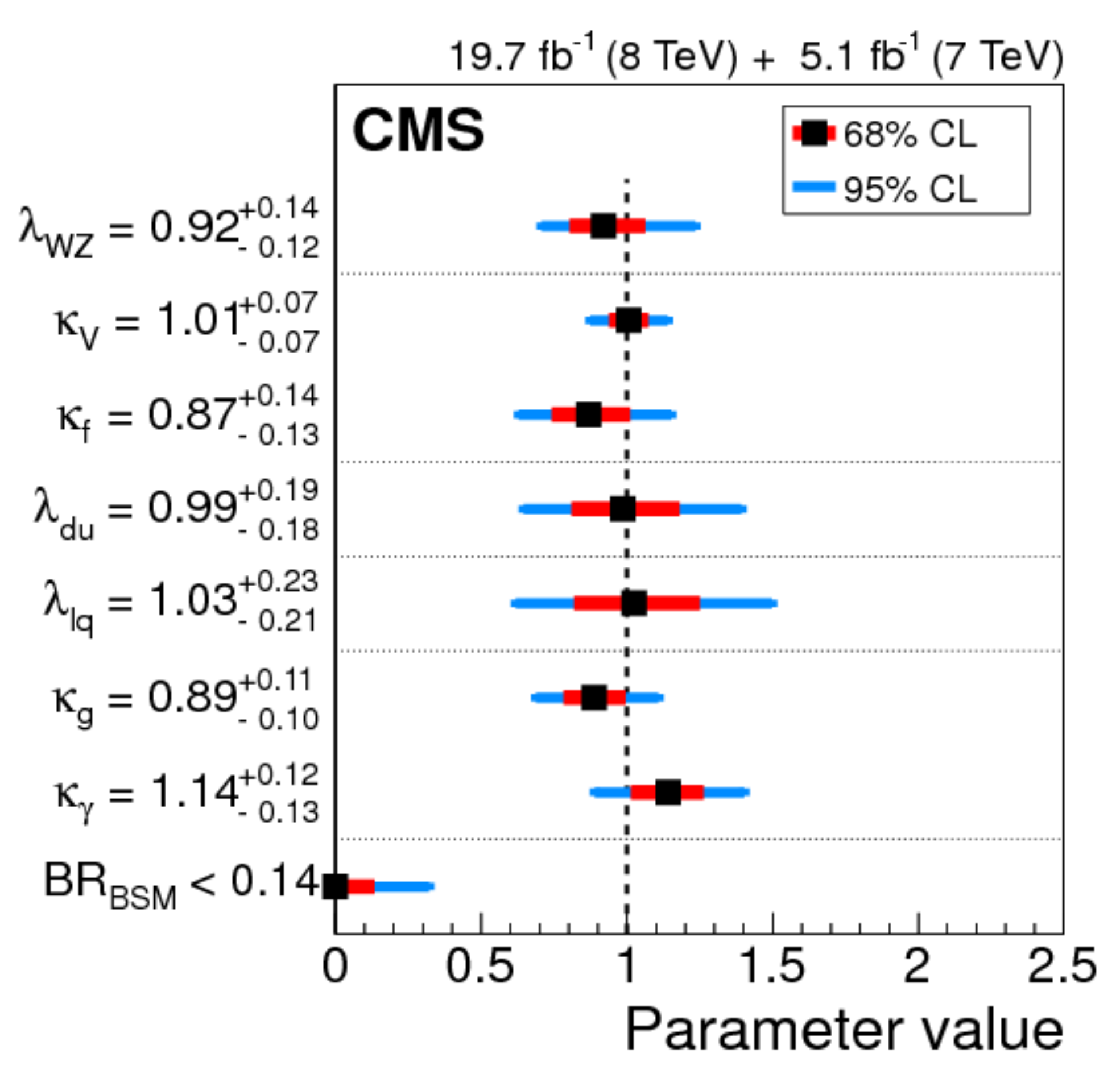}
\caption{
Summary plot of CMS likelihood scan results~\cite{Khachatryan:2014jba} 
for the different parameters of interest in benchmark models documented 
in~\cite{Heinemeyer:2013tqa}.
The inner bars represent the 68\% CL confidence intervals while the outer bars represent the 
95\% C.L.~confidence intervals. 
}
\label{fig:lhc10}        
\end{figure}

Summaries of 
CMS results~\cite{Khachatryan:2014jba}  from such coupling studies are presented in 
Fig.~\ref{fig:lhc10}. Within each of the specific sets of assumptions, consistency 
with the SM expectation is found.  Corresponding studies by CMS~\cite{Khachatryan:2014jba} yield the same 
conclusions. It should be noted, however, that this does not yet constitute a complete, unconstrained analysis of the Higgs boson couplings. 

\begin{figure}[h!]
\centering
\includegraphics[width=0.85\columnwidth]{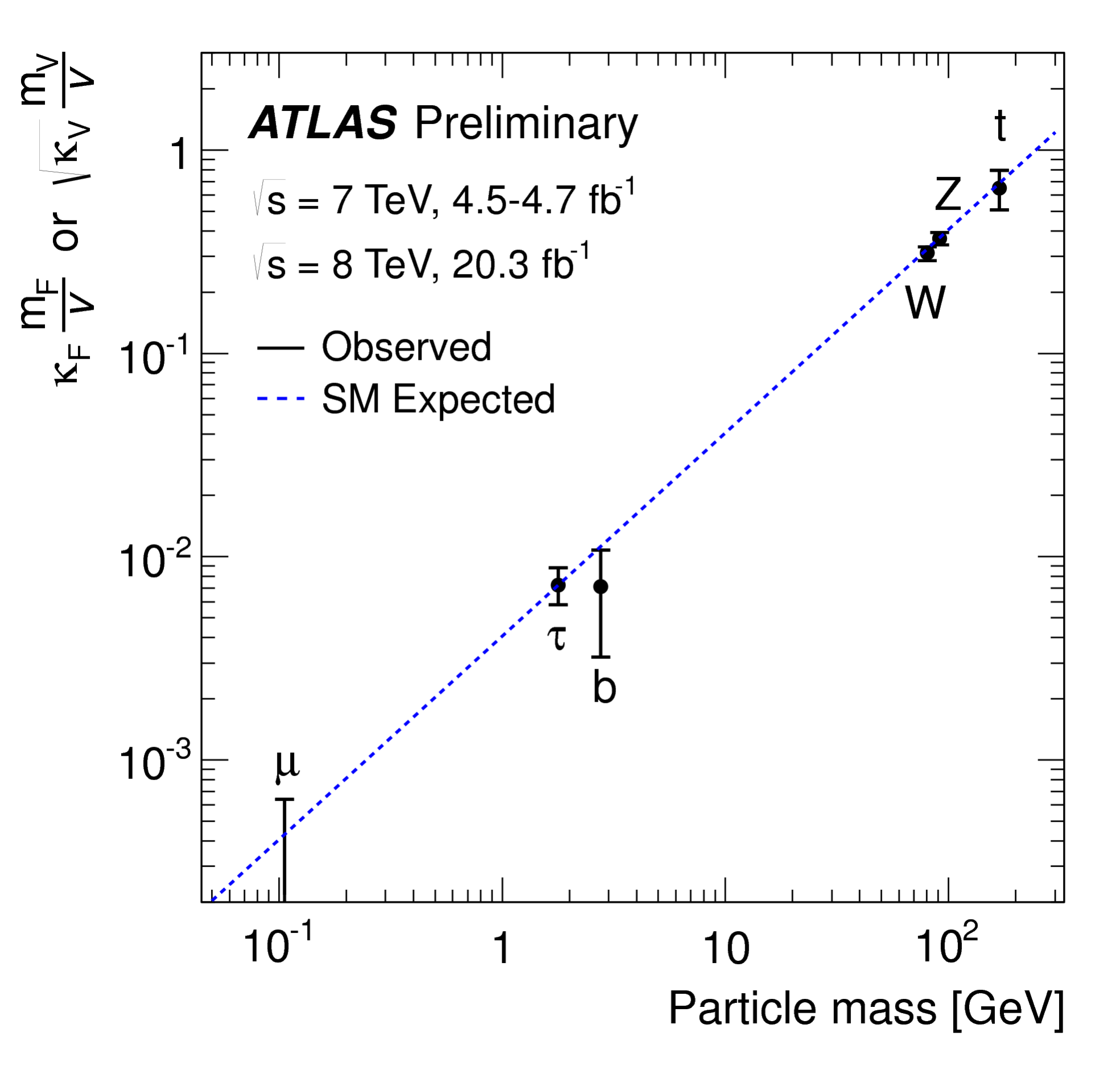}
\caption{
ATLAS summary of the fits for modifications of the SM Higgs boson couplings 
expressed as function of the particle mass. For the fermions, the values of the fitted Yukawa couplings for the $Hf\bar{f}$ vertex are shown, while for vector bosons the square-root of the coupling for the $HVV$ vertex divided by twice the vacuum expectation  value of the Higgs boson field~\cite{ATLAS-CONF-2015-007}.
}
\label{fig:lhc11}        
\end{figure}

For the 
fit assuming that loop-induced couplings follow the SM structure as in~\cite{Heinemeyer:2013tqa} without any BSM contributions to Higgs boson decays or particle loops, ATLAS, see Fig.~\ref{fig:lhc11}, and CMS also demonstrate
that the results follow the predicted relationship between Higgs boson couplings and the SM particle masses.


\subsubsection*{Mass}
\begin{figure}[h!]
\centering
\includegraphics[width=0.8\columnwidth]{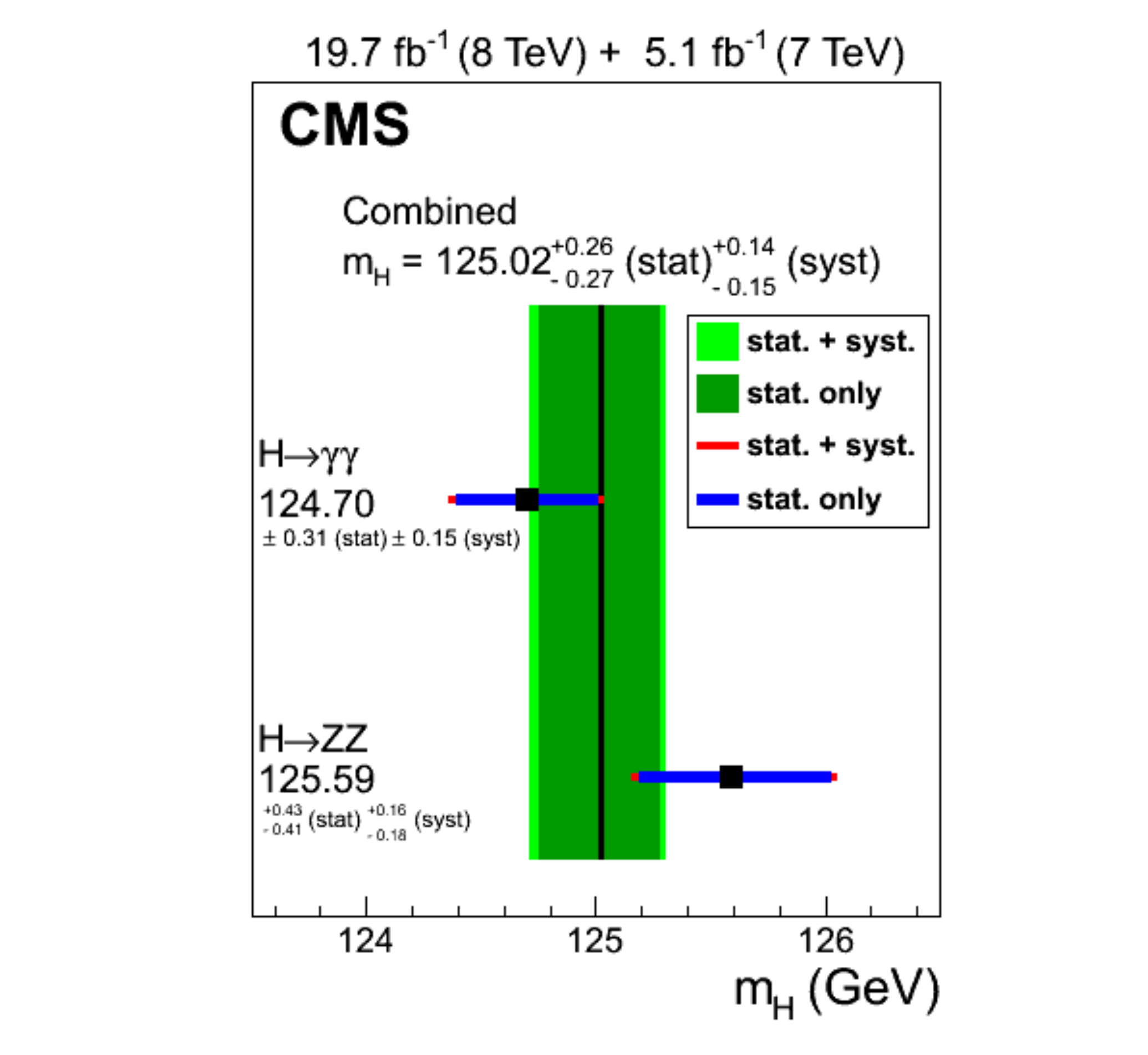}	
\caption{
CMS mass measurements~\cite{Khachatryan:2014jba} in the $\gamma\gamma$ and $ZZ\to4\ell$ final states and their combinations. The vertical band shows the combined uncertainty. The horizontal bars indicate the $\pm1$ standard deviation uncertainties for the individual channels.
%
}
\label{fig:lhc12}        
\end{figure}

\begin{sloppypar}
Current measurements of the Higgs boson mass are based on the two
high-resolution decay channels
\mbox{$H\to\gamma\gamma$} and $H\to ZZ\to 4\ell$. 
Based on fits to the invariant diphoton and four-lepton mass
spectra, \mbox{ATLAS} measures~\cite{Aad:2014aba} $m_H=125.98\pm
0.42\mathrm{(stat)}\pm 0.28\mathrm{(sys)}$ and $m_H=124.51\pm
0.52\mathrm{(stat)}\pm 0.06\mathrm{(sys)}$, respectively. A
combination of the two results, which are consistent within $2.0$
standard deviations, yields
$m_H=125.36\pm0.37\mathrm{(stat)}\pm 0.18\mathrm{(sys)}.$
%
An analysis~\cite{Khachatryan:2014jba} of the same decays by CMS find consistency between the two channels 
at 1.6$\sigma$, see Fig.~\ref{fig:lhc12}. The combined result 
$m_H=125.02^{+0.26}_{-0.27}\mathrm{(stat)}^{+0.14}_{-0.15}\mathrm{(sys)}$
agrees well with the corresponding ATLAS measurement. 

A preliminary combination~\cite{LHCmass} of both experiments gives a measurement of the Higgs boson mass of
$$m_H=125.09 \pm0.21\mathrm{(stat)}\pm 0.11\mathrm{(sys)},$$
with a relative uncertainty of $0.2\%$.
\end{sloppypar} 

Other decay channels currently do not provide any significant contributions to the overall mass precision but they can still be used for consistency tests. For example, CMS obtains $m_H=128^{+7}_{-5}$~GeV and $m_H=122\pm7$~GeV from the analysis of $WW$~\cite{Chatrchyan:2013iaa}  and $\tau\tau$~\cite{Chatrchyan:2014nva} final states  
states, respectively.


\subsubsection*{Width}
Information on the decay width 
of the Higgs boson obtained from the above mass measurements is limited by the experimental resolution to
about 2~GeV, whereas the SM prediction for $\Gamma_H$ is about 4 MeV.

\begin{figure}[t!]
\centering
\includegraphics[width=0.7\columnwidth]{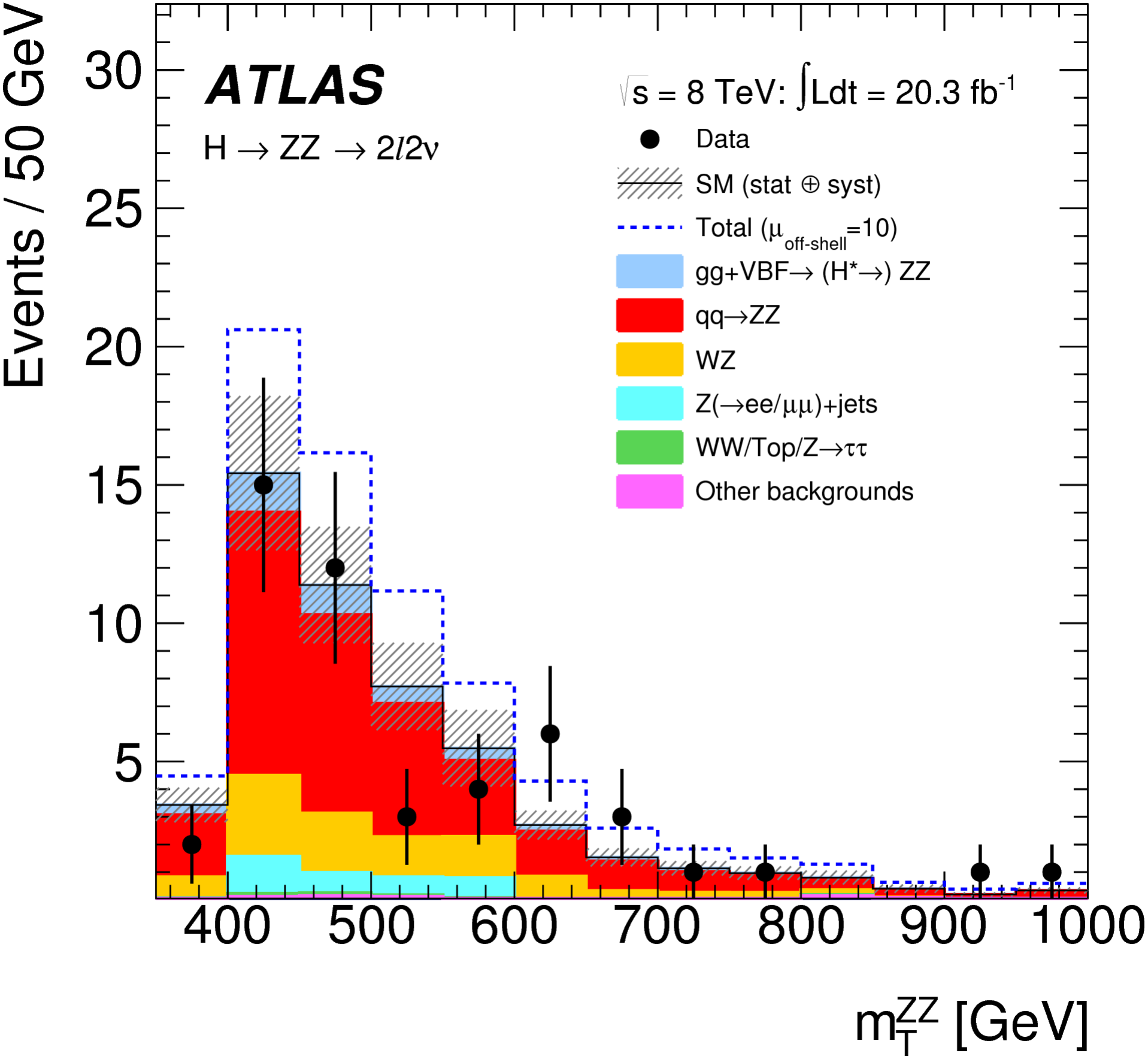}
\caption{
Observed transverse mass distributions for the ATLAS $ZZ\to2\ell2\nu$  analysis~\cite{Aad:2015xua} in 
the signal region compared to the expected contributions from ggF and
VBF Higgs production with the decay $H^*\to ZZ$ SM and with
$\mu_{\mathrm{off-shell}}=10$ 
(dashed) in the $2e2\nu$  channel. 
A relative $gg\to ZZ$ background 
K-factor of 1 is assumed. 
}
\label{fig:lhc13}        
\end{figure}

Analyses of 
$ZZ$ and $WW$ events in the mass range above the 2$m_{Z,W}$ threshold provide an 
alternative approach~\cite{Kauer:2012hd,Caola:2013yja}, which was first pursued 
by CMS~\cite{Khachatryan:2014iha} based on the $ZZ\to4\ell$  and $ZZ\to2\ell2\nu$ 
channels; a later ATLAS analysis~\cite{Aad:2015xua} included also the $WW\to e\nu\mu\nu$ 
final state.
The studied distributions vary between experiments 
and channels; for example, Fig.~\ref{fig:lhc13} 
\begin{figure}
\centering
\includegraphics[width=0.7\columnwidth]{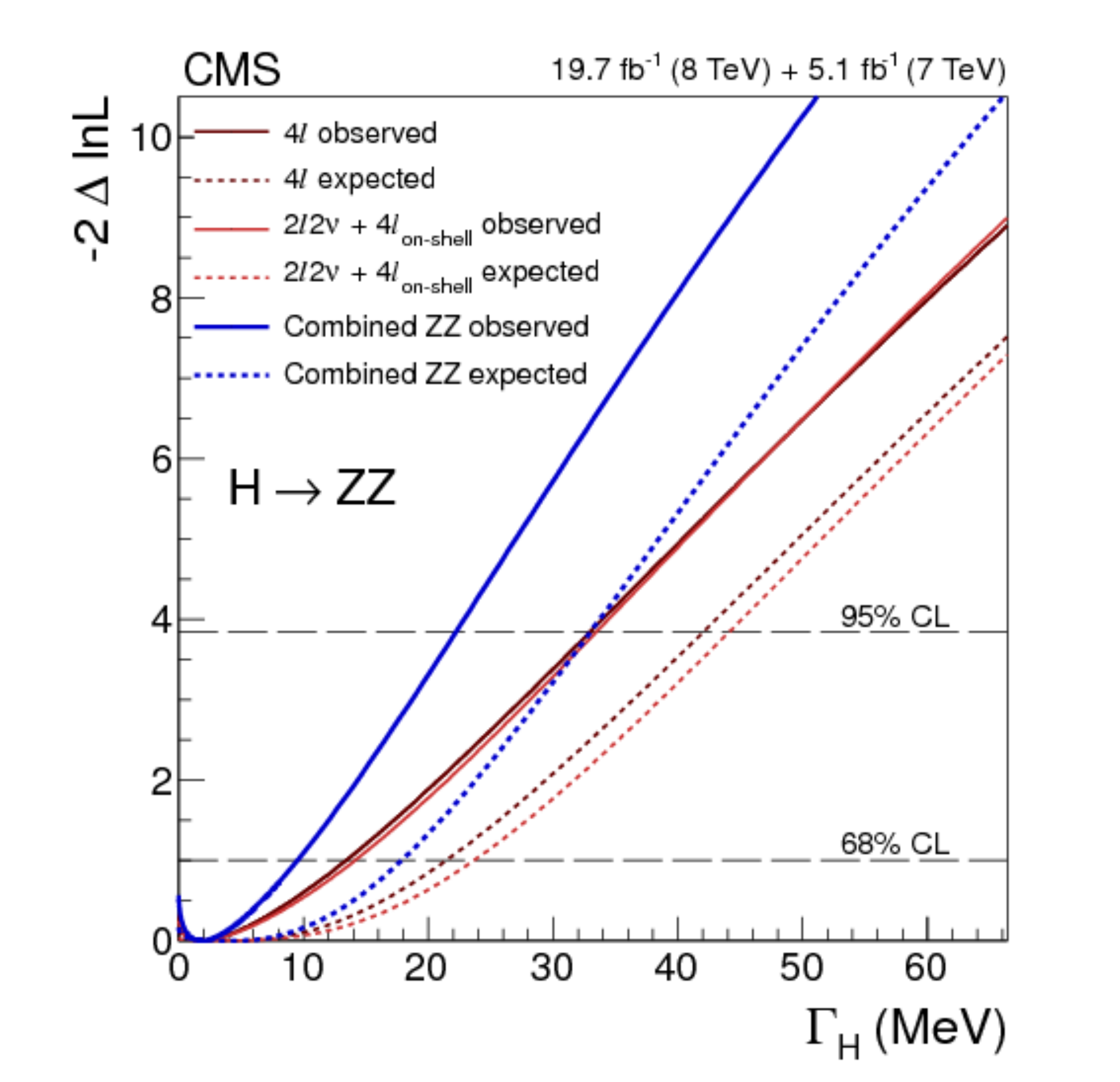}
\caption{
CMS likelihood scan versus $\Gamma_H$. Different colours refer to: combination of $4\ell$ low-mass and 
high-mass (ochre), combination 
of $4\ell$  low-mass and $2\ell2\nu$ high-mass 
and combination of $4\ell$ low-mass and both channels at high-mass (blue). Solid and dashed  lines represent 
observed and expected limits, respectively~\cite{Khachatryan:2014iha}.
}
\label{fig:lhc14}        
\end{figure}
shows the high-mass \mbox{$ZZ\to2\ell2\nu$} transverse mass distribution observed by ATLAS with the expected 
background contributions and the predicted signal for different assumptions for the off-shell $H\to ZZ$ signal 
strength $\mu_{\mathrm{off-shell}}$. The resulting constraints on $\mu_{\mathrm{off-shell}}$, together 
with the on-shell $H\to ZZ\to 4\ell$ $\mu_{\mathrm{on-shell}}$ measurement, can be interpreted as a limit on the 
Higgs boson width if the relevant off-shell and on-shell Higgs couplings are assumed to be equal.\footnote{
However, the relation between the off-shell and on-shell couplings can be severely affected by new-physics 
contributions, in particular via threshold effects.  In fact, such effects may be needed to give rise to a Higgs-boson 
width that differs from the one of the SM by the currently probed amount, see also the discussion in~\cite{Englert:2014aca}. 
In this sense, these analyses currently provide a consistency test of the SM rather than model-independent bounds 
on the total width.
 }

Combining $ZZ$ and $WW$ channels, ATLAS find an observed (expected)  95\% C.L. limit of 
$$5.1(6.7)<\mu_{\mathrm{off-shell}}<8.6(11.0)$$ when varying the unknown K-factor ratio between the $gg\to ZZ$ 
continuum background and the $gg\to H^*\to ZZ$ signal between 0.5 and 2.0. This translates into 
$$4.5(6.5)<\Gamma_H/\Gamma^{\rm SM}_H<7.5(11.2)$$ if identical on-shell and off-shell couplings are assumed. 

Fig.~\ref{fig:lhc14} illustrates the results of a corresponding CMS analysis, yielding observed (expected) 
95\% C.L. limit of $\Gamma_H/\Gamma^{\rm SM}_H<22(33)$~MeV or
$\Gamma_H/\Gamma^{\rm SM}_H<5.4(8.0)$.


\subsubsection*{Spin and parity}
Within the SM, the Higgs boson is a spin-0, $CP$-even particle. Since the decay kinematics 
depend on these quantum numbers, the $J^P=0^+$ nature of the SM Higgs boson can be used 
as constraint to increase the sensitivity of the SM analyses. After dropping such assumptions, 
however, these analyses can also be used to test against alternative spin-parity hypotheses. 
These studies are currently based on one or several of the bosonic decays modes discussed above: 
$H\to\gamma\gamma$, \mbox{$H\to ZZ\to4\ell$}, and $H\to WW\to 2\ell2\nu$.
\begin{figure}[b!]
\centering
\includegraphics[width=0.45\columnwidth]{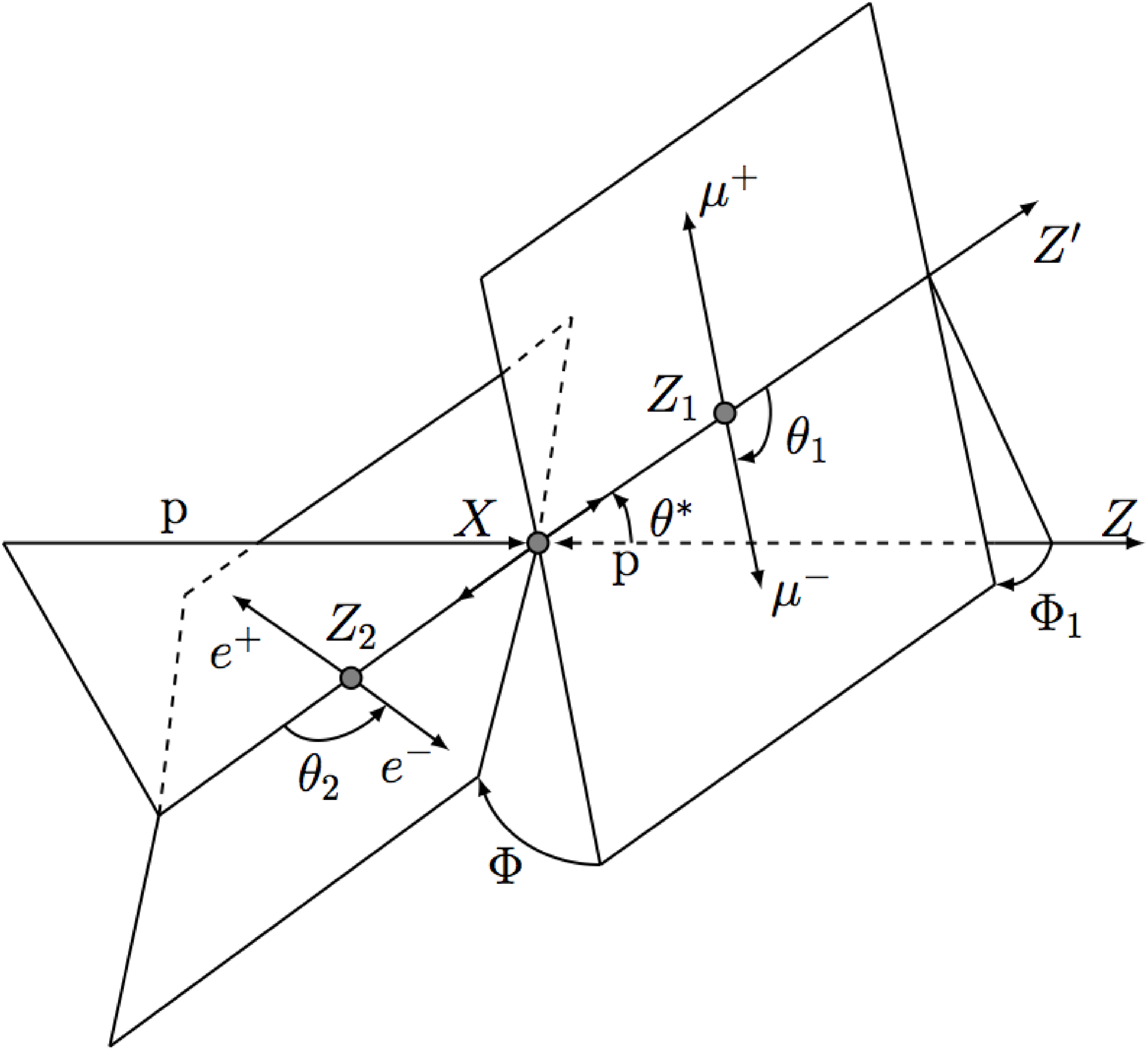}
\includegraphics[width=0.7\columnwidth]{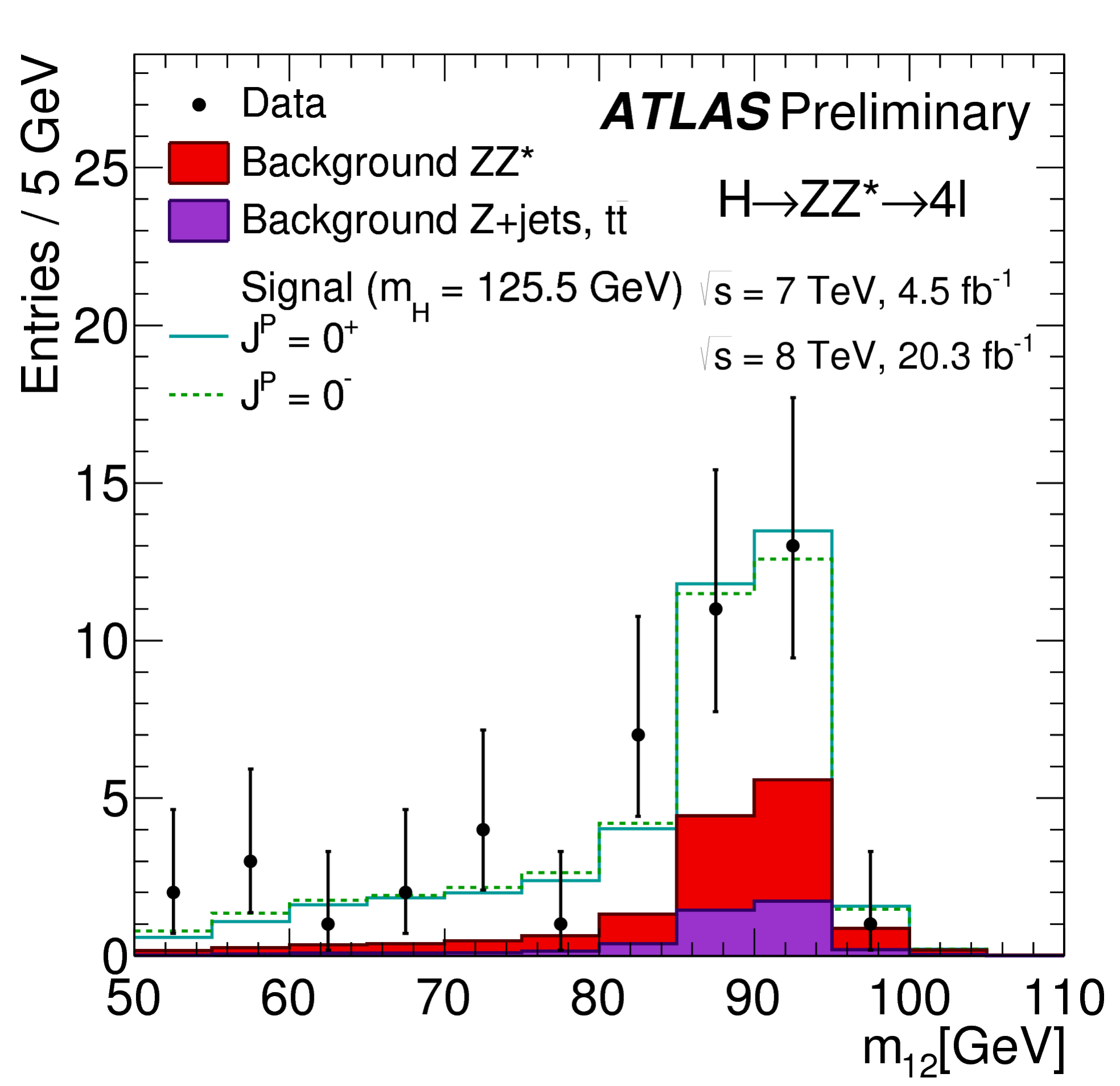}
\caption{
Top: Final state observables sensitive to the spin and parity of the decaying resonance in $ZZ^*\to4\ell$ final states.
Bottom: $\cos\theta_1$ distribution for ATLAS data (point with errors), 
the backgrounds (filled histograms) and several spin hypotheses (SM solid line and alternatives dashed 
lines)~\cite{ATLAS-CONF-2015-008}. 
}
\label{fig:lhc15}        
\end{figure}

In the $H\to\gamma\gamma$ analysis, the $J^P=0^+$ and $J^P=2^+$ hypothesis can be distinguished 
via the Collins-Soper angle $\theta^*$ of the photon system. Since there is a large 
non-resonant diphoton background, the spin information is extracted from a simultaneous fit to the 
$|\cos\theta^*|$ and $m_{\gamma\gamma}$ distributions.
The charged-lepton kinematics and the missing transverse energy in $H\to WW\to e\nu_e\mu\nu_\mu$ 
candidate decays are combined in multivariate analyses to compare the data to the SM and three alternative 
($J^P=2^+,1^+,1^-$) hypotheses.
The $H\to ZZ\to 4\ell$ analysis combines a high signal-to-background ratio with a complete final state reconstruction.
This makes it possible to perform a full angular analysis, c.f.~Fig.~\ref{fig:lhc15}, albeit currently still with a rather limited 
number of events. Here, in addition to the spin-parity scenarios discussed above, also the  $J^P=0^-$ hypothesis is tested.

\begin{figure}[t!]
\centering
\includegraphics[width=1.0\columnwidth]{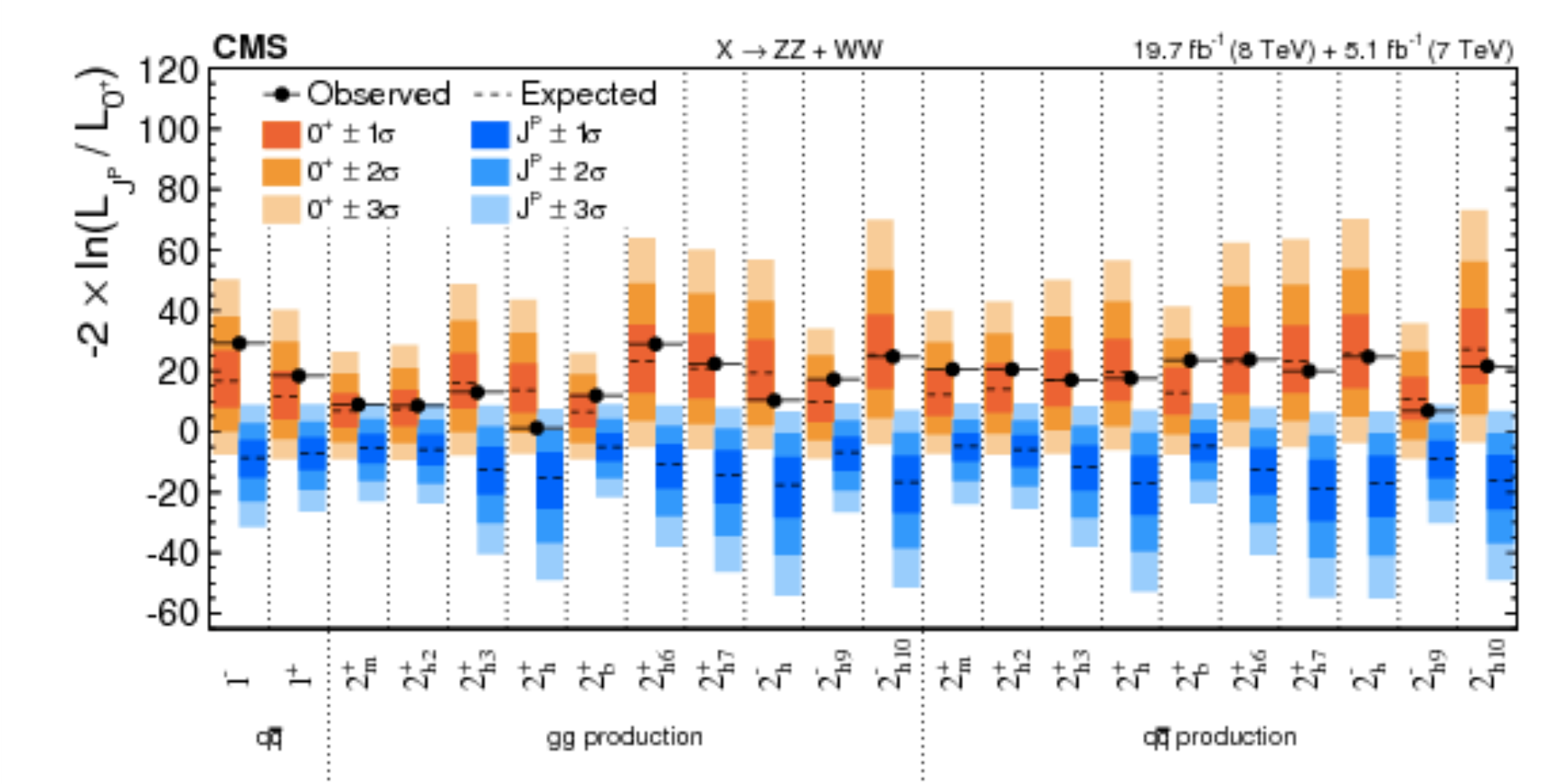}
\caption{
Distributions of the test statistic $q=-2\ln(\mathcal{L}_{J^P}/\mathcal{L}_{0^+})$ for the spin-one and spin-two JP models 
tested against the SM Higgs boson hypothesis in the combined $X\to ZZ$ and $WW$ 
analyses~\cite{Khachatryan:2014kca}. The expected median and the 68.3\%, 95.4\%, and 99.7\% C.L.~regions 
for the SM Higgs boson (orange, the left for each model) and for the alternative $J^P$ hypotheses (blue, right) 
are shown. The observed $q$ values are indicated by the black dots.
}
\label{fig:lhc16}       
\end{figure}

Including the spin-1 hypotheses in the analyses of the decays into vector bosons provides a test independent 
of the $H\to\gamma\gamma$ channel, where  $J = 1$ is excluded by the Landau-Yang theorem, and implies 
the assumptions that the signals observed in the two-photon and $VV$ final states are not originating from a 
single resonance. A representative sample of spin-two alternatives to SM hypothesis is considered, also including 
different assumptions concerning the dominant production mechanisms.

For example, Fig.~\ref{fig:lhc16} shows the results obtained from CMS analyses of the $H\to ZZ\to4\ell$ and  $H\to WW\to2\ell2\nu$ channels~\cite{Khachatryan:2014kca}. Agreement with the SM ($J^P=0^+$) within $1.5\sigma$ and inconsistency with alternative hypotheses at a level of at least $3\sigma$ is found.  
%
%
Corresponding ATLAS studies~\cite{Aad:2013xqa,Aad:2015rwa,ATLAS-CONF-2015-008} yield similar conclusions.

\subsubsection*{Other analyses}
\begin{sloppypar}
In addition to the results discussed above, a number of other analyses
have been performed, making use of the increase in the available data
since the first Higgs boson discovery in different ways. These
include, for example, measurements of differential distributions in
$H\to\gamma\gamma$~\cite{Aad:2014lwa} and $H\to ZZ$~\cite{Aad:2014tca}
events and searches for rarer decays, such as
$H\to\mu\mu$~\cite{Aad:2014xva,Khachatryan:2014aep}, $H\to
ee$~\cite{Khachatryan:2014aep}, $H\to
Z\gamma$~\cite{Aad:2014fia,Chatrchyan:2013vaa}, decays to heavy
quarkonia states and a photon~\cite{Aad:2015sda}, and invisible
modes~\cite{Aad:2014iia,Chatrchyan:2014tja}.  These searches are not
expected to be sensitive to a SM Higgs boson signal based on the
currently available data and thus are as of now mainly relevant for
the preparation for the larger datasets expected from LHC Run2
and/or for using Higgs boson events as a probe for effects beyond the SM. 
\end{sloppypar}

Additional production modes are searched for as well. Here,
top-associated production is of particular interest because it would
provide direct access to the top-Higgs Yukawa coupling.  While the
results from recent
analyses~\cite{Aad:2014lma,Khachatryan:2014qaa,Khachatryan:2015ila,Aad:2015gra}
of these complex final states do not quite establish a significant
signal yet, they demonstrate a lot of promise for LHC Run2, where, in
addition to larger datasets, an improved signal-to-background ratio is
expected due to the increased collision energy.


\subsubsection{Future Projections}
Studies of longer-term Higgs physics prospects currently focus on the scenario of an LHC upgraded during 
a shutdown starting in 2022  to run at a 
levelled luminosity of $5\times10^{34}$ cm$^{-2}$s$^{-1}$,  
resulting in a typical 
average of 140 pile-up events per bunch crossing. This so-called HL-LHC is expected to deliver a total integrated 
luminosity of 3000 fb$^{-1}$ to be compared to a total of 300~fb$^{-1}$ expected by the year 2022.

The following summary of SM Higgs boson analysis prospects for such large datasets is based on preliminary results 
by the ATLAS and CMS collaborations documented in \cite{atlpub14016} and \cite{cmspas13002}, respectively.
While the prospects for measurements of other Higgs boson properties are being studied as well, the discussion below 
focusses on projections concerning signal strength measurements and coupling analyses.

\subsubsection*{Underlying assumptions}
CMS extrapolates the results of current Run1 measurements to $\sqrt{s}=14$~TeV data samples corresponding 
to 300 fb$^{-1}$ and 3000 fb$^{-1}$ assuming 
that the upgraded detector and trigger systems will provide 
the same performance in the high-luminosity environment as the current experiments during 2012, i.e.~the  
signal and background event yields are scaled according to the increased luminosities and cross sections. 
Results based on two different assumptions concerning the systematic uncertainties are obtained: a first scenario 
assumes no changes with respect to 2012, while in a second scenario theoretical uncertainties are reduced by a factor 
of two and other uncertainties scaled according to the square root of the integrated luminosities.

%
\begin{figure}[b!]
\centering
\includegraphics[width=0.7\columnwidth]{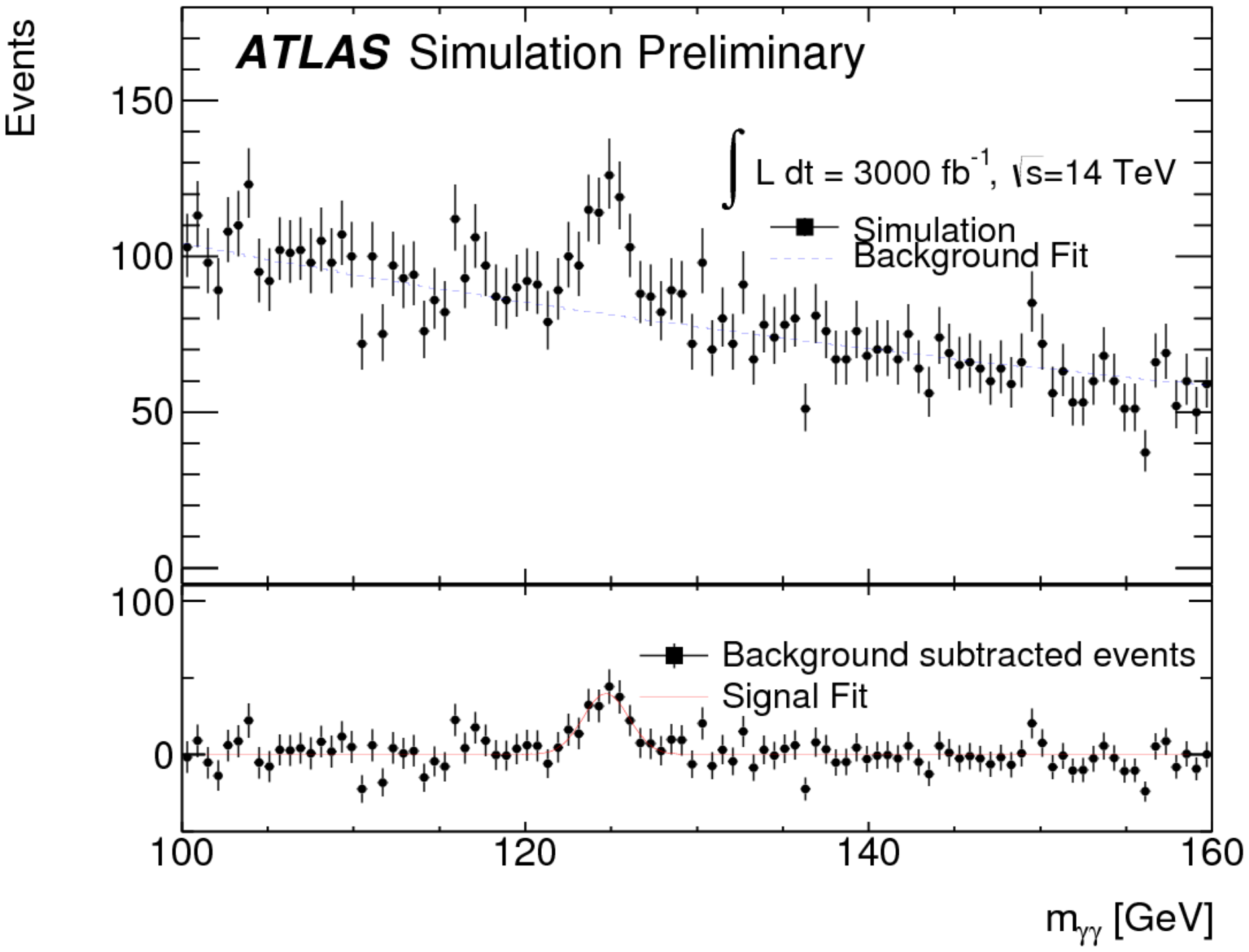}
\includegraphics[width=0.7\columnwidth]{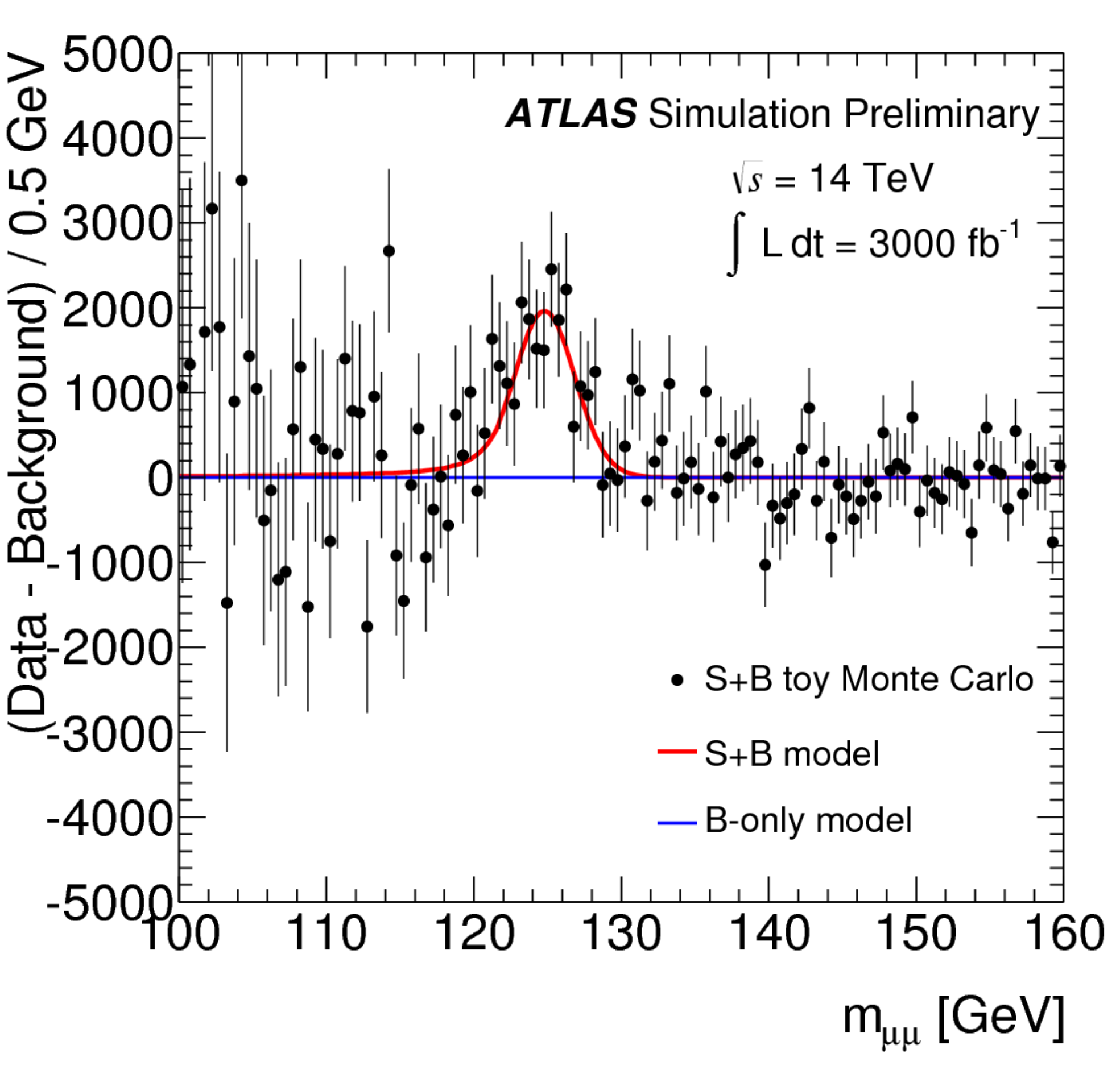}
\caption{Projected (a) diphoton mass distribution for the SM Higgs boson signal and background processes after 
VBF selection and (b) background-subtracted dimuon mass distribution based on ATLAS simulations assuming 
an integrated luminosity of 3000~fb$^{-1}$~\cite{atlpub13014}.
\label{fig:lhc17}  }     
\end{figure}

ATLAS investigates the physics prospects for 14~TeV datasets corresponding to the same integrated luminosities as 
CMS but here the expected detector performance is parameterised based on efficiency and resolution modifications 
at the detector object level. These are  obtained from full simulations corresponding to current and/or upgraded 
ATLAS detector components assuming values for the number of pile-up events per bunch crossing ranging from 40 to 
200. The theoretical uncertainties are assumed to be similar to those used in recent analysis of the Run1 data but some 
of the experimental systematic uncertainties are re-evaluated taking into account, e.g., the expected improved  
background estimates due to an increased number of events in data control regions.

\subsubsection*{Signal strength}
Both experiments study expectations for the experimentally most
significant SM Higgs boson decay modes
$H\to\gamma \gamma$, $H\to ZZ\to4\ell$, $H\to WW\to2\ell2\nu$,
$H\to\tau\tau$, and $H\to bb$ but also include analyses of additional
sub-modes as well as rare decays to $Z\gamma$, $\mu\mu$, and invisible
final states. Fig.~\ref{fig:lhc17} shows two examples for expected
mass signals based on ATLAS simulations of SM Higgs boson decays to
two photons (after a VBF selection) and two muons, respectively.
%
\begin{table}[b!]
\centering
\caption{Relative uncertainty on the determination of the signal strength expected for the CMS experiment for integrated 
luminosities of 300~fb$^{-1}$ and 3000~fb$^{-1}$~\cite{cmspas13002} and the two uncertainty scenarios described in the text.
}
\label{tab:lhc18}
\footnotesize
\vspace{0.2em}
\begin{tabular}{|c||cc|cc|}
\hline \hline
$\mathcal{L}$ &   \multicolumn{2}{|c|}{300~fb$^{-1}$}   & \multicolumn{2}{|c|}{3000~fb$^{-1}$}   \\
 Scenario & 2&1 & 2& 1\\\hline
$\gamma\gamma$ &6\%& 12\% & 4\% & 8\% \\
$WW$  &6\%& 11\% & 4\% & 7\% \\
$ZZ$  &7\%& 11\% & 4\% & 7\% \\
$bb$  &11\%& 14\% & 5\% & 7\% \\
$\tau\tau$  &8\%& 14\% & 5\% & 8\% \\
$Z\gamma$  &62\%& 62\% & 20\% & 24\% \\
$\mu\mu$  &40\%& 42\% & 14\% & 20\% \\
\hline \hline
\end{tabular}
\end{table}
%

%
\begin{figure}[b!]
\centering
\includegraphics[width=0.7\columnwidth]{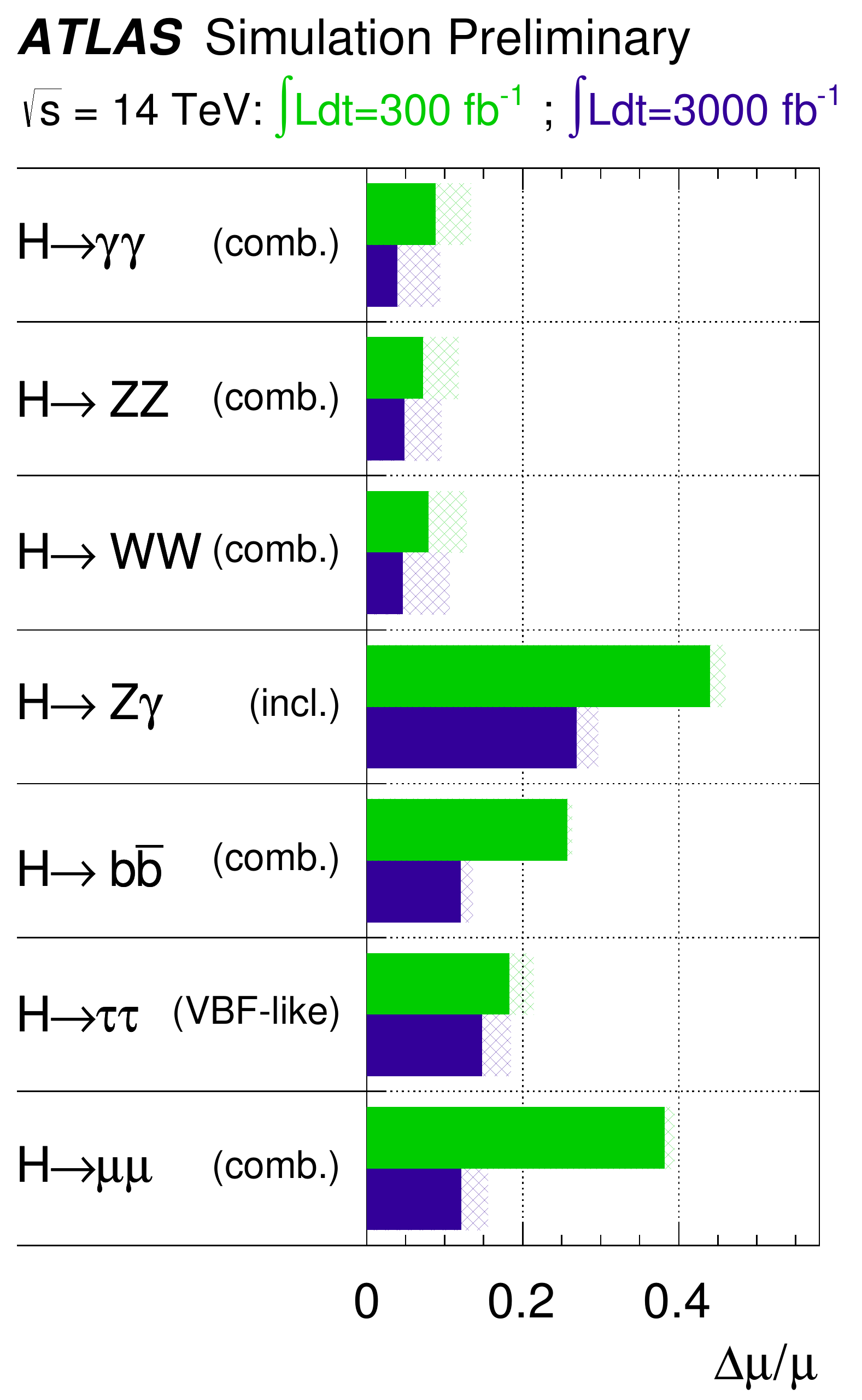}
\caption{
Relative uncertainty on the signal strength determination expected for the ATLAS experiment~\cite{atlpub14016}.
assuming a SM Higgs boson with a mass of 125 GeV and  300~fb$^{-1}$ and 3000~fb$^{-1}$  of 14 TeV data. The
uncertainty pertains to the number of events passing the experimental selection, not to the particular Higgs boson 
process targeted. The hashed areas indicate the increase of the estimated error due to current theory systematic
uncertainties. 
}
\label{fig:lhc19}       
\end{figure}
%
The expected relative uncertainties on the signal strength for CMS and ATLAS are 
shown in Table~\ref{tab:lhc18} and Fig.~\ref{fig:lhc19}, indicating that for the most sensitive channels, experimental uncertainty 
around 5\% should be reachable with 3000~fb$^{-1}$. 
Combining different final states and again assuming SM branching ratios, projections on 
the sensitivity to individual Higgs boson production can be obtained; the corresponding ATLAS results are summarised in 
Table~\ref{tab:lhc20}. For 3000~fb$^{-1}$, the expected experimental uncertainties on the signal strength range from about 
4\% for the dominant ggF production to about 10\% for the rare $t\bar{t}H$ production mode. Fig.~\ref{fig:lhc19} and 
Table~\ref{tab:lhc20} also indicate the contribution of current theoretical uncertainties, showing that reducing them further 
will be important to fully exploit the HL-LHC for Higgs boson precision studies.

%
\begin{table}[h!]
\centering
\caption{
Relative uncertainty on the signal strength projected by ATLAS for different production modes using the combination of Higgs final states 
based on integrated luminosities of 300~fb$^{-1}$ and 3000~fb$^{-1}$~\cite{atlpub14016}, assuming a SM Higgs boson with a mass of 
125~GeV and branching ratios as in the SM. 
}
\label{tab:lhc20}
\footnotesize
\vspace{0.2em}
\begin{tabular}{|c||cc|cc|}
\hline \hline
$\mathcal{L}$ &   \multicolumn{2}{|c|}{300~fb$^{-1}$}   & \multicolumn{2}{|c|}{3000~fb$^{-1}$}   \\
 Uncertainties & All &No theory & All & No theory \\\hline
$gg\to H$ &12\%& 6\% & 11\% & 4\% \\
VBF &18\%& 15\% & 15\% & 9\% \\
$WH$ &41\%& 41\% & 18\% & 18\% \\
$qqZH$ &80\%& 79\% & 28\% & 27\% \\
$ggZH$ &371\%& 362\% & 147\% & 138\% \\
$ttH$ &32\%& 30\% & 16\% & 10\% \\
\hline \hline
\end{tabular}
\end{table}
%

\begin{figure}[b!]
\centering
\includegraphics[width=0.9\columnwidth]{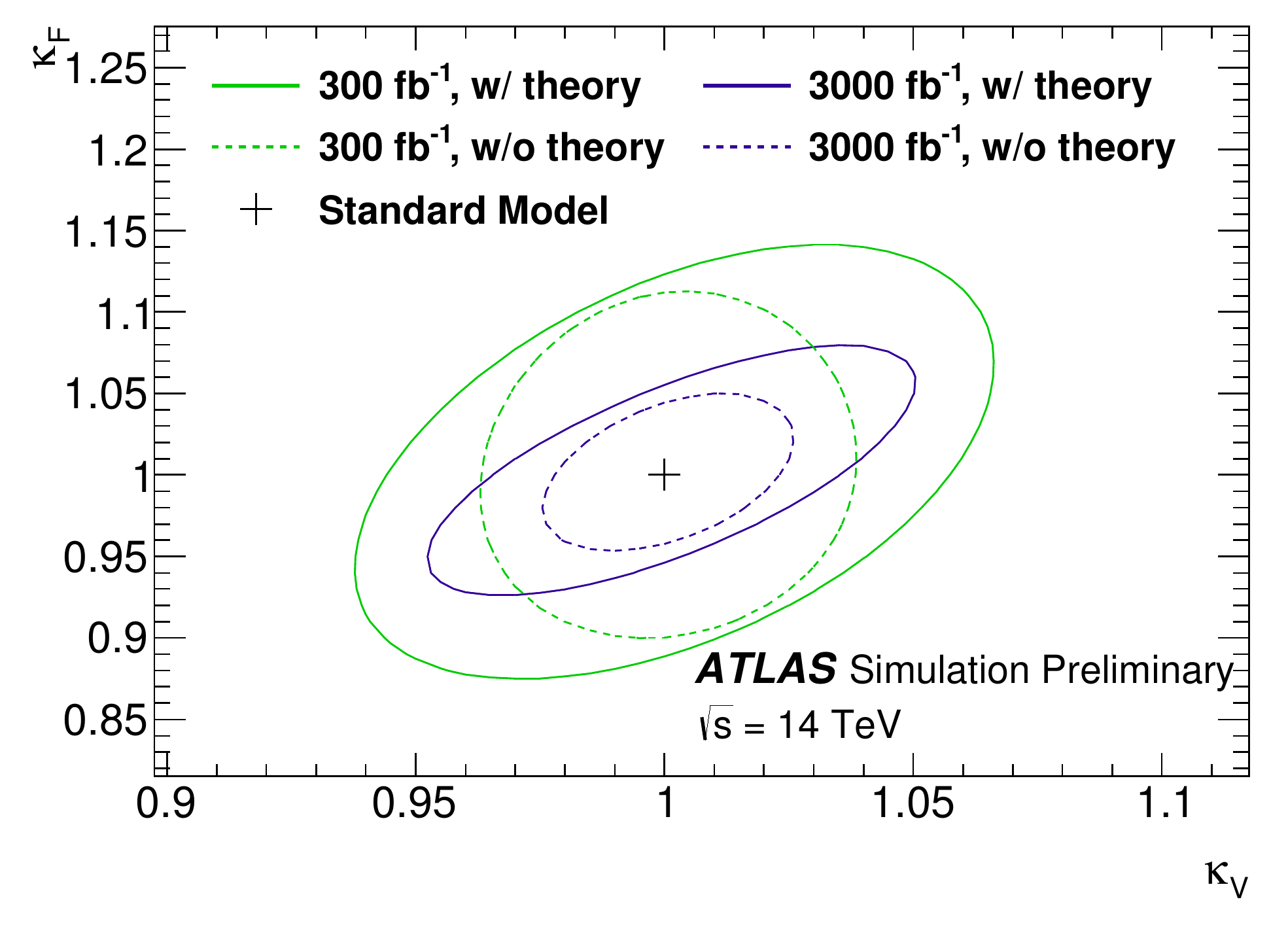}
\caption{
Expected ATLAS 68\% and 95\% C.L.\ likelihood contours for $\kappa_V$ and $\kappa_F$ in a minimal coupling fit 
for an integrated luminosity of 300~fb$^{-1}$ and 3000~fb$^{-1}$~\cite{atlpub14016}.
}
\label{fig:lhc21}       
\end{figure}

\begin{figure}[t!]
\centering
\includegraphics[width=0.8\columnwidth]{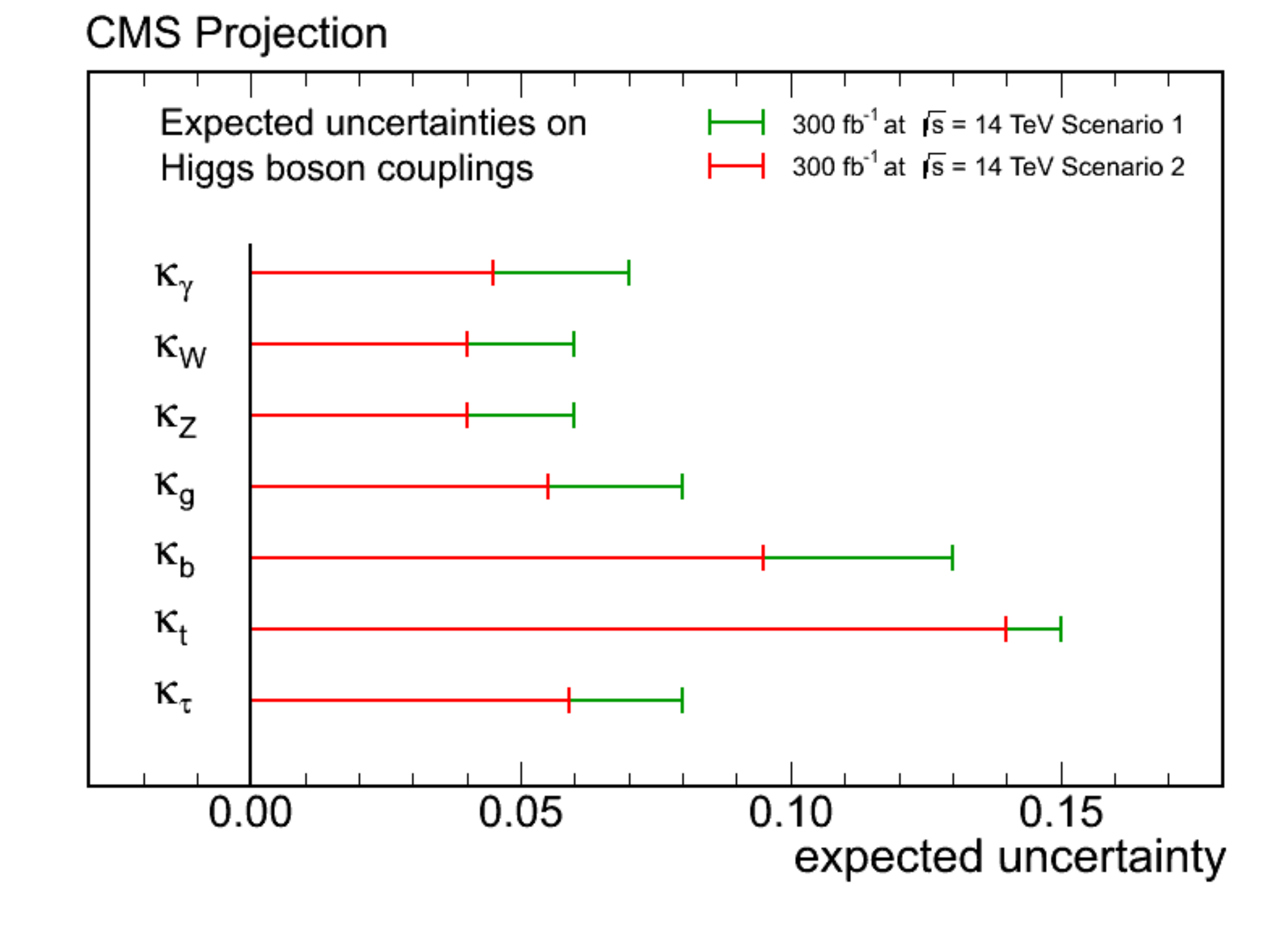}
\includegraphics[width=0.8\columnwidth]{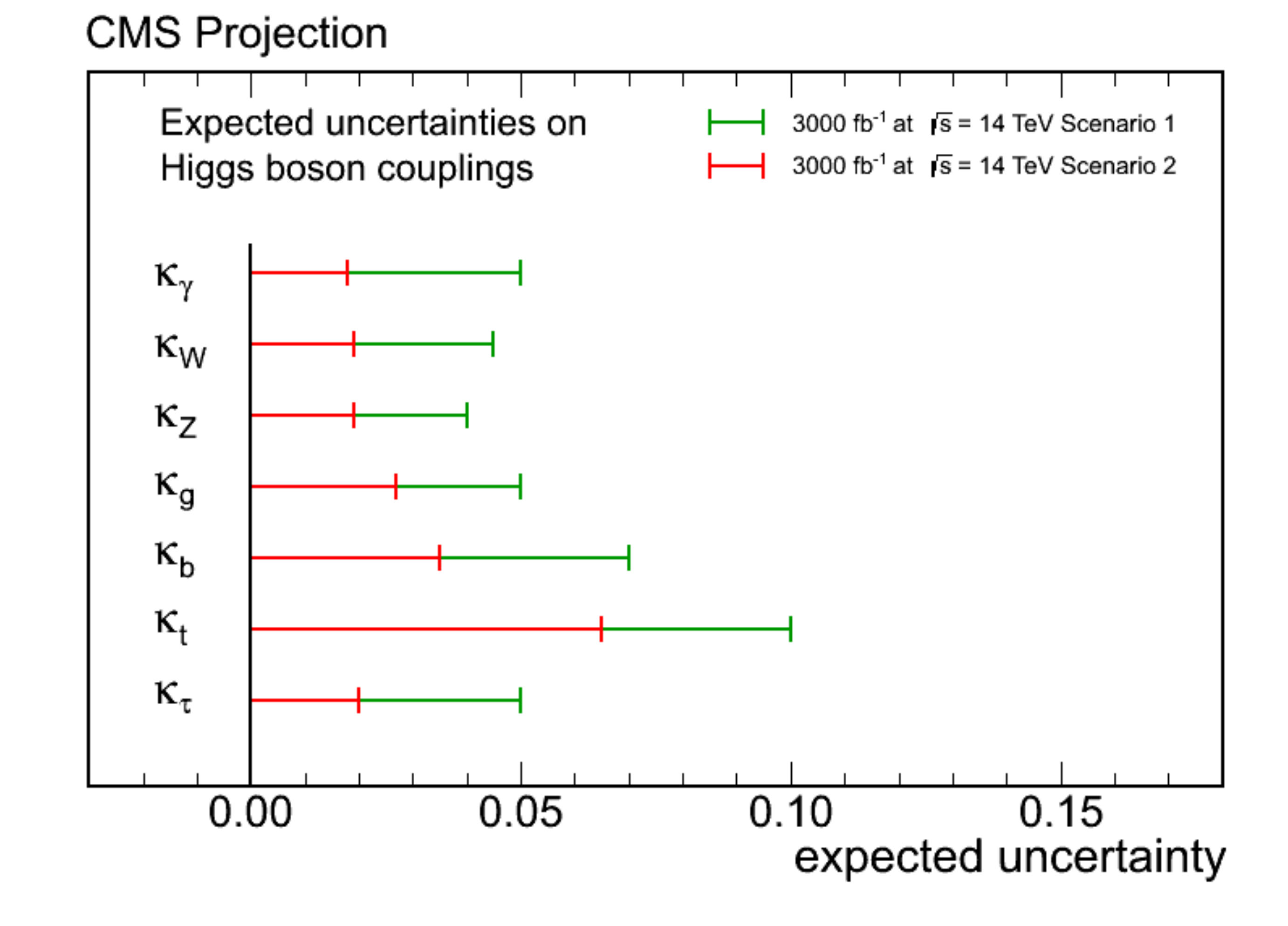}
\caption{
CMS projected relative uncertainty on the measurements of 
$\kappa_\gamma$, 
$\kappa_V$,
$\kappa_g$,
$\kappa_b$,
$\kappa_t$,
and
$\kappa_\tau$ 
assuming $\sqrt{s}$ = 14 TeV and an integrated luminosity 300 and 3000 fb$^{-1}$.
The results are shown for two uncertainty scenarios described in the text~\cite{cmspas13002}.
}
\label{fig:lhc22}       
\end{figure}

\subsubsection*{Couplings to other particles}
The individual channels are combined to obtain projections on the experimental sensitivity concerning 
Higgs boson couplings to other elementary bosons and fermions. 
Following the same formalism and set of assumptions used for the current Run1 
results described above, coupling scale factors $\kappa_X$ are extracted. 
Fig.~\ref{fig:lhc21}, for example, shows the 
projected ATLAS results of the minimal coupling fit constrained to common scale factors $\kappa_F$ and $\kappa_V$ 
for all fermions and bosons, respectively, and assuming SM values for both; cf. Fig.~\ref{fig:lhc7} for the 
corresponding Run1 results. Fig.~\ref{fig:lhc22} gives an overview of the precision on the extraction of 
individual coupling scale factors expected for the CMS experiment. 

\begin{figure}[t!]
\centering
\includegraphics[width=0.75\columnwidth]{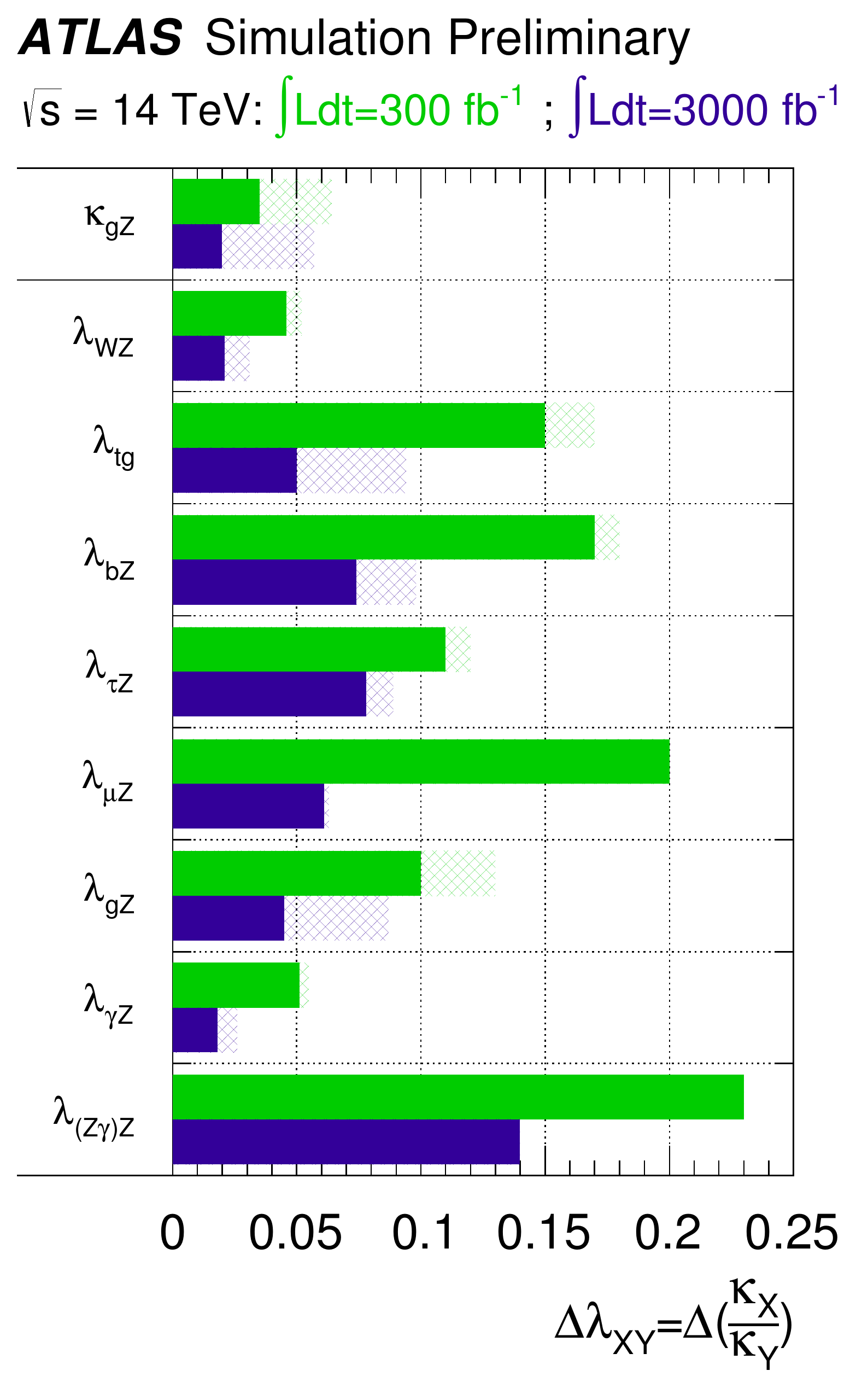}
\caption{
Relative uncertainty expected for the ATLAS experiment on the determination of coupling scale factor ratios 
$\lambda_{XY}=\kappa_X/\kappa_Y$ from a 
generic fit~\cite{atlpub14016}, assuming a SM Higgs boson with a mass of 125 GeV and  300~fb$^{-1}$ and 
3000~fb$^{-1}$ of 14 TeV data. The hashed areas indicate the increase of the estimated error due to current 
theory uncertainties.
}
\label{fig:lhc23}       
\end{figure}

The $\kappa_X$ extraction requires assumptions on the total width of the Higgs boson. Without total width 
information, only ratios of couplings can be studied. As for the current Run1 analyses, results are obtained
for several different sets of assumptions. 
%
\begin{table}[b!]
\centering
\caption{
Relative uncertainty on the determination of the coupling scale factor ratios expected for the CMS experiment for 
integrated luminosities of 300~fb$^{-1}$ and 3000~fb$^{-1}$~\cite{cmspas13002} and the two uncertainty scenarios described in the text.
}
\label{tab:lhc24}
\footnotesize
\vspace{0.2em}
\begin{tabular}{|c||cc|cc|}
\hline \hline
$\mathcal{L}$ &   \multicolumn{2}{|c|}{300~fb$^{-1}$}   & \multicolumn{2}{|c|}{3000~fb$^{-1}$}   \\
 Scenario & 2&1 & 2& 1\\\hline
$\kappa_\gamma\cdot\kappa_Z/\kappa_H$ &4\%& 6\% & 2\% & 5\% \\
$\kappa_W/\kappa_Z$ &4\%& 7\% & 2\% & 3\% \\
$\lambda_{tg}=\kappa_t/\kappa_g$ &13\%& 14\% & 6\% & 8\% \\
$\lambda_{bZ}=\kappa_b/\kappa_Z$ &8\%& 11\% & 3\% & 5\% \\
$\lambda_{\tau Z}=\kappa_\tau/\kappa_Z$ &6\%& 9\% & 2\% & 4\% \\
$\lambda_{\mu Z}=\kappa_\mu/\kappa_Z$ &22\%& 23\% & 7\% & 8\% \\
$\lambda_{Zg}=\kappa_Z/\kappa_g$ &6\%& 9\% & 3\% & 5\% \\
$\lambda_{\gamma Z}=\kappa_\gamma/\kappa_Z$ &5\%& 8\% & 2\% & 5\% \\
$\lambda_{(Z\gamma)Z }=\kappa_{Z\gamma}/\kappa_Z$ &40\%& 42\% & 12\% & 12\% \\
\hline \hline
\end{tabular}
\end{table}
%
An overview of the expected CMS precision for the most generic of these scenarios, still with a single, narrow, 
CP-even scalar Higgs boson but without further assumptions, e.g. on new-particle contributions through loops, is given in  
Table~\ref{tab:lhc24}. Results from corresponding ATLAS analyses are shown in Fig.~\ref{fig:lhc23}, where, 
for an integrated luminosity of 3000~fb$^{-1}$, the experimental uncertainties range from about 2\% 
for the coupling scale factors between the electroweak bosons to 5-8\% for the ratios involving gluons and fermions
outside the first generation.

%

\begin{figure}[h!]
\centering
\includegraphics[width=0.85\columnwidth]{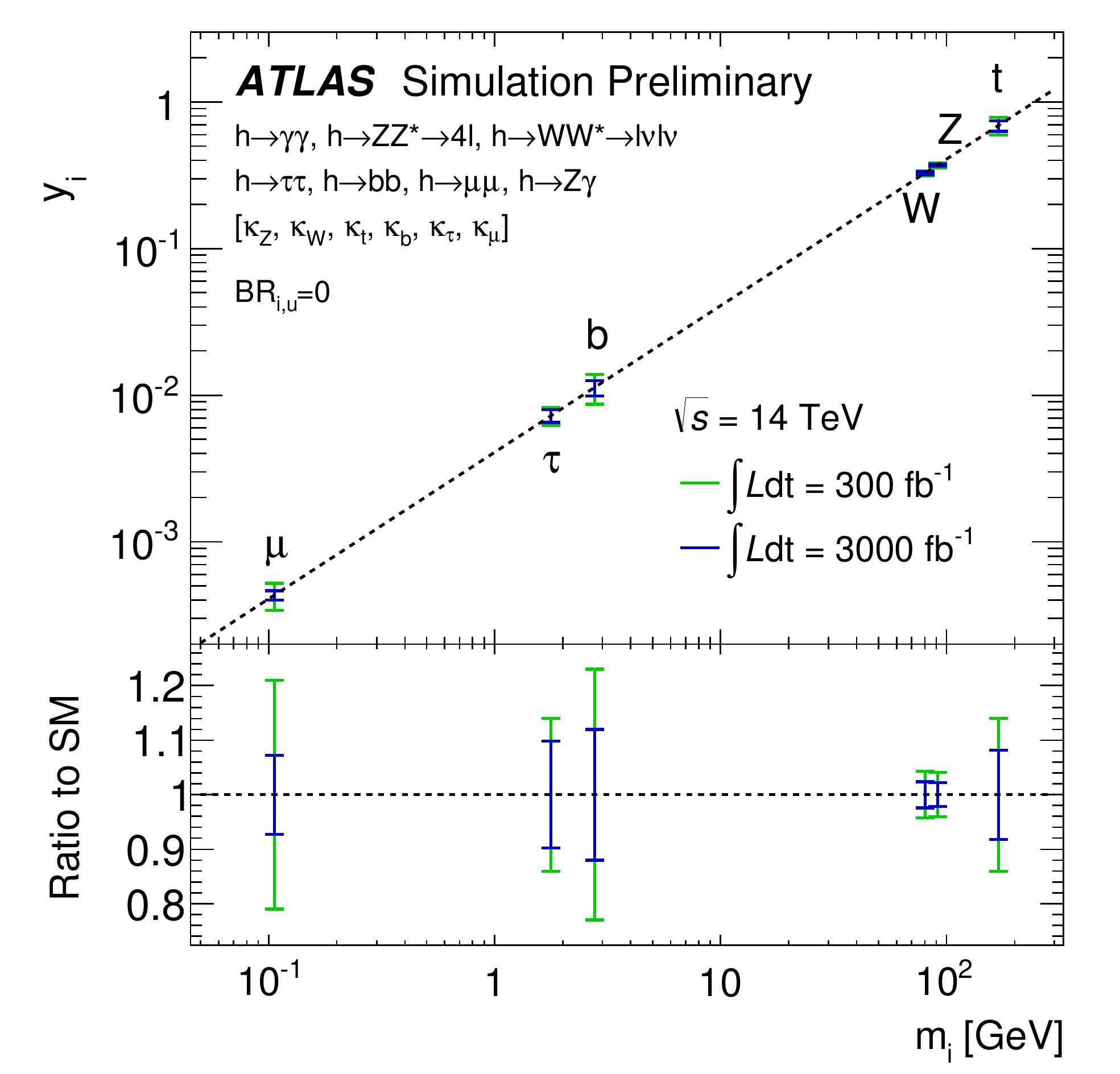}
\caption{Fit results for the reduced coupling scale factors for weak bosons and fermions as a function of the particle mass,  
assuming 300/fb or 3000/fb of 14 TeV data and a SM Higgs boson with a mass of 125 GeV~\cite{atlpub14016}. 
}
\label{fig:lhc25}       
\end{figure}

Fig.~\ref{fig:lhc25} gives the ATLAS projection for the precision of the Higgs boson couplings to other elementary SM 
particles as a function of the particle masses  obtained from fits assuming no BSM contributions to Higgs boson decays 
or particle loops; see Fig.~\ref{fig:lhc11} for  corresponding CMS Run1 results.

\subsubsection*{Higgs self-coupling}
One of the most important long-term goals of the SM Higgs physics
program is the measurement of the trilinear self-coupling
$\lambda_{HHH}$, which requires the study of Higgs boson pair
production.
%
\begin{figure}[b!]
\centering
\includegraphics[width=0.8\columnwidth]{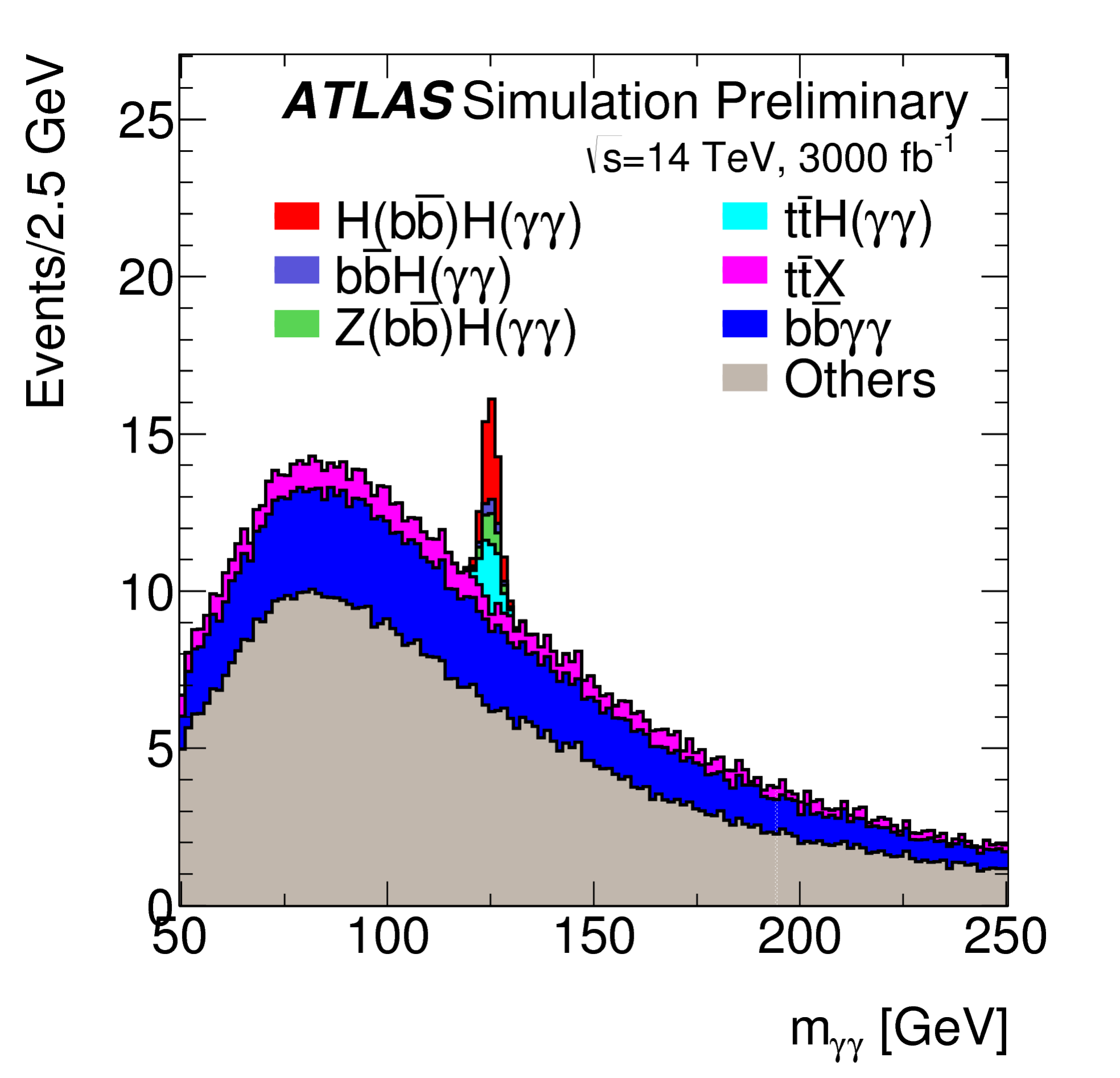}
\caption{Projected diphoton mass distribution for signal and background processes based on ATLAS simulations for a
search for Higgs boson pair production with subsequent decays $H\to b\bar{b}$ and $H\to\gamma\gamma$  assuming 
an integrated luminosity of 3000~fb$^{-1}$~\cite{atlpub14019}. The simulated distributions are scaled to match the expected 
event yields but do not necessarily reflect the corresponding statistical fluctuations. 
}
\label{fig:lhc26}       
\end{figure}
 At the LHC the dominant production mechanism is gluon-gluon fusion with a cross section of about 40 fb at $\sqrt{s}=14$~TeV. 
Several combinations of Higgs decays can be considered. For example, assuming  3000 fb$^{-1}$ of 14 TeV data, \cite{atlpub14019} 
presents the ATLAS prospects for the search for Higgs pair production in the channel $H(\to \gamma\gamma)H(\to bb)$, which 
combines the large $H\to bb$ branching ratio with the good mass resolution of the two-photon final state. 
The projected diphoton mass distribution for simulated ggF-produced signal and background processes after signal 
selection requirements is shown in Fig.~\ref{fig:lhc26}; the statistical analysis gives a 
signal yield of about eight events and 
signal significance of 1.3$\sigma$. Although additional observables, the application of more sophisticated analysis 
techniques and the inclusion of other production modes can be expected to improve on this result, a combination 
with other decay channels will likely be needed to find evidence for SM Higgs pair production (or to exclude that the Higgs 
self-coupling strength is close to its SM expectation) with an integrated luminosity of 3000 fb$^{-1}$.

%

\subsection{Higgs at ILC: prospects\protect\footnotemark}
\footnotetext{Keisuke Fujii\\
The presented materials were prepared for the ILC
  TDR physics chapter and for the Snowmass Higgs white paper in
  collaboration with the members of the ILC physics working group
  \cite{ref:ilcphys, ref:snowmass_higgs_white_paper} and the members
  of the ILC physics panel.  The author would like to thank them for
  useful discussions, especially M.\,Peskin, Y.\,Okada, S.\,Kanemura.,
  H.\,Haber, T.\,Barklow, A.\,Miyamoto, J.\,Tian, H.\,Ono, and
  T.\,Tanabe. 
}
\label{sec:ewsb3}

%

\newcommand{\sla}[1]{\not{\! {#1}}}
\newcommand{\bra}[1]{\left< {#1} \,\right\vert}
\newcommand{\ket}[1]{\left\vert\, {#1} \, \right>}
\newcommand{\braket}[2]{\hbox{$\left< {#1} \,\vrule\, {#2} \right>$}}
\newcommand{\mathbold}[1]{\mbox{\boldmath $#1$}}

\subsubsection{Introduction}
\label{sec:ilc_higgs_intro}

The success of the Standard Model (SM) is a success of gauge principle. 
It is the success of the transverse components of $W$ and $Z$ identified as gauge fields of the electroweak (EW) gauge symmetry. 
Since explicit mass terms for $W$ and $Z$ are forbidden by the gauge
symmetry, it must be spontaneously broken by {\it something condensed in
  the vacuum} which carries EW charges 
($I_3$ and $Y$ denoting the third component of the weak iso-spin
  and the hyper charge, respectively),
\begin{eqnarray}
\bra{0} I_3, Y \ket{0} \ne 0  \mbox{  while } \bra{0} I_3 + Y \ket{0} = 0.
\end{eqnarray}
We are hence living in a weak-charged vacuum.
This {\it something} provides three longitudinal modes of $W$ and~$Z$:
\begin{eqnarray}
  \mbox{Goldstone modes :}  \chi^+, \chi^-, \chi_3 \to W_L^+, W_L^-, Z_L~.
\end{eqnarray}
It should be emphasized that we do not know the nature of these longitudinal modes which stem from the {\it something}.
The gauge symmetry also forbids explicit mass terms for matter fermions, since left- ($f_L$) and right-handed ($f_R$) matter fermions carry different EW charges, hence, as long as the EW charges are conserved, they cannot mix.
Their Yukawa interactions with some weak-charged vacuum can compensate the EW-charge difference and hence allow the $f_L$-$f_R$ mixing.
In the SM, the same {\it something} is responsible for the $f_L$-$f_R$ mixing, thereby generating masses and inducing flavor-mixings among generations.
To form gauge-invariant Yukawa interaction terms, we need a complex doublet scalar field,
which has four real components.
In the SM, three of them are identified with the three Goldstone modes and are used to supply the longitudinal modes of $W$ and $Z$. The remaining one is the physical Higgs boson.
There is no reason for this simplicity of the symmetry breaking sector of the SM.
The symmetry breaking sector (hear after called the Higgs sector) can well be much more complicated.
The {\it something} could be composite instead of being elementary.
We know it's there around us with a vacuum expectation value of 246\,GeV. 
But this was about all we knew concerning the {\it something} until July 4th, 2012.

\begin{figure}[h]
\centering
\includegraphics[width=0.9\hsize]{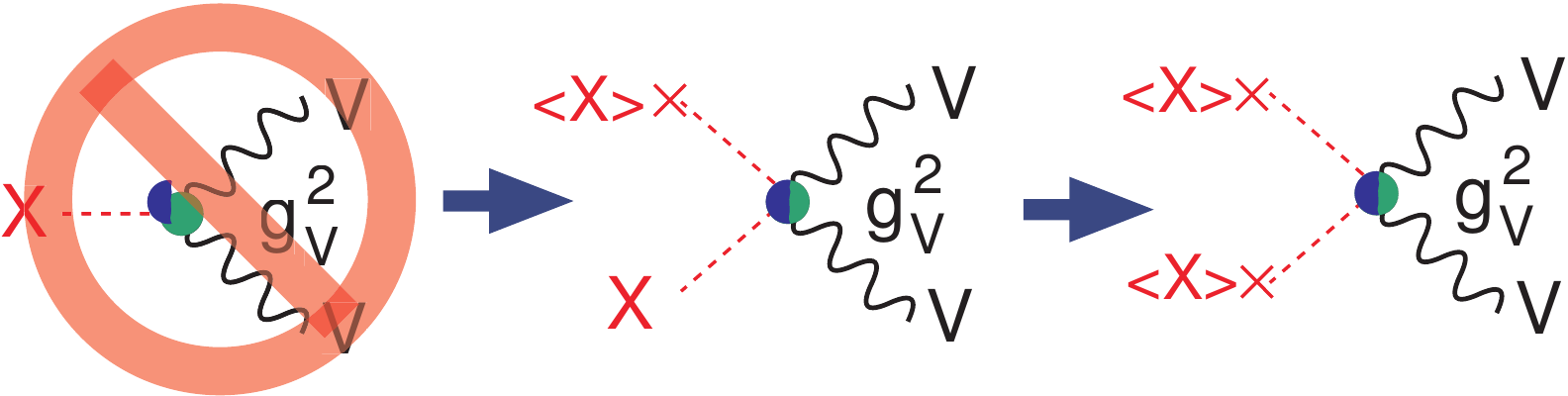}
\caption{The origin of $XVV$ coupling and its relation to the mass term of $V$.} \label{fig:xvv}
\end{figure}

\begin{figure}[h]
\centering
\includegraphics[width=0.9\hsize]{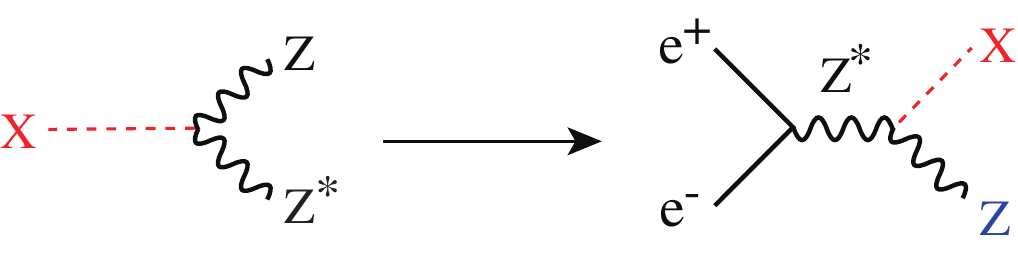}
\caption{$X \to ZZ^*$ decay and $e^+e^- \to ZX$ process.} \label{fig:xzzandzh}
\label{fig:no_loose}
\end{figure}

Since the July 4th, the world has changed! 
The discovery of the 125\,GeV boson ($X(125)$) at the LHC could be called a quantum jump \cite{ref:LHChiggs}.
The observation of $X(125) \to \gamma\gamma$ decay implies $X$ is a neutral boson having a spin not
equal to 1 (Landau-Yang theorem).
We know that the 125\,GeV boson decays also to $ZZ^*$ and $WW^*$, indicating the existence
of $XVV$ couplings, where $V=W/Z$, gauge bosons.
There is, however, no gauge coupling like $XVV$, see Fig.~\ref{fig:xvv}.
There are only $XXVV$ and $XXV$.
The $XVV$ coupling is hence most probably from
$XXVV$ with one $X$ replaced by its vacuum expectation value $\left< X \right> \ne 0$,
namely $\left<X\right>XVV$.
Then there must be $\left<X\right>\left<X\right>VV$, a mass term for $V$, meaning that
$X$ is at least part of the origin of the masses of $V=W/Z$.
This is a great step forward to uncover the nature of the {\it something} in the vacuum but we need to know whether $\left<X\right>$ saturates the
SM vev of 245\,GeV.
The observation of the $X \to ZZ^*$ decay means that $X$ can be produced via $e^+e^- \to Z^* \to ZX$, since by attaching an $e^+e^-$ pair to the $Z^*$ leg and rotate the whole diagram we can get the X-strahlung diagram as shown in Fig.\ref{fig:no_loose}.
By the same token, $X \to WW^*$ means that $X$ can be produced via the $WW$-fusion process:
$e^+e^- \to \nu\bar{\nu}X$.
So we now know that the major Higgs production processes in $e^+e^-$ collisions are indeed
available at the ILC, which can be regarded as a {\it no lose theorem} for the ILC.
The $125\,$GeV is the best place for the ILC, where variety of decay modes are accessible.
We need to check the 125GeV boson in detail to see if it has indeed all the required properties of the {\it something} in the vacuum.

\begin{figure}[h]
\centering
\includegraphics[width=0.9\hsize]{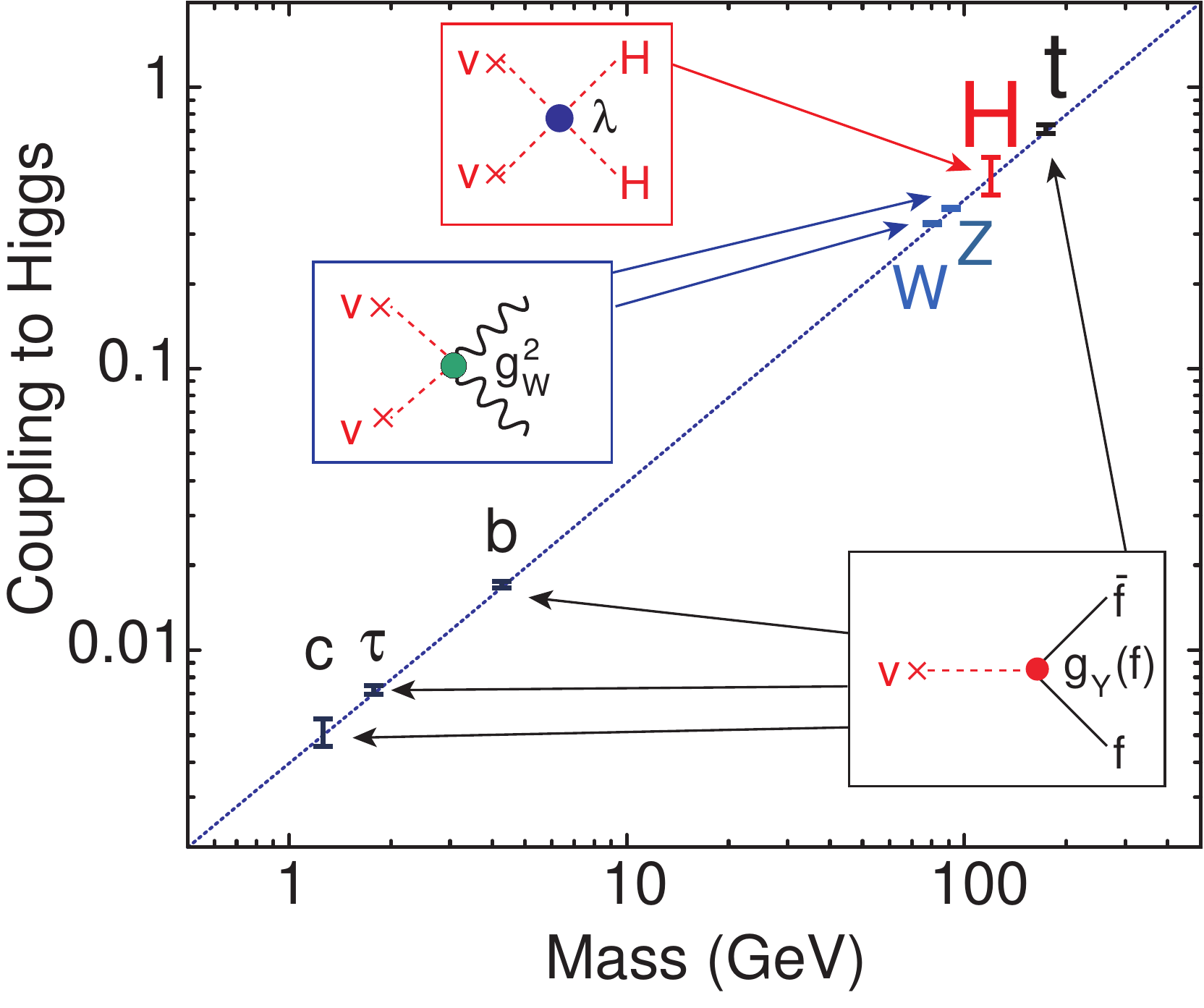}
\caption{Mass-coupling relation \cite{ref:acfa}.} \label{fig:mass-coupling1}
\end{figure}

The properties to measure are the mass, width, and $J^{PC}$, its gauge, Yukawa, and self couplings. 
The key is to confirm {\it the mass-coupling relation}. 
If the 125\,GeV boson is the one to give masses to all the SM particles,
coupling should be proportional to mass as shown in
Fig.\ref{fig:mass-coupling1}. 
Any deviation from the straight line signals physics beyond the Standard Model (BSM).
The Higgs serves therefore as a window to BSM physics. 

Our mission is the bottom-up model-independent reconstruction of the electroweak symmetry 
breaking (EWSB) sector through the coupling measurements.
We need to determine the multiplet structure of the Higgs sector by answering questions like:
Is there an additional singlet or doublet or triplet?
What about the underlying dynamics? 
Is it weakly interacting or strongly interacting?
In other words, is the Higgs boson elementary or composite?
We should also try to investigate its possible relation to other questions of particle physics
such as dark matter, electroweak baryogenesis, neutrino masses, and inflation.

There are many possibilities and different models predict different deviation patterns in the mass-coupling relation.
An example is given in Table \ref{tab:coupling_deviation_pattern}, where a model with an extra singlet and four types of two-Higgs doublet models (2HDM) are compared.
The four types of 2HDMs differ in the assignment of a $Z_2$ charge to the matter fermions, which protects them from inducing dangerous flavor-changing neutral currents\cite{ref:2hdm_fcnc}.

\begin{table}[ht]
\centering
\caption{The expected deviation pattern for various Higgs couplings,
assuming small deviations for $\cos(\beta-\alpha) < 0$.  
The arrows for Yukawa interactions are reversed for 2HDMs 
with $\cos(\beta-\alpha) > 0$.  
}
\label{tab:coupling_deviation_pattern}
\vspace{0.2em}
\begin{tabular}{|l|cccccc|}
\hline \hline
Model &
$\mu$ & $\tau$ & $b$   &
$c$   & $t$    & $g_V$ \\
\hline
Singlet mixing &
$\downarrow$ & $\downarrow$ & $\downarrow$ &
$\downarrow$ & $\downarrow$ & $\downarrow$ \\
2HDM-I &
$\downarrow$ & $\downarrow$ & $\downarrow$ &
$\downarrow$ & $\downarrow$ & $\downarrow$ \\
2HDM-II (SUSY) &
$\uparrow$ & $\uparrow$ & $\uparrow$ &
$\downarrow$ & $\downarrow$ & $\downarrow$ \\
2HDM-X (Lepton-specific) &
$\uparrow$ & $\uparrow$ & $\downarrow$ &
$\downarrow$ & $\downarrow$ & $\downarrow$ \\
2HDM-Y (Flipped) &
$\downarrow$ & $\downarrow$ & $\uparrow$ &
$\downarrow$ & $\downarrow$ & $\downarrow$ \\
\hline \hline
\end{tabular}
\end{table}
Notice that though both singlet mixing and 2HDM-I with $\cos(\beta-\alpha)<0$ give downward deviations,
they are quantitatively different: the singlet mixing reduces the coupling constants universally, while 2HDM-I reduces them differently for matter fermions and gauge bosons.  
In these models, $g_V < 1$ is guaranteed because of the sum rule for the vacuum expectation values of the SM-like Higgs boson and the additional doublet or singlet. 
When a doubly charge Higgs boson is present, however, $g_V >1$ is possible.
The size of any of these deviations is generally written in the following form due to the decoupling theorem:
\begin{equation}
\frac{\Delta g}{g} = \mathcal{O} \left( \frac{v^2}{M^2} \right)
\end{equation}
where $v$ is the SM vev and $M$ is the mass scale for the new physics.
Since there is no hint of new physics beyond the SM seen at the LHC, $M$ should be rather large implying small deviations.
In order to detect possible deviations and to fingerprint the BSM physics from the deviation pattern, we hence need a \% level precision, which in turn requires a 500\,GeV linear collider such as the International Linear Collider (ILC) and high precision detectors that match the potential of the collider.

\begin{figure}[h]
\centering
\includegraphics[width=0.5\hsize]{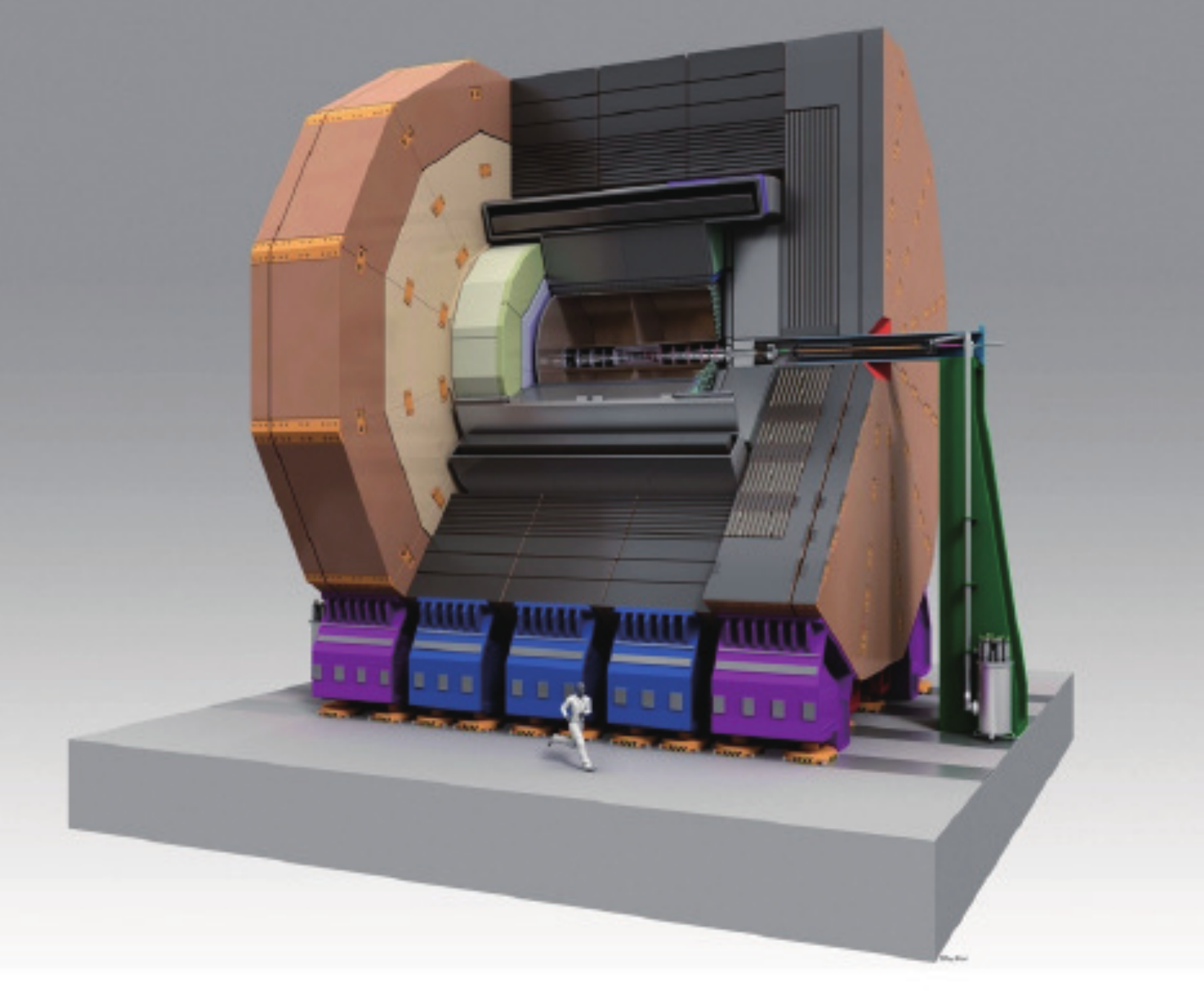} \hspace{0.05\hsize}
\includegraphics[width=0.4\hsize]{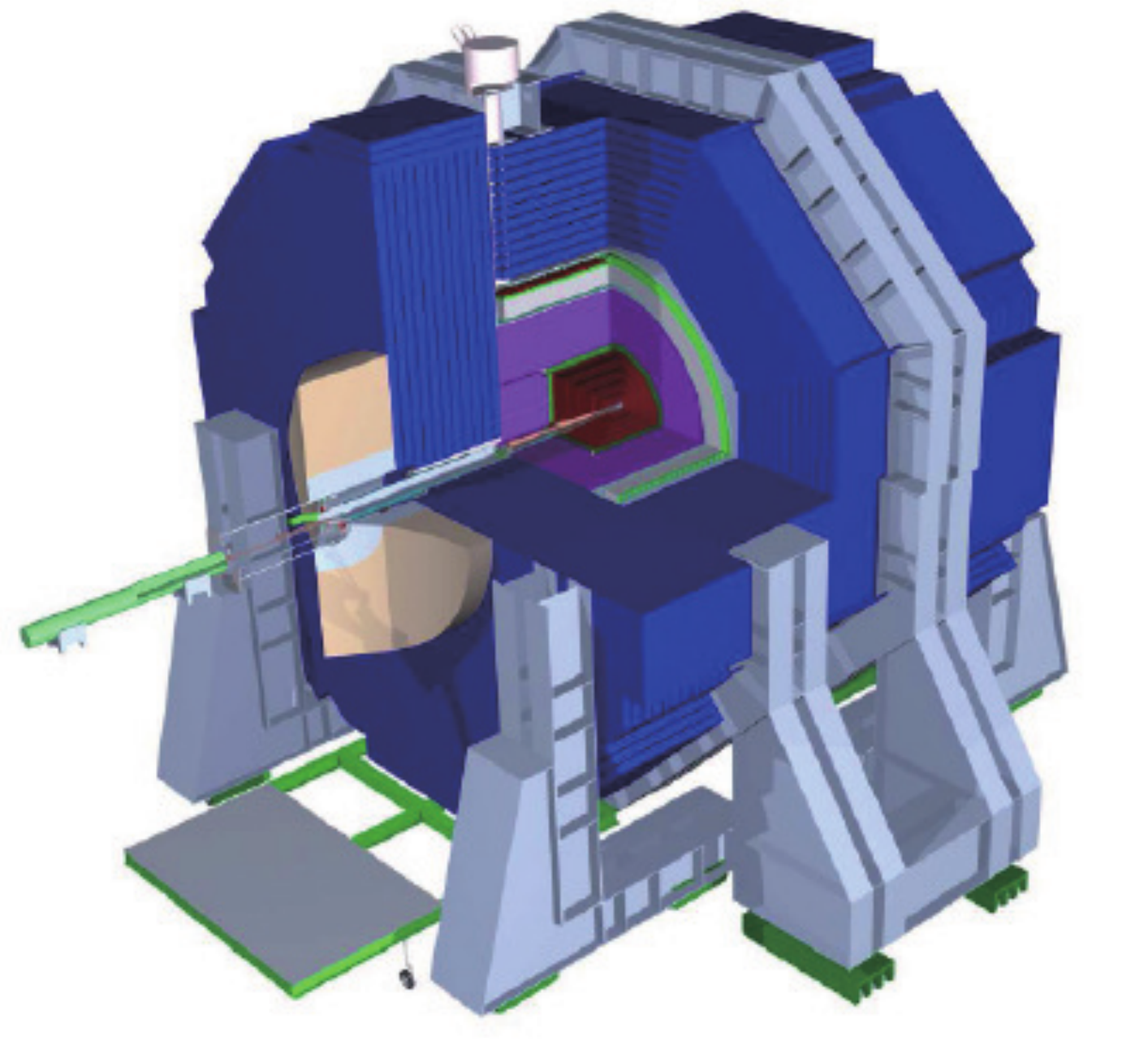}
\caption{Two proposed detector concepts for the ILC: ILD (left) and SiD (right) \cite{ref:dbd}.} \label{fig:ild_sid}
\end{figure}

The ILC, being an $e^+e^-$ collider, inherits all of its traditional merits: cleanliness, democracy, detail, and calculability. 
The two detector concepts proposed for the ILC: ILD and SiD (see Fig.\ref{fig:ild_sid}) take advantage of these merits.
\begin{sloppypar}
Moreover, they are designed with an ambitious goal of reconstructing
all the events in terms of fundamental particles such as quarks,
leptons, gauge bosons, and Higgs bosons, thereby viewing events as
viewing Feynman diagrams.  This requires a thin and high resolution
vertex detector that enables identification of $b$- and $c$-quarks by
detecting secondary and tertiary vertices, combination of a high
resolution charged particle tracker and high granularity calorimeters
optimized for Particle Flow Analysis (PFA) to allow identification of
$W$, $Z$, $t$, and $H$ by measuring their jet invariant masses, and
hermeticity down to $\mathcal{O} \left( 10 {\rm ~mrad} \right)$ or
better for indirect detection of a neutrino as missing momentum.
Notice that both ILD and SiD put all the calorimeters inside the
detector solenoidal magnets to satisfy the requirement of hermeticity
and high performance PFA.  Furthermore, the power of beam
polarizations should be emphasized.  Consider the $e^+e^- \to W^+W^-$
process.  At the energies explored by the ILC, $SU(2)_L \otimes
U(1)_Y$ symmetry is approximately recovered and hence the process can
be regarded as taking place through two diagrams: $s$-channel $W_3$
exchange and $t$-channel $\nu_e$ exchange.  Since both $W_3$ and
$\nu_e$ couple only to a left-handed electron (and right-handed
positrons), right-handed electrons will not contribute to the process.
This is also the case for one of the most important Higgs production
process at the ILC: $e^+e^- \to \nu_e \bar{\nu}_e H$ ($WW$-fusion
single Higgs production).  If we have an 80\% left-handed electron
beam and a 30\% right-handed positron beam the Higgs production cross
section for this $WW$-fusion process will be enhanced by a factor of
2.34 as compared to the unpolarized case.  Beam polarization hence
plays an essential role.\\
\end{sloppypar}

\noindent{\it Why 250 to 500\,GeV?}\\

\begin{figure}[h]
\centering
\includegraphics[width=\hsize]{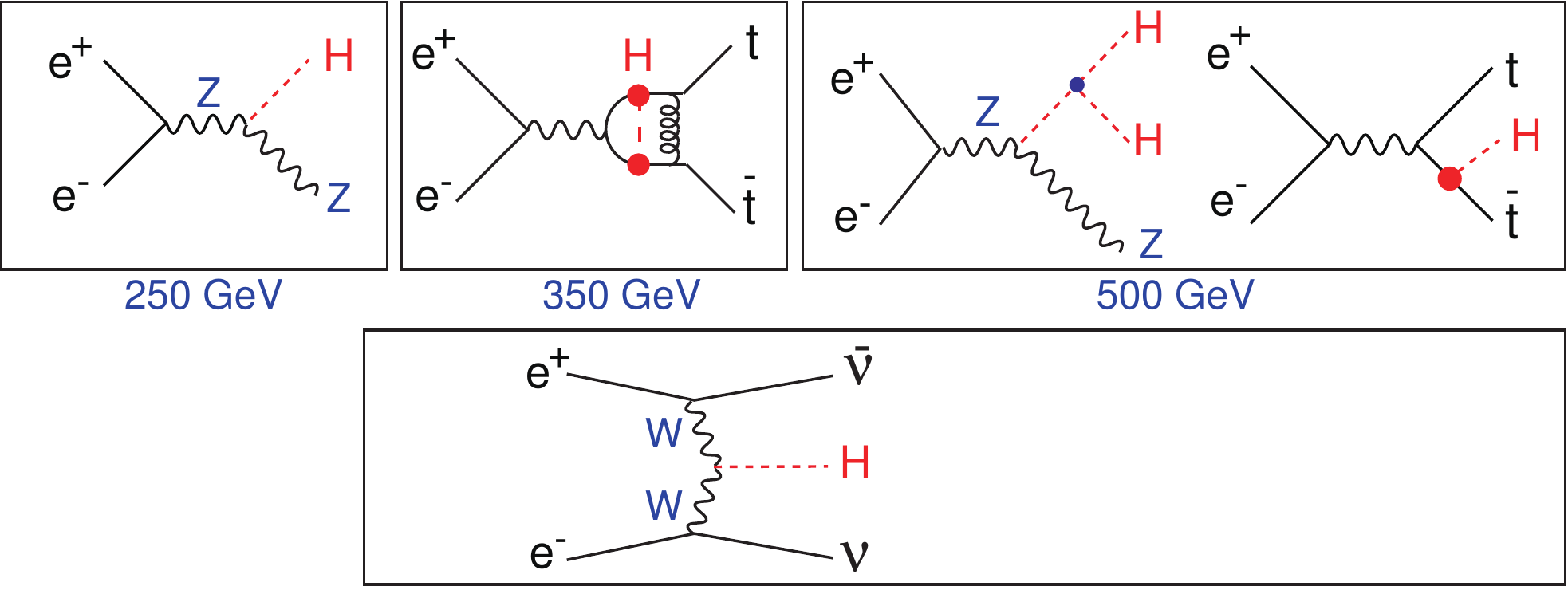}
\caption{Why 250-500\,GeV? The three thresholds.} \label{fig:the3thresholds}
\end{figure}

\noindent The ILC is an $e^+e^-$ collider designed primarily to cover the energy range from $\sqrt{s}=250$ to $500\,$GeV.
This is because of the following three very well know thresholds, see 
Fig.~\ref{fig:the3thresholds}.
\begin{sloppypar}
The first threshold is at around $\sqrt{s}=250\,$GeV, where the
$e^+e^- \to Zh$ process will reach its cross section maximum.  This
process is a powerful tool to measure the Higgs mass, width, and
$J^{PC}$.  As we will see below, this process allows us to measure the
$hZZ$ coupling in a completely model-independent manner through the
recoil mass measurement.  This is a key to perform model-independent
extraction of branching ratios for various decay modes such as $h \to
b\bar{b}$, $c\bar{c}$, $\tau\bar{\tau}$, $gg$, $WW^*$, $ZZ^*$,
$\gamma\gamma$, as well as invisible decays.
\end{sloppypar}
The second threshold is at around $\sqrt{s}=350\,$GeV, which is the well known $t\bar{t}$ threshold.
The threshold scan here provides a theoretically very clean measurement of the top quark mass, which can be translated into $m_t(\overline{\mbox{MS}})$ to an accuracy of $100\,$MeV.
The precise value of the top mass obtained this way can be combined with the precision Higgs mass measurement
to test the stability of the SM vacuum \cite{ref:vacuumstability}.
The $t\bar{t}$ threshold also enables us to indirectly access the top Yukawa coupling through the Higgs exchange diagram.
It is also worth noting that with the $\gamma\gamma$ collider option at this energy the double Higgs production: $\gamma\gamma \to hh$ is possible, which can be used to study the Higgs self-coupling \cite{ref:aahh}.
Notice also that at $\sqrt{s}=350\,$GeV and above, the $WW$-fusion Higgs production process, $e^+e^- \to \nu\bar{\nu}h$, becomes sizable with which we can measure the $hWW$ coupling and accurately determine the total width.

\begin{sloppypar}
The third threshold is at around $\sqrt{s}=500\,$GeV, where the double
Higgs-strahlung process, $e^+e^- \to Zhh$ attains its cross section
maximum, which can be used to access the Higgs self-coupling.  At
$\sqrt{s}=500\,$GeV, another important process, $e^+e^- \to
t\bar{t}h$, will also open, though the product cross section is much
smaller than its maximum that is reached at around
$\sqrt{s}=800\,$GeV.  Nevertheless, as we will see, QCD threshold
correction enhances the cross section and allows us a reasonable
measurement of the top Yukawa coupling concurrently with the
self-coupling measurement.
\end{sloppypar}

By covering $\sqrt{s}=250$ to $500\,$GeV, we will hence be able complete the mass-coupling plot.
This is why the first phase of the ILC project is designed to cover the energy up to $\sqrt{s}=500\,$GeV.

\subsubsection{ILC at 250\,GeV}

\begin{figure}[h]
\centering
\includegraphics[width=0.9\hsize]{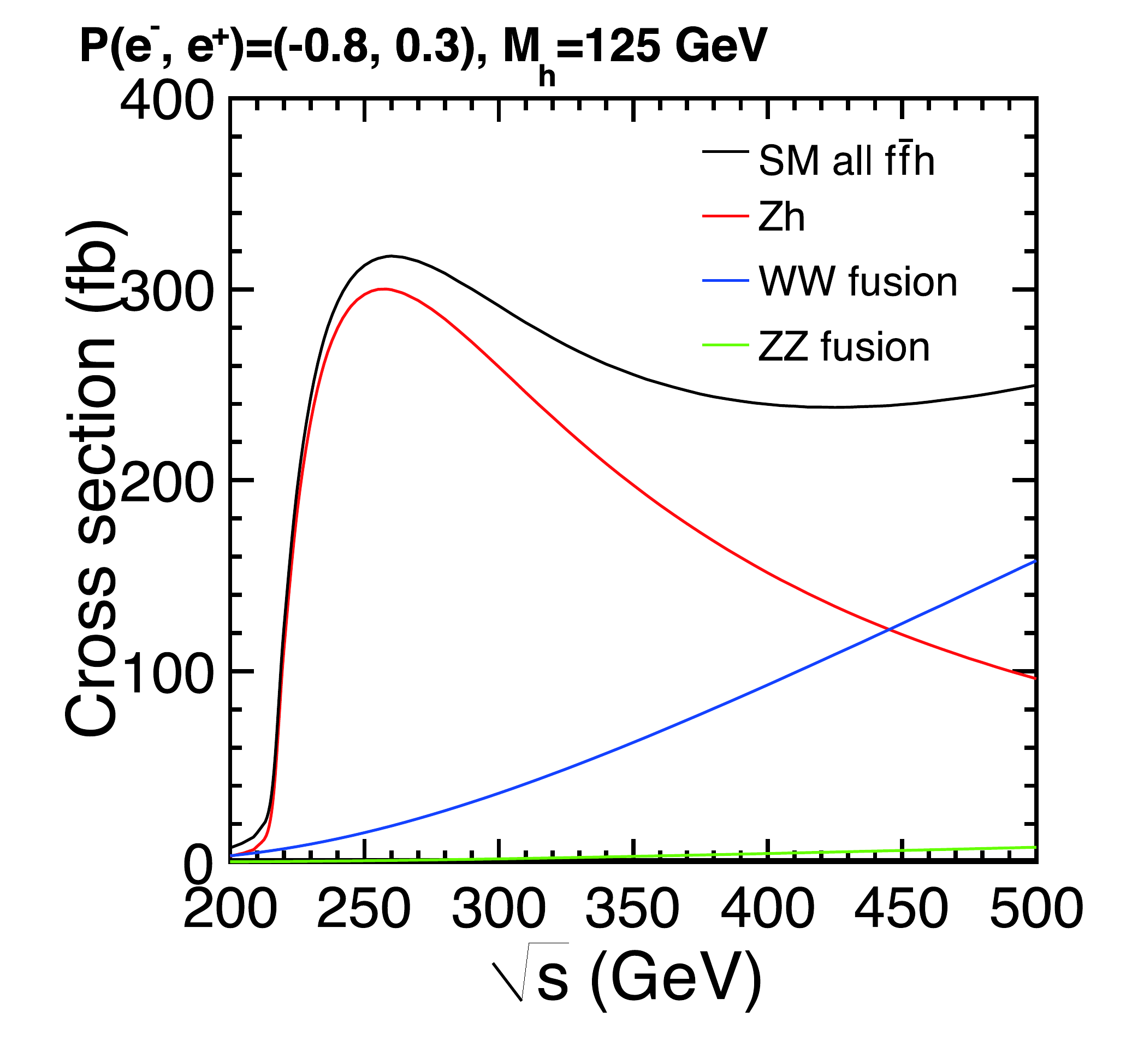}
\caption{Cross sections for the three major Higgs production processes as a function of center of mass energy.} 
\label{fig:sigma_h}
\end{figure}

The first threshold is at around $\sqrt{s}=250\,$GeV, where the $e^+e^- \to Zh$ (Higgs-strahlung) process attains its cross section maximum (see Fig.\ref{fig:sigma_h}).

\begin{figure}[h]
\centering
\includegraphics[width=0.9\hsize]{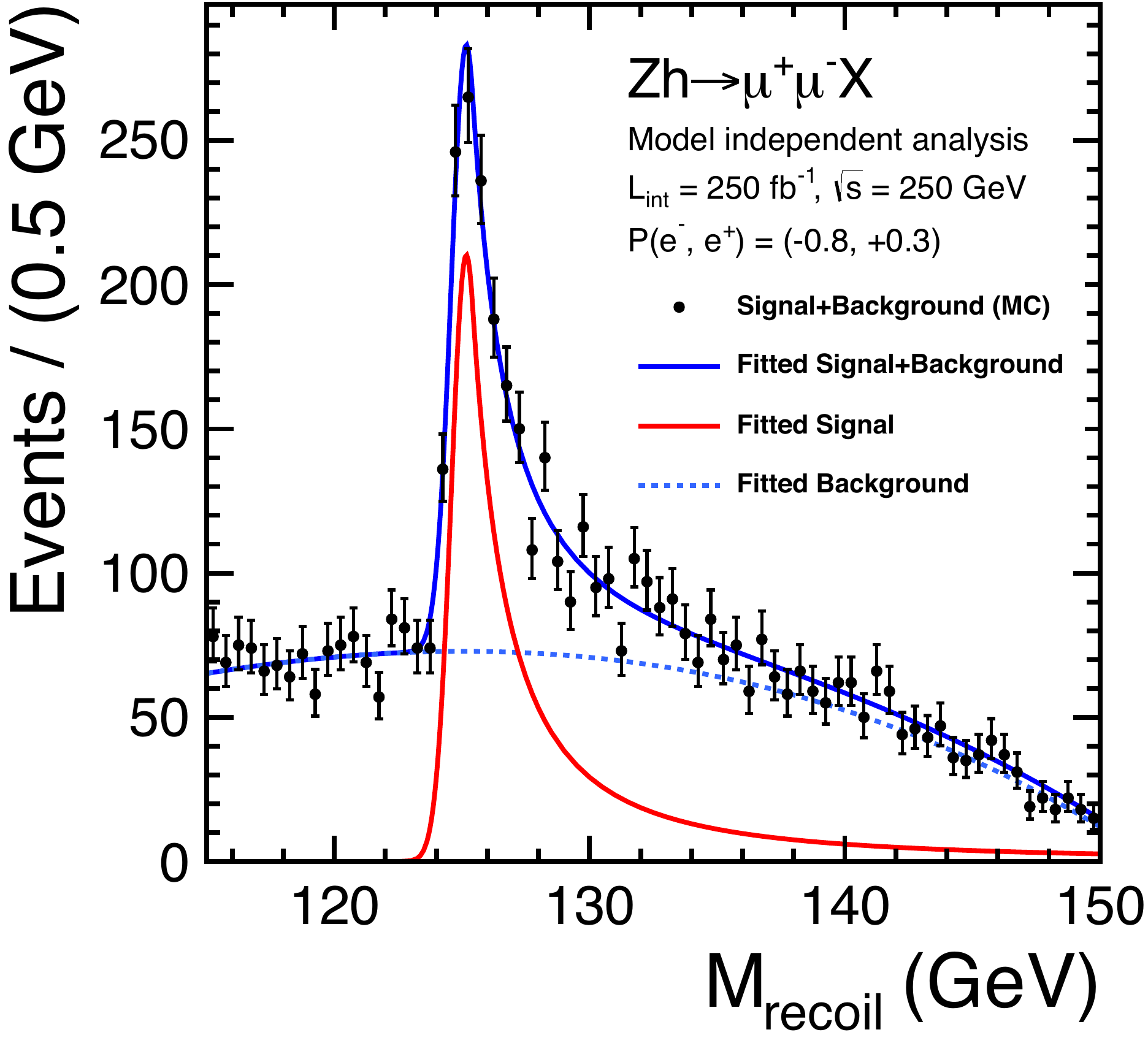}
\caption{Recoil mass distribution for the process: $e^+e^- \to Zh$
  followed by $Z \to \mu^+\mu^-$ decay for $m_h=125\,$GeV with
  $250\,$fb$^{-1}$ at $\sqrt{s}=250\,$GeV \cite{ref:watanuki}.}
\label{fig:mrec}
\end{figure}

The most important measurement at this energy is that of the recoil mass for the process: $e^+e^- \to Zh$ followed by $Z \to \ell^+\ell^- ~(\ell=e,\mu)$ decay.
By virtue of the $e^+e^-$ collider, we know the initial state 4-momentum.
We can hence calculate the invariant mass of the system recoiling against the lepton pair from the $Z$ decay
by just measuring the momenta of the lepton pair:
\begin{eqnarray}
M_X^2 = \left(p_{CM} - (p_{\ell^+} + p_{\ell^-})\right)^2 .
\end{eqnarray}
The recoil mass distribution is shown in Fig. \ref{fig:mrec} for a $m_h=125\,$GeV Higgs boson with 
250\,fb$^{-1}$ at $\sqrt{s}=250\,$GeV.
A very clean Higgs peak is sticking out from small background.
Notice that with this recoil mass technique even invisible decay is
detectable since we do not need to look at the Higgs decay at all \cite{ref:higgsportal2011}.
This way, we can determine the Higgs mass to $\Delta m_h=30\,$MeV and the production cross section to $\Delta \sigma_{Zh} /\sigma_{Zh} = 2.6\,$\%, and limit the invisible branching ratio to $1\%$ at the $95\%$ confidence level
\cite{ref:BRinvOno2012, ref:BRinvIshikawa2013}.
This is the flagship measurement of the ILC at 250\,GeV that allows
absolute measurement of the $hZZ$ coupling thereby unlocking the door to completely model-independent determinations of various couplings of the Higgs boson as well as its total width as we will see below.\\

\begin{figure}
\begin{center}
\includegraphics[width=0.9\hsize]{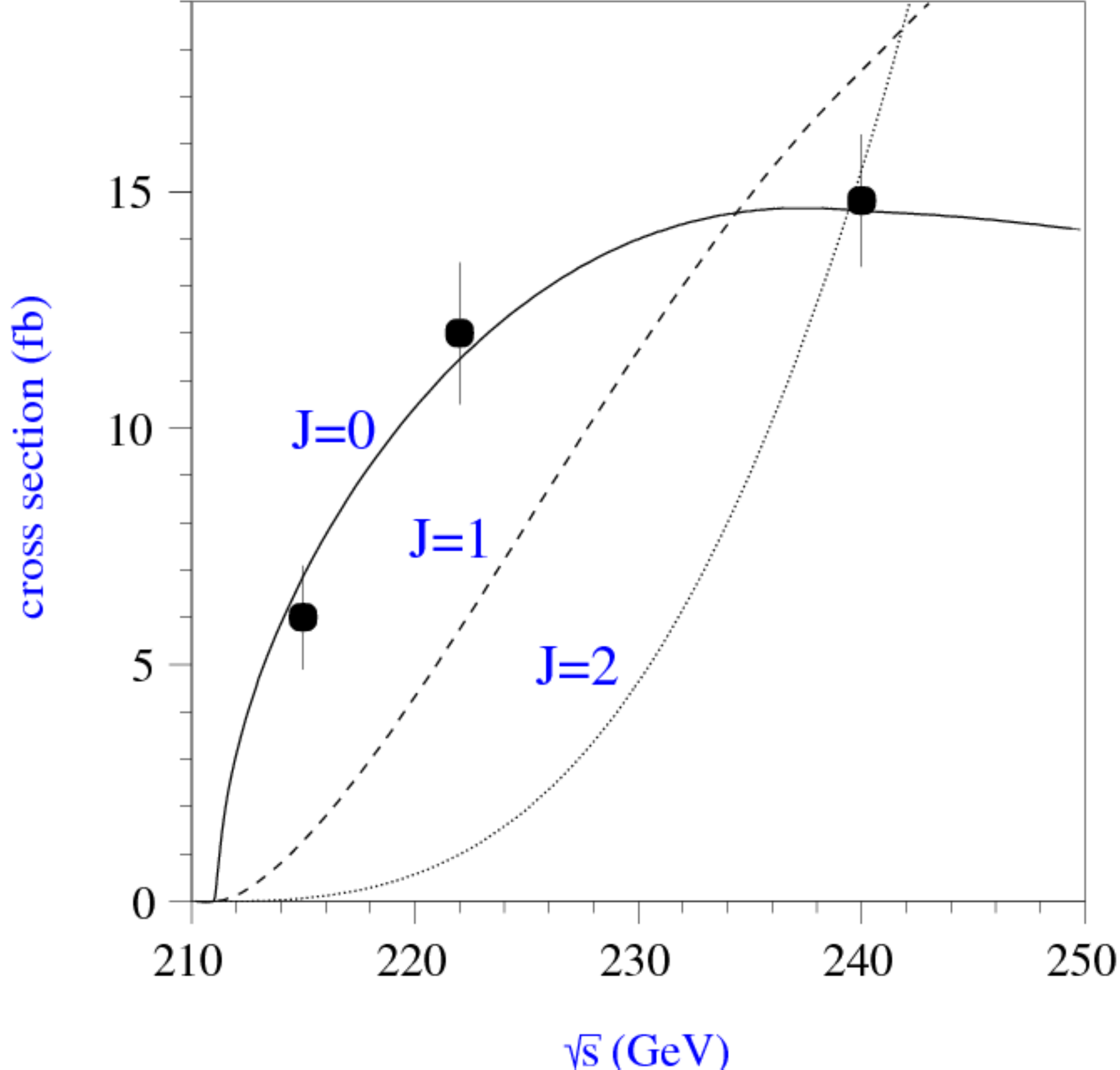}
\end{center}
\caption{ Threshold scan of the $e^+e^- \to Zh$ process for $m_h =
120$\,GeV, compared with theoretical predictions for  $J^{P}= 0^{+}$,
$1^{-}$, and $2^{+}$ \cite{ref:Dova:2003py}.}
\label{fig:ZH:JPC}
\end{figure}

Before moving on to the coupling determinations, let us discuss here the determination of the  spin and CP properties of the 
Higgs boson.
The LHC observed the $h \to \gamma \gamma$ decay, which fact alone rules out the possibility of spin 1 and restricts the charge conjugation C to be positive.
The more recent LHC analysis strongly prefer the $J^P=0^+$ assignment
over $0^-$ or $2^\pm$~\cite{ref:LHC.hspin}.
By the time of the ILC the discrete choice between different spin and 
CP-even or odd assignments will certainly be settled, assuming that the 125\,GeV boson is a CP eigen state.
Nevertheless, it is worth noting that the ILC also offers an additional, orthogonal, and clean test of these assignments.
The threshold behavior of the $Zh$ cross section has a characteristic shape for each spin and each possible CP parity.  
For spin 0, the cross section rises as $\beta$ near the threshold for a 
CP-even state and as $\beta^3$ for a CP-odd state.   
For spin 2,  for the canonical form of the coupling to the energy-momentum tensor, the rise is also $\beta^3$.
If the spin is higher than 2, the cross section will grow as a higher power of $\beta$. 
With a three-$20$\,fb$^{-1}$-point threshold scan of the $e^+e^- \to Zh$ production cross section we can
separate these possibilities~\cite{ref:Dova:2003py} as shown in Fig.~\ref{fig:ZH:JPC}.  
The discrimination of more general forms of the coupling is possible by the use of angular correlations in the boson decay; this is discussed in detail in \cite{ref:Miller:2001bi}.

\begin{figure}
\begin{center}
      \includegraphics[width=0.9\hsize]{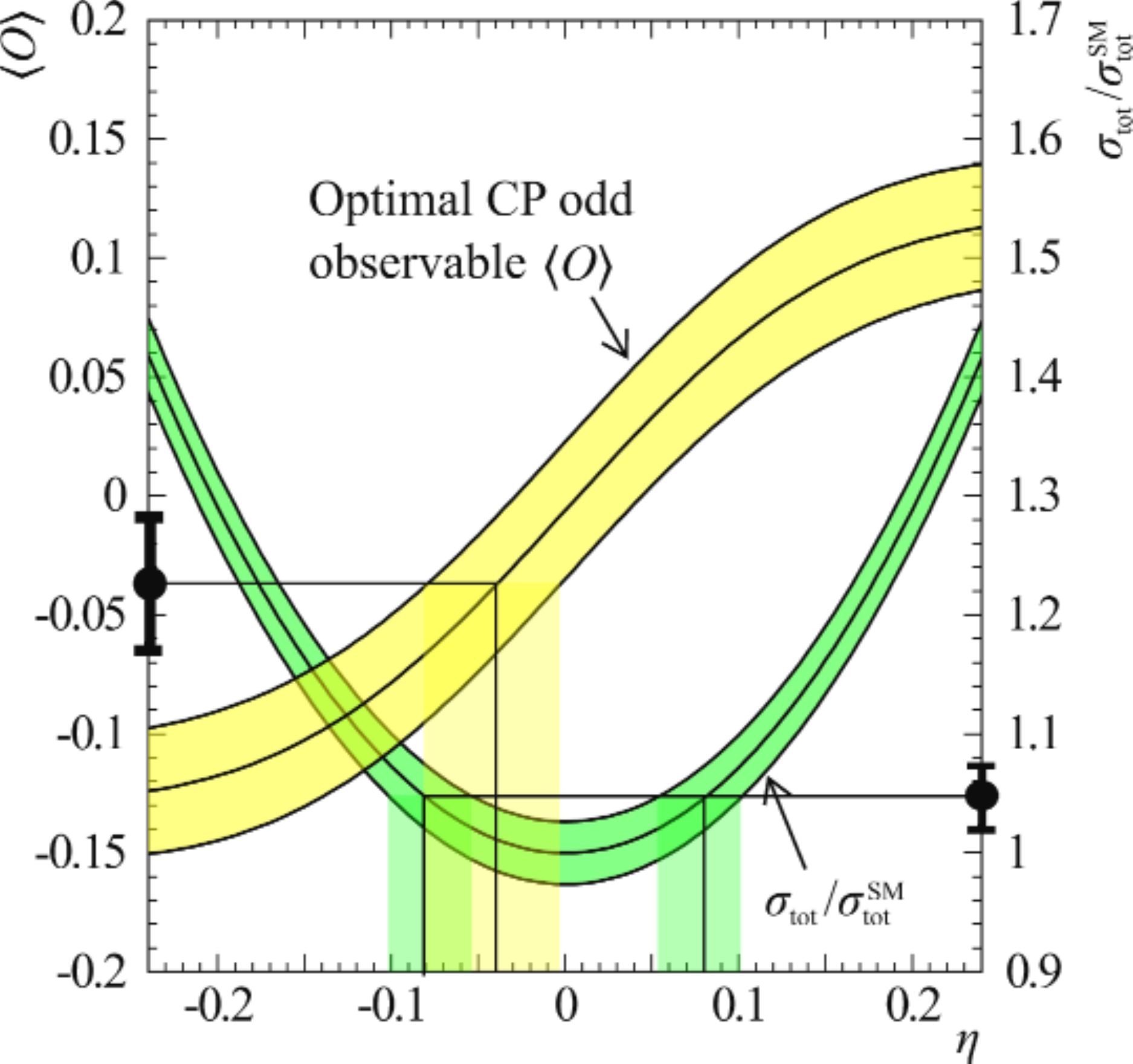}
\end{center} 
\caption{\label{fig:ZH:CPmix}
Determination of $CP$-mixing with $1$-$\sigma$ bands expected at $\sqrt{s}=350$\,GeV and $500$\,fb$^{-1}$ \cite{ref:CPmix:ZH}. }
\end{figure}

The power of the ILC manifests itself when we ask more subtle questions.
There is no guarantee that the $h$ is a CP eigenstate.
It can rather be a mixture of CP-even and CP-odd components.   
This happens if CP is violated in the Higgs sector.
A small CP-odd contribution to the $hZZ$ coupling can affect the threshold behavior.
Figure\,\ref{fig:ZH:CPmix} shows the determination of the small CP-odd
component $\eta$ at $\sqrt{s}=$350\,GeV from the  value of the total cross section and from an appropriately defined optimal observable~\cite{ref:CPmix:ZH}.
The $hZZ$ coupling is probably not the best tool to study possible CP
admixture, since in many scenarios the CP-odd $hZZ$ coupling is only generated through loops.  
It is, hence, more effective to use a coupling for which the CP-even and 
CP-odd components are on the same footing as in the $h$  coupling to $\tau^+\tau^-$, given by 
\begin{equation}
   \Delta {\cal L} =     - {m_\tau\over v} h\  \bar \tau (\cos\alpha + i
   \sin\alpha \gamma^5) \tau
\end{equation}
for a Higgs boson with a CP-odd component.
The polarizations of the final state $\tau$s can be determined from the kinematic distributions of their decay products; the CP-even and odd components interfere in these distributions~\cite{ref:ZPC64-21-1994, Berge:2012wm}.   
In \cite{Desch:2003rw}, it is estimated that the angle $\alpha$ can be determined at the ILC to an accuracy of 6$^\circ$.\\

The $e^+e^- \to Zh$ process can also be used to measure various branching ratios for various Higgs decay modes.
For this purpose $Z \to q\bar{q}$ and $\nu\bar{\nu}$ decays can be included in our analysis to enhance the statistical precision. 
We should stress here that as with similar Higgs-related measurements at the LHC what we can actually measure is {\it NOT} branching ratio ($\br$) itself but the cross section times branching ratio ($\sigma \times \br$).
The crucial difference is the recoil mass measurement at the ILC that provides $\sigma$ to extract $\br$ from $\sigma \times \br$ model-independently.
Table \ref{tab:sigBR250} summarizes the expected precisions for the $\sigma\times BR$ measurements together with those for the extracted $BR$s \cite{ref:BRs250Ono2013, ref:BRs250Banda2010, ref:BRs250Ono2012, ref:BRtau, ref:kawada2013, ref:JCBrient2001, ref:fastMCresults.2007tk, ref:tino2013}.\\

\begin{table}[h]
\begin{center}
\caption{Expected relative errors for the $\sigma\times \br$ measurements at $\sqrt{s}=250\,$GeV with $250\,$fb$^{-1}$ for $m_h=125\,$GeV.}
\begin{tabular}{|c|c|c|c|}
\hline
process &decay mode & $\Delta (\sigma \br)/(\sigma \br)$ &  $\Delta \br/\br$ \\
\hline
Zh & $h \to b\bar{b}$ & 1.2\% & 2.9\% \\
\cline{2-4}
     & $h \to c\bar{c}$ & 8.3\% & 8.7\% \\
\cline{2-4}
     & $h \to gg$ & 7.0\%  & 7.5\% \\
\cline{2-4}
     & $h \to WW^*$ & 6.4\% & 6.9\% \\
\cline{2-4}
     & $h \to \tau\bar{\tau}$ & 4.2\% & 4.9\% \\
\cline{2-4}
     & $h \to ZZ^*$ & 19\% & 19\% \\
\cline{2-4}
     & $h \to \gamma\gamma$ & 34\%  & 34\% \\
\hline
\end{tabular}
\label{tab:sigBR250}
\end{center}
\end{table}
Notice that the cross section error, $\Delta \sigma_{Zh}/\sigma_{Zh}=2.5\%$, eventually limits the precision of the BR measurements.
We hence need more data at $\sqrt{s}=250\,$GeV so as to improve the situation.
We will return to the possible luminosity upgrade scenario later.\\

In order to extract couplings from branching ratios, we need the total
width, since the coupling of the Higgs boson to a particle
  $A$, $g_{hAA}$, squared
is proportional to the partial width which is given by the
total width times the branching ratio:
\begin{eqnarray}
g_{hAA}^2 \propto \Gamma(h \to AA) = \Gamma_h \cdot \br(h \to AA).
\end{eqnarray}
Solving this for the total width, we can see that we need at least one partial width
and corresponding branching ratio to determine the total width:
\begin{eqnarray}
\Gamma_h = \Gamma(h \to AA) / \br(h \to AA) .
\end{eqnarray}
In principle, we can use $A=Z$ or $A=W$, for which we can measure both the $\br$s and the couplings.
In the first case, $A=Z$, we can determine $\Gamma(h \to ZZ^*)$ from the recoil mass measurement
and $\br(h \to ZZ^*)$ from the $\sigma_{Zh} \times \br(h \to ZZ^*)$ measurement together with the $\sigma_{Zh}$ measurement from the recoil mass.
This method, however, suffers from the low statistics due to the small
branching ratio, $\br(h \to ZZ^*)= {\cal O}(1\%)$,
A better way is to use $A=W$, where $\br(h \to WW^*)$ is subdominant and $\Gamma(h \to WW^*)$ can be determined by the $WW$-fusion process: $e^+e^- \to \nu\bar{\nu}h$.
The measurement of the $WW$-fusion process is, however, not easy at $\sqrt{s}=250\,$GeV since the cross section is small. Nevertheless, we can determine the total width to $\Delta \Gamma_h /\Gamma_h = 11\%$ with $250\,$fb$^{-1}$ \cite{ref:Durig, ref:GammaH}.
Since the $WW$-fusion process becomes fully active at
$\sqrt{s}=500\,$GeV, a much better measurement of the total width is
possible there, as will be discussed in the next subsection.

\subsubsection{ILC at 500\,GeV}

At $\sqrt{s}=500\,$GeV, the $WW$-fusion process $e^+e^- \to \nu\bar{\nu}h$ already starts dominating the higgsstrahlung process: $e^+e^- \to Zh$.
We can use this $WW$-fusion process for the $\sigma \times \br$ measurements as well as to determine the total width to $\Delta \Gamma_h / \Gamma_h = 5\%$ \cite{ref:GammaH}.
Table~\ref{tab:sigBR500} summarizes the $\sigma \times \br$ measurements for various modes.
\begin{table}[h]
\begin{center}
\caption{Expected relative errors for the $\sigma\times \br$ measurements at $\sqrt{s}=250\,$GeV with $250\,$fb$^{-1}$ and at $\sqrt{s}=500\,$GeV with $500\,$fb$^{-1}$ for $m_h=125\,$GeV and $(e^{-}, e^{+})=(-0.8, +0.3)$ beam polarization. The last column of the table shows the relative errors on branching ratios. Then numbers in the parentheses are as of $250\,$fb$^{-1}$ at $\sqrt{s}=250\,$GeV alone.}
\begin{tabular}{|l|r|r|r|r|}
   \hline
             & \multicolumn{3}{c|}{$\Delta (\sigma \cdot \br) / (\sigma \cdot \br)$} & $\Delta \br/\br$ \\
   \hline
   energy (GeV) & 250 & \multicolumn{2}{c|}{500} &  250+500 \\
   \hline
   mode & $Zh$  & $Zh$  & $\nu\bar{\nu}h$              & combined \\
   \hline\hline
   $h \to b\bar{b}$       & 1.2\% & 1.8\%  & 0.66\%     & 2.2 (2.9)\% \\
   \hline
   $h \to c\bar{c}$        & 8.3\%  & 13\%    & 6.2\%     & 5.1 (8.7)\%  \\
   \hline
   $h \to gg$                & 7.0\%   & 11\%   & 4.1\%     & 4.0 (7.5)\%  \\
   \hline
   $h \to WW^*$          & 6.4\%   & 9.2\%  & 2.4\%      & 3.1 (6.9)\%  \\
   \hline
   $h \to \tau^+\tau^-$ & 4.2\%   & 5.4\%    & 9.0\%    &  3.7 (4.9)\%  \\
   $h \to ZZ^*$            & 19\%    & 25\%     & 8.2\%    &  7.5 (19)\%   \\
   $h \to \gamma\gamma$ & 29-38\% & 29-38\% & 20-26\% & 17(34)\% \\
   \hline
\end{tabular}
\label{tab:sigBR500}
\end{center}
\end{table}
We can see that the $\sigma_{\nu\bar{\nu}h} \times \br(h \to b\bar{b})$ can be very accurately measured
to better than $1\%$ and the $\sigma_{\nu\bar{\nu}h} \times \br(h \to WW^*)$ to a reasonable precision with $500\,$fb$^{-1}$ at $\sqrt{s}=500\,$GeV.
The last column of the table shows the results of $\Delta \br / \br$ from the global analysis combining all the measurements including the total cross section measurement using the recoil mass at $\sqrt{s}=250\,$GeV (2.6\%) and $500\,$GeV (3\%).
The numbers in the parentheses are with the $250\,$GeV data alone.
We can see that the $\Delta \br(h \to b\bar{b})/\br(h \to b\bar{b})$ is already limited by the recoil mass measurements.\\

\begin{figure}[h]
\centering
\includegraphics[width=0.9\hsize]{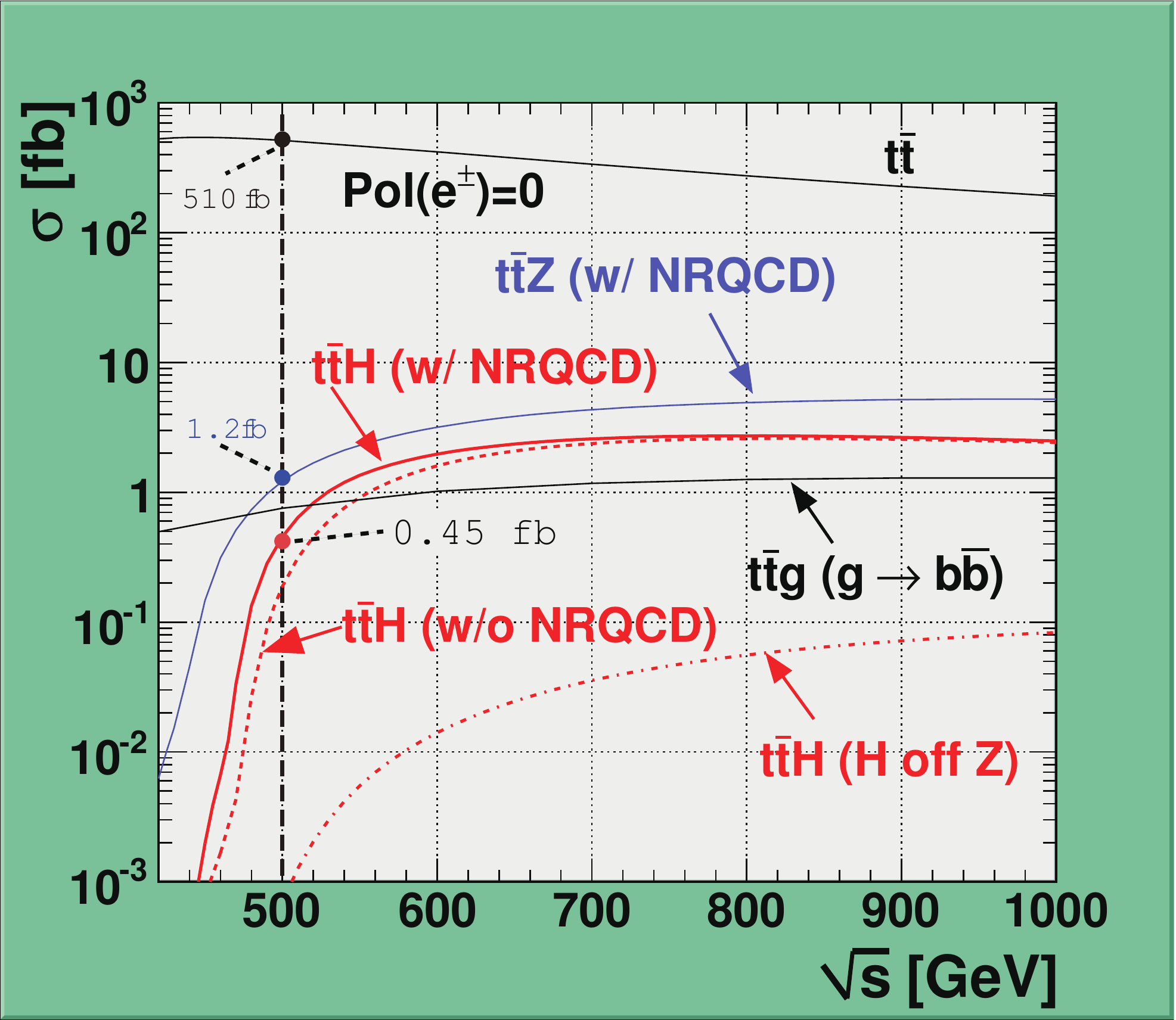}
\includegraphics[width=0.9\hsize]{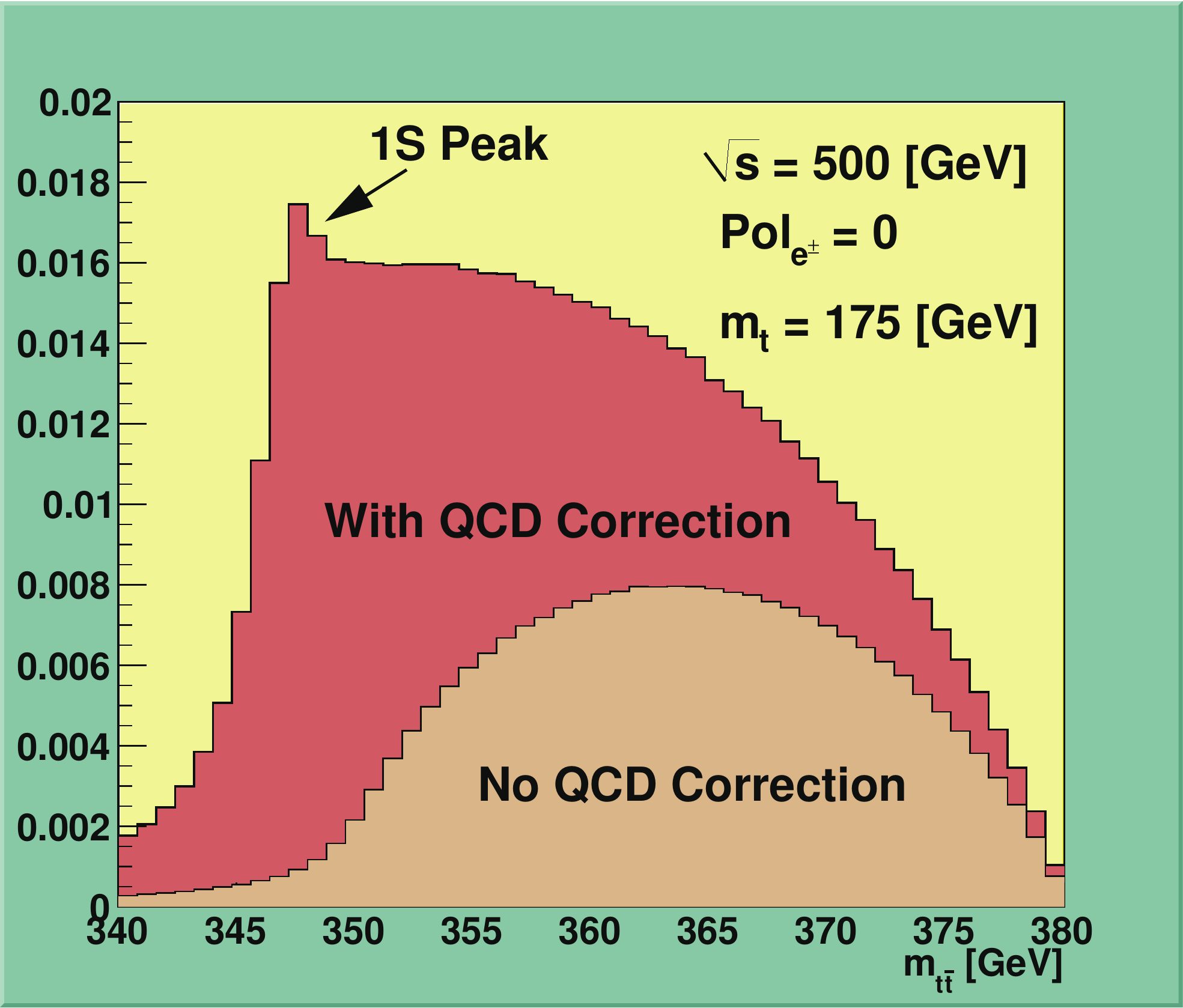}
\caption{Cross sections for the signal $t\bar{t}h$ process with and without the non-relativistic QCD (NRQCD) correction together with those for the background processes: $t\bar{t}Z, t\bar{t}g (g \to b\bar{b})$ and $t\bar{t}$ (upper plot). The invariant mass distribution for the $t\bar{t}$ subsystem with and without the NRQCD correction (lower plot).} 
\label{fig:sigtth}
\end{figure}

\begin{sloppypar}
Perhaps more interesting than the branching ratio measurements is the measurement of the top Yukawa coupling using the $e^+e^- \to t\bar{t}h$ process \cite{ref:tthDjouadi1992, ref:1998dklsz, ref:1999bdr}, since it is the largest among matter fermions and not yet directly observed. 
Although its cross section maximum is reached at around $\sqrt{s}=800\,$GeV as seen in Fig.\ref{fig:sigtth}, 
the process is accessible already at $\sqrt{s}=500\,$GeV, thanks to the QCD bound-state effects (non-relativistic QCD correction) that enhance the cross section by a factor of two\cite{ref:1998dklsz,ref:1999dr,ref:2003gb,ref:2004ddrw,ref:2003yyou,ref:2005fh,ref:2006fh}.
Since the background diagram where a Higgs boson is radiated off the $s$-channel $Z$ boson makes negligible contribution to the signal process, we can measure the top Yukawa coupling by simply counting the number of signal events.
The expected statistical precision for the top Yukawa coupling is then $\Delta g_Y(t) / g_Y(t) = 9.9\%$ for $m_h=125\,$GeV with $1$ab$^{-1}$ at $\sqrt{s}=500\,$GeV \cite{ref:tthRyo2011, ref:tthTabassam2012, ref:tth_DBD, ref:juste1999, ref:gay2007, ref:juste2005}.
Notice that if we increase the center of mass energy by $20\,$GeV, the cross section doubles.
Moving up a little bit hence helps significantly.\\
\end{sloppypar}

\begin{figure}[h]
\centering
\includegraphics[width=0.9\hsize]{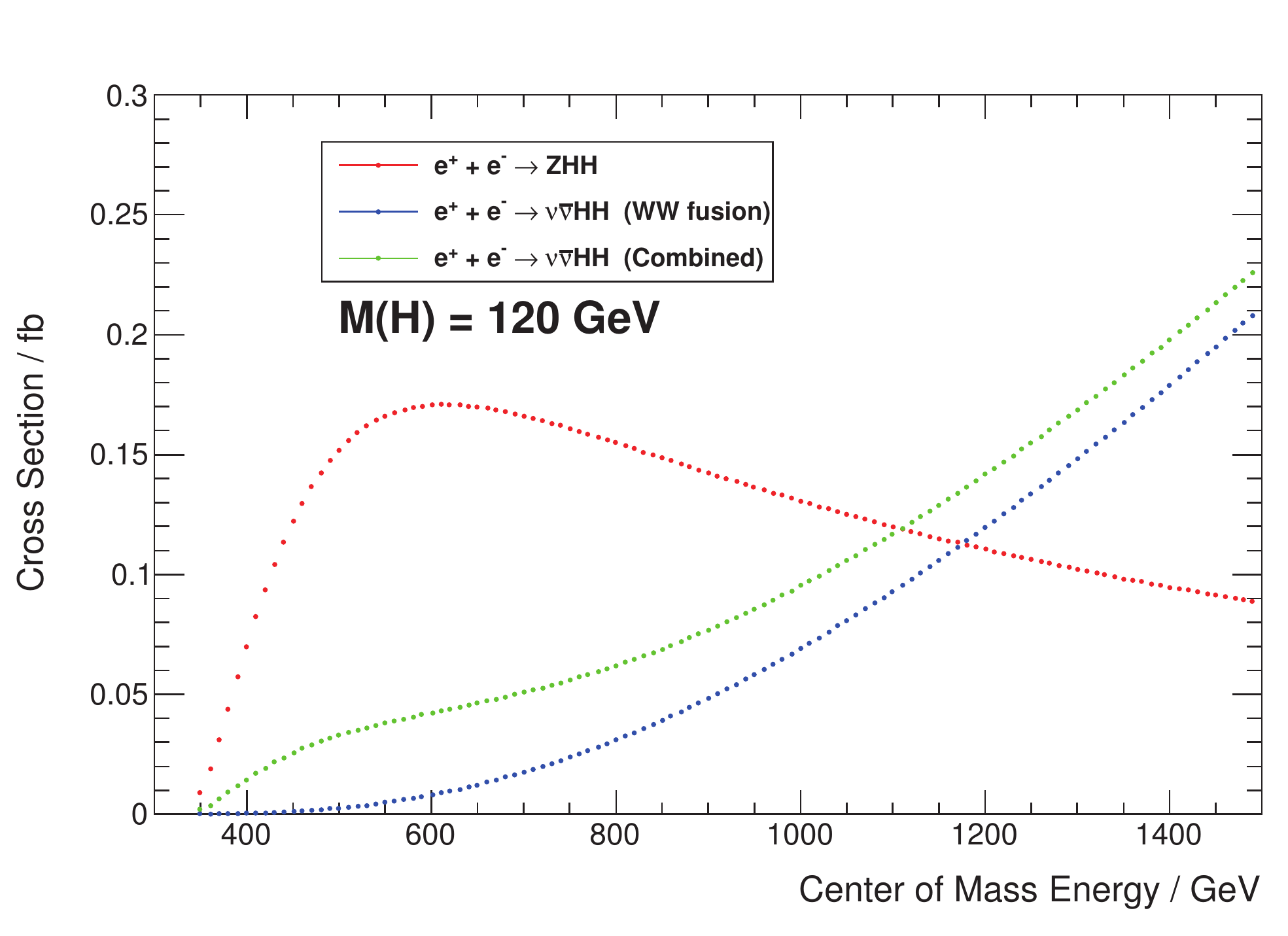}
\caption{Cross sections for the double Higgs production processes, $e^+e^- \to Zhh$ and $e^+e^- \to \nu\bar{\nu}hh$, as a function of $\sqrt{s}$ for $m_h=120\,$GeV.} 
\label{fig:sigzhh_vvhh}
\end{figure}

\begin{figure}[h]
\centering
\includegraphics[width=0.9\hsize]{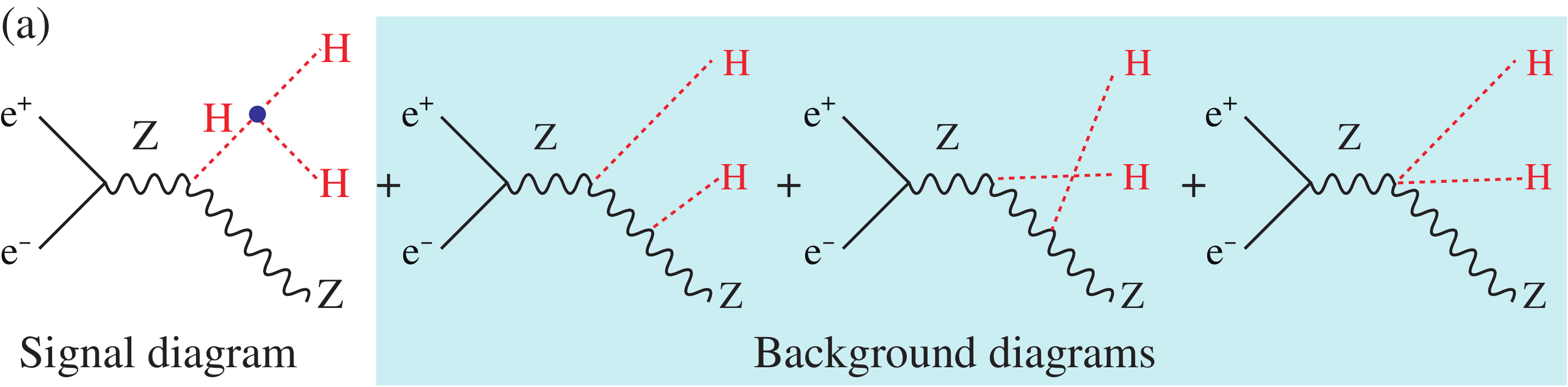}
\includegraphics[width=0.9\hsize]{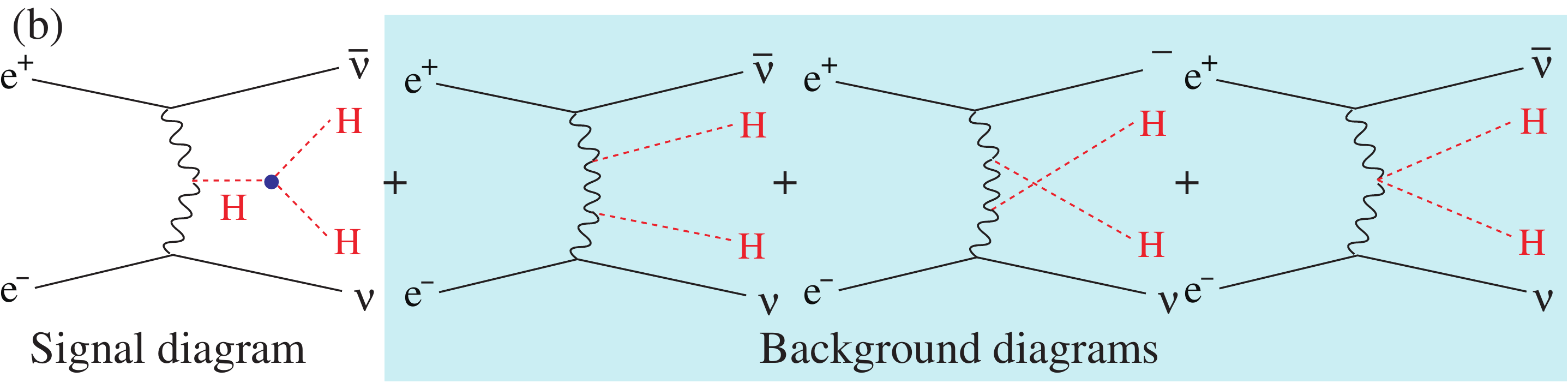}
\caption{Diagrams contributing to (a) $e^+e^- \to Zhh$ and (b) $e^+e^- \to \nu\bar{\nu}hh$.} 
\label{fig:hhdiagrams}
\end{figure}

\begin{figure}[h]
\centering
\includegraphics[width=0.9\hsize]{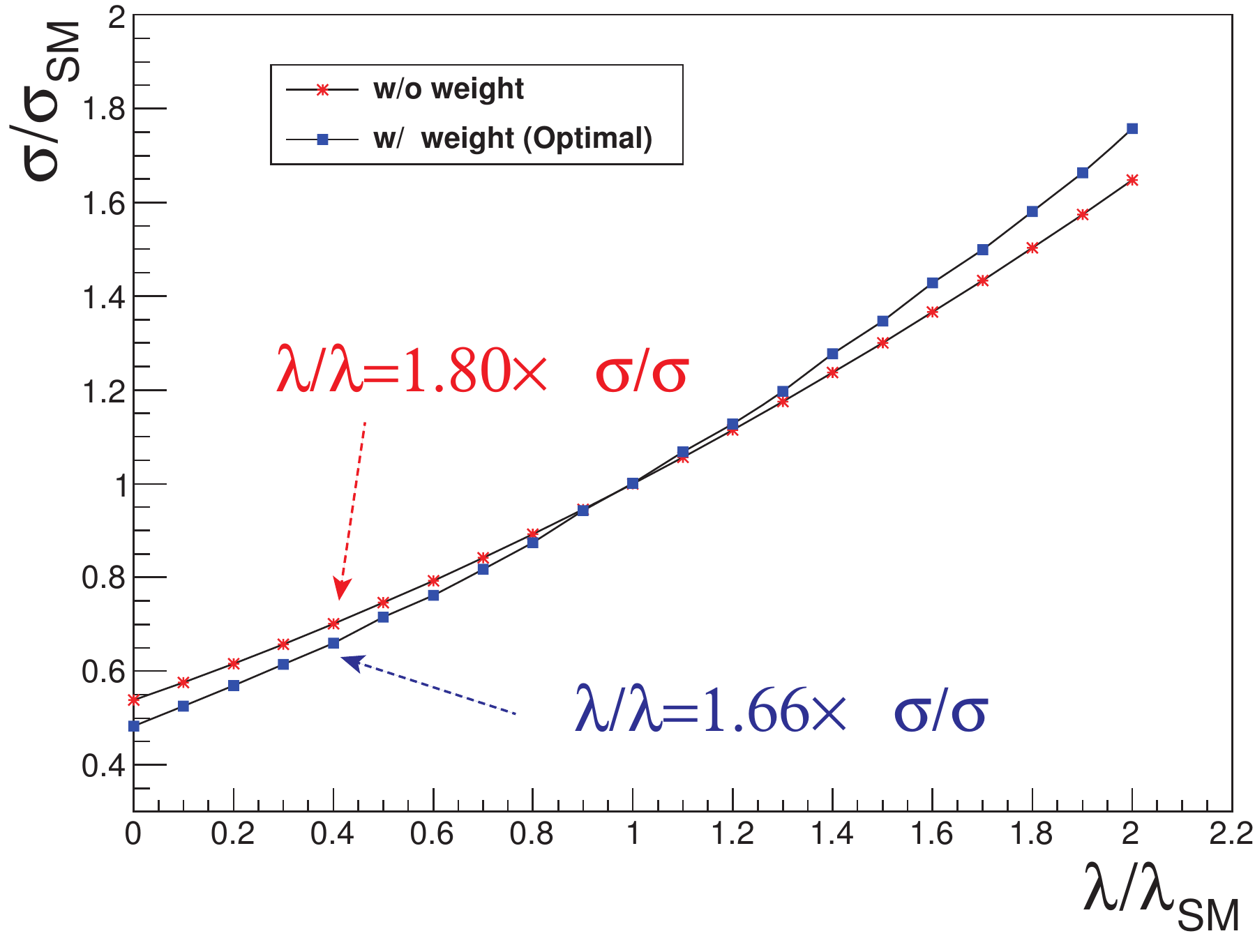}
\includegraphics[width=0.9\hsize]{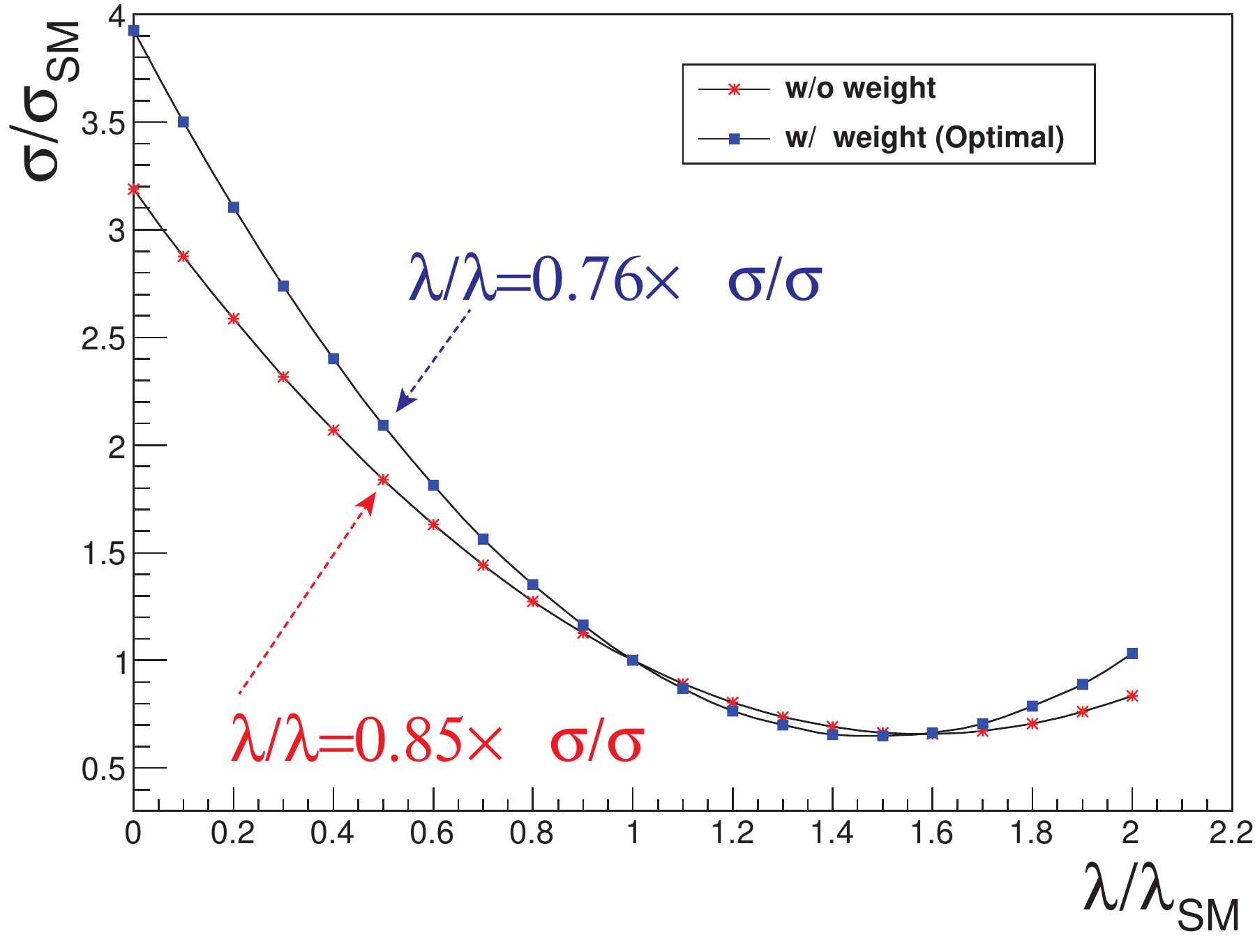}
\caption{ (Upper plot) cross section for $e^+e^- \to Zhh$ at $\sqrt{s}=500$\,GeV normalized by that of the SM as a function of the self-coupling normalized by that of the SM. (Lower plot) a similar plot for $e^+e^- \to \nu\bar{\nu}hh$ at $\sqrt{s}=1$\,TeV.}
\label{fig:sensitivity}
\end{figure}

Even more interesting is the measurement of the trilinear Higgs self-coupling, since it is to observe the force that makes the Higgs boson condense in the vacuum, which is an unavoidable step to uncover the secret of the EW symmetry breaking.
In other words, we need to measure the shape of the Higgs potential.
There are two ways to measure the trilinear Higgs self-coupling.
The first method is to use the double higgsstrahlung process: $e^+e^- \to Zhh$
and the second is by the double Higgs production via $WW$-fusion: $e^+e^- \to \nu\bar{\nu}hh$.
The first process attains its cross section maximum at around $\sqrt{s}=500\,$GeV,
while the second is negligible there but starts to dominate at energies above $\sqrt{s}\simeq 1.2\,$TeV, as seen in Fig.\ref{fig:sigzhh_vvhh}.
In any case the signal cross sections are very small ($0.2\,$fb or less) and as seen in Fig.\ref{fig:hhdiagrams} irreducible background diagrams containing no self-coupling dilute the contribution from the self-coupling diagram, thereby degrading the sensitivity to the self-coupling, even if we can control the relatively huge SM backgrounds from $e^+e^- \to t\bar{t}$, $WWZ$, $ZZ$, $Z\gamma$, $ZZZ$, and $ZZh$.
See Fig.\ref{fig:sensitivity} for the sensitivity factors for $e^+e^- \to Zhh$ at $\sqrt{s}=500\,$GeV and $e^+e^- \to \nu\bar{\nu}hh$ at $\sqrt{s}=1\,$TeV, which are 1.66 (1.80) and 0.76 (0.85), respectively, with (without) weighting to enhance the contribution from the signal diagram. 
Notice that if there were no background diagrams, the sensitivity factor would be $0.5$.
The self-coupling measurement is very difficult even in the clean
environment of the ILC and requires a new flavor tagging algorithm
that precedes jet-clustering, sophisticated neural-net-based data
selection, and the event weighting technique \cite{ref:junping2013,Djouadi:1999gv, ref:gay2001, ref:2001bby, ref:yasui2002,ref:yamashita2003}.  
The current state of the art for the $Zhh$ data
selection is summarized in Table \ref{tab:zhh500}.

\begin{table}[h]
\begin{center}
  \caption{The number of remaining events for the three event
    selection modes: $Zhh \to (\ell\bar{\ell})(b\bar{b})(b\bar{b})$, $
    (\nu\bar{\nu})(b\bar{b})(b\bar{b})$, and $
    (q\bar{q})(b\bar{b})(b\bar{b})$ and corresponding excess and
    measurement sensitivities for $m_h=120\,$GeV at
    $\sqrt{s}=500\,$GeV with $2\,$ab$^{-1}$ and $(e^{-}, e^{+})=(-0.8,
    +0.3)$ beam polarization.}
\begin{tabular}{|c|c|c|c|c|}
   \hline
           mode & signal & BG &
            \multicolumn{2}{c|}{significance} \\
  \cline{4-5}
   & & & excess & meas.\\
  \hline
  \hline
  $Zhh \to (\ell\bar{\ell})(b\bar{b})(b\bar{b})$ 
         & 3.7 & 4.3 & 1.5$\sigma$ & 1.1$\sigma$ \\
  \cline{2-5}
         & 4.5 & 6.0 &  1.5$\sigma$ & 1.2$\sigma$ \\
   \hline
  $Zhh \to (\nu\bar{\nu})(b\bar{b})(b\bar{b})$ 
         & 8.5 & 7.9 & 2.5$\sigma$ & 2.1$\sigma$ \\
   \hline
  $Zhh \to (q\bar{q})(b\bar{b})(b\bar{b})$ 
         & 13.6 & 30.7 & 2.2$\sigma$ & 2.0$\sigma$ \\
  \cline{2-5}
         & 18.8 & 90.6 &  1.9$\sigma$ & 1.8$\sigma$ \\
   \hline
\end{tabular}
\label{tab:zhh500}
\end{center}
\end{table}

Combining all of these three modes, we can achieve $Zhh$ excess
significance of $5\sigma$ and measure the production cross section to
$\Delta \sigma / \sigma = 27\%$, which translates to a relative precision of $44 (48)\%$ for the self-coupling  
with (without) the event weighting for $m_h=120\,$GeV at
$\sqrt{s}=500\,$GeV with $2\,$ab$^{-1}$ and $(e^{-}, e^{+})=(-0.8,
+0.3)$ beam polarization \cite{ref:junping2013}.  The expected
precision is significantly worse than that of the cross section
because of the background diagrams.  Since the sensitivity factor for
the $e^+e^- \to \nu\bar{\nu}hh$ process is much closer to the ideal
$0.5$ and since the cross section for this $WW$-fusion double Higgs
production process increases with the center of mass energy, 
$\sqrt{s} = 1 \tev$ is of particular interest, as will be discussed
  in the next subsection.

\subsubsection{ILC at 1000\,GeV}

As we already pointed out the $WW$-fusion processes become more and
more important at higher energies.  In addition the machine luminosity
usually scales with the center of mass energy.  Together with the
better sensitivity factor we can hence improve the self-coupling
measurement significantly at $\sqrt{s}=1\,$TeV, using the $e^+e^- \to
\nu\bar{\nu}hh$ process.  Table \ref{tab:vvhh1000} summarizes a full
simulation result for the numbers of expected signal and background
events before and after selection cuts with corresponding measurement
significance values.

\begin{table}[h]
\begin{center}
\caption{The numbers of signal and background events before and after selection cuts and  measurement significance for $m_h=120\,$GeV 
at $\sqrt{s}=1\,$TeV with $2\,$ab$^{-1}$ and $(e^{-}, e^{+})=(-0.8, +0.2)$ beam polarization.}
\begin{tabular}{|c|c|c|}
   \hline
           mode & no cut & after cuts \\
  \hline
  \hline
  $\nu\bar{\nu}hh$ ($WW$-fusion) & 272 & 35.7 \\ 
  \hline
  $\nu\bar{\nu}hh$ ($Zhh$) & 74.0 & 3.88 \\
  \hline
  BG ($t\bar{t}/\nu\bar{\nu}Zh$) & $7.86 \times 10^{5}$ & 33.7 \\
  \hline
  meas. significance & 0.30 & 4.29 \\
  \hline 
\end{tabular}
\label{tab:vvhh1000}
\end{center}
\end{table}

With $2\,$ab$^{-1}$ and $(e^{-}, e^{+})=(-0.8, +0.2)$ beam
polarization at $\sqrt{s}=1\,$TeV, we would be able to determine the
cross section for the $e^+e^- \to \nu\bar{\nu}hh$ process to $\Delta
\sigma / \sigma = 23\%$, corresponding to the self-coupling precision
of $\Delta \lambda / \lambda = 18 (20)\%$ with (without) the event
weighting to enhance the contribution from the signal diagram for
$m_h=120\,$GeV \cite{ref:junping2013}.  According to preliminary
results from a on-going full simulation study \cite{ref:kurata2013},
adding $hh \to WW^*b\bar{b}$ would improve the self-coupling
measurement precision by about 20\% relatively, which means $\Delta
\lambda / \lambda = 21\%$ for $m_h=125\,$GeV with the baseline
integrated luminosity of $1\,$ab$^{-1}$ at 1\,TeV.

At $\sqrt{s}=1\,$TeV, the $e^+e^- \to t\bar{t}h$ process is also near
its cross section maximum, making concurrent measurements of the
self-coupling and top Yukawa coupling possible.  We will be able to
observe the $e^+e^- \to t\bar{t}h$ events with $12\sigma$ significance
in 8-jet mode and $8.7\sigma$ significance in lepton-plus-6-jet mode,
corresponding to the relative error on the top Yukawa coupling of
$\Delta g_Y(t) / g_Y(t) = 3.1\%$ with $1\,$ab$^{-1}$ and $(e^{-},
e^{+})=(-0.8, +0.2)$ beam polarization at $\sqrt{s}=1\,$TeV for
$m_h=125\,$GeV \cite{ref:tth_Price2013}.

However, obvious but most important advantage of higher energies in
terms of Higgs physics is its higher mass reach to extra Higgs bosons
expected in extended Higgs sectors and higher sensitivity to $W_LW_L$
scattering to decide whether the Higgs sector is strongly interacting
or not.  In any case thanks to the higher cross section for the
$WW$-fusion $e^+e^- \to \nu\bar{\nu}h$ process at $\sqrt{s}=1\,$TeV,
we can expect significantly better precisions for the $\sigma \times
\br$ measurements (see Table \ref{tab:ilc_obs_baseline}), which also
allows us to access very rare decays such as $h \to \mu^+\mu^-$
\cite{ref:2003tb, ref:2001br}.

\begin{table}[h]
\begin{center}
\caption{Independent Higgs measurements using the higgsstrahlung ($Zh$) and the $WW$-fusion ($\nu\bar{\nu}h$) processes for $m_h=125\,$GeV at three energies: 
$\sqrt{s}=250\,$GeV with $250\,$fb$^{-1}$, $500\,$GeV with $500\,$fb$^{-1}$ both with  $(e^{-}, e^{+})=(-0.8, +0.3)$ beam polarization, $\sqrt{s}=1\,$TeV with $1\,$ab$^{-1}$ and $(e^{-}, e^{+})=(-0.8, +0.2)$ beam polarization.}
\begin{tabular}{|c|c|c|c|c|c|}
   \hline
           $\sqrt{s}$ & \multicolumn{2}{c|}{250\,GeV} & \multicolumn{2}{c|}{500\,GeV} & 1\,TeV \\
  \hline
           lumi.  & \multicolumn{2}{c|}{250\,fb$^{-1}$} & \multicolumn{2}{c|}{500\,fb$^{-1}$} & 1\,ab$^{-1}$ \\
  \hline
           process & $Zh$ & $\nu\bar{\nu}h$ & $Zh$ &  $\nu\bar{\nu}h$ & $\nu\bar{\nu}h$  \\
  \cline{1-6}
                         & \multicolumn{5}{c|}{$\Delta \sigma / \sigma$} \\
  \cline{2-6}
                         & 2.6\% & - & 3.0\% & - & - \\
  \cline{1-6}
             mode    & \multicolumn{5}{c|}{$\Delta (\sigma \cdot \br) / (\sigma \cdot \br)$} \\
  \hline   
           $h \to b\bar{b}$          & 1.2\% & 10.5\% & 1.8\% & 0.66\% & 0.5\% \\
  \hline
           $h \to c\bar{c}$          & 8.3\% &              & 13\% & 6.2\%   & 3.1\% \\
  \hline
           $h \to gg$                  & 7.0\% &              & 11\% & 4.1\%   & 2.3\% \\
  \hline
           $h \to WW^*$             & 6.4\% &              & 9.2\% & 2.4\%   & 1.6\% \\
  \hline
           $h \to \tau^+\tau^-$    & 4.2\% &              & 5.4\% & 9.0\%   & 3.1\% \\
  \hline
           $h \to ZZ^*$             & 18\% &              & 25\% & 8.2\%   & 4.1\% \\
  \hline
           $h \to \gamma\gamma$   & 34\% &              & 34\% & 23\%   & 8.5\% \\
  \hline
           $h \to \mu^+\mu^-$   & 100\% &     -      & -  & -   & 31\% \\
  \hline
\end{tabular}
\label{tab:ilc_obs_baseline}
\end{center}
\end{table}

\subsubsection{ILC 250+500+1000: global fit for couplings}

\begin{figure}[h]
\centering
\includegraphics[width=0.89\hsize]{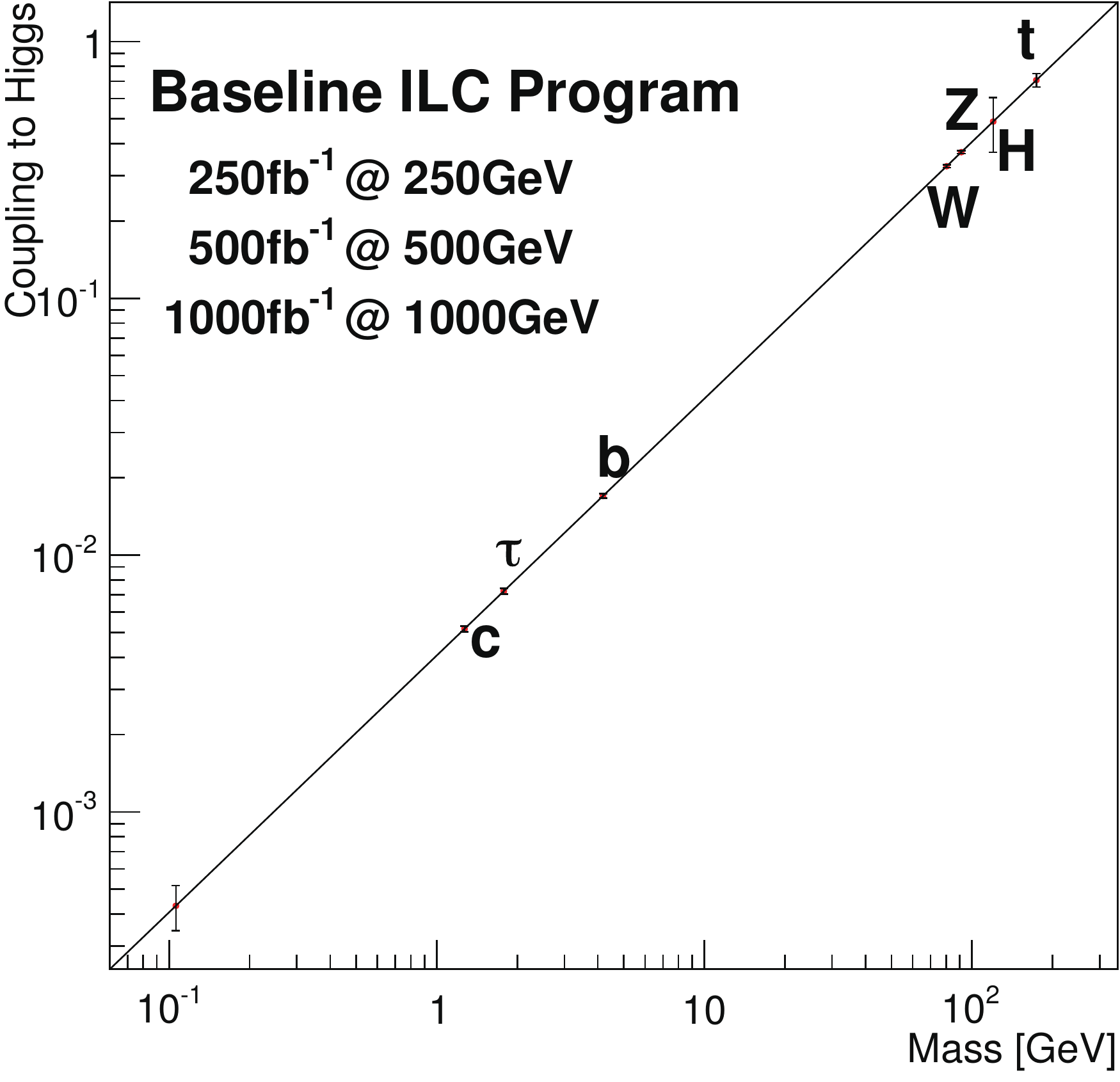}
\caption{Expected mass-coupling relation for the SM case after the full ILC program.} 
\label{fig:mass-coupling-full-ILC}
\end{figure}

The data at $\sqrt{s}=250$, $500$, and $1000\,$GeV can be combined to
perform a global fit to extract various Higgs couplings
\cite{ref:global_fitJunping2013}.  We have 33 $\sigma\times
\br$ measurements: 31 shown in Table \ref{tab:ilc_obs_baseline}
plus two $\sigma(t\bar{t}h) \times \br(h\to b\bar{b})$ measurements at
$\sqrt{s}=500$ and $1000\,$GeV.  The key is the recoil mass
measurement that unlocks the door to a fully model-independent
analysis.  Notice that such a fully model-independent analysis is
impossible at the LHC.  As shown in Table \ref{tab:ilc_obs_baseline},
we can measure the recoil mass cross section at $\sqrt{s}=250$ and
$500\,$GeV.  Altogether we have 35 independent measurements: 33
$\sigma\times \br$ measurements ($Y_i : i=1\cdots 33$) and 2
$\sigma(Zh)$ measurements ($Y_{34,35}$).  We can then define a
$\chi^2$ function:
\begin{eqnarray}
\chi^2 & = & \sum_{i=1}^{35} \left(\frac{Y_i - Y'_i}{\Delta Y_i}\right)^2
\end{eqnarray}
where
\begin{eqnarray}
Y'_i & := & F_i \cdot \frac{g^2_{hA_i A_i} g^2_{hB_i B_i}}{\Gamma_0} ~ (i=1, \cdots, 33)
\end{eqnarray}
with $A_i$ being $Z$, $W$, or $t$, and $B_i$ being $b$, $c$, $\tau$,
$\mu$, $g$, $\gamma$, $Z$, and $W$, $\Gamma_0$ denoting the total
  width and  
\begin{eqnarray}
F_i & = & S_i G_i
\end{eqnarray}
with
\begin{eqnarray}
S_i & = & \left(\frac{\sigma_{Zh}}{g^2_{hZZ}}\right), ~ \left(\frac{\sigma_{\nu\bar{\nu}h}}{g^2_{hWW}}\right), ~{\rm or} ~ \left(\frac{\sigma_{t\bar{t}h}}{g^2_{htt}}\right) \cr\rule{0in}{5ex}
G_i & = & \left( \frac{\Gamma_i}{g^2_i} \right) .
\end{eqnarray}
Cross section calculations ($S_i$) do not involve QCD ISR unlike with the LHC.
Partial width calculations ($G_i$), being normalized by the coupling squared, do not need quark mass as input.
We are hence confident that the goal theory errors for $S_i$ and $G_i$ will be at the 0.1\% level at the time of ILC running.
The free parameters are 9 coupling constants: $g_{hbb}$, $g_{gcc}$, $g_{h\tau\tau}$, $g_{h\mu\mu}$, $g_{hgg}$, $g_{h\gamma\gamma}$, $g_{hZZ}$, $g_{hWW}$, and 1 total width: $\Gamma_0$.
Table \ref{tab:ilc_couplings_baseline} summarizes the expected coupling precisions for $m_h=125\,$GeV with the baseline integrated luminosities of 250\,fb$^{-1}$ at $\sqrt{s}=250\, $GeV, 500\,fb$^{-1}$ at 500\,GeV both with $(e^{-}e^{+})=(-0.8, +0.3)$ beam polarization, and 1\,ab$^{-1}$ at 1\,TeV with $(e^{-}e^{+})=(-0.8, +0.2)$ beam polarization. 
\begin{table}[h]
\begin{center}
\caption{Expected precisions for various couplings of the Higgs boson
  with $m_h=125\,$GeV from a model-independent fit to observables
  listed in Table \ref{tab:ilc_obs_baseline} at three energies:
  $\sqrt{s}=250\,$GeV with $250\,$fb$^{-1}$, $500\,$GeV with
  $500\,$fb$^{-1}$ both with $(e^{-}, e^{+})=(-0.8, +0.3)$ beam
  polarization, $\sqrt{s}=1\,$TeV with $2\,$ab$^{-1}$ and $(e^{-},
  e^{+})=(-0.8, +0.2)$ beam polarization, 
cf.\ \cite{ref:snowmass_higgs} and Scen. `Snow' in \cite{parametergroup}. 
Values with (*) assume
  inclusion of $hh\to WW^*b\bar{b}$ decays.}
\begin{tabular}{|c|c|c|c|}
   \hline
                  & \multicolumn{3}{c|}{$\sqrt{s}$ (GeV)}\\
   \cline{2-4}
    coupling & 250 & 250+500 & 250 + 500 + 1000 \\
   \hline
   $hZZ$ & 1.3\% & 1.0\% & 1.0\% \\
   \hline
   $hWW$ & 4.8\% & 1.1\% & 1.1\% \\
   \hline
   $hbb$ & 5.3\% & 1.6\% & 1.3\% \\
   \hline
   $hcc$ & 6.8\% & 2.8\% & 1.8\% \\
   \hline
   $hgg$ & 6.4\% & 2.3\% & 1.6\% \\
   \hline
   $h\tau\tau$ & 5.7\% & 2.3\% & 1.6\% \\
   \hline
   $h\gamma\gamma$ & 18\% & 8.4\% & 4.0\% \\
   \hline
   $h\mu\mu$ & 91\% & 91\% & 16\% \\
   \hline
   $\Gamma_0$ & 12\% & 4.9\% & 4.5\% \\
   \hline
   $htt$ & - & 14\% & 3.1\% \\
   \hline\hline
   $hhh$ & - & 83\%(*) & 21\%(*) \\
   \hline
\end{tabular}
\label{tab:ilc_couplings_baseline}
\end{center}
\end{table}
The expected coupling precisions are plotted in the mass-coupling plot expected for the SM Higgs sector in Fig.\ref{fig:mass-coupling-full-ILC}.
The error bars for most couplings are almost invisible in this logarithmic plot.

\subsubsection{Synergy: LHC + ILC}

\begin{figure}[h]
\centering
\includegraphics[width=0.9\hsize]{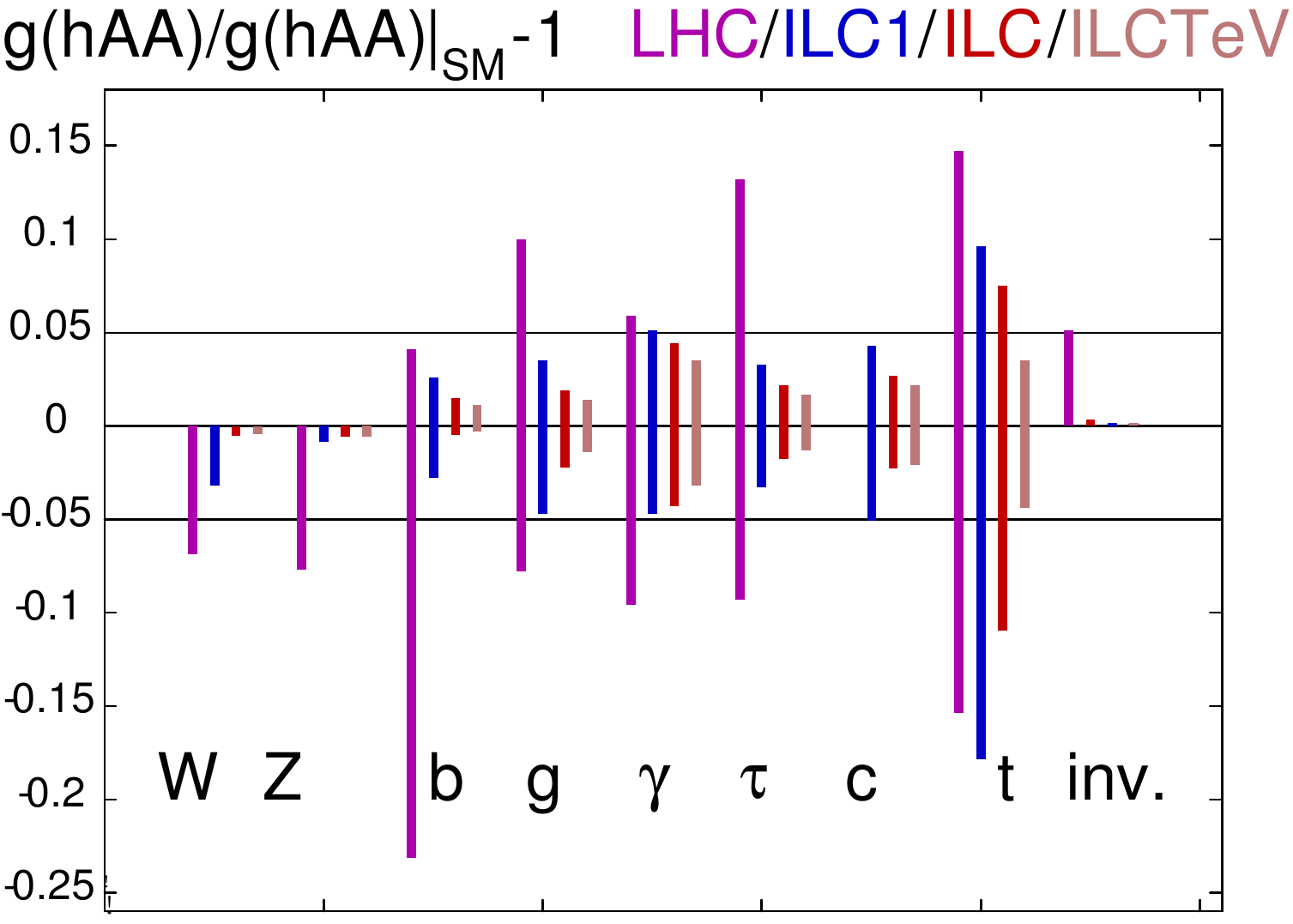}
\caption{Comparison of the capabilities of the LHC and the ILC, when the ILC data in various stages: ILC1 with $250\,$fb$^{-1}$ at $\sqrt{s}=250$, ILC: $500\,$fb$^{-1}$ at $500\,$GeV, and ILCTeV: $1\,$ab$^{-1}$ at $1\,$TeV are cumulatively added to the LHC data with $300\,$fb$^{-1}$ at $14\,$TeV  \cite{ref:peskin}.} 
\label{fig:peskin}
\end{figure}

\begin{sloppypar}
So far we have been discussing the precision Higgs physics expected at
the ILC.  It should be emphasized, however, that the LHC is expected
to impose significant constraints on possible deviations of the
Higgs-related couplings from their SM values by the time the ILC will
start its operation, even though fully model-independent analysis is
impossible with the LHC alone.  Nevertheless, Refs.~\cite{HcoupLHCSM,ref:peskin} demonstrated that with a reasonably weak assumption
such as the $hWW$ and $hZZ$ couplings will not exceed the SM values
the LHC can make reasonable measurements of most Higgs-related
coupling constants except for the $hcc$ coupling.  Figure
\ref{fig:peskin} shows how the coupling measurements would be improved
by adding, cumulatively, information from the ILC with
$250\,$fb$^{-1}$ at $\sqrt{s}=250$, $500\,$fb$^{-1}$ at $500\,$GeV,
and $1\,$ab$^{-1}$ at $1\,$TeV to the LHC data with $300\,$fb$^{-1}$
at $14\,$TeV.
\end{sloppypar}
The figure tells us that the addition of the $250\,$GeV data, the $hZZ$ coupling in particular, from the ILC allows the absolute normalization and significantly improves all the couplings.
It is interesting to observe the synergy for the measurement of the $h\gamma\gamma$ coupling, whose precision significantly exceeds that of the ILC alone. 
This is because the LHC can precisely determine the ratio of the $h\gamma\gamma$ coupling to the $hZZ$ coupling, while the ILC provides a precision measurement of the $hZZ$ coupling from the recoil mass measurement.
The addition of the $500\,$GeV data from the ILC further improves the precisions, this time largely due to the better determination of the Higgs total width.
Finally as we have seen above, the addition of the $1\,$TeV data from the ILC improves the top Yukawa coupling drastically with even further improvements of all the other couplings except for the $hWW$ and $hZZ$ couplings which are largely limited by the cross section error from the recoil mass measurement at $\sqrt{s}=250\,$GeV. 
This way we will be able to determine these couplings to ${\cal O}(1\%)$ or better.
The {\it SFitter} group performed a similar but more model-independent analysis and obtained qualitatively the same conclusions \cite{ref:SFitter}.
This level of precision matches what we need to fingerprint different BSM scenarios, when nothing but the 125~GeV boson would be found at the LHC (see Table \ref{tab:gupta}).
\begin{table}[h]
\begin{center}
\caption{Maximum possible deviations when nothing but the 125~GeV boson would be found at the LHC  \cite{ref:gupta}.}
\begin{tabular}{l c c c}
   \hline
   \hline
   & $\Delta hVV$ & $\Delta h\bar{t}t$ & $\Delta h\bar{b}b$ \\
  \hline
  Mixed-in Singlet & 6\% & 6\% & 6\% \\
  Composite Higgs & 8\% & tens of \% & tens of \% \\
  Minimal SUSY & $<1$\% & 3\% & 10\%$^a$, 100\%$^b$ \\
  LHC {\scriptsize 14\,TeV, 3\,ab$^{-1}$} & 8\% & 10\% & 15\% \\
   \hline
   \hline
\end{tabular}
\label{tab:gupta}
\end{center}
\end{table}
These numbers can be understood from the following formulas for the
three different models in the decoupling limit
(see~\cite{ref:dbd} for definitions and details), 
\begin{eqnarray}
\mbox{~Mixing with singlet: } & & \cr
	\frac{g_{hVV}}{g_{h_{\rm SM} VV}} = \frac{g_{hff}}{g_{h_{\rm SM} ff}}
&=& \cos\theta \simeq 1 - \frac{\delta^2}{2} \cr\rule{0in}{5ex}
\mbox{Composite Higgs: }~~~ & & \cr
	\frac{g_{hVV}}{g_{h_{\rm SM}VV}} &\simeq& 1 - 3\% 
\left( \frac{1~{\rm TeV}}{f} \right)^2 \cr
	\frac{g_{hff}}{g_{h_{\rm SM}ff}} &\simeq& \left\{ 
	\begin{array}{ll}
	1 - 3\% \left( \frac{1~{\rm TeV}}{f} \right)^2 & {\rm (MCHM4)} \cr
	1 - 9\% \left( \frac{1~{\rm TeV}}{f} \right)^2 & {\rm (MCHM5).}
	\end{array} \right. \cr\rule{0in}{3ex}
\mbox{Supersymmetry:}~~ & & \cr
	\frac{g_{hbb}}{g_{h_{\rm SM} bb}} &=& 
\frac{g_{h \tau\tau}}{g_{h_{\rm SM} \tau\tau}} \simeq 1 + 1.7\% \left( \frac{1 \ 
{\rm TeV}}{m_A} \right)^2.
\nonumber
\end{eqnarray}
The different models predict different deviation patterns.
The ILC together with the LHC will be able to fingerprint these models or set the lower limit on the energy scale for BSM physics.

\subsubsection{Model-dependent global fit: example of fingerprinting}

\begin{figure}[h]
\centering
\includegraphics[width=0.9\hsize]{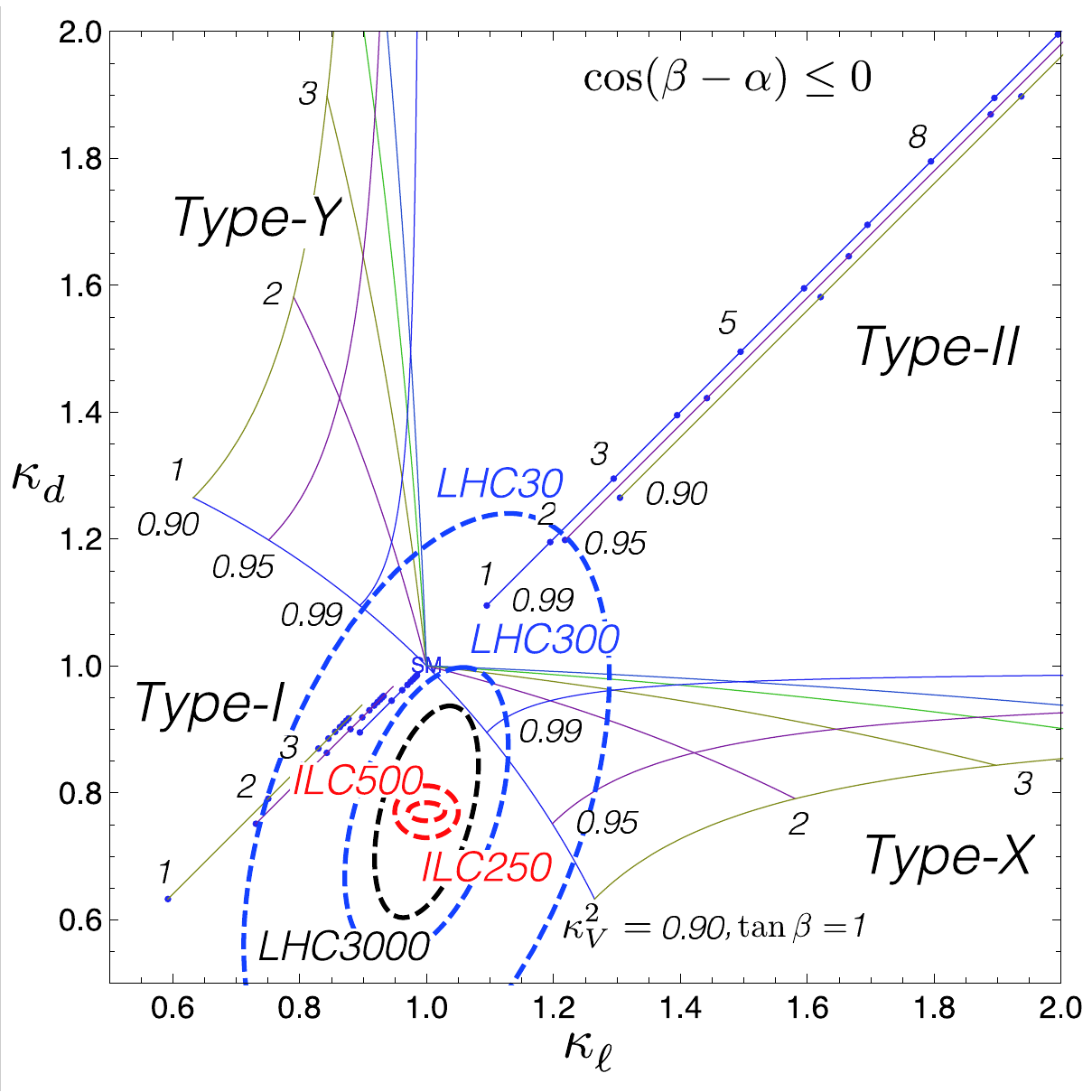}
\caption{Comparison of the model-discrimination capabilities of the LHC and the ILC \cite{ref:kanemura2013}.} 
\label{fig:kl_kd}
\end{figure}

\begin{figure}[h]
\centering
\includegraphics[width=0.9\hsize]{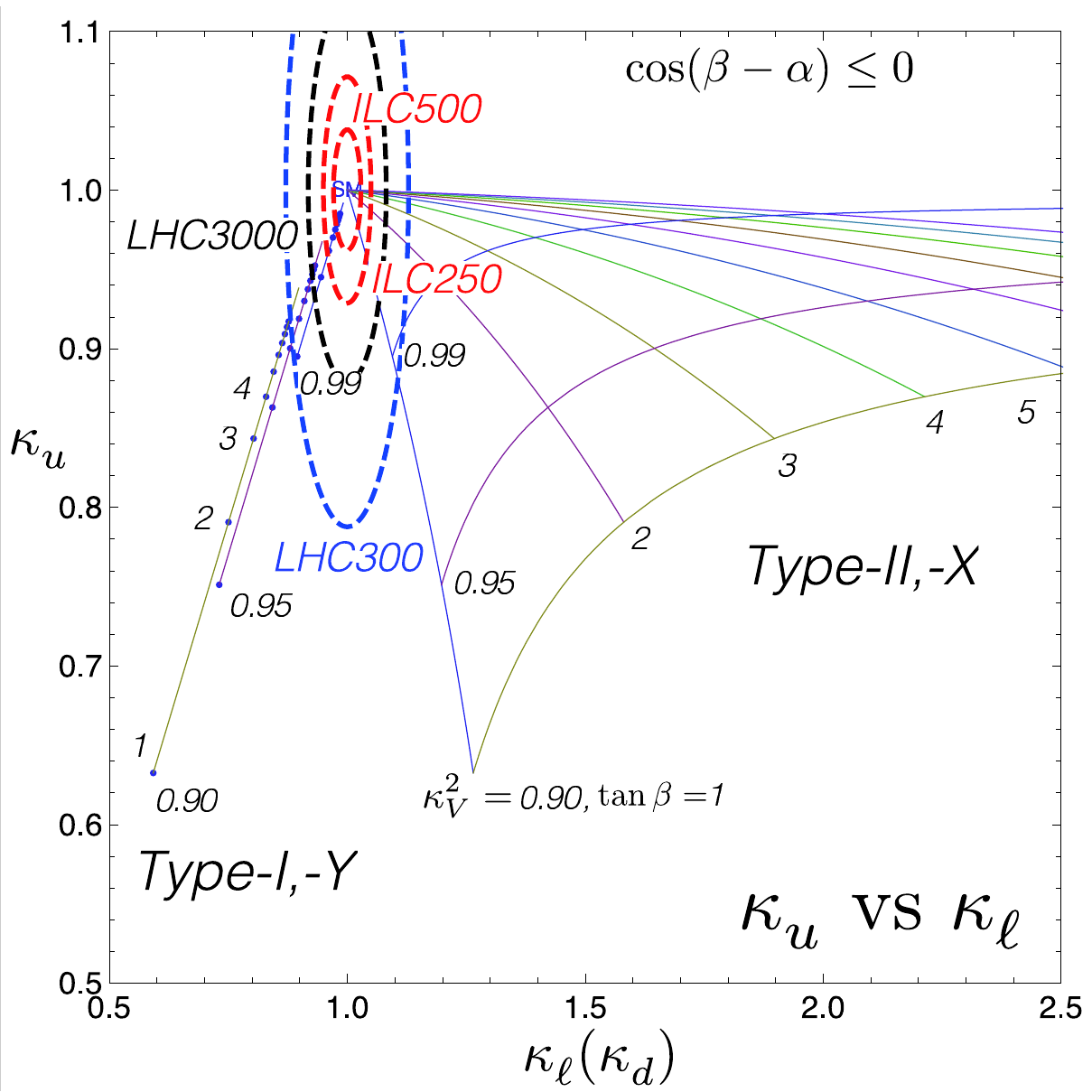}
\caption{Comparison of the model-discrimination capabilities of the LHC and the ILC \cite{ref:kanemura2013}.} 
\label{fig:kl_ku}
\end{figure}

\begin{figure}[h]
\centering
\includegraphics[width=0.9\hsize]{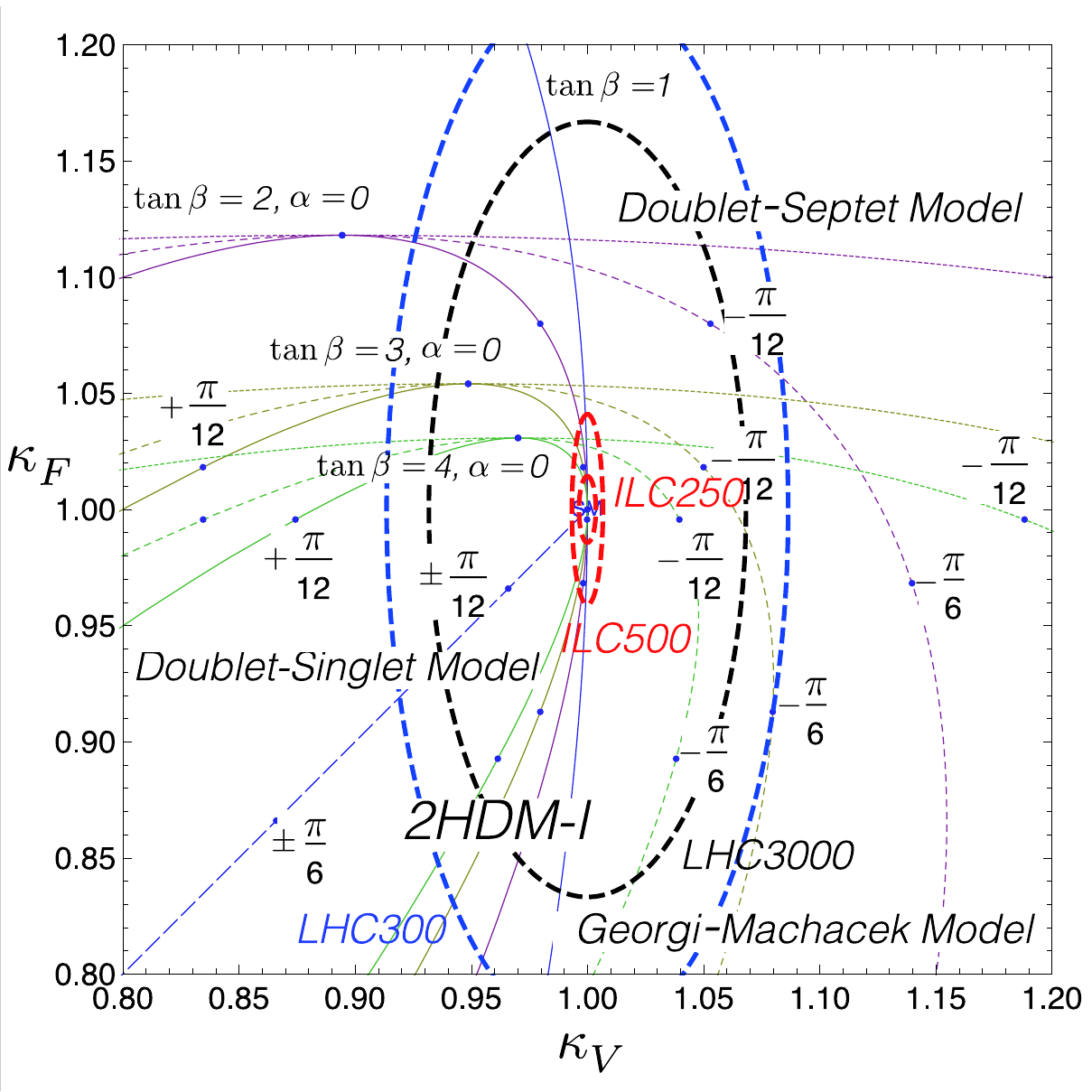}
\caption{Comparison of the model-discrimination capabilities of the LHC and the ILC \cite{ref:kanemura2013}.} 
\label{fig:kv_kf}
\end{figure}

As mentioned above, the LHC needs some model assumption to extract Higgs couplings.
If we use stronger model assumptions we may have higher discrimination power at the cost of loss of generality.
As an example of such a model-dependent analysis, let us consider here a
7-parameter 
global fit with the following assumptions:
\begin{eqnarray}
~~~~~~ \kappa_c &=& \kappa_t  ~=:~  \kappa_u , \cr
~~~~~~ \kappa_s &=& \kappa_b ~=:~  \kappa_d , \cr
~~~~~~ \kappa_\mu &=& \kappa_\tau ~=:~ \kappa_\ell , \cr
\mbox{and} ~~~~~~~ && \cr
~~~~~~ \Gamma_{\rm tot} &=& \sum_{i \in \mbox{SM decays}} \, \Gamma_i^{\rm SM} \, \kappa_i^2  ,
\end{eqnarray}
where $\kappa_i := g_i / g_i ({\rm SM})$ is a Higgs coupling normalized by its SM value.
The first three of these constrain the relative deviations of the up-type and down-type quark Yukawa couplings as well as that of charged leptons to be common in each class, while the last constraint restricts unknown decay modes to be absent.
The results of the global fits assuming projected precisions for the LHC and the ILC are summarized in Table~\ref{tab:model_dep_fit} \cite{ref:global_fitJunping2013}.
\begin{table}[h]
\begin{center}
\caption{Expected Higgs precisions on normalized Higgs couplings ($\kappa_i := g_i / g_i ({\rm SM})$) for $m_h=125\,$GeV from model-dependent 7-parameter fits for the LHC and the ILC, where $\kappa_c = \kappa_t =: \kappa_u$, $\kappa_s = \kappa_b =: \kappa_d$, $\kappa_\mu = \kappa_\tau =: \kappa_\ell$, and $\Gamma_{\rm tot} = \sum \Gamma_i^{\rm SM} \, \kappa_i^2$ are assumed.
}
\begin{tabular}{|c|c|c|c|c|}
   \hline
     Facility & LHC & HL-LHC & ILC500 & ILC1000 \\
     {\scriptsize $\sqrt{s}\,$(GeV)} & {\scriptsize 1,400} & {\scriptsize 14,000} & {\scriptsize 250/500} & {\scriptsize 250/500/1000} \\
     {\scriptsize $\int {\cal L} dt\,$(fb$^{-1}$)} & {\scriptsize300/exp} &{\scriptsize 3000/exp} & {\scriptsize 250+500} & {\scriptsize 250+500+1000} \\
   \hline 
   $\kappa_\gamma$ & 5-7\% & 2-5\% & 8.3\% & 3.8\% \\
   $\kappa_g$            & 6-8\% & 3-5\% & 2.0\% & 1.1\% \\
   $\kappa_W$             & 4-6\% & 2-5\% & 0.39\% & 0.21\% \\
   $\kappa_Z$             & 4-6\% & 2-4\% & 0.49\% & 0.50\% \\
   $\kappa_\ell$           & 6-8\% & 2-5\% & 1.9\%  & 1.3\% \\
   $\kappa_d$           & 10-13\% & 4-7\% & 0.93\%  & 0.51\% \\
   $\kappa_u$           & 14-15\% & 7-10\% & 2.5\%  & 1.3\% \\
   \hline
\end{tabular}
\label{tab:model_dep_fit}
\end{center}
\end{table}
Figures \ref{fig:kl_kd} and \ref{fig:kl_ku} compare the model discrimination power of the LHC and the ILC in the $\kappa_\ell$-$\kappa_d$ and $\kappa_\ell (\kappa_d)$-$\kappa_u$ planes for the four types of two-Higgs-doublet model discussed in \ref{sec:ilc_higgs_intro} \cite{ref:snowmass_higgs_white_paper,ref:kanemura2013}. 
Figure \ref{fig:kv_kf} is a similar plot in the $\kappa_V$-$\kappa_F$
plane showing the discrimination power for four models: doublet-singlet
model, 2HDM-I, Georgi-Machacek model, and doublet-septet model, all of which naturally realize $\rho = 1$ at the tree level \cite{ref:snowmass_higgs_white_paper,ref:kanemura2013}.

\subsubsection{High Luminosity ILC?}

\begin{figure}[h]
\centering
\includegraphics[width=0.9\hsize]{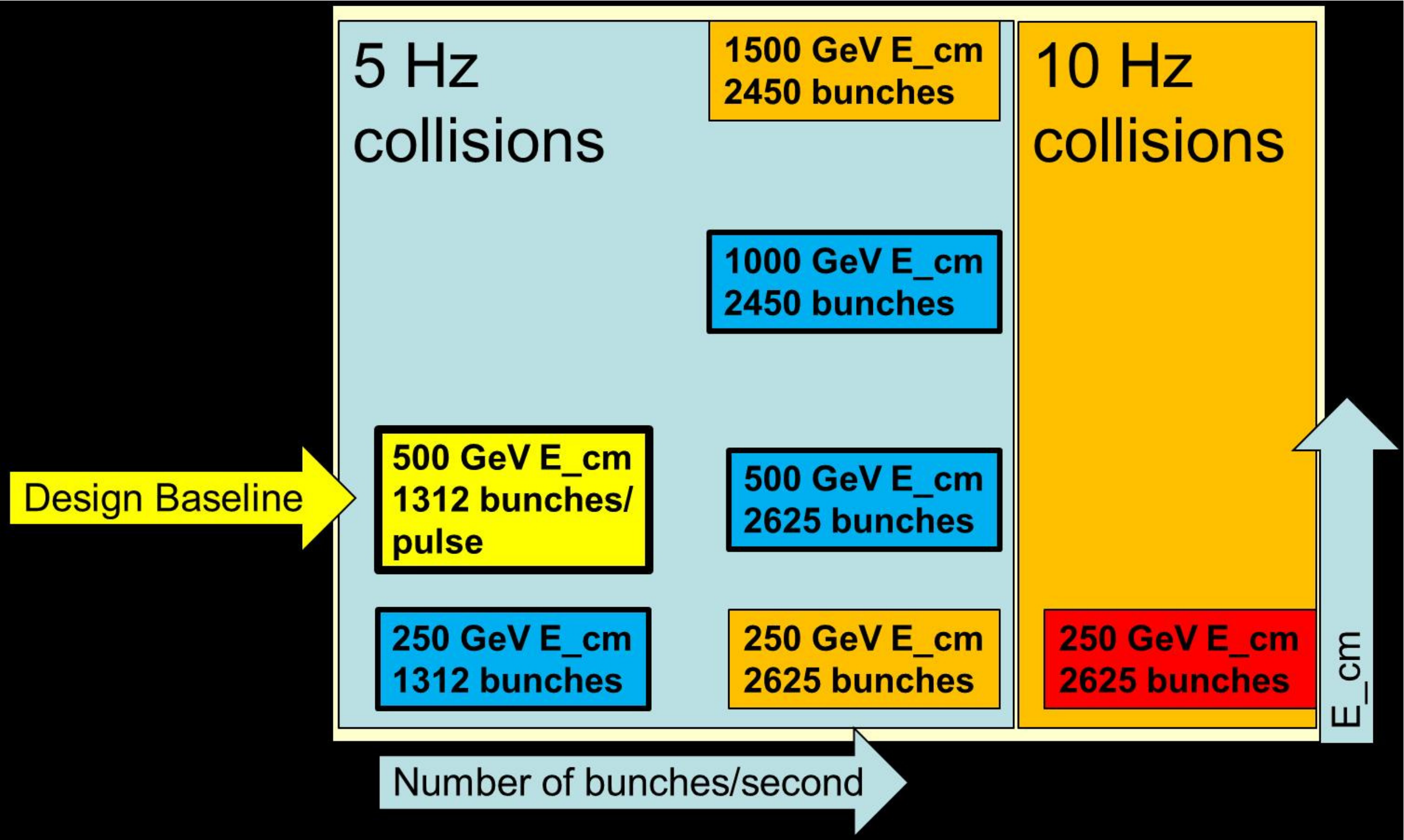}
\caption{Possible machine upgrade scenarios for the ILC \cite{ref:ilc_tdr, ref:snowmass_higgs_white_paper}.} 
\label{fig:ilc_upgrade}
\end{figure}

We have seen the crucial role played by the recoil mass measurement for the model-independent coupling extraction.
We have also pointed out that because of this the recoil mass measurement would eventually limit the coupling precisions achievable with the ILC.
Given the situation, let us now consider the possibility of luminosity upgrade.
As a matter of fact, the ILC technical design report (TDR) \cite{ref:ilc_tdr} describes some possible luminosity and energy upgrade scenarios, which are sketched in Fig.\ref{fig:ilc_upgrade} as blue boxes.
In order to improve the recoil mass measurement significantly a new luminosity upgrade option (doubling of the number of bunches plus 10\,Hz collisions instead of nominal 5\,Hz) was proposed for the 250\,GeV running in the Snowmass 2013 process \cite{ref:snowmass_higgs_white_paper} (see the red box in Fig.\ref{fig:ilc_upgrade}).
It should be noted that the number of bunches was 2625 in the original ILC design given in the reference design report \cite{ref:RDR}, which was reduced to 1312 in the TDR so as to reduce the construction cost.
The 10\,Hz operation is practical at 250\,GeV, since the needed wall plug power is lower at the lower energy.
The upgrade would hence allow a factor of four luminosity upgrade at $\sqrt{s}=250\,$GeV.
Let us now assume that after the baseline program at $\sqrt{s}=250$, $500$, and $1000\,$GeV we will run at the same three energies with the luminosity upgrade, thereby achieving $1150\,$fb$^{-1}$ at 250\,GeV, $1600\,$fb$^{-1}$ at 500\,GeV, and $2500\,$fb$^{-1}$ at 1000\,GeV.\\

\begin{sloppypar}
The expected precisions for the independent Higgs-related measurements
are summarized in Table \ref{tab:ilc_obs_lumiup} for the full data
after the luminosity upgraded running.  Corresponding expected
precisions for various Higgs couplings are tabulated in Table
\ref{tab:ilc_couplings_lumiup}.  The table shows that with the
luminosity upgrade we can achieve sub-\% level precisions for most of
the Higgs couplings even with the completely model-independent
analysis.
\end{sloppypar}

\begin{table}[h]
\begin{center}
\caption{Similar table to Table \ref{tab:ilc_obs_baseline} but with the luminosity upgrade described in the text: 
1150\,fb$^{-1}$ at 250\,GeV, 1600\,fb$^{-1}$ at 500\,GeV, 
and 2500\,fb$^{-1}$ at 1\,TeV.}
\begin{tabular}{|c|c|c|c|c|c|}
   \hline
           $\sqrt{s}$ & \multicolumn{2}{c|}{250\,GeV} & \multicolumn{2}{c|}{500\,GeV} & 1\,TeV \\
  \hline
           lumi.  & \multicolumn{2}{c|}{1150\,fb$^{-1}$} & \multicolumn{2}{c|}{1600\,fb$^{-1}$} & 2.5\,ab$^{-1}$ \\
  \hline
           process & $Zh$ & $\nu\bar{\nu}h$ & $Zh$ &  $\nu\bar{\nu}h$ & $\nu\bar{\nu}h$  \\
  \cline{1-6}
                         & \multicolumn{5}{c|}{$\Delta \sigma / \sigma$} \\
  \cline{2-6}
                         & 1.2\% & - & 1.7\% & - & - \\
  \cline{1-6}
             mode    & \multicolumn{5}{c|}{$\Delta (\sigma \cdot \br) / (\sigma \cdot \br)$} \\
  \hline   
           $h \to b\bar{b}$          & 0.56\% & 4.9\% & 1.0\% & 0.37\% & 0.3\% \\
  \hline
           $h \to c\bar{c}$          & 3.9\% &              & 7.2\% & 3.5\%   & 2.0\% \\
  \hline
           $h \to gg$                  & 3.3\% &              & 6.0\% & 2.3\%   & 1.4\% \\
  \hline
           $h \to WW^*$             & 3.0\% &              & 5.1\% & 1.3\%   & 1.0\% \\
  \hline
           $h \to \tau^+\tau^-$    & 2.0\% &              & 3.0\% & 5.0\%   & 2.0\% \\
  \hline
           $h \to ZZ^*$             & 8.4\% &              & 14\% & 4.6\%   & 2.6\% \\
  \hline
           $h \to \gamma\gamma$   & 16\% &              & 19\% & 13\%   & 5.4\% \\
  \hline
           $h \to \mu^+\mu^-$   & 46.6\% &     -      & -  & -   & 20\% \\
  \hline
\end{tabular}
\label{tab:ilc_obs_lumiup}
\end{center}
\end{table}

\begin{table}[h]
\begin{center}
\caption{Similar table to Table \ref{tab:ilc_couplings_baseline} but with the luminosity upgrade described in the text: 
1150\,fb$^{-1}$ at 250\,GeV, 1600\,fb$^{-1}$ at 500\,GeV, and 2500\,fb$^{-1}$ at 1\,TeV, cf.\ \cite{ref:snowmass_higgs} and Scen. `Snow' 
in \cite{parametergroup}. Values with (*) assume inclusion of $hh\to WW^*b\bar{b}$ decays.}
\begin{tabular}{|c|c|c|c|}
   \hline
                  & \multicolumn{3}{c|}{$\sqrt{s}$ (GeV)} \\
   \cline{2-4}
    coupling & 250 & 250+500 & 250 + 500 + 1000 \\
   \hline
   $hZZ$    & 0.6\% & 0.5\% & 0.5\% \\
   \hline
   $hWW$  & 2.3\% & 0.6\% & 0.6\% \\
   \hline
   $hbb$     & 2.5\% & 0.8\% & 0.7\% \\
   \hline
   $hcc$     & 3.2\% & 1.5\% & 1.0\% \\
   \hline
   $hgg$     & 3.0\% & 1.2\% & 0.93\% \\
   \hline
   $h\tau\tau$ & 2.7\% & 1.2\% & 0.9\% \\
   \hline
   $h\gamma\gamma$ & 8.2\% & 4.5\% & 2.4\% \\
   \hline
   $h\mu\mu$ & 42\% & 42\% & 10\% \\
   \hline
   $\Gamma_0$ & 5.4\% & 2.5\% & 2.3\% \\
   \hline
   $htt$ & - & 7.8\% & 1.9\% \\
   \hline\hline
   $hhh$ & - & 46\%(*) & 13\%(*) \\
   \hline
\end{tabular}
\label{tab:ilc_couplings_lumiup}
\end{center}
\end{table}

\subsubsection{Conclusions}

The primary goal for the next decades is to uncover the secret of the electroweak symmetry breaking.
This will open up a window to BSM and set the energy scale for the energy frontier machine that will follow the LHC and the ILC 500.
Probably the LHC will hit systematic limits at $O$(5-10\%) for most of $\sigma \times \br$ measurements, being insufficient to see the BSM effects if we are in the decoupling regime.
The recoil mass measurements at the ILC unlocks the door to a fully model-independent analysis.
To achieve the primary goal we hence need a 500\,GeV linear collider for self-contained precision Higgs studies to complete the mass-coupling plot, where
we start from $e^+e^- \to Zh$ at $\sqrt{s}=250\,$GeV,
then $t\bar{t}$ at around $350\,$GeV,
and then $Zhh$ and $t\bar{t}h$ at $500\,$GeV.
The ILC to cover up to $\sqrt{s}=500\,$GeV is an ideal machine to carry out this mission (regardless of BSM scenarios) and we can do this {\it completely model-independently} with staging starting from $\sqrt{s}\simeq250\,$GeV.
We may need more data at this energy depending on the size of the deviation, since the recoil mass measurement eventually limits the coupling precisions.
Luminosity upgrade possibility should be always kept in our scope.
If we are lucky, some extra Higgs boson or some other new particle might be within reach already at the ILC 500. 
Let's hope that the upgraded LHC will make another great discovery in the next run from 2015.
If not, we will most probably need the energy scale information from the precision Higgs studies.
Guided by the energy scale information, we will go hunt direct BSM signals, if necessary, with a new machine. Eventually we will need to measure $W_L W_L$ scattering to decide if the Higgs sector is strongly interacting or not.

%
%
%

%





\subsection{Higgs at CLIC: prospects\protect\footnotemark}
\footnotetext{Marcel Stanitzki: The materials presented in this subsection 
were prepared for the CLIC Conceptual
Design Report.}
\label{sec:ewsb4}

\subsubsection{Introduction}
The CLIC accelerator~\cite{Aicheler:2012bya} offers the possibility 
to study $e^+e^-$ collisions at center-of-mass energies from 350~GeV up to 3~TeV.  
The novel CLIC acceleration schemes uses a two-beam acceleration scheme and
normal conducting cavities, which operate at room temperature. A high-intensity
drive beam generates the necessary RF power at 12~GHz, 
which is then used to accelerate the main beam. Compared to the 
ILC~\cite{Behnke:2013xla}, the pulse length is significantly 
shorter (150~ns) with an bunch spacing of just 0.5~ns and a repetition rate of 50~Hz.

\begin{figure}[htbp!]
        \centering

\includegraphics[width=0.49\hsize]{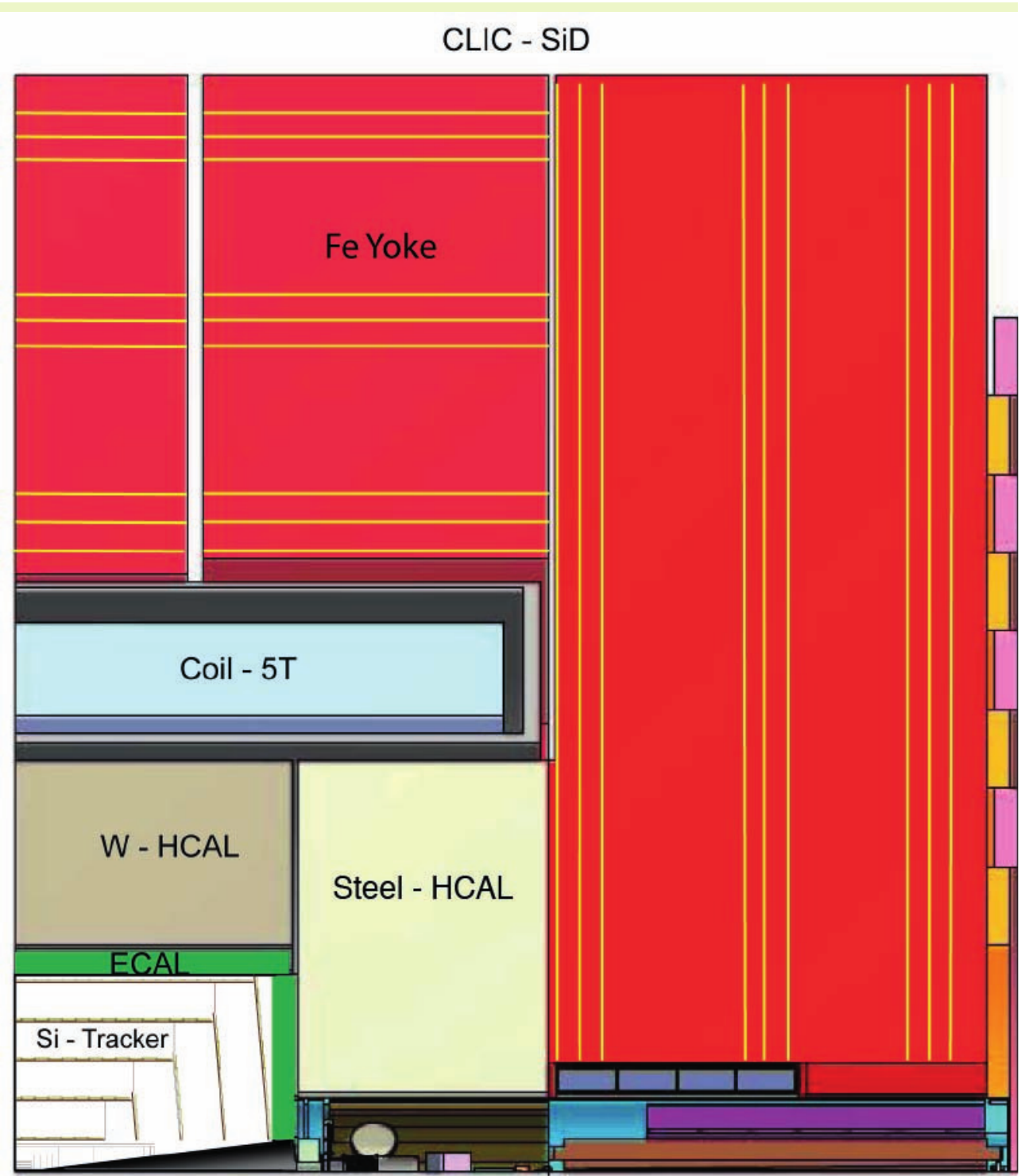}
\includegraphics[width=0.49\hsize]{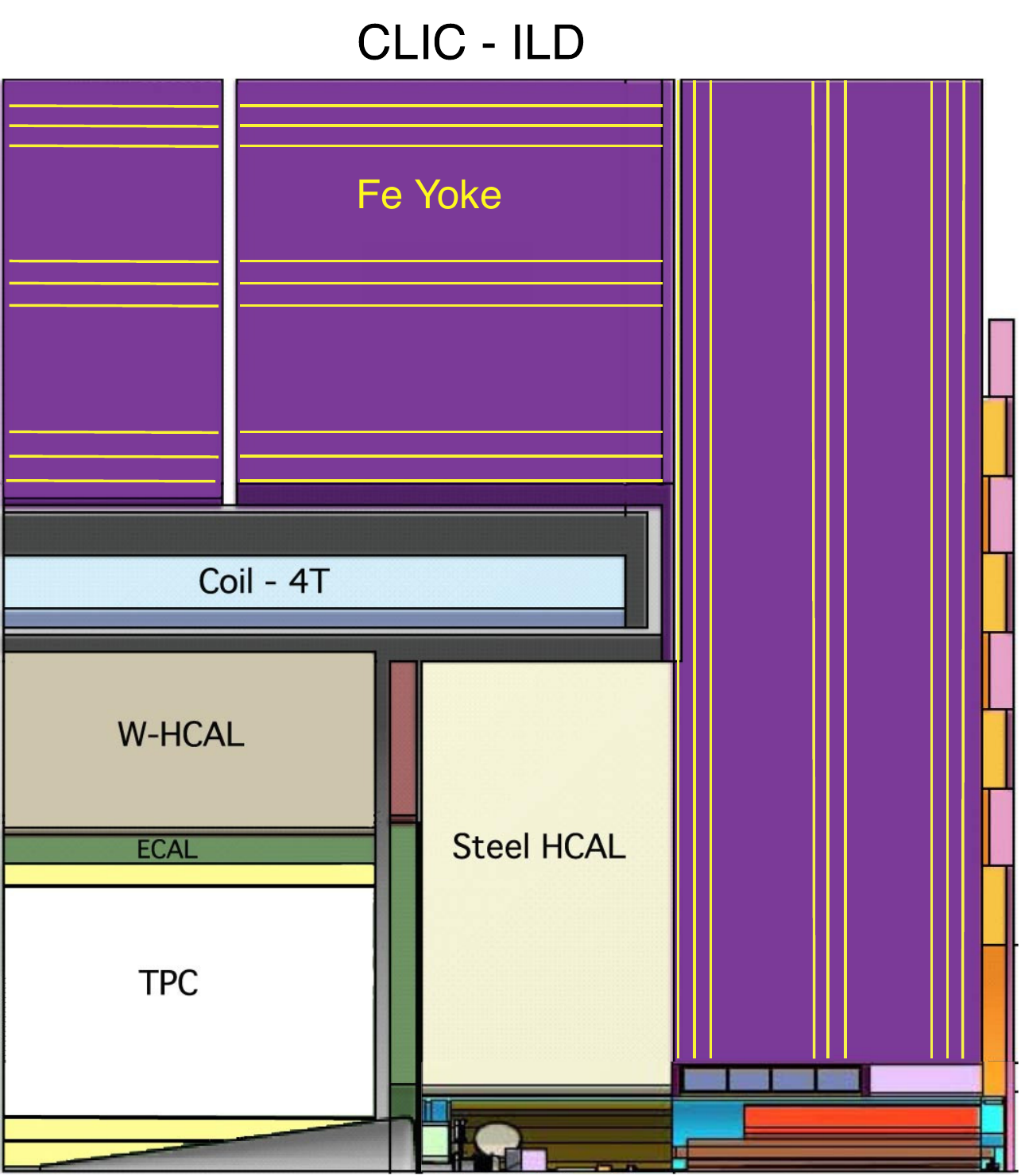}
	\caption{Longitudinal cross section of the top quadrant of CLIC\_SiD (left)
        and CLIC\_ILD (right)~\cite{Linssen:2012hp}.}
        \label{clichiggs_fig1}
\end{figure}

The detectors used for the CLIC physics and detector studies~\cite{Linssen:2012hp}
are based on the SiD~\cite{Aihara:2010zz,Behnke:2013lya} and
ILD~\cite{Abe:2010aa,Behnke:2013lya}
detectors proposed for the ILC. They have been adapted for the more challenging environment of
running at $\sqrt{s}=3$~TeV. The most significant changes for both CLIC\_SID and
CLIC\_ILD (see Fig.~\ref{clichiggs_fig1}) is the use of tungsten in the hadronic
calorimeter and an increase of the depth of hadronic calorimeter to 7.5
$\Lambda_{\mathrm{int}}$. 

Running in the multi-TeV regime and with small intense bunches means, that the
CLIC detectors experience much higher backgrounds from beamstrahlung. This also 
leads to a long tail of the luminosity spectrum. To cope with these harsh backgrounds, the
CLIC detectors plan to use highly granular detectors with time-stamping on the 10~ns level in
for the tracking detectors and 1~ns level for the calorimeters in order to
suppress these backgrounds~\cite{Linssen:2012hp}. 

\begin{figure}[hbt]
\centering
\includegraphics[width=0.49\linewidth]{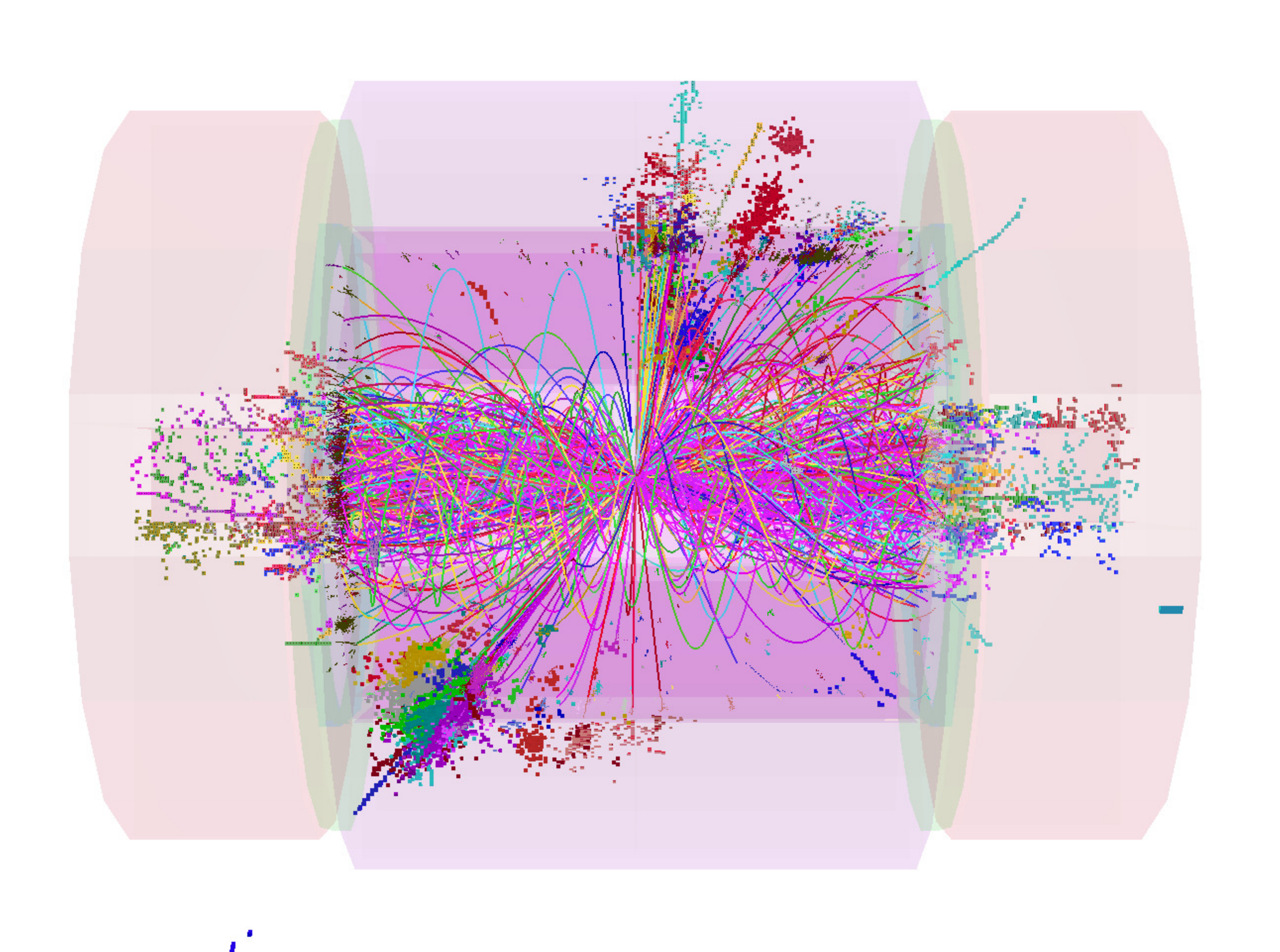}
\includegraphics[width=0.49\linewidth]{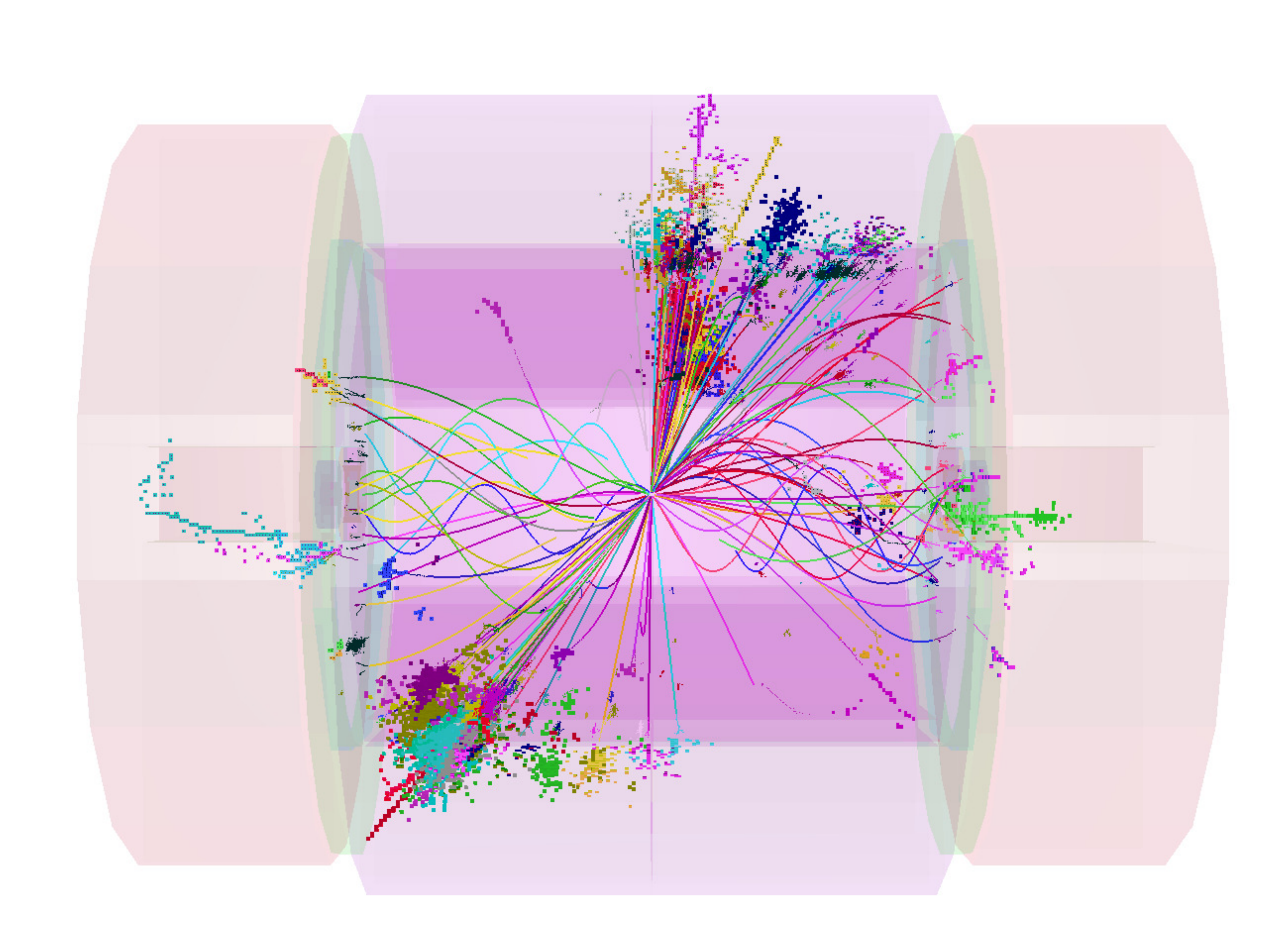}
\caption{Reconstructed particles in a simulated
$e^+e^-\rightarrow H^+H^- \rightarrow t\bar{b}\bar{t}b$ event at $\sqrt{s}$=3~TeV in the CLIC\_ILD
detector including the background from $\gamma\gamma\rightarrow \mbox{hadrons}$ before
(left) and after(right( applying tight timing cuts on the reconstructed cluster
times~\cite{Linssen:2012hp}.
\label{fig:clicbackgrounds}}
\end{figure}

\begin{sloppypar}
An entire bunch train at CLIC roughly deposits around 20~TeV in the detector,
which is predominantly coming from $\gamma\gamma\rightarrow \mbox{hadrons}$ events. By
applying tight cuts on the reconstructed particles this number can be reduced to
about 100~GeV. Using hadron-collider type jet clustering algorithms, which treat
the forward particles in a similar way than an underlying event this can be even
further improved~\cite{Linssen:2012hp}. The impact of this approach is
illustrated with a reconstructed $e^+e^-\rightarrow H^+H^- \rightarrow
t\bar{b}\bar{t}b$ event in the CLIC\_ILD detector (see
Fig.~\ref{fig:clicbackgrounds}).
\end{sloppypar}

This section focuses on the production of heavy Higgs bosons ($H, A, H^\pm$),
which are predicted in extended models like the 2HDM or supersymmetric models.
The CLIC capabilities for studying light, Standard Model (SM)-like
Higgs bosons are 
summarized elsewhere~\cite{Linssen:2012hp,Abramowicz:2013tzc}.

\subsubsection{Searches for heavy Higgs Bosons}

\begin{figure}[htbp]
\centering
\includegraphics[width=\linewidth]{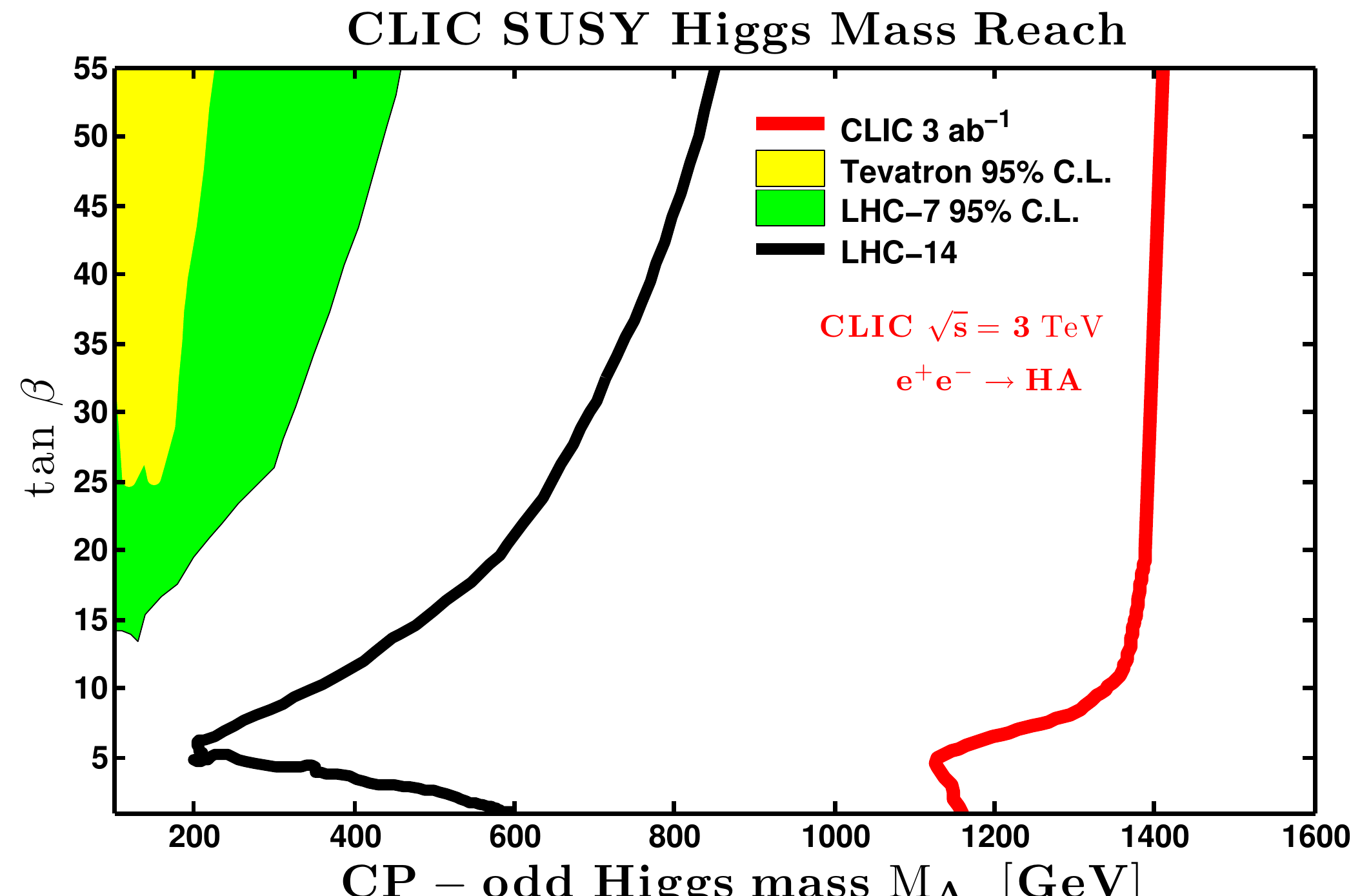}
\caption{Search reach in the $m_{\mathrm{A}}-\tan\beta$ plane for LHC and CLIC\@. The
left-most colored regions are current limits from the Tevatron with $\sim
7.5$~$\mathrm{fb}^{-1}$ of data at $\sqrt{s}=1.96$~TeV and from $\sim
1$~$\mathrm{fb}^{-1}$ of LHC data at $\sqrt{s}=7$~TeV. The black line is
projection of search reach at LHC with $\sqrt{s}=14$~TeV and
300~$\mathrm{fb}^{-1}$ of luminosity~\cite{Gianotti:2002xx}. The right-most red
line is search reach of CLIC in the $\rm HA$ mode with $\sqrt{s}=3$~TeV\@. This
search capacity extends well beyond the LHC~\cite{Linssen:2012hp}.}
\label{clichiggs:matb}
\end{figure}

In many supersymmetric scenarios, the Higgs sector consists of one light
Higgs boson $h$, consistent with a SM Higgs boson while the remaining four Higgs
bosons are almost mass-degenerate and have masses way beyond
500~GeV, see Sect.~\ref{sec:SUSYHiggs}. These
scenarios are consistent with current results from ATLAS and CMS on the Higgs
boson~\cite{ATLAS:2013mma,CMS:yva}.
If this scenario for the Higgs sector has been realized, it will be extremely
challenging to discover these additional final states at the LHC, especially in the low
$\tan\beta$ regime, where e.g. the reach for the pseudoscalar $A$ can be as low
as 200~GeV (See Fig.~\ref{clichiggs:matb}).

\begin{sloppypar}
The pair production processes $e^+e^-\rightarrow H^+H^-$ and 
$e^+e^-\rightarrow HA$ will give access to these heavy Higgs bosons almost up to the kinematic
limit~\cite{Coniavitis:2007me,Battaglia:2008nk}. Two
separate scenarios have recently been studied~\cite{Linssen:2012hp},
with a mass of the pseudoscalar Higgs boson A of $m_A$=902~GeV (Model I) or 
$m_A$=742~GeV (Model II). In both scenarios,
the dominant decay modes are $HA\rightarrow b\bar{b}b\bar{b}$ and 
$H^{+}H^{-}\rightarrow t\bar{b}\bar{t}b$. As already mentioned above, the 
analyses use the anti-$k_T$ algorithm that has been developed for the LHC 
in order to suppress the background originating from
$\gamma\gamma\rightarrow{hadrons}$. 
\end{sloppypar}

\begin{figure}[hbt]
\centering
\includegraphics[width=0.49\linewidth]{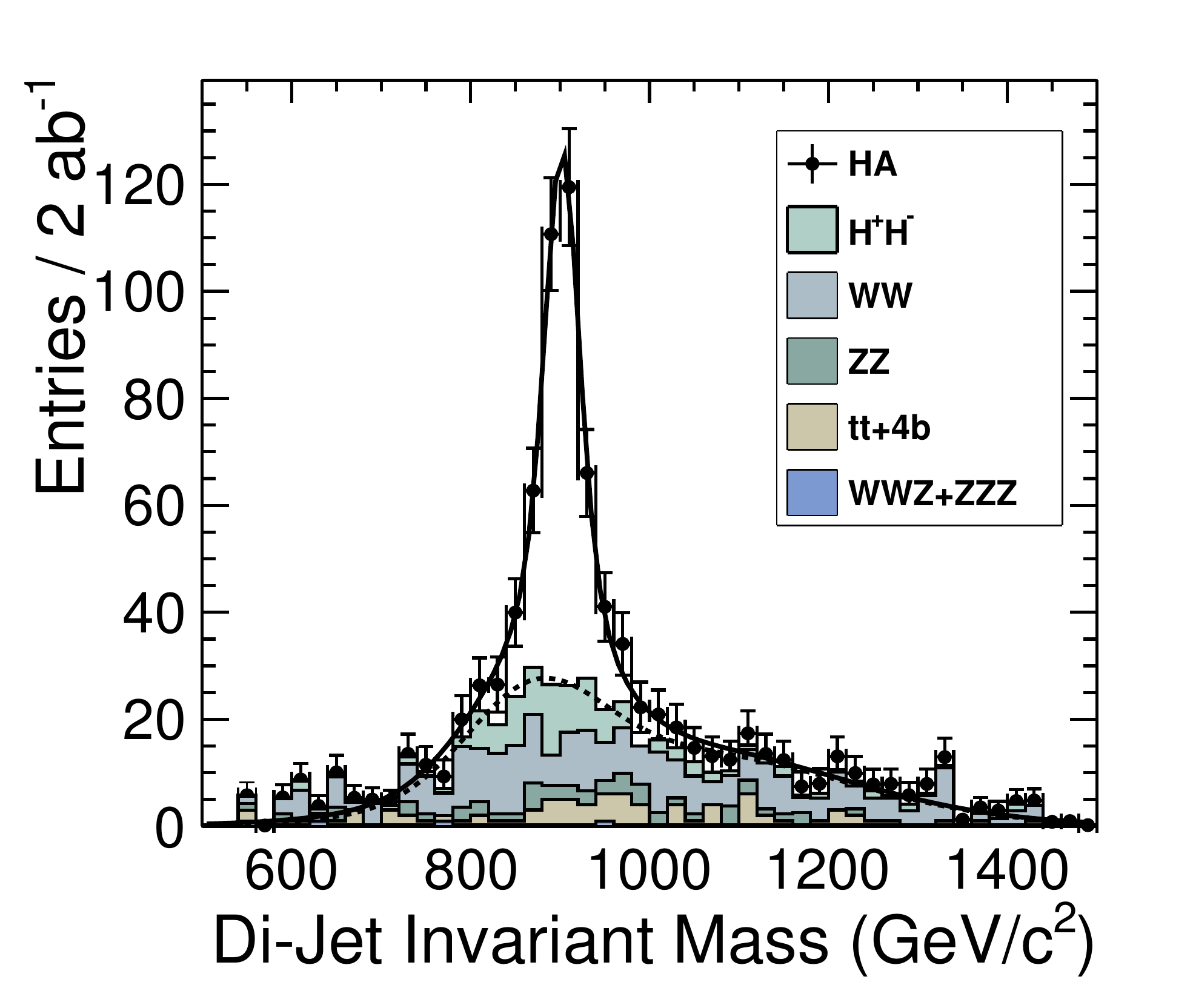}
\includegraphics[width=0.49\linewidth]{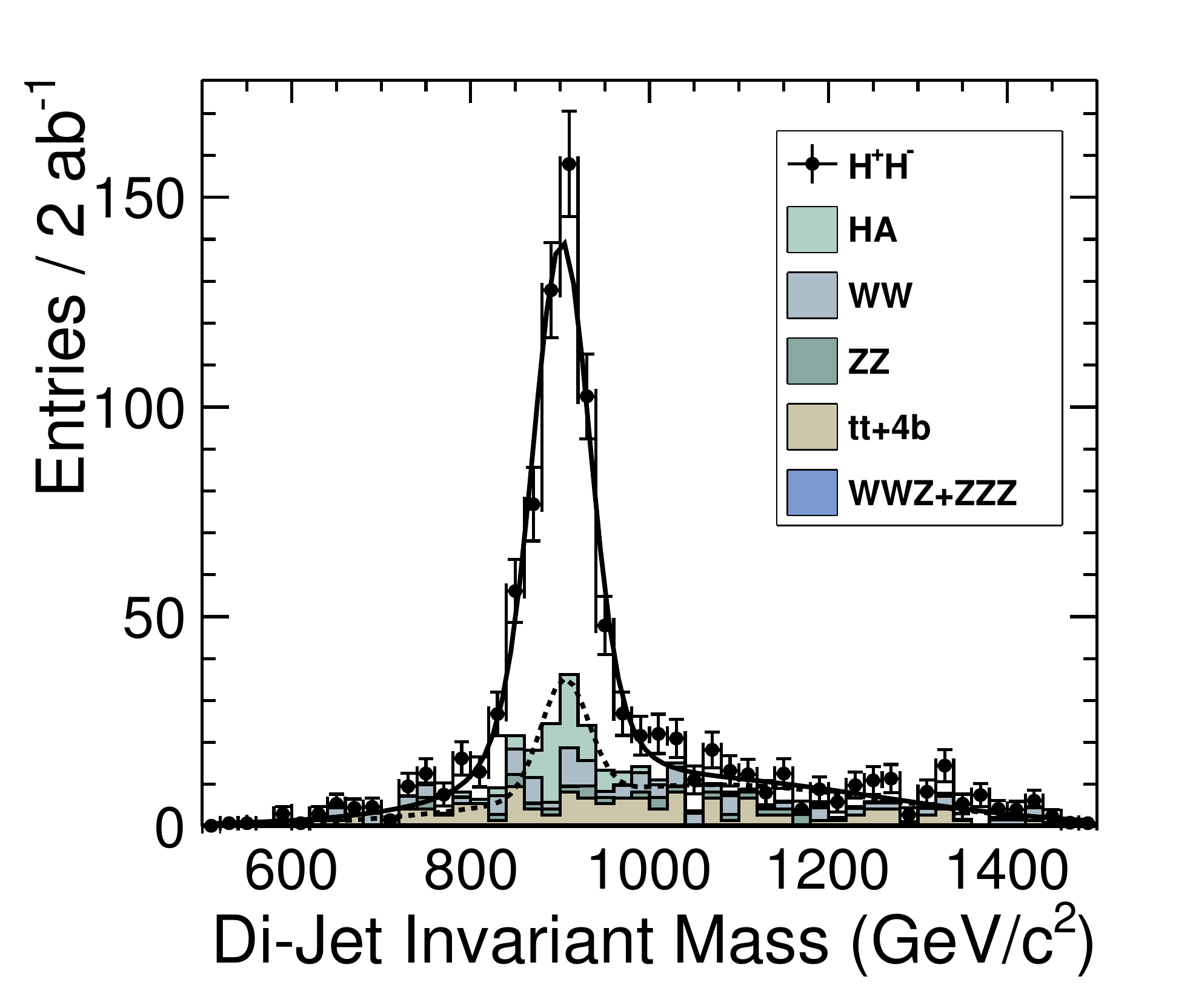}
\caption{Di-jet invariant mass distributions for the $e^+e^-\rightarrow
HA\rightarrow b\bar{b}b\bar{b}$  (left) and the 
$e^+e^-\rightarrow H^{+}H^{-}\rightarrow t\bar{b}\bar{t}b$ (right) 
signal together with the individual background contributions for model I~\cite{Linssen:2012hp}.}
\label{clichiggs:heavyhiggs1}
\end{figure}

\begin{figure}[hbt]
\centering
\includegraphics[width=0.49\linewidth]{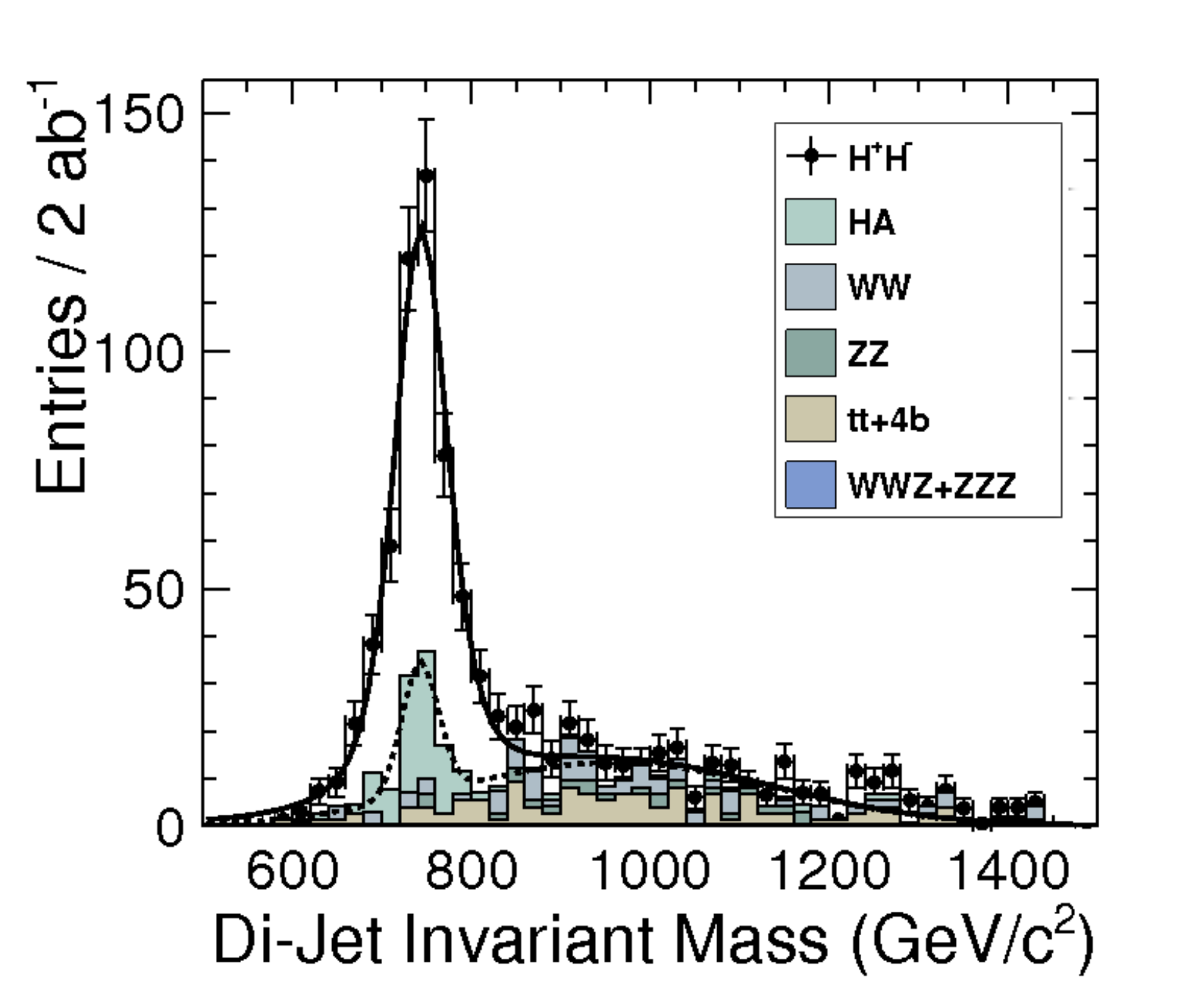}
\includegraphics[width=0.49\linewidth]{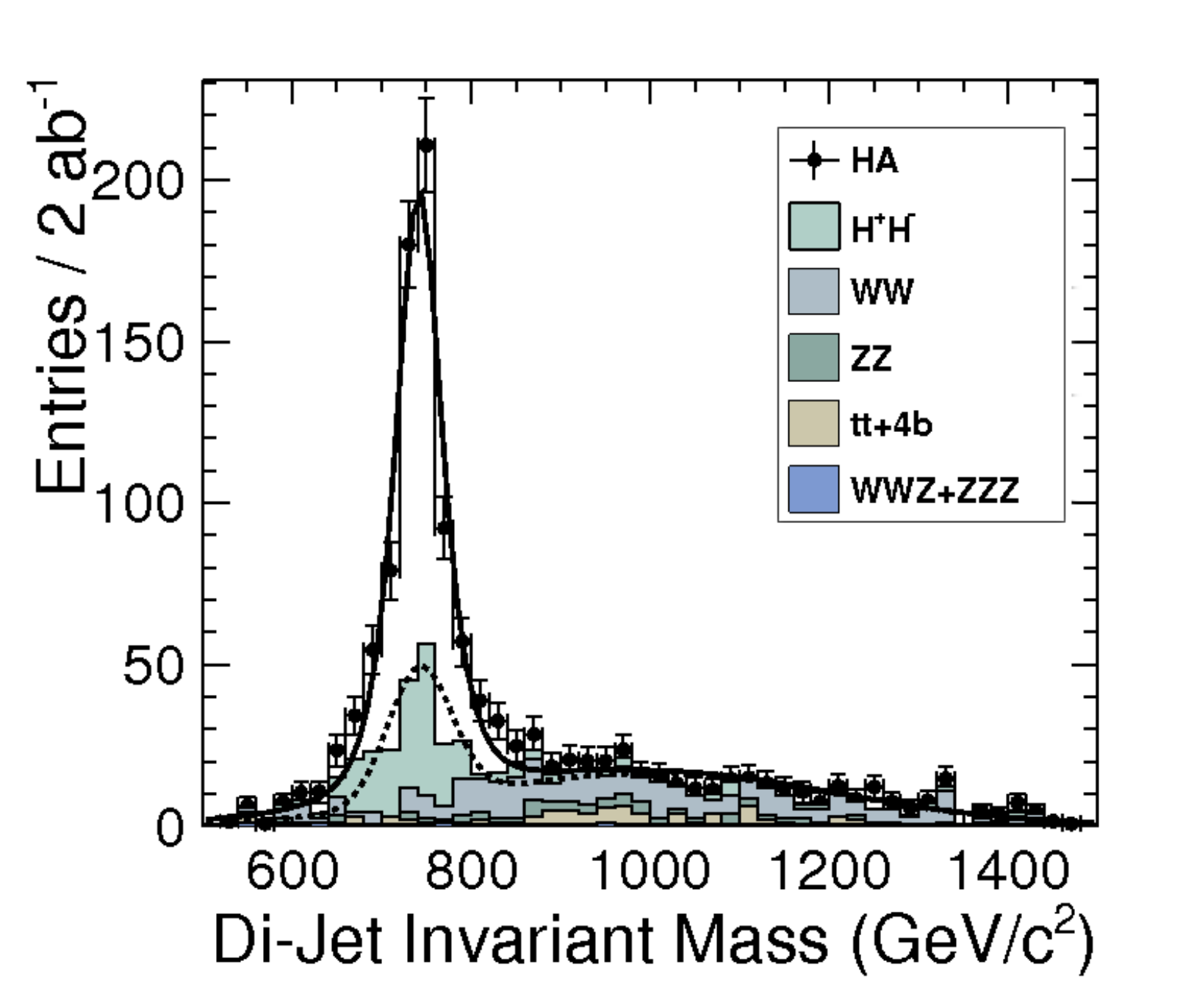}
\caption{Di-jet invariant mass distributions for the $e^+e^-\rightarrow
HA\rightarrow b\bar{b}b\bar{b}$  (left) and the 
$e^+e^-\rightarrow H^{+}H^{-}\rightarrow t\bar{b}\bar{t}b$ (right) 
signal together with the individual background contributions for model II~\cite{Linssen:2012hp}.}
\label{clichiggs:heavyhiggs2}
\end{figure}

The resulting di-jet mass distributions including the background processes
are shown in Fig.~\ref{clichiggs:heavyhiggs1}(Model I) and
and Fig.~\ref{clichiggs:heavyhiggs2}(Model II).
The achievable accuracy on the Higgs boson mass using a dataset of 2~$ab^{-1}$ 
at $\sqrt{s}$=3~TeV is about 0.3\%~\cite{Linssen:2012hp} 
and the width can be determined with an
accuracy of 17-31\% for the $b\bar{b}b\bar{b}$ final state and 23-27\% for the
$t\bar{b}\bar{t}b$ final state, showing the excellent physics capabilities of
CLIC for studying heavy Higgs bosons.

\subsection{Prospects for MSSM Higgs bosons\protect\footnotemark}
\footnotetext{Sven Heinemeyer} 
\label{sec:SUSYHiggs}
%
%
%
%
%




We will briefly review the MSSM Higgs sector, the relevance of
higher-order corrections and the implications of the recent discovery of
a Higgs-like state at the LHC at $\sim \MHexp\ \gev$. Finally we look at
the prospects in view of this discovery for MSSM Higgs physics at the
LC. We will concentrate on the MSSM with real parameters%
\footnote{
Analyses with complex parameters can be found in
\citeres{mhiggsCPXRG1,mhcMSSMlong} and references therein.}%
. The NMSSM will be covered in \refse{sec:ewsb9}


\subsubsection{The Higgs sector of the MSSM at tree-level}
\label{sec:Higgs-MSSM}

Contrary to the SM, in the MSSM~\cite{mssm} two Higgs doublets are
required (since the superpotential is a holomorphic function of the
superfields). The  Higgs potential
\BEA
V &=& m_{1}^2 |\cHe|^2 + m_{2}^2 |\cHz|^2 
      - m_{12}^2 (\epsilon_{ab} \cHe^a\cHz^b + \mbox{h.c.})  \non \\
  & & + \frac{1}{8}(g_1^2+g_2^2) \left[ |\cHe|^2 - |\cHz|^2 \right]^2
        + \frac{1}{2} g_2^2|\cHe^{\dag} \cHz|^2~,
\label{higgspot}
\EEA
contains $m_1, m_2, m_{12}$ as soft SUSY breaking parameters;
$g_2$ and $g_1$ are the $SU(2)$ and $U(1)$ gauge couplings,
respectively, and $\epsilon_{12} = -1$.

The doublet fields $\cHe$ and $\cHz$ are decomposed  in the following way:
\BEA
\cHe &=& \VL \cHe^0 \\[0.5ex] \cHe^- \VR \; = \; \VL v_1 
        + \frac{1}{\sqrt2}(\phi_1^0 - i\chi_1^0) \\[0.5ex] -\phi_1^- \VR~,  
        \non \\
\cHz &=& \VL \cHz^+ \\[0.5ex] \cHz^0 \VR \; = \; \VL \phi_2^+ \\[0.5ex] 
        v_2 + \frac{1}{\sqrt2}(\phi_2^0 + i\chi_2^0) \VR~,
\label{higgsfeldunrot}
\EEA
where $\phi^0_{1,2}$ denote the $\cp$-even fields, $\chi^0_{1,2}$ the
$\cp$-odd fields and $\phi^\pm_{1,2}$ the charged field components.
The potential (\ref{higgspot}) can be described with the help of two  
independent parameters (besides $g_2$ and $g_1$): 
$\tan\beta = v_2/v_1$ (with $v_1^2 + v_2^2 =: v^2 \approx (246\, \gev)^2$) 
and $\MA^2 = -m_{12}^2(\tan\beta+\CTb)$,
where $\MA$ is the mass of the $\cp$-odd Higg boson~$A$.

The diagonalization of the bilinear part of the Higgs potential,
i.e.\ of the Higgs mass matrices, is performed via orthogonal
transformations, introducing the mixing angle $\al$ for the $\cp$-even
part (with $\mh$ denoting the tree-level value of the light $\cp$-even
Higgs, see below),
\BE
\tan\,\al = \KKL 
  \frac{-(\MA^2 + \MZ^2) \Sbe \Cb}
       {\MZ^2 \CQb + \MA^2 \SQb - \mh^2} \KKR~, ~~
 -\frac{\pi}{2} < \al < 0~.
\label{alphaborn}
\end{equation}

One gets the following Higgs spectrum:
\BEA
\mbox{2 neutral bosons},\, {\cal CP} = +1 &:& h, H \non \\
\mbox{1 neutral boson},\, {\cal CP} = -1  &:& A \non \\
\mbox{2 charged bosons}                   &:& H^+, H^- \non \\
\mbox{3 unphysical Goldstone bosons}      &:& G, G^+, G^- .
\EEA

At tree level the masses squares are given by
\begin{align}
m^2_{H, h} &=
 \frac{1}{2} \Big[ \MA^2 + \MZ^2 \nonumber \\
         &\quad \pm \sqrt{(\MA^2 + \MZ^2)^2 - 4 \MZ^2 \MA^2 \CQZb} \Big] \\
\mHp^2 &= \MA^2 + \MW^2~.
\end{align}
In the decoupling limit~\cite{decoupling}, $\MA \gg \MZ$, the light
$\cp$-even Higgs 
becomes SM-like, i.e.\ all its couplings approach their SM value.

\subsubsection{The relevance of higher-order corrections}
\label{sec:HiggsHO}

\begin{sloppypar}
Higher-order corrections give large contributions to the Higgs sector
predictions in the MSSM~\cite{MSSMreviews}.
%
Most prominently, they affect the prediction of
the Higgs boson masses in terms of the other model parameters. In the
MSSM, in particular, the light $\cp$-even Higgs boson mass receives
higher-order contributions up to \order{100\%}~\cite{ERZ}. The very leading
one-loop correction reads
\begin{align}
\De\Mh^2 = \frac{3\, g_2^2\, \mt^4}{8 \,\pi^2\,\MW^2}
           \, \KKL \log \KL \frac{M_S^2}{\mt^2} \KR
                 + \frac{\Xt^2}{M_S^2} \KL 1 - \frac{\Xt^2}{12\,M_S^2} \KR
              \KKR~,
\label{MHmt}
\end{align}
where $M_S = (\mste + \mstz)/2$ denotes the average of the two scalar
top masses, and $\mt\Xt$ is the off-diagonal element in the scalar top
mass matrix.
Via this kind of higher-order corrections the light Higgs mass is
connected to all other sectors of the model and can serve as a precision
observable. 
The missing higher-order uncertainties have been estimated to be at the
level of $\sim 2-3 \gev$~\cite{mhiggsAEC,ehowp}
\end{sloppypar}

\bigskip
\begin{sloppypar}
Higher-order corrections also affect the various couplings of the Higgs
bosons and thus the production cross sections and branching ratios. 
Focusing on the light $\cp$-even Higgs boson, the couplings to down-type
fermions are modified with respect to the SM coupling by an additional
factor $-\Sa/\Cb$, and higher-order corrections can be absorbed into the
$\cp$-even mixing angle, $\al \to \aeff$~\cite{hff}. 
For large higher-order corrections which drive $\aeff \to 0$ the decay
widths $\Ga(h \to b \bar b)$ and $\Ga(h \to \tau^+\tau^-)$ could be
substantially smaller than in the SM~\cite{benchmark2}, altering the
available search modes for such a Higgs boson.
\end{sloppypar}

\medskip
\begin{sloppypar}
The relation between the bottom-quark mass and the Yukawa coupling
$h_b$, which controls also the interaction between the Higgs fields and
the sbottom quarks, is also affected by higher-order corrections,
summarized in the quantity $\db$~\cite{deltamb1,deltamb2,db2l}.
These, often called threshold corrections, are generated either by
gluino--sbottom one-loop diagrams (resulting in \order{\alb\als}
corrections), or by chargino--stop loops (giving
\order{\alb\alt} corrections). Analytically one finds
$\db \propto \mu \tan\beta$. 
The effective Lagrangian is given by~\cite{deltamb2}
\begin{eqnarray}
\label{effL}
\cL =& \frac{g_2}{2\MW} \frac{\mbms}{1 + \db} \Bigg[ 
\tan\beta\; A \, i \, \bar b \ga_5 b 
   + \wz \, V_{tb} \, \tan\beta \; H^+ \bar{t}_L b_R \nonumber \\
&+ \KL \frac{\Sa}{\Cb}
- \db \frac{\Ca}{\Sbe} \KR h \bar{b}_L b_R \nonumber \\
&\mbox{\hspace{1cm}} - \KL \frac{\Ca}{\Cb} + \db \frac{\Sa}{\Sbe} \KR H \bar{b}_L b_R
    \Bigg] + {\rm h.c.} \non
\end{eqnarray}
\end{sloppypar}
\begin{sloppypar}
Large positive (negative) values of $\db$ lead to a strong suppression
(enhancement) of the bottom Yukawa coupling. For large $\MA$ the
decoupling of the light $\cp$-even Higgs boson to the SM bottom Yukawa
coupling is ensured in \refeq{effL}. Effects on the searches for heavy
MSSM Higgs bosons via $\db$ have been analyzed in
\citeres{benchmark3,cmsHiggs}. 
\end{sloppypar}

\medskip
Deviations from the SM predictions can also be induced by the appearance
of light virtual SUSY particles in loop-induced processes. Most
promiently a light scalar top can have a strong impact on the prediction
of $gg \to h$. The additional contributions can interfere negatively
with the top loop contribution, leading to a strong suppression of the
production cross
section~\cite{ggh-djouadi,benchmark2,benchmark4}. Similarly, it 
was shown that light scalar taus can lead to an enhancement of up to
$\sim 50\%$ of the decay width of the light $\cp$-even Higgs to photons,
$\Ga(h \to \ga\ga)$~\cite{Mh125gaga}.


\subsubsection{Implicatios of the discovery at $\sim \MHexp\ \gev$}
\label{sec:Higgs126}

The discovery of a new state 
with a mass around $M_H\simeq \MHexp\ \gev$, which has been announced by
ATLAS~\cite{ATLASdiscovery} and CMS~\cite{CMSdiscovery}, 
marks a milestone of an effort that has been 
ongoing for almost half a century and opens a new era of particle physics. 
Both ATLAS and CMS reported a clear excess around $\sim \MHexp\ \gev$
in the two photon channel as well as in the $ZZ^{(*)}$ channel, 
supported by data in the $WW^{(*)}$ channel.
The combined sensitivity
in each of the experiments reaches by now far beyond $5 \si$. 
Also the final Tevatron results~\cite{TevHiggsfinal} show a broad excess
in the region around $\MH \sim 125 \gev$ that reaches a significance of
nearly $3\,\si$.
Within theoretical and experimental uncertainties the newly observed
boson behaves SM-like~\cite{HiggsMoriond14,HiggsMoriond14-QCD}.

Several types of investigations have analyzed the compatibility of the
newly observed state around $\sim \MHexp\ \gev$ with the MSSM.

\begin{itemize}

\item
Looking into pre-defined benchmark scenarios it was shown that the light
$\cp$-even Higgs boson can be interpreted as the new boson around 
$\MHexp\ \gev$. On the other hand, also the heavy $\cp$-even Higgs boson
can in principle be interpreted as the newly discovered
state~\cite{Mh125}. 
The latter option, however, is challenged by the latest ATLAS results on
charged Higgs boson searches~\cite{ATLAS-charged-Higgs}.

\begin{sloppypar}
Here we briefly discuss the results in two of the
new benchmark scenarios~\cite{benchmark4}, devised for the search for
heavy MSSM Higgs bosons. In the upper plot of \reffi{fig:mhmax} the
\mhmax\ scenario is shown. The red area is excluded by LHC searches for
the heavy MSSM Higgs bosons, the blue area is excluded by LEP Higgs
searches, and the light shaded red area is excluded by LHC searches for
a SM-like Higgs boson. The bounds have been obtained with 
{\tt HiggsBounds}~\cite{higgsbounds} (where an extensive list of
original references can be found). The green area yields 
$\Mh = 125 \pm 3 \gev$, i.e.\ the region allowed by the experimental
data, taking into account the theoretical uncertainty in the $\Mh$
calculation as discussed above. 
The left plot also allows to extract new {\em lower} limits on $\MA$ and
$\tan\beta$. From this analysis it can be concluded that if the light
$\cp$-even Higgs is interpreted as the newly discovered state at 
$\sim \MHexp\ \gev$, that $\tan\beta \gsim 4$, $\MA \gsim 200 \gev$ and
$\MHp \gsim 220 \gev$~\cite{benchmark4}.
\end{sloppypar}

In the lower plot of \reffi{fig:mhmax} we show the $\mh^{\rm mod+}$
scenario that differs from the \mhmax\ scenario in the choice of
$\Xt$. While in the \mhmax\ scenario $\Xt/M_{\rm SUSY} = +2$ had been chosen to
maximize $\Mh$, in the $\mh^{\rm mod+}$ scenario $\Xt/M_{\rm SUSY} = +1.5$ is
used to yield a ``good'' $\Mh$ value over the nearly the entire
$\MA$-$\tan\beta$ plane, which is visible as the extended green region.

\begin{figure}[htb!]
\begin{center}
\includegraphics[width=0.45\textwidth]{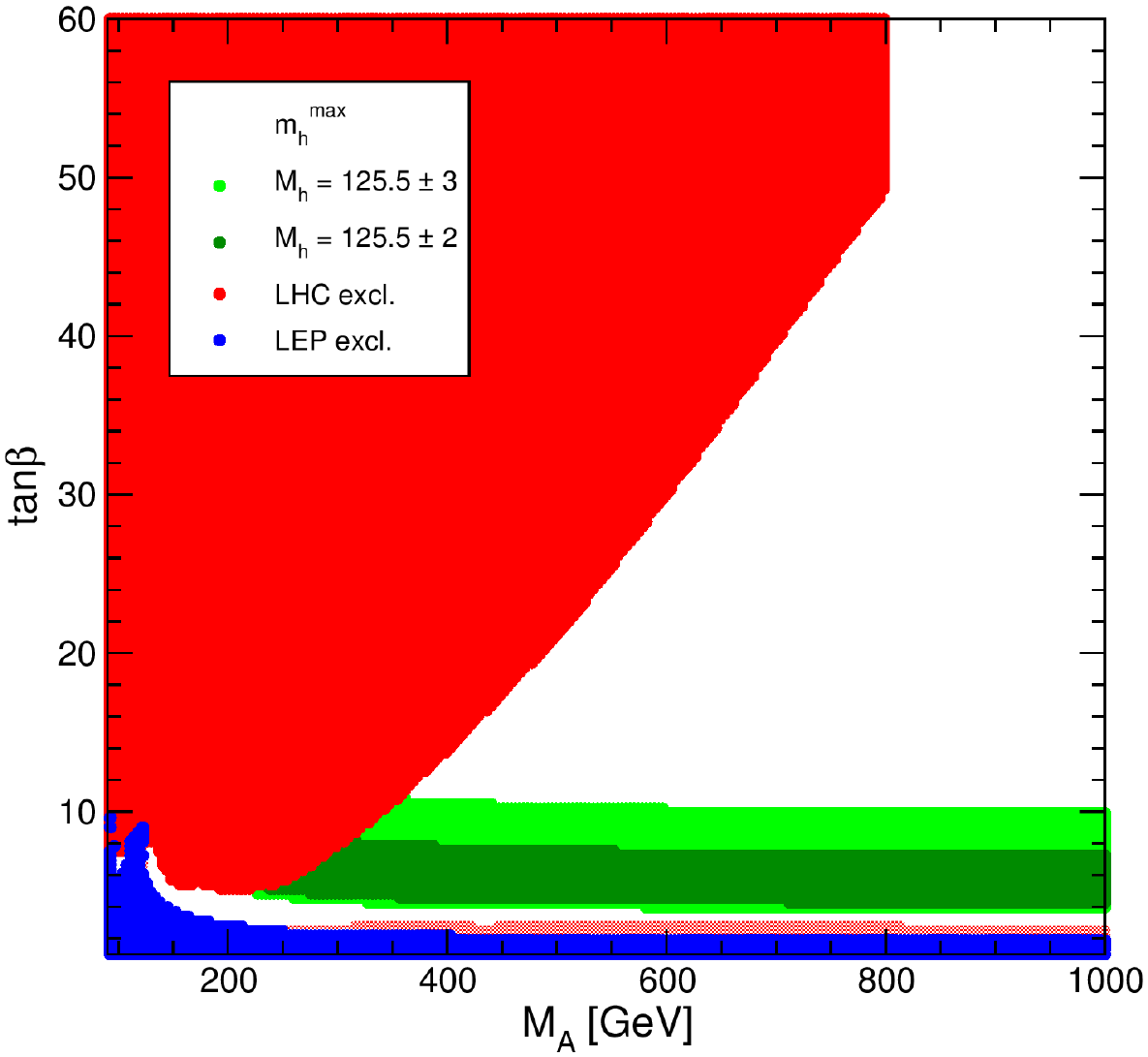}
\includegraphics[width=0.45\textwidth]{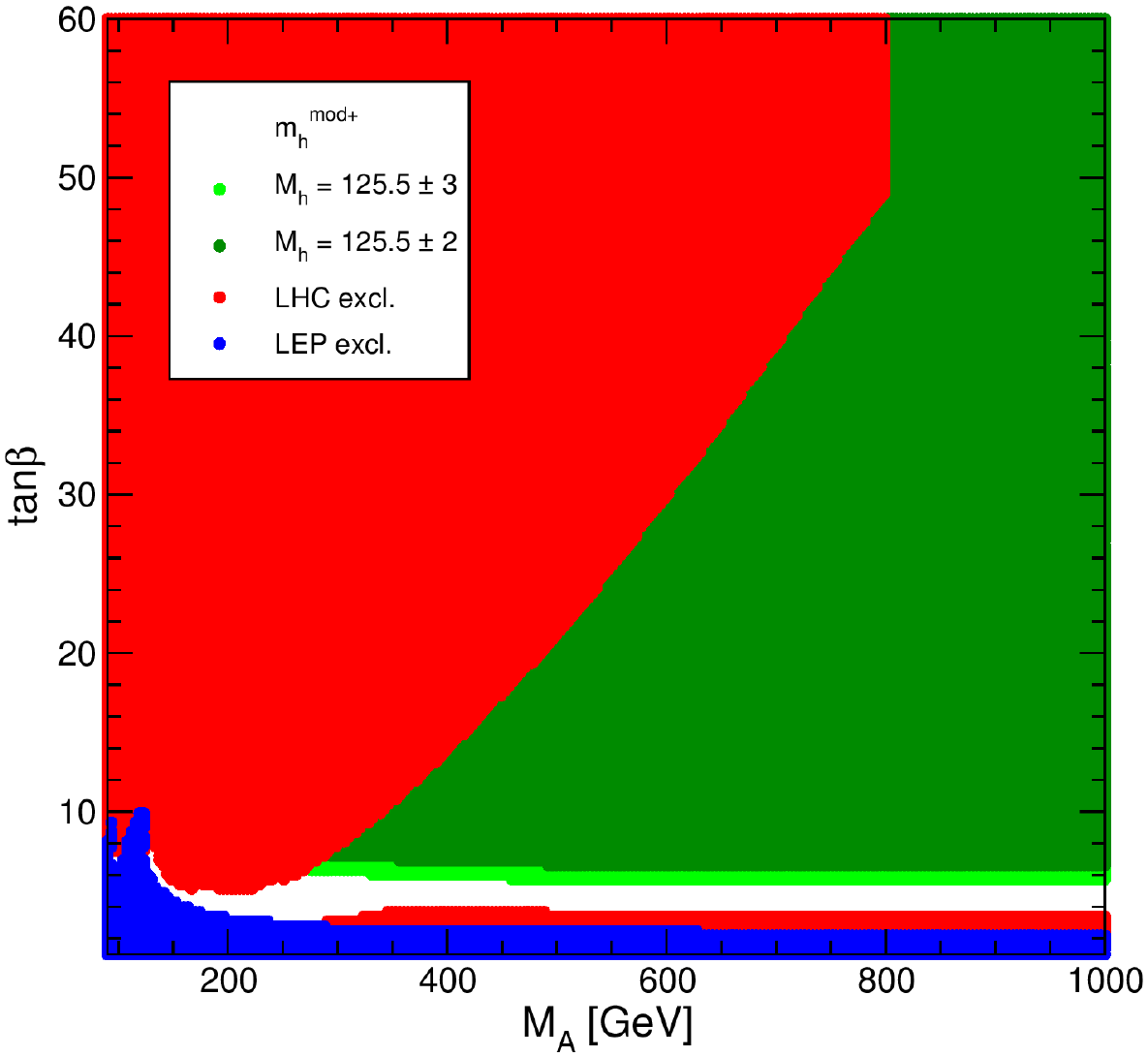}
\caption{
$\MA$-$\tan\beta$ plane in the \mhmax\ scenario (upper) and in the 
$\mh^{\rm mod+}$ scenario (lower plot)~\cite{benchmark4}. 
The green shaded area
yields $\Mh \sim 125 \pm 3 \gev$, the red area at high $\tan\beta$ is excluded
by LHC heavy MSSM Higgs boson searches, the blue area is excluded by LEP Higgs
searches, and the red strip at low $\tan\beta$ is excluded by the LHC SM Higgs
searches.}
\label{fig:mhmax}
\end{center}
\end{figure}

\item
In GUT based scenarios such as the CMSSM and the NUHM1%
\footnote{
In the CMSSM we have four free parameters, $m_0$, $m_{1/2}$ and $A_0$
defined at the GUT scale, as well as $\tan\beta$ defined at the EW
scale. Furthermore the sign of the $\mu$ parameter remains free. In the
NUHM1 in addition the Higgs sector has one free parameter at the GUT
scale, $m_H$. Details on the definition as well as the differences to
mSUGRA scenarios can be found in, e.g., \citere{pMSSM19def} and
references therein.}%
~it was shown that a light 
$\cp$-even Higgs boson around or slightly below $\MHexp\ \gev$ is a
natural prediction of these models~\cite{mc8}. These predictions take
into account 
the current SUSY search limits (but no direct light Higgs search
limits), as well as the relevant electroweak precision observables,
$B$-physics observables and the relic Dark Matter density.
In \reffi{fig:gut} we show the predictions in the CMSSM (upper) and the
NUHM1 (lower plot). The red bands indicate a theory uncertainty of 
$\sim 1.5 \gev$ on the evaluation of $\Mh$. The green columns indicate
the range of the newly discovered particle mass.

\begin{figure}[htb!]
\begin{center}
\includegraphics[width=0.45\textwidth]{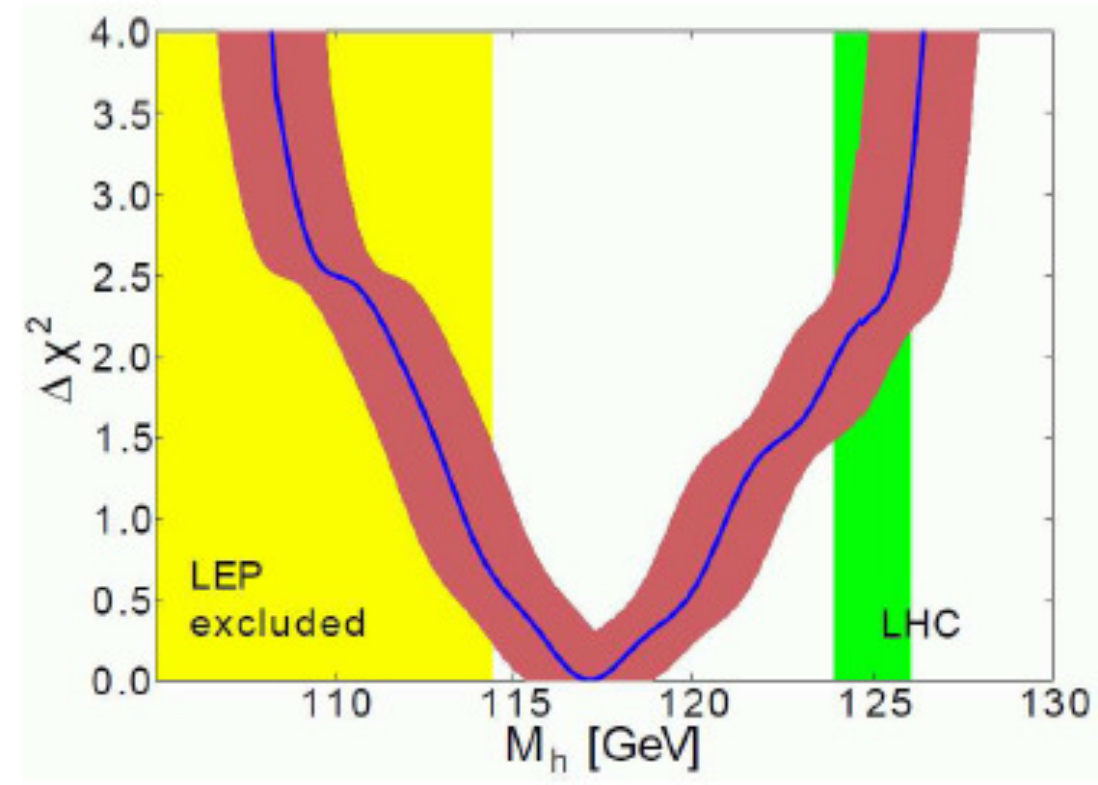}
\includegraphics[width=0.45\textwidth]{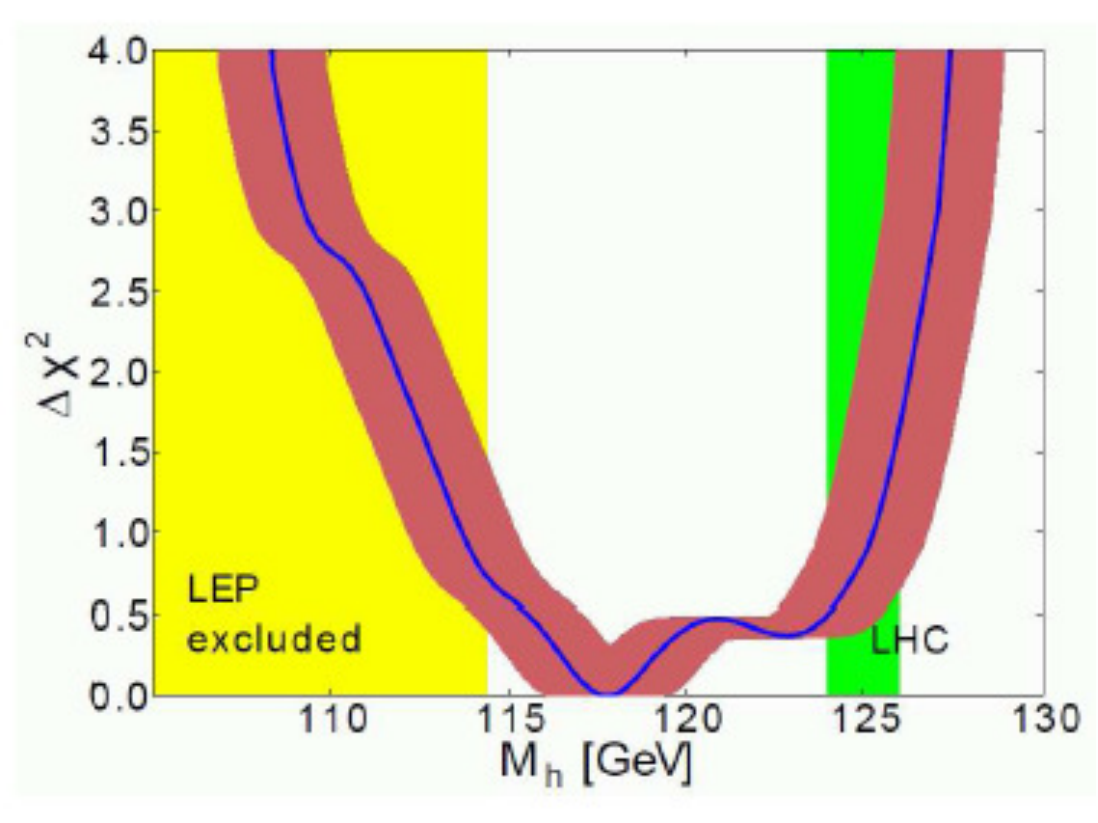}
\caption{
Fit for the light $\cp$-even Higgs mass in the CMSSM (left) and NUHM1
(right)~\cite{mc8}. Direct searches for the light Higgs boson are 
{\em not} included.
}
\label{fig:gut}
\end{center}
\end{figure}

\item
Parameter scans in the MSSM with 19 free parameters
(pMSSM--19~\cite{pMSSM19def}) are naturally compatible with a light Higgs
boson around $\Mh \sim \MHexp\ \gev$, as has been analyzed in
\citere{pMSSM19-Mh125} (see also \citere{Strege:2014ija}
for a more recent analysis in the pMSSM--15 and \citere{Cao:2014rma} 
for an analysis in the pMSSM--19).
Taking into account the available constraints from SUSY searches, Higgs
sear\-ches, low-energy observables, $B$-physics observables and the relic
abundance of Dark Matter viable scenarios can be identified that can be
analyzed in the upcoming LHC runs. Also the effects on the various
production cross sections and branching ratios were analyzed, where it
was confirmed that light particles can modify in particular the decay
rate to photons~\cite{Mh125gaga}. 

\item
\begin{sloppypar}
Parameter scans in the MSSM with seven free parameters (pMSSM--7) in
comparison to the pMSSM--19 have the advantage of a full sampling of the
parameter space with \order{10^7} points; but they have the disadvantage
of potentially not including all relevant phenomenogy of the MSSM.
The pMSSM--7 fits to the full set of Higgs data (and several low-energy
observables)~\cite{hifi} allow to show an enhancement of the 
$\br(h \to \ga\ga)$, correlated to a suppression of the decays to 
$b \bar b$ and $\tau^+\tau^-$ via the mechanisms outlined in
\refse{sec:HiggsHO} (see also \citere{Benbrik:2012rm}). 
%
In particular, these scans (while not incorporating the latest data) 
demonstrate that light scalar top masses are
compatible with $\Mh \sim \MHexp\ \gev$ (see also \citere{Mh125}). 
In \reffi{fig:h_mstop} we show 
$\Xt/\msqd$ vs.\ the light stop mass (left plot, where 
$\Xt = \At - \mu/\tan\beta$ denotes the off-diagonal entry in the scalar top
mass matrix, $\At$ is the trilinear Higgs-stop coupling, and $\msqd$
denotes the (common) diagonal soft SUSY-breaking parameter in the scalar
top and bottom sector) and the light
vs.\ the heavy stop mass (right plot) in the case that the light
$\cp$-even Higgs boson corresponds to the new state at $\sim \MHexp\ \gev$.
The colored points passed the Higgs exclusion bounds (obtained using
{\tt HiggsBounds}~\cite{higgsbounds}). The red (yellow) points
correspond to the best fit points with a $\De\chi^2 < 2.3 (5.99)$, see
\citere{hifi} for details. 
In the left plot one can see that the case of zero stop mixing in
the MSSM is
excluded by the observation of a light Higgs at $\Mh\sim \MHexp\ \gev$ 
(unless $\msqd$ is extremely large, see, e.g., \citere{Mh-logresum}), and
that values of $|\Xt/\msqd|$ between $\sim 1$ and $\sim 2.5$ must be
realised. For the most favoured region we find $\Xt/\msqd = 2 - 2.5$.
Concerning the value of the lightest scalar top mass, the overall smallest
values are found at $\mste \sim 200 \gev$, where also the regions
favored by the fit to the Higgs rates start, in the case of $\Xt$ positive.
Such a light $\Stope$ is accompanied by a somewhat
heavier $\Stopz$, as can be seen in the right plot of
\reffi{fig:h_mstop}. Values of $\mste \sim 200 \gev$ are realized for 
$\mstz \sim 600 \gev$, which would mean that both stop masses are
rather light, 
offering interesting possibilities for the LHC. The highest
favoured $\mste$ values we find are $\sim 1.4 \tev$. These are the
maximal values reached in the scan in \citere{hifi}, but from
\reffi{fig:h_mstop} it 
is obvious that the favoured region extends to larger values of
both stop masses. Such a scenario would be extremely difficult to
access at the LHC. 
\end{sloppypar}

\begin{figure}
\centering
\includegraphics[width=0.45\columnwidth]{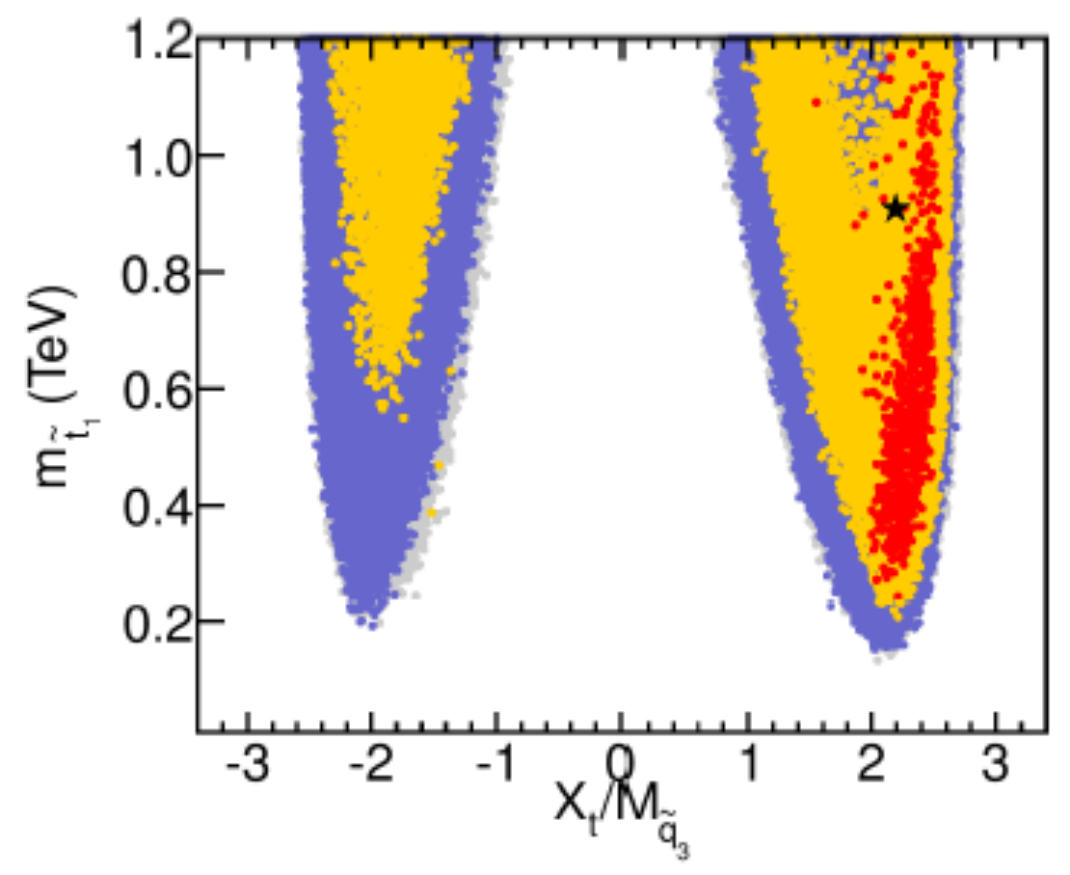}
\includegraphics[width=0.45\columnwidth]{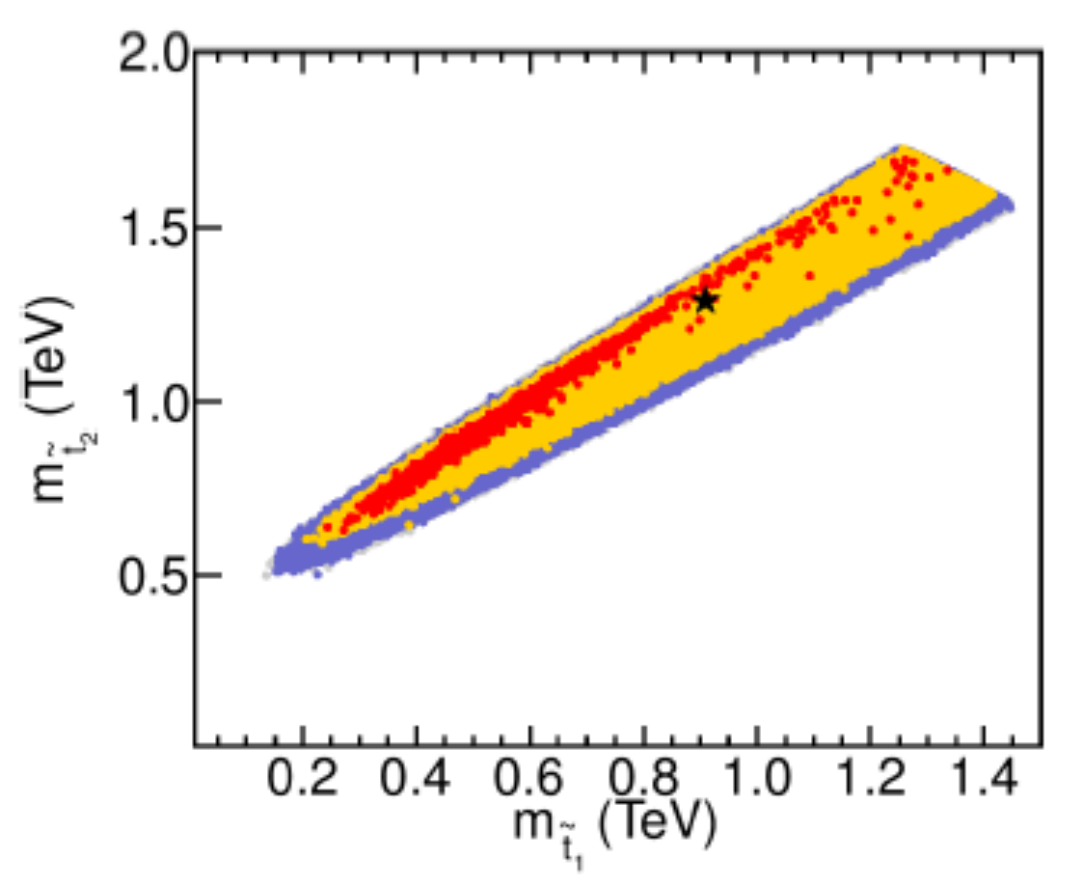}
\caption{Stop mixing parameter $\Xt/\msqd$ vs.\ the light stop mass
  (left), and the light vs.\ heavy stop masses (right), see text.} 
\label{fig:h_mstop}
\end{figure} 

%

\end{itemize}

Searches for the other Higgs bosons of the MSSM have so far not
been successful. This applies to the heavy Higgs bosons of the MSSM
as well as to a potentially light $\cp$-even Higgs bosons in the MSSM in
the case that the new state at $\sim \MHexp\ \gev$ is interpreted as the
heavy $\cp$-even Higgs boson, see Sect.~\ref{sec:ewsb2}.


\subsubsection{Prospects for the MSSM Higgs bosons at the LHC}
\label{sec:HiggsLHC}

The prime task now is to study the properties of the discovered
new particle and in particular to test whether the new particle is
compatible with the Higgs boson of the SM or whether there are
significant deviations from the SM predictions, which would point
towards physics beyond the SM. 
A large part of the current and future LHC physics program is devoted to
this kind of analyses. 

The prospects for the SM Higgs boson in this respect are the
following~\cite{ESPP-ATLAS,ESPP-CMS,LHC2TSP-WG1}:
\begin{itemize}

\item
The Higgs boson mass can be determined down to a level of 
\order{200\, \mev}.

\item
For the coupling determination the following has to be kept in
mind. Since it is not possible to measure the Higgs production cross
sections independently from the Higgs decay (or, equivalently, the Higgs
boson width%
\footnote{
A recent analysis from CMS using the Higgs decays to $ZZ$ far off-shell
yielded an upper limit on the total width about four times larger than the
SM width~\cite{CMS:2014ala}. However, these constraints on the total
width rely on the assumption of the
equality of the on-shell and off-shell couplings of the Higgs boson.
The relation between those couplings can be severely affected by new
physics contributions, in particular via threshold effects, which on the
other hand would be needed to give rise to a Higgs-boson width that
differs from the SM one by the currently probed amount, see the discusson 
in \citere{Englert:2014aca}. 
}), a determination of
couplings is only possible if certain (theory) assumptions on the Higgs
width are made, see, e.g.\ \citere{HcoupSM,LMrec}. 
For instance, it can be assumed that no new particles
contribute to the decay width. Under this kind of assumption, going to
the HL-LHC, precisions on couplings at the $\sim 10\%$ level can be
achieved. Without any assumptions only ratios of couplings can be
determined (see also \citere{Englert:2014uua} for a recent review).

\item
Studies in the context of the HL-LHC indicate that there might be some
sensitivity on the tri-linear Higgs self-coupling; however, this will
require a careful estimate of background contributions. Further studies
to clarify these issues are currently in progress, see \citere{LHC-HHH}
for a discussion.

\item
It can be expected that the spin~2 hypothesis can be rejected using LHC
data. 

\item
A pure $\cp$-even state can be discarded at the $2\,\si$ level already
from current data (assuming that the coupling strength to gauge bosons
is the same one as in the $\cp$-even case). However, the prospects for
the LHC to determine a certain level of
$\cp$-odd admixture to the Higgs state are less clear~\cite{Dawson:2013bba}.

\end{itemize}

In the case that the light $\cp$-even MSSM Higgs boson is identified
with the new state at $\sim \MHexp\ \gev$, as can be seen in
\reffi{fig:mhmax}, the decoupling region, $\MA \gg \MZ$ is a viable
option. In this case the SM Higgs analyses can be taken over directly to
the MSSM case -- and will yield (nearly) identical results. Only light
SUSY particles in the loops mediating the gluon fusion process or the
decay to two photons might result in somewhat different
predictions. However, depending on the actual values of the SUSY mass
scales, these differences might easily remain unobservable with the
anticipated LHC precision. Furthermore, in the decoupling regime the
heavy MSSM Higgs bosons can easily be too heavy to be discovered at the
LHC, in particular for medium or lower values of $\tan\beta$. 

Only in the lower allowed range for $\MA$ in this scenario larger
deviations from the phenomenology of the light $\cp$-even MSSM Higgs
with respect to the SM Higgs can be expected. Depending on the level of
decoupling, the LHC might be able to detect this kind of deviations,
e.g.\ in enhanced rates involving the decay to two photons or in
suppressed rates in the decay to $\tau$~leptons or $b$~quarks.



\subsubsection{Prospects for the MSSM Higgs bosons at the LC}
\label{sec:HiggsLC}

As outlined in the previous subsection, identifying the light $\cp$-even
Higgs with the new state at $\sim \MHexp\ \gev$ can easily result in a
scenario where the LHC can neither distinguish the $h$ from the SM Higgs
boson, nor be able to discover additional Higgs bosons. 
In this case the analyses at an LC offer good prospects to reveal the
non-SM nature of the Higgs particle. The anticipated experimental
precisions for couplings to SM particles, the self-coupling etc., as
given in detail in \refse{sec:ewsb3}. In particular, the following
improvements over the anticipated LHC precision/potential can be expected:

\begin{itemize}

\item
The mass of a SM-like Higgs boson at $\sim \MHexp\ \gev$ can be determined
at the level of $50 \mev$.

\item
Using the $Z$~recoil method the production cross section of a SM-like
Higgs can be determined independently of the decay products, see
\refse{sec:ewsb3}. This allows for a {\em model independent} measurement of
the Higgs couplings at the per-cent level, see
\refta{tab:ILCprecisions}. In particular, a determination of the
tri-linear Higgs self-coupling at the level of $15\%$ can be expected.

\begin{footnotesize}
\begin{table}
\renewcommand{\arraystretch}{1.2}
  \begin{center}
    \begin{tabular}{| l | c     |}
      \hline
      Observable                  & Expected precision \\
      \hline
 $\MH$ [GeV]        & 0.03\,\%  \\ 
 $g_{HWW}      $     & 1.4\,\%   \\
 $g_{HZZ}      $     & 1.4\,\%   \\
 $g_{Hbb}      $     & 1.4\,\%   \\
 $g_{Hcc}      $     & 2.0\,\%   \\
 $g_{H\tau\tau}$     & 2.5\,\%   \\
 $g_{Htt}      $     & 10\,\%   \\
 $g_{HHH}      $     & 40\,\%    \\
 $\br(H \to\ga\ga)$  & 25\,\%   \\
 $\br(H \to gg)$  & 5\,\%       \\
 $\br(H \to \mathrm{invisible})$ & 0.5\,\% \\
 \hline
 \end{tabular}
\end{center}
\caption{\small
Examples of the precision of SM-like Higgs observables
at a $\sqrt{s}=500 \gev$ LC assuming a Higgs boson mass of $125 \gev$. 
The results are based on the ILC set-up.
For the direct measurements, an
integrated luminosity of $\cL^{\rm int} = 500~\mathrm{fb}^{-1}$ is
assumed.
For the indirect measurements at GigaZ, a running time of
approximately one year is assumed, corresponding to 
  $\cL = $~\order{10~\mathrm{fb}^{-1}}. Taken from \citere{Baer:2013cma}.
\label{tab:ILCprecisions} }
\renewcommand{\arraystretch}{1.0}
\end{table}
\end{footnotesize}

\item
The spin can be determined unambigously from a production cross section
threshold scan.

\item
The $\cp$ decomposition can be determined, in particular, using the
channel $e^+e^- \to t \bar t H$~\cite{CPttH}.

\item
The reach for the heavy Higgs bosons can be extended to higher masses in
particular for lower and intermediate values of $\tan\beta$ up to 
$\MA \lsim \sqrt{s}/2$ (and possibly beyond, depending on the SUSY
parameters~\cite{LCnondecoup}). 

An indirect determination of $\MA$ can be performed via a precise
measurement of the Higgs couplings, where a sensitivity up to 
$800 \gev$ was found~\cite{MAindirect}. 

In the $\ga\ga$ option of the LC the Higgs bosons can be produced in the
$s$-channel, and a reach up to $\MA \lsim 0.8 \sqrt{s}$ can be
realized~\cite{MAgaga} (see also \citere{MAgaga2}).
\end{itemize}
%
Another measurement at the LC can turn out to be crucial for Higgs
physics in the MSSM: the determination of $\mt$ from a threshold
scan. As can be seen in \refeq{MHmt}, the theory prediction of $\Mh$
depends strongly on $\mt$. Only the LC determination of a well defined
top quark mass can yield a theory prediction that matches the LHC
precision in $\Mh$. More details can be found in
\refse{sec:quantum-mt}.






\subsection{General multi-Higgs structures\protect\footnotemark}
\footnotetext{Shinya Kanemura}
\label{sec:ewsb6}




 %





 \subsubsection{Introduction}

 We here give a review of extended Higgs sectors and their collider phenomenology.  
 In the Standard Model (SM), one isospin doublet scalar field $\Phi$ is simply introduced 
 as the minimum form.  Under the requirement of the 
 renormalizability its potential can be uniquely written  as 
 \begin{eqnarray}
   V(\Phi) = + \mu^2 |\Phi|^2 + \lambda |\Phi|^4. 
 \end{eqnarray} 
 \begin{sloppypar}
 By putting an assumption of $\mu^2 < 0$ (and $\lambda > 0$), 
 the shape of the potential becomes like a Mexican hat, and the electroweak symmetry is broken 
 spontaneously at the vacuum $\langle \Phi \rangle = (0, v/\sqrt{2})^T$. 
 Consequently, weak gauge bosons, quarks and charged leptons obtain their masses from the unique 
 vacuum expectation value (VEV) $v$ ($=(\sqrt{2}G_F)^{-1/2} \simeq 246$ GeV).
 However, there is no theoretical principle for the SM Higgs sector, 
 and there are many possibilities for non-minimal Higgs sectors. 
 While the current LHC data do not contradict the predictions 
 of the SM, most of the extended Higgs sectors can also satisfy current data. 
 These extended Higgs sectors are often introduced to provide physics sources to solve 
 problems beyond the SM, such as baryogenesis, dark matter and tiny neutrino masses. 
 Each scenario can predict a specific Higgs sector with additional scalars.  
 \end{sloppypar}

 It is also known that the introduction of the elementary scalar field is 
 problematic from the theoretical viewpoint, predicting the quadratic divergence  
 in the radiative correction to the mass of the Higgs boson. 
 Such a quadratic divergence causes the hierarchy problem. 
 There are many scenarios proposed to solve the hierarchy problem such as 
 Supersymmetry, Dynamical Symmetry Breaking, Extra dimensions and so on. 
 Many models based on these new paradigms predict specific Higgs sectors 
 in their low energy effective theories. 

 Therefore, experimental determination of the structure of the Higgs sector is essentially 
 important to deeply understand EWSB and also to find direction to 
 new physics beyond the SM. 
 The discovery of the 125 GeV Higgs boson at the LHC in 2012 is a big step 
 to experimentally investigate the structure of the Higgs sector. 
 From the detailed study of the Higgs sector, we can determine the model of new physics.  

 What kind of extended Higgs sectors we can consider?  
 As the SM Higgs sector does not contradict the current data within the errors, 
 there should be at least one isospin doublet field which looks like the SM Higgs boson. 
 An extended Higgs sector can then contain additional isospin multiplets.  
 There can be infinite kinds of extended Higgs sectors. 
 These extended Higgs sectors are subject to constraints from the current data 
 of many experiments including those of the electroweak $\rho$-parameter and 
 for flavor changing neutral currents (FCNCs). 

 The electroweak $\rho$-parameter is calculated at the tree level 
 for a Higgs sector with $N$ multiplets by  
 \begin{eqnarray}
  \rho = \frac{m_W^2}{m_Z^2 \cos^2\theta_W} = \frac{\sum_i \left\{ 4 T_i (T_i+1)- Y_i^2 \right\} |v_i|^2 c_i}
     {\sum_i 2 Y_i^2 |v_i|^2}, \label{eq:rho_general}
 \end{eqnarray}
 where  $T_i$ and $Y_i$  ($i=1, \cdots , N$) are isospin and hyper charges of the 
 $i$-th multiplet field ($Q_i=T_i+Y_i/2$), and $c_i =1/2$ for real fields ($Y_i=0$) 
 and $1$ for complex fields. 
 The data shows that $\rho=1.0004^{+0.0003}_{-0.0004}$~\cite{PDG}.  
 Higgs sectors with additional doublets $(T_i, Y_i) = (1/2, 1)$  (and singlets with $Y_i=0$) 
 predict $\rho=1$ at the tree level, like the SM Higgs sector. 
 Thus, multi-doublet structures would be a {\it natural} extension of the Higgs sector. 
 The introduction of higher representation fields generally causes a tree-level deviation in 
 the $\rho$- parameter from unity. 
 For example, in the model with a triplet field $\Delta$($1,2$) with the VEV $v_\Delta$, 
 $\rho \sim 1 - 2(v_\Delta/v)^2$ is given, so that in such a model a tuning 
 $(v_\Delta/v)^2 \ll 1$ is required to satisfy the data.   
 We note that there are exceptional Higgs sectors with larger isospin representations  
 which predict $\rho=1$ at the tree level. 
 In the model proposed by Georgi and Machacek~\cite{Georgi:1985nv}, the Higgs sector is composed of 
 an isospin doublet field with additional a complex $(1, 2)$ and a real $(1,0)$ 
 triplet fields, which satisfies $\rho=1$ at the tree level. 
 Addition of the septet field $(3, 2)$ to the SM Higgs sector also predicts $\rho=1$ 
 at the tree level.  

 Extended Higgs sectors with a multi-doublet structure, in general, 
 receive a severe constraint from the results of FCNC experiments. 
 The data show that FCNC processes such as $K^0 \to \mu^+\mu^-$, $B^0-\bar{B}^0$ and so on 
 are highly suppressed~\cite{PDG}.  
 In the SM with a doublet Higgs field, the suppression of FCNC processes is perfectly 
 explained by the so-called Glashow-Illiopoulos-Miani mechanism~\cite{GIM}.
 On the other hand, in general multi Higgs doublet models where multiple Higgs doublets 
 couple to a quark or a charged lepton, Higgs boson mediated FCNC processes 
 can easily occur at the tree level.  
 In these models, in order to avoid such dangerous FCNC processes, 
 it is required that these Higgs doublet fields have different quantum numbers~\cite{Glashow:1976nt}.

 In \refse{2hdm}, we discuss properties of the two Higgs doublet model (2HDM), and its 
 phenomenology at the LHC and the ILC. 
 The physics of the model with the Higgs sector with a triplet is discussed 
 in \refse{htm}. 
 The possibility of more exotic extended Higgs sectors are shortly 
 discussed in \refse{exotic}.

 \subsubsection{Two Higgs doublet models}\label{2hdm}

 The 2HDM is one of the simplest extensions of 
 the standard Higgs sector with one scalar doublet field.  
 The model has many typical characteristics of 
 general extended Higgs sectors, such as the existence of 
 additional neutral Higgs states, charged scalar states, 
 and the source of  CP violation. 
 In fact, the 2HDM often appears in the low energy effective theory of 
 various new physics models which try to solve problems in the SM such as 
 the minimal supersymmetric SM (MSSM), to some models of neutrino masses, 
 dark matter, and electrowak baryogenesis.
 Therefore, it is useful to study properties of 2HDMs with 
 their collider phenomenology.  

 In the 2HDM, two isospin doublet scalar fields $\Phi_1$ and $\Phi_2$ are 
 introduced with a hypercharge $Y=1$.  
 The Higgs potential under the standard gauge symmetry is  given 
 by~\cite{Gunion:1989we}
 \begin{align}
  V& = m_1^2|\Phi_1|^2+m_2^2|\Phi_2|^2
  -  (m_3^2 \Phi_1^{\dagger}\Phi_2 + {\rm h.c.} ) \nonumber \\ &
   + \frac{\lambda_1}{2}|\Phi_1|^4 +\frac{\lambda_2}{2}|\Phi_2|^4
  + \lambda_3|\Phi_1|^2|\Phi_2|^2+\lambda_4|\Phi_1^\dagger\Phi_2|^2 \nonumber \\ &
 \mbox{\hspace{-.5cm}}
   + \left[\frac{\lambda_5}{2}(\Phi_1^\dagger\Phi_2)^2
  + \left\{\lambda_6(\Phi_1^\dagger\Phi_1)
  + \lambda_7(\Phi_2^\dagger\Phi_2)\right\}\Phi^\dagger_1\Phi_2
  + {\rm h.c.} \right], 
 \label{eq:potential-shinya}
 \end{align}
 where $m_1^2$, $m_2^2$ and $\lambda_{1-4}$ are real while
 $m_3^2$ and $\lambda_{5-7}$ are complex. 
 We here discuss the case of CP conservation with taking these complex as real.  
 The doublet fields can be parameterized as 
 \begin{align}
 \Phi_i=\left[\begin{array}{c}
 w_i^+\\
 \frac{1}{\sqrt{2}}(v_i+h_i+iz_i)
 \end{array}\right],\hspace{3mm}(i=1,2), 
 \end{align}
 where $v_1$ and $v_2$ are the VEVs of $\Phi_1$ and $\Phi_2$, 
 which satisfy $v\equiv\sqrt{v_1^2+v_2^2}$. 
 The ratio of the two VEVs is a parameter written as $\tan\beta=v_2/v_1$.  
 The mass eigenstates for the scalar bosons are obtained by  
 \begin{align}
 \left(\begin{array}{c}
 w_1^\pm\\
 w_2^\pm
 \end{array}\right)&=R(\beta)
 \left(\begin{array}{c}
 G^\pm\\
 H^\pm
 \end{array}\right),\quad 
 \left(\begin{array}{c}
 z_1\\
 z_2
 \end{array}\right)
 =R(\beta)\left(\begin{array}{c}
 G^0\\
 A
 \end{array}\right),  \notag\\ 
 \left(\begin{array}{c}
 h_1\\
 h_2
 \end{array}\right)&=R(\alpha)
 \left(\begin{array}{c}
 H\\
 h
 \end{array}\right), 
 \text{with}~R(\theta) = 
 \left(
 \begin{array}{cc}
 \cos\theta & -\sin\theta\\
 \sin\theta & \cos\theta
 \end{array}\right),
 \label{mixing}
 \end{align}
 where $G^\pm$ and $G^0$ are the Nambu-Goldstone bosons absorbed by the longitudinal component of $W^\pm$ and $Z$, respectively.  
 As the physical degrees of freedom,
 consequently, 
 we have two CP-even Higgs bosons $h$ and $H$, a CP-odd Higgs boson $A$ and a pair of singly-charged Higgs boson $H^\pm$. 
 We define $h$ as the SM-like Higgs boson with the mass of about 
 125~GeV.

 \begin{sloppypar}
 As already mentioned, in general 2HDMs,  FCNCs can appear 
 via tree-level Higgs-mediated diagrams, which are not phenomenologically acceptable. 
 The simple way to avoid such dangerous FCNCs is to impose a discrete $Z_2$ symmetry,  
 under which the two doublets are transformed as
 $\Phi_1\to+\Phi_1$ and 
 $\Phi_2\to-\Phi_2$~\cite{Glashow:1976nt,%
 Paschos:1976ay,Haber:1978jt,Donoghue:1978cj}. 
 Then, each quark or lepton can couple with only one of the two doublets,  
 so that the Higgs mediated FCNC processes are forbidden at the tree level.  
 \end{sloppypar}

 We hereafter concentrate on the case with the discrete symmetry.  
 Under this symmetry, $\lambda_6$ and $\lambda_7$ in the Higgs potential 
 in Eq.~(\ref{eq:potential-shinya}) are zero.  
 On the other hand, the soft-breaking mass $m_3^2$ of the discrete symmetry can be 
 allowed, because the discrete symmetry is introduced just to suppress FCNC interactions. 
 As we consider the CP-conserving scenario, $m_3^2$ and $\lambda_5$ are real. 
 Eight parameters in the potential are rewritten as the following eight physical parameters;  
 the masses of $h$, $H$, $A$ and $H^\pm$, 
 two mixing angles $\alpha$ and $\beta$ appearing in Eq.~(\ref{mixing}), 
 the VEV $v$ and the soft-breaking parameter $M^2$ defined by 
 \begin{align}
 M^2=\frac{m_3^2}{\sin\beta\cos\beta}. \label{bigm}
 \end{align}
 In terms of these parameters, the quartic coupling constants
 in the Higgs potential are expressed as~\cite{Kanemura:2004mg}
 \begin{subequations}
 \begin{align}
 \lambda_1 &= \frac{1}{v^2\cos^2\beta}
 (-M^2\sin^2\beta+m_h^2\sin^2\alpha+m_H^2\cos^2\alpha), \\
 \lambda_2 &= \frac{1}{v^2\sin^2\beta}
 (-M^2\cos^2\beta+m_h^2\cos^2\alpha+m_H^2\sin^2\alpha), \\
 \lambda_3 &= \frac{1}{v^2}\left[-M^2
 -\frac{\sin2\alpha}{\sin2\beta}(m_h^2-m_H^2)+2m_{H^{\pm}}^2\right],\\
 \lambda_4 &= \frac{1}{v^2}(M^2+m_A^2-2m_{H^{\pm}}^2),\\
 \lambda_5 &= \frac{1}{v^2}(M^2-m_A^2).
 \end{align}
 \end{subequations}
 \begin{table}[t]
  \begin{tabular}{c|ccccccc}
   \hline
   & $\Phi_1$ & $\Phi_2$ & $u_R$ & $d_R$ & $\ell_R$ &
   $Q_L$ & $L_L$ \\
   \hline
   \hline
   Type-I & $+$ & $-$ & $-$
               & $-$ & $-$ & $+$
                           & $+$  \\
   Type-II & $+ $& $-$ & $-$
   & $+$ & $+$ & $+$
   & + \\
   Type-X & $+$ & $-$ & $-$
   & $-$ & $+$ & $+$
   & $+$  \\
   Type-Y & $+$ & $-$ & $-$
   & $+$ & $-$ & $+$
   & $+$  \\
   \hline
  \end{tabular}
 \caption{Four possible $Z_2$ charge assignments of scalar and fermion
  fields to forbid tree-level Higgs-mediated FCNCs~\cite{Aoki:2009ha}.}
 \label{tab:Z2}
 \end{table}

 \begin{table*}[tbp]
 \begin{center}
  \begin{tabular}{c|ccccccccc}
   \hline
   & $\xi_h^u$ & $\xi_h^d$ & $\xi_h^\ell$ & $\xi_H^u$ & $\xi_H^d$ &
   $\xi_H^\ell$ & $\xi_A^u$ & $\xi_A^d$ & $\xi_A^\ell$ \\
   \hline
   \hline
   Type-I & $c_\alpha/s_\beta$ & $c_\alpha/s_\beta$ & $c_\alpha/s_\beta$
               & $s_\alpha/s_\beta$ & $s_\alpha/s_\beta$ & $s_\alpha/s_\beta$
                           & $\cot\beta$ & $-\cot\beta$& $-\cot\beta$ \\
   Type-II & $c_\alpha/s_\beta$ & $-s_\alpha/c_\beta$ & $-s_\alpha/c_\beta$
   & $s_\alpha/s_\beta$ & $c_\alpha/c_\beta$ & $c_\alpha/c_\beta$
   & $\cot\beta$ & $\tan\beta$& $\tan\beta$ \\
   Type-X & $c_\alpha/s_\beta$ & $c_\alpha/s_\beta$ & $-s_\alpha/c_\beta$
   & $s_\alpha/s_\beta$ & $s_\alpha/s_\beta$ & $c_\alpha/c_\beta$
   & $\cot\beta$ & $-\cot\beta$& $\tan\beta$ \\
   Type-Y & $c_\alpha/s_\beta$ & $-s_\alpha/c_\beta$ & $c_\alpha/s_\beta$
   & $s_\alpha/s_\beta$ & $c_\alpha/c_\beta$ & $s_\alpha/s_\beta$
   & $\cot\beta$ & $\tan\beta$& $-\cot\beta$ \\
   \hline
 \end{tabular}
 \caption{The coefficients for different type of Yukawa
  interactions~\cite{Aoki:2009ha}. 
 $c_\theta=\cos\theta,~{\rm and }~s_\theta=\sin\theta$ for $\theta = 
 \alpha,~\beta$.}
 \label{tab:yukawa}
 \vspace*{1cm}
 \end{center}
 \end{table*}

 Under the softly-broken discrete symmetry, 
 the Yu\-ka\-wa interactions of the 2HDM  can be written as   
 \begin{eqnarray}
  {\mathcal L}^{\rm 2HDM}_{\rm Yukawa} &=& -\bar{Q}_{L}Y_u\tilde\Phi_u u_R
 -\bar{Q}_{L}Y_d\Phi_d d_R \nonumber\\
 &&- \bar{L}_{L}Y_{\ell}\Phi_{\ell}\ell_R
 +{\rm h.c.}, \label{eq:Yukawa}
 \end{eqnarray}
 where $R$ and $L$ are the right-handed and left-handed chirality
 of fermions, respectively, and 
 $\Phi_{f=u,d,\ell}$ are chosen from $\Phi_1$ or $\Phi_2$.
 There are four types of Yukawa interactions  
 depending on the parity assignment of the discrete symmetry for fermions~\cite{Barger:1989fj} 
 shown in Table~\ref{tab:Z2}. 
 Type-I is the case that all the quarks and charged leptons
 obtain the masses from $v_2$, while Type-II is that up-type quark
 masses are generated by $v_2$ but the masses of down-type quarks and
 charged leptons are generated by $v_1$. 
 In Type-X, both up- and down- type quarks couple to $\Phi_2$
 while charged leptons couple to $\Phi_1$. 
 In Type-Y, up-type quarks and charged
 leptons couple to $\Phi_2$ while up-type quarks couple to $\Phi_1$. 
 Because of these variations in types of Yukawa interaction, 
 the 2HDM with the discrete symmetry can provide rich phenomenology.  
 We note that Type-I is for example used in the neutrino-philic
  mode~\cite{neutrinophillic} approximately, 
 Type-II is predicted in the context of the minimal supersymmetric SM 
 (MSSM)~\cite{Haber:1984rc,Gunion:1989we}
 and that Type-X  is used for example in some of radiative seesaw
 models~\cite{Aoki:2008av}. 

 Yukawa interaction in Eq.~(\ref{eq:Yukawa}) is rewritten in terms of the mass eigenstates  
 as 
 \begin{align}
  {\mathcal L}^{\rm 2HDM}_{\rm Yukawa} &= -\sum_{f=u,d,\ell}
 \left[\frac{m_f}{v}\xi_h^f\bar{f}fh+\frac{m_f}{v}\xi_H^f\bar{f}fH \right. \nonumber\\
 &\left.  -i\frac{m_f}{v}\xi_A^f\gamma_5\bar{f}fA\right]\nonumber\\
 &-\left\{
 \frac{\sqrt{2}V_{ud}}{v}\bar{u}\left[m_u\xi_A^uP_L+m_d\xi_A^dP_R
 \right]dH^+ + \right.\nonumber\\
 & \left.\frac{\sqrt{2}m_\ell}{v}\xi_{A}^\ell\bar{v}_{L}\ell_RH^+
 +{\rm h.c.} \right\},
 \end{align}
 where $P_{R,L}$ are the chiral projection operators.
 The coefficients $\xi_\phi^f$ are summarized in
 Table~\ref{tab:yukawa}.

 \begin{sloppypar}
 There are two possibilities to explain the current LHC data, which
 show that the Higgs sector is approximately SM-like.  When $M^2 \gg
 v^2$, the additional Higgs bosons $H$, $A$ and $H^\pm$ are as heavy as
 $\sqrt{M^2}$, and only $h$ stays at the electroweak scale, behaving as
 the SM-like Higgs boson.  The effective Lagrangian is
 \begin{eqnarray}
  {\mathcal{L}}_{\rm eff} = {\mathcal{L}}_{\rm SM} +  \frac{1}{M^2} {\cal O}^{(6)}. 
 \end{eqnarray}
 Another case is for $\sqrt{M^2} \sim v$. In the limit where the $hWW$
 coupling takes the same value as the SM prediction
 $\sin(\beta-\alpha)=1$, all the Yukawa couplings and the self-coupling
 for $h$ take the SM values, while $HWW$ is zero.  In this case, $h$
 behaves as the SM-like Higgs boson.  Contrary, $H$, $A$ and $H^\pm$ do
 not couple to gauge bosons, and they only couple to the SM particles
 via Yukawa interaction.
 When $\sin(\beta-\alpha)$ is slightly smaller than unity, the couplings 
 $hVV$ ($V=W$, $Z$) and $hff$ ($f=t$,$b$,$c$, $\cdots$) 
 deviate from the SM predictions depending on the type of Yukawa interaction. 
 By detecting the pattern of the deviation in each coupling of $h$ at future experimets, 
 we can distinguish the type of Yukawa coupling in the 2HDMs even without directly 
 discovering the additional Higgs bosons.  
 \end{sloppypar}

 \begin{sloppypar}
 The decay widths and branching ratios of additional Higgs bosons can
 be calculated for given values of $\tan\beta$, $\sin(\beta-\alpha)$
 and the masses for each type of Yukawa interaction.  We refer to
 Ref.~\cite{Aoki:2009ha} where the total decay widths are discussed in
 details for $\sin(\beta-\alpha)\simeq1$.  Explicit formulae for all
 the partial decay widths can be found, e.g., in
 Ref.~\cite{Aoki:2009ha}.
 \end{sloppypar}

\begin{figure*}
 \centering
 \includegraphics[width=\textwidth]{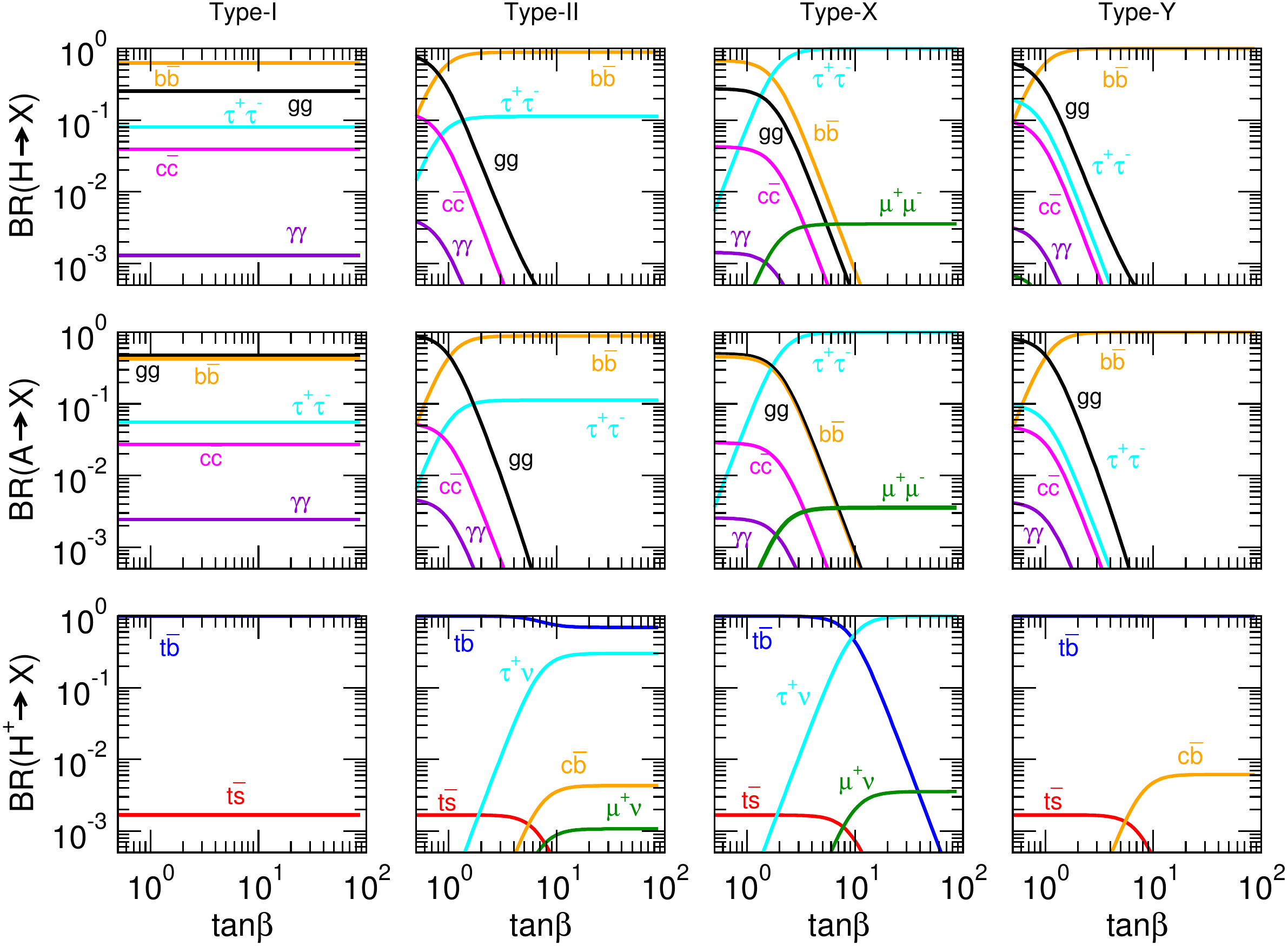} 
 \vspace{.5cm}
 \caption{The decay branching ratios of $H$, $A$ and $H^\pm$ in 2HDMs 
 for Type I, Type II, Type X and Type Y as a function of $\tan\beta$ with 
 $m_H=m_A=m_{H^\pm}=250$~GeV and $\sin(\beta-\alpha)=1$~\cite{Kanemura:2014dea}.  
 }
 \label{fig:Br_250}
 \end{figure*}

 In Fig.~\ref{fig:Br_250}, decay branching ratios of additional Higgs
 bosons $H$, $A$, and $H^\pm$ are plotted in each type of Yukawa interaction 
 as a function of $\tan\beta$ for the masses of $250$~GeV.  %
 For simplicity, the SM-like limit $\sin(\beta-\alpha)=1$ is taken. 
 In this limit, the decay modes of $H\to W^+W^-$, $ZZ$, $hh$ as well as
 $A\to Zh$ are absent. 
 In this limit, decay branching ratios of the SM-like Higgs boson are completely
 the same as those in the SM at the tree level, so that we cannot
 distinguish models by precision measurements of the couplings of the
 SM-like Higgs boson $h$\footnote{%
 The decay branching ratios of $h$ can be different from the SM prediction at
 the next-to-leading order~\cite{Guasch:2001wv,Hollik:2001px,%
 Dobado:2002jz,Kanemura:2004mg,Kanemura:2014dja}.
 }.\\

 \noindent
 {\it Constraints on the Higgs potential from perturbative
   unitarity and vacuum stability}

 The condition of tree-level unitarity requires the scattering amplitudes
 to be perturbative~\cite{Lee:1977yc}; 
 {\it i.e.} $|a^0_i|<1/2$~\cite{Gunion:1989we},
 where $a^0_i$ are the
 eigenvalues of the $S$-wave amplitudes of the elastic
 scatterings of the longitudinal component of weak gauge bosons and the
 Higgs boson.
 In the 2HDM with the softly-broken $Z_2$ symmetry, this condition gives
 constraints on the quartic couplings in the Higgs 
 potential~\cite{Kanemura:1993hm,Akeroyd:2000wc,Ginzburg:2005dt}. 
 The eigenvalues for $14\times14$ scattering matrix for neutral states
 are given as~\cite{Kanemura:1993hm}, 
 \begin{subequations}
 \begin{align}
 &a_1^{\pm} = \frac{1}{16\pi}\left[\frac{3}{2}(\lambda_1+\lambda_2)
 \pm\sqrt{\frac{9}{4}(\lambda_1-\lambda_2)^2+(2\lambda_3+\lambda_4)^2}
  \right],\\
 &a_2^{\pm} = \frac{1}{16\pi}\left[\frac{1}{2}(\lambda_1+\lambda_2)
 \pm\sqrt{\frac{1}{4}(\lambda_1-\lambda_2)^2+\lambda_4^2}
 \right],\\
 &a_3^{\pm} = \frac{1}{16\pi}\left[\frac{1}{2}(\lambda_1+\lambda_2)
 \pm\sqrt{\frac{1}{4}(\lambda_1-\lambda_2)^2+\lambda_5^2}
 \right],\\
 &a_4 = \frac{1}{16\pi}(\lambda_3+2\lambda_4-3\lambda_5),
 a_5 = \frac{1}{16\pi}(\lambda_3-\lambda_5),\\
 &a_6 = \frac{1}{16\pi}(\lambda_3+2\lambda_4+3\lambda_5),
 a_7 = \frac{1}{16\pi}(\lambda_3+\lambda_5),\\
 &a_8 = \frac{1}{16\pi}(\lambda_3+\lambda_4),
 \end{align}
 \end{subequations}
 and for singly charged states, one additional eigenvalue is
 added~\cite{Akeroyd:2000wc}; 
 \begin{subequations}
 \begin{align}
 a_9 = \frac{1}{16\pi}(\lambda_3-\lambda_4).
 \end{align}
 \end{subequations}

 The condition of vacuum stability that the Higgs potential must be 
 bounded from below gives~\cite{Deshpande:1977rw,Nie:1998yn,Kanemura:1999xf} 
 \begin{align}
  &\lambda_1>0,\quad \lambda_2>0,\quad
  \sqrt{\lambda_1\lambda_2}+\lambda_3+{\rm Min}(0,\lambda_4-|\lambda_5|)>0. 
 \end{align}

 The parameter space of the model is constrained by these conditions on
 the coupling constants in the Higgs potential. \\

 \noindent 
 {\it Constraints on the Higgs potential from electroweak
   precision observables}

 Further constraints on the Higgs sector of the 2HDM are from the
 electroweak precision measurements. 
 The $S$, $T$ and $U$ parameters~\cite{Peskin} 
 are sensitive to the loop effects of Higgs bosons~\cite{Toussaint:1978zm,Bertolini:1985ia}. 
 The $T$ parameter corresponds to the electroweak $\rho$ parameter,
 which is severely constrained by experimental observations as has been discussed. 
 The mass splitting between the additional Higgs bosons are strongly 
 bounded~\cite{Haber:2010bw,Kanemura:2011sj}. 
 This implies that the Higgs potential has to respect the custodial $SU(2)$ symmetry 
 approximately. \\

 \noindent
 {\it Flavour constraints on $m_{H^\pm}$ and $\tan\beta$ }

\begin{figure*}
  \begin{center}
 \vspace*{.5cm}
   \includegraphics[width=0.245\textwidth]{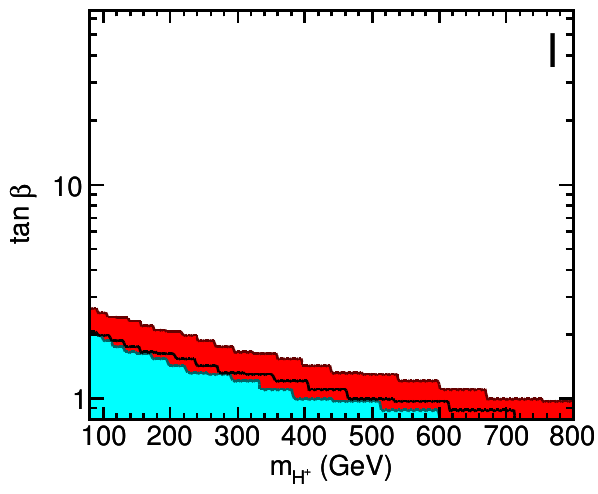}
   \includegraphics[width=0.245\textwidth]{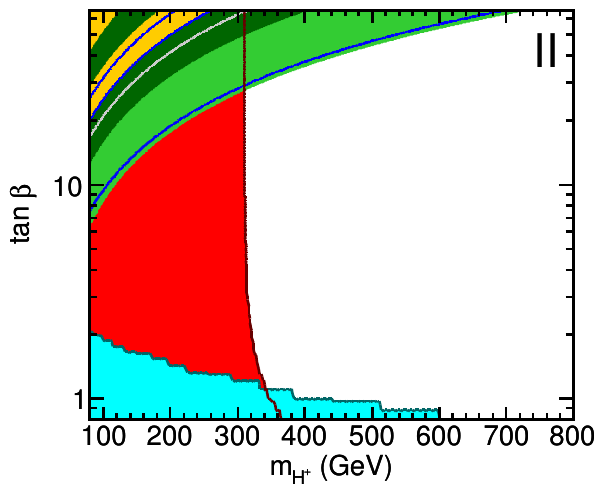}
   \includegraphics[width=0.245\textwidth]{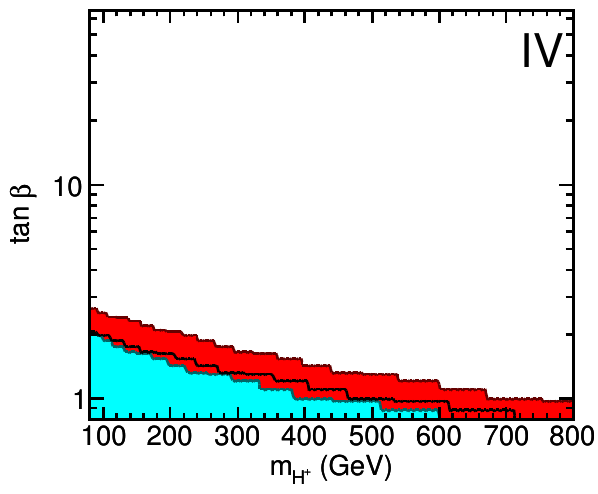}
   \includegraphics[width=0.245\textwidth]{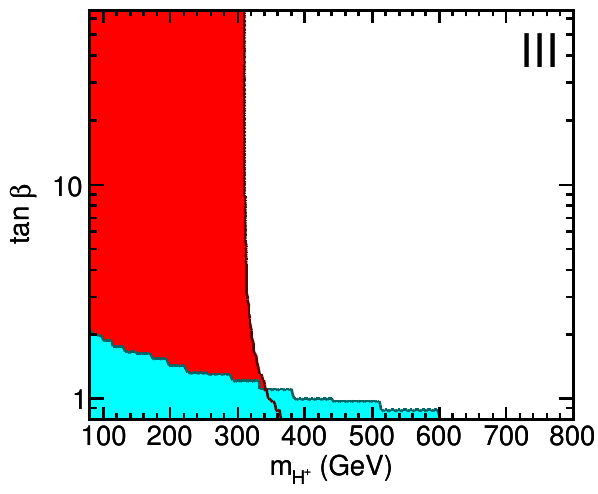}
   \caption{The constraint on the parameter space in the 2HDM for Type I, Type II, Type IV (Type X) and Type III (Type Y) 
   from various flavor experiments~\cite{Mahmoudi:2009zx}. }\label{fig:flavor}
   \end{center}
   \end{figure*}

 \begin{sloppypar}
 Flavour experiments provide strong constraints on the 2HDMs through 
 the $H^\pm$ contribution to the flavour mixing observables at the tree level 
 or at the loop level~\cite{Aoki:2009ha,Logan:2009uf,Su:2009fz}. 
 Because the amplitudes of these processes necessarily contain 
 the Yukawa interaction, constraints on the 2HDM strongly depends on 
 the type of Yukawa interaction. 
 In Ref.~\cite{Mahmoudi:2009zx}, the limits on the general couplings 
 from flavour physics are translated into those on the
 ($m_{H^\pm},\tan\beta$) plane for all four types of Yukawa interaction 
 in the 2HDM: see Fig.~\ref{fig:flavor}, where  
 Type III and Type IV  correspond to Type Y and Type X, respectively.
 See also the more recent studies~\cite{Botella:2014ska,Cheng:2014ova,Bhattacharyya:2014nja}.
 \end{sloppypar}

 A strong exclusion limit is given from the result for the
 branching ratio of the $B\to X_s\gamma$ process~\cite{Amhis}. 
 For Type-II and Type-Y, a $\tan\beta$-independent lower limit of
 $m_{H^\pm}\gtrsim 380$~GeV is obtained~\cite{oai:arXiv.org:1208.2788} 
 by comparing with the NNLO calculation~\cite{Misiak:2006ab}. 
 For Type-I and Type-X, on the other hand, $\tan\beta\lesssim 1$ is
 excluded for $m_{H^\pm}\lesssim 800$~GeV, while no lower bound on
 $m_{H^\pm}$ is obtained. 

 By the results for the  $B_{d}^0$-$\bar{B}^0_{d}$ mixing,  
  lower $\tan\beta$ regions ($\tan\beta\le1$) are excluded for $m_{H^\pm}\lesssim500$~GeV  
  for all types of Yukawa interaction~\cite{Amhis}.

 Constraints in larger $\tan\beta$ regions are obtained only for 
 Type-II, which come from the results for leptonic meson decay 
 processes~\cite{Amhis}, $B\to\tau\nu$~\cite{Hou:1992sy} and
 $D_s\to\tau\nu$~\cite{Akeroyd:2009tn}. 
 Upper bounds on $\tan\beta$ are obtained at around 30 for 
 $m_{H^\pm}\simeq350$~GeV and around 60 for
 $m_{H^\pm}\simeq700$~GeV~\cite{Mahmoudi:2009zx}.
 On the other hand, the other types do not receive 
 any strong constraint for large $\tan\beta$ values, because
 the relevant couplings behave $\xi_A^d\xi_A^\ell=\tan^2\beta$ for Type-II 
 while $\xi_A^d\xi_A^\ell=-1$ ($\cot^2\beta$) for Type-X and Type-Y (Type-I). 
  \\

 \noindent
 {\it Constraint from the data at LEP/SLC, Tevatron and also from the current LHC data}

 \begin{sloppypar}
 At the LEP direct search experiments, lower mass bounds on $H$ and $A$ have been
 obtained as $m_H>92.8$~GeV and $m_A>93.4$~GeV in the CP-conservation
 scenario~\cite{Abdallah:2004wy,Schael:2006cr}. 
 Combined searches for $H^\pm$ give the lower mass bound $m_{H^\pm}>80$~GeV, 
 by assuming ${\mathcal B}(H^+\to\tau^+\nu)+{\mathcal B}(H^+\to
 c\bar{s})=1$~\cite{Achard:2003gt,Abdallah:2003wd,Abbiendi:2013hk}.
 \end{sloppypar}

 At the Fermilab Tevatron, CDF and D0 collaborations have studied 
 the processes of $p\bar{p}\to b\bar{b}H/A$, followed by $H/A\to b\bar{b}$ or
 $H/A\to\tau^+\tau^-$~\cite{Aaltonen:2011nh,Abazov:2011up,Aaltonen:2012zh}. 
 By using the $\tau^+\tau^-$ ($b\bar{b}$) decay mode, which can be
 sensitive for the cases of Type-II (Type-II and Type-Y), upper bounds on 
 $\tan\beta$ have been obtained to be from about 25 to 80 (40 to 90) for $m_A$
 from 100~GeV to 300~GeV, respectively. 
 For the direct search of $H^\pm$, the decay modes of
 $H^\pm\to\tau\nu$ and $H^\pm\to cs$ have been investigated by using 
 the production from the top quark decay  $t\to
 bH^\pm$~\cite{Abazov:2008rn,Abazov:2009aa,Aaltonen:2009ke}. 
 Upper bounds on  ${\mathcal B}(t\to bH^\pm)$ have been obtained, 
 which can be translated into the bound on $\tan\beta$ in various scenarios. 
 For Type-I with $H^\pm$ heavier than the top quark, 
 upper bounds on $\tan\beta$ have been obtained to be from around 20 to 70 for
 $m_{H^\pm}$ from 180~GeV to 190~GeV, respectively~\cite{Abazov:2008rn}. 

 At the LHC, additional Higgs boson searches have been
 performed by using currently accumulated events at the experiments 
 with a center-of-mass energy of 7~TeV with the integrated luminosity of 
 4.9~fb$^{-1}$ in 2011 and also 8~TeV with 19.7~fb$^{-1}$ in 2012. 
 The CMS Collaboration has searched $H$ and $A$ which decay into the
 $\tau^+\tau^-$ final state, and upper limits on $\tan\beta$ have been
 obtained in the MSSM (or in the Type-II 2HDM) from 4 to 60 for $m_A$ from 
 140~GeV to 900~GeV, respectively~\cite{CMS_neutral_new}. 
 By the ATLAS Collaboration similar searches have also been done~\cite{Aad:2012cfr} .
 In the Type-II and Type-Y 2HDMs, CMS has also searched the 
 bottom-quark associated production process of $H$ or $A$
 which decays into the $b\bar{b}$ final state~\cite{CMS_neutral_old}, 
 and has obtained the upper bounds on $\tan\beta$: i.e.,  
 $\tan\beta\gtrsim 16$ (28) is excluded at $m_{A}=100$~GeV (350~GeV). 
 ATLAS has reported the $H^\pm$ searches via the $\tau$+jets
 final state~\cite{Aad:2012tj,ATLAS_charged_constraints}. 
 In the Type-II 2HDM with $m_{H^{\pm}}\lesssim m_{t}$, wide parameter
 regions have been already excluded by the data for $100$~GeV $\lesssim
  m_{H^\pm}\lesssim140$~GeV with $\tan\beta\gtrsim 1$. 
 Moreover,  the parameter regions of $\tan\beta\gtrsim 50$ at $m_{H^\pm}=200$~GeV 
 and $\tan\beta\gtrsim 65$ at $m_{H^\pm}=300$~GeV have been excluded 
 for $m_{H^\pm}\gtrsim 180$~GeV, respectively. 
 The searches for $H^\pm$ in the $cs$ final-state have been
 performed by ATLAS~\cite{Aad:2013hla}, and the upper limit on the
 branching ratio of $t\to bH^\pm$ decay is obtained assuming the 100\%
 branching ratio of $H^\pm\to cs$.
 For $\sin(\beta-\alpha)<1$, searches for $H\to W^+W^-$, $hh$ and $A\to
  Zh$ give constraints on the 2HDMs with Type-I and Type-II
  Yukawa interactions~\cite{ATLAS:2013zla,CMS:2013eua}. \\

 \begin{figure*}[t]
  \begin{center}
   \includegraphics[width=0.245\textwidth]{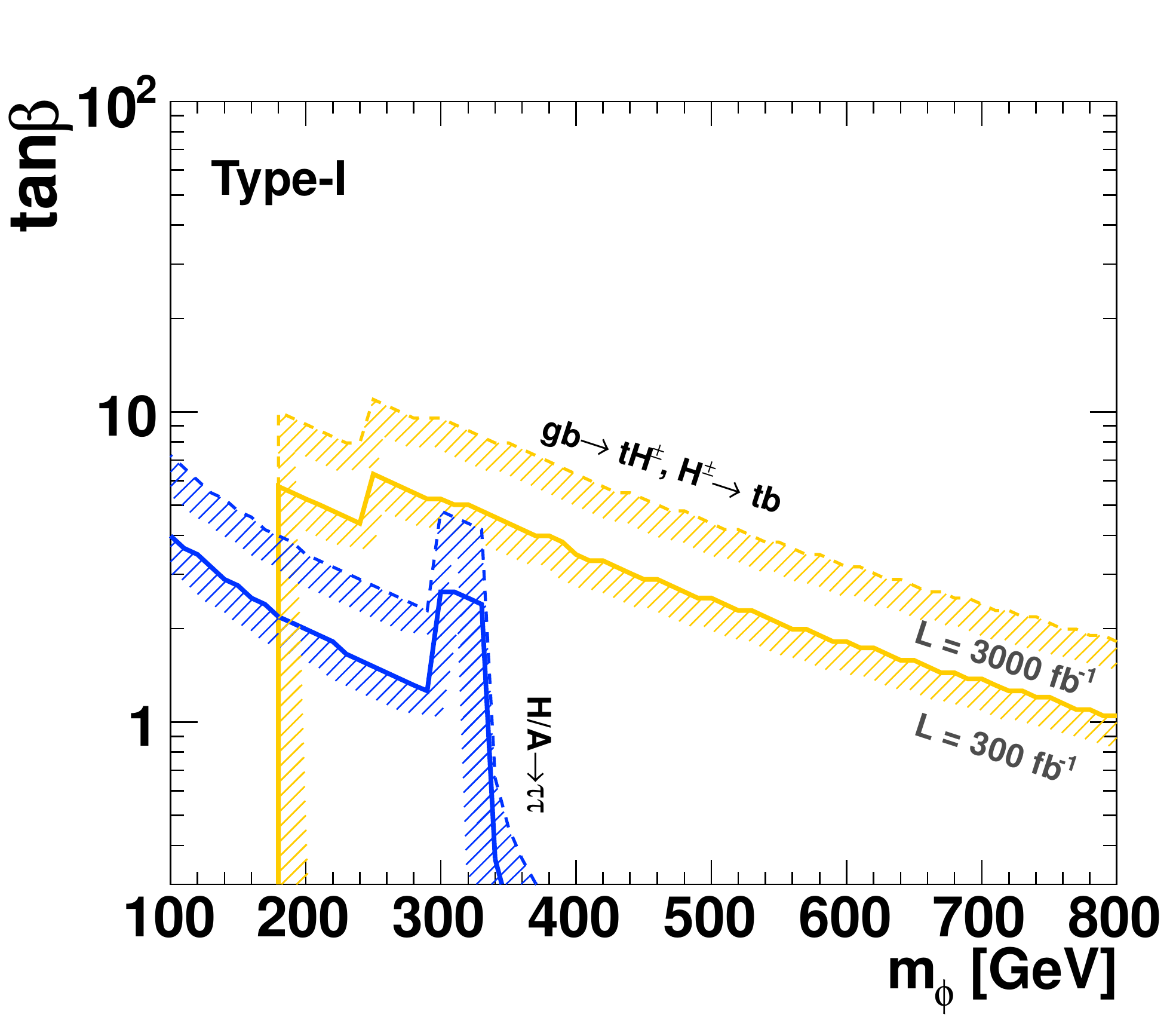}
   \includegraphics[width=0.245\textwidth]{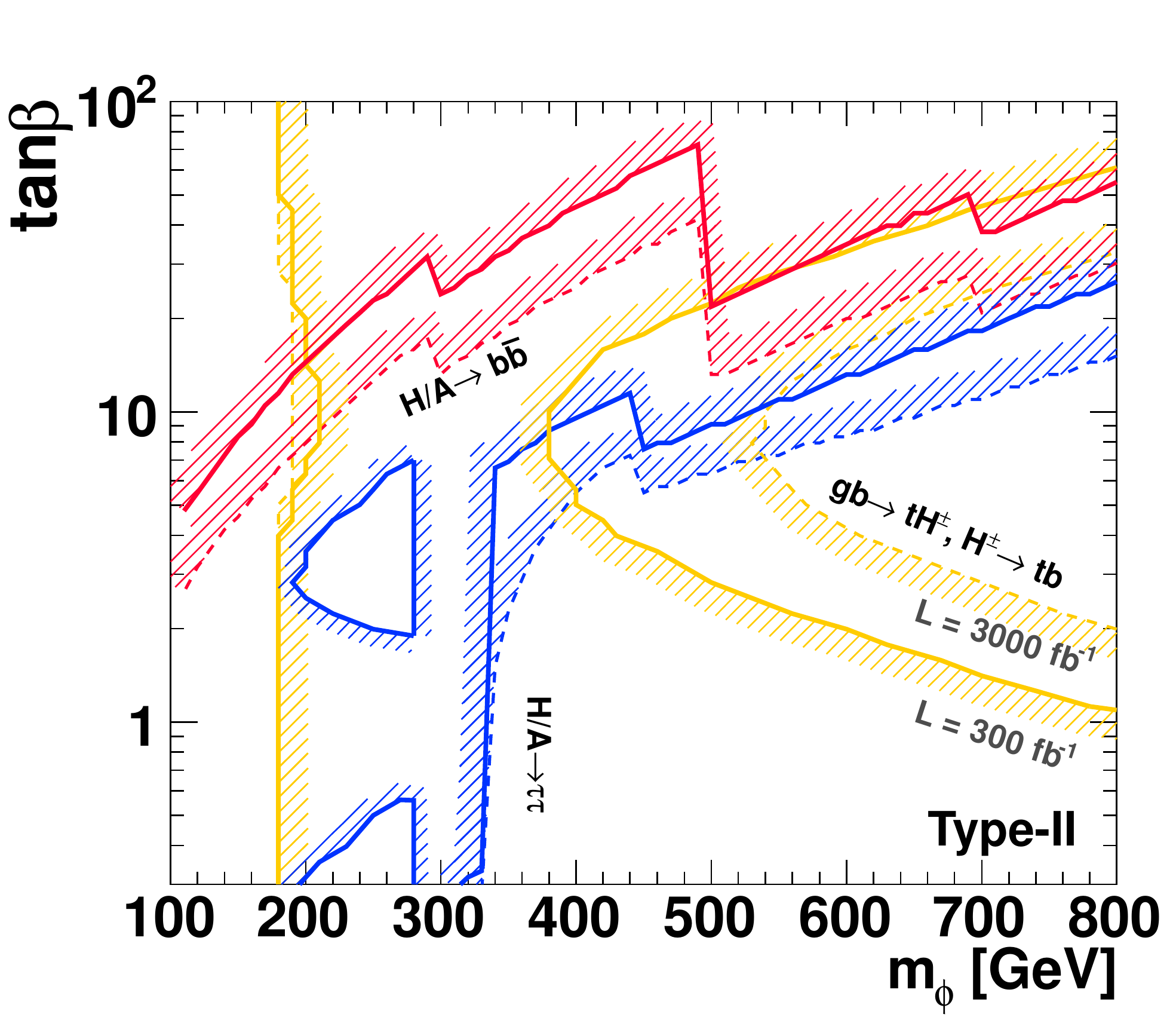}
   \includegraphics[width=0.245\textwidth]{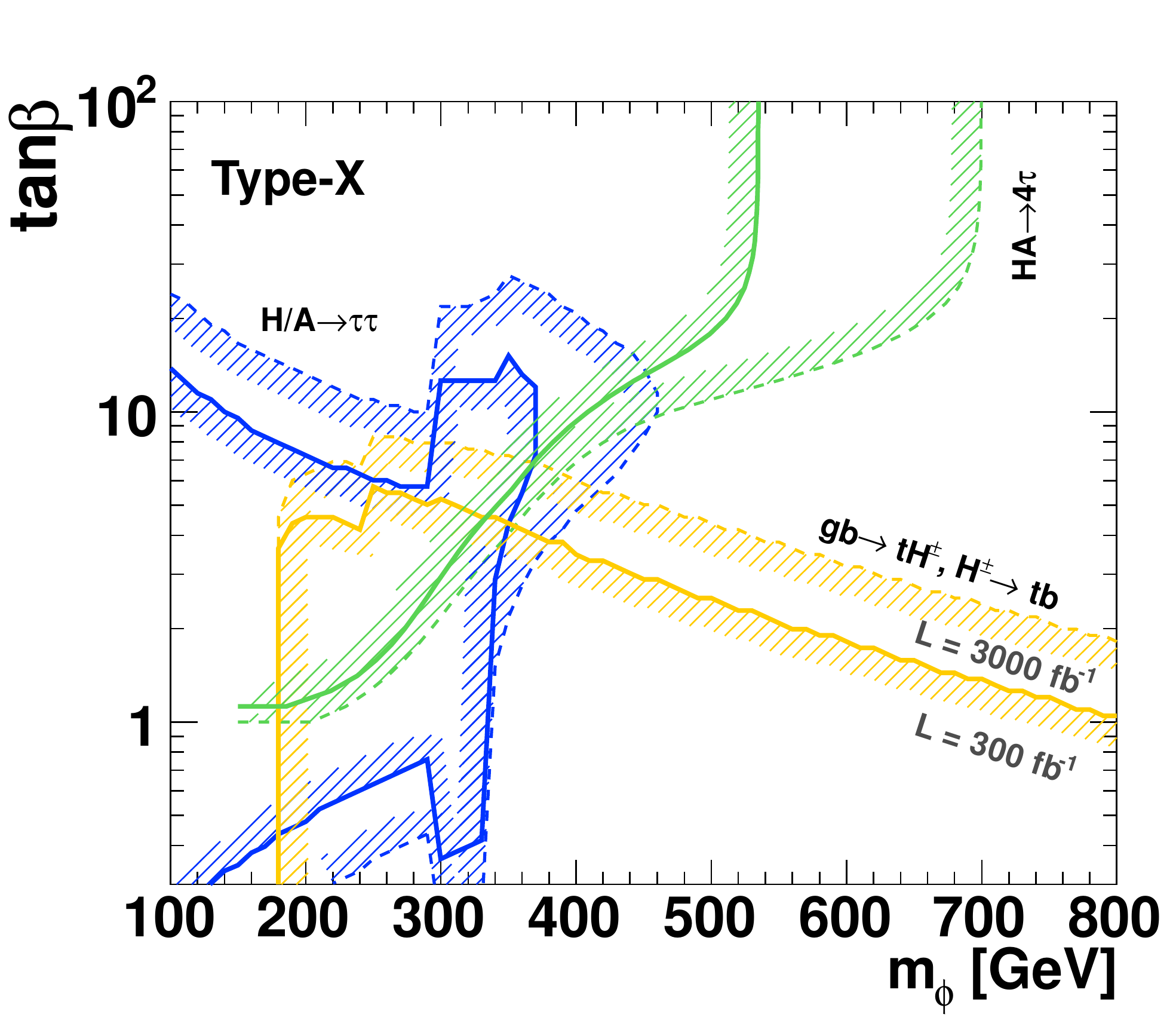}
   \includegraphics[width=0.245\textwidth]{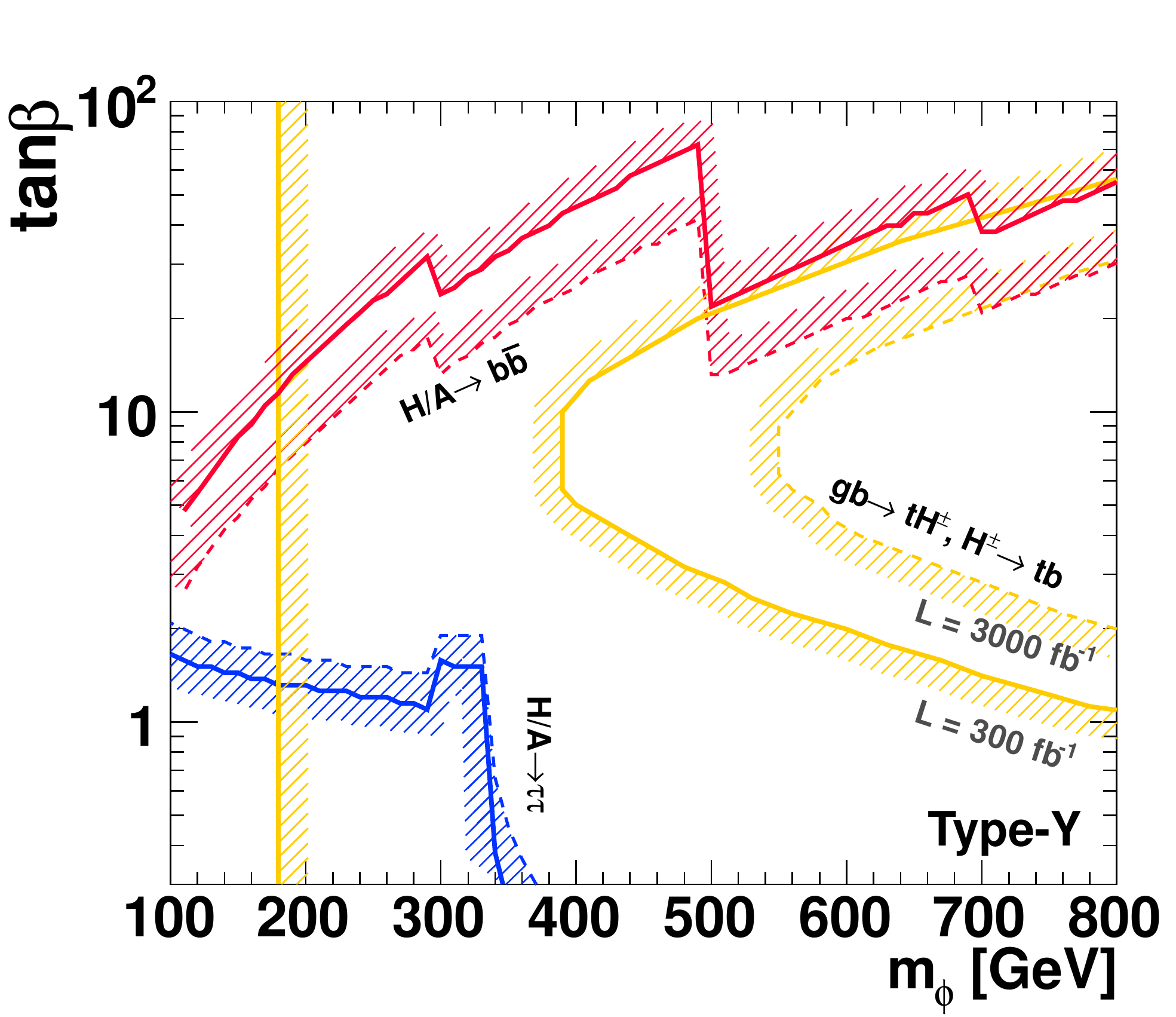}
   \caption{
   Expected exclusion regions ($2\sigma$ CL) in the
   plane of $\tan\beta$ and the mass scale $m_{\phi}$ of the additional
   Higgs bosons at the LHC.
   Curves are evaluated by using the signal and background analysis given
   in Ref.~\cite{ATLAS:1999vwa} for each process, where the signal events
   are rescaled to the prediction in each case~\cite{Asner:2013psa,KTYY},
   except the $4\tau$ process for which we follow the analysis in
   Ref.~\cite{Kanemura:2011kx}. 
   Thick solid lines are the expected exclusion contours by
   $L=300$~fb$^{-1}$ data, and thin dashed lines are for
   $L=3000$~fb$^{-1}$ data. 
   For Type-II, the regions indicated by circles may not be excluded by
   $H/A\to\tau^+\tau^-$ search by using the 300~fb$^{-1}$ data due to the
   large SM background. 
   }\label{fig:LHCreach}
  \end{center}
 \end{figure*}
 %

 \noindent
 {\it Prospect of extra Higgs boson searches at the LHC (13-14 TeV)}

  At the  LHC experiments with the collision energy of $13$-$14$ TeV 
  and the integrated luminosity of $L=300$~fb$^{-1}$ and also 3000~fb$^{-1}$,  
  the expected discovery potential for additional Higgs bosons 
  have been studied in the 2HDM in Refs.~\cite{Asner:2013psa,KTYY,Kanemura:2014dea},  by using the signal 
  and background analysis for various channels given in Ref.~\cite{ATLAS:1999vwa}.
 Processes available for the searches for additional Higgs bosons are~\cite{Kanemura:2014dea} 
 \begin{sloppypar}
 \begin{itemize}
  \item $H/A (+ b\bar{b})$ inclusive and associated production followed
        by the $H/A\to\tau^+\tau^-$ decay~\cite{Baglio:2010ae}.
  \item $H/A+b\bar{b}$ associated production followed by the $H/A\to
        b\bar{b}$ decay~\cite{Dai:1994vu,DiazCruz:1998qc,Baglio:2010ae}.
  \item $gb\to tH^\pm$ production followed by the $H^\pm\to
        tb$ decay~\cite{Borzumati:1999th,Plehn:2002vy}.
  \item $q\bar{q}\to HA\to 4\tau$
        process~\cite{Kanemura:2011kx,Liu:2013gba}.
 \end{itemize}
 \end{sloppypar}
 For the production cross sections, the tree-level cross sections 
 have been convoluted with the CTEQ6L parton distribution
 functions~\cite{Pumplin:2002vw}. 
 The scales of the strong coupling constant and the parton distribution
 function are chosen to the values used in
 Ref.~\cite{Djouadi:2005gi}. 
 For details, see Ref.~\cite{Kanemura:2014dea}, 
 where the latest recommendations from the LHC Higgs Cross Section
 Working Group for 2HDM cross section (and branching ratio) evaluations
 can be found in \citere{2hdm-lhchxswg}.

 In Fig.~\ref{fig:LHCreach}, the contour plots of the expected
 exclusion regions [$2\sigma$ confidence level (CL)] in the
 $(m_\phi,\tan\beta)$ plane are shown 
 at the LHC $\sqrt{s}=14$~TeV with the integrated luminosity of 300~fb$^{-1}$ (thick solid lines) and
 3000~fb$^{-1}$ (thin dashed lines), where $m_\phi$ represents common masses of
 additional Higgs bosons. 
 From the left panel to the right panel, the results for
 Type-I, Type-II, Type-X and Type-Y are shown separately. 
 Following the analysis in Ref.~\cite{ATLAS:1999vwa}, the reference values of the expected 
 numbers of signal and background events are changed at the several values of 
 $m_\phi$~\cite{KTYY, Kanemura:2014dea}, 
 which makes sharp artificial edges of the curves in Figs.~\ref{fig:LHCreach}. 

 For Type-I, $H/A$ production followed by the decay into  $\tau^+\tau^-$ can
 be probed for $\tan\beta\lesssim 3$ and $m_{H,A}\le350$~GeV, 
 where the inclusive production cross section is enhanced by the relatively 
 large top Yukawa coupling with the sizable $\tau^+\tau^-$ branching ratio.  
 The $tH^\pm$ production decaying into $H^\pm\to tb$ can be
 used to search $H^\pm$ in relatively smaller $\tan\beta$ regions. 
  $H^\pm$ can be discovered for $m_{H^\pm} <  800$~GeV and 
 $\tan\beta\lesssim1$ (2) for the integrated luminosity of 300~fb$^{-1}$
 (3000~fb$^{-1}$). 

 \begin{sloppypar}
 For Type-II, the inclusive and the bottom-quark-associated production
 processes of $H/A$ with the decay into $\tau^+\tau^-$ or the $b
 \overline b$ can be used to search $H$ and $A$ for relatively large
 $\tan\beta$.  They can also be used in relatively small $\tan\beta$ regions 
 for $m_{H,A}\lesssim350$~GeV.  
 $H^\pm$ can be searched by the $tH^\pm$ production with 
 $H^\pm\to tb$ decay for $m_{H^\pm}\gtrsim180$~GeV for 
 relatively small and large $\tan\beta$ values. 
 The region of $m_{H^\pm}\gtrsim350$~GeV (500~GeV) could be 
 excluded with the 300~fb$^{-1}$ (3000~fb$^{-1}$) data. 
 \end{sloppypar}

 For Type-X, $H$ and $A$ can be searched 
 via the inclusive production and $HA$ pair production  
 by using the $\tau^+\tau^-$ decay mode, which is dominant.  
 The inclusive production could exclude the region of 
 $\tan\beta\lesssim 10$ with $m_{H,A}\lesssim 350$~GeV.   
 Regions up to $m_{H,A}\simeq500$~GeV (700~GeV) 
 with $\tan\beta\gtrsim 10$
 could be excluded by using the pair production 
 with the 300~fb$^{-1}$ (3000~fb$^{-1}$) data. 
 The search for $H^\pm$ is the similar to that for Type-I. 

 Finally, for Type-Y, the inclusive production of $H$ and $A$ f
 ollowed by $H/A \to \tau^+\tau^-$ can be searched for the regions of
 $\tan\beta\lesssim 2$ and $m_{H,A}\le350$~GeV.
 The bottom-quark associated production of $H$ and $A$ with 
 $H/A\to b\bar{b}$ can be searched for the regions of
 $\tan\beta\gtrsim 30$ up to $m_{H,A}\simeq800$~GeV. 
 The search of $H^\pm$ is similar to that for Type-II. 

 For Type-II and Type-Y (Type-X),  if all the curves are combined by 
 assuming that all the masses of additional Higgs bosons are the same, 
  the mass below 400~GeV~(350~GeV) coud be excluded by the 300~fb$^{-1}$ data  
  for all value of $\tan\beta$, and with 3000~fb$^{-1}$, the mass below 
  550~GeV (400~GeV) could be excluded. 
 On the other hand,  for Type-I, the regions with $\tan\beta\gtrsim 5$~(10) 
 cannot be excluded by 300~fb$^{-1}$~(3000~fb$^{-1}$) data. 
 In the general 2HDM, however, the mass spectrum of additional Higgs
 boson has more degrees of freedom, so that we can still find allowed 
 parameter regions where $m_H$ is relatively light but  
 $m_A (\simeq m_{H^\pm})$ are heavy. 
 Thus, the overlaying of these exclusion curves for different additional
 Higgs bosons may only be applied to the case of $m_H=m_A=m_{H^\pm}$.

 \begin{figure*}[t]
 \begin{center}
 \includegraphics[width=0.24\textwidth]{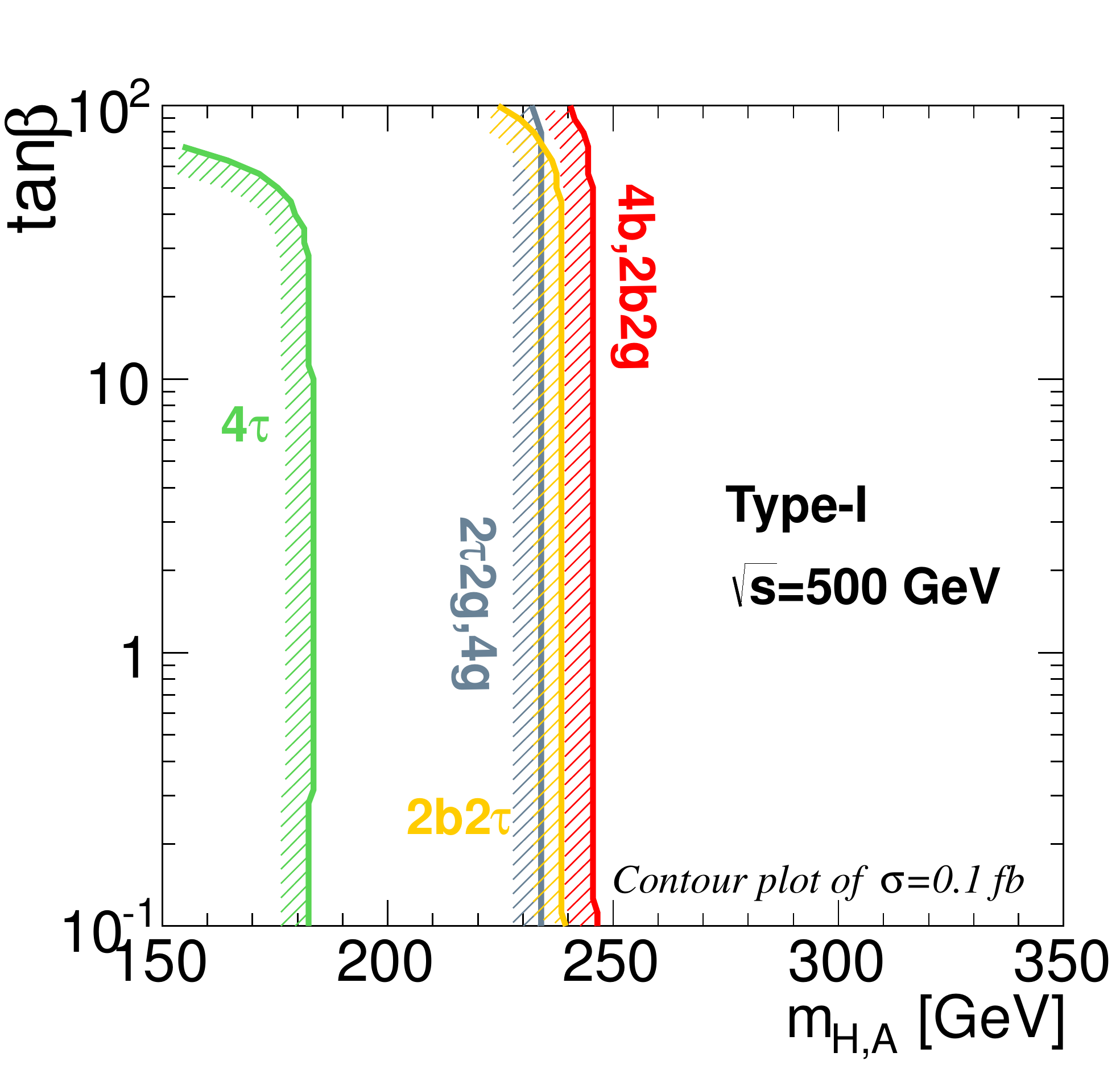} 
 \includegraphics[width=0.24\textwidth]{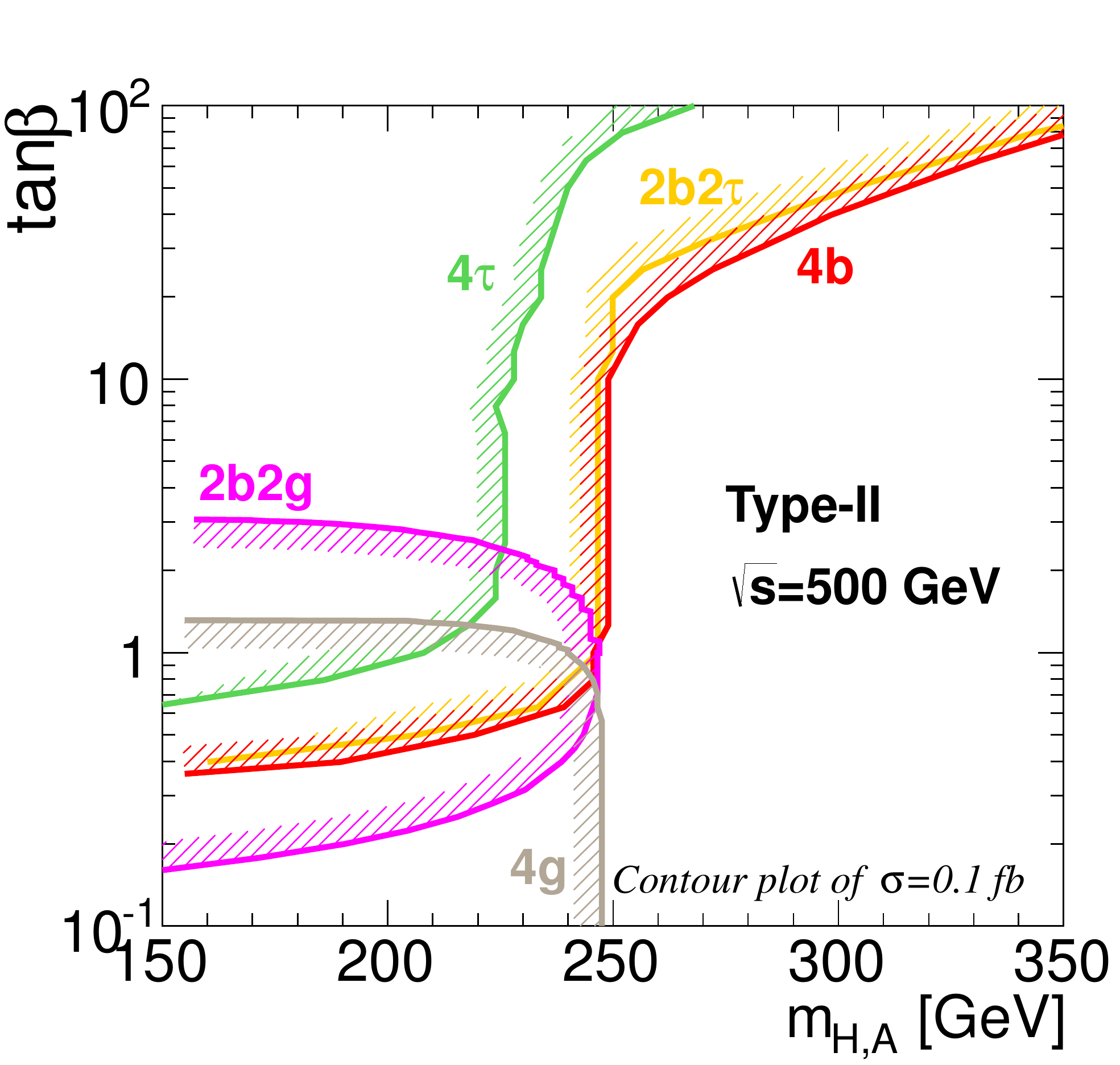} 
 \includegraphics[width=0.24\textwidth]{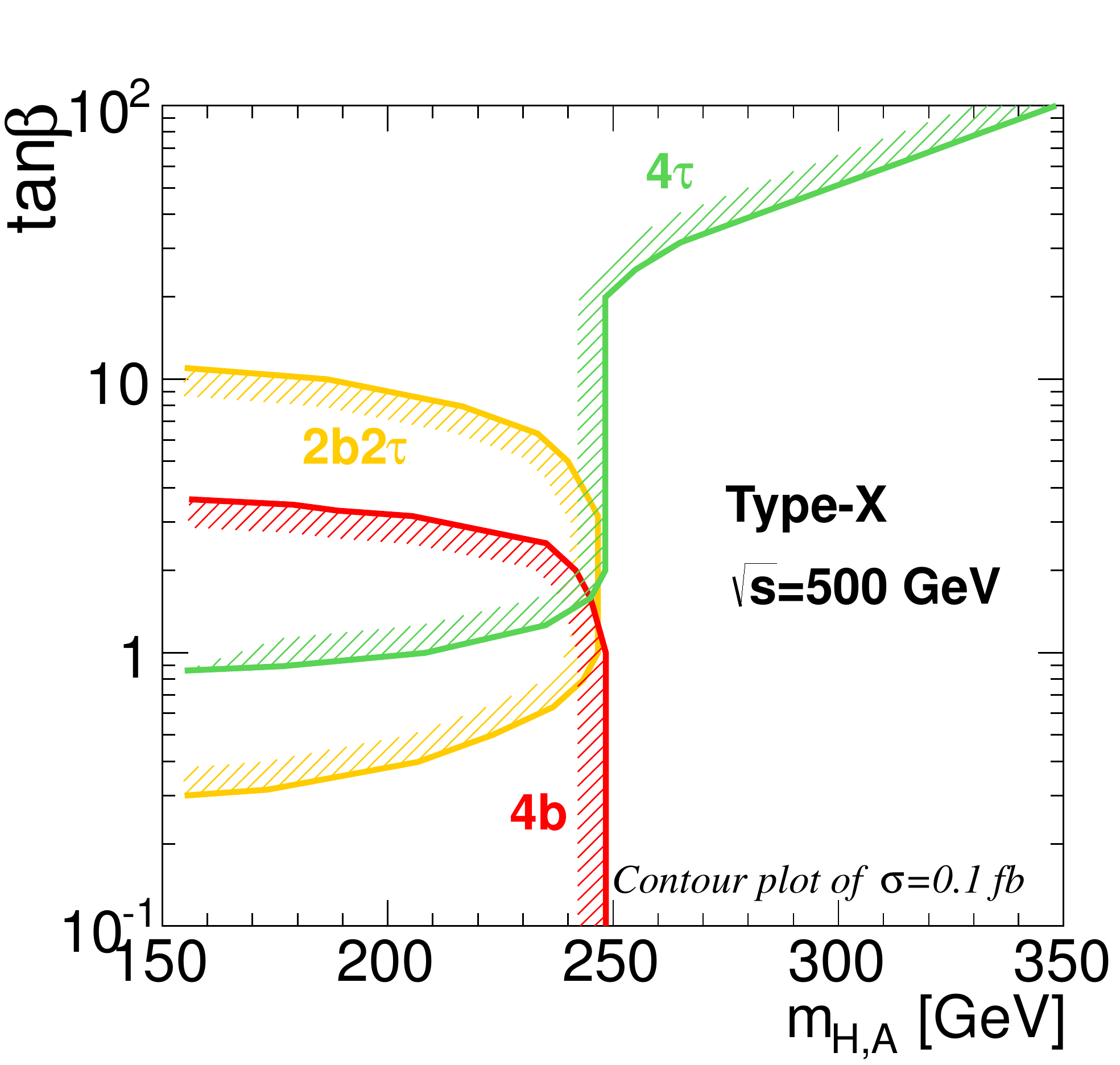} 
 \includegraphics[width=0.24\textwidth]{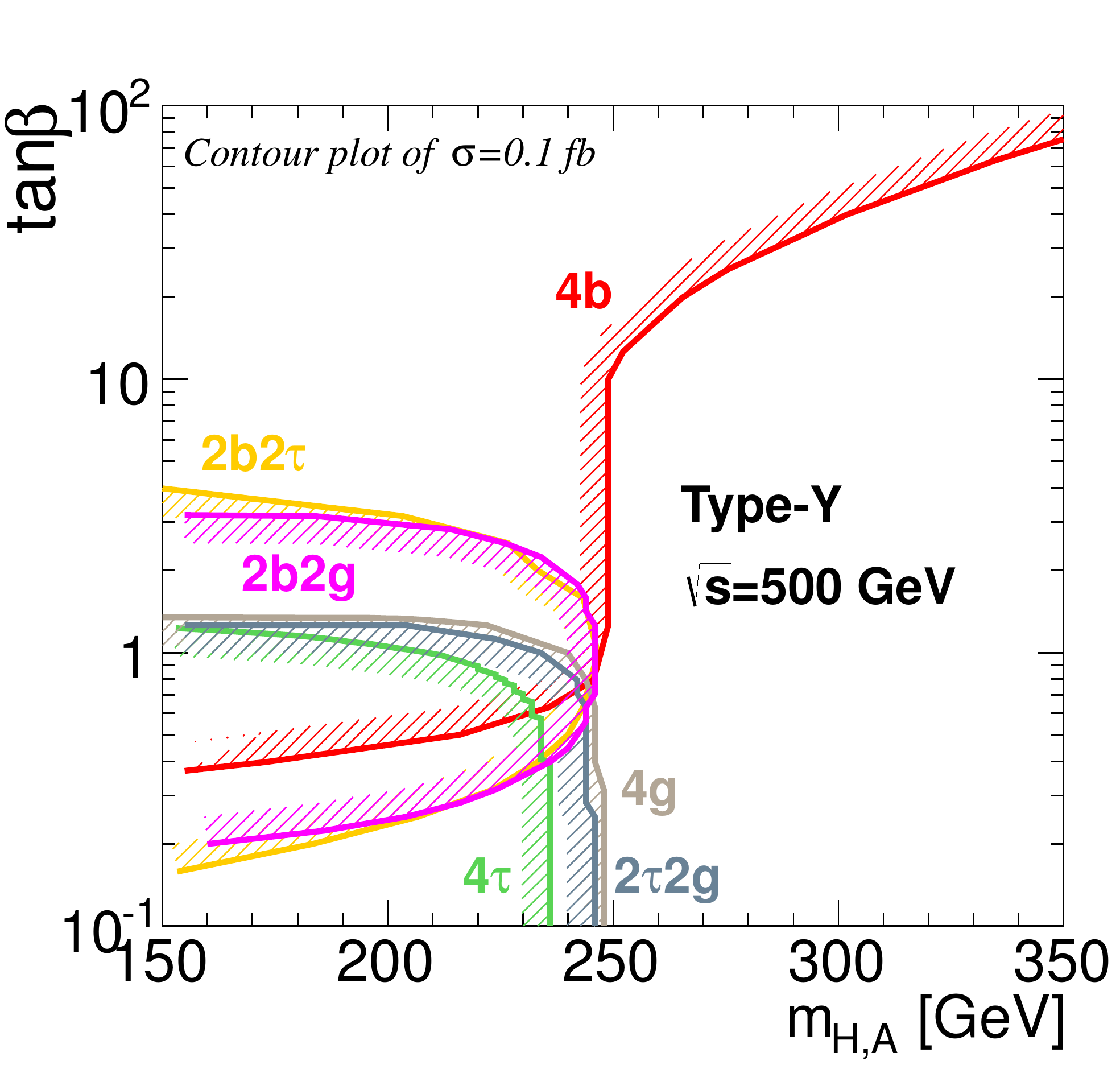} \\
  \includegraphics[width=0.24\textwidth]{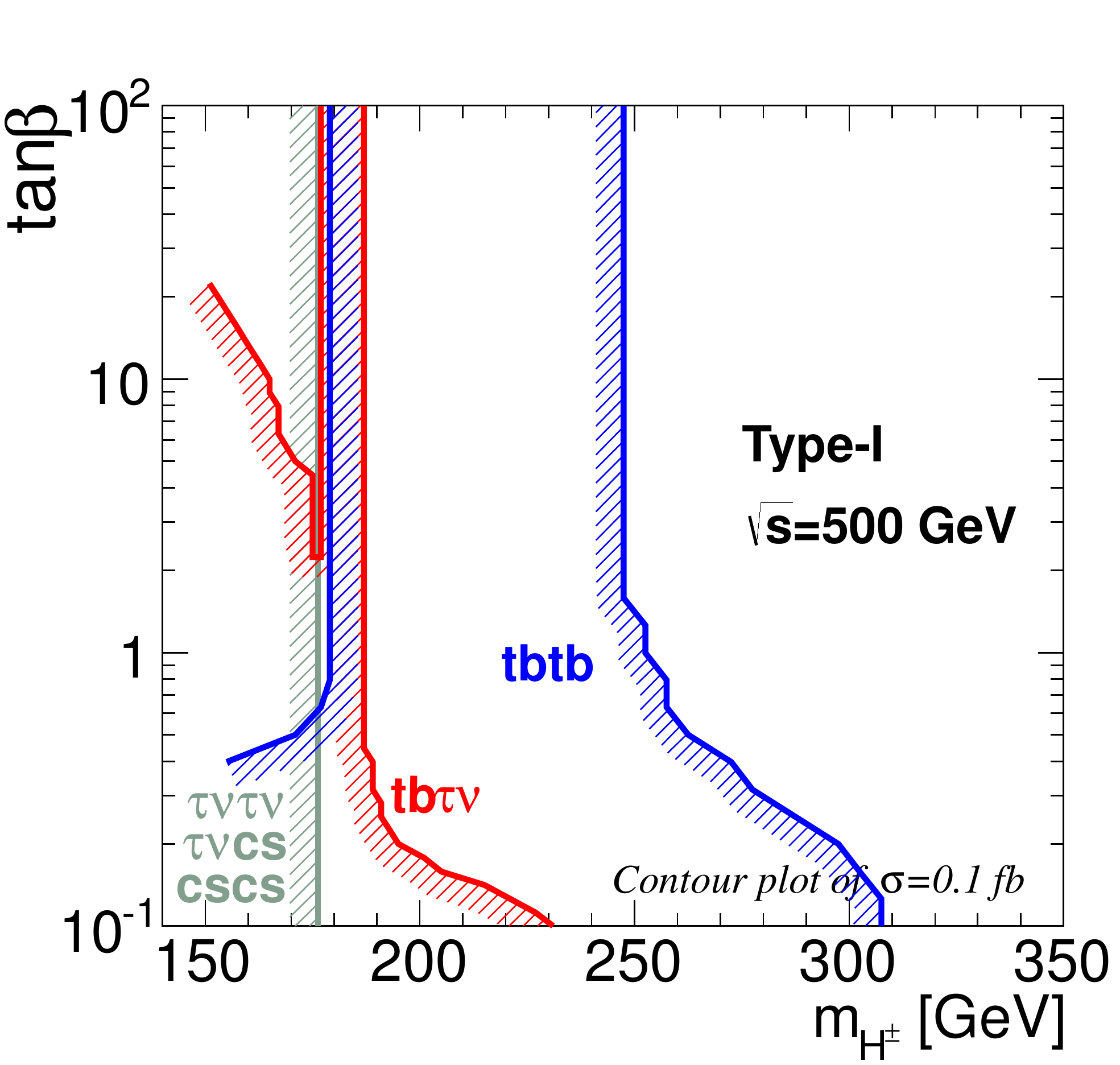} 
   \includegraphics[width=0.24\textwidth]{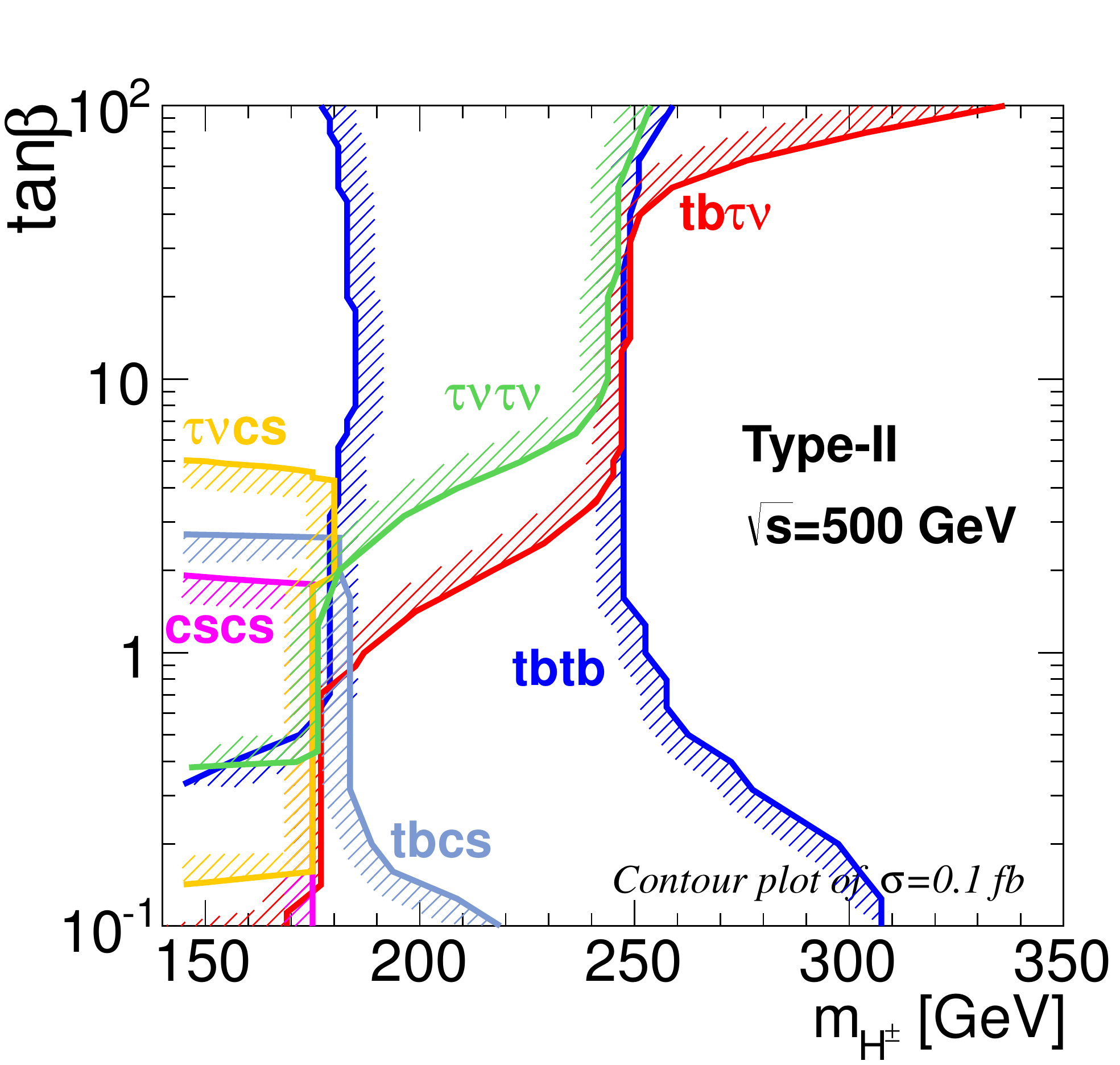} 
    \includegraphics[width=0.24\textwidth]{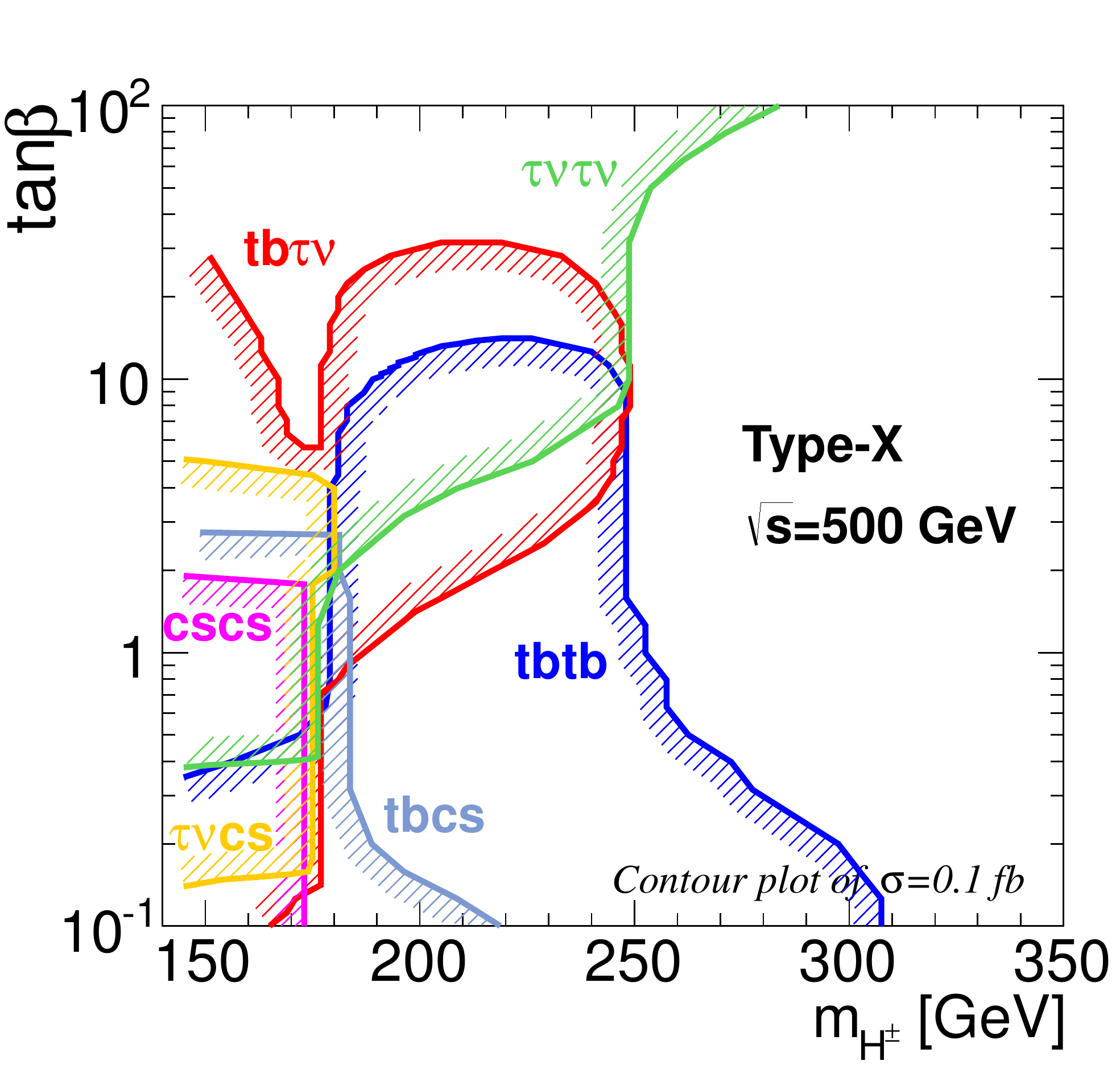}   
   \includegraphics[width=0.24\textwidth]{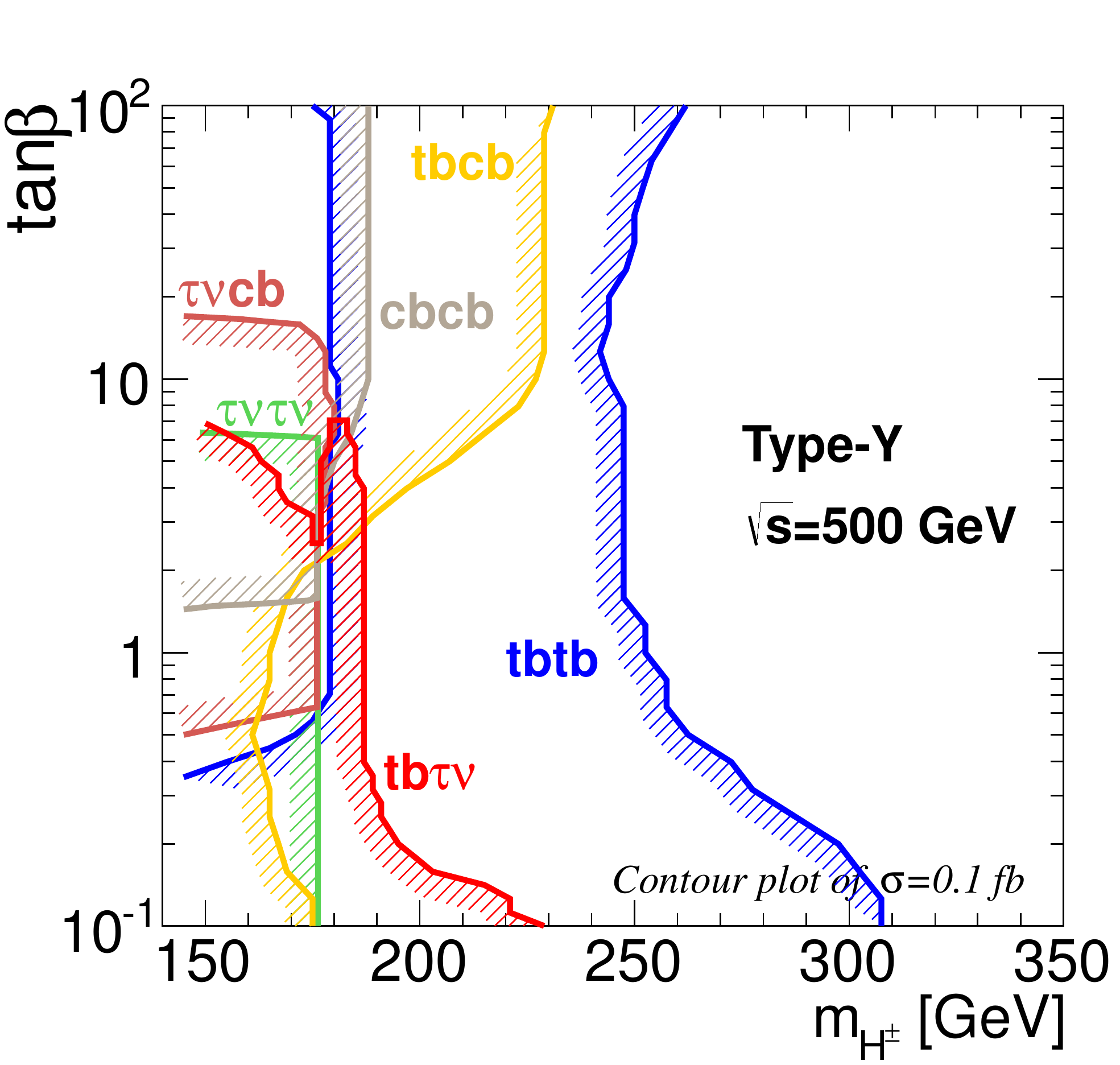} 
  \caption{Contour plots of the four-particle production cross sections
  through the $H/A$ production and $H^\pm$ production process at the ILC with 
  $\sqrt{s}=500$ ~GeV in the $(m_{H^\pm},\tan\beta)$  plane.
  Contour of $\sigma=0.1$~fb is drawn for each signature~\cite{Kanemura:2014dea}..} 
  \label{fig:ILC500GeV}
  \end{center}
 \end{figure*}

 If $H^\pm$ is discovered at the LHC, its mass could be determined 
 immediately~\cite{ATLAS:1999vwa,Rindani:2013mqa}. 
 Then the determination of the type of the Yukawa interaction becomes 
 important. 
 At the LHC, however, we would not completely distinguish the types of Yukawa interaction, 
 because the Type-I and Type-X, or Type-II and Type-Y have  
 a common structure for the $tbH^\pm$ interaction. 
 In addition, as seen in Fig.~\ref{fig:LHCreach}, there can be no
 complementary process for the neutral Higgs boson searches in some
 parameter regions; e.g., $m_{H,A}\gtrsim350$~GeV with relatively small
 $\tan\beta$, depending on the type of the Yukawa interaction.  
 At the ILC, on the other hand, as long as $m_{H,A}\lesssim 500$~GeV, 
 the neutral Higgs bosons can be produced and investigated almost
 independent of $\tan\beta$. 
 Therefore,  it is quite important to search for the
 additional Higgs bosons with the mass of $350$-$500$~GeV, and to
 determine the models and parameters at the ILC, even after the LHC. 

 Notice that the above results are obtained in the SM-like limit,
 $\sin(\beta-\alpha)=1$. 
 A deviation from the SM-like limit causes appearance of additional 
 decay modes such as  $H\to W^+W^-$, $ZZ$, $hh$ as well as $A\to
 Zh$~\cite{Gunion:1989we,Gunion:1990kf,Craig:2013hca,Baglio:2014nea}. 
 Especially, for Type-I with a large value of $\tan\beta$, branching
 ratios of these decay modes can be dominant even with a small deviation
 from the SM-like limit~\cite{Aoki:2009ha,Craig:2013hca}. 
 Therefore, searches for additional Higgs bosons in these decay modes can
 give significant constraints on the deviation of $\sin(\beta-\alpha)$
 from the SM-like limit~\cite{ATLAS:2013zla,CMS:2013eua}, which is
 independent of coupling constants of $hVV$. \\

 \noindent
 {\it Prospect for the searches for the additional Higgs bosons at the
  ILC}\label{sec:ilc} 

 At LC's the main production mechanisms of additional Higgs bosons in the 2HDM 
 are $e^+ e^- \to H A$ and $e^+ e^- \to H^+ H^-$, where a pair of
 additional Higgs bosons is produced via gauge interactions as long as 
 kinematically allowed. 
 For energies below the threshold, the single production processes, 
 $e^+e^- \to H(A) f \bar{f}$ and $e^+ e^- \to H^\pm f \bar{f}'$ are the
 leading contributions~\cite{Kanemura:2000cw}. 
 They are enhanced when the relevant Yukawa
 couplings  $\phi f \bar{f}^{(')}$ are large. 
 The cross sections of these processes have been studied
 extensively~\cite{Kanemura:2000cw,Moretti:2002pa,Kiyoura:2003tg,%
 Behnke:2013lya},
 mainly for the MSSM or for the Type-II 2HDM. 

 Here, we discuss the result in the general 2HDMs but with
 softly-broken discrete symmetry.  The following processes are 
 considered:  
 \begin{subequations}
 \begin{align}
  &e^+e^-\to \tau^+\tau^- H, \quad  \tau^+\tau^- A,\label{eq:tautauH}\\
  &e^+e^-\to b\bar{b} H, \quad\quad b\bar{b} A,\label{eq:bbH}\\
  &e^+e^-\to t\bar{t} H,  \quad\quad t\bar{t} A, \label{eq:ttH}\\
  &e^+e^-\to \tau^-\nu H^+, \quad \tau^+ \bar{\nu} H^-, \label{eq:taunuH}\\
  &e^+e^-\to \bar{t}b H^+, \quad\quad \bar{b}t H^-.\label{eq:tbH}
 \end{align}
 \end{subequations}
 For energies above the threshold of the pair production,
$\sqrt{s}>m_H+m_A$, the contribution from $e^+e^-\to HA$ can be
significant in the processes in Eqs.~(\ref{eq:tautauH}-\ref{eq:ttH}). 
Similarly for $\sqrt{s}>2m_{H^\pm}$, the contribution from $e^+e^-\to
H^+H^-$ can be significant in the processes in
Eqs.~(\ref{eq:taunuH}, \ref{eq:tbH}). 
Below the threshold, the processes including diagrams of 
$e^+e^-\to f \bar{f}^\ast$ and $e^+e^-\to f^\ast\bar{f}$ dominate.

\begin{figure*}[t]
\begin{center}
 \includegraphics[width=0.24\textwidth]{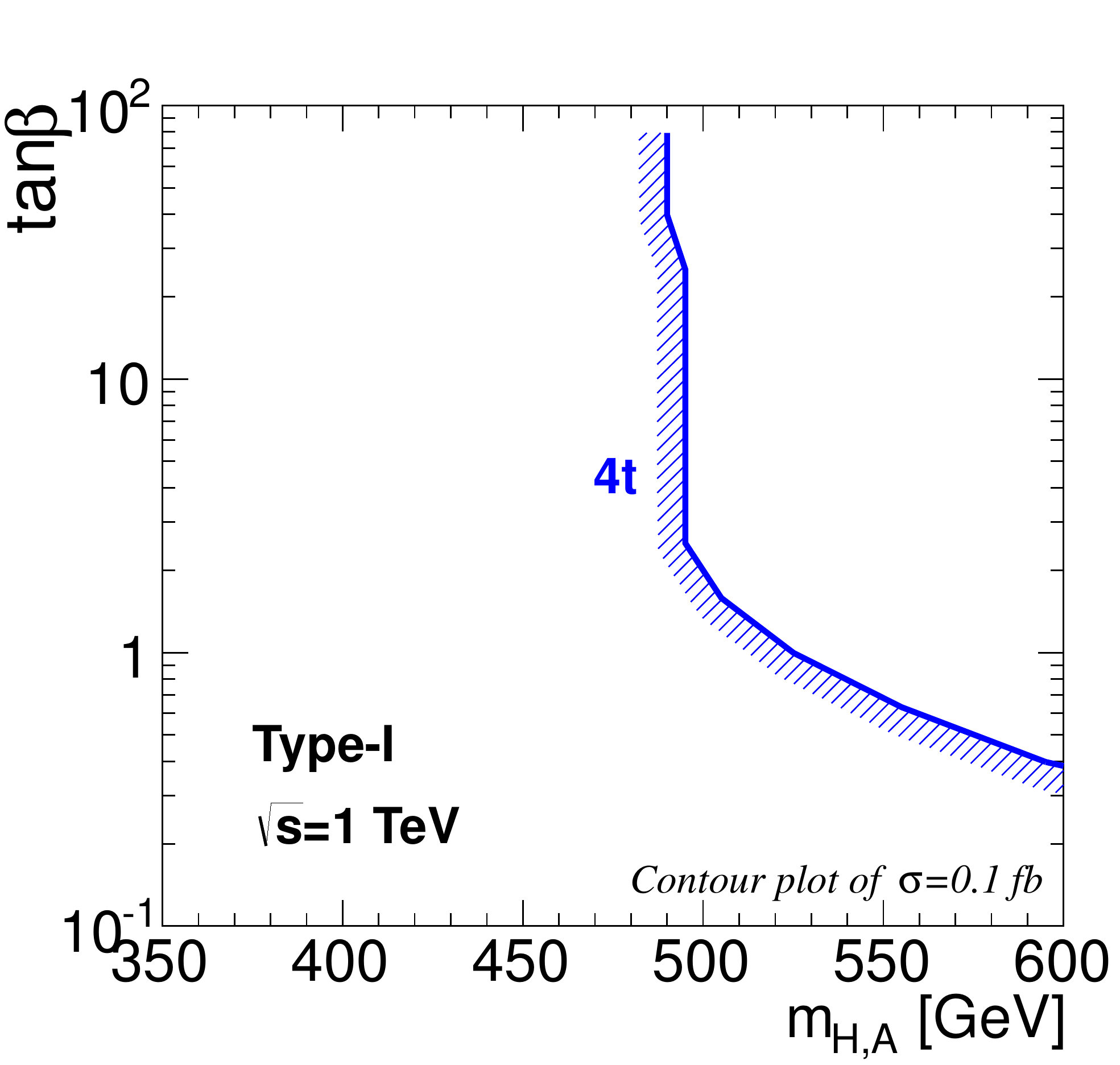} 
 \includegraphics[width=0.24\textwidth]{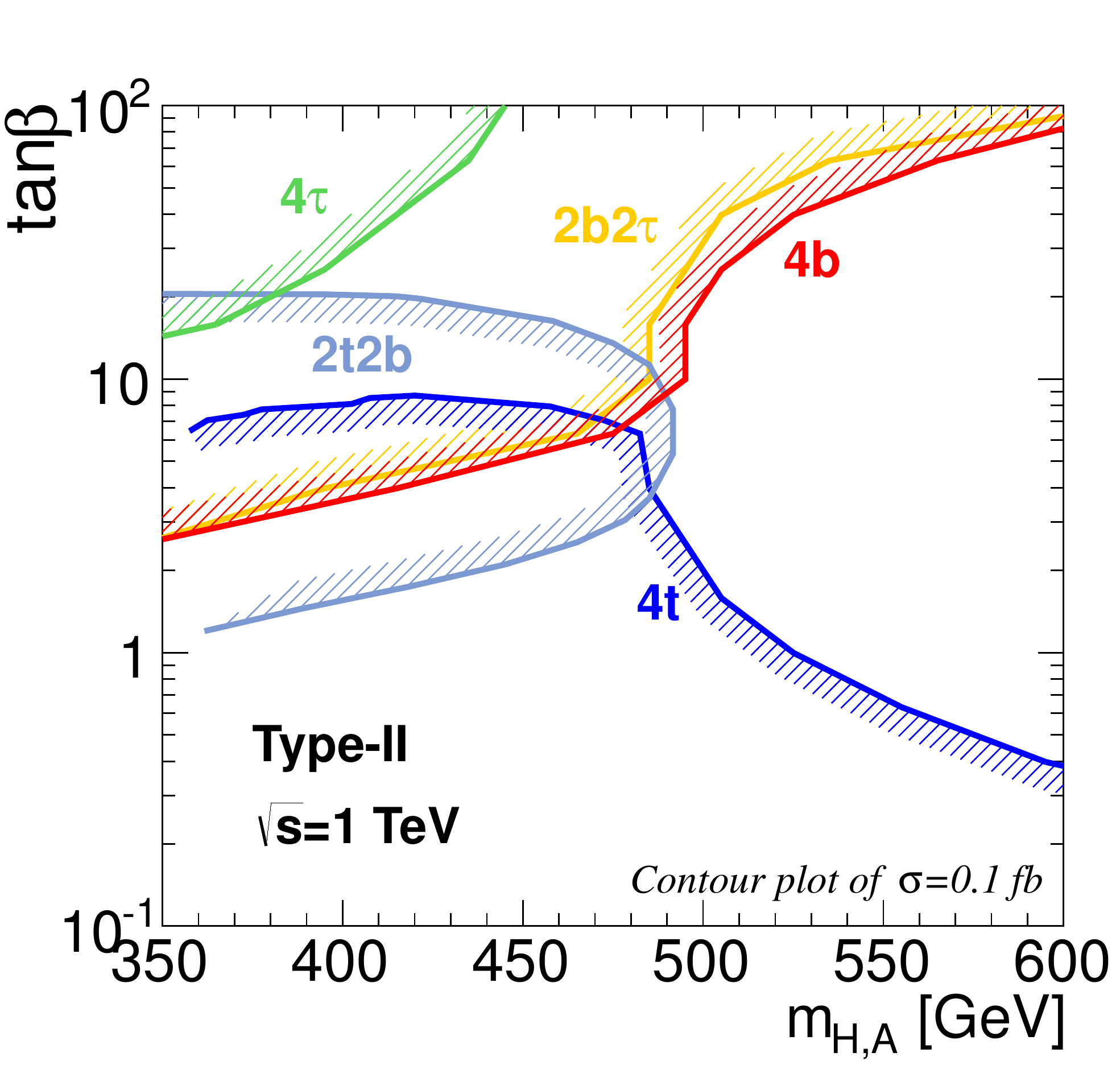} 
 \includegraphics[width=0.24\textwidth]{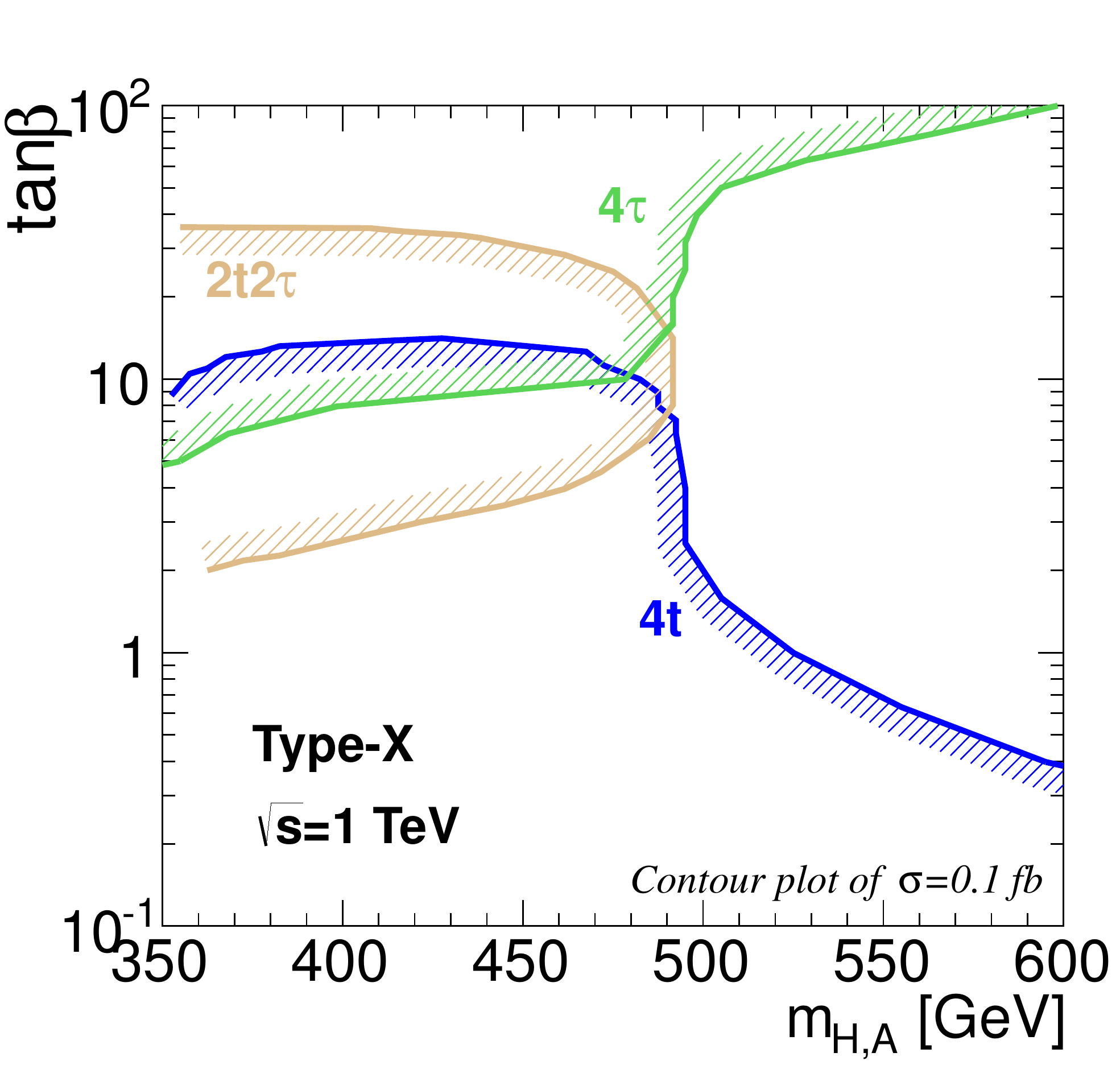} 
 \includegraphics[width=0.24\textwidth]{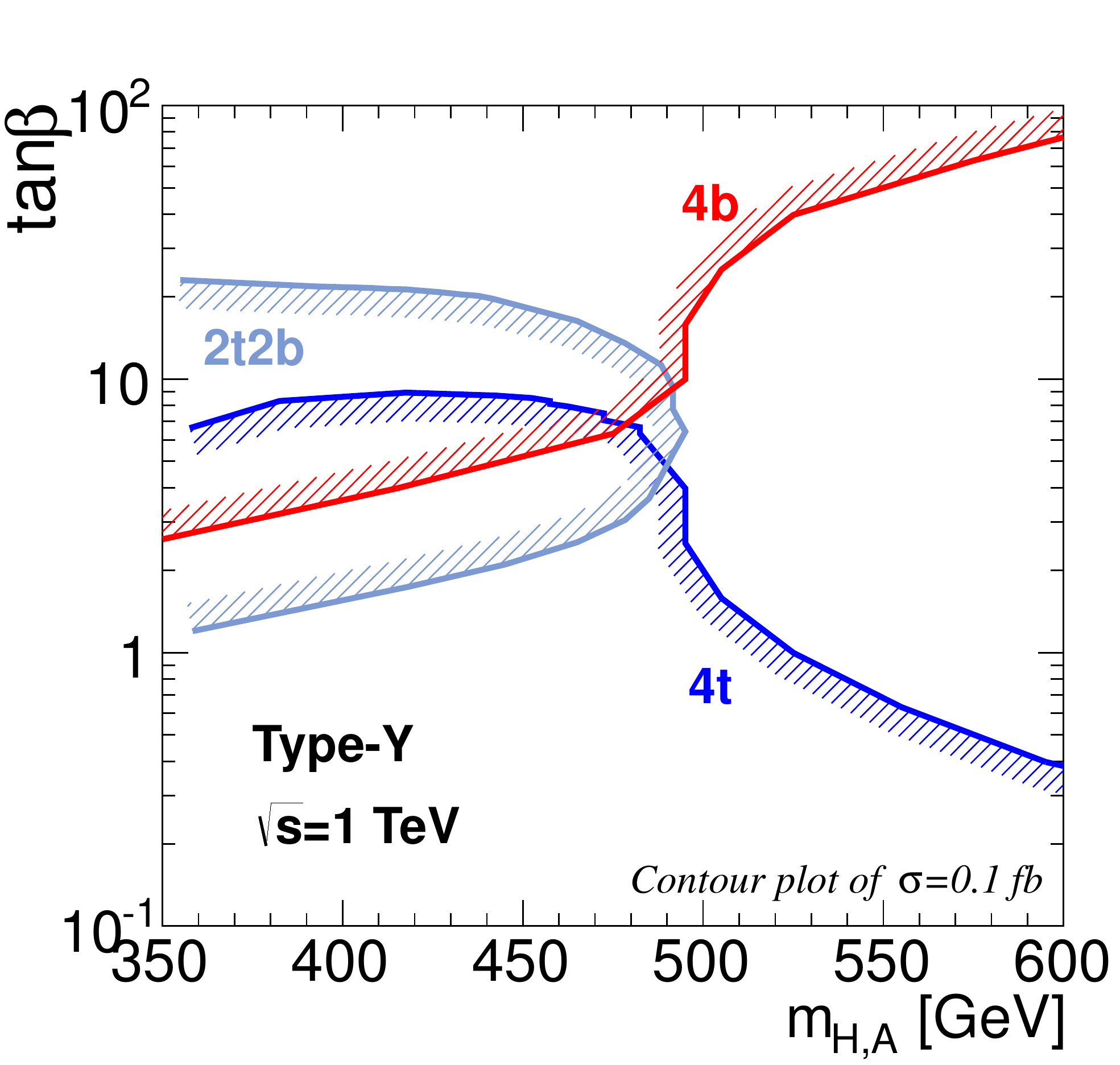} 
   \includegraphics[width=0.24\textwidth]{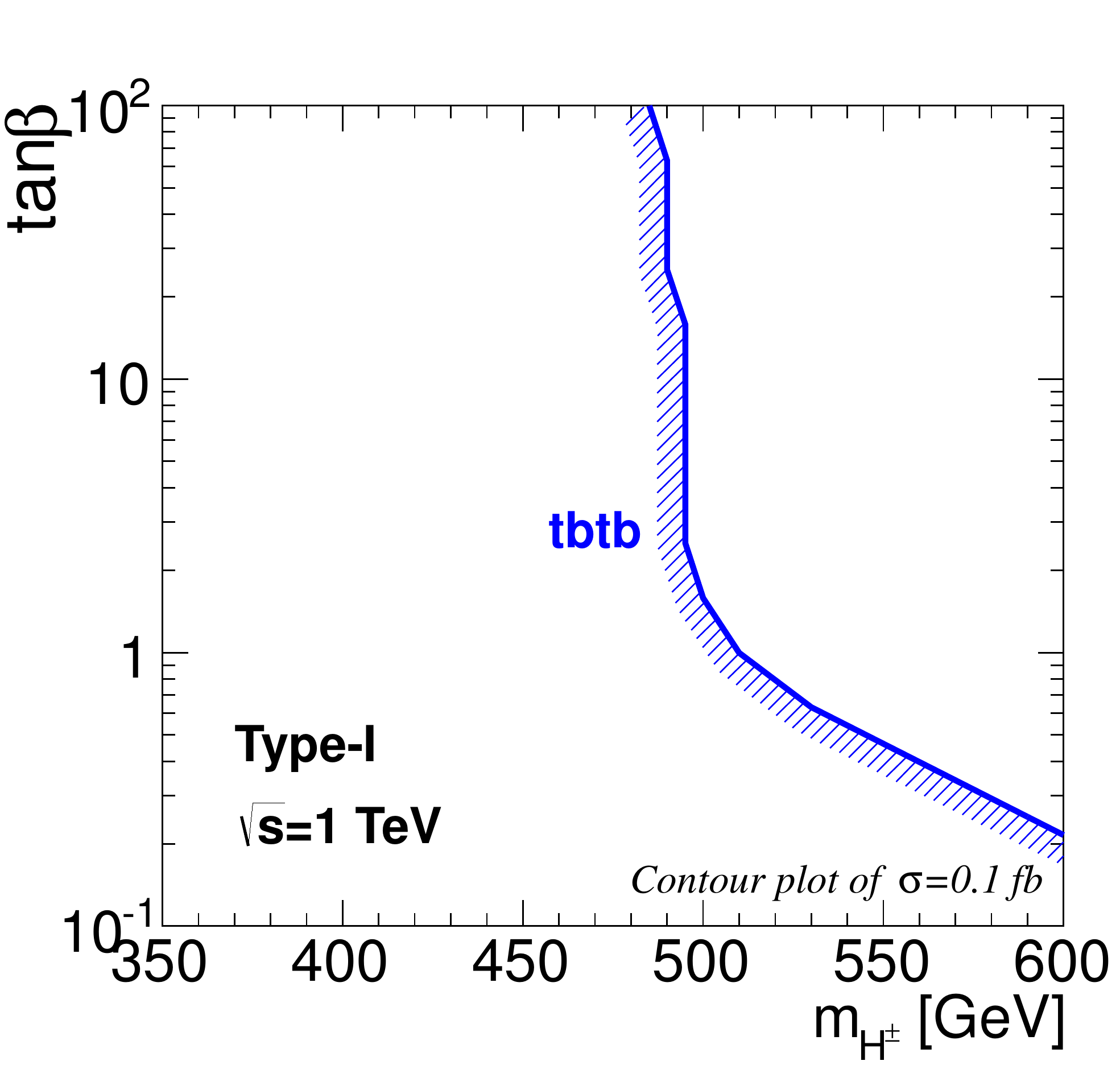} 
 \includegraphics[width=0.24\textwidth]{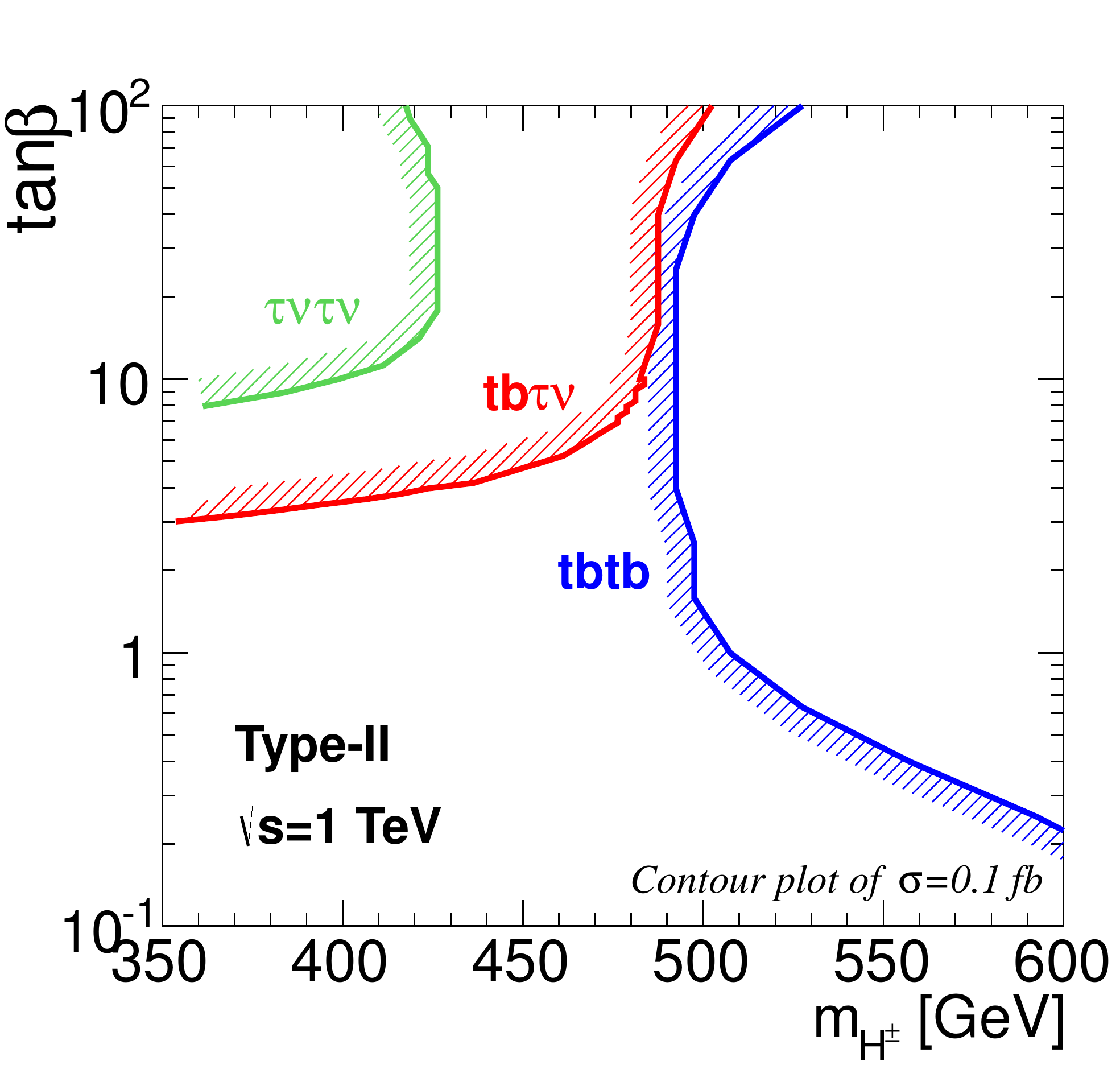} 
 \includegraphics[width=0.24\textwidth]{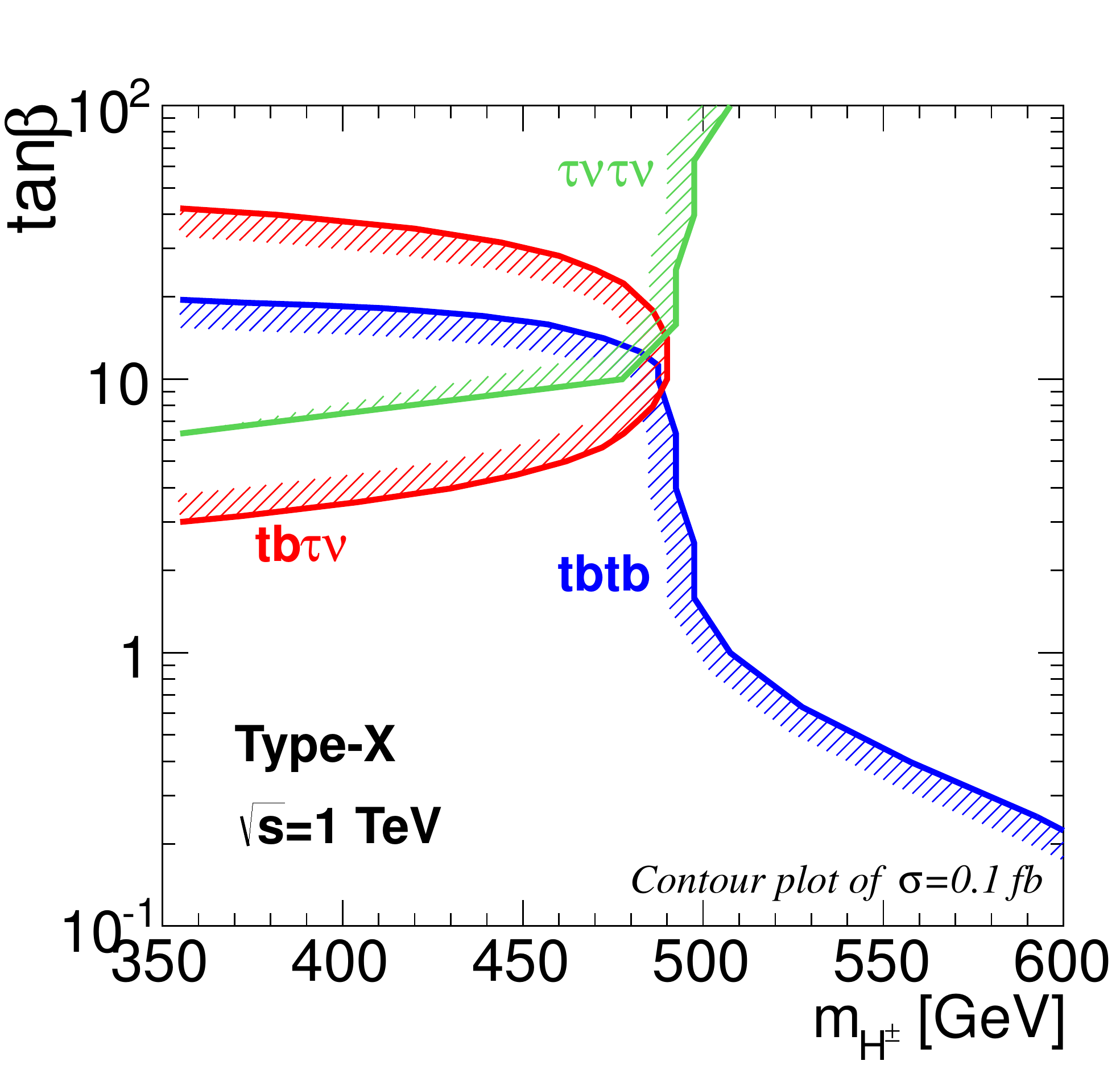} 
 \includegraphics[width=0.24\textwidth]{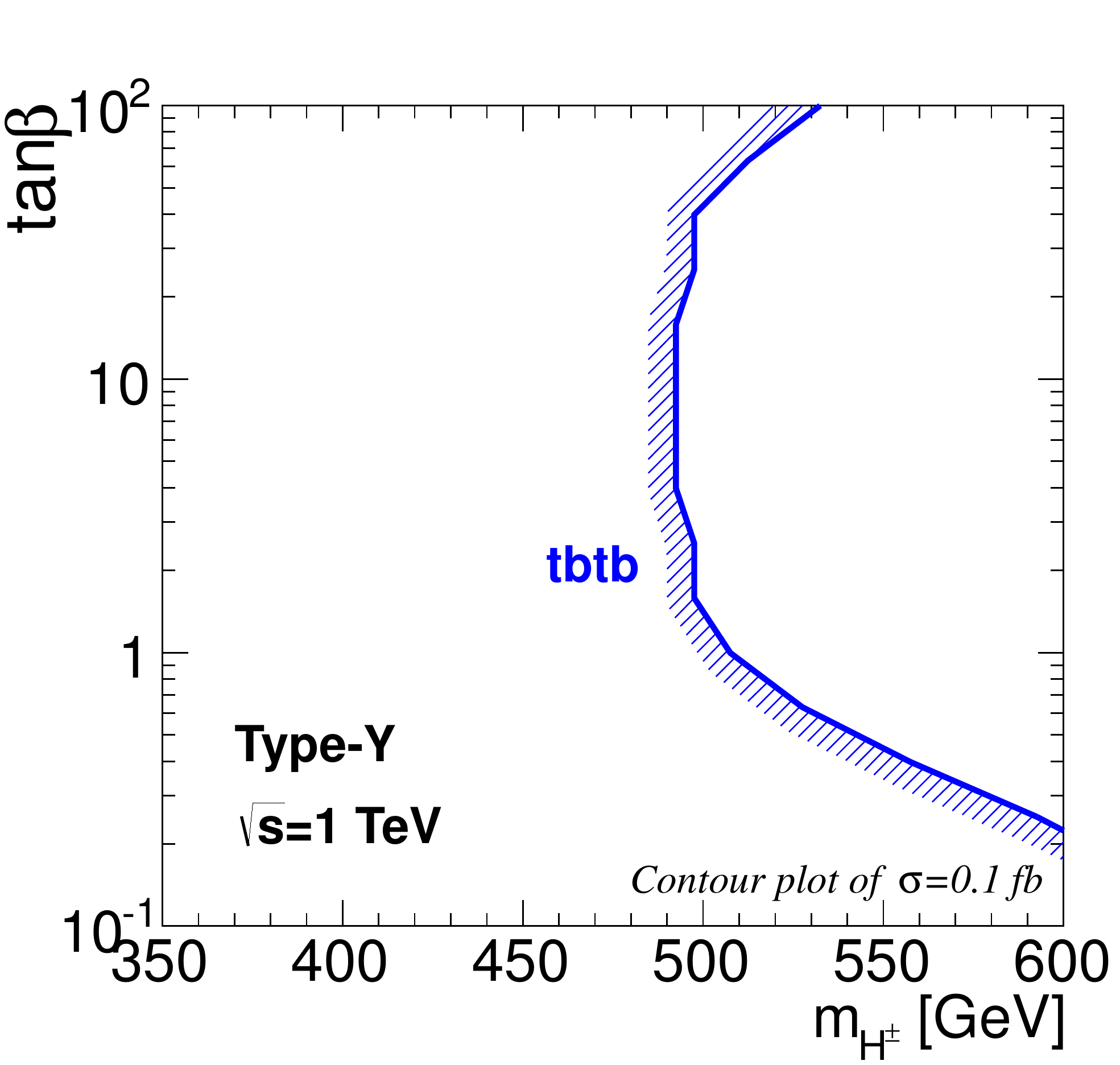} 
 \caption{Contour plots of the four-particle production cross sections
 through the $H/A$ production and $H^\pm$ production process at the ILC with 
 $\sqrt{s}=1$ ~TeV in the $(m_{H^\pm},\tan\beta)$  plane.
 Contour of $\sigma=0.1$~fb is drawn for each signature~\cite{Kanemura:2014dea}. }
 \label{fig:ILC1TeV}
 \end{center}
\end{figure*}

Both the pair and single production processes of additional Higgs bosons mostly result   
in four-particle final-states (including neutrinos).
In Ref.~\cite{Kanemura:2014dea},  the cross sections of various four-particle final-states
are studied for given masses of additional Higgs bosons and $\tan\beta$ with setting
$\sin(\beta-\alpha)=1$, and draw contour curves where the cross sections
are $0.1$~fb. 
This value is chosen commonly for all processes as it could be regarded
as a typical order of magnitude of the cross section of the additional
Higgs boson production~\cite{Kiyoura:2003tg}. 
In addition, this value can also be considered as a criterion for
observation with the expected integrated luminosity at the
ILC~\cite{Djouadi:2007ik,Behnke:2013lya}. 
Certainly, the detection efficiencies are different for different
four-particle final-states, and 
the decay of unstable particles such as tau leptons and
top quarks have to be considered if they are involved. 
We here restrict ourselves to simply compare the various four-particle 
production processes in four types of Yukawa interaction in the 2HDMs
with taking the criterion of 0.1~fb as a magnitude of the cross
sections. 
Expected background processes and a brief strategy of observing the
signatures are discussed in Ref.~\cite{Kanemura:2014dea}. 

In Fig.~\ref{fig:ILC500GeV},  
contour plots of the cross sections of four-particle production processes
through $H$ and/or $A$ are shown in the $(m_{H/A},\tan\beta)$ plane (upper figures),  
 and those through $H^\pm$ are shown in the$(m_{H^\pm},\tan\beta)$ plane (lower figures) 
 for the collision energy to be $\sqrt{s}=500$ GeV.  
 From left to right, the figures correspond to the results in in Type 1, Type II, Type X and Type Y.
 We restrict ourselves to consider the degenerated mass case, $m_H=m_A$.

In Fig.~\ref{fig:ILC1TeV}, contour plots of the cross sections of four-particle production processes
through $H$ and/or $A$ are shown in the $(m_{H/A},\tan\beta)$ plane (upper figures),  
 and those through $H^\pm$ are shown in the$(m_{H^\pm},\tan\beta)$ plane (lower figures) 
 for the collision energy to be $\sqrt{s}=1$~TeV.  
 From left to right, the figures correspond to the results in in Type 1, Type II, Type X and Type Y.
 We restrict ourselves to consider the degenerated mass case, $m_H=m_A$.

We here give a comment on the SM background processes and their cross
sections~\cite{Kanemura:2014dea}. 
In general, for the four-particle production processes, the SM
background cross sections are larger for $\sqrt{s}=250$~GeV, but
decrease with the collision energy. 
The typical orders of cross sections are of the order of 1~fb to 10~fb 
for the $Z/\gamma$ mediated processes,
and of the order of 10 to 100~fb 
for the processes which are also mediated by $W^\pm$.
For the four-quark production processes, gluon exchange diagrams also
contribute.
In order to reduce the background events, efficient kinematical cuts are
required. 

The cross section of the $4t$ production is very small in the SM. 
Therefore, a clean signature can be expected to be detected in this
mode . 
Detailed studies on the signal and background processes for
$tbtb$ production can be found in
Ref.~\cite{Moretti:2002pa}, and the signal-to-background analysis for
the $4\tau$ production can be found in Ref.~\cite{Kanemura:2012az} with
the reconstruction method of the masses of additional Higgs bosons.

Finally, we discuss some concrete scenarios to show the
complementarity of direct searches for the additional Higgs bosons 
in the 2HDMs at the LHC and the ILC. 
As benchmark scenarios,  three cases $\tan\beta=2$, 7 and 20 
are considered for $m_\phi=220$~GeV and $\sin(\beta-\alpha)=1$, 
where $m_\phi$ represents the common mass of $H$, $A$ and
$H^\pm$. 
In Table~\ref{tab:Benchmark1}, the expected signatures 
of $H/A$ and $H^\pm$ are summarized to be observed at the LHC with 300~fb$^{-1}$, 
3000~fb$^{-1}$ and at the ILC with $\sqrt{s}=500$~GeV. 
\begin{table*}[t]
 \begin{tabular}{c||l|cc|cc|cc|cc}
  $(m_\phi, \tan\beta)$ & & \multicolumn{2}{c|}{Type-I} &
  \multicolumn{2}{c|}{Type-II} & \multicolumn{2}{c|}{Type-X} &
  \multicolumn{2}{c}{Type-Y} \\ \hline\hline
  & & $H, A$ & $H^\pm$ & $H, A$ & $H^\pm$ & $H, A$ & $H^\pm$ & $H, A$ &
  $H^\pm$ \\ \hline  \hline
  & LHC300 & $-$ & $-$ & $\tau\tau$, $bb$ & $tb$ & $4\tau$ & $-$ &
  $bb$ & $tb$ \\ 
  (220~GeV, 20) & LHC3000 & $-$ & $-$ & $\tau\tau$, $bb$  & $tb$ &
  $4\tau$ & $-$ & $bb$ & $tb$ \\[1mm] \cline{2-10}
  & ILC500 & \shortstack{{}\\$4b,2b2\tau,4g$,\\$2b2g,2\tau2g$} & $tbtb$ &
  \shortstack{$4b,2b2\tau$,\\$4\tau$} &
  \shortstack{$tbtb,tb\tau\nu$,\\$\tau\nu\tau\nu^{}$} & $4\tau$ &
  \shortstack{$tb\tau\nu$,\\$\tau\nu\tau\nu$} & $4b$ & $tbtb,tbcb$ \\
  \hline \hline
  & LHC300 & $-$ & $-$ & $\tau\tau$ & $tb$ & $4\tau$ & $-$ &
  $-$ & $tb$ \\
  (220~GeV, 7) & LHC3000 & $-$ & $tb$ & $\tau\tau$ & $tb$ &
  $\tau\tau,4\tau$ & $-$ & $-$ & $tb$ \\[1mm] \cline{2-10}
  & ILC500 & \shortstack{{}\\$4b,2b2\tau,4g$,\\$2b2g,2\tau2g$} & $tbtb$ &
  \shortstack{$4b,2b2\tau$,\\$4\tau$} & 
  \shortstack{$tbtb,tb\tau\nu$,\\$\tau\nu\tau\nu$} &
  $2b2\tau,4\tau$ & 
  \shortstack{$tbtb,tb\tau\nu$,\\$\tau\nu\tau\nu$} & $4b$ &
  $tbtb,tbcb$ \\ \hline \hline
  & LHC300 & $-$ & $tb$ & $\tau\tau$ & $tb$ &
  $\tau\tau,4\tau$ & $tb$ & $-$ & $tb$ \\
  (220~GeV, 2) & LHC3000 & $\tau\tau$ & $tb$ & $\tau\tau$ &
  $tb$ & $\tau\tau,4\tau$ & $tb$ & $-$ & $tb$\\[1mm] \cline{2-10}
  & ILC500 & \shortstack{{}\\$4b,2b2\tau,4g$,\\$2b2g,2\tau2g$} & $tbtb$ &
	  \shortstack{$4b,2b2\tau$,\\
  $4\tau,2b2g$} & \shortstack{$tbtb$,\\$tb\tau\nu$} &
  \shortstack{$4b,2b2\tau$,\\$4\tau$} &
  \shortstack{$tbtb$,\\$tb\tau\nu$} &
  \shortstack{$4b,2b2\tau$,\\$2b2g$} & $tbtb$ \\
 \hline \hline
 \end{tabular}
\caption{Expected signatures to be observed at the LHC and ILC
 for the benchmark scenarios with $m_\phi=220$~GeV~\cite{Kanemura:2014dea}. 
 Observable final-states are listed as the signatures of additional
 Higgs bosons, $H$, $A$ and $H^{\pm}$. 
 LHC300, LHC3000, ILC500 represent the LHC run of 300~fb$^{-1}$,
 3000~fb$^{-1}$ luminosity, ILC run of 500~GeV, respectively.}
 \label{tab:Benchmark1}
\end{table*}

First,  for the case of  $(m_\phi, \tan\beta) = (220~{\rm GeV}, 20)$.  
no signature is predicted for Type-I, while different signatures
are predicted for Type-II, Type-X and Type-Y at the LHC with 300~fb$^{-1}$ and 3000~fb$^{-1}$. 
Therefore those three types could be discriminated at the LHC. 
On the other hand, at the ILC with $\sqrt{s}=500$~GeV, all the four
types of the Yukawa interaction including Type-I predict signatures 
which are different from each other. 
Therefore, complete discrimination of the type of Yukawa
interaction could be performed at the ILC. 

\begin{sloppypar}
Next, we turn to the second case with $(m_\phi, \tan\beta) = (220~{\rm
GeV}, 7)$. 
At the LHC with 300~fb$^{-1}$, Type-I cannot be observed, while Type-II,
Type-X and Type-Y are expected to be observed with different signatures. 
At the LHC with 3000~fb$^{-1}$, the signature of Type-I can also be
observed with the same final state as Type-Y.
Type-I and Type-Y can be basically separated, because for Type-Y the
signals can be observed already with 300~fb$^{-1}$ while for Type-I that
can be observed only with 3000~fb$^{-1}$. 
Therefore, at the LHC with 3000~fb$^{-1}$, the complete discrimination
can be achieved. 
At the ILC, the four types of Yukawa interaction can also be separated
by a more variety of the signatures for both channels with the neutral
and charged Higgs bosons. 
\end{sloppypar}

Finally, for the case of $(m_\phi, \tan\beta) = (220~{\rm
GeV}, 2)$, signals for all the four types of Yukawa
interaction can be observed at the LHC with 300~fb$^{-1}$, . 
However, the signatures of Type-I and Type-Y are identical, so that the
two types cannot be discriminated. 
With 3000~fb$^{-1}$, the difference between the
Type-I and Type-Y emerges in the $H/A$ signature.
Therefore the two types can be discriminated at this stage. 
Again, at the ILC, the four types can also be separated with a more
variety of the signatures for both channels with the neutral and charged
Higgs bosons. \\

\noindent
{\it Fingerprinting the type of the 2HDM by precision measurement of 
the Higgs couplings at the ILC}

Extra Higgs bosons in extended Higgs sectors 
can be discovered as long as their masses are 
not too large as compared to the electroweak scale. 
On the other hand, at the ILC~\cite{ILC_TDR}, 
these extended Higgs sectors can also be tested by accurately 
measuring the coupling constants with the discovered Higgs bosons $h$. 
This is complementary with the direct searches at the LHC. 

\begin{figure}[t]
\begin{center}
\includegraphics[width=55mm]{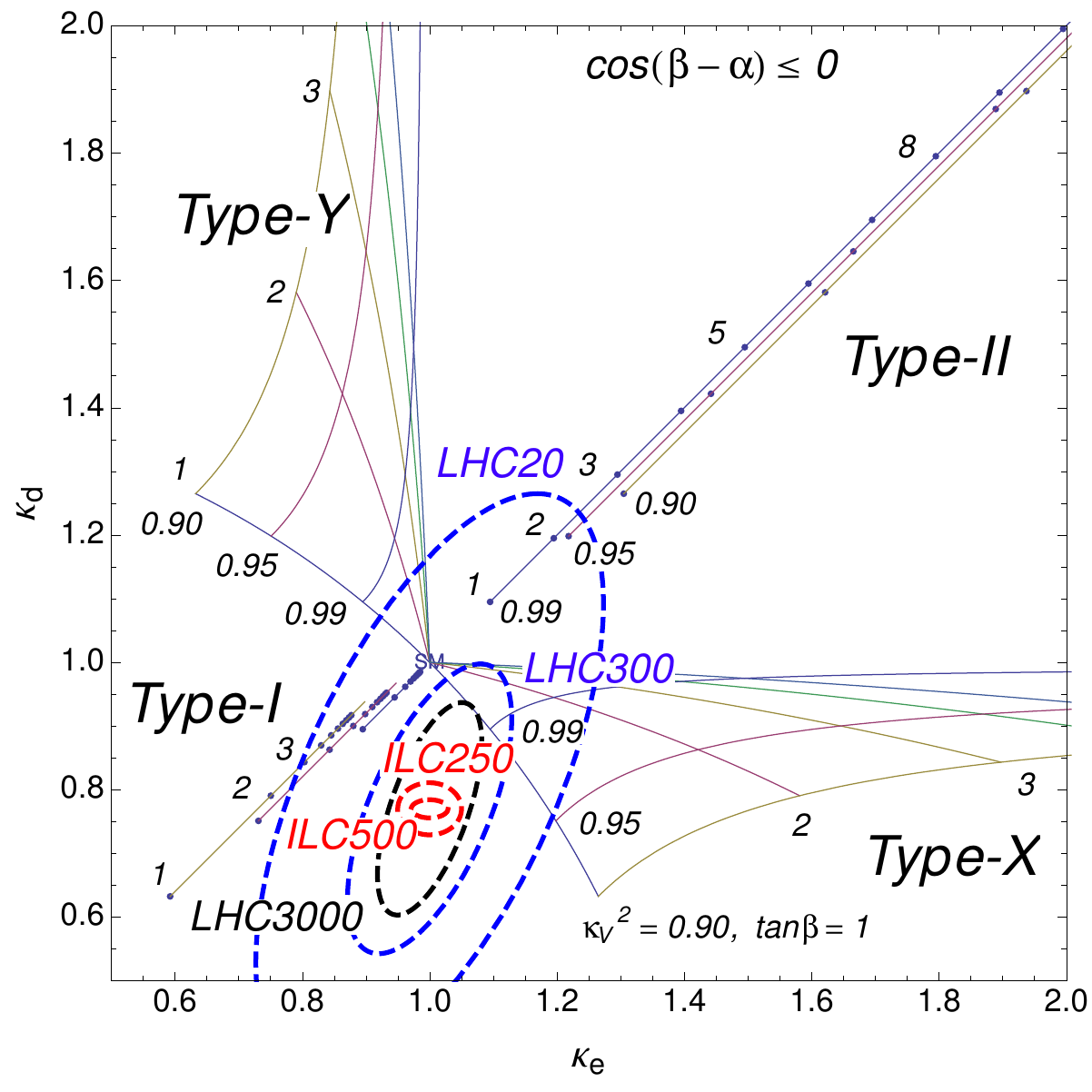}     
\hspace{1cm}
\includegraphics[width=55mm]{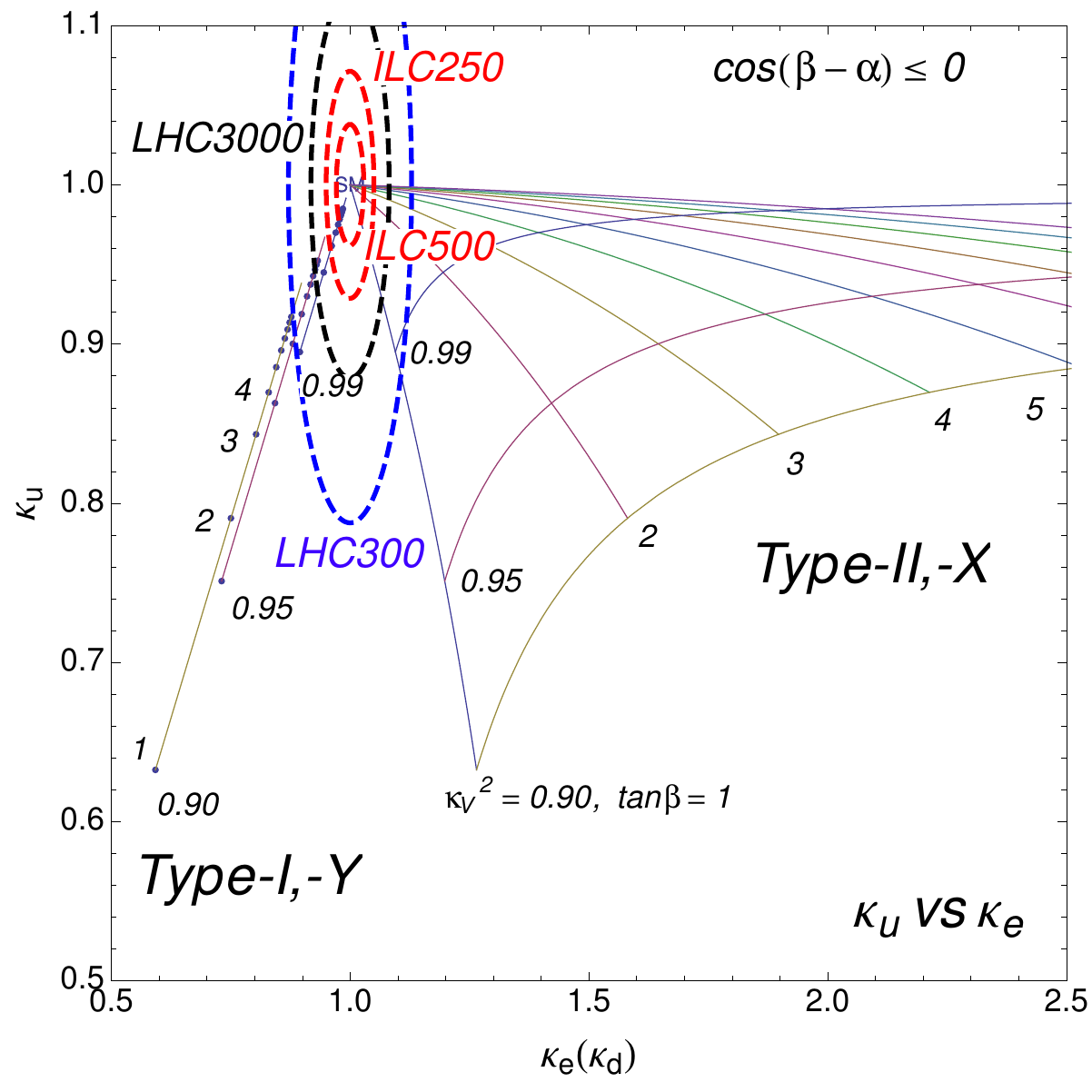}  
\caption{Left: The scaling factors in 2HDM with four types of Yukawa interactions. 
Right: The scaling factors in models with universal Yukawa couplings. The current LHC 
bounds and the expected LHC and ILC sensitivities are also shown at the 68.27 \%  C.L.. 
For details, see Ref.~\cite{Asner:2013psa, KTYY}}
\label{finger_print}
\end{center}
\end{figure}

In the extended Higgs sectors, the gauge couplings and Yukawa interactions of $h$ are 
parameterized by
\begin{eqnarray}
{\mathcal L}^{\rm int}
&=& +\kappa_W \frac{2m_W^2}{v} hW^{+\mu}W^-_\mu + \kappa_Z \frac{m_Z^2}{v} hZ^\mu Z_\mu \nonumber \\
&& -\sum_f\kappa_f\frac{m_f}{v} {\overline f}fh + \cdots , 
\end{eqnarray}
where $\kappa_V$ ($V=W$ and $Z$) and $\kappa_f$ ($f=t,b,c, \cdots$) are the scaling factors measuring 
the deviation from the SM predictions. In the SM, we have $\kappa_V=\kappa_f=1$.
According to Refs.~\cite{ILC_TDR,Asner:2013psa,Dawson:2013bba}, 
the $hVV$ couplings are expected to be measured with about 4$\%$
accuracy at the LHC with 300~fb$^{-1}$ (although requiring some
theory input). 
The accuracy for the $ht\bar{t}$, $hb\bar{b}$ and $h\tau\tau$ couplings are supposed to be about 16\%, 14\% and 11\%, respectively. 
At the ILC250 (ILC500) where the collision energy and the integrated luminosity are 250 GeV (500 GeV) and 
250 fb$^{-1}$ (500 fb$^{-1}$) combining with the results assuming 300~fb$^{-1}$ at the LHC, 
the $hWW$ and $hZZ$ couplings are expected to be measured by about 1.9\% (0.2\%) and about 0.4\% (0.3\%), respectively. 
The $hc\bar{c}$, $hb\bar{b}$ and $h\tau\tau$ couplings are supposed to be measured by
about 5.1\% (2.6\%), 2.8\% (1.0\%) and 3.3\% (1.8\%) at the ILC250 (ILC500).  
For the $ht\bar{t}$ coupling, it will be measured with 12.0\% and 9.6\% at the ILC250 and ILC500, respectively. 

In the 2HDM, the scaling factors $\kappa_V$ are  given by
$\kappa_V=\sin(\beta-\alpha)$, while those for the Yukawa interactions are
given depending on the type of Yukawa interaction~\cite{Aoki:2009ha}.
For the SM-like limit $\kappa_V^{}=1$, all the scaling factors  $\kappa_f$ become unity. 
In Fig.~\ref{finger_print} (left), the scale factors $\kappa_f$ in the 2HDM 
with the softly-broken symmetry are shown on the  $\kappa_\ell$-$\kappa_d$ plane 
for various values of $\tan\beta$ and $\kappa_V^{}$ ($=\sin(\beta-\alpha)$). 
The points and the dashed curves denote changes of $\tan\beta$ by steps of one.
$\kappa_V$ ($=\kappa_W=\kappa_Z$) is taken as $\kappa_V^2 = 0. 99, 0.95$ and $0.90$.
The current LHC constraints as well as the expected LHC and ILC sensitivities 
for $\kappa_d$ and $\kappa_\ell$  are also shown at the 68.27 \%  Confidence Level (C.L.).
For the current LHC constraints (LHC30), we take the numbers
from the universal fit in Eq.~(18) of Ref.~\cite{Giardino:2013bma}. 
For the future LHC sensitivities (LHC300 and LHC3000), 
the expectation numbers are taken from the Scenario 1 in Tab.~1 of Ref.~\cite{CMS:2012zoa}. 
The central values and the correlations are assumed to be the same as in LHC30. 
The ILC sensitivities are taken from Table. 2.6 in Ref.~\cite{ILC_TDR}. 
The same central value without correlation is assumed for the ILC sensitivity curves.
For more details see Refs.~\cite{Asner:2013psa}, and for some revisions see Ref.~\cite{KTYY}.

\begin{figure}[!t]
\begin{center}
\includegraphics[width=80mm]{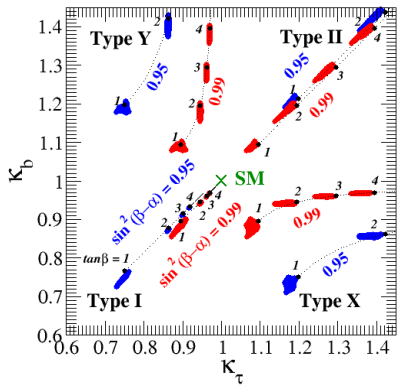}
\includegraphics[width=80mm]{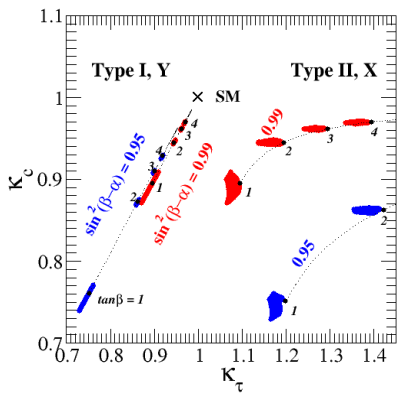}
\caption{Predictions of various scale factors on the $\kappa_\tau$ vs $\kappa_b$ (upper panel),  
and $\kappa_\tau$ vs $\kappa_c$ (bottom panel) in four types of Yukawa interactions
in the cases with $\cos(\beta-\alpha)<0$~\cite{Kanemura:2014dja}.
Each black dot shows the tree level result with $\tan\beta$=1,~2,~3 and 4. 
One-loop corrected results are indicated by red for $\sin^2(\beta-\alpha)=0.99$ and 
blue for $\sin^2(\beta-\alpha)=0.95$ regions where 
$m_\Phi$ and $M$ are scanned over from 100 GeV to 1 TeV and 0 to $m_\Phi$, respectively. 
All the plots are allowed by the unitarity and vacuum stability bounds.  
}
\label{fing_loop}
\end{center}
\end{figure}

\begin{sloppypar}
The analysis including radiative corrections has been done recently~\cite{Kanemura:2014dja}. 
We show the one-loop results for the Yukawa couplings in the planes of fermion scale factors.
In Fig.~\ref{fing_loop}, predictions of various scale factors are shown 
on the $\kappa_\tau$ vs $\kappa_b$ (upper panels), 
and $\kappa_\tau$ vs $\kappa_c$ (bottom panels) planes. 
When we consider the case with $\sin(\beta-\alpha)\neq 1$, the sign dependence of $\cos(\beta-\alpha)$ to 
$\kappa_f$ is also important. We here show the both cases with $\cos(\beta-\alpha)<0$.  
The value of $\tan\beta$ is discretely taken as $\tan\beta$=1,~2,~3 and 4. 
The tree level predictions are indicated by the black dots, while 
the one-loop corrected results are shown by the red for $\sin^2(\beta-\alpha)=0.99$ and 
blue for $\sin^2(\beta-\alpha)=0.95$ regions where 
the values of $m_\Phi$ and $M$ are scanned over from 100 GeV to 1 TeV and 0 to $m_\Phi$, respectively. 
All the plots are allowed by the unitarity and vacuum stability bounds. 
\end{sloppypar}

Even when we take into account the one-loop corrections to the Yukawa couplings,   
this behavior; i.e., predictions are well separated among the four types of THDMs, does not so change as we see the 
red and blue colored regions. 
Therefore, we conclude that 
all the 2HDMs can be distinguished from each other by measuring the charm, bottom and tau Yukawa couplings precisely when 
the gauge couplings $hVV$ are deviated from the SM prediction with $\mathcal{O}(1)$\%
\footnote{We here give a  comment on the radiative 
correction to the $hVV$ couplings in the THDMs. 
Although the tree level deviations in the $hVV$ couplings are described by the factor $\sin(\beta-\alpha)$, 
these values can be modified at the one-loop level.  
In Ref.~\cite{KOSY}, the one-loop corrected $hZZ$ vertex has been calculated 
in the softly-broken $Z_2$ symmetric 2HDM.  It has been found that 
for the fixed value of $\sin(\beta-\alpha)$, 
the one-loop corrections to the $hZZ$ vertex are less than a few $\%$. }.  

\begin{sloppypar}
The Higgs boson couplings $h\gamma\gamma$ and $hgg$ are absent 
at the tree level but are produced at the one-loop level via 
the higher dimensional operators 
\begin{eqnarray}
  \frac{1}{M^2} |\Phi|^2 F^{\mu\nu} F_{\mu\nu}, \quad \frac{1}{M^2}
  |\Phi|^2 G^{(a)\mu\nu} G^{(a)}_{\mu\nu}, 
\end{eqnarray} 
where
  $F^{\mu\nu}$ and $G^{(a)\mu\nu}$ are the field strength tensors of
  $U(1)_{\rm EM}$ and $SU(3)_C$, and $M$ is a dimensionful parameter.
  In the 2HDM, the coupling can deviate from the SM due to the mixing
  effect of neutral scalar bosons and, for $h\gamma\gamma$, also due
  to the loop contributions of additional Higgs bosons $H$, $A$ and
  $H^\pm$.  The latter effect can be significant even in the SM-like
  limit where $\sin(\beta-\alpha)=1$ as long as $M$ is not too large.
  At the LHC (300\,fb$^{-1}$), the HL-LHC (3000\,fb$^{-1}$), and the
  ILC (1\,TeV-up)~\cite{Dawson:2013bba, Asner:2013psa},
  $\kappa_\gamma$ is expected to be measured with 5-7 \%, 2-5 \% and
  2.4 \%, respectively.  If deviations in $\kappa_\gamma$ and
  $\kappa_g$ are detected in future precision measurements at the LHC
  and the ILC, we can directly extract information of new particles in
  the loop such as their mass scales.
\end{sloppypar}
 
\begin{figure}[t]
\begin{center}
\includegraphics[width=55mm]{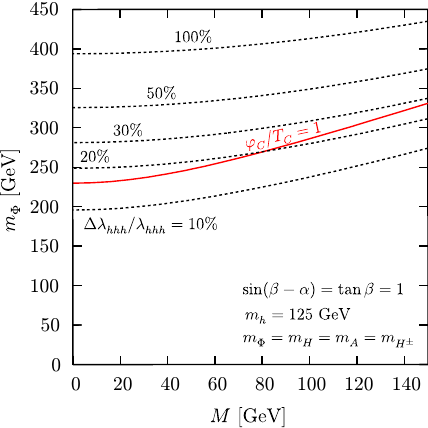} 
 \caption{
 Contour plots of the deviation in the $hhh$ coupling in the $(m_\Phi^{}, M)$  plane for $m_h=125$ GeV 
 and $\sin(\beta-\alpha)=1$. The red line indicates $\varphi_c/T_c=1$, above which 
  the strong first order phase transition occurs ($\varphi_c/T_c >1$)~\cite{senaha}. } 
 \label{fig:hhh_ewbg}
 \end{center}
\end{figure}

The triple Higgs boson coupling $hhh$ is essentially important to be measured to 
obtain the information of the Higgs potential. 
The tree-level behavior of the $hhh$ coupling constant has been discussed 
in the 2HDM in Ref.~\cite{hhh_2hdm}.  The deviation from the SM predictions are sensitive 
to the mixing parameters $\tan\alpha$ and $\sin(\beta-\alpha)$. 
In the SM-like limit $\sin(\beta-\alpha)=1$, the value of the $hhh$ coupling 
coincide with that in the SM. 
At the one-loop level, even when the SM-like limit, the $hhh$ coupling can 
deviate from the SM prediction due to the quantum loop effects of 
$H$, $A$ and $H^\pm$~\cite{KKOS,KOSY}.  
For the SM-like limit $\sin(\beta-\alpha)=1$, the one-loop corrected 
effective $hhh$ coupling in the 2HDM can be expressed as    
\begin{eqnarray}
&& \lambda_{hhh}^{eff}  = \frac{3 m_h^2}{v}
      \left\{ 1  
              + \frac{m_{H}^4}{12 \pi^2 m_h^2 v^2} 
                         \left(1 - \frac{M^2}{m_H^2}\right)^3 \right.\nonumber\\
&& 
\left.
              + \frac{m_{A}^4}{12 \pi^2 m_h^2 v^2} 
                         \left(1 - \frac{M^2}{m_A^2}\right)^3               + \frac{m_{H^\pm}^4}{6 \pi^2 m_h^2 v^2} 
                         \left(1 - \frac{M^2}{m_{H^\pm}^2}\right)^3\right.\nonumber\\
&& \left.
              - \frac{N_{c_t} m_t^4}{3 \pi^2 m_h^2 v^2} + 
              {\cal O} \left(\frac{p^2_i m_\Phi^2}{m_h^2 v^2},
                           \;\frac{m_\Phi^2}{v^2},
                           \;\frac{p^2_i m_t^2}{m_h^2 v^2},  
                           \;\frac{m_t^2}{v^2}  \right)
      \right\}, \label{m4THDM}    
\end{eqnarray}
where $m_\Phi^{}$ and $p_i$ represent the mass of $H$, $A$ or $H^\pm$ 
and the momenta of external Higgs lines, respectively.     
The deviation from the SM prediction can be ${\cal O}(100)$ \% under the  constraint from 
perturbative unitarity and vacuum stability as well as the current  LHC results , 
in the non-decoupling case $v^2 \sim M^2$.
For $M^2 \gg v^2$, such a large quantum effect decouples in the $hhh$ coupling 
because of the decoupling theorem.

It is well known that such a large non-decoupling loop effect on the triple Higgs 
boson coupling is related to the strong first-order phase transition of the electroweak gauge symmetry~\cite{Kanemura:2004ch},   
which is required for successful electroweak baryogenesis~\cite{EWBG,EWBG2}\footnote{
See also Ref.~\cite{Grojean:2004xa}. }.  
In the scenario of electroweak baryogenesis, one of the Sakharov's  conditions of the departure from thermal 
equilibrium is satisfied when $\varphi_c/T_c >1$, where $T_c$ is 
the critical temperature and $\varphi_c$ is the order parameter at $T_c$.    
With the mass of the discovered Higgs boson to be 125 GeV, the SM cannot satisfy this condition. 
On the other hand, in the extended Higgs sector, the condition $\varphi_c/T_c >1$ can be satisfied 
without contradicting the current data. 
In Fig.~\ref{fig:hhh_ewbg}, the correlation between the large deviation in the $hhh$ coupling 
and the first order phase transition is shown~\cite{Kanemura:2004ch,senaha}. 
This results show that we may be able to test the scenario of electroweak baryogenesis 
by measuring the $hhh$ coupling by the 13\% accuracy~\cite{Asner:2013psa}.  
Such a precision measurement can be achieved at the ILC.

\begin{figure*}[t]
\vspace*{.2cm}
\begin{center}
\includegraphics[width=70mm]{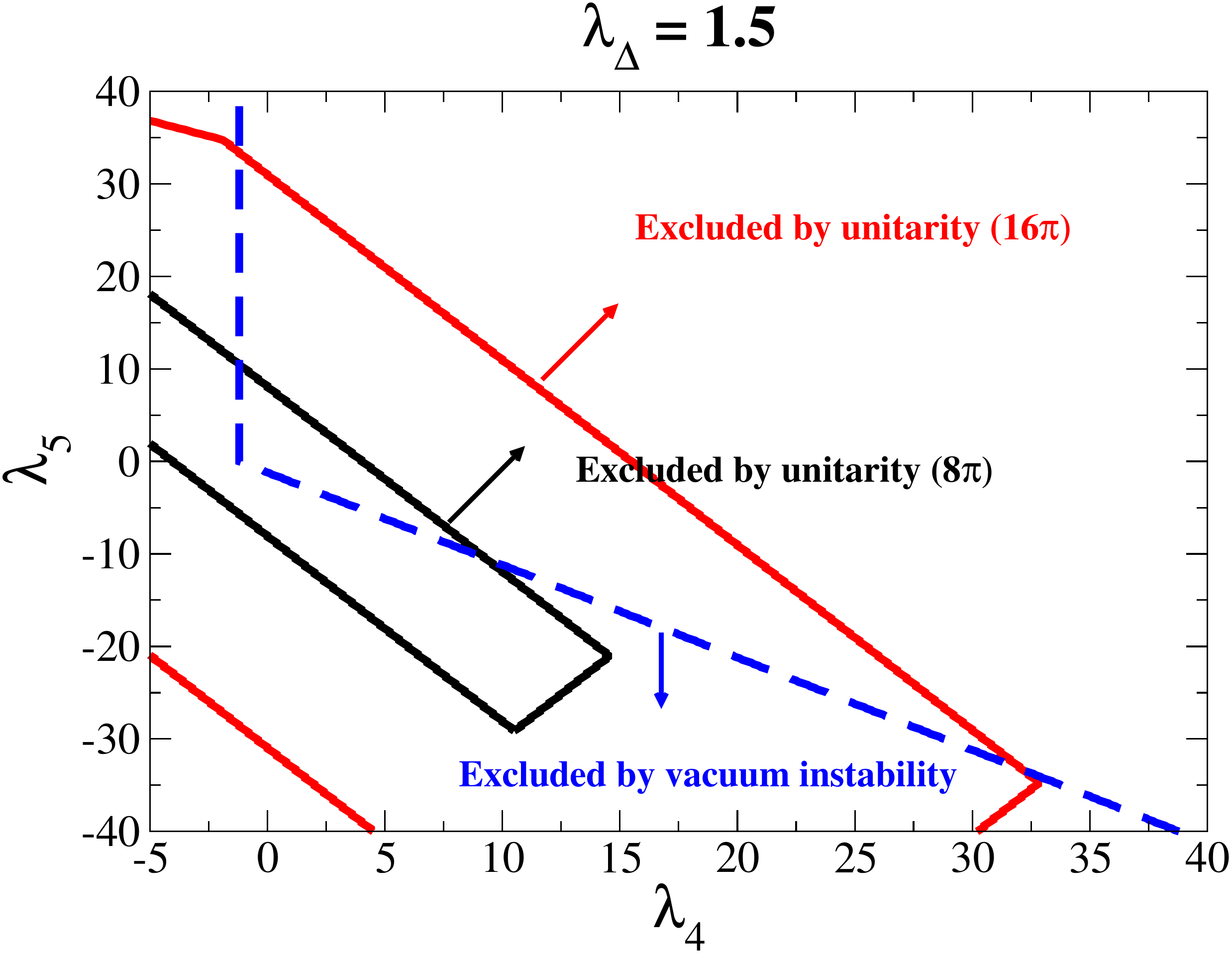}\hspace{3mm}
\includegraphics[width=70mm]{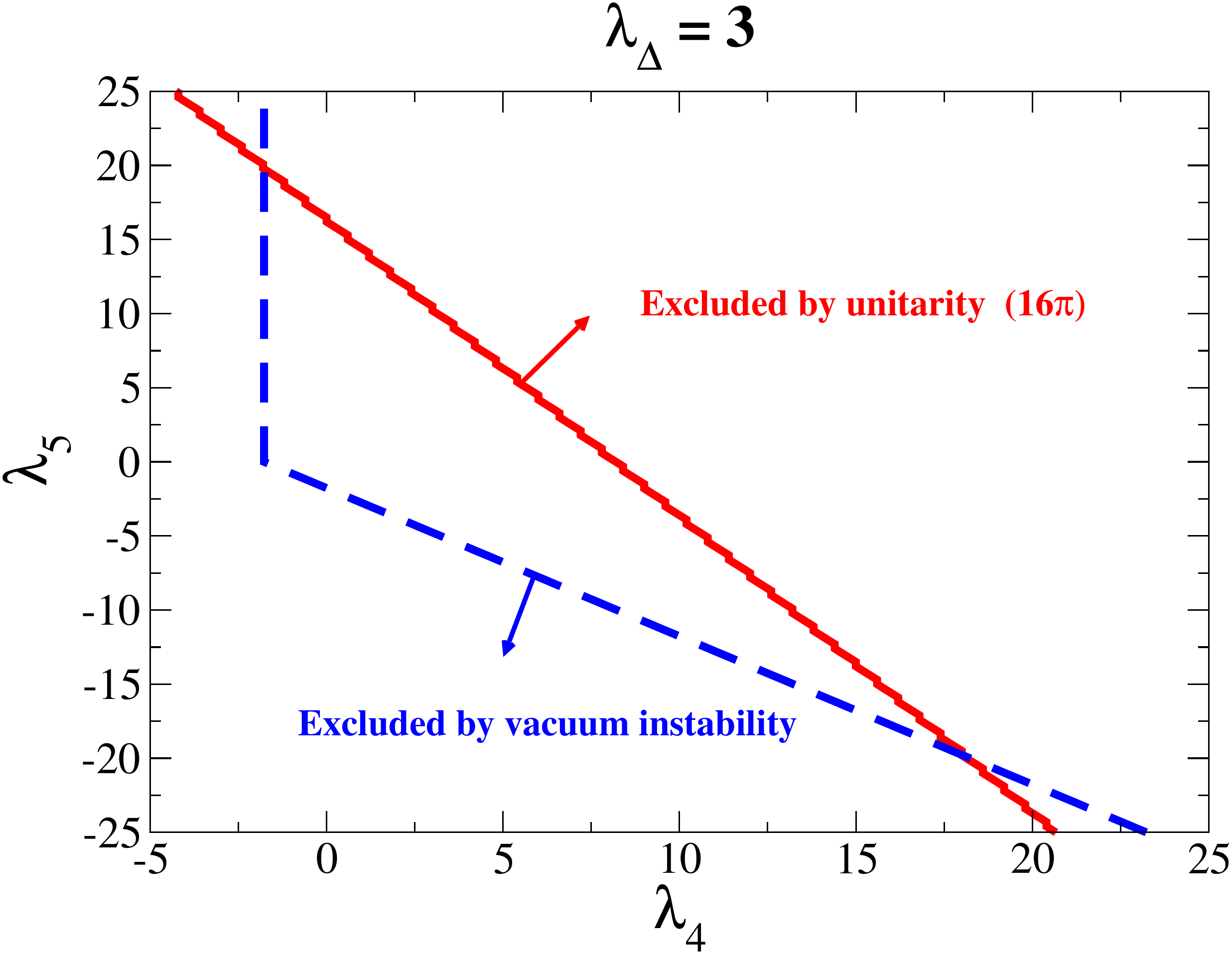}
\caption{Constraints from the unitarity and vacuum stability bounds for $\lambda_1=m_h^2/(2v^2)\simeq 0.13$
in the $\lambda_4$-$\lambda_5$ plane. 
We take $\lambda_\Delta=1.5$ for the left panel and $\lambda_\Delta=3$ for the right panel with 
$\lambda_\Delta=\lambda_2=\lambda_3$~\cite{yagyu}. 
}
\label{FIG:pu_vs2}
\end{center}
\end{figure*}

\subsubsection{Higgs Triplet models}\label{htm}
  
We here discuss the Higgs boson properties in the minimal Higgs triplet model (HTM). 
A motivation to study this model is that tiny neutrino masses can be explained 
via the so-called type-II seesaw mechanism~\cite{typeII}. 
The Higgs sector of the HTM is composed of one isospin doublet field $\Phi$ with 
hypercharge $Y=1$ and the triplet field $\Delta$ with $Y=2$. 
The Higgs fields can be parameterized by
\begin{align}
\Phi&=\left[
\begin{array}{c}
\phi^+\\
\frac{1}{\sqrt{2}}(\phi+v_\phi+i\chi)
\end{array}\right],\quad \Delta =
\left[
\begin{array}{cc}
\frac{\Delta^+}{\sqrt{2}} & \Delta^{++}\\
\Delta^0 & -\frac{\Delta^+}{\sqrt{2}} 
\end{array}\right] \nonumber \\
&\text{ with } \Delta^0=\frac{1}{\sqrt{2}}(\delta+v_\Delta+i\eta), 
\end{align}
where $v_\phi$ and $v_\Delta$ are the VEVs of the neutral components of 
doublet Higgs field $\phi^0$ and the triplet Higgs field $\delta^0$, respectively, which satisfy 
$v^2\equiv v_\phi^2+2v_\Delta^2\simeq$ (246 GeV)$^2$. 
The masses of the $W$~boson and the $Z$~boson are obtained at the tree level as
\begin{align}
m_W^2 = \frac{g^2}{4}(v_\phi^2+2v_\Delta^2),\quad m_Z^2 =\frac{g^2}{4\cos^2\theta_W}(v_\phi^2+4v_\Delta^2). \label{mV}
\end{align}

One of the striking features of the  HTM is the prediction that 
the electroweak $\rho$- parameter $\rho$ deviates from unity at the tree level 
due to the non-zero VEV of the triplet field $v_\Delta$. From Eq.~(\ref{eq:rho_general}), 
we obtain 
\begin{align}
\rho \equiv \frac{m_W^2}{m_Z^2\cos^2\theta_W}=\frac{1+\frac{2v_\Delta^2}{v_\phi^2}}{1+\frac{4v_\Delta^2}{v_\phi^2}}. \label{rho_triplet}
\end{align}
The experimental value of the $\rho$-parameter is quite close to unity, 
so that $v_\Delta$ has to be less than about 8 GeV from the tree level formula given in Eq.~(\ref{rho_triplet}). 

\begin{figure*}[t]
\centering
\includegraphics[width=48mm]{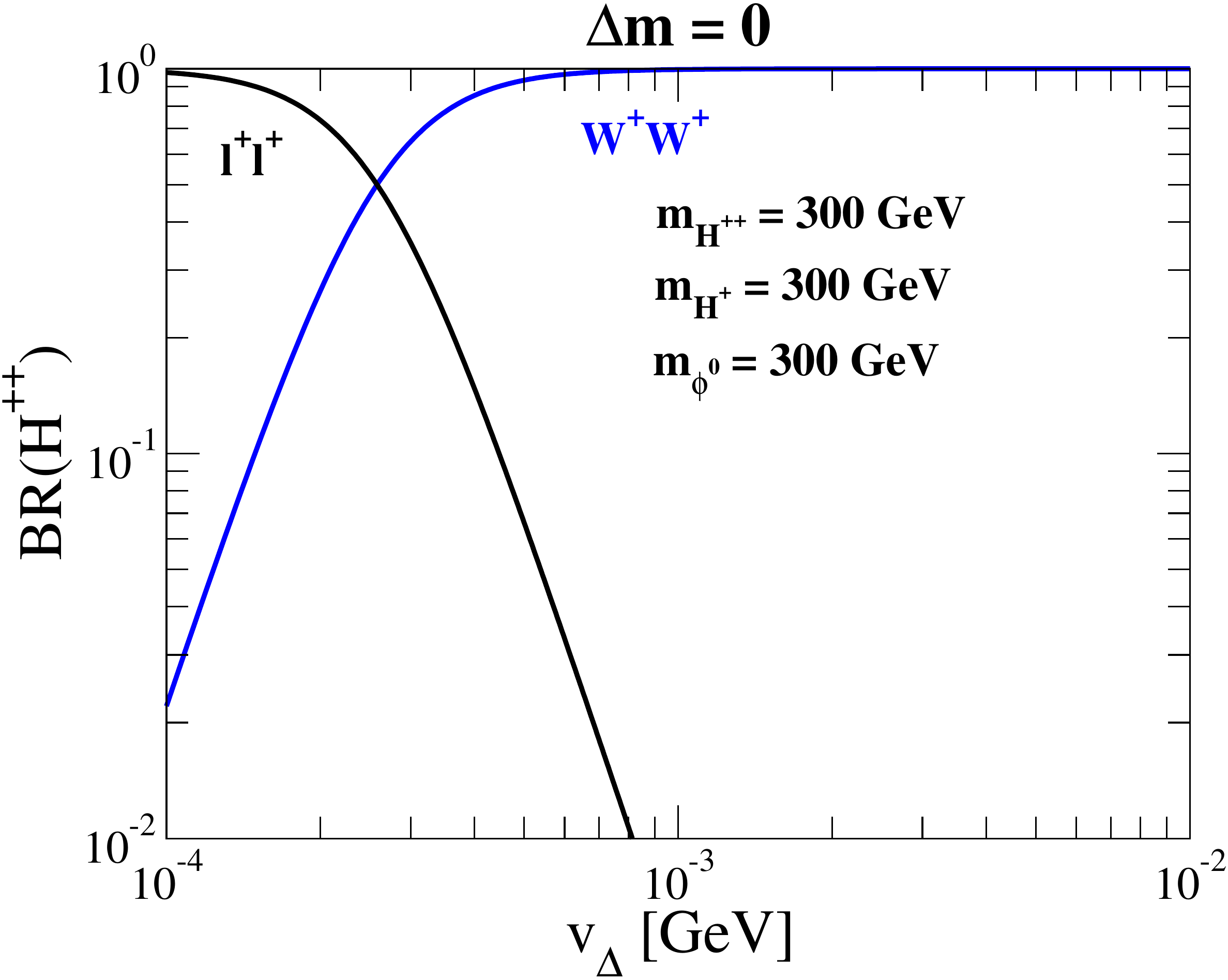}
\includegraphics[width=48mm]{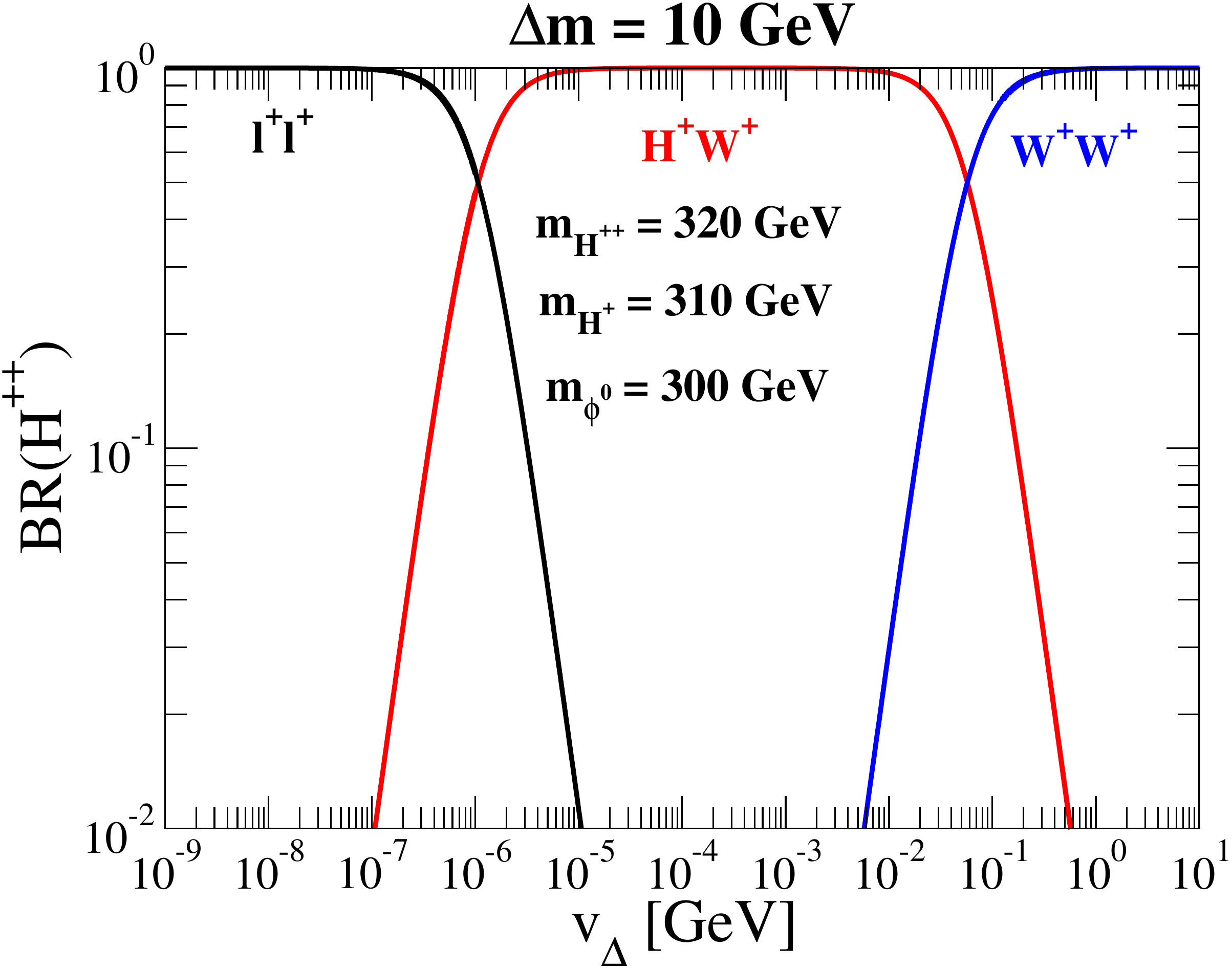}
\includegraphics[width=48mm]{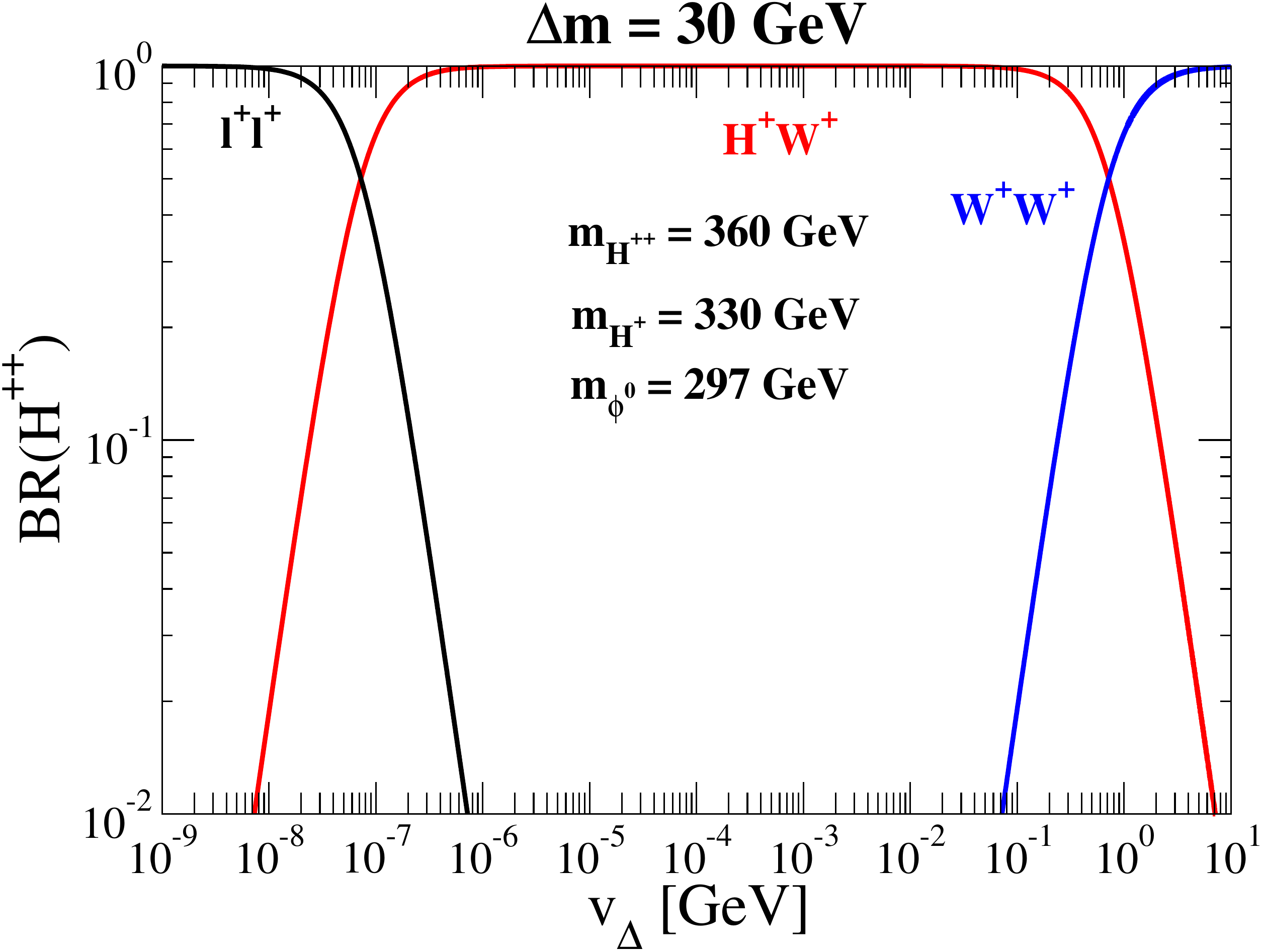}
\caption{Decay branching ratio of $H^{++}$ as a function of $v_\Delta$.
In the left figure, $m_{H^{++}}$ is
fixed to be 300 GeV, and $\Delta m$ is taken to be zero.
In the middle  figure, $m_{H^{++}}$ is
fixed to be 320 GeV, and $\Delta m$ is taken to be 10 GeV.
In the right figure, $m_{H^{++}}$ is
fixed to be 360 GeV, and $\Delta m$ is taken to be 30 GeV.}
\label{FIG:BR1_HTM}
\centering
\end{figure*}

The Yukawa interaction for neutrinos~\cite{typeII} is given by 
\begin{align}
\mathcal{L}_Y&=h_{ij}\overline{L_L^{ic}}i\tau_2\Delta L_L^j+\text{h.c.}, \label{nu_yukawa}
\end{align}
where $h_{ij}$ is the $3\times 3$ complex symmetric Yukawa matrix. 
Notice that the triplet field $\Delta$ carries the lepton number of $-2$.  
The mass matrix for the left-handed neutrinos is obtained as 
\begin{align}
(\mathcal{M}_\nu)_{ij}=\sqrt{2}h_{ij}v_\Delta.  \label{eq:mn}
\end{align}
Current neutrino oscillation data  can be explained in the HTM~\cite{lplp2,Han,HTM_pheno_w1}. 
It is seen from Eq.~(\ref{eq:mn}) that the neutrino mixing pattern is simply determined by the $h_{ij}$ matrix. 
Since the decay rate of $H^{\pm\pm}$ into the same-sign dilepton is proportional to $|h_{ij}|^2$, 
the type-II seesaw scenario can be tested  by looking at the same-sign dilepton decay mode of $H^{\pm\pm}$~\cite{lplp2,Han,HTM_pheno_w1}. 

The Higgs potential of the HTM is given by 
\begin{align}
V(\Phi,\Delta)&=m^2\Phi^\dagger\Phi+M^2\text{Tr}(\Delta^\dagger\Delta)+\left[\mu \Phi^Ti\tau_2\Delta^\dagger \Phi+\text{h.c.}\right]\notag\\
&+\lambda_1(\Phi^\dagger\Phi)^2+\lambda_2\left[\text{Tr}(\Delta^\dagger\Delta)\right]^2+\lambda_3\text{Tr}[(\Delta^\dagger\Delta)^2] 
\nonumber \\
&+\lambda_4(\Phi^\dagger\Phi)\text{Tr}(\Delta^\dagger\Delta)+\lambda_5\Phi^\dagger\Delta\Delta^\dagger\Phi, \label{pot_htm}
\end{align}
where $m$ and $M$ are the dimension full real parameters, $\mu$ is the dimension full complex parameter 
which violates the lepton number, and 
$\lambda_1$-$\lambda_5$ are the coupling constants which are real. 
We here take $\mu$ to be real.

The potential respects additional global symmetries in some limits. 
First, there is the global $U(1)$ symmetry in the potential in the limit of $\mu = 0$, which conserves the lepton number. 
As long as we assume that the lepton number is not spontaneously broken, 
the triplet field does not carry the VEV; i.e., $v_\Delta=0$. 
Next,  an additional global $SU(2)$ symmetry appears in the limit where $\mu = \lambda_5 =0$ . 
Under this $SU(2)$ symmetry, $\Phi$ and $\Delta$ can be transformed with the different $SU(2)$ phases.  
All the physical triplet-like Higgs bosons are then degenerate in mass. 

The mass matrices for the scalar bosons can be diagonalized by rotating the 
scalar fields as 
\begin{align}
\left(
\begin{array}{c}
\phi^\pm\\
\Delta^\pm
\end{array}\right)&=
\left(
\begin{array}{cc}
\cos \beta & -\sin\beta \\
\sin\beta   & \cos\beta
\end{array}
\right)
\left(
\begin{array}{c}
G^\pm\\
H^\pm
\end{array}\right),\nonumber\\
\left(
\begin{array}{c}
\chi\\
\eta
\end{array}\right)&=
\left(
\begin{array}{cc}
\cos \beta' & -\sin\beta' \\
\sin\beta'   & \cos\beta'
\end{array}
\right)
\left(
\begin{array}{c}
G^0\\
A
\end{array}\right),\notag\\
\left(
\begin{array}{c}
\phi\\
\delta
\end{array}\right)&=
\left(
\begin{array}{cc}
\cos \alpha & -\sin\alpha \\
\sin\alpha   & \cos\alpha
\end{array}
\right)
\left(
\begin{array}{c}
h\\
H
\end{array}\right), \label{mixing1}
\end{align}
with the mixing angles
\begin{align}
\tan\beta&=\frac{\sqrt{2}v_\Delta}{v_\phi},\quad \tan\beta' = \frac{2v_\Delta}{v_\phi}, \nonumber\\
\tan2\alpha &=\frac{v_\Delta}{v_\phi}\frac{2v_\phi^2(\lambda_4+\lambda_5)-4M_\Delta^2}{2v_\phi^2\lambda_1-M_\Delta^2-2v_\Delta^2(\lambda_2+\lambda_3)}. \label{tan2a}
\end{align}
In addition to the three Nambu-Goldstone  bosons $G^\pm$ and $G^0$ which are absorbed by the longitudinal components 
of the $W$ boson and the $Z$ boson, 
there are seven physical mass eigenstates; i.e., 
a pair of doubly-charged (singly-charged) Higgs bosons $H^{\pm\pm}$ ($H^\pm$), a CP-odd Higgs boson $A$ 
and CP-even Higgs boson $H$ and $h$, where $h$ is taken as the SM-like Higgs boson. 
The six parameters $\mu$ and $\lambda_1$-$\lambda_5$ in the Higgs potential in Eq.~(\ref{pot_htm}) 
can be written in terms of the physical scalar masses, the mixing angle $\alpha$ and VEVs $v_\phi$ and $v_\Delta$.

As required by the $\rho$- parameter data, when the triplet VEV $v_\Delta$ is much less than the doublet VEV $v_\phi$,  
there is relationships among the masses of the triplet-like Higgs bosons by neglecting $\mathcal{O}(v_\Delta^2/v_\phi^2)$ 
terms as 
\begin{align}
m_{H^{++}}^2-m_{H^{+}}^2&=m_{H^{+}}^2-m_{A}^2~~\left(=-\frac{\lambda_5}{4}v^2\right), \label{eq:mass_relation1}\\
m_A^2&=m_{H}^2~~(=M_\Delta^2). \label{eq:mass_relation2}
\end{align}
In the limit of $v_\Delta/v_\phi\to 0$, the four mass parameters of the triplet-like Higgs bosons are determined by two parameters. 
Eqs.~(\ref{eq:mass_relation1}) and (\ref{eq:mass_relation2}) can be regarded as the consequence of the global symmetries 
mentioned above. 

\begin{sloppypar}
The condition for the vacuum stability bound has been derived in Ref.~\cite{Arhrib}, where 
we require that 
the Higgs potential is bounded from below in any direction of the large scalar fields region. 
The unitarity bound in the HTM has been discussed in Ref.~\cite{Arhrib}. 
In Fig.~\ref{FIG:pu_vs2}, the excluded regions by the unitarity bound and the vacuum stability condition 
are shown for $\lambda_1 = m_h^2/(2v^2)\simeq 0.13$ in the $\lambda_4$-$\lambda_5$ plane~\cite{yagyu}. 
We take $\lambda_\Delta=1.5$ $(3)$ in the left (right) panel. 
Excluded regions by the unitarity and vacuum stability bounds are shown.  
\end{sloppypar}

\begin{figure}[t]
\begin{center}
 \includegraphics[width=80mm]{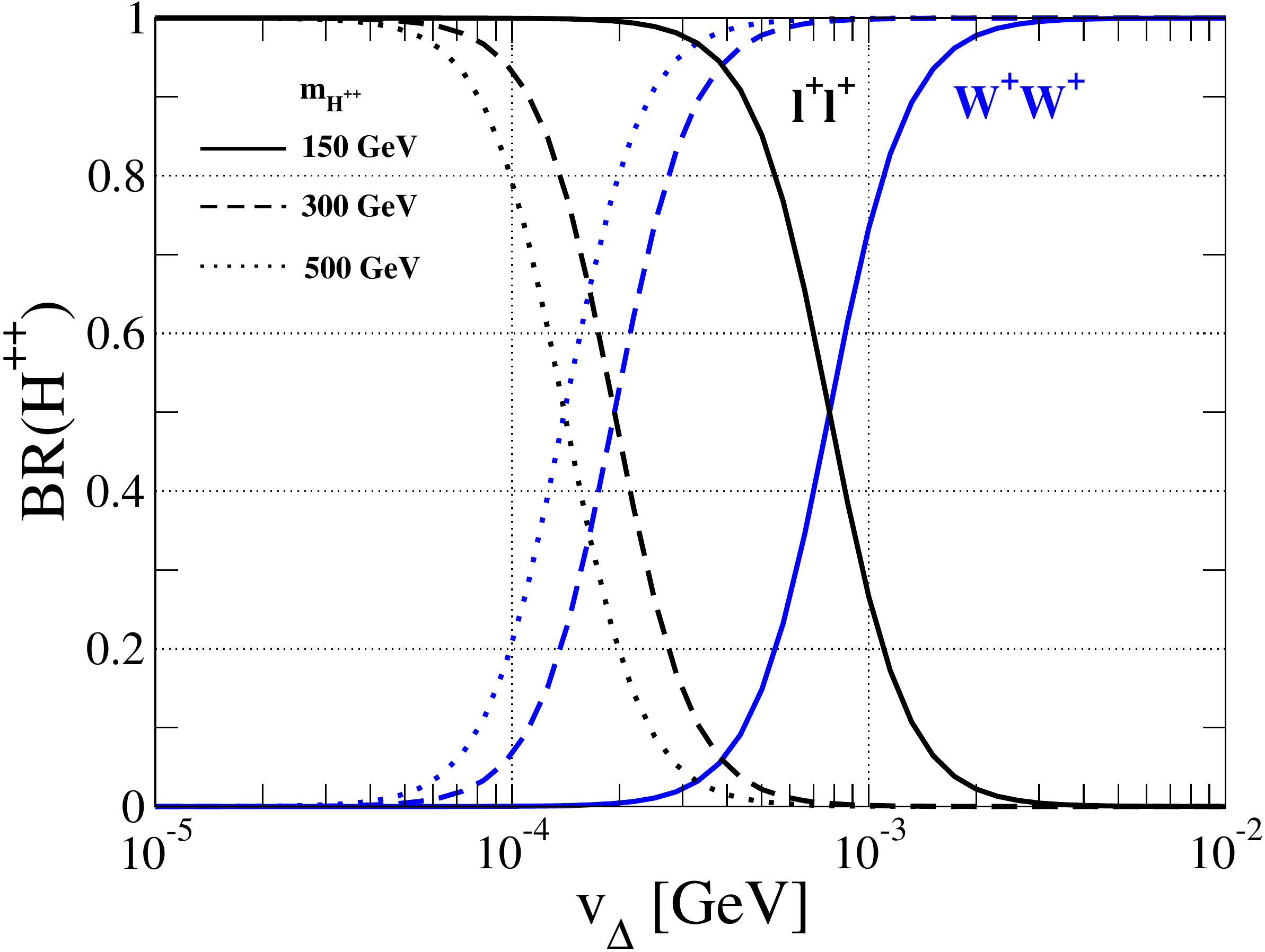}
   \caption{
Decay branching ratio of $H^{++}$ as a function of $v_\Delta$ with $m_{H^+}=m_{H^{++}}$. 
The solid, dashed and dotted curves respectively show the results in the case of $m_{H^{++}}=150$, 300 and 500 GeV~\cite{yagyu}. }
   \label{Fig:BR}
\end{center}
\end{figure}

The most interesting feature of the HTM is the existence of doubly charged Higgs bosons $H^{\pm\pm}$.
Their discovery at colliders can be a direct probe of the exotic Higgs sectors.
The doubly charged Higgs bosons $H^{\pm\pm}$ can decay into $\ell^\pm \ell^\pm$, $H^\pm W^\pm$ and $W^\pm W^\pm$
depending on the magnitude of $v_\Delta$~\cite{Perez:2008ha}.
In Fig.~\ref{FIG:BR1_HTM} , the branching ratios are shown as a function of
the vacuum expectation value of the triplet field, 
$v_\Delta$, for the cases with the mass difference $\Delta m=m_{H^{++}}-m_{H^+}=0$, $10$ GeV and $30$ GeV~\cite{Aoki:2011pz}.
The decay branching ratio of $H^{\pm\pm}$ is shown in Fig.~\ref{Fig:BR} assuming all the elements in $(M_\nu)_{ij}$ to be 0.1 eV. 
The dominant decay mode changes from the same-sign dilepton mode to the same-sign diboson mode at $v_\Delta=0.1$-1 MeV. 

\begin{sloppypar}
When the triplet-like Higgs bosons are degenerate in mass or
$H^{\pm\pm}$ is the lightest of all of them, the main decay mode of
$H^{\pm\pm}$ is the same-sign dilepton (diboson) in the case where
$v_\Delta$ is less (larger) than about 1 MeV.  The signal directly
shows the existence of the doubly charged scalar boson with lepton
number 2, which can be a strong evidence for the neutrino mass
generation via Eq.~(\ref{nu_yukawa}).
At the LHC, $H^{\pm\pm}$ are produced by the Drell-Yan process $pp\to
Z^*/\gamma^* \to H^{++}H^{--}$ and the associated process $pp\to
W^* \to H^{\pm\pm}H^{\mp}$.  The search for $H^{\pm\pm}$ in the
dilepton decay scenario has been performed at the LHC.  The scenario
based on the same-sign dilepton decay of $H^{\pm\pm}$ has been studied
in Refs.~\cite{lplp2,Han,HTM_pheno_w1}.  The strongest lower limit on
$m_{H^{++}}$ has been given by 459 GeV~\cite{400GeV_CMS} at the 95\%
CL assuming the 100\% decay of $H^{\pm\pm}\to \mu^\pm \mu^\pm$ from
the 7 TeV and 4.9 fb$^{-1}$ data.  This bound becomes weaker as 395
GeV~\cite{400GeV_CMS} when we only use the pair production process.
However, when $H^{\pm\pm}$ mainly decay into the same-sign diboson,
this bound can no longer be applied.
\end{sloppypar}

\begin{figure}[t]
\begin{center}
 \includegraphics[width=60mm]{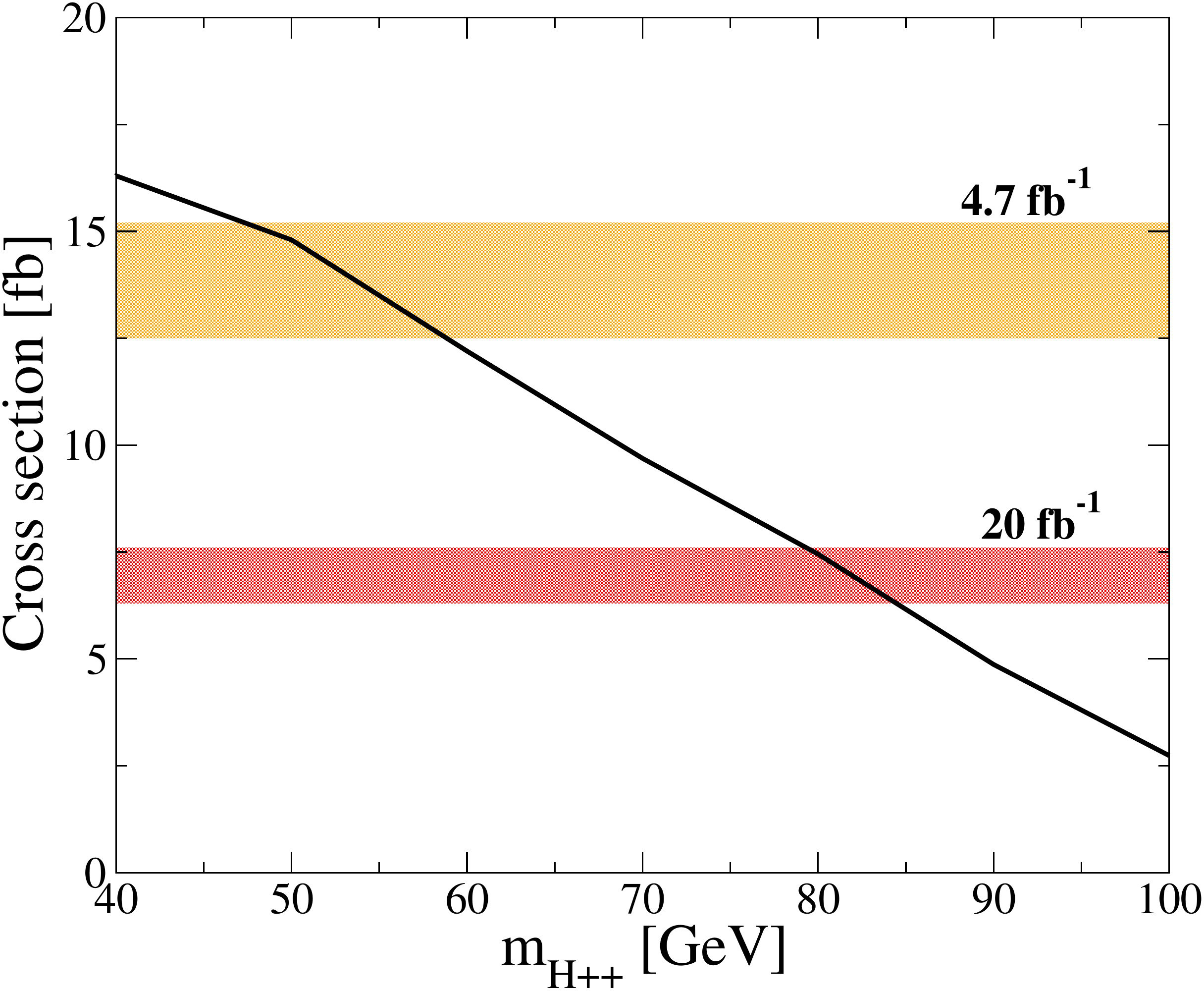}
\caption{The signal cross section as a function of $m_{H^{++}}$ with the collision energy to be 7 TeV from Ref.~\cite{KYY}. 
The light (dark) shaded band shows the 95\% CL (expected) upper bound for the cross section from the data with the integrate luminosity to be 4.7 fb$^{-1}$ (20 fb$^{-1}$). }
   \label{Fig:LHC}
\end{center}
\end{figure}

When $v_\Delta$ is sufficiently larger than $10^{-3}$ GeV,  the diboson decay $H^{\pm\pm} \to W^\pm W^\pm$
becomes dominant. In this case, the signal can also be same sign four leptons, but its rate is reduced by
the branching ratios of leptonic decays of $W$s.
The scenario for the same sign diboson decay of $H^{\pm\pm}$ has been 
discussed in Refs.~\cite{Han,Chiang_Nomura_Tsumura}. 
The discovery potential of $H^{\pm\pm}$ at the LHC has also been investigated in 
Ref.~\cite{Chiang_Nomura_Tsumura} in the HTM and also the Georgi-Machacek model~\cite{Georgi:1985nv}.  
 In Ref.~\cite{KYY}, the lower bound on $m_{H^{++}}$ has been obtained 
 by using the same-sign dilepton event measured at the LHC with 
7 TeV and 4.7 fb$^{-1}$ data~\cite{ATLAS_SS}. 
In Fig.~\ref{Fig:LHC}, the sum of the cross sections of the processes 
\begin{align}
pp\to H^{++}H^{--}\to W^{+(*)}W^{+(*)}H^{--}\to \mu^+\mu^+ E_{\text{miss}}H^{--},\notag\\
pp\to H^{++}H^{-}\to W^{+(*)}W^{+(*)}H^{-}\to \mu^+\mu^+ E_{\text{miss}}H^{-}, \label{Eq:signal}
\end{align}
are shown as a function of $m_{H^{++}}$ assuming $m_{H^+}=m_{H^{++}}$.
We can see that $m_{H^{++}}$ smaller than about 60 GeV is excluded at
the 95\% CL.  The bound is much relaxed as compared to that in the
dilepton decay scenario.  By the extrapolation of the data to
20~fb$^{-1}$ with the same collision energy, the lower limit is
obtained as 85 GeV.  Therefore, a light $H^{\pm\pm}$ such as around
100 GeV is still allowed by the current data at the LHC, and in this
case the ILC may be able to discover the doubly charged Higgs boson.
See also recent progress in Ref.~\cite{Kanemura:2014goa}.
 
\begin{figure}[t]
\begin{center}
 \includegraphics[width=80mm]{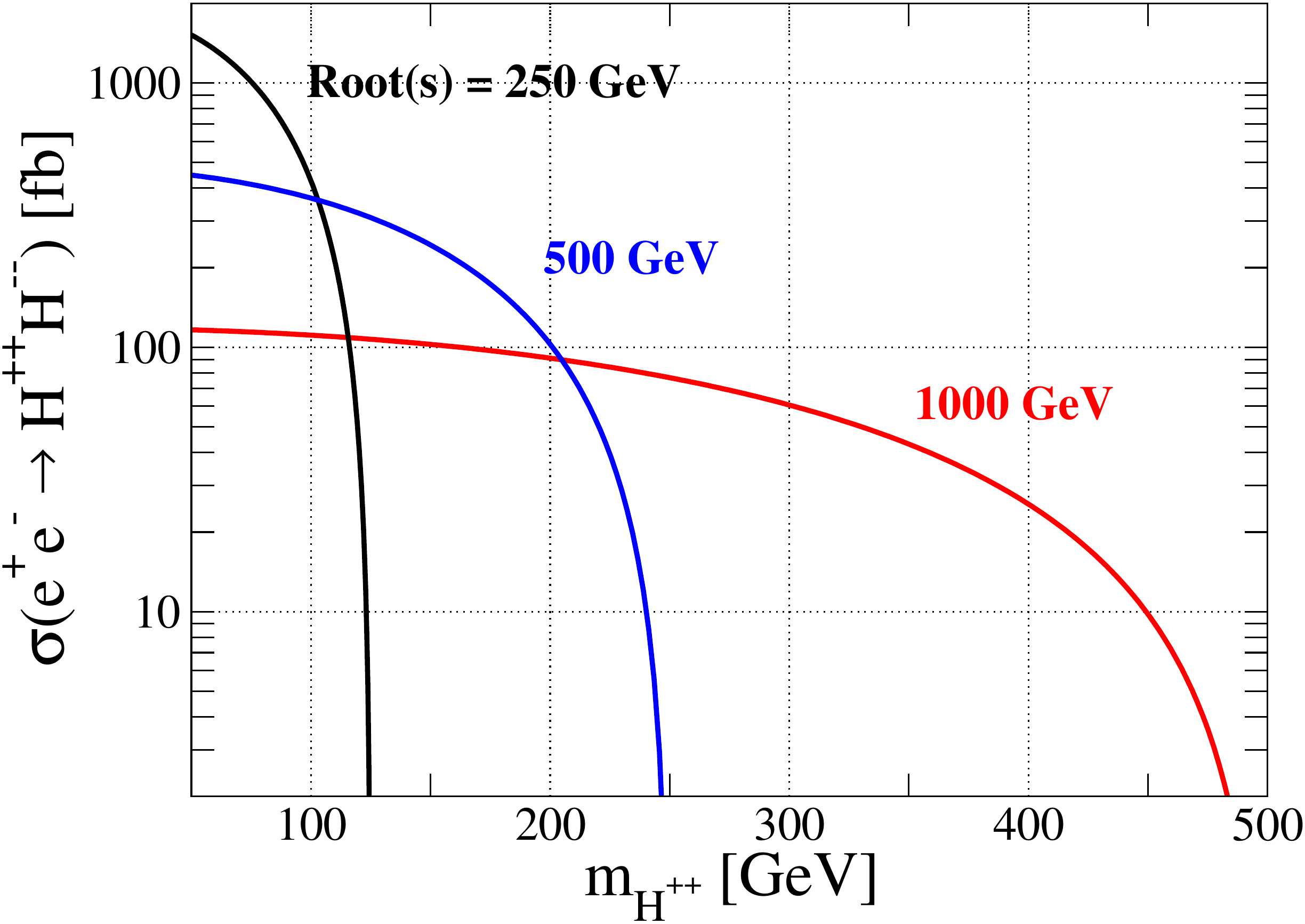}
   \caption{
Production cross section of the $e^+e^-\to H^{++}H^{--}$ process as a function of $m_{H^{++}}$. 
The black, blue and red curves are respectively the results with the collision energy $\sqrt{s}=$250, 500 and 1000 GeV.  }
   \label{Fig:sigma_Hpp}
\end{center}
\end{figure}

\begin{sloppypar}
 At the ILC, doubly charged Higgs bosons are produced via the pair
production $e^+e^-\to H^{++}H^{--}$.  In the diboson decay scenario,
the final state is the same-sign dilepton, missing energy and
multi-jets; i.e., $e^+e^-\to H^{++}H^{--}\to \ell^+\ell^+
E_\text{miss}jjjj$, where $\ell=e,\mu$~\cite{yagyu}.  The background
comes from the four W bosons production; i.e., $e^+e^-\to
W^+W^+W^-W^-\to \ell^\pm\ell^\pm E_\text{miss}jjjj$.  For example,
when $\sqrt{s}=500$ GeV and the $m_{H^{++}}=230$~GeV is taken, the
signal (background) cross section of the final state 
$\ell^\pm\ell^\pm E_\text{miss}4j$ is obtained to 
be 1.07 fb (2.37$\times 10^{-3}$
fb) (Fig.~\ref{Fig:sigma_Hpp})~\cite{yagyu}.  
The above numbers are obtained after taking the
following basic kinematic cuts
\begin{align}
 p_T^\ell \geq 15 ~\text{GeV},\quad |\eta^\ell| \leq  2.5, 
\end{align}
where $p_T^\ell$ and $\eta^\ell$ are the transverse momentum and pseudo rapidity for $\ell$, respectively. 
Therefore, this process is almost background free. 
\begin{figure*}[t]
\vspace{.6cm}
\begin{center}
 \includegraphics[width=60mm]{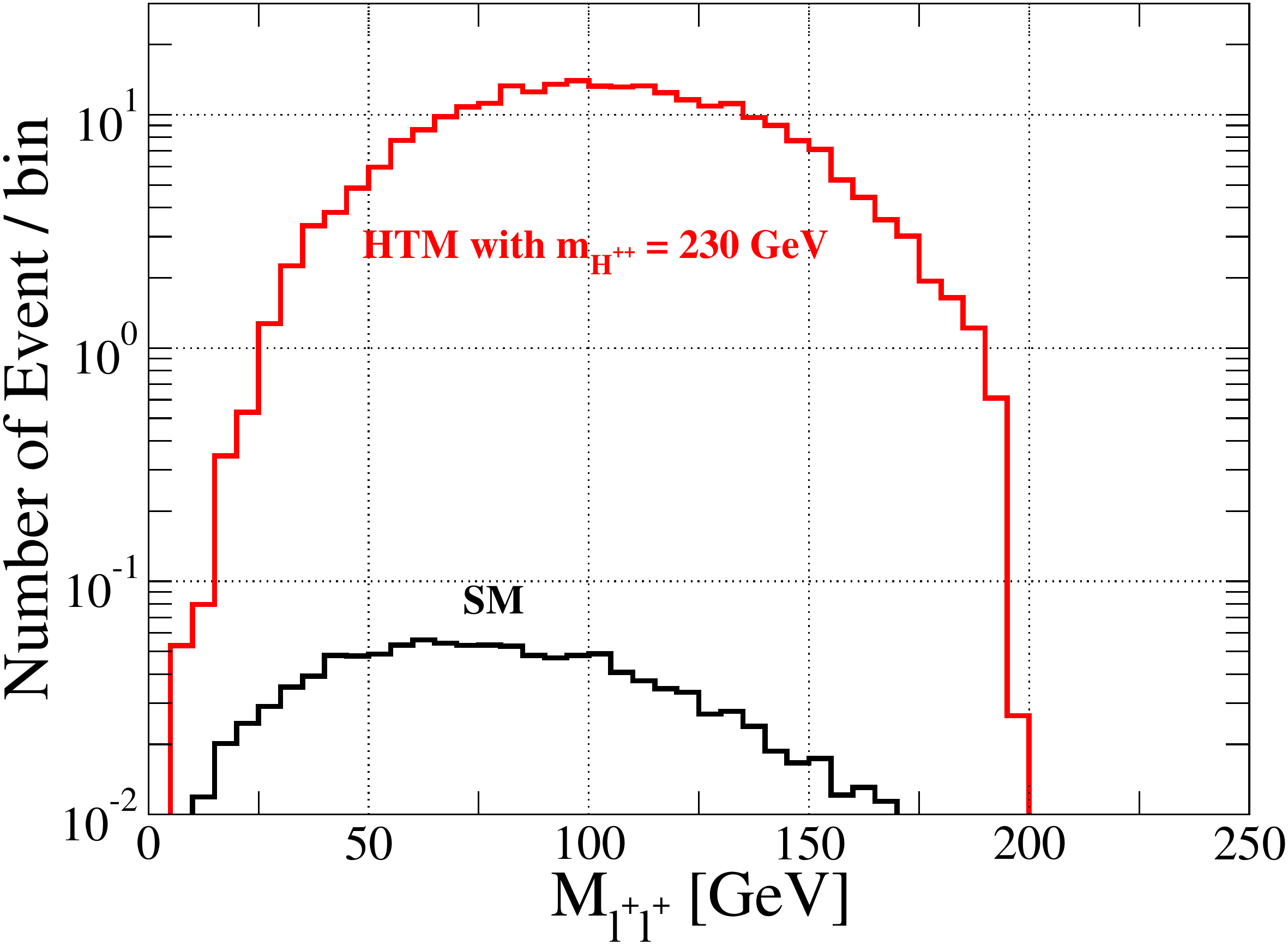}\hspace{5mm}
 \includegraphics[width=60mm]{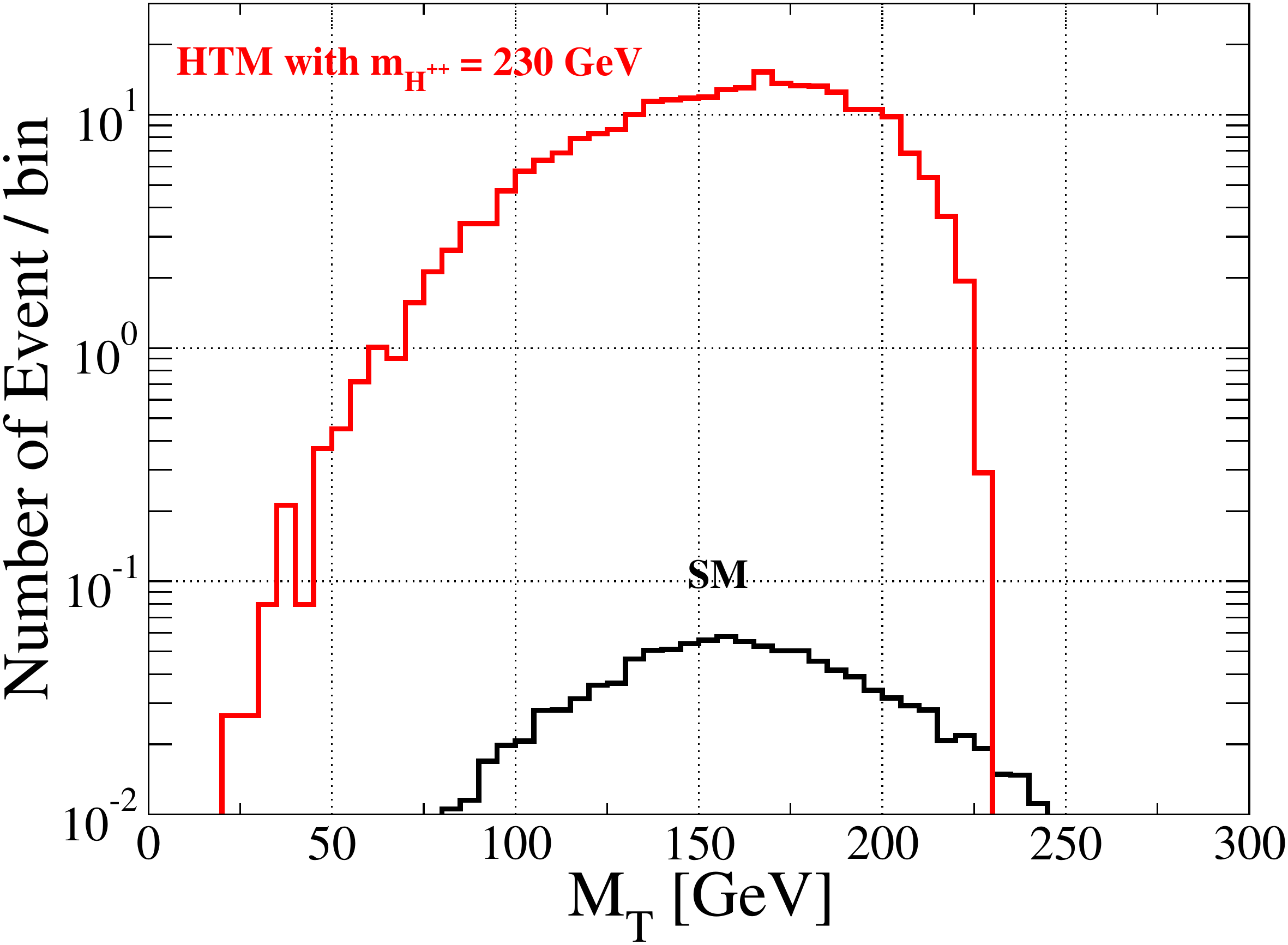}
   \caption{
The invariant mass distribution (left panel) and the transverse mass distribution (right panel) for the $\ell^+\ell^+$ and $\ell^+\ell^+E_{\text{miss}}$ systems, respectively, in the case of $m_{H^{++}}=230$ GeV and $\sqrt{s}=500$ GeV~\cite{yagyu}. 
The integrated luminosity is assumed to be 500 fb$^{-1}$.  }
   \label{Fig:distr}
\end{center}
\end{figure*}
In Fig.~\ref{Fig:distr}, the invariant mass $M_{\ell^+\ell^+}$ for the $\ell^+\ell^+$ system (left panel) and 
the transverse mass $M_T$ (right panel) distributions for $\ell^+\ell^+E_{\text{miss}}$ system are shown.
The red and black curves denote the distribution from the signal and background, respectively. 
Around 230 GeV,  there is an endpoint in the $M_T$ distribution that corresponds to $m_{H^{++}}$. 
The $M_T$ distribution is useful to measure $m_{H^{++}}$. \\
\end{sloppypar}

If the triplet-like Higgs bosons are light enough, 
the direct detection of them at the LHC and the ILC is the most important probe of the HTM as already discussed. 
On the other hand, they can also be indirectly tested by measuring 
the deviations from the SM in the Higgs boson couplings for the SM-like Higgs boson $h$,  such as 
the coupling constants with the weak gauge bosons $hVV$, the Yukawa couplings $hf\bar{f}$ and 
the triple Higgs boson coupling $hhh$, where $V$ represents gauge bosons, and $f$ does 
quarks and leptons. The indirect searches can be useful even when no new particles is directly 
found. 
At the ILC, the Higgs boson couplings are expected to be precisely measured. 
For example, the Higgs boson couplings with the weak gauge bosons ($hZZ$ and $hWW$) 
and the Yukawa couplings ($hb\bar{b}$, $h\tau\bar{\tau}$ and $ht\bar{t}$) are expected to be measured with 
$\mathcal{O}(1)\%$ accuracy~\cite{Higgs_couplings_LHC,ILC_TDR,Asner:2013psa,Dawson:2013bba}. 
In the HTM, the loop induced $h\gamma\gamma$ coupling has been calculated in Refs.~\cite{hgg_HTM}. 
The one-loop corrections to the $hWW$, $hZZ$ and $hhh$ vertices have also been calculated in Refs.~\cite{AKKY,AKKY_full}. 
In Ref.~\cite{AKKY_full}, it has been found that there is a correlation among the deviation in the Higgs boson couplings. 
For example, when the decay rate of $h\to \gamma\gamma$ deviates by 30\% (40\%)
from the SM prediction, deviations in the one-loop corrected $hVV$ and $hhh$ vertices are
predicted to be about $-0.1\%$ ($-2\%$) and $-10\%$ ($150\%$), respectively\footnote{
In the HTM, deviations in $h\bar f f $ couplings are small because of $v_\Delta \ll v_\phi$.}.
By comparing these deviations with the precisely measured value at the ILC, 
we can discriminate the HTM from the other models.

\subsubsection{Other exotic models}\label{exotic}

\begin{table*}[t]
\begin{center}
\begin{tabular}{c||l|l|l}\hline\hline 
& $\tan\beta$ & $\kappa_f$ &$\kappa_V^{}$   \\ \hline 
Doublet-Singlet Model & --- & $\cos\alpha$ & $\cos\alpha$ \\ \hline 
Type-I THDM  &$v_0/v_\text{ext}^{}$ &$\cos\alpha/\sin\beta=\sin(\beta-\alpha)+\cot\beta\cos(\beta-\alpha)$ &$\sin(\beta-\alpha)$  \\ \hline 
GM Model &$v_0/(2\sqrt{2}v_\text{ext}^{})$& $\cos\alpha/\sin\beta$ &   $\sin\beta \cos\alpha -\tfrac{2\sqrt6}3 \cos\beta \sin\alpha$ \\  \hline 
Doublet-Septet Model &$v_0/(4v_\text{ext}^{})$&$\cos\alpha/\sin\beta$&$\sin\beta \cos\alpha 
-4 \cos\beta \sin\alpha$\\\hline\hline
\end{tabular}
\end{center}
\caption{The fraction of the VEVs $\tan\beta$ and the scaling factors $\kappa_f$ and $\kappa_V$ in the extended Higgs sectors with 
universal Yukawa couplings~\cite{KTYY}. }
\label{Tab:ScalingFactor}
\end{table*}

Precision measurements for the couplings of the SM-like Higgs boson $h$ at the ILC 
can also discriminate exotic Higgs sectors.  Accrding to Refs.~\cite{Asner:2013psa,KTYY}, 
we here consider various extended Higgs sectors which satisfy $\rho=1$ at the tree level; i.e., 
the model with an additional singlet scalar field with $Y=0$, 
 the 2HDM (Type~I), the model with a septet scalar field with $Y=4$~\cite{Hisano:2013sn}, 
and the Georgi Machacek model where a complex ($Y=2$) and a real ($Y=0$) 
triplet scalar fields are added to the SM-like Higgs doublet~\cite{Georgi:1985nv}.  
In these models, all quark and leptons receive their masses from only one scalar doublet. 
Consequently,  the Yukawa coupling constants with respect to the SM-like Higgs boson 
$h \bar{f} f$ from the SM values are commonly suppressed due to the mixing between 
the two (or more) neutral states.
In a, we have a universal suppression on 
the coupling constants, $\kappa_F^{} = \kappa_V^{} = \cos\theta$ with $\theta$ 
being the mixing angle between the doublet field and the singlet field.
However, $\kappa_F^{} \neq \kappa_V^{}$ is usually predicted in more complicated 
Higgs sectors such as the 2HDM (Type I), the Georgi-Machacek model~\cite{Georgi:1985nv} and 
the doublet-septet model~\cite{Hisano:2013sn}.  
Notice that in exotic models with higher representation scalar fields such as  the Georgi-Machacek model 
and doublet-septet model, $\kappa_V$ can be greater than 1.  
This can be a signature of exotic Higgs sectors.
From Eq.~(\ref{eq:rho_general}), a VEV from these additional scalar multiplets do not change $\rho=1$ at the tree level. 
All the VEVs  $v_\text{ext}$ of  these additional Higgs multiplets except for that of the singlet 
partially contribute to the spontaneous breaking of the electroweak gauge symmetry.  
The VEVs satisfy $v^2= v_0^2 + (\eta_\text{ext}\, v_\text{ext})^2$, where 
$v_\phi$ is the VEV of the SM-like Higgs doublet $\Phi$ 
and $\eta_{\text{ext}}$ = 1 and 4 in the Type-I THDM and the model with the septet, 
respectively. 
It is convenient to define the ratio of the VEVs as $\tan\beta = v_0/(\eta_\text{ext}\, v_\text{ext})$~\cite{KTYY}. 
In tab.~\ref{Tab:ScalingFactor},  the scaling factors $\kappa_f$ and $\kappa_V$ are listed  
in terms of $\alpha$ and $\beta$ in the four models. 

\begin{figure}[t]
\begin{center}
\includegraphics[width=70mm]{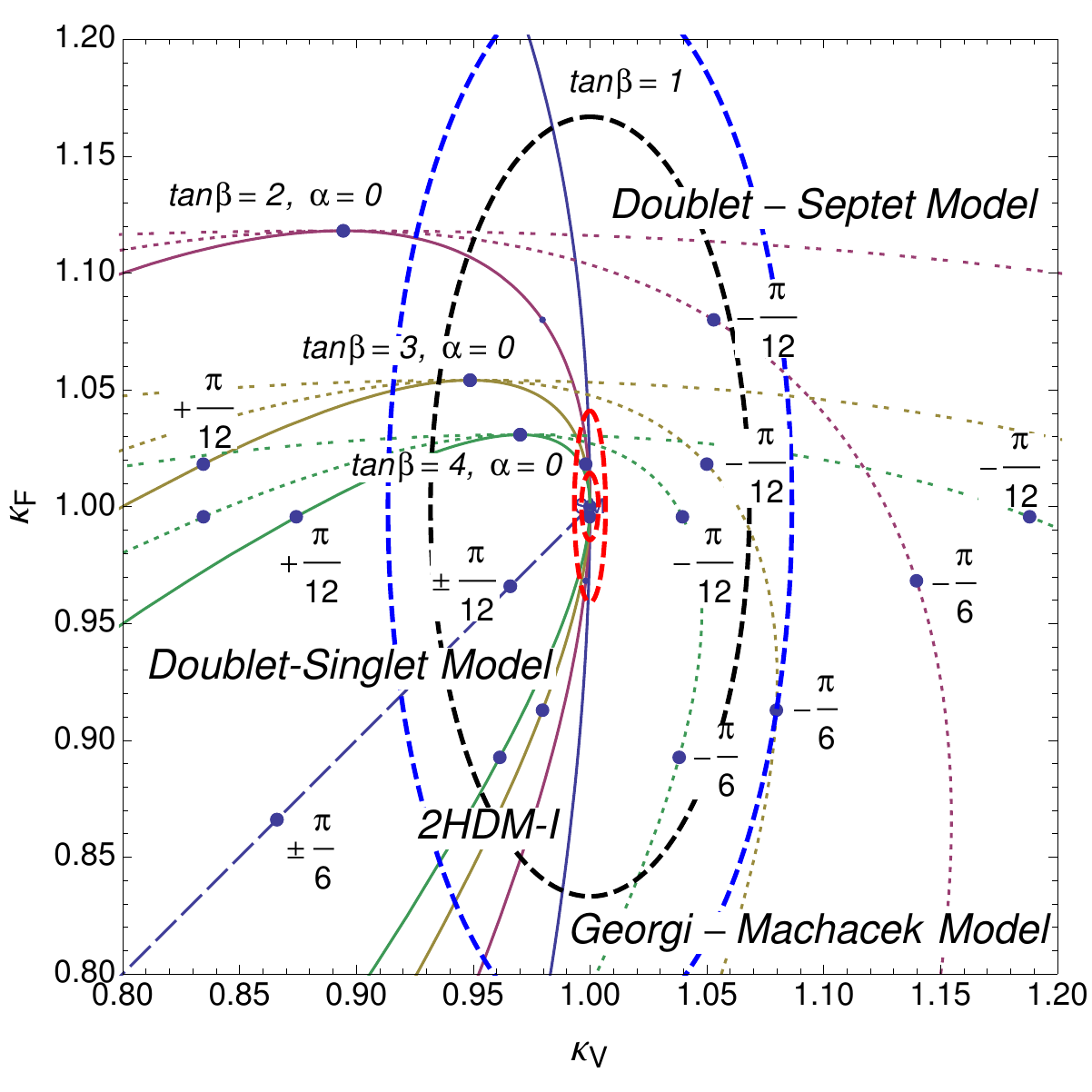}  
\caption{The scaling factors in models with universal Yukawa couplings. The current LHC 
bounds and the expected LHC and ILC sensitivities are also shown at the 68.27 \%  C.L.. 
For details, see Ref.~\cite{KTYY}}
\label{finger_print2}
\end{center}
\end{figure}

In Fig.~\ref{finger_print2}, the predictions for the scale factors of the universal 
Yukawa coupling $\kappa_F$ and the gauge coupling $\kappa_V$ are plotted 
in exotic Higgs sectors for each set of mixing angles.
The current LHC bounds, expected LHC and ILC sensitivities 
for $\kappa_F$ and $\kappa_V$ are also shown at the 68.27 \%  C.L..
Therefore, exotic Higgs sectors can be discriminated by measuring $\kappa_V$ and $\kappa_F$ 
precisely. For details, see Refs.~\cite{Asner:2013psa, KTYY}.

\subsubsection{Summary}

Although the Higgs boson with the mass 125 GeV was found at the LHC, 
knowledge about the structure of the Higgs sector is very limited.
Since there are no theoretical principle for the minimal Higgs sector with 
one Higgs doublet, there are many possibilities of non-minimal Higgs sectors. 
Such extended Higgs sectors appear in many new physics models beyond the SM. 
Therefore, the Higgs sector is a window to new physics, and we can explore new physics 
from clarifying the structure of the Higgs sector by coming collider experiments.  
At the LHC, direct discovery of additional Higgs bosons can be expected as long as 
they are not too heavy. 
On the other hand, the Higgs sector can also be explored by precisely measuring the 
properties of the discovered Higgs boson $h$ accurately.
The precision measurements will be performed partially at the high luminosiity
 LHC with 3000 fb$^{-1}$.  
Using high ability of the ILC for measuring the Higgs boson couplings,   
we can further test extended Higgs sectors,  
and consequently narrow down the new physics models.






\subsection{Higgs physics in strong-interaction scenarios\protect\footnotemark}
\footnotetext{Christoph Grojean, M.\ Margarete M\"uhlleitner}
\label{sec:ewsb7}
%

\def\MySubsection#1{\vspace{.2cm} \noindent {\it  #1} \vspace{.2cm}}
\def\overleftrightarrow#1{\buildrel\hbox{$\scriptstyle \leftrightarrow$}\over{#1}\!}

\noindent
The Higgs mechanism~\cite{Englert:1964et,Higgs:1964ia,Higgs:1964pj,Guralnik:1964eu}, which has
been introduced to provide masses for the fermions and gauge bosons
without violating gauge principles, can describe electroweak
symmetry breaking (EWSB) but fails to explain it. Within the Standard Model
(SM) there is no dynamics leading to the typical mexican hat shape of the
Higgs potential. Moreover, in order to keep the Higgs boson mass at the
experimentally measured value of $125$~GeV~\cite{Aad:2012tfa,Chatrchyan:2012ufa} in the
presence of high scales at which the SM will eventually has to be amended, a
substantial amount of finetuning is necessary unless the mass is
protected from higher order corrections due to some symmetry. Such a
symmetry must act non-linearly on the Higgs field. Besides
supersymmetry a prominent example is given by a global symmetry when
the Higgs boson appears as a pseudo Nambu--Goldstone boson. 
A Higgs boson is needed to ensure the proper decoupling of the longitudinal polarizations of the massive EW gauge bosons at high energy.
Indeed, these
longitudinal modes of  $W^\pm$ and $Z$ can
be described by Nambu--Goldstone bosons associated
to the coset $SU(2)_L\times SU(2)_R / SU(2)_{\mbox{\scriptsize
    isospin}}$. Their kinetic term corresponds to the gauge boson mass terms,
\begin{eqnarray}
\frac{1}{2}m_Z^2 Z_\mu Z^\mu + m_W^2 W_{\mu}^+ W^{-\mu}=
\frac{v^2}{4} \mbox{Tr} ( D_\mu \Sigma^\dagger D^\mu \Sigma)
\label{eq:kin}
\end{eqnarray}
with $\Sigma = e^{i\sigma^a \pi^a/v}$, where $\sigma^a$ ($a=1,2,3$) are the usual Pauli matrices. Due to the
Goldstone boson equivalence theorem the non-trivial scattering of the 
longitudinal gauge bosons $V$ ($V=W^\pm,Z$) is controlled by the contact
interactions among four pions from the expansion of the Lagrangian
Eq.~(\ref{eq:kin}), leading to amplitudes growing with the energy,
\begin{eqnarray}
{\cal A} (V^a_L V^b_L \to V^c_L V^d_L) &=& {\cal A}(s) \delta^{ab}
\delta^{cd} + {\cal A} (t) \delta^{ac} \delta^{bd} \nonumber \\
&& \hspace*{-0.7cm} + {\cal A} (u)
\delta^{ad} \delta^{bc} \quad \mbox{with} \quad {\cal A} (s) \approx
\frac{s}{v^2} \:.
\end{eqnarray}
\begin{sloppypar}
Here $s,t,u$ denote the Mandelstam variables, and $v$ represents the vacuum
expectation value (VEV) with $v \approx 246$~GeV. The amplitude grows with the
center-of-mass (c.m.) energy squared $s$, and therefore perturbative unitarity
will be lost around $4 \pi v \sim 3$~TeV, unless there is a new weakly coupled
elementary degree of freedom. The simplest realization of new dynamics
restoring perturbative unitarity is given by a single scalar field
$h$, which is singlet under $SU(2)_L\times SU(2)_R / SU(2)_{\mbox{\scriptsize
    isospin}}$ and
couples to the longitudinal gauge bosons and fermions
as~\cite{Giudice:2007fh,Contino:2010mh,Grober:2010yv}, 
\end{sloppypar}
\begin{eqnarray}
{\cal L}_{EWSB} &=& \frac{1}{2} (\partial_\mu h)^2 - V(h) +
 \nonumber \\
&&\hspace*{-1.7cm} \frac{v^2}{4} \,\mbox{Tr} (D_\mu \Sigma^\dagger D^\mu \Sigma) \left(
  1+2a \frac{h}{v} + b \frac{h^2}{v^2} + \sum_{n\geq 3} b_n \frac{h^n}{v^n}+\ldots
  \right) \nonumber \\
&&\hspace*{-1.7cm} - \frac{v}{\sqrt{2}}
(\bar{u}^i_L \bar{d}^i_L) \Sigma \left( 1 + c \frac{h}{v} + \sum_{n\geq 2}c_n
  \frac{h^n}{v^n} + ... \right) \left(\begin{array}{c} y^u_{ij} u^j_R
    \\ y^d_{ij} d^j_R \end{array}\right) \nonumber \\
&&\hspace*{-1.7cm} +{\rm h.c.}\label{eq:lewsb}
\end{eqnarray}
with
\begin{eqnarray}
V(h) &=& \frac{1}{2} m_h^2 h^2 + \frac{d_3}{6} \left(\frac{3m_h^2}{v}\right) h^3 + \frac{d_4}{24}
\left(\frac{3m_h^2}{v^2}\right) h^4 \nonumber \\
+ ... \label{eq:vewsb}
\end{eqnarray}
For $a=1$ the scalar exchange cancels the piece growing with the
energy in the $V_L V_L$ amplitude. If in addition $b=a^2$ then also in
the inelastic amplitude $V_L V_L \to hh$ unitarity is maintained,
while for $ac=1$ the $V_L V_L \to f \bar{f}'$ amplitude remains
finite. The SM Higgs boson is defined by the point $a=b=c=1$ and
$d_3=d_4=1$, $c_{n\geq2}=b_{n\geq3}=0$. The scalar resonance and the pions then
combine to from a doublet which transforms linearly under $SU(2)_L
\times SU(2)_R$. The Lagrangian Eq.~(\ref{eq:lewsb}) describes either
an elementary or a composite Higgs boson. For $a\ne 1$ the Higgs boson
alone cannot fully unitarize the $V_L V_L$ scattering, with the
breakdown of perturbative unitarity pushed to a higher scale now,
which is of the order $4\pi v/ \sqrt{1-a^2}$. The residual growth of the
scattering amplitude ${\cal A}(s) \approx (1-a^2) s/v^2$ will finally
be cancelled by the exchange of other degrees of freedom. The
Lagrangian Eqs.~(\ref{eq:lewsb}), (\ref{eq:vewsb}) introduces deviations in
the Higgs boson phenomenology~\cite{Giudice:2007fh,Espinosa:2010vn} away from the SM point by rescaling all
Higgs couplings through the modifiers $a,b$ and $c$,
\begin{eqnarray}
g_{hVV} = a g_{hVV}^{SM} \; , \; g_{hhVV}= b g_{hhVV}^{SM} \; , \;
g_{hf\bar{f}'} = c g_{hf \bar{f}'}^{SM} \; ,
\end{eqnarray}
while keeping the same Lorentz structure.
With $c$ being flavor-universal, minimal flavor violation is built  in and the only
source of flavor violation are the usual SM Yukawa couplings. There are
additional new couplings as, {\it e.g.}, the $c_2$ coupling between two
Higgs bosons and two fermions, which contributes to multi-Higgs
production~\cite{Giudice:2007fh,Contino:2010mh,Grober:2010yv}. 

In composite Higgs models, the deviations from the SM point $a=b=1$ are
controlled by the ratio of the weak scale over the compositeness scale
$f$. In these models the Higgs boson is a composite bound state which
emerges from a strongly interacting sector~\cite{Kaplan:1983fs,Dimopoulos:1981xc,Banks:1984gj,Georgi:1984ef,Georgi:1984af,Dugan:1984hq}. The good agreement
with the electroweak precision data is achieved by a mass gap that
separates the Higgs scalar from the other resonances of the strong
sector. This mass gap arises dynamically in a natural way if the
strongly interacting sector has a global symmetry $G$, which is
spontaneously broken at a scale $f$ to a subgroup $H$ so that the
coset $G/H$ contains a fourth Nambu-Goldstone boson which is
identified with the Higgs boson. Composite Higgs models can be viewed
as a continuous interpolation between the SM and technicolor type
models. With the compositeness scale of the Higgs boson given by the
dynamical scale $f$, the limit $\xi \equiv v^2/f^2 \to 0$ corresponds to the
SM where the Higgs boson appears as an elementary light particle and
the other resonances of the strong sector decouple. In the limit $\xi \to
1$ the Higgs boson does not couple to the $V_L$'s any longer and other (heavy) resonances
are necessary to ensure unitarity in the gauge boson scattering. The $\xi\to 1$ limit corresponds to the technicolor paradigm~\cite{Weinberg:1975gm,Susskind:1978ms} where the strong dynamics directly breaks the electroweak symmetry down to the electromagnetism subgroup. 

\subsubsection{Effective Lagrangian and Higgs couplings}

\begin{sloppypar}
Independently of its dynamical origin, the physics of a strongly
interacting light Higgs (SILH) boson can be captured in a
model-independent way by an effective Lagrangian which involves two
classes of higher dimensional operators: (i) those being genuinely
sensitive to the new strong force and which will qualitatively affect
the Higgs boson phenomenology and (ii) those being sensitive only to
the spectrum of the resonances and which will simply act as form
factors. The size of the various operators is controlled by simple
rules and the effective Lagrangian can be cast into the generic
form~\cite{Giudice:2007fh}
\begin{eqnarray}
\mathcal{L}_{\rm SILH} & = & \frac{c_H}{2f^2} \left( \partial_\mu |H|^2 
\right)^2
+ \frac{c_T}{2f^2}  \left(   H^\dagger  {\overleftrightarrow { D^\mu}}  H\right)^2 
- \frac{c_6\lambda}{f^2} |H|^6 \nonumber \\
&& \hspace*{-1.3cm} + \left( \frac{c_yy_f}{f^2} |H|^2 {\bar f}_L Hf_R +{\rm h.c.}\right) 
+\frac{ic_Wg}{2m_\rho^2}\left( H^\dagger  \sigma^i
  {\overleftrightarrow { D^\mu}}
H \right ) \times \nonumber \\
&& \hspace*{-1.3cm}
( D^\nu  W_{\mu \nu})^i +\frac{ic_Bg'}{2m_\rho^2}\left( H^\dagger  {\overleftrightarrow { D^\mu}} H 
\right )( \partial^\nu  B_{\mu \nu})  +\ldots 
\label{eq:lsilh}
\end{eqnarray}
with the SM electroweak (EW) couplings $g,g'$, the SM Higgs quartic
coupling $\lambda$ and the SM Yukawa coupling $y_f$ to the fermions
$f_{L,R}$. The coefficients in Eq.~(\ref{eq:lsilh}) are expected to be
of order one unless protected by some symmetry. The SILH Lagrangian
gives rise to oblique corrections at tree-level. 
 The coefficient $c_T$  vanishes in
case the strong sector is assumed to respect custodial symmetry. 
The
form factor operators induce a contribution to the $\hat S$ parameter,
$\hat{S} = (c_W + c_B)m_W^2/m_\rho^2$, where $m_\rho$ denotes the mass
scale of the heavy strong sector resonances, which imposes a lower
bound $m_\rho \ge 2.5$~TeV. Since the Higgs couplings to the SM vector
bosons receive corrections of the order $v^2/f^2$ the cancellation
between the Higgs and the gauge boson contributions taking place in
the SM, is spoiled and the $\hat S$ and $\hat T$ parameters become
logarithmically divergent~\cite{Barbieri:2007bh} when all the low energy degrees of freedom are considered. This infrared (IR) contribution
imposes an upper bound of $\xi \raisebox{-0.13cm}{~\shortstack{$<$ \\[-0.07cm] $\sim$}}~0.1$~\cite{Agashe:2005dk,Gillioz:2008hs,Anastasiou:2009rv,Ciuchini:2013pca} which can be relaxed by a
factor of 2 if a partial cancellation of ${\cal O}(50\%)$ with
contributions from other states is allowed. Light top partners, as required to generate the Higgs mass, also contribute to the EW oblique parameters and can change the range of value of $\xi$ preferred by EW precision data~\cite{Grojean:2013qca}.
 The Higgs kinetic term,
which receives a correction from the operator $c_H$, can be brought
back to its canonical form by rescaling the Higgs field. This induces
in the Higgs couplings a universal shift by a factor $1-c_H
\xi/2$. For the fermions, it adds up to the modified Yukawa
interactions. 
\end{sloppypar}

\begin{sloppypar}
The effective Lagrangian Eq.~(\ref{eq:lsilh}) represents the
first term in an expansion in $\xi = v^2/f^2$. For large values of
$\xi \sim {\cal O}(1)$ the series has to be resummed, examples of
which have been given in explicit models such as those constructed in
5D warped space based on the coset $SO(5)/SO(4)$~\cite{Contino:2003ve,Agashe:2004rs,Contino:2006qr}. 
In the MCHM4~\cite{Agashe:2004rs}, where the SM fermions transform as spinorial 
representations of $SO(5)$, all SM Higgs couplings are suppressed by
the same modification factor as function of $\xi$, so that the
branching ratios are unchanged and only the total width is
affected. In the MCHM5~\cite{Contino:2006qr} with the fermions in the
fundamental representation of $SO(5)$ on the other hand the Higgs
couplings to gauge bosons and to fermions are modified differently
inducing non-trivial changes both in the branching ratios and the
total width. The relations between the couplings in the effective
Lagrangian Eq.~(\ref{eq:lewsb}), the SILH Lagrangian
Eq.~(\ref{eq:lsilh}) and the MCHM4 and MCHM5 models is summarized in
Table~\ref{tab:coupl}, see also Ref.~\cite{Contino:2013kra}.
\end{sloppypar}
\begin{table}
\caption{Higgs coupling values of the effective
  Lagrangian Eq.~(\ref{eq:lewsb}), in the SILH set-up Eq.~(\ref{eq:lsilh}) and
  in explicit SO(5)/SO(4) composite Higgs models built in warped 5D
  space-time, MHCM4 and MHCM5. From Ref.~\cite{Espinosa:2012qj}.}
\begin{tabular}{cccc}
\hline\noalign{\smallskip}
\hspace{-.2cm} Parameters & \hspace{-1.cm} SILH & \hspace{-.1cm} MCHM4 & MCHM5  \\
\noalign{\smallskip}\hline\noalign{\smallskip}
\hspace{-1cm} $a$ &  \hspace{-1.cm}$1-c_H\xi/2$ & \hspace{-.1cm}$\sqrt{1-\xi}$ & \hspace{-.1cm}$\sqrt{1-\xi}$\\[.2cm]
\hspace{-1cm} $b$ &  \hspace{-1.cm}$1-2 c_H \xi$ & \hspace{-.1cm}$1-2\xi$ & \hspace{-.1cm}$1-2\xi$\\[.2cm]
\hspace{-1cm} $b_3$ &  \hspace{-1.cm} $-\frac{4}{3}\xi$ & \hspace{-.1cm}$-\frac{4}{3} \xi \sqrt{1-\xi}$ & \hspace{-.1cm}$-\frac{4}{3} \xi \sqrt{1-\xi}$ \\[.3cm]
\hspace{-1cm} $c$ &  \hspace{-1.cm}$1-(c_H/2+c_y) \xi$ &  \hspace{-.1cm}$\sqrt{1-\xi}$ & \hspace{-.1cm}$\frac{1-2\xi}{\sqrt{1-\xi}}$  \\[.3cm]
\hspace{-1cm} $c_2$ &  \hspace{-1.cm} $-(c_H+3c_y)\xi/2$ &  \hspace{-.1cm}$-\xi/2$ & \hspace{-.1cm}$-2\xi$  \\[.2cm]
\hspace{-1cm} $d_3$ &  \hspace{-1.cm}$1+(c_6 - 3 c_H/2) \xi$ & \hspace{-.1cm}$\sqrt{1-\xi}$ & \hspace{-.1cm}$\frac{1-2\xi}{\sqrt{1-\xi}}$\\[.3cm]
\hspace{-1cm} $d_4$ &  \hspace{-1.cm}$1+(6 c_6 - 25 c_H/3) \xi$ & \hspace{-.1cm}$1-7 \xi/3$ & \hspace{-.1cm}$\frac{1-28\xi(1-\xi)/3}{1-\xi}$\\
\noalign{\smallskip}\hline
\label{tab:coupl}    
\end{tabular}
\end{table}

\begin{figure}[hb]
\begin{center}
\includegraphics[width=0.4\textwidth]{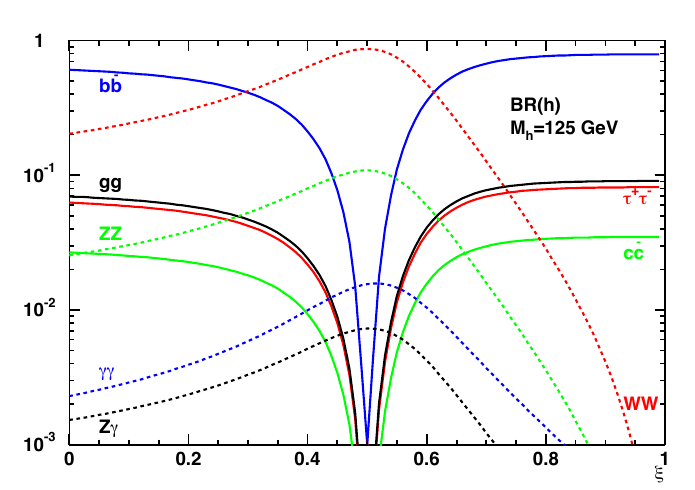}
\caption{\label{fig:125bran}
 Higgs boson branching ratios in MCHM5 as a function of $\xi$
  for $M_h=125$~GeV.}
\end{center}
\end{figure}

The Higgs anomalous couplings affect both the Higgs production and
decay processes. The Higgs boson bran\-ching ratios of a 125~GeV Higgs
boson are shown in Fig.~\ref{fig:125bran} for MCHM5. For $\xi=0.5$ the
Higgs boson becomes fermiophobic and the branching ratios into fer\-mions and
gluons vanish while the ones into gauge bo\-sons become enhanced. As
explained above, in MCHM4 the branching ratios are unchanged. The
modified production cross sections can easily be obtained from the
corresponding SM results by rescaling with the appropriate coupling
modification factors squared. As the QCD couplings are not affected  
the higher order QCD corrections can be taken over from the SM, while
the EW corrections would change and have to be omitted as they are
not available so far. 

The anomalous couplings can be tested by a
measurement of the Higgs interaction strengths. In case of a
universal coupling modification as, {\it e.g.}, in MCHM4 the production
rates and the total width have to be tested. At an $e^+e^-$ linear
collider an accuracy of a few percent can be achieved in the measurement
of the SM Higgs couplings to gauge bosons and fermions~\cite{Djouadi:2007ik}. For an investigation of the prospects for
the determination of $\xi$ at the LHC, see Ref.~\cite{Bock:2010nz}.
In Ref.~\cite{Barger:2003rs} a study of Higgs
couplings performed in the context of genuine dimension-six
operators showed that a sensitivity of up to $4\pi f \sim 40$~TeV can
be reached for a 120~GeV Higgs boson already at 500~GeV with
1~ab$^{-1}$ integrated luminosity. At the high-energy phase of the CLIC
project, i.e., at 3~TeV with 2~ab$^{-1}$ integrated luminosity, the
compositeness scale of the Higgs boson will be probed up to 60 - 90~TeV~\cite{Contino:2013gna}.
Also the total width of a 125~GeV
Higgs boson can be measured at a few percent precisely already at the low-energy phase of the  ILC programme.

\subsubsection{Strong processes}

\begin{sloppypar}
If no new
particles are discovered at the LHC, deviations from the SM predictions
for production and decay rates can point towards models with strong
dynamics. It is, however, only the characteristic signals of a composite
Higgs boson in the high energy region which unambiguously imply the
existence of new strong 
interactions. Since in the composite Higgs scenario the $V_L V_L$
scattering amplitude is not fully unitarized the related interaction
necessarily becomes strong and eventually fails tree-level unitarity at
the cutoff scale. The $VV$ scattering therefore becomes strong at high
energies. As the transversely polarized vector boson scattering is
numerically large in the SM, the test of the energy growth in
longitudinal gauge boson scattering is difficult at the LHC~\cite{Contino:2010mh}. 
Another probe of the strong dynamics at the
origin of EWSB is 
provided by longitudinal vector boson fusion in Higgs pairs which also
grows with the energy. For the test of strong double Higgs production
the high luminosity upgrade of the LHC would be needed, however~\cite{Contino:2010mh}. 
Besides testing the high-energy behaviour in
strong double Higgs production, new resocances unitarizing the
scattering amplitudes can be searched for. The ILC has been shown to
be able to test anomalous strong gauge couplings up to a scale $\sim
3$~TeV and exclude $\rho$-like resonances below 2.5~TeV~\cite{Djouadi:2007ik}.  
\end{sloppypar}

\subsubsection{Non-linear Higgs couplings}

\begin{sloppypar}
Vertices involving more than one Higgs boson could also provide a way to test the composite nature of the Higgs.
Double Higgs production  is a process that depends on the Higgs self-coupling and on the coupling between two Higgs bosons and two massive gauge bosons.
At a low energy $e^+e^-$ collider, double Higgs production proceeds mainly via  double
Higgs-strahlung off $Z$ bosons, $e^+e^- \to ZHH$,  
\begin{figure}[h]
\begin{center}
\includegraphics[width=0.45\textwidth]{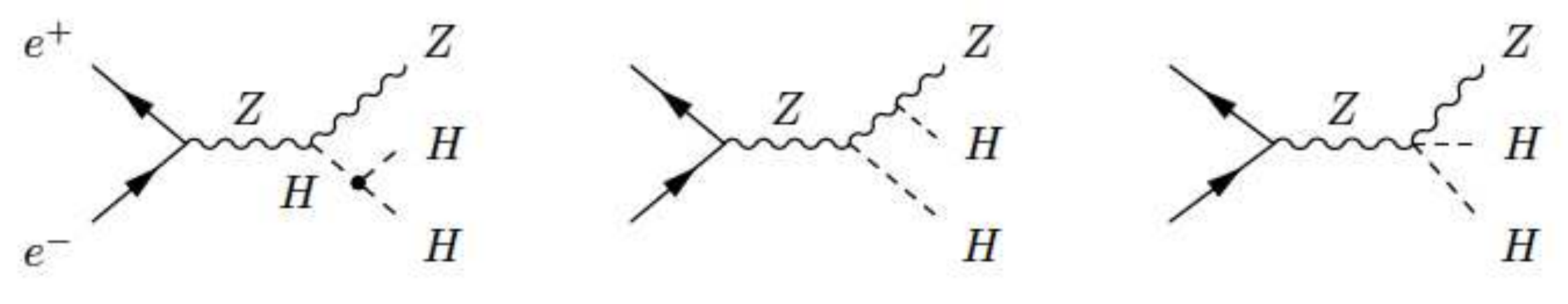}
\caption{\label{fig:hrad}Generic Feynman diagrams contributing to Higgs pair
  production via Higgs-strahlung off $Z$ bosons. }
\end{center}
\end{figure}
and $WW$ boson fusion
to Higgs pairs, $e^+e^- \to HH \nu\bar{\nu}$~\cite{Djouadi:1999gv}. Generic diagrams are
shown in Fig.~\ref{fig:hrad} for double Higgs-strahlung and
Fig.~\ref{fig:wwfus} for $WW$ boson fusion.
\end{sloppypar}
\begin{figure}[hb]
\begin{center}
\includegraphics[width=0.45\textwidth]{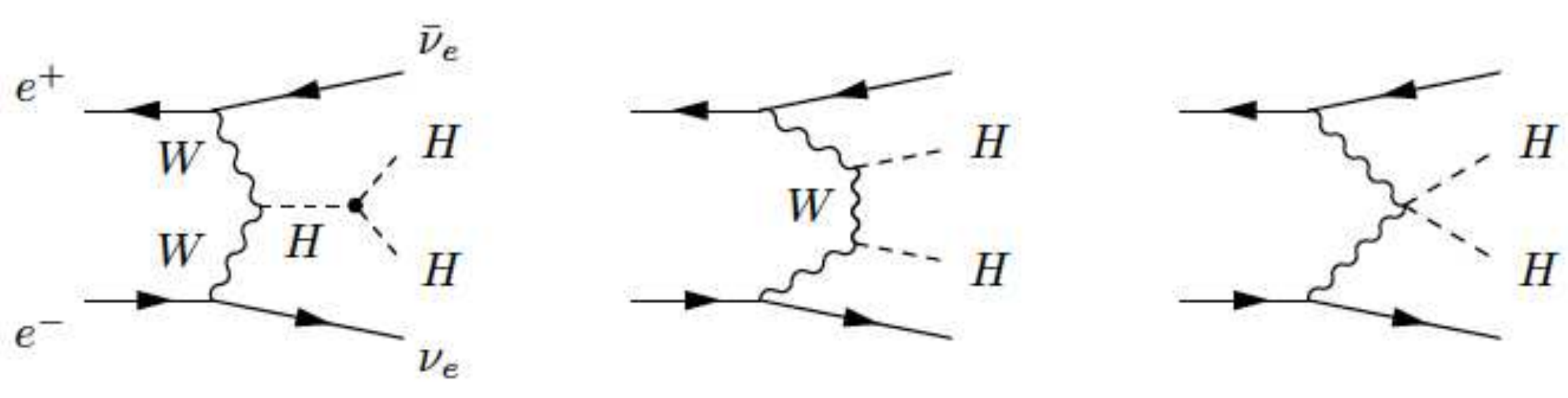}
\caption{\label{fig:wwfus}Generic Feynman diagrams contributing to Higgs pair
  production via $W$ boson fusion. }
\end{center}
\end{figure}
\begin{figure}[ht]
\begin{center}
\includegraphics[width=0.35\textwidth]{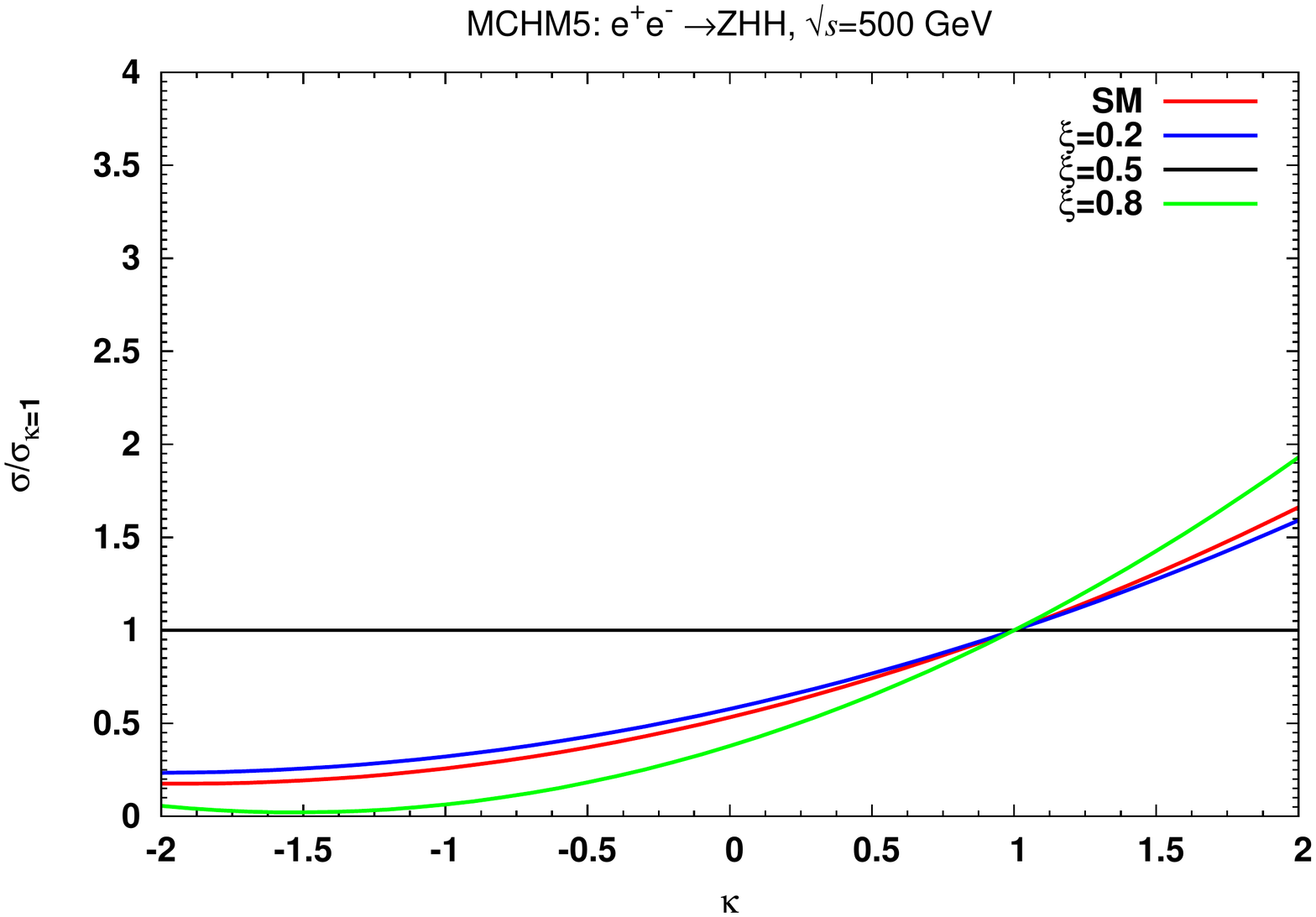}
\\
\includegraphics[width=0.35\textwidth]{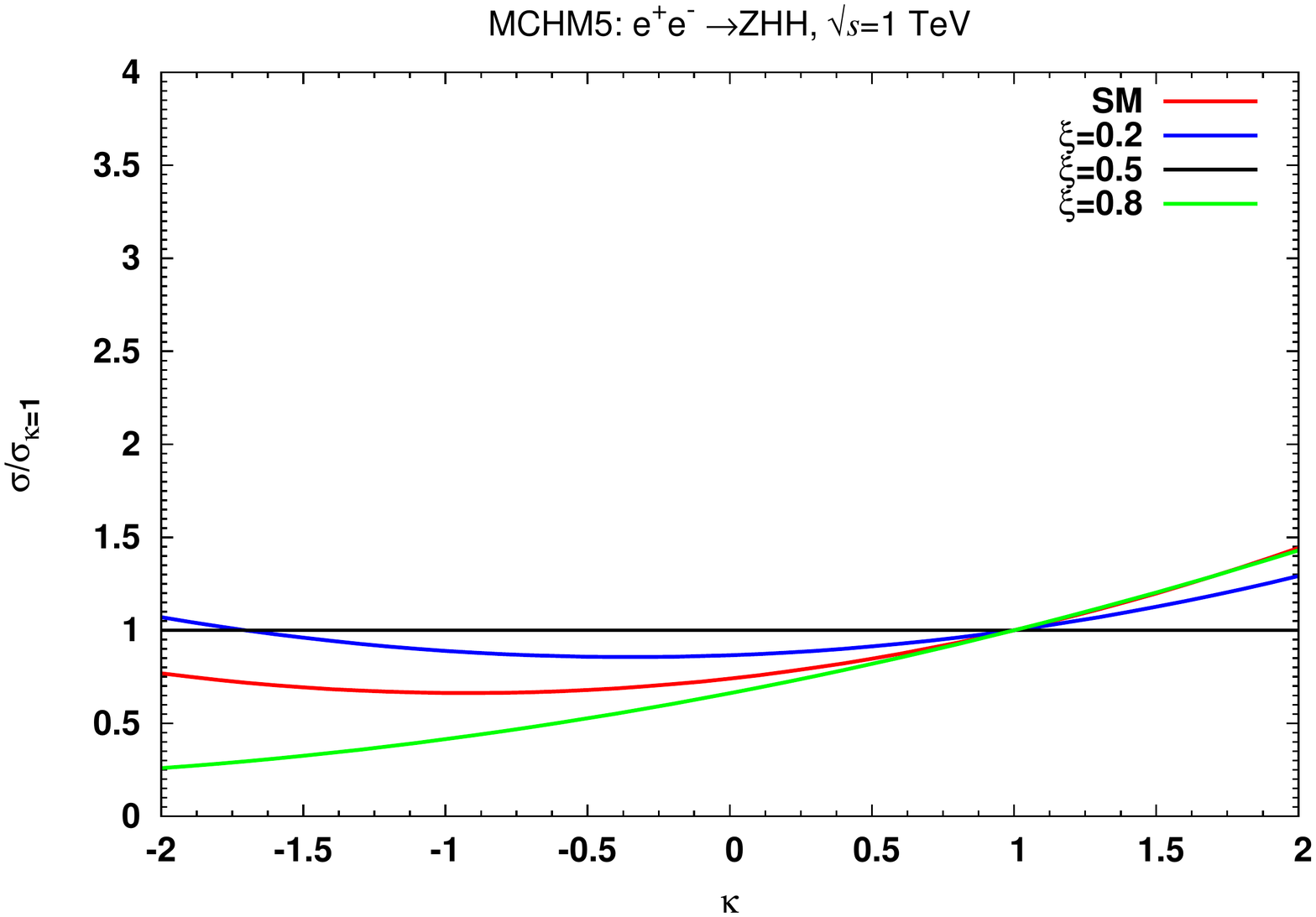}
\\
\includegraphics[width=0.35\textwidth]{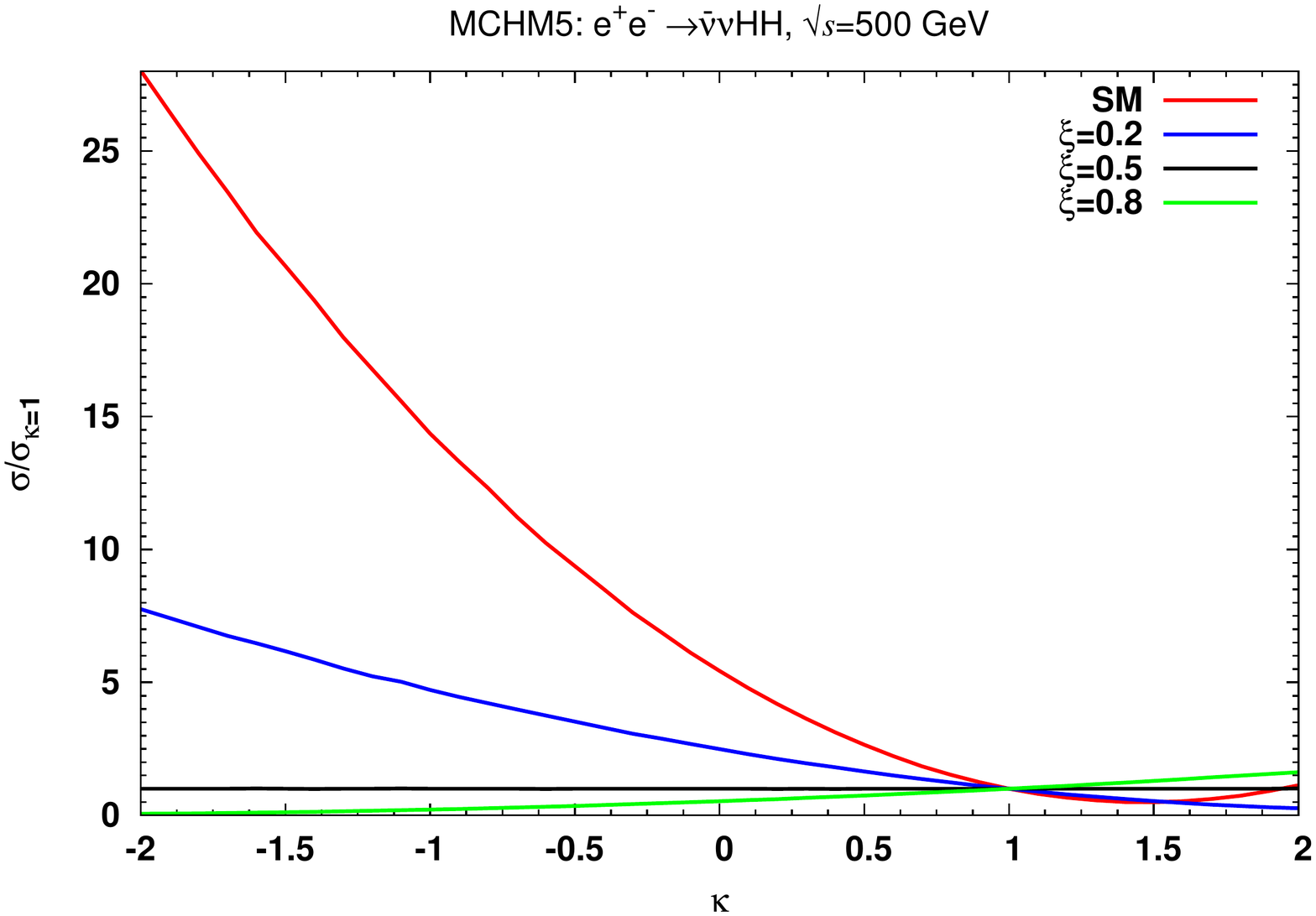}
\\
\includegraphics[width=0.35\textwidth]{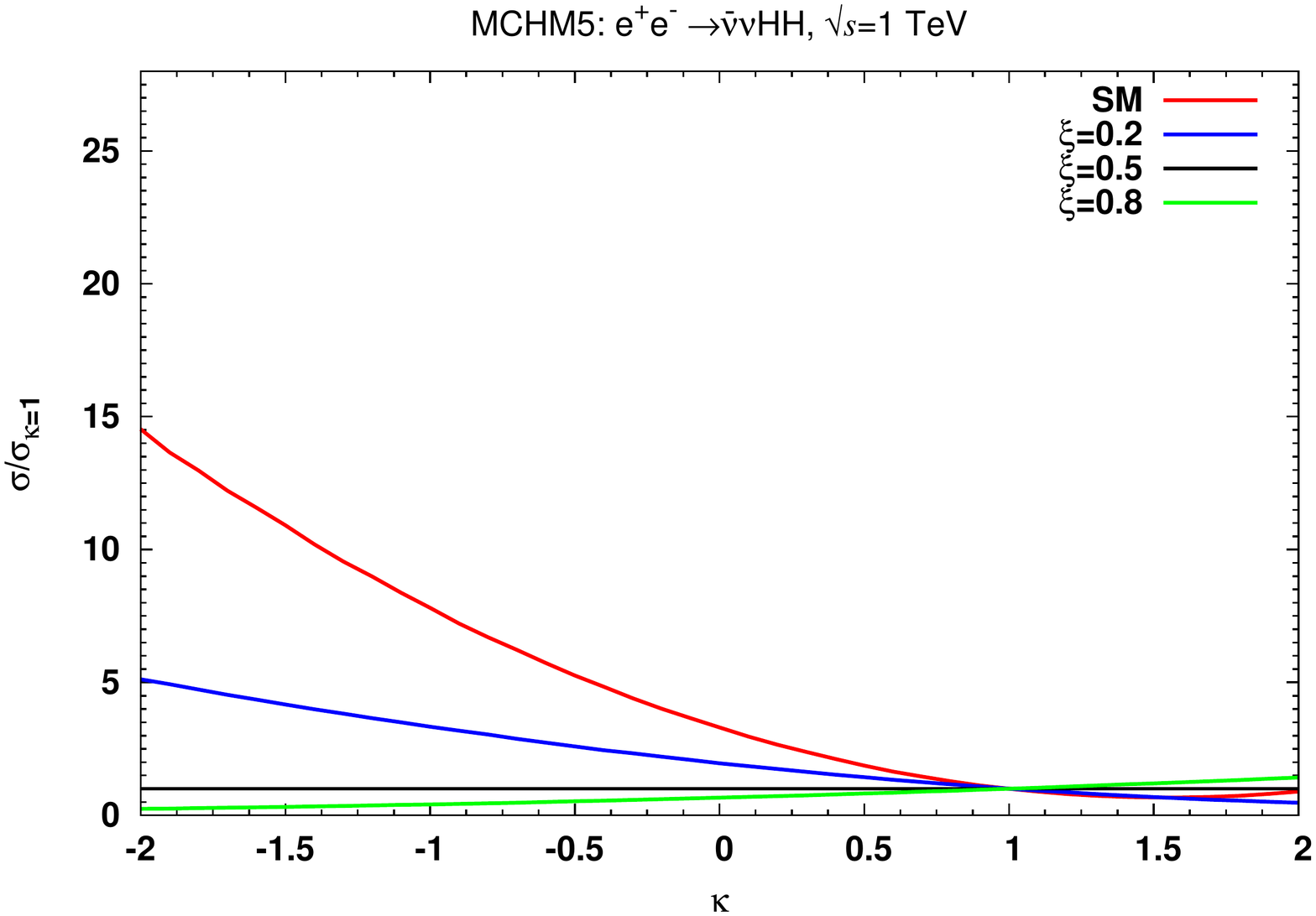}
\caption{\label{fig:ilchh} The $ZHH$ (upper two) and $WW$ fusion
  (lower two)
  cross sections in the SM (red) and the MCHM5 for $\xi=0.2$
  (blue), $\xi=0.5$ (black) and $\xi=0.8$ (green) divided by the cross
  section of the corresponding model at $\kappa$=1, as a function of
  $\kappa$,  for $\sqrt{s}=500$~GeV and $\sqrt{s}=1$
  TeV.}
\end{center}
\end{figure}
\begin{sloppypar}
The double Higgs-strahlung process dominates at low energies, and in
the MCHM4 and MCHM5 it is always smaller than in the SM, which is due
to the suppressed Higgs-gauge couplings. On the other hand, the $WW$
fusion process, which becomes important for higher c.m. energies, is 
enhanced compared to the SM for non-vanishing values of $\xi$~\cite{rgrober1,rgrober2}. 
This is due to interference effects related to the
anomalous Higgs couplings. Furthermore, 
the amplitude growths like the  c.m.~energy
squared contrary to the SM where it remains constant. The sensitivity of double
Higgs-strahlung and gauge boson fusion processes to the trilinear Higgs
self-coupling of the corresponding model can be studied by varying the
Higgs trilinear coupling in terms of the respective
self-interaction of the model in consideration, hence $\lambda_{HHH}
(\kappa)= \kappa \, \lambda_{HHH}^{\mbox{\scriptsize{MCHM4,5}}}$. 
This
gives an estimate of how accurately the Higgs pair production process
has to be measured in order to extract $\lambda_{HHH}$ within in the
investigated model with a certain precision. Note, however, that this
does not represent a test of models beyond the actually investigated theory.
Figure~\ref{fig:ilchh} shows for the SM and for the MCHM5 with three
representative $\xi$ values ($\xi=0.2,0.5,0.8$) 
the normalized double
Higgs production cross sections for Higgs-strahlung and gauge boson fusion,
respectively, at two c.m.~energies, $\sqrt{s}=500$~GeV and
1~TeV, as a function of the modification factor $\kappa$. The cross
sections are normalized with respect to the double Higgs production
cross sections at $\kappa=1$ of the respective model. As can be
inferred from the figure, both Higgs-strahlung and double Higgs
production are more sensitive to $\lambda_{HHH}$ at lower
c.m. energies. This is due to the suppression of the propagator in the
diagrams which contain the trilinear Higgs self-coupling with
higher energies. In addition in $WW$ fusion the $t$- and $u$-channel
diagrams, insensitive to this coupling, become more important with
rising energy. A high energy $e^+e^-$ collider can exploit the $WW$ fusion process to study
the deviations in the coupling between two Higgs bosons and two gauge  bosons by looking at the large $m_{HH}$ invariant mass distribution~\cite{Contino:2013gna}. The sensivity obtained on $\xi$ via this process is almost an order of magnitude better than the one obtained from the study of double Higgs-strahlung~\cite{Contino:2013gna}.
\end{sloppypar}

The parton level analysis in Refs.~\cite{rgrober1,rgrober2} showed that both
double Higgs-strahlung and $WW$ fusion have, in the $4b$ final state
from the decay of the two 125~GeV Higgs bosons, sensitivity to a
non-vanishing $\lambda_{HHH}$ at the 5$\sigma$ level in almost the
whole $\xi$ range, with the exception of $\xi=0.5$ in MCHM5, where the
trilinear Higgs coupling vanishes, {\it cf.}~Table~\ref{tab:coupl}. 

\subsubsection{Top sector}

The fermionic sector of composite Higgs models, in particular the top
sector, also shows an interesting phenomenology. With the fermion coupling
strengths being proportional to their masses the top quark has the
strongest coupling to the new sector and is most sensitive to new
physics. It is hence natural to consider one of the two top helicities
to be partially composite. The top quark mass then arises through
linear couplings to the strong sector. 
ATLAS and CMS already constrained the top partners to be heavier than
600 - 700~GeV at  95\% confidence level~\cite{CMS:2013tda}.
 The associated new heavy top
quark resonances have been shown to influence double Higgs production
through gluon fusion~\cite{Gillioz:2012se,Dawson:2012mk}. At $e^+e^-$
colliders these new resonances can 
be searched for either in single or in pair production~\cite{Agashe:2013hma}.

\subsubsection{Summary}

\begin{figure}[ht]
\begin{center}
\includegraphics[width=0.4\textwidth,angle=0]{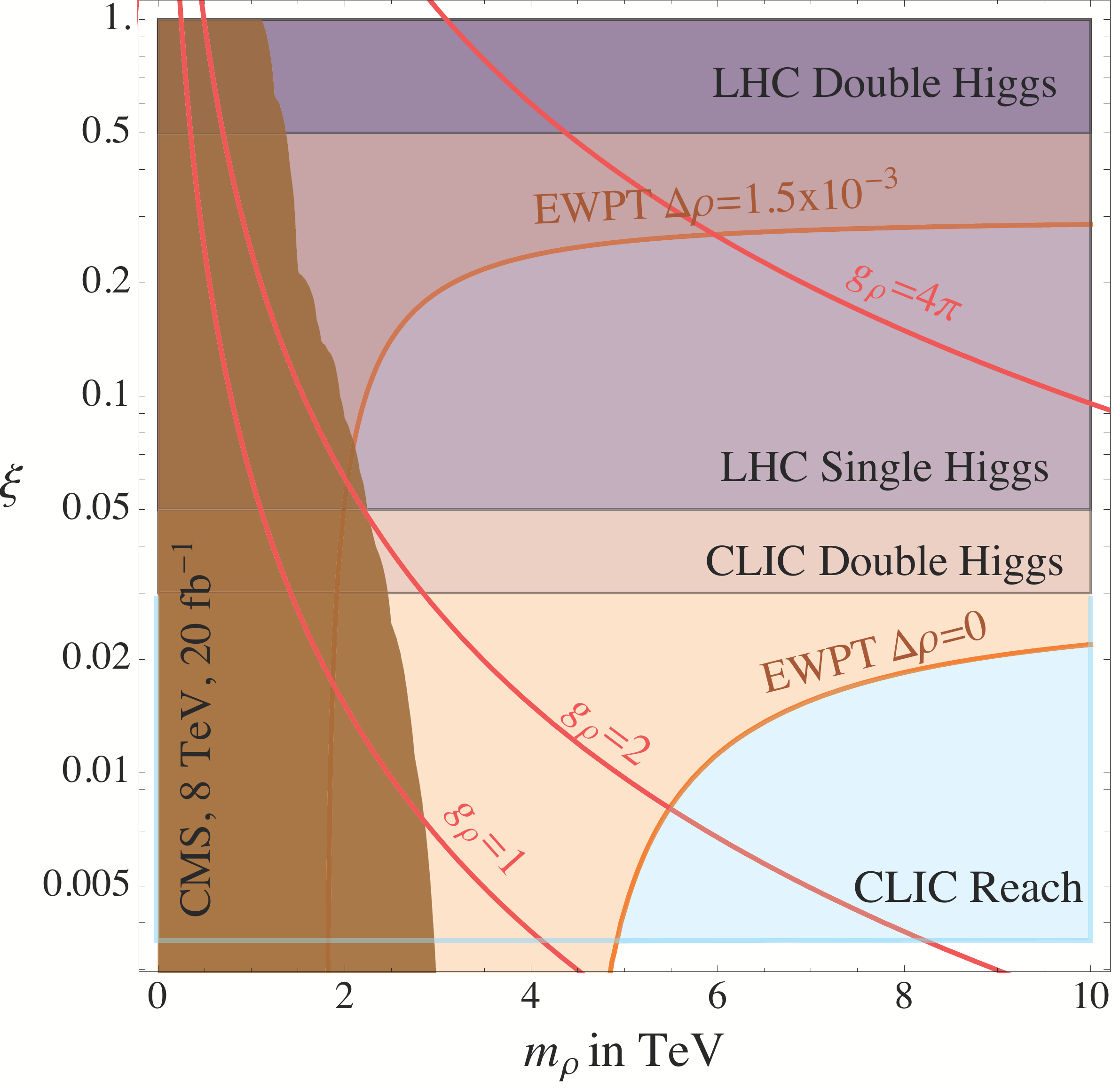}
\caption{\label{fig:xi-mrho} Summary plot of the current constraints
  and prospects for direct and indirect probes of Higgs compositeness.
  The dark brown region shows the current LHC limit from direct search
  for vector resonance. The dark (medium light) horizontal purple
  bands indicate the sensitivity on $\xi$ expected at the LHC from
  double (single) Higgs production with 300~fb$^{-1}$ of integrated
  luminosity. The pink horizontal band reports the sensitivity reach
  on $\xi$ from the study of double Higgs processes alone at CLIC with
  1 ab$^{-1}$ of integrated luminosity at 3 TeV while the light blue
  horizontal band shows the sensitivity reach on $\xi$ when
  considering single Higgs processes. Finally, experimental
  electroweak precision tests (EWPT) favor the region below the orange
  thick line with and without additional contributions to $\Delta
  \rho$. From Ref.~\cite{Contino:2013gna}.}
\end{center}
\end{figure}

Composite Higgs models offer a nice possibility to
solve the hierarchy problem by introducing a Higgs boson which emerges
as pseudo Nambu--Goldstone boson from a strongly interacting
sector. The phenomenology of these models is characterized by a light
Higgs resonance which is separated through a mass gap from the other
resonances of the strong sector, and which has modified couplings to
the SM fermions and gauge bosons. At an $e^+e^-$ collider these
couplings can be tested at high accuracy, and  interactions with more than one Higgs boson, 
among which the Higgs self-interactions, will alse be accessible. Genuine
probes of the strong sector are provided by strong double Higgs
production through gauge boson fusion and longitudinal gauge boson
scattering, which both rise with the energy. A high-energy $e^+e^-$ collider like CLIC 
can also become sensitive to the  tails of the spin-1 resonance contributions to
the $WW\to WW$ and $WW\to HH$ amplitudes.
Assuming partial compositeness
in the top sector, new top resonances arise which can be also searched for
at a future linear collider above the current LHC bound around 700~GeV.
Figure~\ref{fig:xi-mrho} summarizes  the sensitivities at the LHC and CLIC for observing non-SM signatures from the composite nature of the Higgs boson in the plane of $\xi$ and $m_\rho$, the typical mass scale of the strong sector resonances.

%
%


\subsection{The Higgs portal\protect\footnotemark}
\footnotetext{Christoph Englert}
\label{sec:ewsb8}



A large fraction of matter in the Universe is dark and not
incorporated in the Standard Model (SM). Nevertheless, this new kind of 
invisible matter is expected to interact with the SM fields,
naturally by gravitational interaction. However, another path
could be opened by a Higgs portal which connects the SM Higgs
field with potential Higgs fields in the dark sector, respecting
all symmetry principles and well founded theoretical SM concepts 
like renormlizability.

Even though the particles of the novel sector are invisible, the
portal nevertheless induces observable signals in the SM,
in the Higgs sector in particular. Mixings among Higgs bosons of the
SM and of the dark sector modify Higgs couplings to the SM particles
and give rise to invisible Higgs decays (beyond the cascades to
neutrinos).

Crucial to an extraction of the $m_H\simeq 125~{\text{GeV}}$ Higgs boson
candidate's couplings to known matter is a good understanding of Higgs
production $p$ and decay mechanisms $d$, which can be constrained by
measuring
\begin{equation}
  \label{decaywidth}
  \sigma_p \times \text{BR}_d \sim {\Gamma_p\,\Gamma_q\over
    \Gamma_{\text{tot}}}\sim {g_p^2\,g_d^2\bigg/ \bigg(
    \sum_{\text{modes}} g_i^2 \bigg)} \,,
\end{equation}
where $\sigma_p$, BR$_d$, and $g_i$ denote the involved production
cross sections, branching ratios and couplings, as usual. Precisely
reconstructing these underlying parameters is systematically hindered
by the unavailable measurement of the total decay width
$\Gamma_{\text{tot}}$. As a matter of fact, un-adapted search
strategies at LHC miss certain non-SM decay modes, which naturally arise in
models beyond the SM
\cite{Falkowski:2010hi,Chen:2010wk,Englert:2011iz,Englert:2012wf,Englert:2011us}
and which would then manifest as an invisible branching ratio
\cite{Shrock:1982kd} in global fits. The expected constraint on such
an invisible Higgs boson decay at the LHC is
${\text{BR}}(H\to\text{invisible})\simeq 10\%$ \cite{Englert:2011aa},
a bound too loose to efficiently constrain physics beyond the SM,
especially models where the Higgs field provides a portal to a
hidden sector~\cite{Binoth:1996au,Patt:2006fw,Schabinger:2005ei},
which can provide a viable dark matter candidate
\cite{Kanemura:2010sh}.

At a linear collider (LC) it is straightforward to derive the total width of the Higgs boson
by combining the model-independent measurement of the partial width 
$\Gamma(ZZ^\ast)$ in semi-inclusive Higgs-strahlung with the measurement 
of the branching ratio ${\text{BR}}(ZZ^\ast)$:
\begin{equation}
\Gamma_{\text{tot}}(H) = \Gamma(ZZ^\ast)/{\text{BR}}(ZZ^\ast) \,.
\end{equation}
Subsequently ${\text{BR}}(H\to \text{invisible})$ can be determined in
a model-independent way \cite{Schumacher:2003ss}.

\begin{figure}[!t]
  \begin{center}
    \includegraphics[width=0.35\textwidth]{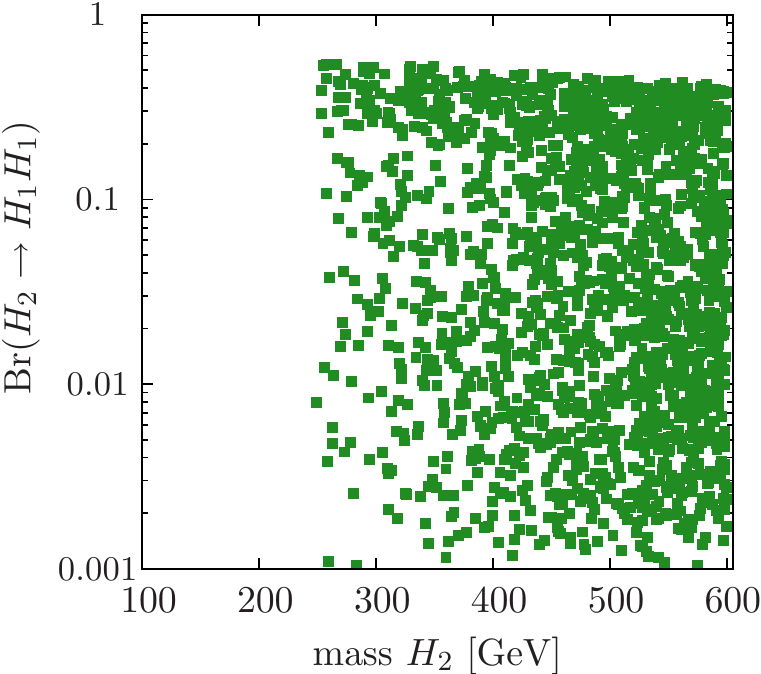}
    \caption{\label{cascade} Scan over the Higgs portal potential
      Eq.~\ref{eq:potential}. We include the constraints from
      electroweak precision measurements.}
  \end{center}
\end{figure}

From Eq.~\eqref{decaywidth}, we need to interpret the strong Higgs
exclusion for heavy Higgs masses as a sign of a highly suppressed
production cross section for heavier Higgs-like resonances. That heavy
Higgs copies need to be weakly coupled in simple model-building
realizations is already known from the investigation of electroweak
precision measurements performed during the LEP era. This complements
the requirement to include unitarizing degrees of freedom for
longitudinal gauge boson scattering $V_LV_L\to V_L V_L$ $(V=W^\pm,
Z)$, and, to less constraint extent, massive quark annihilation to
longitudinal gauge bosons $q\bar q\to V_LV_L$. Saturating all three of
these requirements fixes key characteristics of the phenomenological
realization of the Higgs mechanism, and does not allow dramatic
modifications of the couplings $\{g_i\}$ in Eq.~\eqref{decaywidth} away
from the SM expectation of a light Higgs --- the common predicament of
electroweak scale model building. In this sense gaining additional
sensitivity to invisible Higgs decays (or the Higgs total width in
general) beyond the limitations of the LHC's hadronic environment is
crucial to the understanding of electroweak physics at the desired
level, before the picture will be clarified to the maximum extent
possible at a LC.

\begin{figure*}[!t]
  \begin{center}
    \includegraphics[width=0.49\textwidth]{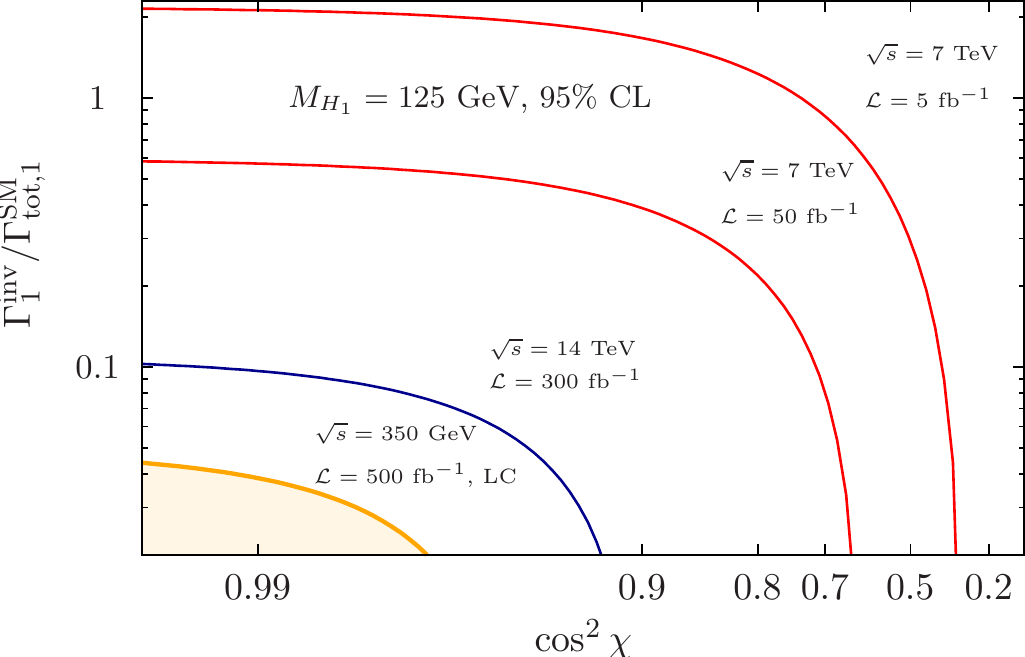}
    \caption{\label{fig5} 95\% confidence level contours for a
      measurement of $\Gamma_1^{\text{hid}}/\Gamma_1^{\text{SM}}$ at the
      LHC and a $350~{\text{GeV}}$ LC. We use {\sc{Sfitter}}
      \cite{Lafaye:2009vr} for the LHC results and we adopt the linear
      collider uncertainties of reference~\cite{Schumacher:2003ss}.}
  \end{center}
\end{figure*}

\begin{sloppypar}
The aforementioned Higgs portal model
\cite{Binoth:1996au,Patt:2006fw,Schabinger:2005ei} provides a
theoretically well-defined, renormalizable, and yet minimal framework
to explore both effects in a consistent way \cite{Englert:2011yb}: the
influence of $\Gamma_{\text{inv}}$ on the Higgs phenomenology is
captured while heavier Higgs boson-like particles with suppressed
couplings are naturally incorporated. Therefore, the Higgs portal
model not only provides a well-motivated SM Higgs sector extension in
the context of dark matter searches\footnote{In fact, there are only
  two other possibilities to couple the SM to a hidden sector: U(1)
  mixing \cite{Okun:1982xi,Holdom:1985ag} and mixing with a
  right-handed sterile neutrino
  \cite{Dodelson:1993je,MarchRussell:2008yu}. The Higgs portal model
  is least constrained amongst these possibilities.} and current data,
but it represents an ideal model to generalize the SM in its
phenomenologically unknown parameters to facilitate the SM's
validation by constraining the additional portal parameters beyond
introducing biases ({\it{e.g.}}  $\Gamma_H^{\text{tot}}\equiv
\Gamma_H^{\text{SM}}$).
\end{sloppypar}

In its simplest form, leading to both a modified electroweak
phenomenology and an invisible Higgs decay channel, the Higgs
portal is given by the potential
\begin{multline}
  \label{eq:potential}
  \mathcal{V} =
  \mu^2_s |\phi_s|^2 + \lambda_s |\phi_s|^4 
  \; + \;
  \mu^2_h |\phi_h|^2 + \lambda_h |\phi_h|^4\\
  \; + \;
  \eta_\chi |\phi_s|^2 |\phi_h|^2 \,,
\end{multline}
where $\phi_{s,h}$ are the SM and the hidden Higgs doublet fields,
respectively, {\it i.e.} the Higgs sector is mirrored
\cite{Barbieri:2005ri}. The visible sector communicates to the hidden
world via the additional operator $\eta_\chi |\phi_s|^2 |\phi_h|^2$,
which exploits the fact that both $|\phi_s|^2$ and $|\phi_h|^2$ are
singlet operators under both the SM and the invisible gauge groups.

After symmetry breaking which is triggered by the Higgs fields
acquiring vacuum expectation values $|\phi_{s,h}|=v_{s,h}/\sqrt{2}$,
the would-be-Nambu Goldstone bosons are eaten by the $W^\pm$, $Z$ fields,
and correspondingly in the hidden sector. The only
effect (formulated here in unitary gauge) is a two-dimensional isometry 
which mixes the visible and the hidden Higgs bosons $H_{s,h}$:
\begin{equation}
  \begin{split}
    H_1 =&\phantom{-} \cos\chi \, H_s + \sin\chi \, H_h\,,     \\
    H_2 =& -\sin\chi \, H_s + \cos\chi \, H_h  \,,
    \label{eq:mixi}
  \end{split}
\end{equation}
with the mixing angle 
\begin{equation}
  \tan \, 2\chi = \frac{\eta_\chi v_s v_h}{\lambda_s v^2_s -
    \lambda_h v^2_h} \,. 
\end{equation}
The masses of the two Higgs fields are given by
\begin{multline}
  \label{eq:masses}
  M^2_{1,2} = [\lambda_s v^2_s + \lambda_h v^2_h] \\ \hspace{-1cm} \mp | \lambda_s
  v^2_s - \lambda_h v^2_h | \,
  \sqrt{1+\tan^2 2\chi}     \,.
\end{multline}
We assume $M_1\simeq 125~{\text{GeV}}$ in the following. The inverse
phenomenological situation $M_1<M_2\simeq 125~{\text{GeV}}$, {\it i.e.} a
Higgs field hiding below the upper LEP2 bound, is obviously reconciled
by $\chi\to \pi-\chi$ since the potential has a ${\mathbb{Z}}_2$
symmetry. Consistency with electroweak precision data and an
efficient unitarization of the $V_LV_L$ scattering amplitudes relies
in this case 
on $\cos^2\chi$ being close to unity.

\begin{figure*}[!t]
  \begin{center}
    \includegraphics[width=0.49\textwidth]{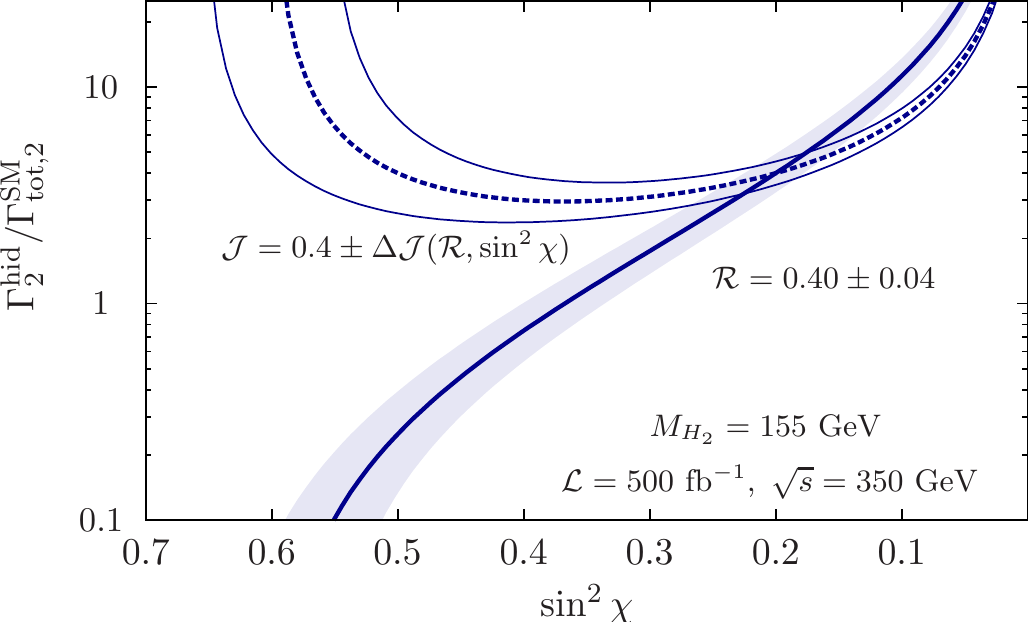}
    \caption{\label{fig:sad} Measurement of a hypothetical portal
      model at a 350 GeV linear collider, uncertainties are adopted
      from Ref.~\cite{Schumacher:2003ss}. A measurement of
      ${\cal{R}}_2$ at the LHC, with only an upper 95\% confidence
      level bound on ${\cal{J}}_2$ does not constrain the region
      $\Gamma_2^{\text{hid}}/\Gamma_{{\text{tot}},2}^{\text{SM}}$
      below the ${\cal{J}}_2$ curve. This degeneracy is lifted with a
      measurement at a linear collider.}
  \end{center}
\end{figure*}

As a consequence of the mixing we have universally suppressed cross
sections of the SM-Higgs
\begin{subequations}
  \label{eq:1}
  \begin{equation}
    \begin{split}
      \sigma_{1}  = \cos^2\chi \, \sigma^\text{SM}_1 \,\phantom{,} \\
      \sigma_{2}  = \sin^2\chi \, \sigma^\text{SM}_2 \,,
    \end{split}
  \end{equation}
  and 
  \begin{equation}
    \begin{split}
      \label{eq:2}
      \Gamma^\text{\rm vis}_{1,2} = \cos^2\chi \, \{\sin^2\chi\} \,
    \Gamma^\text{SM}_{1,2}
    + \Delta^{\rm vis}_2 \, \Gamma^\text{HH}_2 \,\phantom{,}     \\
    \Gamma^\text{\rm inv}_{1,2} = \sin^2\chi \, \{\cos^2\chi\} \,
    \Gamma^\text{hid}_{1,2} + \Delta^{\rm inv}_2 \, \Gamma^\text{HH}_2\,,
      \end{split}
    \end{equation}
\end{subequations}
where $\Delta^{vis\{inv\}}_2 = \zeta^2 \, \{[1-\zeta]^2\} \neq 0$ and 
$\zeta = 1/[1+\tan^2\chi $\, $\Gamma_1^\text{hid}/
\Gamma^\text{SM}_{\text{tot},1}]$. We understand the index in
$\Delta_2$ such that this contribution only arises for the heavier
state labelled with index=2.

We have also included cascade decays $\Gamma^\text{HH}_2$ (if they are
kinematically allowed for $M_2\geq 2M_1$) and the possibility for a
hidden partial decay width in Eq.~\eqref{eq:2}. The latter naturally
arise if the hidden sector has matter content with $2m\leq m_{H_1}$,
{\it{i.e.}} in models with light dark matter candidates. Weak coupling
of the heavier Higgs-like state is made explicit when correlating the
Higgs portal model with electroweak precision constraints
\cite{Englert:2011yb}.

Generically, the branching ratio of the heavier Higgs boson to two
light Higgs states is small (Fig.~\ref{cascade}) and kinematically
suppressed, so that a direct measurement of the cascade decay at the
LHC is challenging. Measurement strategies targeting invisible
Higgs boson decays at the LHC \cite{DeRoeck:2009id} are based on
measurements in weak boson fusion \cite{Eboli:2000ze} and associated
production \cite{Godbole:2003it,Davoudiasl:2004aj}. Recent re-analysis
of the monojet+Higgs production \cite{Englert:2011us,Djouadi:2012zc},
however, suggest that additional sensitivity can be gained in these
channels, at least for the 7 and 8 TeV data samples.

The production of multiple final state Higgs particles is another
strong test of this model, since it predicts resonant contributions
which can be large, see Fig.~\ref{cascade}. A measurement of the involved
trilinear coupling $H_2H_1H_1$ is challenging at the LHC
\cite{Baur:2002qd,Dolan:2012rv} and can be achieved more
straightforwardly at a high luminosity LC
\cite{Djouadi:1999gv}. Especially because we can separate the
different final states of the light Higgs decay at the latter
experiment, we can use the prediction of the various trilinear
couplings that arise from Eq.~\eqref{eq:potential} to reconstruct the
potential.

The precision to which invisible decays can be studied at the LHC is
ultimately limited by the machine's systematics which will saturate at
luminosities ${\cal{L}}\simeq 300~{\text{fb}}^{-1}$,
see Fig.~\ref{fig5}. Bounds on visible decays are typically expressed as
ratios to the SM expectation, which, for the lighter $M_1\simeq
125~{\text{GeV}}$ state, can be rephrased in the portal model for
either $i = pp$ or $e^+e^-$ initial stares
\begin{equation}
  \label{eq:excl}
  \dfrac{\sigma[i \to H_1 \to F]}{\sigma[i \to H_1 \to F]^\text{SM}} 
  = \dfrac{\cos^2\chi}{1 + \tan^2\chi \, 
    [{\Gamma^\text{hid}_1}/{\Gamma^\text{SM}_{\text{tot},1}}]}
  \leq \mathcal{R}_1 \,,
\end{equation}
\begin{sloppypar}
where ${\cal{R}}_1$ denotes the observed exclusion limit (signal
strength). An identical quantity can be derived from future
constraints on invisible decays
\end{sloppypar}
\begin{equation}
  \label{eq:inv}
  \dfrac{\sigma[i \to H_1 \to inv]}{\sigma[i \to H_1]^\text{SM}}
  = \dfrac{\sin^2\chi \, [\Gamma^\text{hid}_1 / \Gamma^\text{SM}_{\text{tot},1}]}
  {1 + \tan^2\chi \, [{\Gamma^\text{hid}_1}/{\Gamma^\text{SM}_{\text{tot},1}}]}
  \leq \mathcal{J}_1                                                            \,.
\end{equation}
Similar relations hold for $H_2$, and there are portal-specific sum
rules which facilitate the reconstruction of the mixing angle from
measurements of ${\cal{J}}_{1,2}$ and ${\cal{R}}_{1,2}$,
\begin{equation}
  \begin{split}
    \label{eq:sumrules}
    {\cal{R}}_1+{\cal{J}}_1=&\cos^2\chi\,,\\
    {\cal{R}}_2+{\cal{J}}_2=&\sin^2\chi\,.
  \end{split}
\end{equation}
While the LHC running at 14~TeV will eventually probe small visible
production cross sections ${\cal{R}}_{2}$ (Eq.~\eqref{eq:inv} becomes
an equality), the invisible decay searches at the LHC will most likely
yield a 95\% confidence level bound \cite{Read:2002hq} on
${\cal{J}}_{1,2}$ \cite{DeRoeck:2009id} rather than a statistically
significant observation. The bounds can be vastly improved by
performing by performing precision spectroscopy of the 125 GeV Higgs
candidate in the associated production channel $e^+e^-\to HZ$ at, {\it
  e.g.}, a 350~GeV LC (see also Ref.~\cite{wells}). Still,
invisible Higgs searches that solely provide upper limits on both
${\cal{J}}_{1,2}$ are not enough to fully reconstruct the portal model
if a second Higgs-like state is discovered as a result of
Eq.~\eqref{eq:sumrules}. Only the precise {\it{measurement}}, which is
impossible at the LHC, solves this predicament, but an LC is the
perfect instrument to pursue such an analysis in the associated
production channel.

In Fig.~\ref{fig:sad} we show a hypothetical situation, where $H_2$ is
discovered at the LHC with ${\cal{R}}_2=0.4$; the error is given by a
more precise measurement at a 350~GeV LC,
see Fig.~\ref{cascade}. The measurement of ${\cal{J}}_2=0.4$ allows to
reconstruct $\sin^2\chi$, which can be seeded to a reconstruction
algorithm \cite{Englert:2011yb} that yields the full Higgs portal
potential Eq.~\eqref{eq:potential}.

From Eq.~\eqref{eq:sumrules} we also obtain the sum rule
\begin{equation}
  \label{eq:sum}
  {\cal{R}}_1+{\cal{J}}_1+  {\cal{R}}_2+{\cal{J}}_2=1.
\end{equation}
which provides a strong additional test of the portal model
Eq.~\eqref{eq:sumrules} when a measurement of the invisible branching
ratios via ${\cal{J}}_{1,2}$ becomes available at a future linear
collider.

To summarize, the Higgs portal can open the path to the
dark sector of matter and can allow crucial observations on this
novel kind of matter in a global way. While first hints may be 
expected from LHC experiments, high-precision analyses of Higgs 
properties and the observation of invisible decays at LC can give 
rise to a first transparent picture of a new world of matter.


\subsection{The NMSSM\protect\footnotemark}
\footnotetext{Ulrich Ellwanger}
\label{sec:ewsb9}




\begin{sloppypar}
In the Next-to-Minimal Supersymmetric Standard Mo\-del (NMSSM) the Higgs
sector of the MSSM is extended by an additional gauge singlet superfield
$\widehat{S}$ \cite{Ellwanger:2009dp}. It is the simplest supersymmetric
extension of the Standard Model with a scale invariant superpotential;
the $\mu$-term $\mu \widehat{H_u} \widehat{H_d}$ in the superpotential
$W_{\mathrm{MSSM}}$ of the MSSM is replaced by
\end{sloppypar}
\begin{equation}\label{eq:nmssm1}
W_{\mathrm{NMSSM}} = \lambda \widehat{S} \widehat{H_u} \widehat{H_d}
+ \frac{\kappa}{3} \widehat{S}^3\; .
\end{equation}
Once the scalar component $S$ of the superfield $\widehat{S}$ assumes a
vacuum expectation value $s$, the first term in the superpotential
(\ref{eq:nmssm1}) generates an effective $\mu$-term with
\begin{equation}\label{eq:nmssm2}
\mu_{\mathrm{eff}} = \lambda s\; .
\end{equation}
In addition to the NMSSM-specific Yukawa couplings $\lambda$ and
$\kappa$, the parameter space of the NMSSM contains soft supersymmetry
breaking trilinear couplings $A_\lambda$, $A_\kappa$ and soft
supersymmetry breaking mass terms $m_{H_u}^2$, $m_{H_d}^2$ and $m_S^2$.
The order of $s$ and hence of $\mu_{\mathrm{eff}}$ is essentially
determined by $A_\kappa$ and $m_S^2$, hence $\mu_{\mathrm{eff}}$ is
automatically of the order of the soft supersymmetry breaking terms.

\begin{sloppypar}
The physical states in the Higgs sector of the NMSSM (assuming
CP-conservation) consist in three neutral CP-even states $H_i$ (ordered
in mass), two neutral CP-odd states $A_i$ and charged Higgs bosons
$H^\pm$. The CP-even states $H_i$ are mixtures of the real components of
the weak eigenstates $H_u$, $H_d$ and $S$:
\begin{equation}\label{eq:nmssm3}
H_i = S_{1,d}\ H_d + S_{1,u}\ H_u +S_{1,s}\ S\; ,
\end{equation}
where the mixing angles $S_{i,j}$ depend on the a priori unknown
parameters in the Higgs potential. Similarly, the two CP-odd states
$A_i$ are mixtures of the imaginary components of the weak eigenstates
$H_u$, $H_d$ and $S$ without the Goldstone boson. In addition, the
fermionic component of the superfield $\widehat{S}$ leads to a fifth
neutralino, which mixes with the four neutralinos of the MSSM.
\end{sloppypar}

\begin{sloppypar}
In view of the mass of 125-126~GeV of the at least approximately
Standard Model-like Higgs boson $H_{\mathrm{SM}}$ measured at the LHC,
the NMSSM has received considerable attention: In contrast to the MSSM,
no large radiative corrections to the Higgs mass (implying a fine-tuning
in parameter space) are required in order to obtain
$M_{H_{\mathrm{SM}}}$ well above $M_Z$, the upper bound on
$M_{H_{\mathrm{SM}}}$ at tree level in that model. In the NMSSM, additional tree level
contributions to $M_{H_{\mathrm{SM}}}$ originate from the superpotential
Eq.~(\ref{eq:nmssm1}) \cite{Ellwanger:2009dp}. Also a mixing with a
lighter mostly singlet-like Higgs boson can increase the mass of the
mostly Standard-Model-like Higgs boson \cite{Ellwanger:1999ji}, in which
case one has to identify $H_{\mathrm{SM}}$ with $H_2$. Both effects
allow to obtain $M_{H_{\mathrm{SM}}}\sim$~125-126~GeV without
fine-tuning and, moreover, such a mixing could easily explain an
enhanced branching fraction of 
this Higgs boson (from now on denoted as $H_{125}$) into $\gamma\gamma$
\cite{Ellwanger:2010nf,Hall:2011aa,Ellwanger:2011aa,King:2012is,Cao:2012fz,Vasqu
ez:2012hn,Ellwanger:2012ke,Benbrik:2012rm,Cao:2012yn,Gunion:2012gc,SchmidtHoberg
:2012yy}.
\end{sloppypar}

\begin{sloppypar}
Depending on the mixing angles, on the masses of the additional Higgs
bosons and on their branching fractions, the LHC can be blind to the
extended Higgs sector of the NMSSM beyond the mostly Standard Model-like
state. Then the detection of the additional states will be possible
only at a LC. Also if hints for such an extended Higgs sector are
observed at the LHC, only a LC will be able to study its properties in
more detail. Earlier studies of the detection of NMSSM Higgs bosons at
$e^+ e^-$ colliders can be found in
\cite{Kamoshita:1994iv,Ham:1996dc,Espinosa:1998xj,Gunion:2003fd,Ellwanger:2003jt
,Weiglein:2004hn}.
\end{sloppypar}

The dominant production modes of CP-even Higgs bosons at a LC (associate
$ZH$ production and VBF) depend on the Higgs couplings to the
electroweak gauge bosons. Denoting the coupling of $H_{\mathrm{SM}}$ to
electroweak gauge bosons by $g_\mathrm{SM}$, the couplings $g_i$ of the
CP-even states $H_i$ satisfy the sum rule
\begin{equation}\label{eq:nmssm4}
\sum_i g_i^2 = g_\mathrm{SM}^2\; .
\end{equation}

If a measurement of the coupling $g_i$ of the 125~GeV Higgs boson at the
LC gives a value significantly below $g_\mathrm{SM}$, one can deduce the
presence of additional Higgs states. The scenario where $H_{125}=H_2$ is particularly natural in the parameter space of the
NMSSM. Then the coupling $g_1$ of the lightest Higgs boson $H_1$ must satisfy
constraints from LEP~II, if its mass is below $\sim 114$~GeV.

\begin{figure}
  \includegraphics*[width=0.40\textwidth]{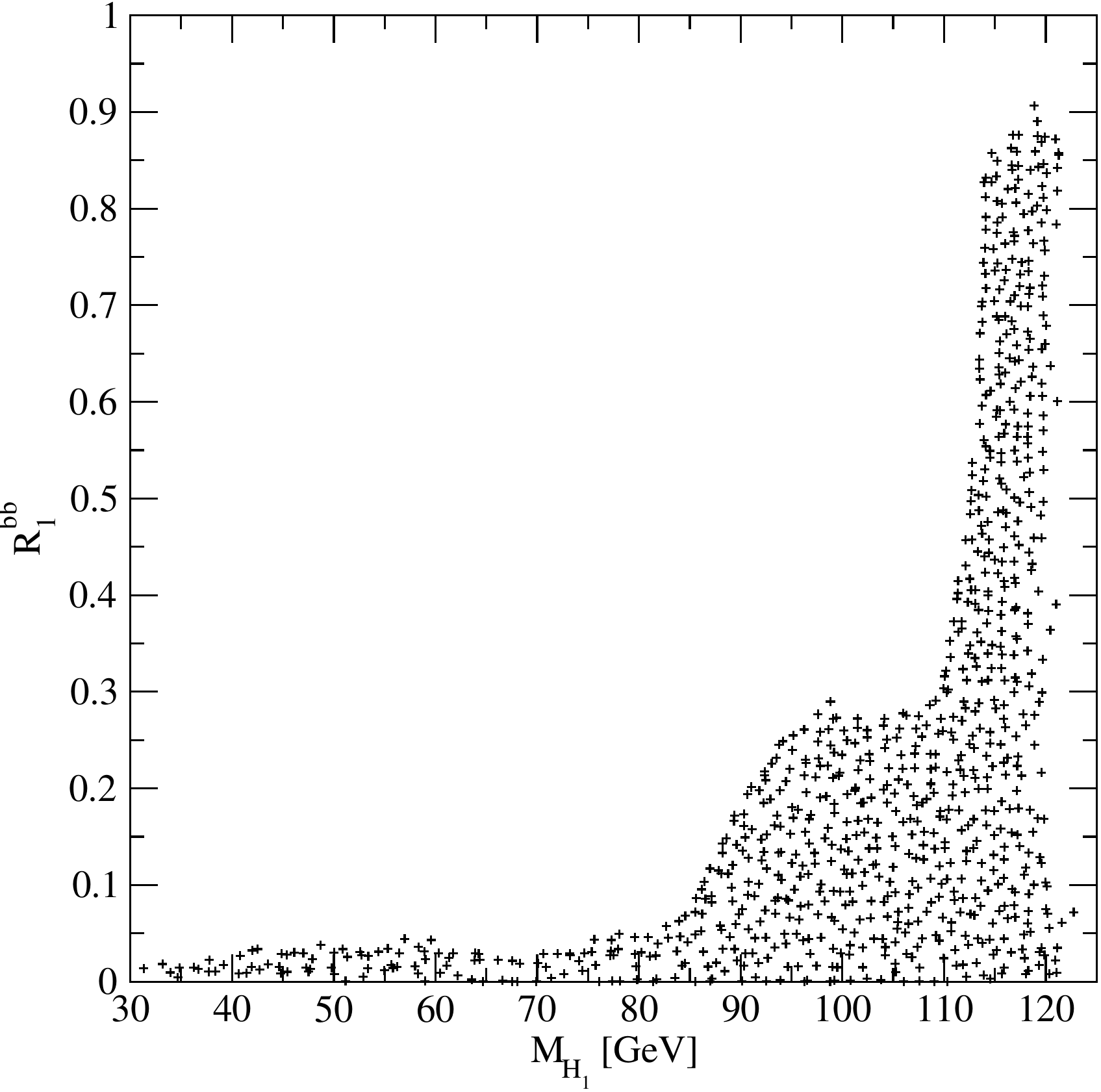}
\caption{The reduced signal cross section $R_1^{bb}$ at a $e^+ e^-$
collider as defined in the text, as function of $M_{H_1}$ in
the semi-constrained NMSSM (from \cite{Ellwanger:2012ke}).}
\label{fig:nmssm1}
\end{figure}

\begin{figure}
  \includegraphics*[width=0.40\textwidth]{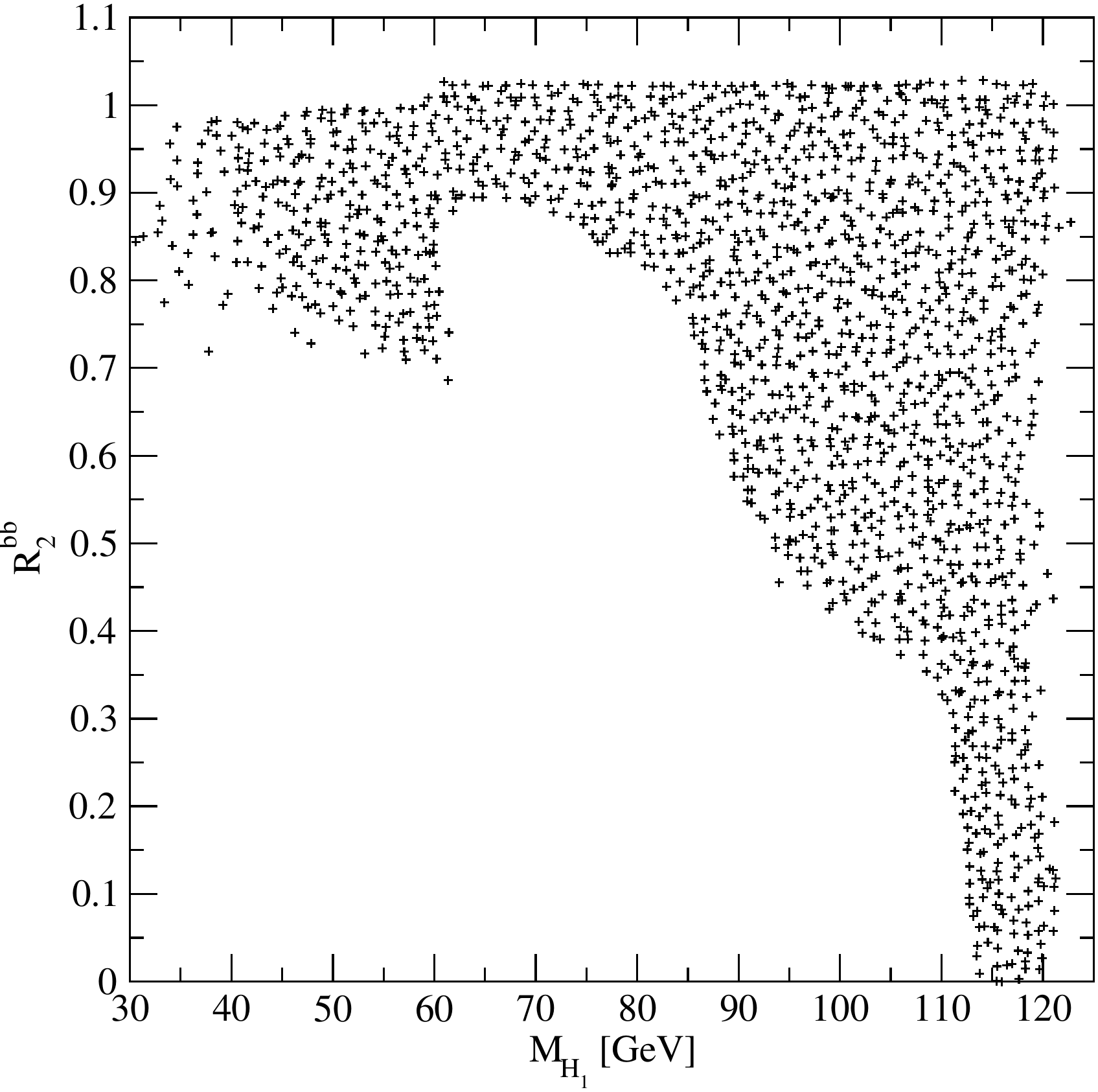}
\caption{The reduced signal cross section $R_2^{bb}$ as function of
$M_{H_1}$ in the semi-constrained NMSSM (from \cite{Ellwanger:2012ke}).}
\label{fig:nmssm2}
\end{figure}

The allowed gauge couplings$^2$ $\times$ branching fractions into $bb$
of $H_1$ and $H_2$ have been studied as function of $M_{H_1}$, once
$M_{H_2}\sim 125$~GeV is imposed, in the parameter space of the
semi-constrained NMSSM in \cite{Ellwanger:2012ke}. (In the
semi-constrained NMSSM, squark and slepton masses at the GUT scale are
given by a common value $m_0$, gaugino masses by a common value
$M_{1/2}$, but the NMSSM-specific soft Higgs masses and trilinear
couplings are left free.) The results for the allowed values of
$R_i^{bb} = \frac{g_i^2 \times BR(H_i \to bb)} {g_\mathrm{SM}^2 \times
BR(H_\mathrm{SM} \to bb)}$ are shown in Figs.~\ref{fig:nmssm1} and
\ref{fig:nmssm2}. Since here $BR(H_i \to bb) \approx
BR(H_\mathrm{SM} \to bb)$, one has $R_i^{bb} \approx \frac{g_i^2}
{g_\mathrm{SM}^2 }$.

For $M_{H_1} < 114$~GeV, the upper bounds on $R_1^{bb}$ in
Fig.~\ref{fig:nmssm1} follow from the LEP~II constraints in
\cite{Schael:2006cr}. Still, even for $M_{H_1} < 110$~GeV, a detection
of $H_1$ at a LC is possible (but difficult at the LHC within the
semi-constrained NMSSM). From Fig.~\ref{fig:nmssm2} one finds that, if
$M_{H_1} > 114$~GeV, $R_2^{bb}$ can assume all possible values from 0 to
1. Note that $R_1^{bb}$ and $R_2^{bb}$ satisfy approximately
$R_1^{bb}+R_2^{bb}\sim 1$.

For $M_{H_1} \sim 100$~GeV and $R_1^{bb} \sim 0.1-0.25$, $H_1$ can
explain the $\sim 2 \sigma$ excess in the $bb$ final state for this
range of Higgs masses at LEP~II \cite{Schael:2006cr}. Properties of such
points in the parameter space of the semi-constrained NMSSM have been
studied in \cite{Belanger:2012tt}, amongst others the production cross
sections of the various Higgs bosons in various channels at a LC.

\begin{figure}
  \includegraphics*[width=0.54\textwidth]{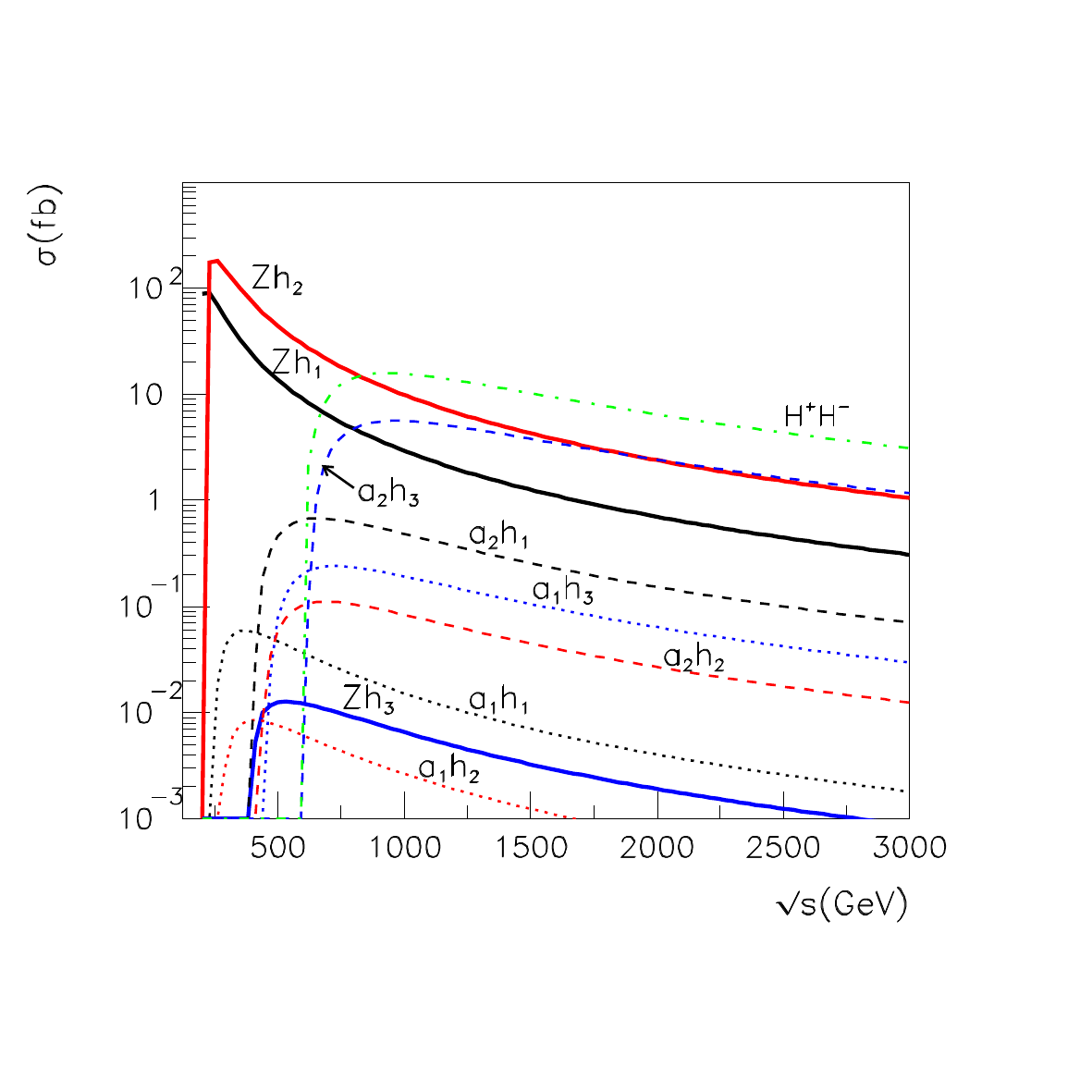}
\vspace{-1.6cm}
\caption{Higgs production cross sections at a $e^+ e^-$ collider in the
channels $Z H_1$, $Z H_2$, $H^+ H^-$ and $H_i A_j$ for a point in the
parameter space of the semi-constrained NMSSM with Higgs masses as
indicated in the text, from \cite{Belanger:2012tt}.}
\label{fig:nmssm3}
\end{figure}

For a typical point with $M_{H_1} \sim 99$~GeV,
$M_{H_2} \sim 124$~GeV (and an enhanced signal rate in the
$\gamma\gamma$ final state at the LHC), $M_{H_3} \sim 311$~GeV,
$M_{A_1} \sim 140$~GeV, $M_{A_2} \sim 302$~GeV and $M_{H^\pm} \sim
295$~GeV, the production cross sections in the channels $Z H_1$, $Z
H_2$, $H^+ H^-$ and $H_i A_j$ are shown in Fig.~\ref{fig:nmssm3} as
function of $\sqrt{s}$ of a LC (from \cite{Belanger:2012tt}). Note that,
for suitable mixing angles of $H_i$ and $A_j$, also $H_i A_j$ production
via  $e^+ + e^- \to H_i A_j$ is possible as in the MSSM.
However, an additional CP-even Higgs boson with sizeable coupling $g_i$
can also be heavier than 125~GeV; such a scenario is motivated by best
fits to present LHC and Tevatron data \cite{Belanger:2012sd}.

Other NMSSM-specific scenarios are possible Higgs-to-Higgs decays (see,
e.g., \cite{Dermisek:2008uu}). For the 125~GeV Higgs boson, the measured
Standard Model-like decay modes at the LHC indicate that Higgs-to-Higgs
decays are not dominant for this state, but branching fractions of
${\cal O}(10\%)$ are allowed. In the NMSSM, $H_{125}$
could decay into pairs of lighter CP-even or CP-odd states (if
kinematically possible). If these states are heavier than $\sim 10$~GeV
and decay dominantly into $bb$, such decay modes of
$H_{125}$ 
into $4b$ (or $2b2\tau$) would be practically invisible
at the LHC. At a LC, using the leptonic decays of $Z$ in the $ZH$ Higgs
production mode and/or VBF, such unconventional decays can be discovered
\cite{Ellwanger:2003jt}.

In addition, more Higgs-to-Higgs decays involving all three CP-even
states $H$ and both CP-odd states $A$ (omitting indices for simplicity)
like $H\to HH$, $H\to AA$, $H\to ZA$, $A\to AH$, $A\to ZH$, $H^\pm \to
W^\pm H$ and $H^\pm \to W^\pm A$ are possible whenever kinematically
allowed, and visible whenever the ``starting point'' of the cascade has
a sufficiently large production cross section (see, e.g.,
Fig.~\ref{fig:nmssm3}) and the involved couplings are not too
small. Even if a mostly Standard Model-like Higgs boson at 125~GeV is
imposed, the remaining unknown parameters in the Higgs sector of the
NMSSM allow for all of these scenarios.

The relevance of a $\gamma\gamma$ collider for the study of
Higgs-to-Higgs decays in the NMSSM has been underlined
in~\cite{Gunion:2004si}. Astonishingly, also pure singlet-like states
$H$ and $A$ can be produced in the $\gamma\gamma$ mode of a LC. In the
Standard Model, a $H\gamma\gamma$-vertex is loop-induced with mainly
$W^\pm$ bosons and top-quarks circulating in the loops. In the case of
the NMSSM and dominantly singlet-like states $H_S$ and $A_S$ (without
couplings to $W^\pm$ bosons or top-quarks), higgsino-like charginos can
circulate in the loops. The corresponding couplings of $H_S$ and $A_S$
to higgsino-like charginos originate from the term $\lambda \widehat{S}
\widehat{H_u} \widehat{H_d}$ in the superpotential (\ref{eq:nmssm1}) and
are absent for the MSSM-like CP-even and CP-odd Higgs states.

\begin{figure}
  \includegraphics*[angle=0,width=0.5\textwidth]{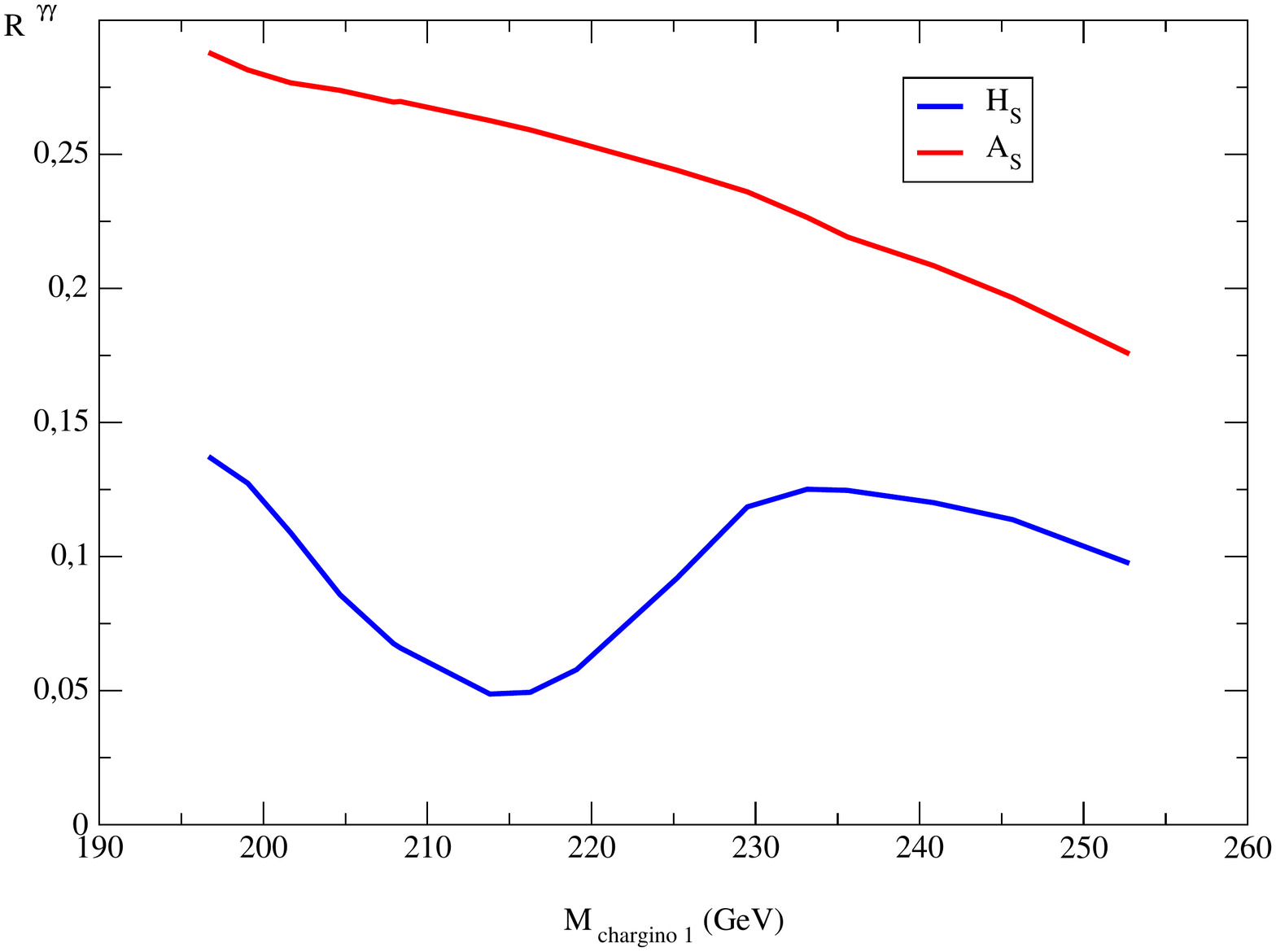}
  \vspace{-.5cm}
\caption{The reduced coupling $R^{\gamma\gamma}$, as defined in
Eq.~(\ref{eq:nmssm5}), as function of $M_{\mathrm{chargino}_1}$ for
$M_{A_S} \sim M_{H_S} \sim 260$~GeV, for a scenario explaining a 130~GeV
photon line from dark matter annihilation in the galactic center.}
\label{fig:nmssm4}
\end{figure}

Possible values of the reduced couplings $R^{\gamma\gamma}$ of such
nearly pure singlet-like states $H_S$ and $A_S$ are shown in
Fig.~\ref{fig:nmssm4}, where we define
\begin{equation}\label{eq:nmssm5}
R^{\gamma\gamma} = \frac{\Gamma(H/A \to \gamma\gamma)}
{\Gamma(H_\mathrm{SM} \to \gamma\gamma)}
\end{equation}
for a Standard Model-like $H_\mathrm{SM}$ of the same mass as $H_S$ or
$A_S$. The production cross sections of these states in the
$\gamma\gamma$ mode of a LC are given by the  production cross section
of $H_\mathrm{SM}$ multiplied by same ratio $R^{\gamma\gamma}$.

\begin{sloppypar}
The values of $R^{\gamma\gamma}$ shown in Fig.~\ref{fig:nmssm4}
correspond to a region in the parameter space of the NMSSM where the
Standard Model-like $H_\mathrm{SM}$ has a mass of $\sim 125$~GeV and,
simultaneously, dark matter annihilation in the galactic center can give
rise to a 130~GeV photon line~\cite{Das:2012ys}. Hence the LSP mass is
130~GeV, $M_{A_S} (\equiv M_{A_1}) \sim 260$~GeV in order to produce two
photons from LSP annihilation with $A_S$ exchange in the s-channel, and
$M_{H_S} (\equiv M_{H_2}) \approx 260$~GeV such that $H_S$ exchange in
the s-channel gives a relic density compatible with WMAP. $\lambda$
varies between 0.6 andd 0.65, the wino mass parameter is fixed to
$M_2=300$~GeV, but $\mu_\mathrm{eff}$ varies from $250 - 350$~GeV. The
nature of the chargino$_1$ varies slightly with $\mu_\mathrm{eff}$, but
is always $\approx 50\%$ wino and higgsino-like. The values shown in
Fig.~\ref{fig:nmssm4} have been obtained using the code
NMSSMTools~\cite{Ellwanger:2004xm,Ellwanger:2005dv}.
We see in Fig.~\ref{fig:nmssm4} that notably $R^{\gamma\gamma}(A_S)$ can
assume values close to 0.3, leading to a significant production cross
section in the $\gamma\gamma$ mode of a LC.
\end{sloppypar}

\begin{figure}
  \includegraphics*[width=0.45\textwidth]{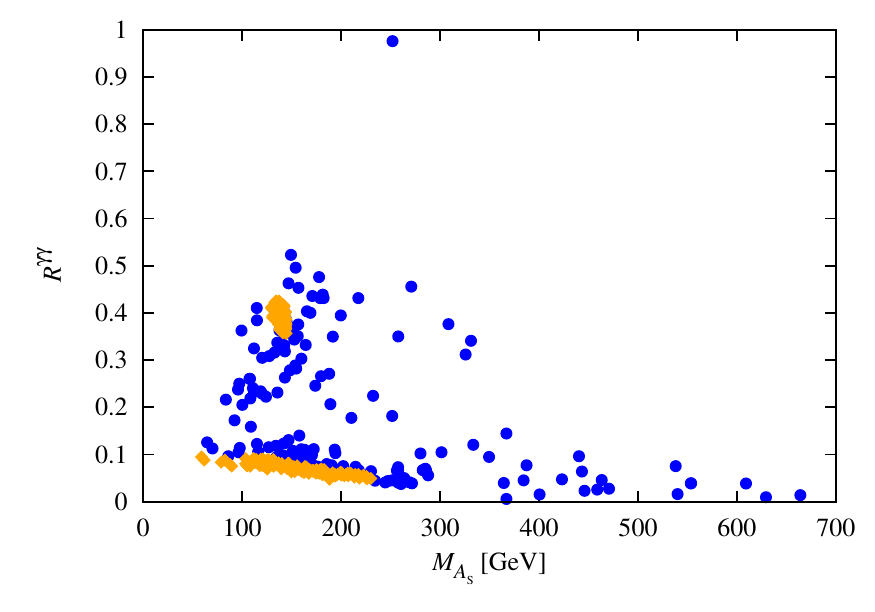}
  \vspace{0mm}
\caption{The reduced coupling $R^{\gamma\gamma}$ as function of
$M_{A_S}$, for points in the semi-constrained NMSSM where $H_S$ with
$M_{H_S}\sim 100$~GeV explains the excess in $bb$ at LEP~II (from
\cite{Belanger:2012tt}; orange diamonds satisfy the WMAP constraint on
the dark matter relic density).}
\label{fig:nmssm5}
\end{figure}

\begin{figure}
  \includegraphics*[width=0.45\textwidth]{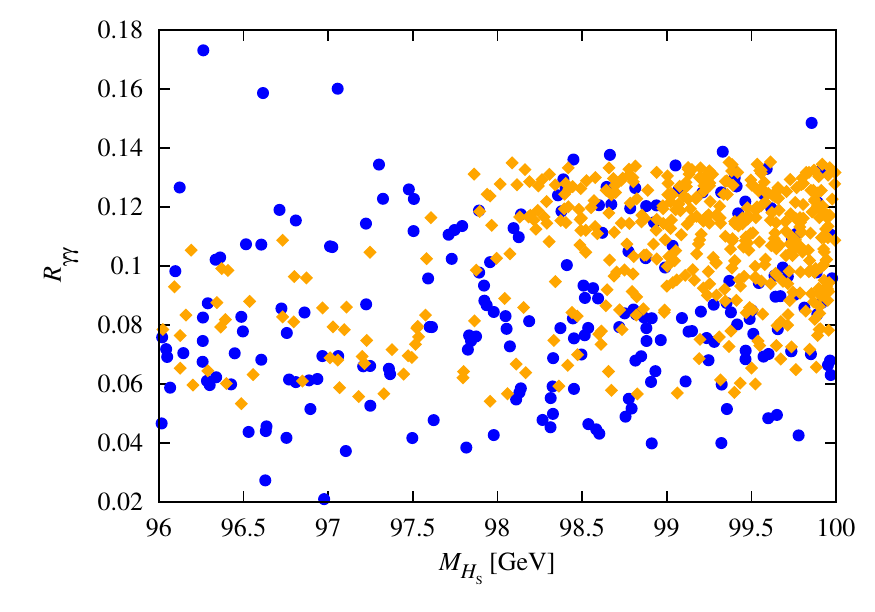}
  \vspace{0mm}
\caption{The reduced coupling $R^{\gamma\gamma}$ as function of
$M_{H_S}$, for points in the semi-constrained NMSSM where $H_S$ with
$M_{H_S}\sim 100$~GeV explains the excess in $bb$ at LEP~II (from
\cite{Belanger:2012tt}; orange diamonds satisfy the WMAP constraint on
the dark matter relic density).}
\label{fig:nmssm6}
\end{figure}

Returning to the semi-constrained NMSSM with\break $M_{H_1} \equiv
M_{H_S} \sim 100$~GeV and $M_{H_2} \sim 125$~GeV, scatter plots for
$R^{\gamma\gamma}(A_S)$ and $R^{\gamma\gamma}(H_S)$ as function of
$M_{A_S}$ and $M_{H_S}$ are shown in Figs.~\ref{fig:nmssm5} and
\ref{fig:nmssm6} (from \cite{Belanger:2012tt}). Again we see that the
prospects for $A_S$/$H_S$ discovery are quite promising for sufficiently
large  luminosity, since the production cross sections are typically
about 10\% (possibly larger) than those of a SM-like Higgs boson of a
corresponding mass.

Finally the NMSSM differs from the MSSM also due to the presence of a
fifth neutralino, the fermionic component of the superfield
$\widehat{S}$. Phenomenological analyses of pair production of
neutralinos in the NMSSM at $e^+\,e^-$ colliders at higher energies have
been performed in
\cite{Franke:1995tf,Choi:2001ww,Franke:2001nx,Choi:2004zx,MoortgatPick:2005vs,Ba
su:2007ys}
. Since the information on the neutralino sector from the LHC will be
quite limited, a $e^+\,e^-$ collider can be crucial to distinguish the
NMSSM neutralino sector from the one of the MSSM
\cite{MoortgatPick:2005vs}, although it cannot be guaranteed that the
difference is visible if one is close to the decoupling limit $\lambda,\
\kappa \to 0$. This question has also been addressed in the radiative
production of the lightest neutralino pair, $e^+\,e^- \to
\tilde\chi^0_1\,\tilde\chi^0_1\,\gamma$, at a LC with $\sqrt{s} = 500$~GeV in
\cite{Basu:2007ys}.

To summarize, the NMSSM is a well motivated supersymmetric extension of
the Standard Model, notably in view of the discovery of a Higgs boson at
125~GeV and a potentially enhanced branching fraction into
$\gamma\gamma$. Due to their reduced couplings to electroweak gauge
bosons it is not clear, however, whether the LHC will be able to verify
the extended Higgs and neutralino sectors of the NMSSM. Only a LC will
be able to perform measurements of such reduced couplings,
correspondingly reduced production cross sections, and possible
unconventional decay modes. These incompass both possible Higgs-to-Higgs
cascade decays, as well as cascades in the neutralino sector.


\subsection{Little Higgs\protect\footnotemark}
\footnotetext{Masaki Asano and Shigeki Matsumoto\\ Both M.A. and S.M. 
would like to thank all the members of 
  the ILC physics subgroup \cite{Ref:subgroup} for useful discussions. }
\label{sec:ewsb10}

%

The Litte Higgs (LH)
model~\cite{ArkaniHamed:2001nc,ArkaniHamed:2002qx,ArkaniHamed:2002qy}
is known to be one of the attractive scenarios for physics beyond the
standard model (SM). In this subsection, we review the physics of the
model at future linear collider experiments by referring to several
studies reported so far.

\subsubsection{About the LH model}
\label{sec:LH1}

The cutoff scale of the standard model (SM) is constrained by 
electroweak precision measurements: If we assume the existence of a
$\sim 125$~GeV SM Higgs-like resonance, the cutoff scale should be higher 
than roughly $5$ TeV
\cite{Barbieri:1998uv,Barbieri:2000gf}. However, such a relatively
high cutoff scale requires a fine-tuning in the Higgs potential 
because the Higgs potential receives the quadratic divergent radiative 
correction. 

\begin{sloppypar}
In LH models, the Higgs boson is regarded as a pseudo Nambu-Goldsone
(NG) boson which arises from a global symmetry breaking at high
energy, $\sim 10$ TeV. 
Although Yukawa and gauge couplings break the global
symmetry explicitly, some global symmetry is not broken
by one of these couplings: in LH models, the breaking of
such a symmetry is achieved only by two or more couplings, 
which is called ``collective'' symmetry breaking.
Because of the collective symmetry breaking,
the quadratic divergence from SM loop diagrams is canceled
by new particle diagrams at the one-loop level.
\end{sloppypar}

\begin{sloppypar}
As a bottom-up approach, specifying a coset group, we investigate the
phenomenology of such a scenario by a non-linear sigma model. In
particular, the littlest Higgs (LLH) model \cite{ArkaniHamed:2002qy}
described by an SU(5)$/$SO(5) symmetry breaking and the simplest
little Higgs (SLH) model \cite{Schmaltz:2004de} described by an
[SU(3)$\times$U(1)]$^2/$ [SU(2)$\times$U(1)]$^2$ symmetry breaking
have been studied about its expected phenomenology well so far. Here
we review the ILC physics mainly focusing on the LLH model.
\end{sloppypar}

The LLH model is based on a non-linear sigma model describing an
SU(5)$/$SO(5) symmetry breaking with the vacuum expectation value $f
\sim \mathcal{O}$(1) TeV. An [SU(2) $\times$U(1)]$^2$ subgroup of 
the SU(5) is gauged and broken down to the SM 
SU(2)$_L\times$U(1)$_Y$. Fourteen NG bosons arise and it can be 
decomposed into ${\bf 1}_0 \oplus {\bf 3}_0 \oplus {\bf 2}_{\pm
  1/2} \oplus {\bf 3}_{\pm 1}$ under the electroweak gauge group. 
The ${\bf 1}_0 \oplus {\bf 3}_0$ are eaten by heavy gauge bosons 
$A_H, Z_H, W_H^\pm$, and ${\bf 2}_{\pm 1/2} \oplus {\bf
  3}_{\pm 1}$ are the SM Higgs field $h$ and new triplet Higgs 
field $\Phi$, respectively. To realize the collective symmetry breaking, 
SU(2) singlet vector-like top quark partners, $T_L$ and $T_R$, are also 
introduced. These heavy particles have masses which are proportional 
to $f$ and depend also on the gauge coupling, charges and Yukawa couplings. 
The Higgs potential is generated radiatively and it depends also 
on parameters of UV theory at the cutoff scale $\Lambda \sim 4 \pi f$.

Even in the model, the new particle contributions are strongly
constrained at precision measurements.\\ Pushing new particle masses up
to avoid the constraint, the fine-tuning in the Higgs potential is
reintroduced. To avoid the reintroducing the fine-tuning, implementing of
the $Z_2$ symmetry called T-parity has been proposed
\cite{Cheng:2003ju,Cheng:2004yc,Low:2004xc}.  \footnote{ As the other
  possibility, for example, the model decoupling the new gauge bosons
  have been also proposed \cite{Schmaltz:2010ac,Katz:2005au}. }

\begin{sloppypar}
In the LLH model, the T-parity is defined as the invariance under the
exchanging gauged $[SU(2) \times U(1)]_1$ and $[SU(2) \times
U(1)]_2$. Then, for all generation of lepton and squark sector, new
heavy fermions are introduced to implement this symmetry. Under the
parity, the new particles are assigned to be a minus charge (T odd),
while the SM particles have a plus charge (T even). Thus, heavy
particles are not mixing with SM particles. Then, the tree level new
particle contribution to electroweak precision measurements are
forbidden and the new particle masses can be light.
\end{sloppypar}

It has been suggested that the T-parity is broken by anomalies in the typical strongly coupled UV theory \cite{Hill:2007nz,Hill:2007zv} and the possibilities of the conserved T-parity scenario and another parity are also studied \cite{Krohn:2008ye,Csaki:2008se,Freitas:2009jq,Pappadopulo:2010jx,Brown:2010ke}. If the T-parity is an exact symmetry, the lightest T-odd  particle, heavy photon in the LLH model, is stable and provides a dark matter candidate. Even if the T-parity  is broken by anomalies, contribution to electroweak precision measurements are still suppressed while the lightest T-odd particle would decay at colliders \cite{Barger:2007df,Freitas:2008mq}.

As described above, top quark partner, new gauge bosons and additional
scalar bosons are expected in LH models while its details strongly
depend on models. In the model with T-parity, T-odd quark partners and
lepton partners are introduced additionally. The Higgs boson
phenomenology would be different from the SM prediction due to the
new particle contributions as well as deviations from the SM coupling
which would be appeared from higher dimensional operators.

\subsubsection{Higgs phenomenology in LH}
\label{sec:LH2}

In LH models, parameters of the Higgs potential cannot be estimated
without calculating the contribution of a specifying UV theory. As a
phenomenological approach, we consider these parameters as free
parameters and these are determined by observables, e.g., Higgs mass. 
As described here, there are possibilities to change the Higgs
boson phenomenology from the SM prediction and it may be checked at
the ILC.

\paragraph{Higgs decay from loop diagram}
One of the possibility to change the Higgs phenomenology is
contributions from top partner as well as the deviation from the SM
couplings. It leads to deviations in the decay branching ratios of the
$h \to gg$ (also indicating deviations in the main Higgs production
channel at the LHC) and $h \to \gamma \gamma$ modes, via the top partner-loop
diagrams. The extra gauge bosons and charged scalar bosons also
contribute to the $h \to \gamma \gamma$ decay.

\begin{figure}
  \includegraphics[width=80mm]{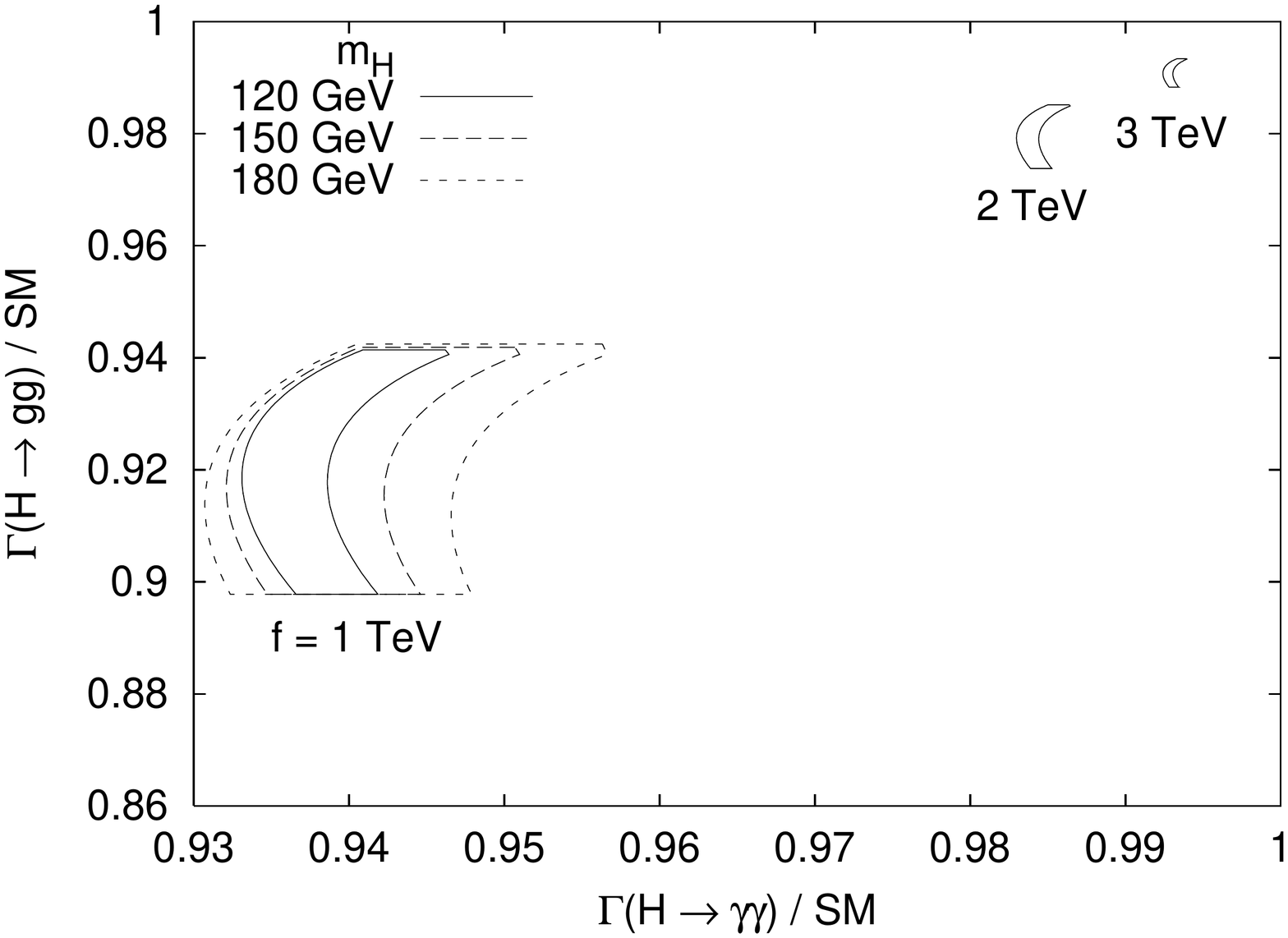}
\caption{Accessible range of $\Gamma(h \to gg)$ and $\Gamma(h \to \gamma \gamma)$ normalized to the SM value in the LLH model (from \cite{Han:2003gf}).}
\label{fig:loop_higgs_decay_LH}       
\end{figure}

Fig.\ref{fig:loop_higgs_decay_LH} shows the range of partial decay
widths, $\Gamma(h \to gg)$ and $\Gamma(h \to \gamma \gamma)$, in the
LLH model varying model parameters \cite{Han:2003gf}. In the model,
the deviation of the top Yukawa coupling suppresses the $\Gamma(h \to
gg)$ while contributions from top partner and mixing 
in the top sector enhance the partial decay width. 
Totally, these additional top sector contributions
suppresses the $\Gamma(h \to gg)$ in
Fig.\ref{fig:loop_higgs_decay_LH}. On the other hand, it enhances the
$\Gamma(h \to \gamma \gamma)$ because the $W$~boson loop contribution is
dominant in the SM and the fermion-loop contributions have a minus
sign. The contribution from the heavy gauge bosons suppresses the
$\Gamma(h \to \gamma \gamma)$ as well as the deviation of the gauge
boson coupling and mixing in the gauge boson sector due to the
sign of the $W_H W_H h$ coupling. The charged Higgs contribution leads
to an enhancement. The doubly-charged Higgs contribution is small
because the coupling to the Higgs boson is suppressed, then, it is
neglected here \cite{Han:2003gf}.  
In a similar way the $\gamma Z$ decay would be
affected~\cite{GonzalezSprinberg:2004bb}.

\begin{figure}
  \includegraphics[width=80mm]{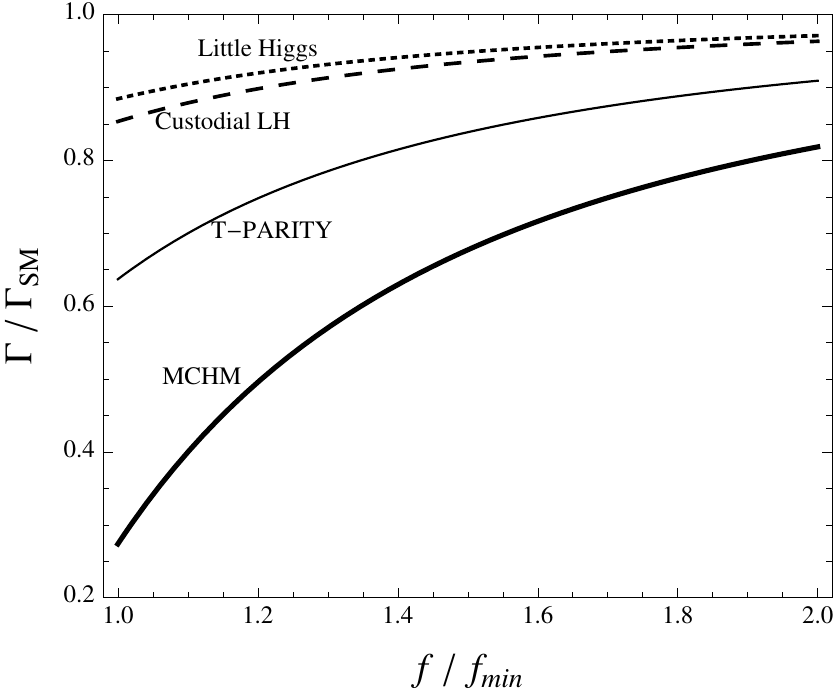}
  \caption{The $\Gamma(h \to gg)$ normalized to the SM value (from
    \cite{Low:2010mr}). The $f_{min}$ is defined as the smallest value
    allowed by electroweak precision measurements and the values are
    $1.2$ TeV for the LLH model, $500$ GeV for T-parity case, 700 GeV
    for custodial littlest Higgs model and $500$ GeV for minimal
    composite Higgs model, respectively (for details, see
    \cite{Low:2010mr}).}
\label{fig:loop_higgs_decay2_LH}       
\end{figure}

In the model with T-parity, there is also the contribution from T-odd
heavy fermions and the contribution is negative to $\Gamma(h \to gg)$
and positive to $\Gamma(h \to \gamma \gamma)$ \cite{Chen:2006cs}.
Furthermore, in the model with T-parity case (and also in a decoupling
gauge partner case, e.g., \cite{Berger:2012ec}), the new particle can
be light consisting with electroweak precision measurements, thus, the
deviation could be greater than the case without T-parity. For
example, in the littlest Higgs model with T-parity (LHT), the
$\Gamma(h \to gg)$ normalized to the SM value can be around $60 \%$ at
$f = 500$ GeV case \cite{Low:2010mr} (see Fig.
\ref{fig:loop_higgs_decay2_LH}).

The expected precision for measurements of the Higgs coupling
including $h \to\gamma \gamma$ and $h \to gg$ 
branch at ILC are 
summarized in Sect.\ref{sec:ewsb3}.
One of the
possibilities to measure the deviation of the $\Gamma(h \to \gamma
\gamma)$ is the $\gamma \gamma \to h \to b \bar{b}$ mode in photon
collider option \cite{Logan:2004hj,Wang:2009qs}.

\paragraph{Higgs decay at tree level} 
The deviation of the SM coupling and new particles would also change the Higgs phenomenology at tree level. The deviation of $ht\bar{t}$ and top partner change the cross section of $ht\bar{t}$ production \cite{Yue:2005db,Wang:2006ke,Kai:2007ji}. In LHT, production cross section of the $e^+e^- \to ht\bar{t}$ normalized to the SM value is about $90 \%$ at $f=1$ TeV \cite{Wang:2006ke}.

The deviation of $hWW$ and $hZZ$ couplings (e.g. \cite{Han:2003wu} in LLH model) also change the cross sections of the Higgs boson production as well as the decay branching ratio \footnote{ For the deviation of vector boson fusion process at ILC, see \cite{Yue:2005av,Wang:2006ui,Liu:2006xa}.}. In some case, the deviation rates of partial decay widths are the same, then, the branching ratio of the Higgs decay can be close to the SM prediction \cite{Chen:2006cs}.

\begin{figure*}
  \includegraphics[width=\textwidth]{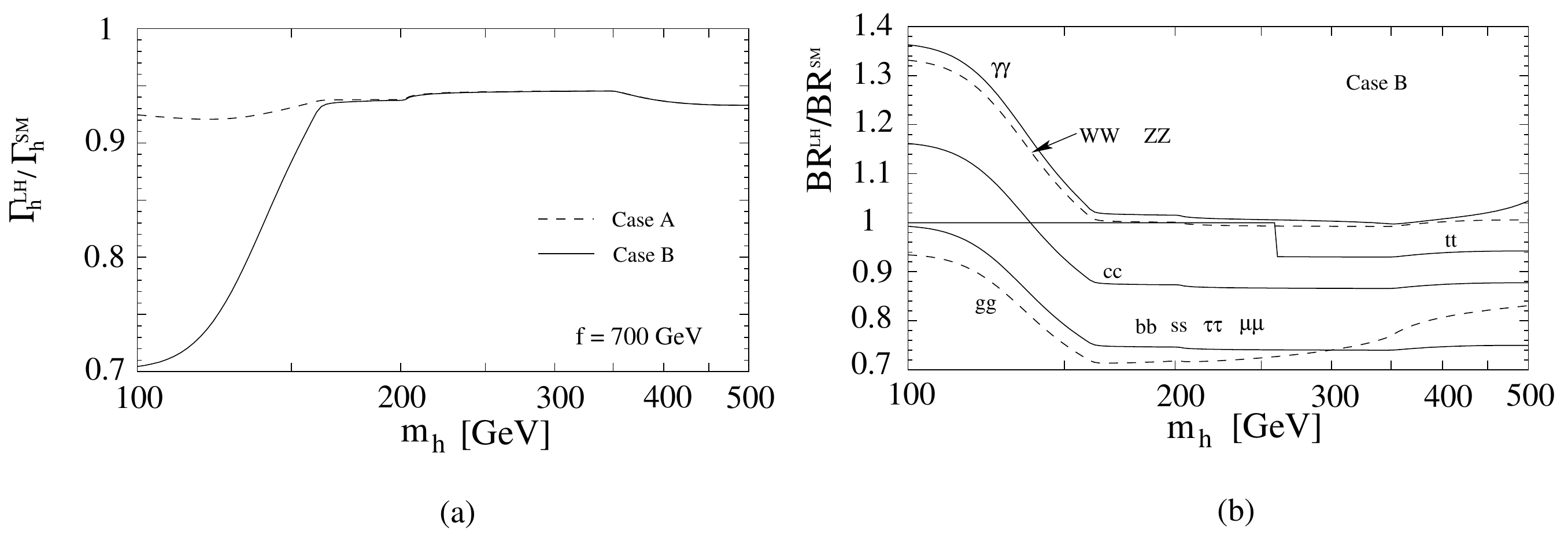}
\caption{The (a) shows the total decay width normalized to the SM value in the LHT (from \cite{Chen:2006cs}). The difference between case A and case B comes from the definition of the down-type Yukawa term (for details, see \cite{Chen:2006cs}). The (b) shows the partial Higgs branching ratios normalized to the SM value (from \cite{Chen:2006cs}).}
\label{fig:higgs_decay_Br}       
\end{figure*}

However, the down-type Yukawa coupling has model dependence and the couplings could be significantly suppressed in some case of the LHT \cite{Chen:2006cs}. Thus, the decay branching ratio of a light Higgs boson ($m_h < 2 m_W$) could significantly change because the dominant decay width, $h \to b\bar{b}$ is suppressed. Fig. \ref{fig:higgs_decay_Br} shows the correction of the branching ratio from the SM prediction \cite{Chen:2006cs}.

\paragraph{Higgs decay to new particles}

Another possibility is additional decay branches of Higgs boson into new
particles. For example, the lightest new particle in the LHT is the
heavy photon which mass is $\sim 60$ GeV with $f = 400$ GeV. If it
kinematically possible, the Higgs boson also decays into two heavy
photons and the value of the branching ratio could be large ($> 80 \%$)
in the $125$~GeV Higgs boson case because it decays via the gauge
coupling \cite{Asano:2006nr,Hundi:2006rh}. If the T-parity is an exact
symmetry, it is the invisible decay. On the other hand, 
the produced heavy photon decays mainly into SM fermions in such a light Higgs boson case if the T-parity is broken by anomaly. The decay width is about $10^{-1}$-$10^{-2}$ eV \cite{Barger:2007df,Freitas:2008mq}.

\paragraph{Additional scalar bosons} 
In some models, e.g., simple group models, there could be a pseudo
scalar, $\eta$,  although the mass depends on the models. The Higgs
boson could also decay into $\eta \eta$ and
$Z \eta$ \cite{Cheung:2006nk} if it is kinematically
possible. Furthermore, because the $Z$-$h$-$\eta$ coupling cannot appear
in product group models, the measurement at ILC helps to distinguish the
kind of LH models \cite{Kilian:2006eh}. Other phenomenology studies for
$\eta$ can be found in Refs.~\cite{Kilian:2004pp,Han:2011zzg}. As another example of additional scalars, there is the triplet Higgs boson in the LLH model, although these mass is proportional $f$ \cite{Yue:2007ym,Cagil:2009fa,Cagil:2010hr,Cagil:2010mk}.

\paragraph{Higgs self coupling} 
The measurement of Higgs self coupling is one of the important test for the Higgs boson. In the LH models, the triplet and quartet coupling could slightly change from the SM expectation. Study for $Zhh$ process in LLH \cite{Liu:2006rc} and the one-loop correction to the $hhh$ coupling from vector-like top quarks \cite{Asakawa:2010xj} have been studied.

\subsubsection{Other direct LH signals}
\label{sec:LH3}

Since the LH model is discussed only in this subsection, we also mention
here other signals of the model at future liner collider
experiments. The signals can be divided into two categories; direct and
indirect signals. The direct signals means the direct productions of new particles predicted by the LH model. The indirect signals are, on the other hand, the LH contributions to the processes whose final states are composed only of SM particles. We consider only the direct signals, while we omit to discuss the indirect ones for want of space. Please see references~\cite{Han:2003wu,Conley:2005et,Yue:2005dd,Yue:2005pa,Choudhury:2006xa,Batra:2006iq,Yue:2007kv,HongSheng:2007ve,Wang:2008qna,Bernreuther:2008us,Huang:2009nv,Berger:2009hi,Riemann:2010zz,Moyotl:2010ss,Zhang:2010wq,Han:2011zza,Han:2011xd,Yang:2011zze,Yang:2011xk,Wang:2011bb,Li:2011ak,Wang:2012zza,Yang:2012tj} for the indirect signals.

The direct signals can future be divided into two subcategories; the direct productions of colored particles and non-colored ones. This is because the LH model requires the cancellation of quadratically divergent corrections to the Higgs mass term from top loop and those of electroweak gauge bosons at one-loop level, and thus the model inevitably predicts both colored and non-colored new particles. When the T-parity (or some other Z$_2$-symmetry distinguishing SM and new particles) is not imposed on the model like the littlest or the simplest Higgs model, non-colored new particles will be produced by following two processes; single productions (i.e., $e^+ e^- \to V_H$)~\cite{Langacker:2008yv,Park:2003sq,Cho:2004xt,Yue:2005nc,Yue:2008jt,Liu:2008bs,Ananthanarayan:2009dw,Liu:2009zza,Aranda:2010cy} and associate productions (i.e., $e^+ e^- \to V_H + \gamma (Z)$)~\cite{Yue:2004fv,Wang:2006pa,Wang:2007um,Wang:2007pt,Yue:2008zp}, where $V_H$ is the LH partner of the weak gauge boson (heavy gauge boson). On the other hand, when the T-parity is imposed like the case of the LHT, non-colored new particles must be produced in pair (i.e., $e^+ e^- \to V_H V_H$)~\cite{Yue:2007yu,Cao:2007pv,Asakawa:2009qb,Asano:2011aj,Zeng:2012tw,Kato:2012jf}. For the productions of colored new particles, associate productions (i.e., $e^+ e^- \to T + t$) and pair productions (i.e. $e^+ e^- \to f_H f_H$) are frequently considered to find LH signals~\cite{Liu:2006zk,Kong:2007uu,Harigaya:2011yg}, where $T$ is the LH partner of the top quark (top partner) and $f_H$ is the new colored fermion like the top partner or heavy fermions which are introduced by imposing the T-parity on the model.

\begin{sloppypar}
We first consider the productions of non-colored new particles. Among
several relevant studies reported so far, the most comprehensive one
involving realistic numerical simulations has been performed in
reference~\cite{Kato:2012jf}. They have considered following five pair
production processes in the framework of the LHT; $e^+ e^- \to Z_H Z_H$,
$Z_H A_H$, $W_H^+ W_H^-$, $e_H^+ e_H^-$, and $\nu_{eH} \bar{\nu}_{eH}$,
which are followed by the decays $Z_H \to A_H h$, $W_H^\pm \to A_H
W^\pm$, $e_H^\pm \to Z_H e^\pm$, $\nu_{eH} \to W_H^+ e^-$
($\bar{\nu}_{eH} \to W_H^- e^+$), where $e_H^-$ ($e_H^+$) and $\nu_{eH}$
($\bar{\nu}_{eH}$) are the T-parity partners of electron (positron) and
electron neutrino (anti-neutrino), respectively. The mass spectrum of
the non-colored new particles used in this study is the following 
(to be taken as a representative example), 
\end{sloppypar}
\begin{center}
\begin{tabular}{c|ccccc}
& $M_{A_H}$ & $M_{W_H}$ & $M_{Z_H}$ & $M_{e_H}$ & $M_{\nu_{eH}}$ \\
\hline
Mass (GeV) & 81.9 & 368 & 369 & 410 & 400 \\
\hline
\end{tabular}
\end{center}
The above mass spectrum has been obtained by choosing the vacuum expectation value of the global symmetry $f$ and the Yukawa coupling of the heavy electron $\kappa_e$ to be 580 GeV and 0.5, respectively.\footnote{The Higgs mass is assumed to be 134 GeV, because this analysis has been performed before the discovery of the Higgs-like boson. The result of the analysis is not changed significantly even if the Higgs mass is set to be around 125 GeV.} Flavor-changing effects caused by the heavy lepton Yukawa couplings are implicitly assumed to be negligibly small.

By measuring the energy distribution of visible (SM) particles emitted in each production process, the masses of the non-colored new particles can be precisely extracted. This is because the initial energy of electron (positron) is completely fixed at the $e^+ e^-$ colliders and thus measuring the energy distribution allow us to reconstruct the process accurately without any assumption of the LHT model. With assuming the integrated luminosity of 500 fb$^{-1}$ at $\sqrt{s} =$ 1 TeV running and use of the four processes, $e^+ e^- \to Z_H Z_H$, $W_H^+ W_H^-$, $e_H^+ e_H^-$, and $\nu_{eH} \bar{\nu}_{eH}$, the resultant accuracies of the mass extractions turns out to be as follows~\cite{Kato:2012jf}.
\begin{center}
\begin{tabular}{c|ccccc}
& $M_{A_H}$ & $M_{W_H}$ & $M_{Z_H}$ & $M_{e_H}$ & $M_{\nu_{eH}}$ \\
\hline
Accuracy & 1.3\% & 0.20\% & 0.56\% & 0.46\% & 0.1\% \\
\hline
\end{tabular}
\end{center}
\begin{sloppypar}
Since the relevant physics of the LHT model is described with only two
model parameters $f$ and $\kappa_e$, the masses of non-colored new
particles are also given by the parameters. Performing these
model-independent mass measurements therefore provides strong evidence
that the discovered new particles are indeed LHT particles. The
parameters $f$ and $\kappa_e$ are eventually extracted from the
measurements very accurately; $f$ and $\kappa_e$ are extracted at
accuracies of 0.16\% and 0.01\%.
\end{sloppypar}

\begin{sloppypar}
More interestingly, by assuming the vertex structures of the LHT model
(i.e. the Lorentz structure, the ratio of right- and left-handed
couplings, etc.), it is possible to extract the couplings concerning
heavy gauge bosons/heavy leptons through cross section
measurements. There are a total of eight vertices concerning the five
pair production processes. Extracting all the couplings is therefore
possible by measuring the total cross sections of the five processes
and the angular distribution (the difference cross section) of the
produced heavy gauge boson for appropriate three processes. See
reference~\cite{Kato:2012jf} for more detailed strategy to extract the
couplings. Though numerical simulations for the three differential
cross sections are not performed yet, the measurement accuracies for
the five total cross sections have already been obtained as follows.
\end{sloppypar}
\begin{center}
{\scriptsize
\begin{tabular}{c|ccccc}
$e^+ e^- \to$
& $A_H Z_H$ & $Z_H Z_H$ & $e_H^+ e_H^-$ & $\nu_{eH} \bar{\nu}_{eH}$ & $W_H^+ W_H^-$ \\
\hline
Accuracy & 7.70\% & 0.859\% & 2.72\% & 0.949\% & 0.401\% \\
\hline
\end{tabular}
}
\end{center}
Only $Z_H A_H$ process has been analyzed with 500 fb$^{-1}$ data at
$\sqrt{s} =$ 500 GeV running, while others have been done with the
same luminosity at 1 TeV running.

We next consider the direct productions of colored new particles.
Among several colored new particles, the most important one is the top
partner $T$ (and its T-parity partner $T_-$), because it is
responsible for the cancellation of the quadratically divergent
correction to the Higgs mass term from top loop. Since the top partner
has a color-charge, it is expected to be constrained by the LHC
experiment when its mass is not heavy. Thus we summarize the current
status of the constraint before going to discuss the physics of the
top partner at future linear collider experiments.

The most severe limit on the mass of the top partner comes from its
pair production process followed by the decay $T \to b
W$~\cite{ATLAS:2012qe}. The limit is $m_T >$ 650 GeV at 95\% C.L. with
assuming BR($T \to bW$) = 1. Since the top partner has other decay
channels like $T \to tZ/T \to th$ and the branching fraction to $bW$
is typically about 40\%, the actual limit on the mass is $m_T >$ 500
GeV. On the other hand, the T-parity partner of the top partner $T_-$
decays into $tA_H$ with BR($T_- \to t A_H$) $\simeq$ 1. The most severe
limit on its mass again comes from its pair production process, which
gives $m_{T_-} >$ 420 GeV at 95\% C.L. when $A_H$ is light
enough~\cite{Aad:2011wc}.

The physics of the top partner at future linear collider experiments
has been discussed in some details in
reference~\cite{Harigaya:2011yg}. When $m_T \simeq$ 500 GeV, the cross
section of its pair production process ($e^+ e^- \to T \bar{T}$) is
${\cal O}(100)$ fb, while that of the associate production process
($e^+ e^- \to t \bar{T} + \bar{t} T$) is ${\cal O}(1-10)$ fb with
appropriate center of mass energy. It has been shown that the Yukawa
coupling of the top partner and the coupling of the interaction
between $h$, $t$, and $T$ can be precisely measured with use of the
threshold productions of these processes. Since these couplings are
responsible for the cancellation of the quadratically divergent
correction to the Higgs mass term from top loop, these measurements
will give a strong test of the LH model.

The physics of the T-parity partner $T_-$ at future linear collider
experiments has been discussed in some details in
reference~\cite{Kong:2007uu}. When $m_{T_-} \simeq$ 500 GeV, the cross
section of its pair production process ($e^+ e^- \to T_- \bar{T}_-$)
is ${\cal O}(100)$ fb with appropriate center of mass energy. Since
$T_-$ decays into $t A_H$, the masses of both $T_-$ and $A_H$ can be
precisely measured using the energy distribution of reconstructed top
quarks, which will provide an excellent test of the LHT model by
comparing this signal with those of non-colored new particles.
Furthermore, it has also been pointed out that the process can be used
to discriminate new physics models at the TeV scale. This is because
many new physics models predict similar processes, a new colored
particle decaying into $t$ and an invisible particle like a squark
decaying into $t$ and a neutralino in the MSSM.

As a recent review and recent studies for current status of new particles
and dark matter in LHT, please see 
also~\cite{Baer:2013cma,Reuter:2013iya,Wang:2013yba,Chen:2014wua}.





\subsection{Testing Higgs physics at the photon linear collider
\label{sec:ewsb12}\protect\footnotemark}

\footnotetext{Maria Krawczyk and  Ilya Ginzburg: 
We are grateful to Filip \.Zarnecki for clarification of old analyses 
as well as to  Jan Kalinowski.
}


%

\newcommand{\bu}{$\bullet$\ }
\newcommand\blue{\color{blue}}
\newcommand\red{\color{red}}
\newcommand{\bes}{\begin{subequations}}
\newcommand{\ees}{\end{subequations}}
\newcommand{\bear}{\begin{equation}\begin{array}}
\newcommand{\eear}[1]{\end{array}\label{#1}\end{equation}}

\newlength{\figwidth}
\newlength{\figheight}
\newlength{\twofigheight}
\newlength{\twofigwidth}
\setlength{\figheight}{0.5\textwidth}
\setlength{\figwidth}{0.7\textwidth}
\setlength{\twofigheight}{0.35\textwidth}
\setlength{\twofigwidth}{0.45\textwidth}
 \newcommand{\fn}[1]{\footnote{ #1}}




%



\begin{sloppypar}
A photon collider (hereafter we use abbreviation PLC -- photon linear
collider) is based on photons obtained from laser light back-scattered
from high-energy electrons of Linear Collider (LC). Various high
energy gamma-gamma and electron-gamma processes can be studied
here. With a proper choice of electron beam and laser polarization,
the high-energy photons with high degree polarization (dependent on
energy) can be obtained. The direction of this polarization can be
easily changed by changing the direction of {electron and} laser
polarization.  By converting both electron beams to the photon beams
one can study $\gamma \gamma$ interactions in the energy range up to
$\sqrt{s_{\gamma\gamma}}\sim 0.8 \cdot \sqrt{s_{ee}}$, whereas by
converting one beam only the $e\gamma$ processes can be studied up to
$\sqrt{s_{e\gamma}}\sim 0.9 \cdot \sqrt{s_{ee}}$~\cite{GKST}.
\end{sloppypar}

\begin{figure}[htb!]
\begin{center}
  \includegraphics[width=0.6\figwidth,clip=]{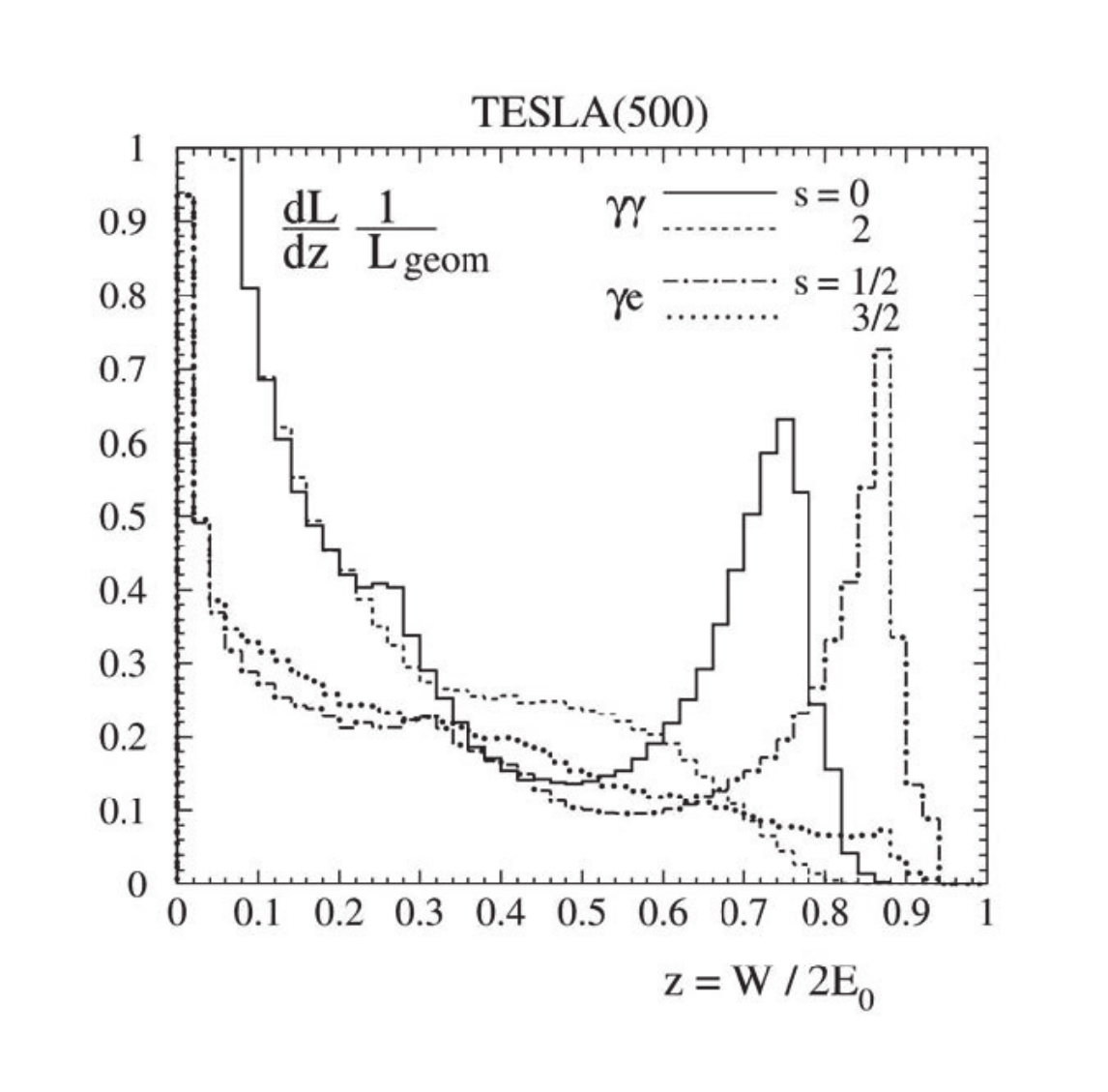}
\vspace{-.5cm}
  \caption{\it The distribution of $\gamma \gamma$ and $e \gamma $ center-of-mass energy
           $W$ with respect to the $e^+e^-$ energy (2$E_0$) from  simulation of
           the PLC luminosity spectra~\cite{Telnov:1999vz}.  Contributions of various spin states of produced system are shown.}
  \label{fig:spectra}
\end{center}
\end{figure}

\begin{sloppypar}
In a nominal LC option, i.e. with the electron-beam energy of 250~GeV,
the geometric luminosity $L_{geom}=12\cdot 10^{34} cm^{-2}s^{-1}$ can
be obtained, which is about four times higher than the expected $e^+
e^-$ luminosity.  Still, the luminosity in the high energy $\gamma
\gamma$ peak (see Fig.~\ref{fig:spectra}) corresponds to about
$\frac{1}{3}$ of the nominal $e^+ e^-$ luminosity -- so we expect
$L_{\gamma \gamma}(\sqrt{s_{\gamma\gamma}}> 0.65 \cdot \sqrt{s_{ee}})$
equal to about 100${\rm fb}^{-1}$ per year (400 ${\rm fb}^{-1}$ for a
whole energy range)~\cite{ilc-tdr,Telnov:1999vz}.
Adjusting the initial electron beam energy and direction of
polarizations { of electrons and laser photons at fixed laser photon
  energy} one can vary a shape of the $\gamma\gamma$ effective mass
spectrum.
\end{sloppypar}

At a $\gamma \gamma$ collider the neutral C-even  resonance  with spin 0  can be produced, in contrast to C-odd  spin 1 resonances in the $e^+e^-$ collision. Simple change of signs of polarizations of incident electron and laser photon for one beam transforms PLC to a mode with  total helicity 2 at its high-energy part. It allows to determine degree of possible admixture of state with spin 2 in the  observed  Higgs state.  The  s-channel resonance production of  $J^{PC}=0^{++}$ particle  allows
to perform precise measurement of its properties at PLC.

\bigskip In summer 2012  a Higgs boson  with mass about  125 GeV  has been discovered at LHC~\cite{Aad:2012tfa}. We will denote this particle as $\cal H$.  The collected data~\cite{Aad:2013wqa,CMS2013} allow to conclude  that {\it {the SM-like scenario}},  suggested e.g. in~\cite{Ginzburg:2001ph,Ginzburg:2003sp},  is realized~\cite{Belanger:2012gc}: all measured  $\cal {H}$  couplings are close to their SM-values in their {\it {absolute value}}.  Still     following interpretations of these data are  discussed:  A) ${\cal H}$ is Higgs boson of the SM. B) We deals with phenomenon beyond SM, with
${\cal H}$  being some other scalar  particle (e.g. one of neutral  Higgs bosons of Two Higgs Doublet Model (2HDM) -- in particular MSSM, in the CP conserving 2HDM that are $h$ or $H$). In this approach { following} opportunities are possible:
1)   Measured  couplings  are close to SM-values, however some of them (especially the $ttH$
 coupling) with  a "wrong" sign. 2) In addition some new heavy charged particles, like $H^\pm$ from 2HDM,  can contribute to the loop couplings.
3) The observed signal is not due to {\it one} particle but it is an effect of two or more particles, which were not resolved  experimentally -- {\it the degenerated Higgses.}
 {  Each of these} { opportunities} can lead to  the enhanced or suppressed,    as compared to the SM predictions,   ${\cal H}\gamma\gamma$, ${\cal H}gg$ and ${\cal H}Z\gamma$  loop-coupling.

\bigskip \ The case with the observed Higgs-like signal being due to
degenerated Higgses $h_i$ demands a special effort to diagnose it. In
this case the numbers of events with production of some particle $x$
are proportional to sums like $\sum_i
(\Gamma^x_i/\Gamma^{tot}_i)\Gamma^{gg}_i$.  Data say nothing about
couplings of the individual Higgs particles and there are no
experimental reasons in favor of the SM-like scenario for {\it one} of
these scalars.  In such case each of degenerated particles have low
total width, and there is a hope that the forthcoming measurements at
PLC can help to distinguish different states due to much better
effective mass resolution. The comparison of different production
mechanisms at LHC, $e^+e^-$ LC and PLC will give essential impact in
the problem of resolution of these degenerated states. Below we do not
discuss the case with degenerated Higgses with masses $\sim$125 GeV in
more details, concentrating on the case when observed is one Higgs
boson $\cal H$, for which the SM-like scenario is realized.

\bigskip \ In the discussion  we  introduce useful  {\it relative couplings}, defined as ratios of the couplings of each neutral Higgs boson $h^{(i)}$ from the considered model, to the gauge bosons $W$ or $Z$ and to the quarks or leptons ($j=V (W,Z),u,d,\ell...$), to the corresponding SM couplings:
$ \chi_j^{(i)}=g_j^{(i)}/g_j^{\rm SM}$.  Note that all couplings to EW gauge bosons $\chi_V^{(i)}$ are real, while the couplings to fermions are generally complex. For CP-conserving case of 2HDM we have in particular $\chi_j^h$, $\chi_j^H$, $\chi_j^A$ (with $\chi_V^A=0$), where couplings of fermions to $h$ and $H$ are real while couplings to $A$ are purely imaginary.

{\it The SM-like scenario}  for the observed  Higgs  ${\cal H}$, to be identified with some neutral $h^{(i)}$,  corresponds to $|\chi_j^{\cal H}|\approx 1$.  Below we assume this scenario is realized at present.

\bigskip It is known already from a long time that the PLC is  very good observatory of the scalar sector of the SM and beyond SM, leading to important  { and in many cases} complementary to the $e^+e^-$ LC case tests of the EW symmetry breaking mechanism~\cite{GinHiggs}-\cite{Tel2012}. The $e^+e^-$ LC, together with its  PLC  options ($\gamma \gamma$ and $e \gamma$),  is very well suited for the precise study of properties of this newly discovered ${\cal H}$ particle, and other scalars. In particular, the
PLC offers a unique opportunity to study resonant production of Higgs bosons  in the process $\gamma \gamma \rightarrow {\rm  Higgs}$ which is sensitive to  charged fundamental particles of the theory.  In principle, PLC
allows  to study also resonant production of heavier neutral Higgs particles from the extension of the SM. Other physics topic which could be studied  well at PLC is the CP property of Higgs bosons. Below
we discuss the most important aspects of the Higgs physics  which can be investigated at PLC. Our discussion is based on analyses  done during last two decades  and takes into account also some recent "realistic" simulations supporting those results.\\

\subsubsection{Studies of 125 GeV Higgs ${\cal H}$}

 The  discussion in this section is related to the case when $\cal H$ is one of the Higgs bosons $h^{(i)}$ of 2HDM.  In the CP conserving case of 2HDM it can be either~$h$ or~$H$.

\begin{figure}[htb!]
\begin{center}
\includegraphics[height=\twofigheight]{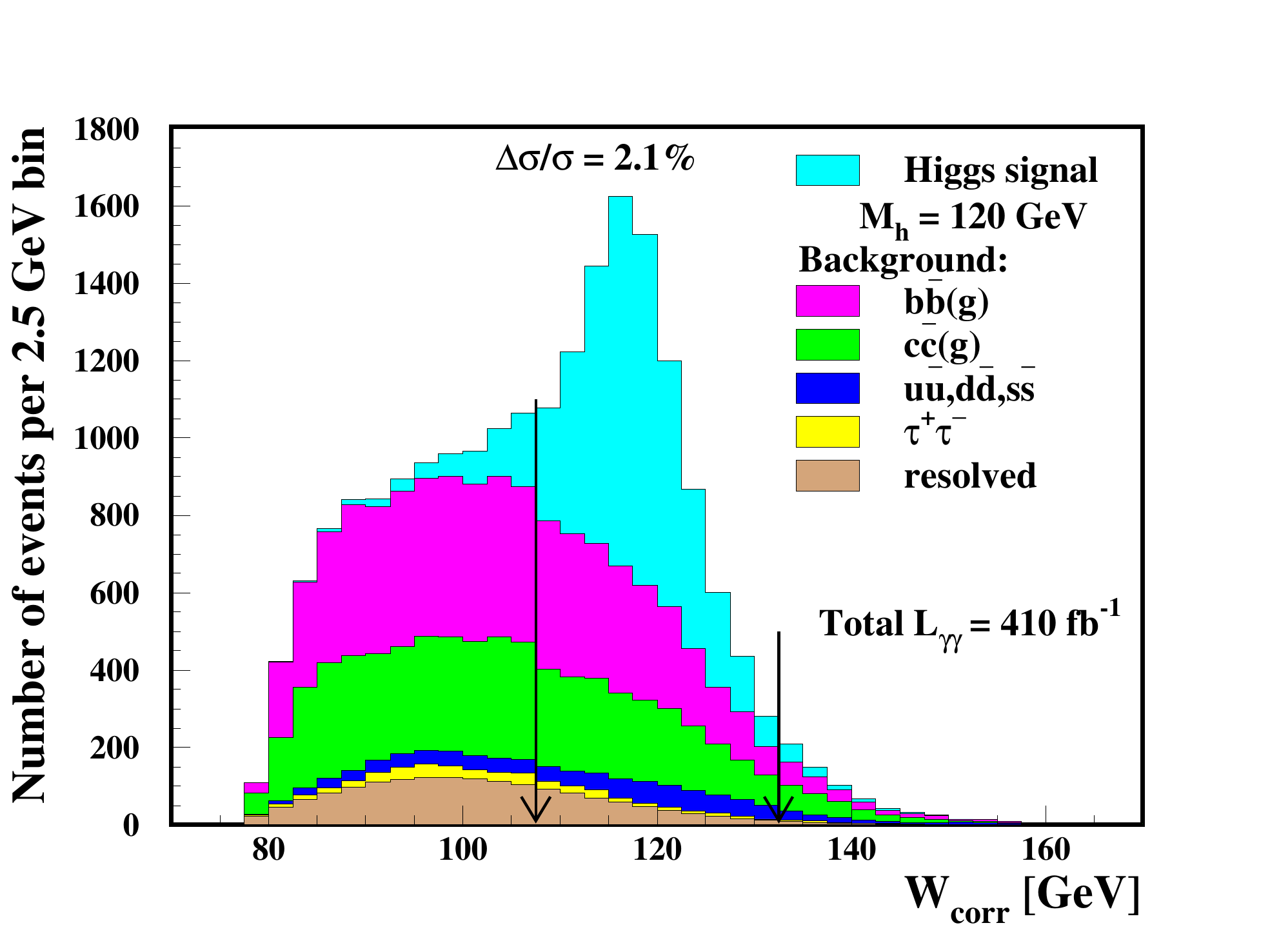}
 \caption{\it
Distributions of the corrected invariant mass, $W_{corr}$, for
selected $b \bar{b}$ events; contributions of the signal, for $M_{H_{SM}} =
$ 120~GeV, and of the different  background processes, are shown separately~\cite{Niezurawski:2005cp}.}
 \label{fig:phys_h1}
\end{center}
\end{figure}

\begin{sloppypar}
  Several NLO analyses of the production at the PLC of a light
  SM-Higgs boson $H_{SM}$ decaying into $b \bar b$ final state were
  performed, including the detector simulation,
  eg.~\cite{JikS}--\cite{SM-Mayda}. These analyses demonstrate a high
  potential of this collider to measure accurately the Higgs
  two-photon width. By combining the production rate for $\gamma
  \gamma \rightarrow H_{SM}\rightarrow b \bar b$
  (Fig. \ref{fig:phys_h1}), to be measured with 2{\blue.1} \%
  accuracy, with the measurement of the $\br(H_{SM}\rightarrow bb)$ at
  $e^+e^-$ LC, with accuracy~$\sim$ 1~\%, the width $\Gamma(H_{SM}
  \rightarrow\gamma \gamma)$ for $H_{SM}$ mass of 120 GeV can be
  determined with precision $\sim$ 2 \%. This can be compared to the
  present value of the measured at LHC signal strength for 125 GeV
  $\cal H$ particle, which ratio to the expected signal for SM Higgs
  with the same mass (approximately equal to the ratio of $|g_{\gamma
    \gamma {\cal H }}|^2/|g_{\gamma \gamma {H_{SM} }}|^2$), are
  1.17$\pm$0.27 and 1.14$^{+0.26}_{-0.23}$ from
  ATLAS~\cite{Aad:2014eha} and CMS~\cite{Khachatryan:2014ira},
  respectively.
\end{sloppypar}

\bigskip \ The process $\gamma\gamma\to {\cal H}\to \gamma\gamma$ is also observable at the PLC with reasonable rate~\cite{SM-Mayda}. This measurement allows to measure directly two-photon width of Higgs without assumptions about unobserved channels, couplings, etc.

\bigskip \  Neutral Higgs resonance  couples to photons via loops with
charged particles. In the  Higgs $\gamma \gamma$ coupling  the heavy
charged particles, with masses generated by the
Higgs mechanism, do not decouple. Therefore the
${\cal H}\to\gamma \gamma$ partial width  is sensitive to the
contributions of charged particles with masses even far beyond the
energy of the $\gamma \gamma$ collision.  This allows to recognize
which type of extension of the  minimal SM is realized.  The $H^+$ contribution to the ${\cal H} \gamma \gamma$ loop coupling   is proportional to  ${\cal H} H^+H^-$ coupling,   which value and sign
can be treated as
free parameters of model\footnote{Except if some additional symmetry is
  present in the model.}. 
The simplest  example gives a 2HDM with type~II Yukawa interaction
(2HDM II). For a small $m_{12}^2$ parameter, see \refse{sec:ewsb6},
the contribution of the charged Higgs boson $H^+$ with mass  larger than 400 GeV  leads to 10\% suppression in the ${\cal H}\to\gamma \gamma$ decay width as compare to the SM one, for $M_{\cal H}$ around 120 GeV~\cite{Ginzburg:2003sp,Ginzburg:2001ph}, Table \ref{table1} (solution A).
The enhancement or decreasing of  the ${\cal H}  \gamma \gamma$ coupling is possible, as discussed for 2HDM with various Yukawa interaction models in~\cite{Bernal:2009rk}--\cite{posch} as well in the  Inert Doublet Model\fn{ That is
the $Z_2$ symmetric 2HDM where one  Higgs doublet  plays a role of SM Higgs field $\phi_S$, interacting with fermions as in Model I, with  the  SM-like Higgs boson $h$ and another Higgs doublet  $\phi_D$, having no v.e.v.. The latter one  contains  four  scalars $D, \,D^A,\,D^\pm $, the lightest among them $D$ (analog of  $H$ of 2HDM) can be DM particle,  scalars $D^A$ and   $D^\pm$ (analog of $A$ and $H^\pm$, respectively).}~\cite{arh,bs}.

\begin{figure}[htb!]
\begin{center}
\includegraphics[height=0.3\textheight,width=0.45\textwidth,angle=0]{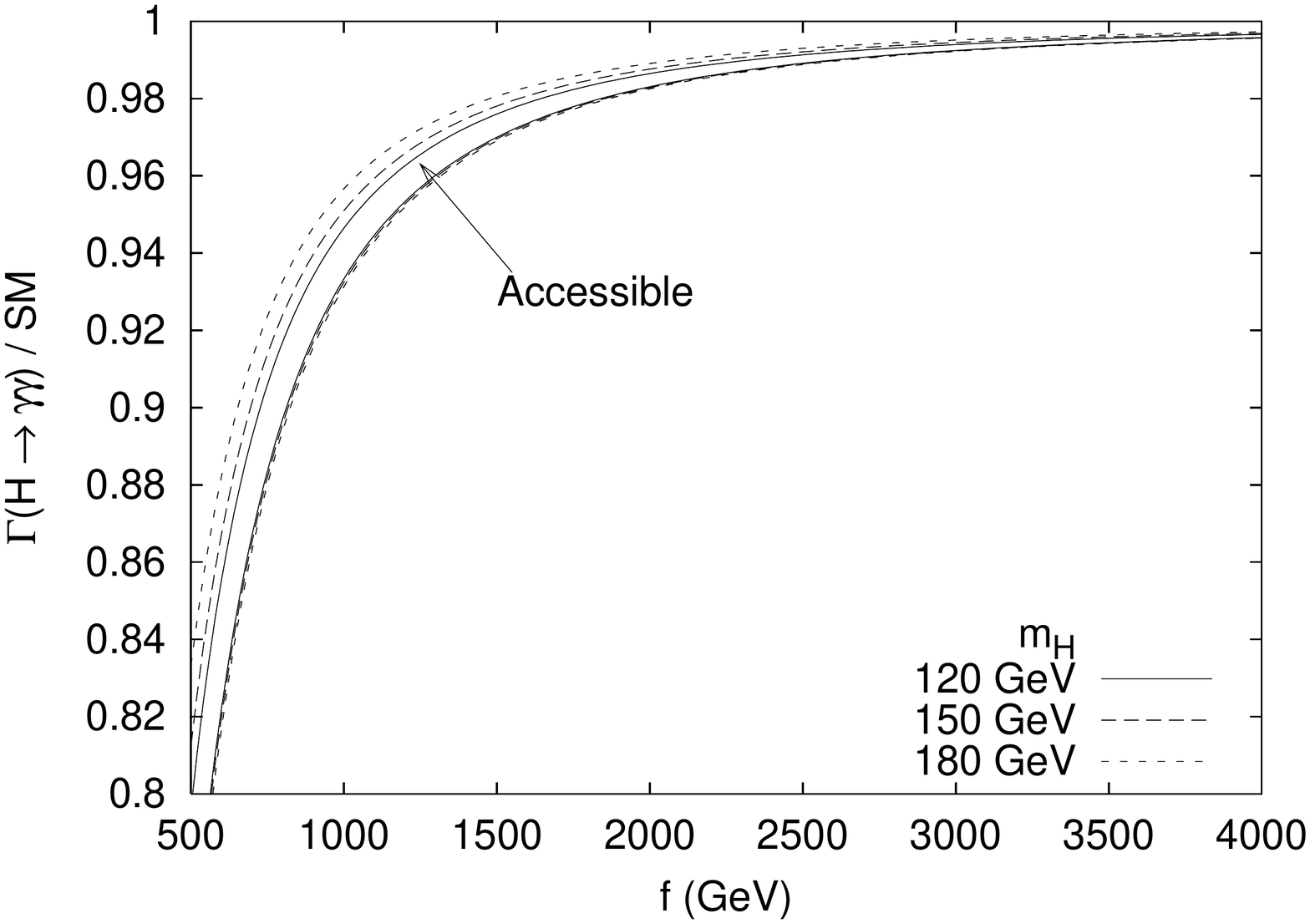}
\vspace{-.5cm}
 \caption{\it
Ratio   $\dfrac{\Gamma(h \rightarrow \gamma \gamma)}{\Gamma(h\to \gamma\gamma)^{SM}}$
as a function of the mass scale of the new physics
$f$ in the Littlest  Higgs model~\cite{Han:2003gf}, for different Higgs
boson masses. ``Accessible'' indicates the possible variation of the
rate for fixed~$f$.
}
 \label{fig:little}
\end{center}
\end{figure}

In the Littlest Higgs model a 10\% suppression of the $\gamma \gamma$ decay
width for $M_{\cal H}\approx 120$~GeV is expected due to the  new heavy
particles with mass around 1 TeV  at the suitable scale of couplings for these
new particles~\cite{Han:2003gf,Huang:2009nv}, see
 Fig.~\ref{fig:little}.

\bigskip \ The  Higgs $\gamma \gamma$ loop coupling is sensitive to the relative signs of various contributions. For example,  in 2HDM II  sign of some  Yukawa couplings  may differ from the SM case, still strength (ie. absolute value) of all squared direct Higgs couplings to $WW/ZZ$ and fermions  being as in the SM. This  may lead to the enhancement of the ${\cal H} \to \gamma \gamma$ decay-width with respect to the SM predictions, up to 2.28 for a "wrong" sign of the ${\cal H} tt$ for $M_{\cal H}=120$  GeV  (1.28 for  ${\cal H} \to gg$ and 1.21 for ${\cal H}  \to Z\gamma$, respectively)  coupling, Table \ref{table1}  (solution $B_{{\cal H} t}$),~\cite{Ginzburg:2001ph}\fn{The recent analysis of the LHC data     leads to constraints of the relative ${\cal{H}}tt$ coupling $\chi_t^{\cal{H}}$
\cite{mele}.}.
The "wrong" sign of ${\cal H}bb$ coupling (solution $B_{{\cal H} b}$ in Table \ref{table1}) could lead to a enhancement in the ${\cal H} \to gg$,  and in the corresponding rate for gluon fusion of Higgs at LHC, similarly as  the "wrong" sign of ${\cal H}tt$ coupling.  Such solution is still considered as a possible for 125 GeV ${\cal H}$ particle~\cite{Aad:2013wqa}.

\begin{table}[ht]
\begin{center}
\begin{tabular}{|c|c|c|c|c|c|}
\noalign{\vspace{-9pt}}
\hline
solution&basic couplings&$|\chi_{gg}|^2$ &
$|\chi_{\gamma\gamma}|^2$ & $|\chi_{Z\gamma}|^2$\\ \hline\hline
$A_{{\cal H}}$
&$\chi_V\approx\chi_b\approx\chi_t\approx \pm1$
& 1.00 & 0.90 & 0.96 \\ \hline
$B_{{\cal H} b}$
&$\chi_V\approx-\chi_b\approx\chi_t\approx \pm1$
& 1.28 & 0.87 & 0.96 \\ \hline
$B_{{\cal H} t}$
& $\chi_V\approx\chi_b\approx-\chi_t\approx \pm1$
& 1.28 & 2.28 & 1.21 \\ \hline
\end{tabular}
\end{center}
\vspace*{-3mm}
\caption{\it SM-like realizations in the 2HDM~II~\cite{Ginzburg:2001ph},\cite{Ginzburg:2003sp} together
with ratios of loop-induced partial widths to their SM values at
$M_{\cal H}=120$~GeV, $M_{H^{\pm}}=
 $800~GeV, $|m_{12}^2|\le 40$~GeV$^2$. }
\vspace*{-2mm}
\label{table1}
\end{table}

\bigskip   The observed Higgs particle can have definite CP parity or
can be admixture of states with different CP parity  ({\it
  {CP-mixing}}). In the latter case the PLC provides the best among  all
colliders place for the study of such mixing. Here, the  opportunity to
simply vary polarization of photon beam allows to study this mixing via
dependence of the production cross section on the incident photon
polarization~\cite{Grzadkowski:1992sa,choi,asakawa,HCPH,gii,Zerwas-CP,Ellis:2004hw}. In
particular, the change of sign of circular polarization ($++
\leftrightarrow --$) results in variation of production cross section of
the 125 GeV Higgs in 2HDM  by up to about 10\%, depending on a degree of
CP-admixture. Using mixed circular and linear polarizations of photons
gives opportunity  for more detailed investigations~\cite{nzk0710}. \\

The important issue is to measure a Higgs selfcoupling, ${\cal H}{\cal
  H}{\cal H}$. In the SM this selfcoupling is precisely fixed via
Higgs mass (and v.e.v. $v=246$~GeV), while deviations from it's SM
value would be a clear signal of more complex Higgs sector. Both at
the $e^+e^-$ collider and at the $\gamma\gamma$ collider the two
neutral Higgs bosons are produced in processes both with and without
selfinteraction, namely
{\small
$$\begin{array}{l}
e^+e^-\to Z\to {{\cal H}(Z\to Z{\cal H})} \oplus e^+e^-\to Z\to {Z({\cal H}\to {\cal H}{\cal H})};\\[2mm]
\gamma\gamma\to \mbox{\,loop\,} \to { {\cal H}{\cal H}}\oplus
\gamma\gamma\to \mbox{\,loop\,} \to  {{\cal H}\to {\cal H}{\cal H}}.\end{array}
$$}
In the SM case  the cross sections for above processes are rather low  but measurable, so that coupling under interest can be extracted,  both in the $e^+e^-$ and $\gamma\gamma$, modes of $e^+e^-$ LC, see~\cite{Belusevic:2004pz}-\cite{Tsumura:2011zz}.
The feasibility of this measurement  at a PLC has been performed recently in~\cite{Kawada:2012uy}. For Higgs mass of 120 GeV and the integrating luminosity 1000 fb$^{-1}$ the  statistical sensitivity as a function of the $\gamma \gamma$ energy for measuring the deviation from the SM Higgs selfcoupling $\lambda=\lambda_{SM} (1+\delta \kappa)$ has been estimated. The optimum  $\gamma \gamma$ collision  energy  was found  to be around 270 GeV for a such Higgs mass,   assuming that  large  backgrounds due to $WW/ZZ$ and $bbbb$ production  can  be suppressed for correct  assignment of tracks.   As a result, the  Higgs  pair  production  can  be  observed  with  a statistical significance  of  5 $\sigma$ by operating the PLC  for 5 years.

\bigskip \ The smaller but interesting effects are expected in $e\gamma\to e{\cal H}$ process with $p_{\bot e}> 30$~GeV, where ${\cal H}Z\gamma$ vertex can be extracted with reasonable accuracy~\cite{Gvych}.\\

\subsubsection{Studies of  heavier Higgses, for  125 GeV ${{\cal H}=h^{(1)}}$}

A direct discovery of other Higgs bosons and measurement of their couplings to gauge bosons and fermions is necessary for clarification the { way the SSB is realized}. In this section we consider the case when observed 125 GeV Higgs is the lightest neutral Higgs, ${\cal H}=h^{(1)}$ (in particular in the CP-conserving case this means ${\cal H}=h$).  A single Higgs production at $\gamma\gamma$  
collider allows to  explore roughly the same  mass region for neutral Higgs bosons at the parent $e^+e^-$ LC but with higher cross section and lower background. The $e\gamma$ collider allows in principle to test wider mass region in the process $e\gamma\to eH, eA$  however  with a  lower cross section.

\bigskip {Before general discussion, we  present some properties of one of the simplest Higgs {} { model} beyond the minimal SM, namely  2HDM (in particular, also the Higgs sector of MSSM), having in mind that the modern data are in favour of a SM-like scenario.
Let us enumerate here some important properties of 2HDM for each neutral Higgs scalar $h^{(i)}$ in the CP conserving case $h^{(1)}=h$, $h^{(2)}=H$, $h^{(3)}=A$:
\begin{itemize}
\item[(i)]  {\it For an arbitrary Yukawa interaction} there are sum rules for coupling of different neutral Higgses to gauge bosons $V=W,\,Z$ and to each separate fermion $f$ (quark or lepton)
\be
\sum\limits_{i=1}^{3} (\chi_V^{(i)})^2=1\,.\qquad\sum\limits_{i=1}^{3}(\chi_f    ^{(i)})^2=1\,.
\label{srW}
\ee
The first  sum rule (to the gauge bosons) was discussed  e.g. in~\cite{gunion-haber-wudka}--\cite{GK05}. The second one was obtained only for Models I and II of Yukawa interaction~\cite{Grzadkowski:1999wj}, however in fact it holds for any Yukawa sector~\cite{GKr2013}.

In the first sum rule all quantities  $\chi_V^{(i)}$ are real. Therefore,  in SM-like case  (i.e. at $|\chi_V^{(1)}|\approx 1$) both    couplings $|\chi_V^{2,3}|$  are small. The couplings entering the second sum rule (for fermions) are generally complex. Therefore this sum rule  shows that for $|\chi_f^{(1)}|$ close to 1, {either  $ \left|\chi_f^{(2)}\right|^2$ and $\left|\chi_f^{(3)}\right|^2$ are simultaneously small, or  $ \left|\chi_f^{(2)}\right|^2 \approx\left|\chi_f^{(3)}\right|^2$}.

\item
[(ii)]  For the 2HDM I   there are  simple relations, which  in the CP conserved case  are
as follows
\be
\chi_u^{(h)}=\chi_d^{(h)}\,,\qquad \chi_u^{(H)}=\chi_d^{(H)}\,.
\label{VodIeq}
\ee
\item
[(iii)]  In the 2HDM II following relations hold:\\
a) {\em The pattern relation} among the relative couplings
for  {\it each neutral Higgs particle $h^{(i)}$}
\cite{Ginzburg:2001ss,Ginzburg:2002wt}:
 \bes\label{ModIIeq}\be\label{2hdmrel}
(\chi_u^{(i)} +\chi_d^{(i)})\chi_V^{(i)}=1+\chi_u^{(i)}
\chi_d^{(i)}\,.
 \end{equation}
b) For each neutral Higgs boson $h^{(i)}$ one can write a horizontal  sum rule
\cite{Grzadkowski:1999ye}:
\begin{equation}
|\chi_u^{(i)}|^2\sin^2\beta+|\chi_d^{(i)}|^2\cos^2\beta=1\,.\label{srules}
\end{equation}\ees
\end{itemize}

\bigskip   Below, in Table~25, we present benchmark points for
the SM-like $h$ scenario in the CP conserving 2HDM II. The total widths for H and A  for various $\chi_t^A=1/\tan\beta$ are shown assuming
with  $\chi_V^h\approx 0.87 $, $|\chi_V^H|=0.5$ and  $|\chi_t^h|=1$ for $H$ and $A$ .   \footnote{The total width  $\Gamma_H$ differs from
the total width  $\Gamma_A$ by the $W/Z$ contribution,  since $\chi^A_V=0$.}
\begin{table}[htb]
\begin{center}
\begin{tabular}{|c|c||c|c|c|}\hline
&$\Gamma_H, \qquad \Gamma_A$&$\Gamma_H, \;\; \Gamma_A$&$\Gamma_H, \;\;
\Gamma_A$\\\hline
$M_{H,A}$& $\tan\beta=1/7$ &$\tan\beta=1$ &
$\tan\beta=7$ \\\hline
200& $0.35\qquad 8\cdot 10^{-5}$&$0.35\qquad 4\cdot 10^{-3}$&$0.4\qquad 0.2$\\\hline
300 &$2.1\qquad 1.2\cdot 10^{-4}$&$2.1\qquad 6\cdot 10^{-3}$&$0.75\qquad 0.3$\\\hline
400& $138\qquad 132$&$8.8\qquad 2.7$&$2.5\qquad
0.45$\\\hline
500& $537\qquad 524$&$22.8\qquad 10.7$&$6.1\qquad
0.7$\\\hline
\end{tabular}
\end{center}
\label{benchtab}
\caption{\it Total width (in MeV) of $H$, $A$ in some benchmark points for the SM-like $h$ scenario ($M_h$=125 GeV)  in the 2HDM  ($\chi_V^h\approx 0.87 $, $|\chi_V^H|=0.5$ and  $|\chi_t^h|=1$).
Results for  $\tan \beta=1/7, \, 1 \ \  and \ \  7$ are shown.
}
\end{table}

In the SM-like $h$  scenario it follows from sum rule \eqref{srW} that the $W$-contribution to the $H\gamma \gamma$
 width is much smaller than that  of would-be heavy SM Higgs, with the same mass, $M_{H_{SM}}\approx M_H$. At the
large $\tan\beta$ also $H\to tt$, $A\to tt$ decay widths  are extremely
small, so that the total widths of $H$, $A$ become very small \fn{At  $\tan\beta\ll 1$ we obtain the  strong interaction in the Higgs sector mediated by $t$-quarks, what is signalizing by the  fact that the  calculated in standard approach total widths of heavy $H$, $A$ is becoming  close to or even higher than the corresponding masses.
Of course,  in this case such  tree-level estimates become inadequate. In the same manner at $\tan\beta>70$ corresponds to the region of a strong interaction in the Higgs sector mediated by $b$-quarks. We don't consider such scenarios.}.

\bigskip \ Let us compare properties of heavy $H$, $A$  in 2HDM  with a  would-be heavy SM Higgs-boson with the same mass.
The cross section for production of such particles  in the main
gluon-gluon fusion channel,  being $\propto
\Gamma_{H,A}^{gg}\Gamma_{H,A} /M_H^3$,   is lower than that in SM. At
large $\tan\beta$ resonances $H,A$ become very narrow, as discussed
above,  besides, the two-gluon decay width become  about $1/\tan^2\beta$
smaller. Consequently, these main at LHC  production channels
cross section are suppressed by roughly $1/\tan^4\beta$
w.r.t.\ the would-be SM Higgs boson with the same mass and   $H$
and $A$ can escape observation in these channels at the LHC. (The same is valid for $e^+e^-$ LC due to small value of $\chi_V^H$ for $H$ and $\chi_V^A=0$.)

\begin{figure}[htb]
 \includegraphics[height=.4\textheight,width=0.45\textwidth]{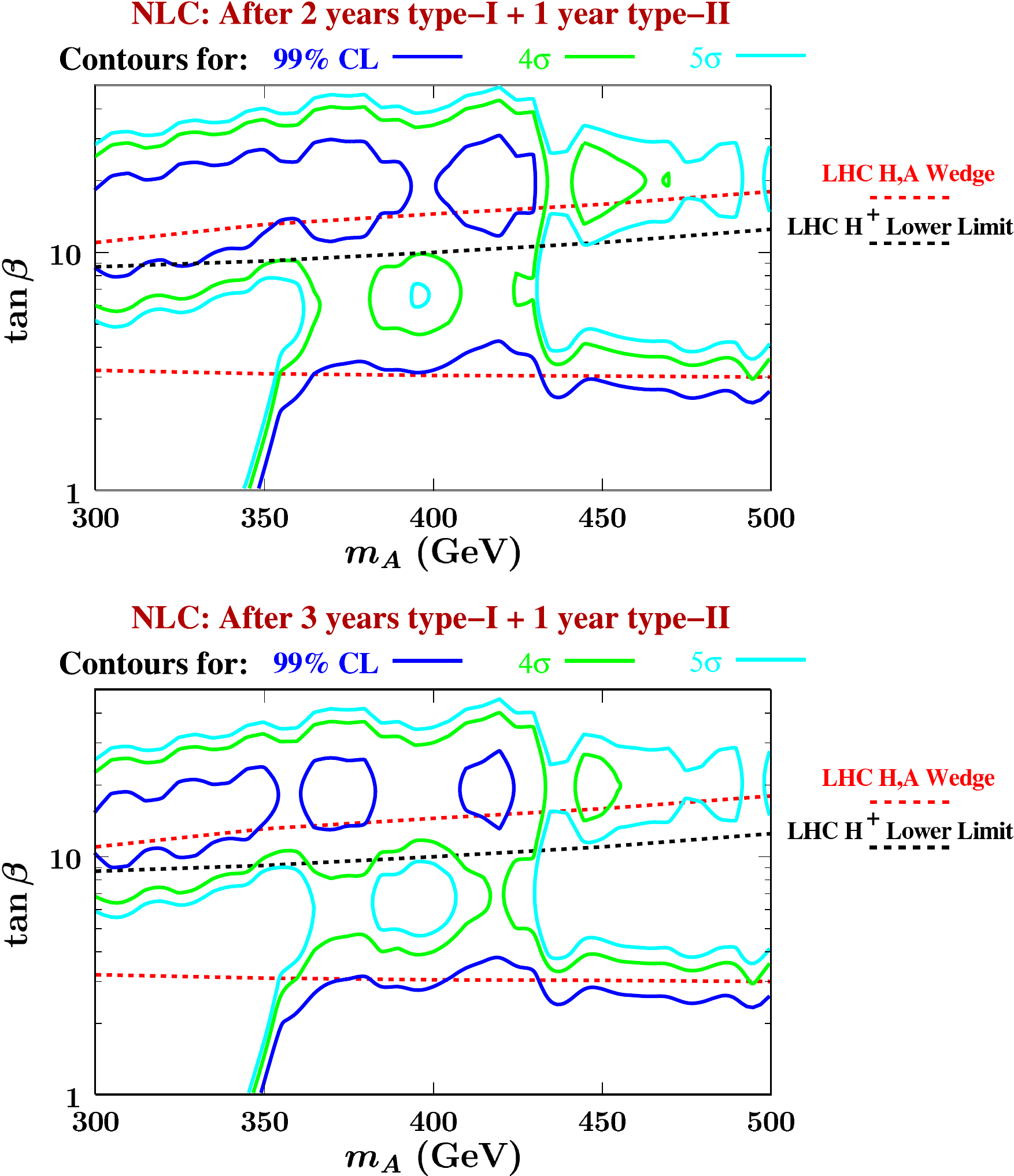}\hspace{7mm}
  \includegraphics[width=0.5\textwidth,height=0.3\textheight]{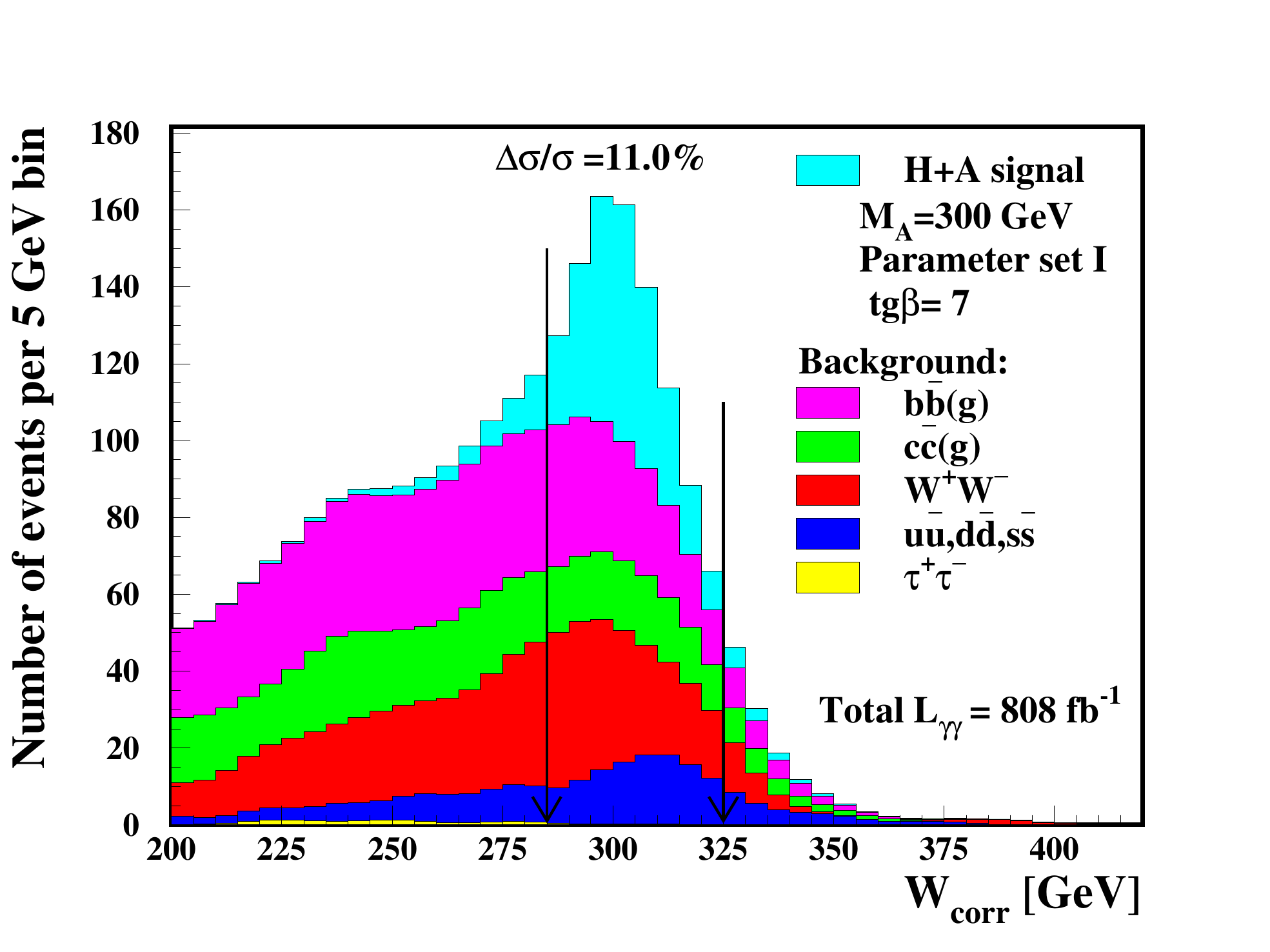}
  \caption{\it
Top: Production of  A and H, with  parameters  corresponding to
 the LHC wedge, at the $\gamma \gamma$ collider.
 Exclusion and discovery limits obtained for NLC collider for
$\sqrt {ee} =$630 GeV, after 2 or 3 years of operation~\cite{Asner:2001ia},
 Bottom: The case $M_H=M_A=300$ GeV at $\chi_V^H\approx 0$ in the MSSM. Distributions of the corrected invariant mass $W_{corr}$ for selected $b \bar{b}$ events  at   $\tan \beta=7$~\cite{nzk2005}.}
   \label{fig:wedge}
\end{figure}

Moreover, in MSSM  with $M_h=125$~GeV we { can } have heavy and
degenerate $H$ and $A$, $M_H\approx M_A$. At large $\tan\beta$  the
  discovery channel  of $H/A$ at LHC is $gg\to b\bar{b}\to b\bar{b}H/A$. Nevertheless,  in some region of parameters, at intermediate $\tan\beta$},  { these $H{\, {\rm and}\,}\,A$ are elusive at LHC.  That is so called {\it LHC wedge region}~\cite{wedgeHein}, see the  latest analysis~\cite{Carena:2013qia}.
The PLC allows to diminish this region of elusiveness, since  here the $H$ and $A$ production is generally not { strongly} suppressed and the $b\bar {b}$ background is under control~\cite{MAgaga,Asner:2001ia,nzk2005,Spira:2006aa}.  Figure~\ref{fig:wedge} show that PLC allows to observe joined effect of $H,\,A$ within this wedge region. Precision
between 11\% to 21 \% for $M_A$ equal to 200-300 GeV,  $\tan\beta$ = 7
of the Higgs-boson production measurement ($\mu$ =200 GeV (the
  Higgs mixing parameter) and $A_f$=1500~GeV (the trilinear
  Higgs-sfermion couplings)) can be reached after one year~\cite{nzk2005}.  To separate these resonances even in the limiting case $\chi_V^H=0$ is a difficult task,
since the total number of expected events  is small.

\bigskip  At $\chi_V^H\neq 0$, taking $\chi_V^H \sim  0.3-0.4$ as
  an example (what is  allowed by current  LHC measurement of  couplings of ${\cal H} = h$ to $ZZ$),   an  observation of $H\to ZZ$ decay channel can be good method for the  $H$ discovery in  2HDM.
The signal $\gamma\gamma\to H\to WW, ZZ$ interferes with background of $\gamma\gamma\to WW, ZZ$, what results in irregular structure in the effective-mass distribution of products of reaction $\gamma\gamma\to WW, ZZ$ (this interference is constructive and destructive below and above resonance, respectively). The study of this irregularity seems to be the   best method for discovery of heavy Higgs, decaying  to $WW,\, ZZ$~\cite{GIvWWH},  and to  measure the  corresponding $\phi_{\gamma\gamma}$ phase, provided it couples to $ZZ/WW$ reasonably strong\fn{Similar calculations given in~\cite{Niezurawski:2002jx} demonstrate this opportunity for a  2HDM version $B_{hu}$. 
}.

\begin{figure}
\begin{center}
\includegraphics[width=0.25\textwidth,height=0.2\textheight]{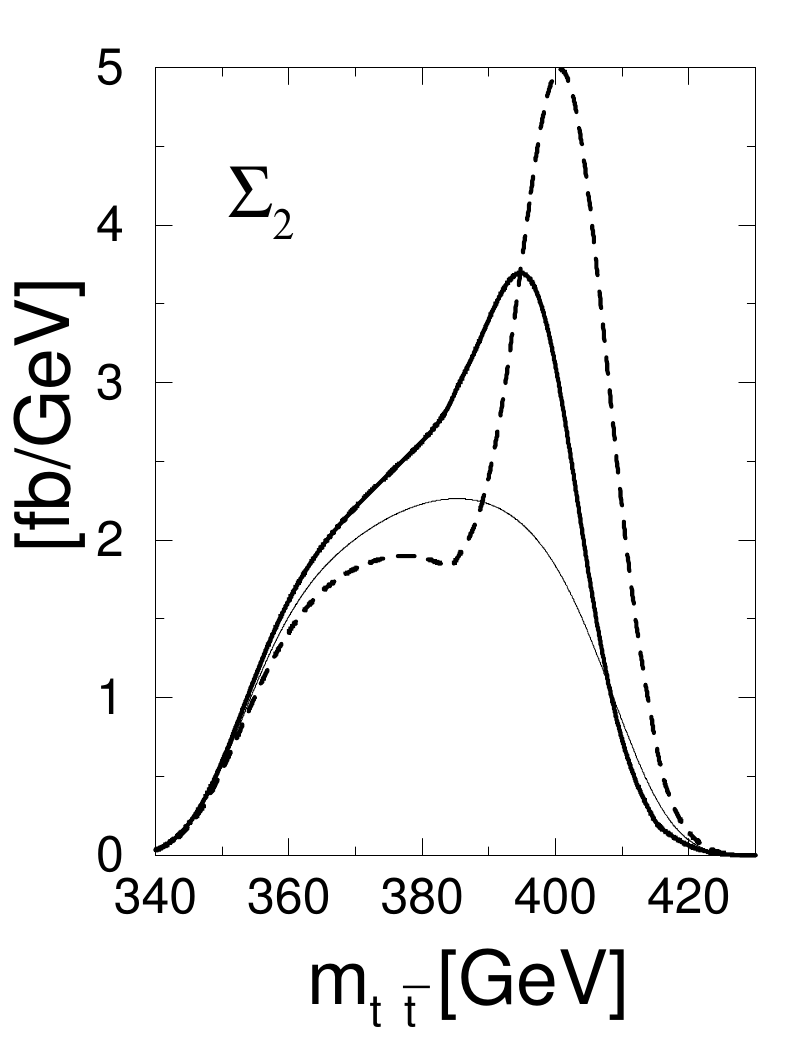}
 \caption{\it The specific  decay angular distributions $\Sigma_i$ in the $\gamma \gamma \rightarrow h^{(i)}\to t\bar{t}$ process in dependence on the $t\bar{t}$ invariant mass for the scalar (dashed)   and pseudoscalar (thick solid) $h^{(i)}$ with $M_H= 400$~GeV~\cite{Asakawa:2003dh}.}
 \label{fig:phys_h2}
\end{center}
 \end{figure}

\bigskip \   Just as it was described above for the observed 125 GeV  Higgs, PLC provides the best among  colliders place for the study of spin and the CP properties of heavy $h^{(2)}$, $h^{(3)}$. That are CP parity in the CP conserved case  (with ($h^{(2)}$, $h^{(3)}$ = ($H,\,A$)), and (complex) degree of   the admixtures of states with different CP parity, if CP is violated. This admixture determines dependence  on the Higgs production cross section  on direction of incident photon polarization~\cite{HCPH,gii,choi,Zerwas-CP,Choi:2002jk,Ellis:2004hw}. These polarization measurements are useful in the study of the case when the heavy states $h^{(2)}$, $h^{(3)}$ ($H,\,A$) are degenerated in their masses.
A study~\cite{nzk0710} shows that the 3-years operation of PLC with linear
polarization of photons,  the production cross-section of the $H$ and $A$
corresponding to the LHC wedge for MSSM (with mass $\sim 300$~GeV) can be separately measured with precison 20\%.
Pure scalar versus  pure pseudoscalar states can be distinguished at
the $\sim 4.5\,\sigma$ level.

We point out on important difference between the $CP$ mixed and the mass-degenerate states.  In the degeneracy of some resonances $A$ and $B$ one should distinguish two opportunities:
\begin{itemize}
\item[a)] instrumental degeneracy when  $|M_B-M_A|>\Gamma_B+\Gamma_A$, with mass difference  within a mass resolution of detector. This effect can be resolved with improving of a resolution of the detector
\item[b)] physical degeneracy when  $|M_B-M_A|<\Gamma_B+\Gamma_A$.
\end{itemize}
In the CP conserving case for both types of degeneracy the overlapping of $H,\,A$ resonances does not result in their mixing, and
the production of a resonante   state cannot vary with change  of sign of photon beam polarization. In the CP violating  case,  the overlapping of resonances results in additional mixing of incident $h^{(2)}$, $h^{(3)}$ states, and the production cross-section varies with  the change of  polarization direction of incident photons.

\bigskip \ Another method for study of CP content of a produced particle provides the measurement of angular distribution of  decay 
products~\cite{Niezurawski:2004ga,Grzadkowski:1992sa,Godbole:2006eb}.
In the  $t \bar t$ decay mode one can perform a study of the CP-violation, exploiting fermion polarization.
The interference between the Higgs exchange and the continuum amplitudes can be sizable for the polarized photon beams, if helicities of the top and anti top quarks are measured. This enables  to determine the CP property of the Higgs boson completely~\cite{Asakawa:2000jy,Asakawa:2003dh},  Fig. \ref{fig:phys_h2}.\\

\begin{sloppypar}
The discovery of charged Higgses $H^\pm$ will be a crucial
signal of the BSM form of Higgs sector. These particles can be produced
both at  the LC ($e^+e^-\to H^+H^-$) and at  the  PLC ($\gamma\gamma \to
H^+H^-$). These processes  are described well by QED. The $H^+H^-$
production process at PLC has worse energy-threshold behaviour than the
corresponding process at the LC, but higher cross section. On the other
hand, the process $e^+e^-\to H^+H^-$ can be analysed at LC better by
measurements  of decay products due to known  kinematics. At the PLC the
variation of a initial-beam polarization could be  used for checking up
spin of $H^\pm$~\cite{ilya2}. See also analysis for flavour violation
models  in~\cite{Martinez:2008hu,Cannoni:2008bg}. \\
\end{sloppypar}

\begin{sloppypar}
After a  $H^\pm$ discovery,  the observation of processes $e^+e^-\to H^+H^-h$ and $\gamma \gamma \to H^+H^-h$, $H^+H^-H$, $H^+H^-A$  may provide direct  information on  a triple Higgs ($H^+H^-h$) coupling $\lambda$,  with  cross sections in both cases  $\propto \alpha^2\lambda^2$.
The $\gamma\gamma$ \,\, collisions are preferable here due to a substantially higher cross section and opportunity of study   polarization effects in the production process via variation of initial photon polarizations.
\end{sloppypar}

\bigskip \  Synergy of  LHC, LC and PLC  colliders may be useful in
determination of  Higgs couplings, as different production processes
dominating at these colliders  lead to different sensitivity to gauge and
Yukawa couplings. For example  LC Higgstrahlung leads to  large sensitivity to
the Higgs coupling to the EW gauge bosons, while at  PLC $\gamma \gamma$ and
$Z\gamma$  loop couplings depend both on the Higgs gauge and Yukawa couplings,
as well as on coupling with $H^+$,  see results both for  CP conservig/CP
violating in e.g.~\cite{Godbole:2006eb,Heinemeyer:2005uf, Niezurawski:2006hy}.
\vskip 1cm

%






\def\had{\mbox{\scriptsize had}}
\def\as{\ensuremath{\alpha_s}}
\def\SM{Standard Model}
\section[Top and QCD]{Top and QCD\protect\footnotemark}
\footnotetext{Authors: Frank Simon, Peter Uwer, Kiyo Yuichiro}
\label{Top-QCD}

\subsection{Introduction}
\label{sec:Top-QCD-Intro}
The experimental studies of electron-positron annihilation into hadrons
were historically essential to establish Quantum Chromodynamics (QCD)
as {\it the} theory of strong interaction: from the measurement of
the $R$-ratio ${\sigma_{\had}/\sigma_t}$ the number of colors could
be determined, the discovery of three-jet events at PETRA provided the
first direct indication of the gluon, and the measurement of the
Bengtson-Zerwas and Nachtmann-Reiter angles illustrated the non-abelian
gauge structure of QCD ---to name only a few milestones on the road to
develop the theory of strong interactions.

At the Large Electron Positron Collider (LEP) the experimental tests
of QCD were further refined. Three, four, and even five-jet rates
were measured with unprecedented accuracy. These measurements provided
important input to constrain the structure constants of the underlying
non-abelian gauge group and to determine the QCD coupling constant
$\as$ with high precision. The $R$-ratio and the forward-backward
asymmetry were studied in detail including precise investigations of the
flavor (in-)dependence. At SLD the measurements were extended to
polarized electrons in the initial state. The tremendous experimental
effort has been complemented over the time by a similar effort on the
theory side: Next-to-leading order (NLO) calculations have been
performed for event-shape observables and jet-rates involving jets
originating from massless as well as massive quarks.  New
jet-algorithms with an improved theoretical behavior were developed.
Very recently theoretical predictions for three-jet rates have been
extended to next-to-next-to-leading order (NNLO) accuracy.  For
inclusive hadron production the theoretical predictions have been
extended to N$^3$LO accuracy in QCD. Beyond fixed order perturbation
theory also power corrections and soft gluon resummation have been
considered. All this effort has paved the way to establish QCD as the
accepted theory of strong interaction.

Today QCD is a mature theory and no longer the primary target of
experimental studies. Assuming QCD as the underlying theory of strong
interaction the precision measurements possible in $e^+e^-$
annihilation can be used to determine fundamental parameters like
coupling constants and particle masses. For example three-jet rates at
LEP have been used to measure the QCD coupling constant and the
$b$-quark mass. Since the small $b$-quark mass leads only to effects
of the order of 5\% at the $Z$-resonance (compared to massless
$b$-quarks), this example nicely illustrates the impressive
theoretical and experimental precision reached. The steadily
increasing experimental accuracy together with LHC as a ``QCD
machine'' and the perspective of a future linear collider have kept
QCD a very active field, where significant progress has been achieved
in the last two decades.  Conceptually effective field theories have been
further developed with specific realizations for dedicated
applications. For example, soft collinear effective theory (SCET) is
nowadays used to systematically improve the quality of the
perturbative expansion through the resummation of logarithmically
enhanced contributions. SCET may also help to deepen our current
understanding of factorization of QCD amplitudes.  Applications to the
production of top-quark pair production have also demonstrated the
power of this approach to assess the impact of non-perturbative
corrections. Non-relativistic QCD (NRQCD) provides the well
established theoretical framework to analyze the threshold production
of top-quark pair production where binding effects between top quarks
are important. The theoretical description of unstable particles in
the context of effective field theories have demonstrated another
successful application of effective field theories.  Theoretical
predictions for a future Linear Collider will profit from the improved
theoretical understanding in terms of an increased precision. Recently
we have witnessed a major breakthrough in the development of
technologies for one-loop calculations. One-loop calculations
involving multiplicities of five or even more particles in the final
state ---which were a major bottleneck over several years in the past---
are today regularly performed for a variety of different
processes. The new techniques have also led to an increased automation
of the required calculations. Various programs are now publicly
available to generate NLO matrix elements.
Furthermore a standardized interface allows the phase space
integration within MC event generators like for example Sherpa.  Also
the two-loop technology has seen important progress and is now a
continuously growing field. The description of threshold effects in
the production of heavy particles notably heavy quarks has been
further improved to include higher order corrections in the
perturbative expansion.

The detailed understanding of QCD achieved today has been proven
essential for the current interpretation of LHC results and the very
precise measurements performed so far. Evidently LHC data can also be
used for QCD studies in the TeV regime. However owing to the
complicated hadronic environment it will be difficult to reach
accuracies at the per cent level or even below. In contrast $e^+e^-$ 
Linear Colliders allows to test QCD at the sub per cent level at
energies above the $Z$ resonance. The reachable precision of any
measurement involving strongly interacting particles will depend on
the ability of making accurate predictions within QCD. QCD studies
will thus continue to play an important role at a future Linear
Collider.  Since non-perturbative effects are intrinsically difficult
to assess, the highest accuracy ---and thus the most precise tests of
the underlying theory--- can be reached for systems, where these
effects are believed to be small or even negligible. A particular
interesting example is provided by top-quark physics. With a mass
almost as heavy as a Gold atom the top quark is the heaviest
elementary fermion discovered so far.

\begin{sloppypar}
Top-quarks have unique properties making them a highly interesting
research topic on their own right. The large mass leads to an extremely
short life time such that top quarks decay before they can form
hadronic bound states. This simple observation has several important
consequences.  First of all the finite width essentially cuts off
non-perturbative physics such that top-quark properties can be
calculated with high accuracy in perturbative QCD.  Top-quark physics
thus allows to study the properties of a `bare quark'.  In the \SM \ 
top quarks decay almost exclusively through electroweak interactions
into a $W$-boson and a $b$-quark. The parity violating decay offers
the possibility to study the polarization of top-quarks through the
angular distribution of the decay products.  Polarization studies,
which are difficult in the case of the lighter quarks since
hadronization usually dilutes the spin information, offer an
additional opportunity for very precise tests of the underlying
interaction.  This is of particular interest since top-quark physics
is controlled in the \SM\ by only `two parameters': The top-quark mass
and the relevant Cabbibo-Kobayashi-Mas\-ka\-wa matrix elements. Once these
parameters are known top-quark interactions are predicted through the
structure of the \SM. In particular all the couplings are fixed
through local gauge invariance. Top-quark physics thus allows to test the
consistency of the \SM\ with high precision. A prominent example is the
relation between the top-quark mass and the mass of the $W$-boson.
Obviously the accuracy of such tests is connected to the precision
with which the top-quark mass ---as most important input parameter--- can
be determined.  While the LHC achieved already an uncertainty in the
mass measurements of one GeV, it is expected that a Linear Collider
will improve this accuracy by an order of magnitude down to 100 MeV or
even below.
Using top-quarks to test the \SM\ with high precision and search for
new physics is very well motivated. In addition to the high
experimental and theoretical accuracy achievable in top-quark
measurements, top-quarks provide a particular sensitive probe to search
for \SM\ extensions.  Due to its large mass top-quarks are very
sensitive to the mechanism of electroweak symmetry breaking. In many
extensions of the \SM\ which aim to present an alternative mechanism
of electroweak symmetry breaking top-quarks play a special role. It
is thus natural to ask whether the top-quark mass, being so much
larger than the masses of the lighter quarks, is indeed produced by
the Englert-Brout-Higgs-Guralnik-Hagen-Kibble mechanism.  
A detailed measurement of the top-quark Yukawa
coupling to the Higgs boson, which is very difficult to assess at a
hadron collider, will provide a crucial information to answer this
question. In the past top-quarks have been extensively studied at the
Tevatron and the LHC.  With exception of the forward-backward charge
asymmetry studied at the Tevatron the measurements are in very good
agreement with the \SM\ predictions.  However it should be noted that due to the
complex environment at a hadron collider the accuracy is often
limited. The top-quark mass which is now measured with sub per cent
accuracy represents an important exception. While the measurements at
the Tevatron and the LHC are perfectly consistent the precise
interpretation of the measured mass value in terms of a renormalized
parameter in a specific scheme is still unclear. The mass which is
determined from a kinematical reconstruction of the top-quark decay
products is assumed to be close to the pole mass. Since precise
theoretical predictions for the measured observable are lacking the
exact relation between the measured mass and the pole mass has not been
quantified so far. An alternative method in which the mass is
determined from cross section measurements where the renormalization is
uniquely fixed through a higher order calculation gives consistent
results. However the experimental uncertainties of this method are
quite large owing to the weak sensitivity of the total cross section
with respect to the top-quark mass. A new method using top-quark pair
production in association with an additional jet represents an
interesting alternative but will most likely also be limited in
precision to one GeV. Although not better in precision the advantage
of this method lies in the fact that the method gives a clear
interpretation of the measured value in a specific renormalization
scheme. Given the importance of a precise determination of the
top-quark mass, going significantly below one GeV may remain the task of a
future Linear Collider.
\end{sloppypar}

In the following we shall briefly describe in Section
\ref{sec:Top-QCD-QCD} recent progress in QCD with a special emphasis
on $e^+e^-$ annihilation. In Section \ref{sec:Top-QCD-Threshold} we
summarize new developments in top-quark physics in particular
concerning the theoretical understanding of top-quark production at
threshold. In the last Section we briefly comment
on the physics potential of a future linear collider with respect to
QCD and top-quark physics. In particular the prospects of a precise
measurement of the top-quark mass are discussed.

\subsection{Recent progress in QCD} \label{sec:Top-QCD-QCD}
\subsubsection{Inclusive hadron production}
The inclusive cross section for the production of hadrons in $e^+e^-$
annihilation or alternatively the $R$-ratio is a fundamental
observable to be studied at any $e^+e^-$ collider. For hadrons
originating from the fragmentation of massless quarks substantial
progress has been obtained over the last 10 years. Starting from the
$n_f^2\as^4$ contribution presented in Ref.~\cite{Baikov:2001aa} more
than ten years ago the full N$^3$LO result including all color structures have been
derived over the last decade in a ground breaking calculation
\cite{Baikov:2002va,Baikov:2008jh,Baikov:2012er}. 
Using $\sin^2\theta_W=0.231$ for the sine squared of the weak mixing angle
the result for the hadronic decay width of the Z-boson reads \cite{Baikov:2012er}:
\begin{eqnarray}
  \Gamma_Z &=& \frac{G_F M_Z^3}{24\pi\sqrt{2}} R^{nc}\\
  R^{nc} &=&  20.1945 + 20.1945\, a_s \nonumber \\
  &+& (28.4587 - 13.0575 + 0)\, a_s^2 \nonumber \\
  &+& (-257.825 - 52.8736 - 2.12068)\, a^3_s \nonumber \\
  &+& (-1615.17 + 262.656 - 25.5814)\, a^4_s,
\end{eqnarray}
with $a_s=\as(M_Z)/\pi$. The three terms inside the brackets display the
non-singlet, axial singlet and vector singlet contributions.  An
important application of the improved theoretical description is the
determination of the QCD coupling constant. It is thus interesting to
investigate the impact of the newly calculated correction on the
determined \as\ value. For $\as(M_Z)=0.1190$ the impact of the 4-loop
correction on the extracted \as\ value is found to be very small. A
shift $\delta\as=-0.00008$ in the $\as$ value when extracted from the
hadronic cross section is expected.  For the quality of the
perturbative expansion not only the size of the corrections is
important but also the residual renormalization scale dependence. In
Ref.~\cite{Baikov:2012er} it has been shown that the scale dependence
is also improved by including the four-loop contributions.  As far as
the order in the QCD coupling constant is concerned the $R$ ratio is
certainly one of the best known QCD observables. 

\subsubsection{Three-jet production at NNLO}
Jet production in $e^+e^-$ annihilation is another classical QCD observable.
The underlying physical picture explaining the outgoing bundles of
hadrons called {\it jets}, is the production of colored high energetic
partons in a short distance process. The partons are then assumed to
fragment into uncolored hadrons.  As a consequence, the naive
expectation is, that the fragmentation products somehow share the
momentum of the mother parton. This simple picture is reflected in
iterative jet algorithms which try to bridge the gap between the
experimentally observed hadrons and the partonic final states
used in the theoretical predictions. To make contact between theory
and experiment, in both analysis the same jet algorithms are applied
and the results are compared. In Born approximation the number of
partons is equal to the number of jets. In this case each jet is thus
`modeled' by a single parton. Including additional real radiation in
higher order predictions allows for the recombination of two or even
more partons
into one jet and gives thus an improved theoretical description of the 
jets. Three jet production in $e^+e^-$ annihilation is of particular
interest since the three jet rate is directly proportional to
the coupling constant of strong interaction.  Until recently the
precision of \as\ extracted from three jet rates was limited due to
the unknown NNLO corrections. The main problems which had to be
overcome were the evaluation of the two-loop amplitudes for the
process $e^+e^-\to(Z^\ast,\gamma^\ast)\to q\bar q g$ and the
systematic cancellation of mass and infrared singularities present in
individual contributions. The former problem was solved in
Refs.~\cite{Garland:2001tf,Garland:2002ak,Moch:2002hm}. The highly
non-trivial combination of virtual corrections, real emission at
one-loop order, and double real emission took another five years until
completion. Predictions for different observables at NNLO accuracy in
QCD have been
presented in Refs.~\cite{GehrmannDeRidder:2007bj,GehrmannDeRidder:2007hr,%
GehrmannDeRidder:2008ug,Weinzierl:2008iv,GehrmannDeRidder:2009dp,%
Weinzierl:2009ms,Weinzierl:2009yz,Weinzierl:2010cw} by two competing
groups. The fixed order NNLO calculation lead to a 10 \% smaller
central value for $\as$ \cite{Dissertori:2007xa}. In addition the
inclusion of the NNLO corrections reduce the variation in \as\ 
extracted from different event-shape observables. The NNLO corrections
thus lead to a more coherent description of the data. Furthermore the
scale uncertainty is reduced by a factor of two compared to the NLO
calculation. However the scale uncertainty still dominates the
extraction of \as\ when compared to uncertainties due to finite
statistics and hadronization. The scale uncertainty is roughly three
times larger than the uncertainty due to hadronization.  In
Ref.~\cite{Dissertori:2009ik} the fixed order NNLO predictions have
been extended by resumming large logarithmic corrections due to
multiple soft gluon emission at next-to-leading logarithmic accuracy (NLLA).
It turns out, that the resummation has very little impact on the
central value of \as\ determined from different event shapes. However
the theoretical uncertainties due to scale variations are slightly
increased.  As a final result
\begin{eqnarray}
  \alpha_s(M_Z) &=& 0.1224 \pm 0.0009
  \mbox{ (stat.) } \pm 0.0009 \mbox{ (exp.) } \nonumber\\
  &\pm& 0.0012 \mbox{ (had.)
    } \pm 0.0035 \mbox{ (theo.) } 
\end{eqnarray}
is quoted \cite{Dissertori:2009ik}.  Evidently the NNLO
predictions will also find their application at a future linear
collider. Even with a limited statistics a future measurement above
the Z resonance will be interesting due to the possibility to further
constrain \as\ at high scale. It is also conceivable that the
theoretical uncertainties are slightly reduced at higher energies due to the
smaller value of \as.

\def\ycut{y_{\mbox{\scriptsize cut}}}
\subsubsection{NLO QCD corrections to 5-jet production and beyond}
At the LEP experiments exclusive production of jet multiplicities up
to five jets were studied experimentally. However until very recently
only NLO results for four-jet production were available due to the
tremendous growth in complexity of the theoretical calculations.  In
Ref.~\cite{Frederix:2010ne} the NLO QCD corrections to five-jet
production are presented.
\begin{sloppypar}
The virtual corrections were calculated using generalized unitarity
(for more details about this method we refer to the next Section
\ref{sec:NLOProgress}), relying to a large extend on amplitudes
calculated in Ref.~\cite{Ellis:2008qc} where one-loop corrections to
$W^+ + \mbox{3-jet}$ production in hadronic collisions were studied.
The real corrections are calculated using MadFKS
\cite{Frederix:2009yq}---an implementation of the Frixione-Kunszt-Signer
(FKS) subtraction scheme \cite{Frixione:1995ms} into Madgraph.  The
Durham jet algorithm is used to define the jets. Results for the
five-jet rate, differential with respect to the parameter $y_{45}$,
which determines the $\ycut$-value at which a five-jet event becomes a
four-jet event, are shown. Furthermore the five-jet rate as function
of the jet resolution parameter $\ycut$ is presented. In addition
hadronization corrections are analyzed using the Sherpa event
generator. At fixed order in perturbation theory it is found that
the scale uncertainty is reduced from
about $[-30\%,+45\%]$ in LO to about $[-20\%,+25\%]$ in
NLO. In this analysis the renormalization scale has been chosen to be
$\mu=0.3\sqrt{s}$ and variations up and down by a factor of two were
investigated. The central scale is chosen smaller than what is usually
used for lower jet multiplicities. The reasoning behind this is that
for increasing multiplicities the average transverse momentum per jet
becomes smaller. This is taken into account by using $\mu=0.3\sqrt{s}$
instead of the more common setting $\mu=\sqrt{s}$. It would be interesting to compare with a dynamical
scale like $H_T$, the sum of the `transverse energies', which has been
proven in four- and five-jet production at hadron colliders to
be a rather useful choice
\cite{Bern:2011ep,Badger:2012pf,Badger:2013yda}.  Using in LO \as=0.130
and in NLO \as=0.118 NLO corrections of the order of 10--20\% are found.
It is noted that using the same value of \as\ in LO and NLO would
amount to corrections at the level of 45--60\%. Including
hadronization corrections through Sherpa the theoretical results are
used to extract \as\ from the experimental data. As final result
$\alpha_s(M_Z) = 0.1156^{+0.0041}_{-0.0034}$ is quoted which is well
consistent with the world average and also shows the large potential
of \as\ measurements using jet rates for high multiplicities: The
uncertainty is similar to the \as\ determinations from three-jet rates
using NNLO+NLLA predictions \cite{Dissertori:2009ik}.  As an
interesting observation it is also pointed out in
Ref.~\cite{Frederix:2010ne} that hadronization corrections calculated
with standard tools like HERWIG, PYTHIA and ARIADNE are typically
large and uncertain unless the tools are matched/tuned to the
specific multi-jet environment. It is suggested to use in such cases event
generators like SHERPA which incorporates high-multiplicity matrix
elements through CKKW matching.\\
\end{sloppypar}
Recently an alternative method to
calculate one-loop corrections has been used to calculate the
NLO corrections for six- and seven-jet production. The method
developed in
\cite{Soper:1999xk,Soper:2001hu,Nagy:2003qn,Nagy:2006xy,Anastasiou:2007qb,Gong:2008ww,Assadsolimani:2009cz,Becker:2010ng}
combines the loop integration together with the phase space
integration. Both integrations are done together using Monte Carlo
integration. Since the analytic structure of the one-loop integrand is
highly non-trivial special techniques have to be developed to enable a
numerical integration. In Ref.~\cite{Becker:2011vg} this technique has
been applied to the NLO calculation of the six- and five-jet rate in
leading color approximation. No phenomenological studies are
presented. It is however shown that the method offers a powerful 
alternative to existing approaches.

\subsubsection{Progress at NLO}
\label{sec:NLOProgress}
An essential input for NLO calculations are the one-loop corrections.
Four momentum conservation at each vertex attached to the loop does
not fix the momentum inside the loop. As a consequence an additional
integration over the unconstrained loop momentum is introduced. Since
the loop momenta appears not only in the denominator through the
propagators but also in the numerator in general tensor integrals have
to be evaluated. The traditional method to deal with these tensor
integrals is the so-called Passarino-Veltman reduction which allows to
express the tensor integrals in terms of a few basic scalar one-loop
integrals \cite{Passarino:1978jh}. All relevant scalar integrals have
been calculated and can be found for example in
Refs.~\cite{vanOldenborgh:1989wn,Ellis:2007qk,vanHameren:2010cp}.  In
practical applications the Passarino-Veltman reduction procedure may
lead to large intermediate expressions when applied analytically to
processes with large multiplicities or many different mass scales.
An alternative to overcome this problem is to apply the reduction
procedure numerically. In this case however numerical instabilities
may appear in specific phase space regions where the scalar one-loop
integrals degenerate for exceptional momentum configurations.
Approaching these exceptional momentum configurations the results
behave as ``0/0''. Evaluating the limit
analytically one finds a well defined result. The numerical evaluation
however will typically lead to instabilities unless special precautions are
taken to deal with these configurations.
In the past various approaches have been developed to
stabilize the numerical evaluation of exceptional momentum
configurations. Details can be found for example in
Refs.~\cite{Binoth:1999sp,Giele:2004iy,Giele:2004ub,Denner:2005nn,Binoth:2008uq,Fleischer:2010sq,Heinrich:2010ax,Fleischer:2011nt,Fleischer:2011hc,Guillet:2013mta}
and references therein. With the steadily increasing computing power
of modern CPU's today an alternative approach is frequently
used: Instead of stabilizing the numerical evaluation it is checked
during the numerical evaluation whether instabilities were
encountered. If this is the case the numerical evaluation of the
respective phase space point is repeated using extended floating point 
precision. The price to pay in this approach is a slightly increase of
computing time which is however affordable as long as the fraction of points
needed to be recomputed remains small.  

Beside the numerical evaluation of tensor integrals the significant
increase in complexity when studying virtual corrections for processes
with large multiplicities is another major bottleneck of one-loop
calculations. Here the recently developed method of generalized
unitarity may provide a solution. The starting point of this method is
the observation that any one-loop amplitude can be written in terms of
scalar one-point, two-point, three-point and four-point one-loop
integrals. No higher point scalar integrals are required. This
observation is a direct consequence of the Passarion-Veltman reduction
procedure. Starting from this observation one can
reformulate the problem of one-loop calculations: How do we calculate
most efficiently the coefficients in this decomposition? One answer to
this question is the method proposed by Ossola, Papadopoulos, Pittau
(OPP) \cite{Ossola:2006us}. The idea of this method is to perform a
decomposition at the integrand level: The integrand is decomposed into
contributions which integrate to zero or lead to scalar integrals. To
derive the decomposition at integrand level internal propagators are
set on-shell. As a consequence the integrand factorizes into a product
of on-shell tree amplitudes. For more details about the method of
generalized unitarity we refer to the recent review of Ellis, Kunszt,
Melnikov and Zanderighi \cite{Ellis:2011cr}. From the practical point
of view the important result is that the algorithm can be implemented
numerically and requires as input only on-shell tree amplitudes. For
on-shell tree amplitudes very efficient methods to calculate them,
like for example the Berends-Giele recursion, exist
\cite{Berends:1987me}. In principle it is also possible to use
analytic results for the tree-level amplitudes or apply on-shell
recursions \`a la Britto, Cachazo, Feng, and Witten ((BCFW) see for
example Ref.~\cite{Britto:2005fq}). Using tree amplitudes instead of
individual Feynman diagrams helps to deal with the increasing
complexity of processes for large multiplicities.  It may also lead to
numerically more stable results since the tree amplitudes are gauge
invariant and gauge cancellation ---usually occuring in Feynman
diagramatic calculations--- are avoided.  The enormous progress made
recently is well documented in the increasing number of publicly
available tools to calculate one-loop amplitudes, see for example
Refs.~\cite{Badger:2010nx,Hirschi:2011pa,Bevilacqua:2011xh,Cullen:2011ac,Badger:2012pg,Bern:2013pya}.
As can be seen from recent work e.g.
Refs.~\cite{Cascioli:2011va,Becker:2011vg,Actis:2012qn} further
progress can be expected in the near future (for the method discussed
in Ref.~\cite{Becker:2011vg} see also the discussion at the end of the
previous section).  As mentioned already the calculation of real
emission processes can be considered as a solved problem since very
efficient algorithms to calculate the required Born matrix elements
are available. In principle also the cancellation of the infrared and
collinear singularities appearing in one-loop amplitudes as well as in
the real emission processes can be considered as solved.  General
algorithms like Catani-Seymour subtraction method \cite{Catani:1996jh}
or FKS subtraction \cite{Frixione:1995ms} exist to perform the
required calculation. Also here significant progress has been obtained
in the recent past towards automation. The required subtractions can
now be calculated with a variety of publicly available tools
\cite{Frederix:2009yq,Gleisberg:2007md,Seymour:2008mu,Frederix:2008hu,Hasegawa:2009tx}.
While most of the aforementioned tools have been applied recently to
LHC physics it is evident that an application to $e+e-$ annihilation
is also possible. It can thus be assumed that for a future Linear
Collider all relevant NLO QCD corrections will be available.

\subsection{Recent progress in top-quark physics}\label{sec:Top-QCD-Threshold}
\def\thetaw{{\theta_W}} In the Standard Model the top quark appears in
the third family as up-type partner of the bottom quark. As missing
building block of the third family the existence of the top quark was
predicted long before its discovery in 1994. Top-quark interactions
are fixed through the gauge structure of the Standard Model. The
coupling strengths follow from the local $SU(3)\times SU(2)_L\times
U(1)_Y$ gauge invariance. In particular the QCD coupling to the gluons
is the same as for the lighter quarks. The coupling to the $Z$-boson
involves vector and axial vector couplings while the coupling to the
$W$-boson is of $V-A$ type. The couplings can be expressed in terms of
the third component of the weak iso-spin $T_3$, the hyper charge $Y$
(or alternatively the electric charge $Q$) and the weak mixing angle
$\thetaw$. For example the coupling to the $Z$-boson reads:
\begin{equation}
  -i {e\over \sin \thetaw
    \cos\thetaw}\left(T_3\gamma^\mu{1\over2}(1-\gamma_5)
  -\sin^2\thetaw Q\gamma_\mu\right).
\end{equation}
\begin{sloppypar}
As a matter of fact top-quark specific aspects or more general flavor
dependencies enter only through the top-quark mass and the
Cabbibo-Kobayashi-Maskawa (CKM) matrix which relates the mass eigenstates
and the eigenstates of the weak interaction. Assuming three
families and unitarity the CKM matrix elements are highly constrained
from indirect measurements. A global fit of available flavor data
gives \cite{PDG2012}:
\end{sloppypar}
\begin{eqnarray}\small
 && \mbox{\hspace{-.3cm}}V =\nonumber\\
  &&\mbox{\hspace{-.4cm}}\small\left(
    \begin{array}[h]{ccc}
      0.97427 \pm 0.00015 & 0.22534 \pm 0.00065 & 
      0.00351^{+0.00015}_{-0.00014}\\[0.2cm]
      0.22520 \pm 0.00065 & 0.97344 \pm 0.00016 &
      0.0412^{+0.0011}_{-0.0005}
      \\[0.2cm]
      0.00867^{+0.00029}_{-0.00031}& 0.0404^{+0.0011}_{-0.0005} &
      0.999146^{+0.000021}_{-0.000046}
    \end{array}
\right)\nonumber\\
\end{eqnarray}
\begin{sloppypar}
Very recently $V_{tb}$ has been determined also in direct
measurements using single--top-quark production at Tevatron and
LHC. Combining the various measurements the Particle Data Group quotes
\cite{PDG2012}:
\end{sloppypar}
\begin{equation}
   |V_{tb}|= 0.89 \pm 0.07.
\end{equation}
\begin{sloppypar}
The result is consistent with the indirect measurements. However, the
complicated experimental environment leads to large
uncertainties.
Further improvements can be expected from future measurements at the LHC.\\
The top-quark mass has been measured at the Tevatron and the LHC with
various techniques. At the Tevatron a combination
\cite{Aaltonen:2012ra} of various D0 and
CDF measurements gives 
\end{sloppypar}
\begin{equation}
 M_t = 173.18\pm 0.56\mbox{ (stat.) } \pm 0.75
\mbox{ (syst.)  }\mbox{GeV} .
\end{equation}
The
measurements performed at the LHC are in perfect agreement with the Tevatron
results. For example CMS \cite{Chatrchyan:2012cz} finds using lepton+jets final states
\begin{equation}
  173.49 \pm 0.43 \mbox{ (stat.+JES) } \pm 0.98 \mbox{ (syst.) } 
  \mbox{GeV}. 
\end{equation}
Strictly speaking the renormalization scheme of the experimentally
determined mass parameter is not properly fixed using a kinematic
reconstruction of the top-quark mass. Nevertheless it is usually assumed that
the aforementioned mass values correspond to the
so-called on-shell/pole mass. \\
From the precise knowledge of the CKM matrix elements and the
top-quark mass all other properties can be predicted within the \SM.
Given the large value of $V_{tb}$ the dominant decay of the top-quark
assuming the Standard Model is the decay into a $W$-boson and a
$b$-quark.  In LO the top-quark decay width
is given by
\begin{eqnarray}
\Gamma(t\to bW)&=& \frac{G_F|V_{tb}|^2 M_t^3  }{8\sqrt{2}\pi}
\left(1-\frac{M_W^2}{M_t^2} \right)^2
\left(1+\frac{2M_W^2}{M_t^2}\right).\nonumber\\
\end{eqnarray}
Higher order electroweak and QCD corrections to the width have been
calculated as detailed in the following. In
Refs.~\cite{Denner:1990ns,Eilam:1991iz} the electroweak one-loop
corrections have been calculated. The NNLO
QCD corrections are known for $M_W = 0$ \cite{Czarnecki:1998qc} and
$M_W\not=0$ \cite{Chetyrkin:1999ju}. Including the radiative
corrections the top quark decay width is approximately
$\Gamma_t\approx 1.4 \, {\rm GeV}$. As mentioned earlier the life time
is thus almost an order of magnitude smaller than the typical time
scale for hadronization. The top-quark thus decays without forming hadrons.

\subsubsection{Top-quark decays at next-to-next-to-leading order QCD}
\label{sec:top-quark-decay}
\begin{sloppypar}
In Refs. \cite{Czarnecki:1998qc,Chetyrkin:1999ju} only the NNLO QCD
corrections to the inclusive decay width were calculated. The
calculation for massless $W$-bosons of Ref.~\cite{Czarnecki:1998qc}
has been extended in Ref. \cite{Chetyrkin:1999ju} to include also the
effects of the finite $W$-boson mass through an expansion in
$M_W^2/M_t^2$. These results have been extended recently in various
directions. In Ref.~\cite{Czarnecki:2010gb} the partial decay widths
for top-quarks decaying into polarized $W$-bosons is investigated.
The partial decay widths are particular interesting since the
polarization of the $W$-boson allows to test the $tWb$ vertex
independently from the top-quark production mechanism. Assuming
massless $b$-quarks the $V-A$ nature of the charged currents forbids
the decay into right-handed $W$-bosons in LO. The measurement of the
$W$-polarization in top-quark decays thus provides a sensitive tool to
test the $V-A$ structure and to search for possible extensions of the
Standard Model.  Obviously a finite $b$-quark mass leads to calculable
corrections.  Evidently also higher order corrections which include in
general also real emission processes can alter the LO predictions. It
is thus very important to calculate the branching fractions
\end{sloppypar}
\begin{equation}
  \label{sec:Top:PartialWidth}
  f_{\pm} = {\Gamma_\pm\over \Gamma(t\to Wb)},
  \quad f_L = {\Gamma_L\over \Gamma(t\to Wb)} 
\end{equation}
where $\Gamma(t\to Wb)$ denotes the inclusive top-quark decay width
and $\Gamma_{-/+}$ ($\Gamma_L$) denote the decay width into left/right handed
(longitudinally) polarized $W$-bosons. Similar to what has been done
in previous work an expansion in $x=M_W/M_t$ is used in 
Ref.~\cite{Czarnecki:2010gb} to calculate the partial decay width in
NNLO QCD. For $\as(M_Z) = 0.1176$ and
$M_Z=91.1876$ GeV the results read
\begin{eqnarray}
  F_L & = & 0.6978 - 0.0075 - 0.0023, \\
  F_+ &=& 0 + 0.00103 + 0.00023, \\
  F_- &=& 0.3022 + 0.0065 + 0.0021,
\end{eqnarray}
where the individual terms correspond to the LO, NLO and NNLO
prediction. Note that the ratios in Eq.~(\ref{sec:Top:PartialWidth}) 
for the fractions are not
expanded in $\as$. The sum of $F_L$, $F_+$ and $F_-$ is thus equal to
one which does not hold anymore if the ratios are expanded in \as.
As one can see the NNLO corrections are about one third of the NLO
corrections. Since $F_+$ is non-zero only in NLO the evaluation of the
NNLO corrections are very important to test the reliability of the
theoretical predictions. We observe that $F_+$ remains very small
even after the inclusion of the NNLO corrections. Any observation of
$F_+$ significantly larger than 0.001 would thus signal New Physics.
In Ref.~\cite{Bernreuther:2008us} the impact of various Standard Model
extensions on the $tWb$ vertex have been investigated. In particular
the MSSM, a generic 2-Higgs doublet model (2HDM) and a top-color
assisted Technicolor model are investigated. In top-color assisted
Technicolor models a modification of the left chiral couplings by
several per cent is possible.
In Ref.~\cite{AguilarSaavedra:2010nx} a more detailed analysis of the
$W$-boson polarization which goes beyond the study of helicity
fractions has been proposed. 
\\
The fact that the top-quark decays before hadronization plays a
major role. Since the dominant decay is parity violating the top-quark
polarization of an ensemble of top-quarks is accessible through the
angular distribution of the decay products. In Born approximation a
straight forward calculation leads to
\begin{equation}
  {1\over \Gamma} {d\Gamma\over d\cos{\vartheta}} = 
  {1\over 2} (1+\alpha_f\cos\vartheta)
\end{equation}
where $\vartheta$ denotes the angle between the direction of flight of
the respective top-quark decay product $f$ and the top-quark spin in
the top-quark rest frame. The parameter $\alpha_f$ measures how
efficient a specific decay product analyses the top-quark
polarization. For the $b$-quark one finds $\alpha_b=-0.423$ while for
the charged lepton from $W$-boson decay a value of $\alpha_\ell=1$ is
found. The NLO corrections are also known and turn out to be small. In
Refs.~\cite{Gao:2012ja,Brucherseifer:2013iv} the NNLO corrections for
the fully differential decay width have been calculated.  The NNLO
corrections to differential distributions are found to be small.  In
Ref.~\cite{Brucherseifer:2013iv} also the $W$-boson helicity fractions
have been calculated. The results agree with the aforementioned
results of Ref.~\cite{Czarnecki:2010gb}.

\subsubsection{Two-loop QCD corrections to heavy quark form factors
  and the forward-backward asymmetry for heavy quarks} 
\label{sec:Top:2loop-FormFactor}
The measurements of the forward-backward asymmetry $A_{FB}^b$ for
$b$-quarks differ significantly from the Standard Model predictions
\cite{lepewwg2003}. The theoretical predictions take into account NNLO
QCD corrections, however the $b$-quark mass has been neglected at
NNLO.  The forward-backward asymmetry for massive quarks may be
calculated from the fully differential cross section.  As far as the
2-loop QCD corrections are concerned this requires the calculation of
the two-loop form factor for heavy quarks. These corrections have been
calculated recently. In Ref.~\cite{Bernreuther:2004ih} the NNLO QCD
corrections for the vector form factor are calculated. In
Ref.~\cite{Bernreuther:2004th} the results are extended to the
axial-vector form factor. The anomaly contribution has been studied in
Ref.~\cite{Bernreuther:2005rw}. The two-loop corrections need to be
combined with the one-loop corrections for real emission and the Born
approximation for double real emission.  All individual contributions
are of order $\as^2$ and thus contribute.  The cancellation of the
collinear and soft singularities encountered in the different
contributions is highly non-trivial. In
Refs.~\cite{Bernreuther:2011jt,Bernreuther:2013uma} `antenna
functions' are derived, which match the singular contributions in the
double real emission processes. As an important result also the
integrated antenna functions are computed in
Refs.~\cite{Bernreuther:2011jt,Bernreuther:2013uma}. In principle all
building blocks are now available to calculate the differential cross
section for heavy quark production in NNLO accuracy in QCD. Evidently
these results, once available, can also be applied to top-quark pair
production.

\subsubsection{Threshold cross section}

Threshold production of top-quark pairs in electron positron
annihilation is an unique process where one can extract the top-quark
mass through a threshold scan by measuring the total cross section
$\sigma(e^+e^-\to t\bar{t})$.  It is a counting experiment of the
production rate of the color singlet $t\bar{t}$ bound state. Therefore the
measurement of the threshold cross section for $e^+ e^-\to t\bar{t}$
is very clean experimentally as well as theoretically concerning QCD
non-perturbative effects.

\begin{sloppypar}
The $t\bar{t}$ cross section normalized to the point particle cross section 
 near threshold \cite{Fadin:1987wz,Fadin:1988fn} can be written at LO as
\begin{eqnarray}
R_{t\bar{t}} =
\left(\frac{6\pi N_c e_t^2}{m_t^2}\right)
{\rm Im}  \, G_c(\vec{0}, \vec{0}; E+i\Gamma_t),
\end{eqnarray}
where $E=\sqrt{s} -2m_t$ 
and $G_c(\vec{r}^{\, '}, \vec{r}; E+i\Gamma_t)$ is the non-relativistic 
Coulomb Green function. 
The Green function contains 
resonances at energies $$E_n=-m_t (C_F \alpha_s)^2/(4n^2)$$ corresponding to
Coulomb boundstates, and its residue is given by the Coulomb wave function 
$|\psi_n(0)|^2=(m_t \alpha_s C_F)^3/(8\pi n^3)$:
\begin{eqnarray}
G_c(\vec{0}, \vec{0}; E+i\Gamma_t) 
=  
\sum \hspace{-5mm} \int_n
\,\frac{ \psi_n(\vec{0})\psi_n^\ast(\vec{0})}{E_n - E-i\Gamma_t}.
\end{eqnarray}
Thus the peak position and the magnitude of the cross section is
determined by the Coulomb energy levels $E_n$ and the wave-functions
$|\psi_n(\vec{0})|^2$, respectively. In practice the resonance
structure of $G_c$ is smeared due to the large top quark width
$\Gamma_t \sim 1.4\, {\rm GeV}$.  In Fig.\ref{fig:xs01} the threshold
cross section is shown for $m_t=170 {\rm GeV}$ varying the top-quark
width. Only the $n=1$ ground state peak can been seen for $\Gamma_t
=1.0-1.5\, {\rm GeV}$ as rather wide prominence of the cross
section, and the resonance states are completely smeared out creating
a flat plateau for $\Gamma_t=2\, {\rm GeV}$. Although the resonant
structures are washed out for a large top-quark width, it is still
possible to extract top-quark parameters, $m_t, \Gamma_t$ and also
$\alpha_s$ by performing a threshold scan, provided a precise theory
prediction for the total cross section is at hand.
\end{sloppypar}
 \begin{figure}[htbp]
 \begin{center}
  \includegraphics[width=0.49\textwidth]{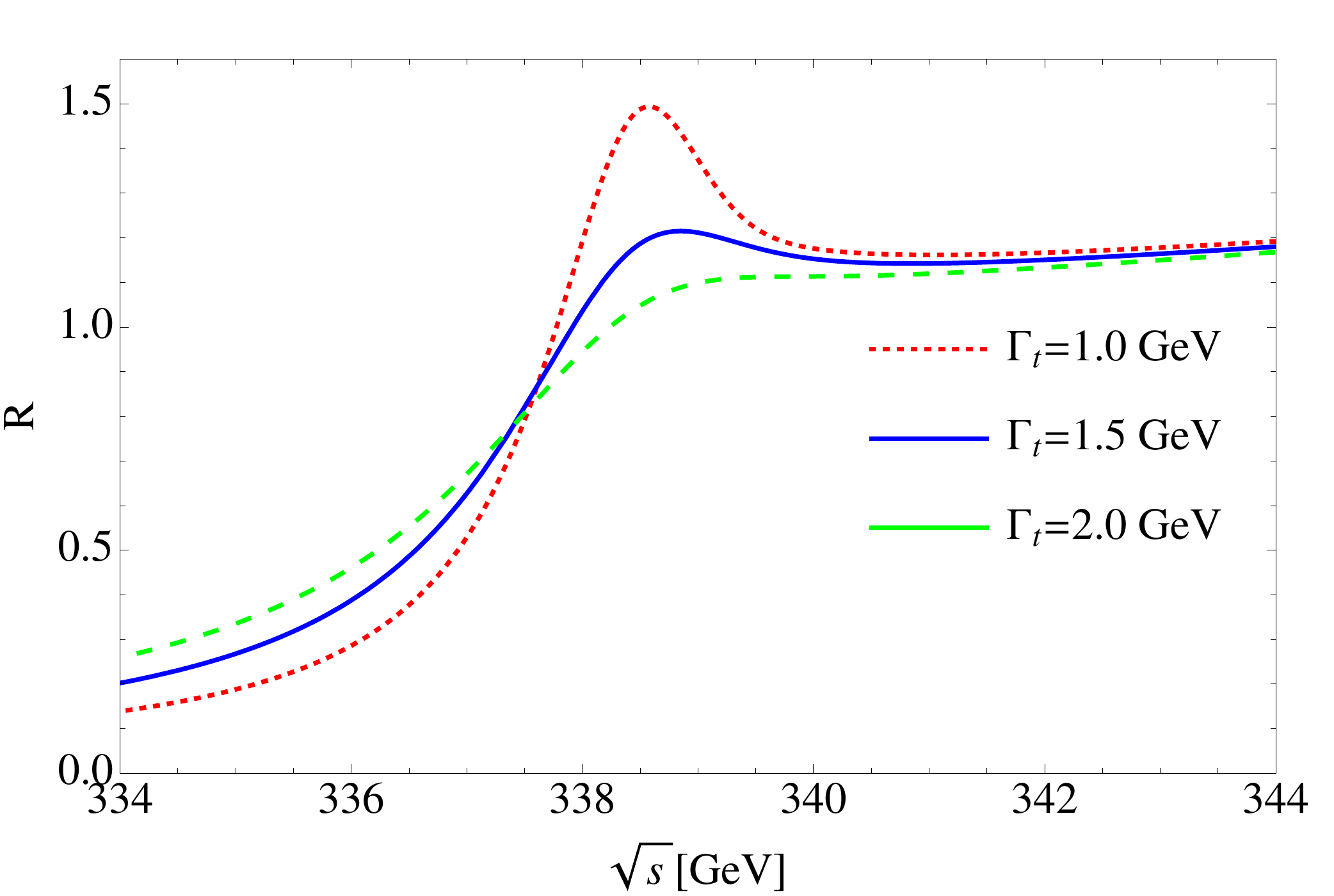}
 \end{center}
  \vspace{-5mm}
 \caption{The top quark production cross section $R$ for $m_t=170\,{\rm GeV}$
  and three values for top quark width. The LO formula for the cross section
  and $\alpha_s(30{\rm GeV})=0.142$ is used}
 \label{fig:xs01}
\end{figure}

\paragraph{QCD corrections}
%
Studies of top quark production near threshold 
\cite{Strassler:1990nw,Sumino:1992ai,Jezabek:1992np,Fujii:1993mk}  
at linear colliders were started several decades ago, and NNLO 
QCD corrections  were completed by several groups 
\cite{Hoang:1998xf,Hoang:1999zc,Melnikov:1998pr,Yakovlev:1998ke,Penin:1998ik,Beneke:1999qg,Nagano:1999nw} and summarized in Ref. \cite{Hoang:2000yr}.  
 One main achievement there was the stabilization of the peak position 
 against QCD corrections taking into account of renormalon cancellation
 using short distance masses like 1S-, kinetic-, PS- masses.
However, despite the completion of the second order QCD corrections
 the normalization of the total cross section still suffers from an
 uncertainty of about 20 percent. 
 
\begin{sloppypar}
 There are efforts to improve the accuracy of the NNLO total cross
 section. These include the resummation of potentially large
 logarithms by renormalization group (RG) methods
 \cite{Hoang:2000ib,Hoang:2001mm,Pineda:2006ri,Hoang:2011gy,Hoang:2013uda} 
and by brute-force
 computations of NNNLO corrections
 \cite{Beneke:2005hg,Beneke:2007gj,Beneke:2008ec,Beneke:2007pj,Marquard:2014pea} to
 increase the precision of the cross section.  Fig.~\ref{fig:nnnlo}
 shows the NNNLO result (using an adhoc estimate of some third order matching
coefficients)~\cite{Beneke:2008ec}
compared to the NNLO cross section.  The
 colored bands correspond to the uncertainty originating from a QCD
 renormalization scale variation between $25 \,{\rm GeV}$ and
 $80\,{\rm GeV}$. A significant reduction of the scale dependence is
 observed when going to NNNLO comparing with the NNLO result.  In
 Fig.\ref{fig:PStop} the RG improved total cross section
 \cite{Hoang:2000ib,Hoang:2001mm,Pineda:2006ri,Hoang:2011gy,Hoang:2013uda} 
is shown, where the
 uppper/lower pannels show the result with fixed order/RG improvement,
 respectively.  Two curves at each order are obtained by varying the
 soft scale $\mu_s$ between $(30-80)\,{\rm GeV}$.  The large scale
 dependence of the fixed order curves is improved by RG resummation in
 the lower pannel. The plot shows that the cross section at the peak
 position has scale dependence of order $2\%$.  
The most complete analysis in RG approach has been performed 
in~\cite{Hoang:2013uda}, where new ultra-soft NNLL 
contributions~\cite{Hoang:2011gy} 
are included. These two approaches,
 NNNLO computation and RG improvement to NNLL, are complementary to
 each other.  The fixed order computation provides the non-logarithmic
 contributions, while the RG improvement reveals the structure of the
 potentially large logarithmic terms to all orders. Therefore it is
 expected that the theory prediction of $t\bar{t}$ cross section with
 $\delta \sigma_{t\bar{t}}/\sigma_{t\bar{t}} = 2-3 \%$ will be
 possible by a combination of the two approaches as far as QCD
 corrections are concerned.  For such a high precision more dedicated
 theoretical studies will be needed, for instance, the calculation of
 electroweak effects and final state interactions in top-quark decays.
\end{sloppypar}
 \begin{figure}[htbp]
 \begin{center}
  \includegraphics[width=0.49\textwidth]{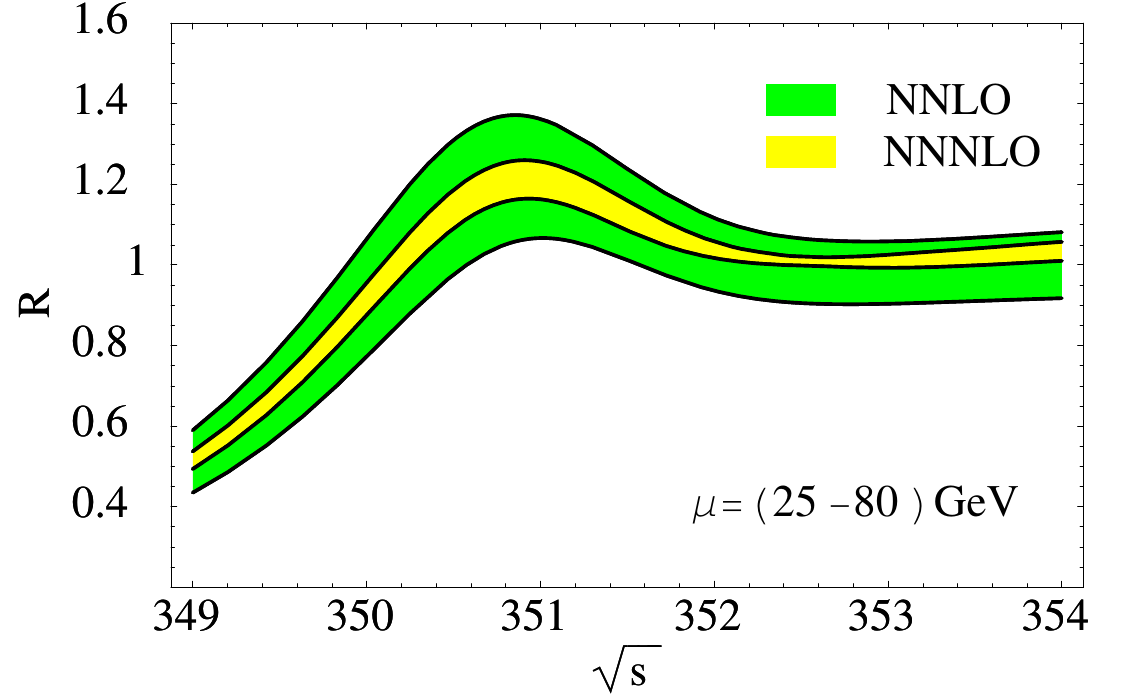}
 \end{center}
 \vspace{-5mm}
 \caption{Total cross section for top quark production near threshold
 at NNNLO (with an estimated third order matching coefficients)
 and NNLO  from \cite{Beneke:2008ec}, where a scale variation
 of $(20- 80)\, {\rm GeV}$ is shown by the colored bands. 
 A top quark PS mass $m_{\rm PS}(20{\rm GeV})=175\, {\rm  GeV}$ is used.  }
 \label{fig:nnnlo}
\end{figure}
%
 \begin{figure}[htbp]
 \begin{center}
  \includegraphics[width=0.49\textwidth]{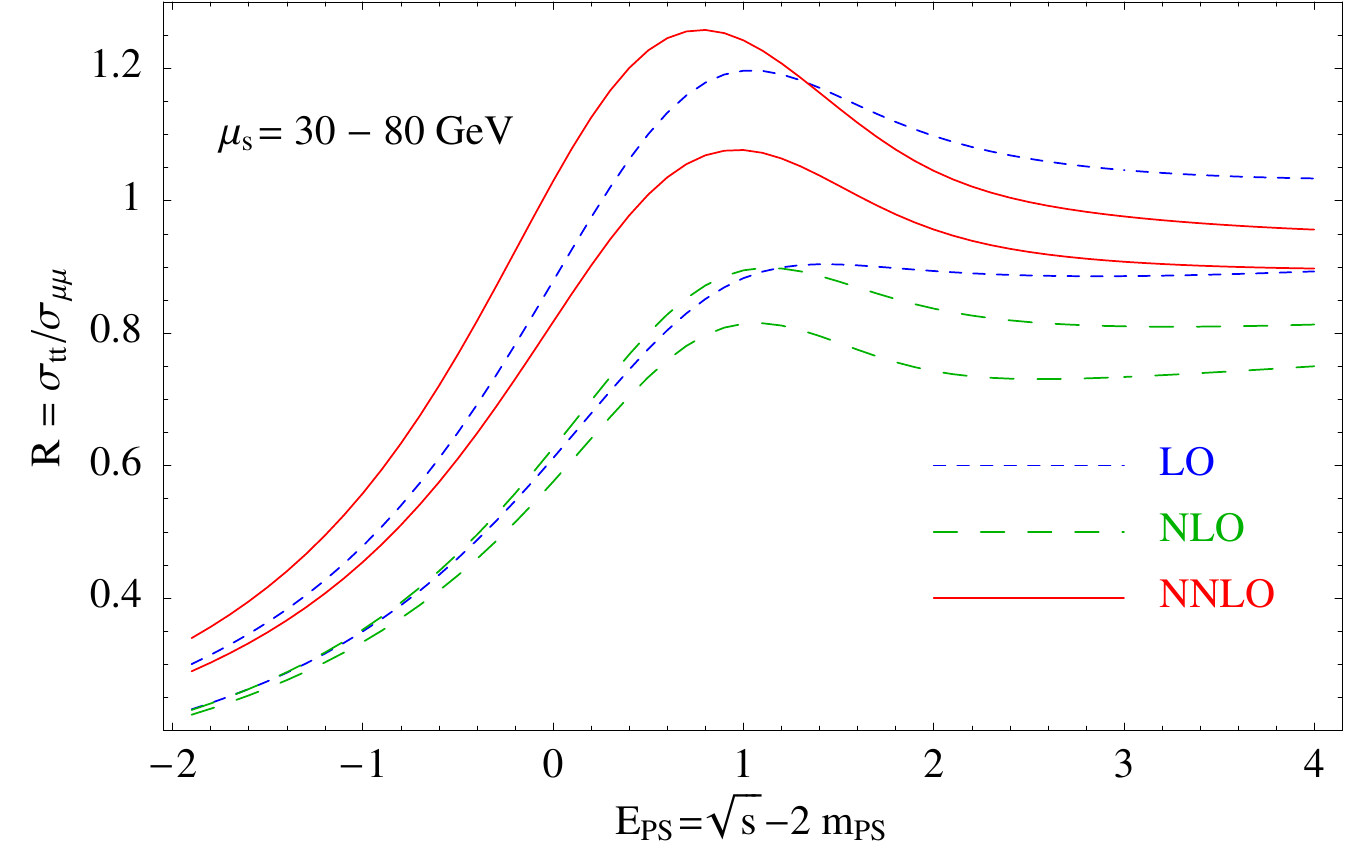}
   \includegraphics[width=0.49\textwidth]{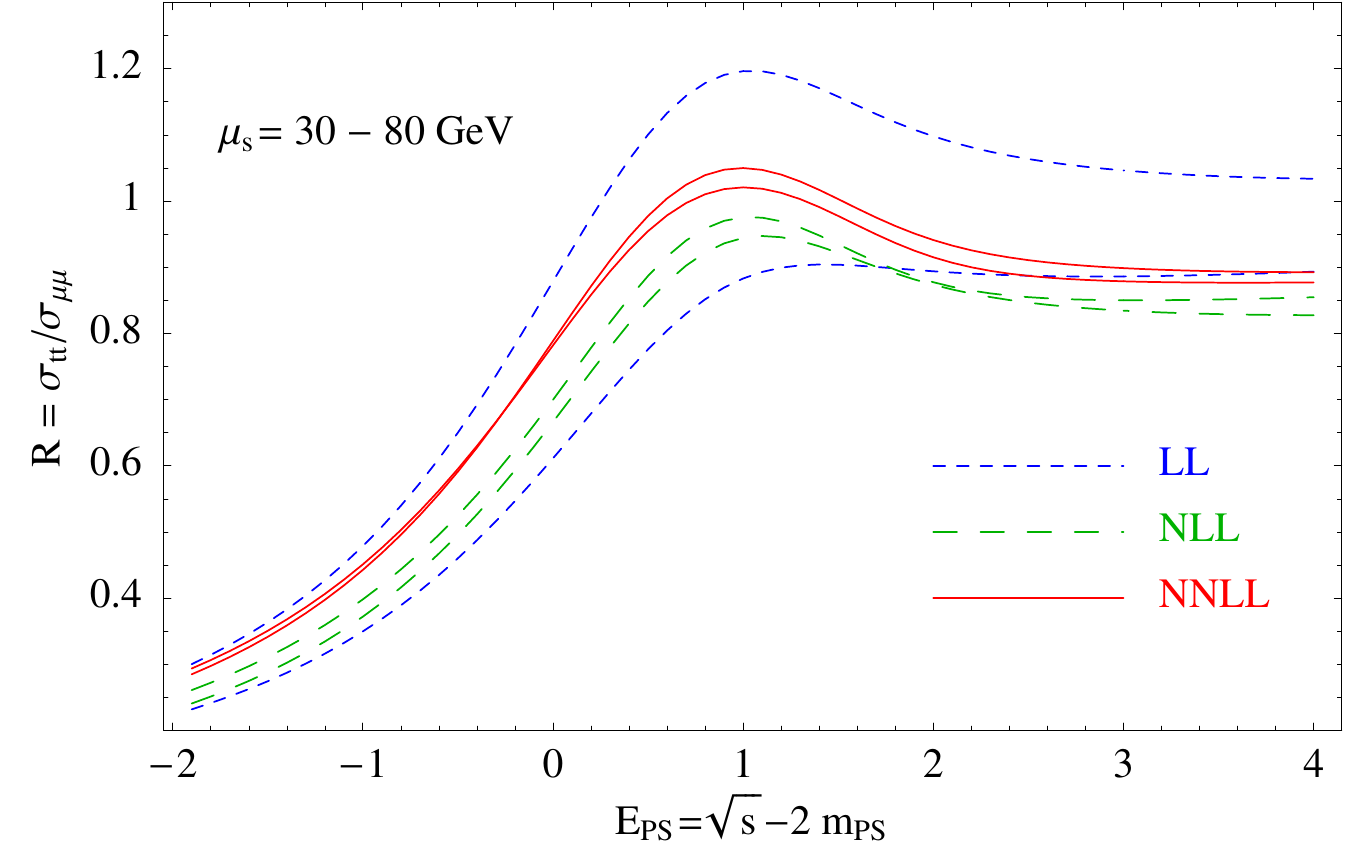}
 \end{center}
  \vspace{-5mm}
 \caption{The threshold cross section at fixed order (upper pannel) and 
 renormalization group improvement (lower pannel) is shown from Ref. \cite{Pineda:2006ri}. 
 The bands between two colored lines at each orders show the scale dependence
 of the results.  The RG improved cross sections are stable against scale variation
 while fixed order result suffers from large dependence on values of $\mu_s$.  }
 \label{fig:PStop}
\end{figure} 

\paragraph{Electroweak corrections and effect of unstable top}
%

In early studies of the $e^+ e^- \to t\bar{t}$ threshold it was
recognized \cite{Fadin:1988fn,Fadin:1987wz} that the effect of the top
quark width can be consistently incorporated into the computation of
the total cross section by the replacement $E\to E+i\Gamma_t$. This prescription
works well up to NLO, but it turns out that in NNLO an
uncanceled ultraviolet divergence appears, which is proportional to
the top-quark
width (in dimensional regularization an example of such a term is
$R_{t\bar{t}}\sim \alpha_s \Gamma_t/\epsilon$).  This is a signal of an
improper treatment of electroweak effects, and the solution of this
problem is to abandon the amplitude $e^+ e^- \to t\bar{t}$ where the
unstable $t\bar{t}$ is treated as a final state of the S-matrix.  Physical
amplitudes should treat stable particles as final states of S-matrix,
i.e. $e^+ e^- \to t\bar{t} \to (bW^-) (\bar{b}W^+)$ \footnote{
  assuming the $W$-boson and $b$-quarks as stable or long-lived
  particles. } and the unstable particles can appear only as
intermediate states.

Electroweak corrections to the production vertex $t\bar{t}-\gamma/Z$ 
 were first described in \cite{Guth:1991ab} and re-derived in 
 \cite{Hoang:2006pd,Hoang:2004tg}. In the later refence 
 it is readily realized that amplitudes for single top production, 
 e.g. $e^+ e^- \to t b W$,
 and even no-top quark production $e^+ e^- \to b W^+ \bar{b} W^-$ 
 can contribute to (or mix with) the top-pair production because 
 the physical final state is the same. 
 
The top-quark width is generated by 
the EW interaction, $t\to b W$, therefore the effects of the top-quark
finite width 
are intimately related to the EW corrections of the process.
To take into account certain electroweak non-resonant effects 
a method referred to as {\it phase-space matching} was introduced in 
\cite{Hoang:2008ud,Hoang:2010gu}. 

\begin{sloppypar}
This idea has been further developped and rephrased in the framework
of an effective theory for unstable particle
\cite{Beneke:2003xh,Beneke:2004km}.  (See
Refs.~\cite{Beneke:2007zg,Actis:2008rb} for an application of the
method to $W$-pair production in $e^+ e^-$ annihilation.)  A systematic
analysis of the electroweak effects in top-quark pair production has
started rather recently, and NLO electroweak non-resonant
contributions were computed \cite{Beneke:2010mp}, e.g.  
$R(e^+e^-\to t\bar{b}W^-)\sim \alpha_{\rm EW}$, where resonant (onshell) top quarks
decay and the final state $(b W^+) (\bar{b} W^+)$ is measured assuming
stable $W$-bosons and $b$-quarks.  In this work invariant mass cuts on the
top-quark and anti--top-quark decay products are implemented. It is
found that the non-resonant correction results in a negative 5\% shift
of the total cross section which is almost energy independent, in
agreement with Ref\cite{Hoang:2010gu}.  The dominant NNLO non-resonant
corrections were computed in Ref. \cite{Penin:2011gg,Jantzen:2013gpa}
and it was shown that the single resonant amplitudes (e.g. $e^+ e^-
\to t (\bar{b} W^-) g$) provide the counter terms for the uncanceled
ultraviolet divergence $\alpha_s \Gamma/\epsilon$ discussed previously
for the double resonant $e^+ e^- \to t\bar{t}$ amplitude at NNLO QCD.
Therefore, the non-resonant corrections provide together with NNLO QCD
a consistent treatment of top quark width effects.
\end{sloppypar} 

It is also known that the final state corrections
 \cite{Melnikov:1993np,Peter:1997rk} between top quarks 
and decay products have to be considered for observables
other than the total cross section. 
A systematic analysis of these effects is still missing beyond NLO.  
Dedicated studies of the electroweak corrections to the threshold cross section
have started rather recently.

\paragraph{Influence of the Higgs boson on the total cross section}

In the SM the large top-quark mass leads to a large top-quark Yukawa coupling
to the Higgs boson, therefore it is expected that Higgs boson exchange 
in top-quark production may lead to observable corrections. 
 \begin{figure}[htbp]
 \begin{center}
  \includegraphics[width=0.49\textwidth]{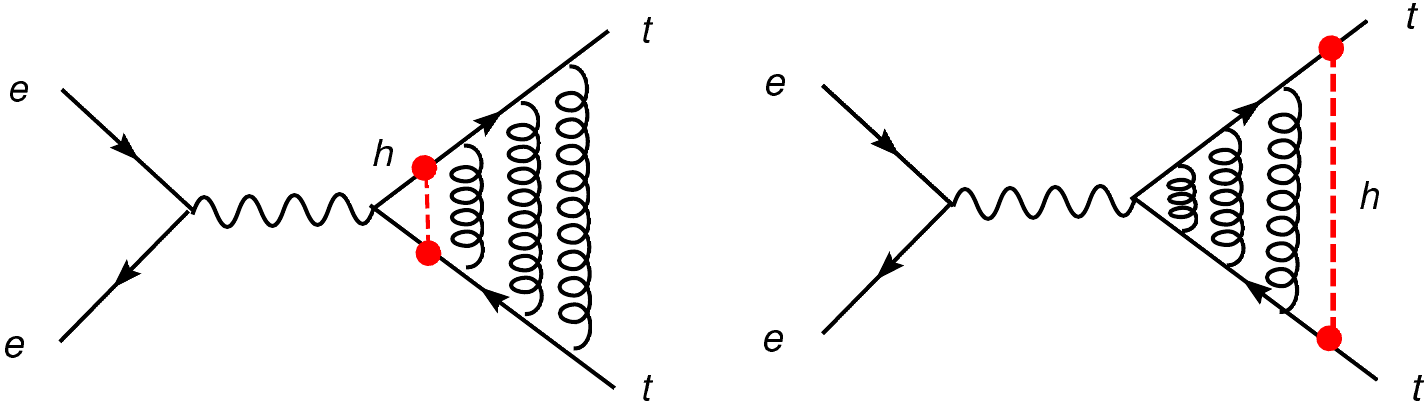}
 \end{center}
  \vspace{-5mm}
 \caption{Corrections due to Higgs exchange in $e^+ e^-\to t\bar{t}$.
   In the left diagram the Higgs exchange contributes to the
   production vertex for $\gamma t\bar{t}, Zt\bar{t}$, which ocurrs at
   short distance when the $t\bar{t}$-pair is separated by $r\sim
   1/m_t$.  
   In the right diagram Higgs exchanges occurs after
   bound-state formation between top and anti-top quarks separated by
   the scale of the bound-state $r\sim 1/(m_t \alpha_s)$.  }
 \label{fig:xs03}
\end{figure} 
Such a Higgs exchange effect appears in two different ways in top and anti-top
production near threshold (see Fig.\ref{fig:xs03}). One is a short distance contribution 
which enhances the top quark production vertex as 
$\bar{t} \gamma^\mu t \to (1+c_h) \bar{t} \gamma^\mu t$. 
The one-loop Higgs correction $c_h^{(1)}$ was determined
 in Refs. \cite{Guth:1991ab}, and Higgs and EW mixed two-loop
correction $c_{h}^{(2)}$ in Ref. \cite{Eiras:2006xm}.
The enhancement factor for the cross section is given by 
\begin{eqnarray}
\delta R/R_{\rm LO}\approx 2 c_h^{(1)} = 6.7/3.4/0.9\times 10^{-2}
\end{eqnarray} 
using $m_h=120/200/500\, {\rm GeV}$. 

In addition, there is a long-distance effect described by the Yukawa
potential $V_h(r)$ for the top quark pair:
\begin{eqnarray}
V_h(r)&=& -\frac{y_t^2}{8\pi} \frac{e^{-m_h r}}{r}
\simeq -\frac{y_t^2}{2m_h^2}\delta(\vec{r}),
\end{eqnarray}
where the second expression is a good approximation for $m_h r  \gg 1$
assuming $m_h\sim 125\,{\rm GeV}$ and $r\sim (m_t \alpha_s)^{-1}$.
In the SM the Yukawa coupling is related to the top-quark mass by 
$y_t=\sqrt{2} e m_t/(s_W M_W)$. 

 \begin{figure}[htbp]
 \begin{center}
  \includegraphics[width=0.49\textwidth]{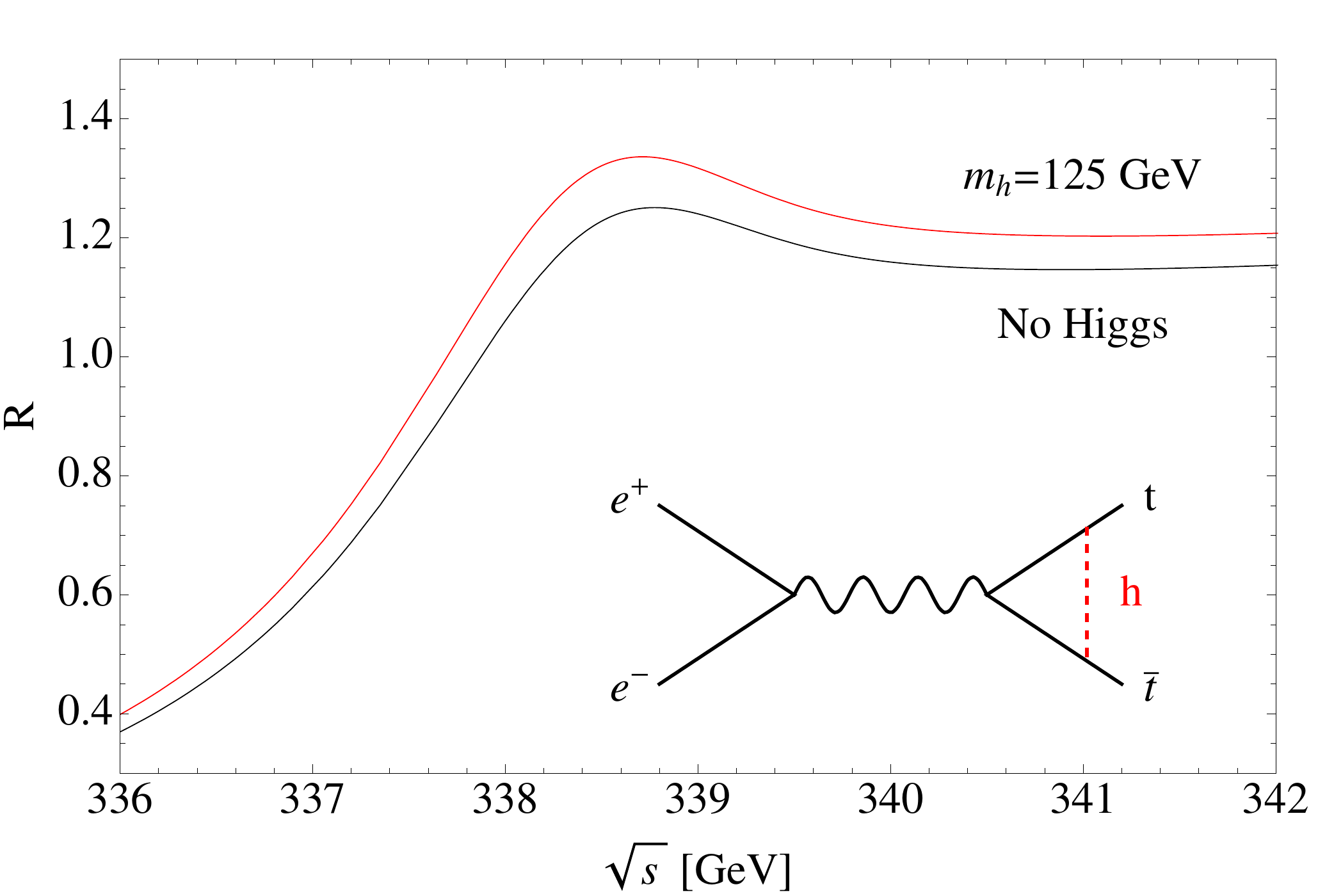}
 \end{center}
  \vspace{-5mm}
 \caption{Cross section for $e^+ e^- \to t\bar{t}$ for $m_t=170\, {\rm GeV}$
   with/without 1-loop Higgs boson corrections. A Higgs boson mass of
   $m_h=125\, {\rm GeV}$ is used.}
 \label{fig:xs03a}
\end{figure} 
In Fig.\ref{fig:xs03a} the threshold cross section is shown taking into
account of Higgs loop effects through $c_h$ and $V_h$. One can see 
that the threshold cross section gets an almost energy independent enhancement.
The Higgs potential $V_h$ produces corrections to the energy and to the 
wave function as
\begin{eqnarray}
\delta E_1/E_{\rm LO}
&=& 
3.2/1.2/0.2\times 10^{-2},
\nonumber\\
\delta |\Psi_{1}(0)|^2/ |\Psi_{\rm LO}(0)|^2
&=& 
4.6/1.6/0.3\times 10^{-2},
\end{eqnarray}
\begin{sloppypar}
using $m_t=175\, {\rm GeV},$ $\mu=30\,$ ${\rm GeV}$
and $m_h=120/200/500\,{\rm GeV}$, respectively. 
The above value for  $\delta E_1$ can be translated into 
a shift  $\delta m_t=25/9/1 \, {\rm MeV}$
of the top-quark mass determined in a threshold scan.
\end{sloppypar}
\paragraph{Distribution and Asymmetry}

In the threshold production, the top-quark momentum $\vec{p}_t$ can be
reconstructed from its decay products. Therefore the top-quark
momentum distribution
\cite{Strassler:1990nw,Sumino:1992ai,Jezabek:1992np} provides complementary
information.  Theoretically it is given
by
 \begin{eqnarray}
 \frac{1}{\sigma_0}
 \frac{d\sigma_{LO}}{d p_t} (e^+ e^- \to t\bar{t})
 &=&
\frac{p_t^2}{2\pi^2} \left( \frac{6\pi N_c e_t^2}{m_t^2} \right) \nonumber\\
&\times& 
\Gamma_t \, | \tilde{G}_c( \vec{p}, \vec{r}=\vec{0}; E+i \Gamma_t)|^2,
%
%
 \end{eqnarray}
 where $\tilde{G}_c(\vec{p},\vec{r};E+i\Gamma_t)$ is the Fourier
 transformation of the Coulomb Green function. For the momentum
 distribution NNLO QCD results \cite{Hoang:1999zc,Nagano:1999nw} are
 available in the literature.  
 
 Fig.\ref{fig:xs02} shows the momentum distribution at specific energy
 points $\Delta E=0, 2, 5{\rm GeV}$ (left panel) and for different
 top-quark masses.  In the lower panel the bands  correspond
 to the uncertainty of the QCD coupling constant assuming $\alpha_s=0.118\pm
 0.003$. As the Green function $\tilde{G}_c(\vec{p}, \vec{r}; E+i\Gamma)$
 is essentially the momentum space wave function averaged over the 
 resonances, a measurement of the top-quark momentum distribution
 gives information on the bound state wave function
 $\tilde{\phi}(\vec{p})$. Therefore the momentum space distribution
 gives independent information on the bound-state and can be used to test 
 the understanding of the QCD dynamics.
 
 \begin{figure}[htbp]
 \begin{center}
  \includegraphics[width=0.49\textwidth]{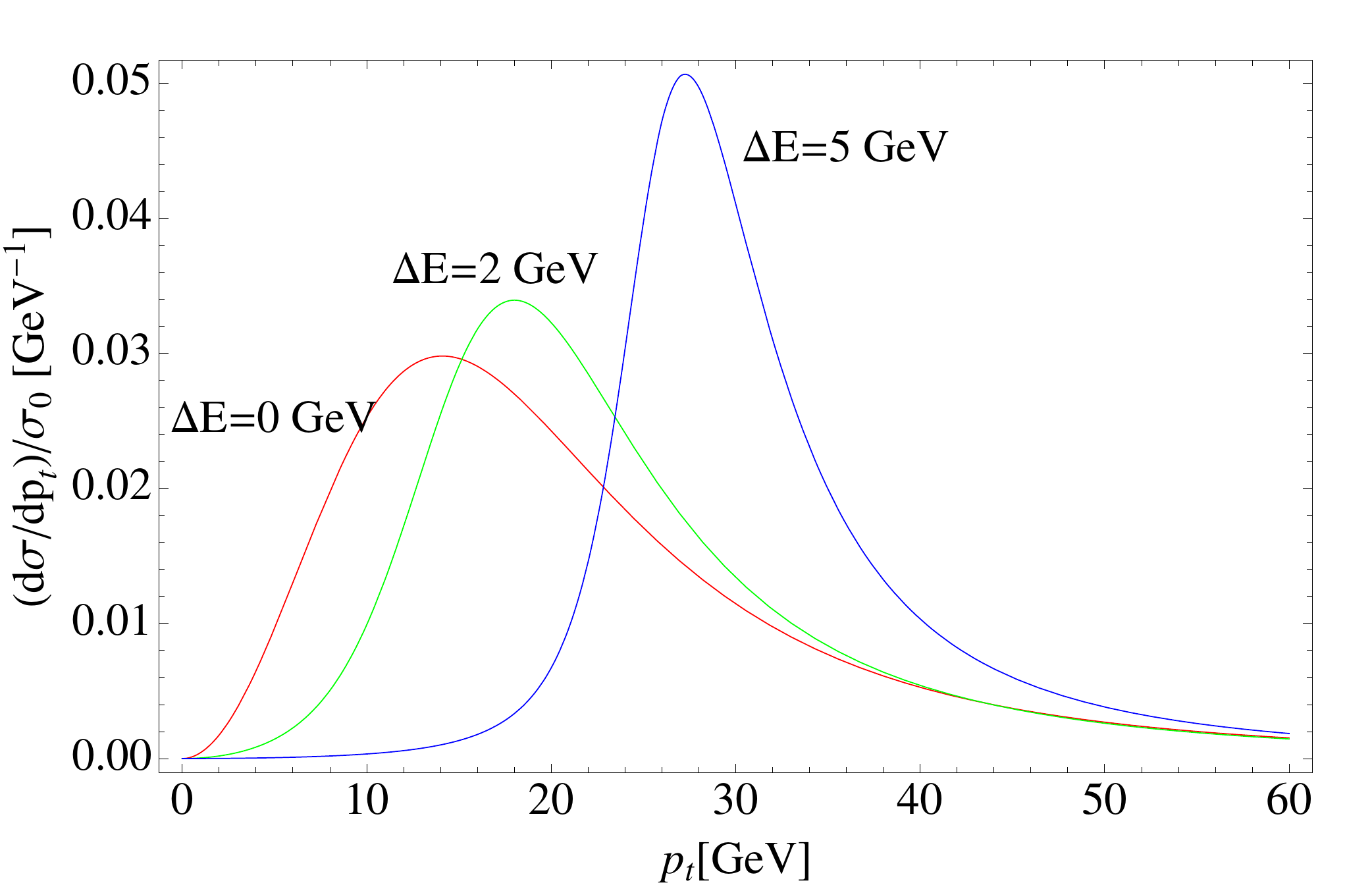}
  \includegraphics[width=0.47\textwidth]{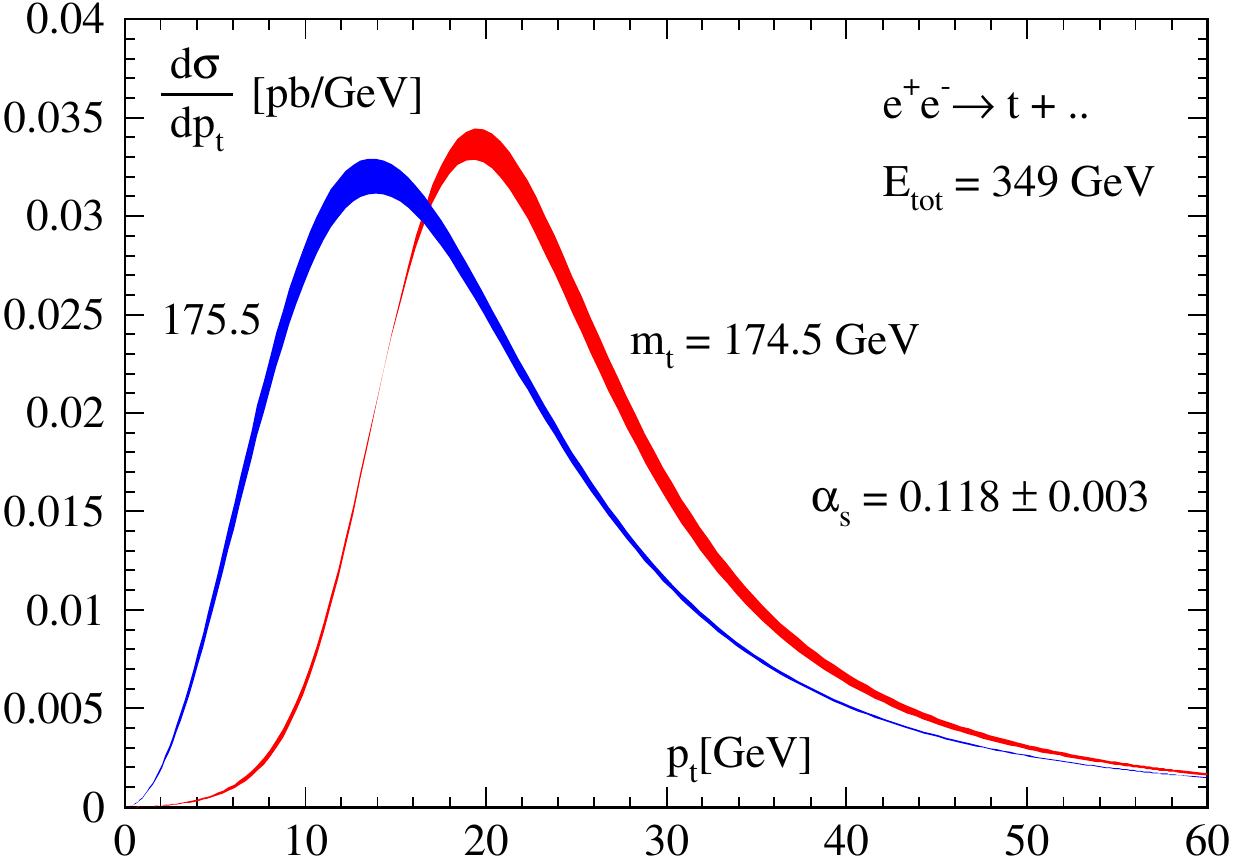}
 \end{center}
 \caption{Top quark momentum distribution at $\Delta E= E- E_{1}= 0, 2, 5$ GeV 
 (top)  for $m_t=170\, {\rm GeV}$ and top quark mass dependence (bottem)
  on the momentum distribution.}
 \label{fig:xs02}
\end{figure} 

Another useful observable which can be measured in top-quark production 
near threshold is the forward-backward asymmetry defined as
\begin{eqnarray}
A_{FB} &=& 
\frac{1}{\sigma_{t\bar{t}}}
\bigg\{ \int_{0}^1 d\cos\theta -\int_{-1}^0 d\cos\theta \bigg\} 
\frac{d\sigma (e^+e^-\to t\bar{t})}{d\cos\theta}.\nonumber \\
&&
\end{eqnarray} 
At lepton colliders top-quark pair production occurs through  
$e^+e^-\to \gamma^\ast/Z^\ast \to t\bar{t}$ and 
the forward-backward asymmetry receives a non-zero contribution 
from the interference of vector and axial-vector couplings.
Vector and axial-vector 
interactions produces S-wave and P-wave bound-states, respectively, 
due to angular momentum conservation. 
 Therefore the for\-ward-back\-ward asymmetry is sensitive to the interference
between S-wave and P-wave top-quark production. 
The asymmetry is sensitive to the top-quark width $\Gamma_t$
because the S-wave and P-wave overlap is non-zero due 
to $\Gamma_t$. 
\begin{figure}[htb]
\begin{center}
\includegraphics[width=0.49\textwidth]{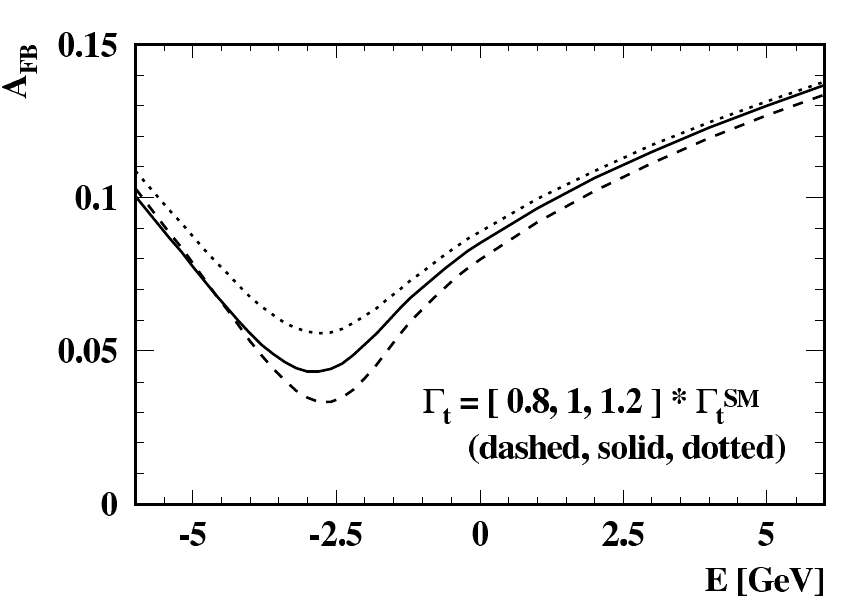}
\includegraphics[width=0.49\textwidth]{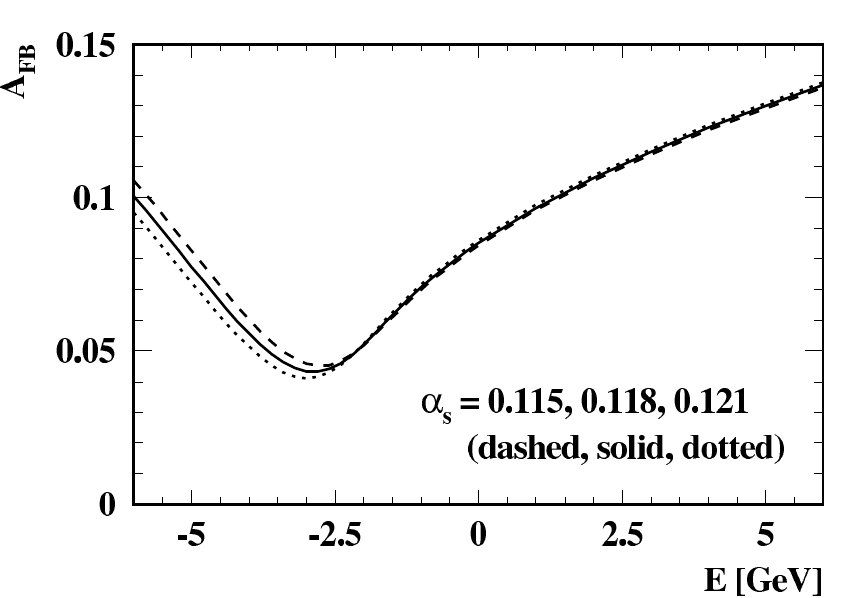}
\caption{Dependence of the forward-backward asymmetry $A_{\rm FB}$ on
  the top quark width (upper plot) and the strong coupling $\alpha_s$
  (lower plot).
  Figures are taken from Ref. \cite{Juste:2006sv}.} 
\label{fig:afb}
\end{center}
\end{figure}
In Fig.\ref{fig:afb} the forward-backward asymmetry is plotted as a
function of energy $E$. Top and bottem panels show the dependence on
$\Gamma_t$ and $\alpha$, respectively. As discussed above the
asymmetry $A_{FB}$ is an effect of $\gamma$ and $Z$-boson interference.
Therefore, the asymmetry provides useful information on the mechanism of
top-quark production near threshold.

\subsubsection{Top-quark production in the continuum}
The total cross section for the production of heavy quarks in
electron-positron annihilation has been calculated in
Refs. \cite{Chetyrkin:1997mb,Hoang:1997ca,Harlander:1997kw,Chetyrkin:1997pn}
at order $\as^2$ in QCD. The results are not applicable very close to
the threshold since in that region Coulomb effects lead to $1/\beta$
corrections where $\beta$ denotes the velocity of the top-quark. For
reliable predictions in the threshold region these contributions need
to be resummed (see also the discussion in the previous section).  In
Ref.~\cite{Chetyrkin:1997pn} it has been estimated that the fixed
order results should be applicable in the case of top-quark pair
production, provided that the center-of-mass energy is about 12 GeV above
threshold.  In Ref.~\cite{Chetyrkin:2000zk} the results have been
extended to order $\as^3$. In particular the quartic mass corrections
with respect to the massless calculation were calculated. Using the
minimal subtraction scheme (MS) to renormalize the mass parameters, sizeable
corrections were found in order $\as^3$. However it is also shown in 
Ref.~\cite{Chetyrkin:2000zk} that using the invariant mass $\hat m$ defined
through
\begin{equation}
  m(\mu) = \hat m \exp\left\{\int da {\gamma_m(a)\over a\beta(a)}\right\},
\end{equation}
where $m(\mu)$ denotes the running mass, $\gamma_m$ the anomalous
dimension of $m(\mu)$ and $\beta(a)$ the QCD beta function in terms of
$a=\as/\pi$, the convergence of the perturbative expansion can be
improved. As discussed in Section \ref{sec:Top:2loop-FormFactor} the
work on the differential cross section at order $\as^2$ is still ongoing.
In Refs. \cite{Fleming:2007qr,Fleming:2007xt} jet observables in
top-quark pair production at high energy have been investigated. The
process is characterized by different mass scales: the center-of-mass
energy $\sqrt{s}$, the top-quark mass $M_t$, the top-quark width
$\Gamma_t$ and $\Lambda_{QCD}$. Large logarithmic corrections
connected with the different mass scales are
resummed in Ref. \cite{Fleming:2007qr} at next-to-leading logarithmic
accuracy. This requires the introduction of soft functions capturing
non-perturbative soft QCD effects. The soft functions can be
obtained from massless dijet events. In Ref. \cite{Fleming:2007xt} the
application to top-quark mass measurements is discussed. In particular
it is demonstrated that a top-quark mass measurement with a precision
of $\Lambda_{QCD}$ is possible significantly above the production threshold. 

\subsection{Physics Potential}

The excellent possibilities for precision top-quark measurements at
$e^+e^-$ colliders have been confirmed by experimental studies of the
physics potential of linear colliders, which, in particular in the
framework of recent reports of the CLIC and ILC physics and detector
projects, often are based on full detector simulations. Particular
emphasis has been placed on the measurement of the top-quark mass, which
has been studied both at and above threshold, and on the study of the
$t\bar{t}Z/\gamma^*$ vertex through the measurement of asymmetries. For
all of these measurements, precise flavor tagging and excellent jet
reconstruction is crucial to identify and precisely reconstruct
top-quark pair events. The detectors being developed for linear colliders
provide these capabilities, and, together with the rather modest
background levels in $e^+e^-$ collisions, allow to acquire
high-statistics high-purity top-quark samples. In the following, the most recent published
results from simulation studies of top-quark mass measurements are discussed. The studies of
top-quark couplings, which make use of the possibilities for polarized beams at linear colliders, are still ongoing.  Preliminary results indicate  
a substantially higher precision than achievable at hadron colliders. 

\subsubsection{Top-quark mass measurement at threshold}
\label{sec:thresholdmeasurements}
\begin{sloppypar}
The measurement of $t\bar{t}$ production cross section in a scan around
the threshold provides direct access to the top quark, as discussed
above. In the experiment, the calculated cross section is modified by initial state radiation and by the luminosity
spectrum of the collider. These two effects are illustrated in Figure
\ref{fig:Top:Exp:CrossSectionISRLS} \cite{Seidel:2013sqa}, where the
pure $t\bar{t}$ cross section is calculated with TOPPIK at NNLO \cite{Hoang:1998xf,Hoang:1999zc}
for a top-quark mass of 174 GeV in the 1S mass scheme, and the luminosity
spectrum of CLIC at 350 GeV is assumed. Both lead to a smearing of the
cross section, resulting in a substantial reduction of the prominence of
the cross section peak, and to a overall reduction of the cross section
due to the lowering of the luminosity available above the production
threshold. Since the beam energy spread at ILC is smaller than at CLIC,
the threshold turn-on is slightly steeper, as visible in Figure
\ref{fig:Top:Exp:Threshold}.
\end{sloppypar}

 \begin{figure}[htbp]
 \begin{center}
  \includegraphics[width=0.49\textwidth]{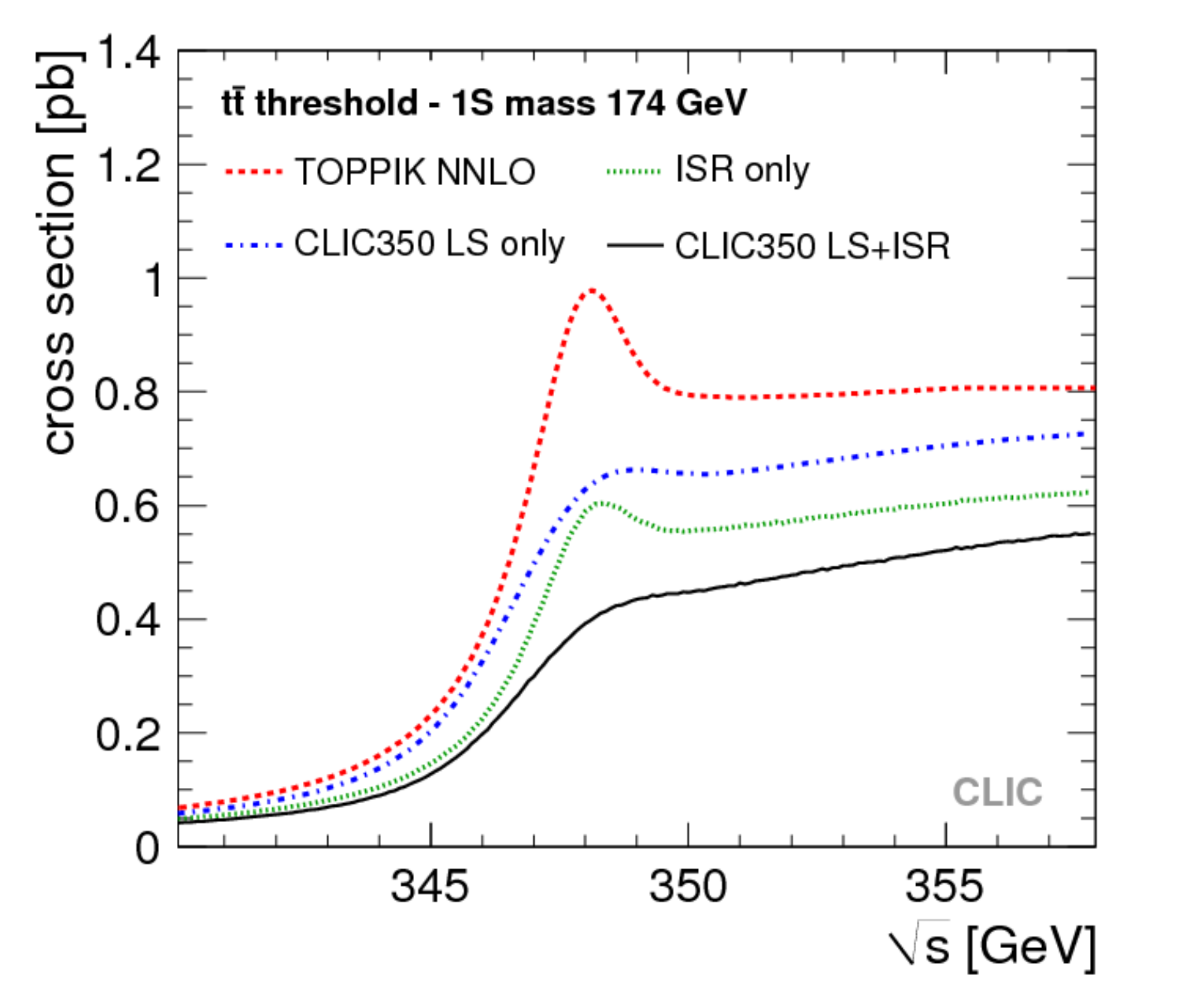}
 \end{center}
  \vspace{-5mm}
 \caption{The top-quark production cross section calculated with TOPPIK for a top mass of 174 GeV in the 1S mass scheme, showing the effects of initial state radiation and of the luminosity spectrum of CLIC. Figure taken from Ref.~\cite{Seidel:2013sqa}.}
 \label{fig:Top:Exp:CrossSectionISRLS}
\end{figure}

Recently, an experimental study has been performed in which the NNLO
cross section shown in Fig.~\ref{fig:Top:Exp:CrossSectionISRLS} was
used, together with signal efficiencies and background contamination
determined with full Geant4 simulations of a CLIC variant of the ILD
detector, including the use of the full reconstruction chain. In the
context of a threshold scan, where the focus is on the efficient
identification of $t\bar{t}$ events, the difference in performance
between the ILC and CLIC detector concepts is expected to be negligible,
allowing to apply this study to both accelerator concepts by using the
appropriate luminosity spectra. The experimental precision of a
threshold scan with a total integrated luminosity of 100 fb$^{-1}$
spread over ten points spaced by 1 GeV for the ILC case is illustrated
in Fig.~\ref{fig:Top:Exp:Threshold}.

\begin{figure}[htbp]
 \begin{center}
  \includegraphics[width=0.49\textwidth]{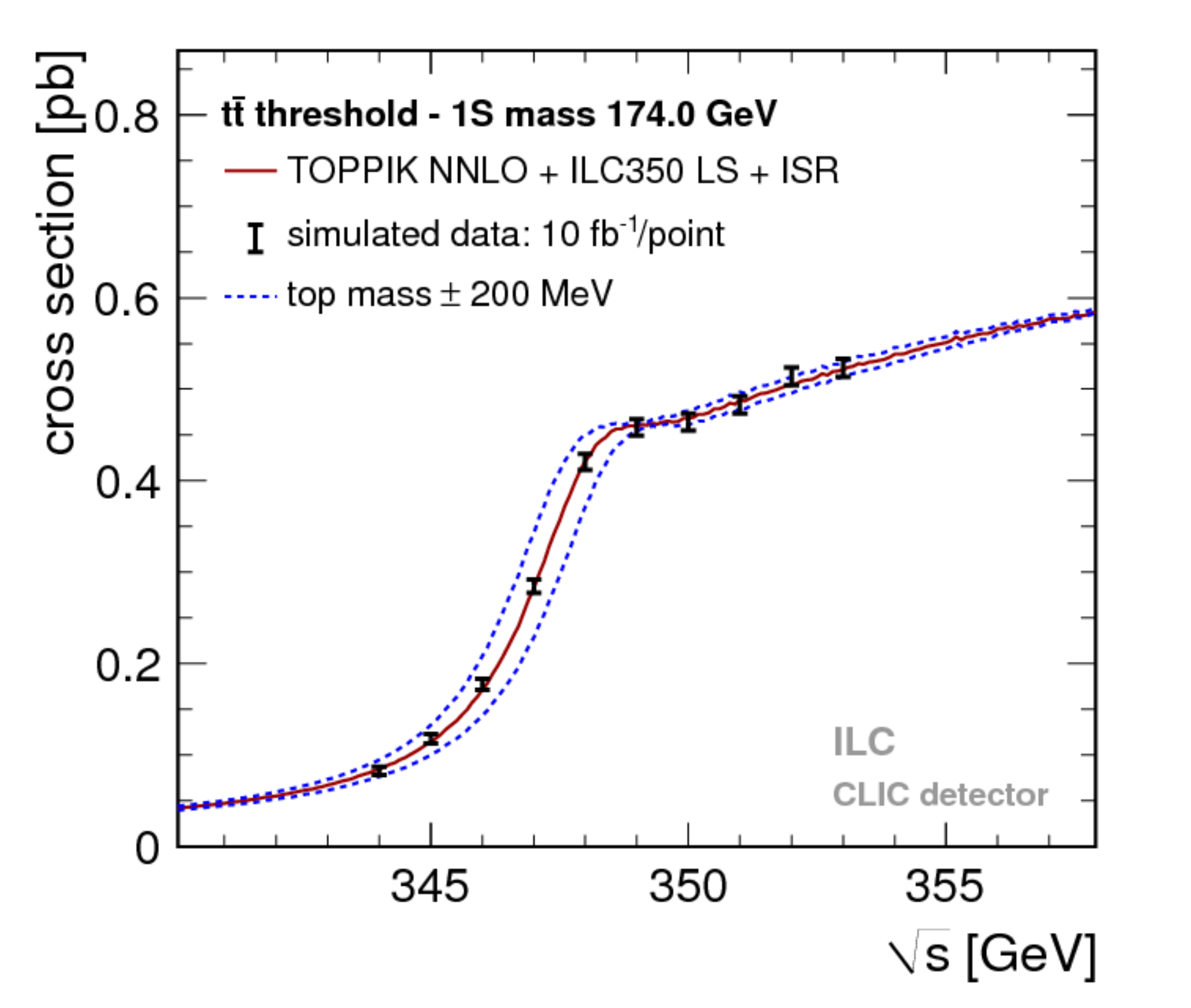}
 \end{center}
  \vspace{-5mm}
 \caption{Simulated measurement of the background-subtracted $t\bar{t}$ cross section with 10 fb$^{-1}$ per data point, assuming a top-quark mass of 174 GeV in the 1s scheme with the ILC luminosity spectrum for the CLIC\_ILD detector. Figure taken from Ref.~\cite{Seidel:2013sqa}.}
 \label{fig:Top:Exp:Threshold}
\end{figure}

Since the cross section depends not only on the top-quark mass, but also on
$\alpha_s$, those two values are determined simultaneously with a
two-dimensional fit, resulting in a statistical uncertainty of 27 MeV on
the mass and 0.0008 on $\alpha_s$. Assuming the CLIC luminosity
spectrum, which is characterized by a somewhat more pronounced
beamstrahlung tail and a larger energy spread, the uncertainties
increase to 34 MeV and 0.0009, respectively. Systematic uncertainties
from the theoretical cross-section uncertainties, from the precision of
the background description and the understanding of the detector
efficiency as well as from the absolute knowledge of the beam energy are
expected to be of similar order as the statistical uncertainties. Thus,
the differences between different linear collider concepts for a top
threshold scan are negligible, and total uncertainties of below 100 MeV
on the mass are expected \cite{Seidel:2013sqa}. For a phenomenological
interpretation, the measured 1S mass typically has to be converted into the 
standard $\overline{\mbox{MS}}$ mass. This incurs additional uncertainties of the order
of 100 MeV, depending on the available precision of $\alpha_s$ \cite{Hoang:1999zc}.

As discussed in detail in Ref.~\cite{Martinez:2002st}, in addition to the
mass and the strong coupling constant, also the top-quark width can be
determined in a threshold scan. The use of additional observables such
as the top-quark momentum distribution and the forward-backward
asymmetry has the potential to further reduce the statistical
uncertainties. The cross-section around threshold is also sensitive to the top-quark Yukawa coupling, as discussed above. However, its effect on
the threshold behavior is very similar to that of the strong coupling constant, so an extraction will only be possible with a substantially
improved knowledge of $\alpha_s$ compared to the current world average uncertainty of 0.0007, and with reduced theoretical uncertainties on the overall cross section.

\subsubsection{Top-quark mass measurement in the continuum}

In the continuum above the $t\bar{t}$ threshold, the top-quark mass is
measured experimentally by directly reconstructing the invariant mass
from the measured decay products, a $W$ boson and a $b$ quark. This is
possible with high precision both in fully hadronic (e.g. both $W$
bosons produced in the $t\bar{t}$ decay decaying into hadrons) and
semi-leptonic (e.g. one $W$ boson decaying into hadrons, one into an
electron or muon and a neutrino) top-quark pair decays. Due to the
well-defined initial state in $e^+e^-$ collisions, full
three-dimensional kinematic constraints can be used for kinematic
fitting, substantially improving the invariant mass resolution compared
to a free measurement.

\begin{figure}[htbp]
 \begin{center}
  \includegraphics[width=0.49\textwidth]{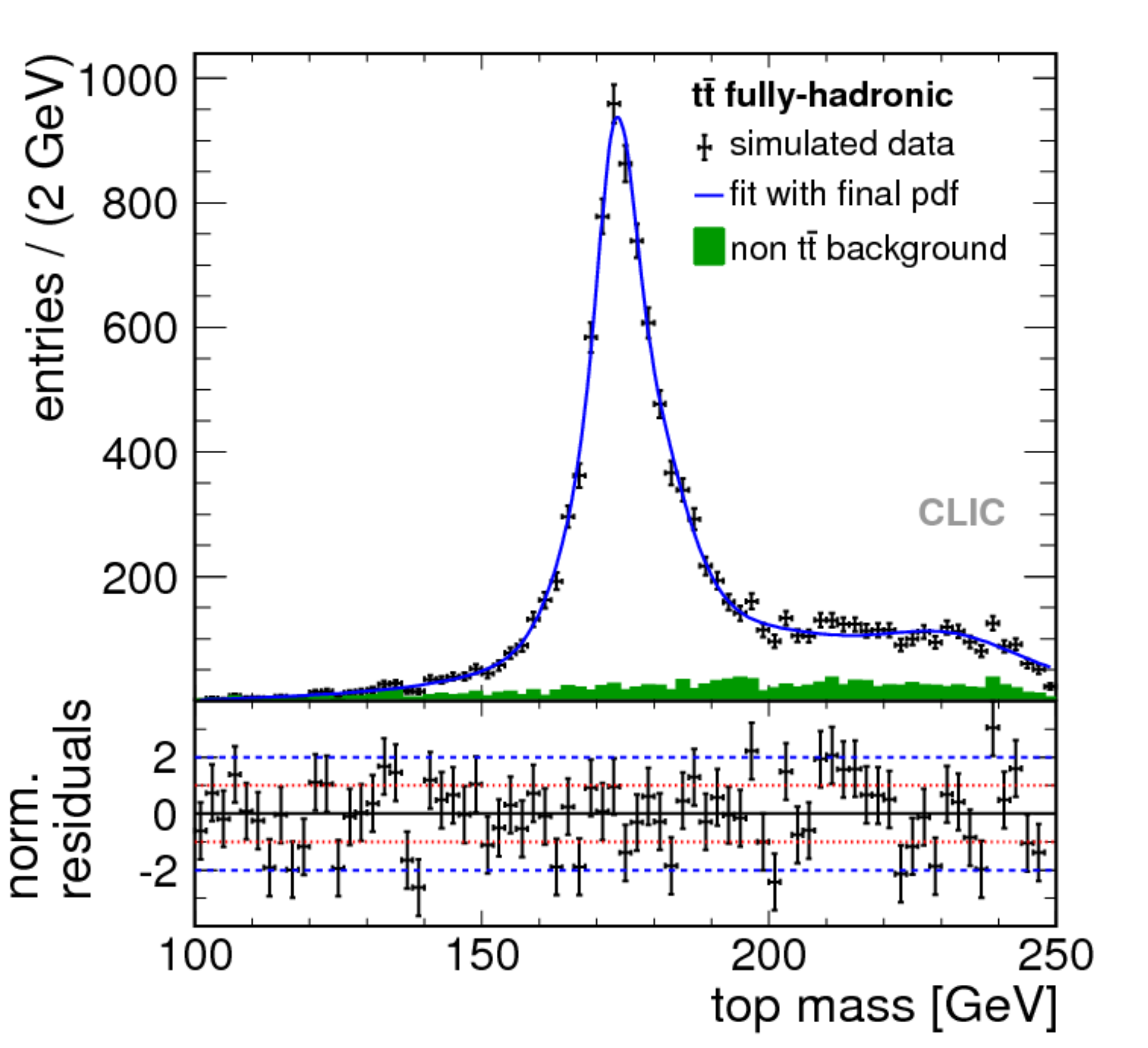}
 \end{center}
  \vspace{-5mm}
 \caption{Simulated measurement of the top-quark invariant mass in the all-hadronic decay channel of top-quark pairs for an integrated luminosity of 100 fb$^{-1}$ at CLIC in the CLIC\_ILD detector at a center-of-mass energy of 500 GeV. The solid green histogram shows the remaining non $t\bar{t}$ background in the data sample. The mass is determined with an unbinned maximum likelihood fit to the distribution. Figure taken from Ref.~\cite{Seidel:2013sqa}.}
 \label{fig:Top:Exp:InvMass}
\end{figure}

For both CLIC and ILC this measurement has been studied using full
detector simulations with all relevant physics backgrounds at an energy
of 500 GeV. In the case of the CLIC study, also the influence of
background from hadron production in two-photon processes was included,
which is more severe at CLIC than at ILC due to the very high
bunch-crossing frequency. The reconstructed invariant mass after
background rejection and kinematic fitting for the fully ha\-dronic
final state at CLIC is shown in Fig.~\ref{fig:Top:Exp:InvMass}. The
figure also illustrates the high purity achievable for top quarks at
linear colliders. For an integrated luminosity of 100 fb$^{-1}$,
combined statistical precisions of 70 MeV and 80 MeV are obtained for
ILC \cite{Abe:2010aa} and CLIC \cite{Seidel:2013sqa}, respectively. The
CLIC study showed that it is expected that systematic uncertainties due
to the jet energy scale can be limited to below the statistical
uncertainty by constraining the light jet energy scale through the
direct reconstruction of the $W$ bosons in the top-quark decay. The $b$ jet
energy scale in turn can be determined in a similar way from
$Z\rightarrow b\bar{b}$ decays. Also other experimental systematics, such as the knowledge
of the beam energy, which enters in the kinematic fit, and uncertainties from color reconnection
effects are expected to be small.

However, in contrast to the measurement via a threshold scan, the mass determined by direct reconstruction is theoretically not well defined. 
Rather, it is obtained in the context of the event generator used to determine the detector and reconstruction effects on the measured invariant
mass. At present, no conversion of this invariant mass value to the 
$\overline{\mbox{MS}}$ mass exists. This leads to additional uncertainties in the interpretation of the result, which potentially far exceed the experimental accuracy of the invariant mass measurement.

\subsubsection{Measurement of coupling constants}
\begin{sloppypar}
For precise test of the Standard Model and New Physics searches a
precise determination of the Standard Model couplings together with
the search for anomalous couplings is important. In the following we
try to review the prospects of a future Linear Collider and compare
where possible with the LHC.  From top-quark pair production at hadron
collider the top-quark coupling to gluons is already constrained. As
mentioned in Section \ref{sec:thresholdmeasurements} the threshold
studies can be used to measure the top-quark mass together with \as.
Top-quark pairs produced in association with an additional jet can be
used to search directly for anomalous top-gluon couplings. This can be
done independent from the production mechanism in hadronic collisions
as well as in electron--positron annihilation.  For hadronic $t\bar t
+ \mbox{1-Jet}$ production dedicated NLO calculations are available
\cite{Dittmaier:2007wz,Dittmaier:2008uj,Bevilacqua:2011aa,Melnikov:2011qx}.
For electron positron annihilation the corresponding calculations for
massive $b$-quarks
\cite{Rodrigo:1997gy,Bernreuther:1997jn,Brandenburg:1997pu,Nason:1997nw,Rodrigo:1999qg}
can be applied by adjusting the coupling constants. A dedicated
analysis of top-quark pair + 1-Jet production at a future Linear
Collider can be found in Ref.~\cite{Brandenburg:1999fv}.  Since
anomalous couplings will show up more likely in the couplings to the
weak gauge bosons no detailed study of the sensitivity to anomalous
top-gluon couplings has been performed so far for a future Linear Collider.\\
\end{sloppypar}
\begin{sloppypar}
The $Wtb$-coupling can be probed through top-quark decay and
single--top-quark production. A detailed measurement of this coupling
is interesting because the $V-A$ structure of the vertex can be
tested.  Furthermore the existence of a fourth family ---if not yet
ruled out by other measurements--- could significantly change the SM
predictions for the respective coupling.  Tevatron and LHC
measurements constrain the coupling already through the measurement of
the top-quark width \cite{Abazov:2010tm} and the measurements of the
$W$-boson helicity fractions
\cite{Aaltonen:2012lua,Aad:2012ky,CMS:ywa}.  A measurement of the
top-quark width from threshold studies can be used to indirectly
constraint the coupling in electron-positron annihilation. A direct
measurement of the $Wtb$ coupling at a Linear Collider is difficult
\cite{Juste:2006sv}.  In top-quark pair production close to the
threshold the coupling enters only through the branching ratio for
$t\to Wb$ which is expected to be very close to one and thus does not
lead to a strong dependence on the $Wtb$ coupling.  Measurements
using single--top-quark production are difficult owing to sizeable
backgrounds.  In Ref.~\cite{Batra:2006iq} it has been argued that
using $e^+e^-\to W^+bW^-\bar b$ events below the $t\bar t$ threshold
the coupling can be measured at ILC with an accuracy of about $3\%$
using an integrated luminosity of about 100/fb.\\
The top-quark coupling to the photon or more precisely the top-quark
charge is constrained through indirect measurements at hadron
colliders.  Using the charge of the top-quark
decay products reconstructed from top-candidate events the top-quark
charge has been measured in Ref.~\cite{Aad:2013uza} to be 
\begin{equation}
  Q = 0.64 \pm 0.02 (\mbox{stat.}) \pm
  0.08 (\mbox{syst.}) 
\end{equation}
\begin{sloppypar}
\noindent in units of the electron charge. A direct
measurement of $t\bar t +\gamma$ production is difficult at the LHC
due to the small cross sections although a measurement with an
uncertainty of 10\% might nevertheless be feasible
\cite{Juste:2006sv}. (First results have been presented already by CDF
\cite{Aaltonen:2011sp} and ATLAS \cite{ATLAS:2011nka}.) At the Linear
Collider the analysis of the SM couplings is usually combined with the
search for anomalous couplings.  As a staring point one may use a form
factor decomposition of the form \cite{Juste:2006sv}:
\end{sloppypar}
\begin{eqnarray}
  \Gamma_\mu^{ttX}(k^2,q,\bar q) &=& i e 
  \bigg\{
    \gamma_\mu\left(\tilde F^X_{1V}(k^2)+\gamma_5 \tilde
  F^{X}_{1A}\right)\nonumber \\ 
&& + {(q-\bar Q)_\mu\over 2M_t}\left(\tilde F^X_{2V}(k^2)+\gamma_5 \tilde
  F^{X}_{2A}\right)
  \bigg\}
\end{eqnarray}
\end{sloppypar}
\begin{sloppypar}
where $X$ can be a photon as well as a $Z$ boson. In Refs.
\cite{AguilarSaavedra:2001rg,Baer:2013cma,Amjad:2013tlv} it has been
shown, that the precision, with which the various couplings can be
determined, can be improved at a Linear Collider by about a factor of
ten compared to what is possible at the LHC. At the LHC the precision
for $\tilde F_{1V}^\gamma$ and $\tilde F_{1A}^\gamma$ is at the level
of 10\% \cite{Amjad:2013tlv} and much larger for the remaining couplings. \\
Given that the top-quark is so much heavier than the next heavy quark
it seems reasonable to question whether the mechanism to generate the
top-quark mass is the same as for the lighter quarks. In this context
the measurement of the $t\bar tH$ Yukawa coupling is of great importance.
At the LHC this coupling can be accessed through the measurement of
top-quark pair production in association with a Higgs boson.  A recent
study of the sensitivity where the subsequent decay $H\to b\bar b$ has
been used can be found for example in Ref.~\cite{Goncalo:2013oma}.  In
Ref.~\cite{Dawson:2013bba} it has been estimated that the $ttH$
coupling can be measured at the LHC with an accuracy of about $15\%$
assuming an integrated luminosity of 300/fb at 14 TeV center of mass
energy. With an increased luminosity of 3000/fb a measurement at the
level of $7-14$\% may become feasible. 
Due to the large mass of the
final state it is difficult to improve this measurement significantly
at a linear collider operating at 500 GeV. For an integrated
luminosity of 1000/fb at 500 GeV center-of-mass energy an uncertainty
of 10 \% has been estimated \cite{Dawson:2013bba}. Increasing the
energy to 1 TeV (ILC) or even 1.4 TeV (CLIC) will help to improve the
situation: In both cases a precision of 4\% seems to be feasible. Using
the ILC design at 1 TeV would require 1000/fb of integrated
luminosity while at 1.4 TeV 1500/fb would be required.
\end{sloppypar}

Very recently it has been argued in Ref.~\cite{Farina:2012xp} that the
$ttH$ coupling could also be inferred at the LHC from single--top-quark
production in association with an additional Higgs. Since the cross
section of this process is below 100 fb such a measurement will be
challenging. In the Standard Model the cross section is reduced
through an accidental cancellation. As a consequence BSM models may
show sizeable deviations compared to the Standard Model prediction.

\subsubsection{The top-quark polarization}
Top-quarks produced in electron-positron annihilation are polarized.
Furthermore the spin of the top-quark is also correlated with the spin
of the anti--top-quark. As mentioned in Section
\ref{sec:top-quark-decay} the top-quark polarization can be inferred
from the angular distributions of the decay products. The top-quark
polarization thus provides an additional observable which allows a
more detailed test of the top-quark interactions. The top-quark
polarization and spin-correlations in electron-positron annihilation
have been studied in detail for example in
Refs.~\cite{Korner:1993dy,Tung:1994fe,Bernreuther:1992,Groote:1995ky,Parke:1996pr,Tung:1997ur,Groote:1997su,Harlander:1994ac,Brandenburg:1998xw}. 
In Ref.~\cite{Groote:2010zf} the impact of the beam polarization on the
polarization of the produced top-quarks has been investigated.  In
difference to the production rate the observables sensitive to the
top-quark polarization depend only on the effective beam polarization
\begin{equation}
  P_{eff} = {P_{e^-} -P_{e^+}\over 1-P_{e^-}P_{e^+}}
\end{equation}
where $\lambda_-$ $(\lambda_-)$ denotes the longitudinal polarization
of the incoming electrons (positrons). While the top-quark polarization depends
strongly on the $P_{eff}$ the longitudinal spin correlation depends
only weakly on $P_{eff}$. At a center-of-mass energy of 500 GeV the
polarization is close to maximal for $|P_{eff}|=1$. For higher
energies the polarization is reduced. However for $|P_{eff}|=1$ a
polarization above 85\% is still possible.




\section[Exploring the Quantum Level:\\ Precision Physics in the SM and BSM]{Exploring the Quantum Level:\\ Precision Physics in the SM and BSM\protect\footnotemark}
\footnotetext{Editors: S.~Heinemeyer, D.~Wackeroth\\
Contributors: A.~Denner,
S.~Dittmaier,
A.~Freitas,
S.~Godfrey,
N.~Greiner,
M.~Gr\"unewald,
A.~Hoecker,
R.~Kogler,
K.~M\"onig,
M.~Schmitt, 
D. St\"ockinger,
G.~Weiglein,
G.~Wilson,
L.~Zeune}
\label{sec:quantum}

\begin{sloppypar}
We review the LC capabilities to explore the electroweak (EW) sector
of the SM at high precision and the prospects of unveiling signals of
BSM physics, either through the presence of new particles in
higher-order corrections or via direct production of extra EW gauge
bosons.  We discuss the experimental and theory uncertainties in the
measurement and calculation of electroweak precision observables
(EWPO), such as the $W$ boson mass, $Z$ pole observables, in
particular the effective weak mixing angle, $\sweff$, and the
anomalous magnetic moment of the muon, $a_\mu$.  We concentrate on the
MSSM to illustrate the power of these observables for obtaining
indirect information on BSM physics. In particular, we discuss the
potential of two key EWPOs at a LC, $M_W$ and $\sweff$, to provide a
stringent test of the SM and constraints on the MSSM parameter
space. Naturally, the recent discovery of a Higgs-like particle at the
LHC has a profound impact on EW precision tests of the SM.  We present
a study of the impact of this discovery on global EW fits, and also
include a discussion of the important role of the top quark mass in
performing these high precision tests of the SM.  Finally, we review
the anticipated accuracies for precision measurements of triple and
quartic EW gauge boson couplings, and how deviations from SM gauge
boson self interactions relate to different BSM scenarios.  These
observables are of special interest at a LC, since they have the
potential of accessing energy scales far beyond the direct kinematical
reach of the LHC or a LC.  We conclude with a discussion of the LC
reach for a discovery of extra EW gauge bosons, $Z'$ and $W'$, and the
LC's role for pinning down their properties and origin, once
discovered.
\end{sloppypar}


\subsection{The role of precision observables}
\label{sec:quantum-intro}

\begin{sloppypar}
The SM cannot be the ultimate fundamental theory of particle
physics. So far, it succeeded in describing direct experimental data
at collider experiments exceptionally well with only a few notable
exceptions, e.~g., the left-right ($A_{\rm LR}^e$(SLD)) and
forward-backward ($A_{\rm FB}^b$(LEP)) asymmetry (see
Section~\ref{sec:sw2eff-constraints}), and the muon magnetic moment
$g_\mu-2$ (see Section~\ref{sec:quantum-amu}).  However, the SM fails to
include gravity, it does not provide cold dark matter, and it has no
solution to the hierarchy problem, i.e.\ it does not have an
explanation for a Higgs-boson mass at the electroweak scale.  On wider
grounds, the SM does not have an explanation for the three generations
of fermions or their huge mass hierarchies.  In order to overcome (at
least some of) the above problems, many new physics models (NPM) have
been proposed and studied, such as
supersymmetric theories, in particular the Minimal Supersymmetric
Standard Model (MSSM), Two Higgs Doublet Models (THDM), Technicolor,
little Higgs models, or models with (large, warped, or universal)
extra spatial dimensions.
So far, the SM has withstood all experimental tests at past and
present collider experiments, such as the LEP and SLC $e^+ e^-$
colliders, the HERA $ep$, Tevatron $p \bar p$, and LHC $pp$
collider. Even the recently discovered Higgs-like particle at the LHC,
after analyzing the 2012 data agrees with the SM Higgs boson
expectation, albeit more precise measurements of its properties will
be needed to pin down its identity. Measurements of precision
observables and direct searches for NPM particles succeeded to exclude
or set stringent bounds on a number of these models. The direct search
reach is going to be significantly extended in the upcoming years,
when the LHC is scheduled to run at or close to its design energy of
14 TeV.  Future $e^+e^-$ colliders, such as the ILC or CLIC, have good
prospects for surpassing the LHC direct discovery reach, especially in
case of weakly interacting, colorless NPM particles (see, e.~g.,
Section~\ref{sec:quantum-bsm-gb}).
\end{sloppypar}

Even if a direct discovery of new particles is out of reach, precision
measurements of SM observables have proven to be a powerful probe of
NPM via virtual effects of the additional NPM particles.  
In general, precision observables (such as particle masses, mixing
angles, asymmetries etc.)  that can be predicted
within a certain model, including higher order corrections in
perturbation theory, and thus depending sensitively on the other model
parameters, and that can be measured with equally high precision,
constitute a test of the model at the quantum-loop level. Various
models predict different values of the same observable due to their
different particle content and interactions. This permits to
distinguish between, e.~g., the SM and a NPM, via precision
observables. Naturally, this requires a very high precision of both
the experimental results and the theoretical predictions.  The wealth
of high-precision measurements carried out at the $Z$ pole at LEP and
SLC, the measurement of the $W$ boson at LEP and the
Tevatron~\cite{LEPEWWG,Group:2012gb,Schael:2013ita}, as well as measurements at low-energy
experiments, such as $a_\mu=(g_\mu-2)/2$ at the ``Muon $g-2$ Experiment''
(E821)~\cite{Bennett:2006fi}, are examples of EWPOs that probe indirect effects
of NPM particles. These are also examples where both experiment and theory
have shown that they can deliver the very high precision needed to fully
exploit the potential of these EWPOs for 
detecting minute deviations from the SM.
The most relevant EWPOs in which the LC plays a key role are the
$W$~boson mass, $\MW$, and the effective leptonic weak mixing angle,
$\sweff$.
In the MSSM, the mass of the lightest $\cp$-even MSSM Higgs boson, $\Mh$,
constitutes another important EWPO~\cite{Heinemeyer:2004gx}.  Note that
in these examples, the top quark mass plays a crucial role as input
parameter.

Also EWPOs that cannot be measured at a LC can be very
relevant in the assessment of its physics potential. A
prominent role in this respect plays the muon magnetic moment,
$(g_\mu-2)$. It already provides some experimental indication for NPM
particles in reach of a LC, and its role in constraining NPM and its
complementarity to the LC is summarized in
Section~\ref{sec:quantum-amu}.

Another type of PO is connected to the self interactions of EW gauge
bosons in multiple EW gauge boson production, i.e.\ they directly probe
the triple and quartic EW gauge boson couplings.  Deviations from SM
predictions would indicate new physics, entering either through loop
contributions or are due to new heavy resonances, which at low energy
manifest themselves as effective quartic gauge boson
couplings. Precision measurements of these POs could provide
information about NPM sectors far beyond the kinematic reach of the
LHC and LC.

\medskip
As discussed above, in this report we focus our discussion on the
EWPO, i.e.\ (pseudo-~)~observables like the $W$-boson mass, $\MW$, the
effective leptonic weak mixing angle, $\sweff$, and the anomalous
magnetic moment of the muon.  Since in the literature virtual effects of NPM
particles are often discussed in terms of {\em effective} parameters instead
of the EWPO we briefly discuss this approach in the following.

A widely used set of effective parameters are the $S$, $T$,
$U$~parameters~\cite{Peskin:1991sw}. They are defined such that they
describe the effects of new physics contributions that enter only via
vacuum-polarization effects (i.e.\ self-energy corrections) to the
vector boson propagators of the SM (i.e.\ the new physics
contributions are assumed to have negligible couplings to SM
fermions).  The $S$, $T$, $U$ parameters can be computed in different
NPM's as certain combinations of one-loop self-energies, and then
can be compared to the values determined from a fit to EW precision data,
i.e. mainly from $M_W, M_Z$ and $\Gamma_Z$ (see, e.g., the review
in~\cite{Beringer:1900zz}).  A non-zero result for $S$, $T$, $U$
indicates non-vanishing contributions of new physics (with respect to
the SM reference value).
According to their definition, the $S$, $T$, $U$ parameters are
restricted to leading order contributions of new physics. They should
therefore be applied only for the description of {\em small}
deviations from the SM pre\-dic\-tions, for which a restriction to the
leading order is permissible.  Examples of new physics contributions
that can be described in the framework of the $S$, $T$, $U$ parameters
are contributions from a fourth generation of heavy fermions or
effects from scalar quark loops to the $W$- and $Z$-boson observables.
A counter example, i.e. where the $S$, $T$, $U$ framework is not
sufficicent, are SUSY corrections to the anomalous magnetic moment of
the muon.  Due to these restrictions of this {\em effective}
description of BSM effects in $W$ and $Z$ boson observables, in this report we
decided to only present investigations of these effects in the EWPO
themselves.

\medskip
This review of precision physics in the SM and BSM at the LC is
organized as follows: in Section~\ref{sec:quantum-MW} we concentrate
on $M_W$ from both the experimental and theoretical view point, and
then turn to a discussion of $Z$ pole observables, in particular
$\sweff$, in Section~\ref{sec:quantum-sw2eff}. The relevance of the top
quark mass in EW precision physics is briefly summarized in
Section~\ref{sec:quantum-mt}, before we present the prospects of
extracting information about the SM Higgs boson mass from a global EW fit in
Section~\ref{sec:quantum-fit}. We close our discussion of EWPOs with
an overview of predictions for the muon magnetic moment in NPM in
Section~\ref{sec:quantum-amu}. An overview of possible parametrizations
of non-standard EW gauge boson couplings, available calculations and
the experimental prospects for precision measurements of these
couplings is presented in Section~\ref{sec:quantum-agc}. And finally,
in Section~\ref{sec:quantum-bsm-gb} we present an overview of studies
of new gauge bosons at the LC.


\subsection{The \boldmath{$W$}~boson mass}
\label{sec:quantum-MW}

The mass of the $W$ boson is a fundamental parameter of the
electroweak theory and a crucial input to electroweak precision tests.
The present world average for the W-boson mass~\cite{Group:2012gb},
\begin{align}
\MW &= 80.385 \pm 0.015 \gev~,
\label{mwexp}
\end{align}
is dominated by the results from the Tevatron, where the $W$ boson
mass has been measured in Drell--Yan-like single-$W$-boson production.
At LEP2, the $W$-boson mass had been measured in $W$-pair production
with an error of $33\mev$ from direct reconstruction and $\sim200\mev$
from the cross section at threshold~\cite{Alcaraz:2006mx,Schael:2013ita}.  In
this 
section we will review the prospects for the $\MW$ measurements at the
LC from the experimental and theoretical side, as well as the
possibility to constrain indirectly parameters of new physics models
using a precise $\MW$ measurement and prediction.

\subsubsection{Experimental prospects for a precision measurement of
  \boldmath{$\MW$} a the ILC \protect\footnotemark}
\footnotetext{Graham Wilson}
\label{sec:quantum-MWexp}

%
%
%
The ILC facility%
\footnote{
We refer in this section particularly to the ILC which has a number of 
advantages over other proposed facilities, notably
the ability to polarize both beams, to run in an optimized fashion at 
a variety of center-of-mass energies, and with
a good quality luminosity spectrum.
}%
~can contribute decisively by making 
several complementary measurements of 
the $W$~mass using $e^{+}e^{-}$ collisions at 
centre-of-mass energies spanning from near $WW$ threshold to as high as 1 TeV.
Data samples consisting of between 10 and 100 million $W$ decays can be
produced, corresponding to an integrated luminosity of about $250$fb$^{-1}$ at 
$\sqrt{s} = 250$~GeV (and correspondingly lower integrated luminosity at
higher energies).

\begin{sloppypar}
The main production channels of W bosons at ILC are pair production, 
$e^{+} e^{-} \to W^+ W^-$ 
and single-$W$ production, $e^{+} e^{-} \to W e \overline{\nu}_e$ which 
proceeds 
mainly through $\gamma-W$ fusion.
Pair production dominates at lower center-of-mass energies while single-$W$ 
production dominates over other $e^+e^-$ sources of hadronic events at 
the higher energies.
\end{sloppypar}

The three most promising approaches to measuring the $W$~mass are: 
\begin{itemize}
\item Polarized threshold scan of 
the $W^+W^-$ cross-section as discussed in~\cite{Wilson:2001aw}.
\item Kinematically-constrained reconstruction of $W^+W^-$ using 
constraints from four-momentum conservation and optionally mass-equality 
as was done at LEP2.\footnote{The 
literature from the LEP2 era usually refers to these methods 
as ``direct reconstruction''.}
\item Direct measurement of the hadronic mass. This can be applied 
particularly to single-$W$ events decaying hadronically 
or to the hadronic system in semi-leptonic $W^+W^-$ events.
\end{itemize}

Much of the existing literature on $\MW$ measurement from LEP2 is still very relevant, 
but one should be aware of a number of LC features which make the LC 
experimental program for $\MW$ measurements qualitatively different. 
Notable advantages are: availability of longitudinally polarized beams, 
energy and luminosity reach, and much better detectors. Notable concerns are related to 
potential degradation of the precision knowledge of the initial state.

We first give an outline of statistical considerations 
for $\MW$ measurements and then outline the strategies considered 
for being able to make use of this considerable statistical power 
in experimentally robust ways.

\bigskip 

\begin{sloppypar}
The {\it statistical} errors on a $W$~mass determination at ILC 
are driven by the cross-sections, the intrinsic width 
of the $W$ ($\Gamma_W \approx 2.08 \gev$), the potential integrated luminosity,
the availability of polarized beams, and where appropriate 
the experimental di-jet mass resolution, event selection efficiencies and
backgrounds. 
The width is the underlying fundamental issue. 
This broadens the turn-on of the $W$-pair cross-section near threshold, 
decreasing its dependence on $\MW$. It also broadens 
the $W$ line-shape, diluting the statistical power of mass measurements for 
both kinematically-constrained reconstruction and direct mass reconstruction.
For the detectors envisaged at ILC, hadronically decaying $W$'s should be
measured with mass resolutions in the $1-2 \gev$ range.
\end{sloppypar}

We have estimated the statistical sensitivity dependence on experimental mass 
resolution quantitatively using a fit to the simulated measured line-shape for 
one million $W$ decays while varying the assumed experimental mass resolution
(per decay). 
Results of a fit with a (non-relativistic) Breit-Wigner convolved with 
a Gaussian of known width (Voigtian fit) are shown in
Figure~\ref{fig:mwstat}. 
One sees from this that statistical sensitivities  
of around $2.5 \mev$ per million $W$ decays 
are to be expected for mass resolutions in 
the $1-2 \gev$ range. In practice experiments will use a variety of analysis 
techniques such as convolution fits where 
one takes into account the mass resolution on an event-by-event basis 
maximizing the statistical power of well-measured events and de-weighting 
events with worse resolution. With a data-sample with several 
tens of millions of W decays, the end result will be 
statistical sensitivity on $\MW$ below $1 \mev$ and potentially in the 
$0.5 \mev$  range.

\begin{figure}
\centering
\includegraphics[width=0.45\textwidth]{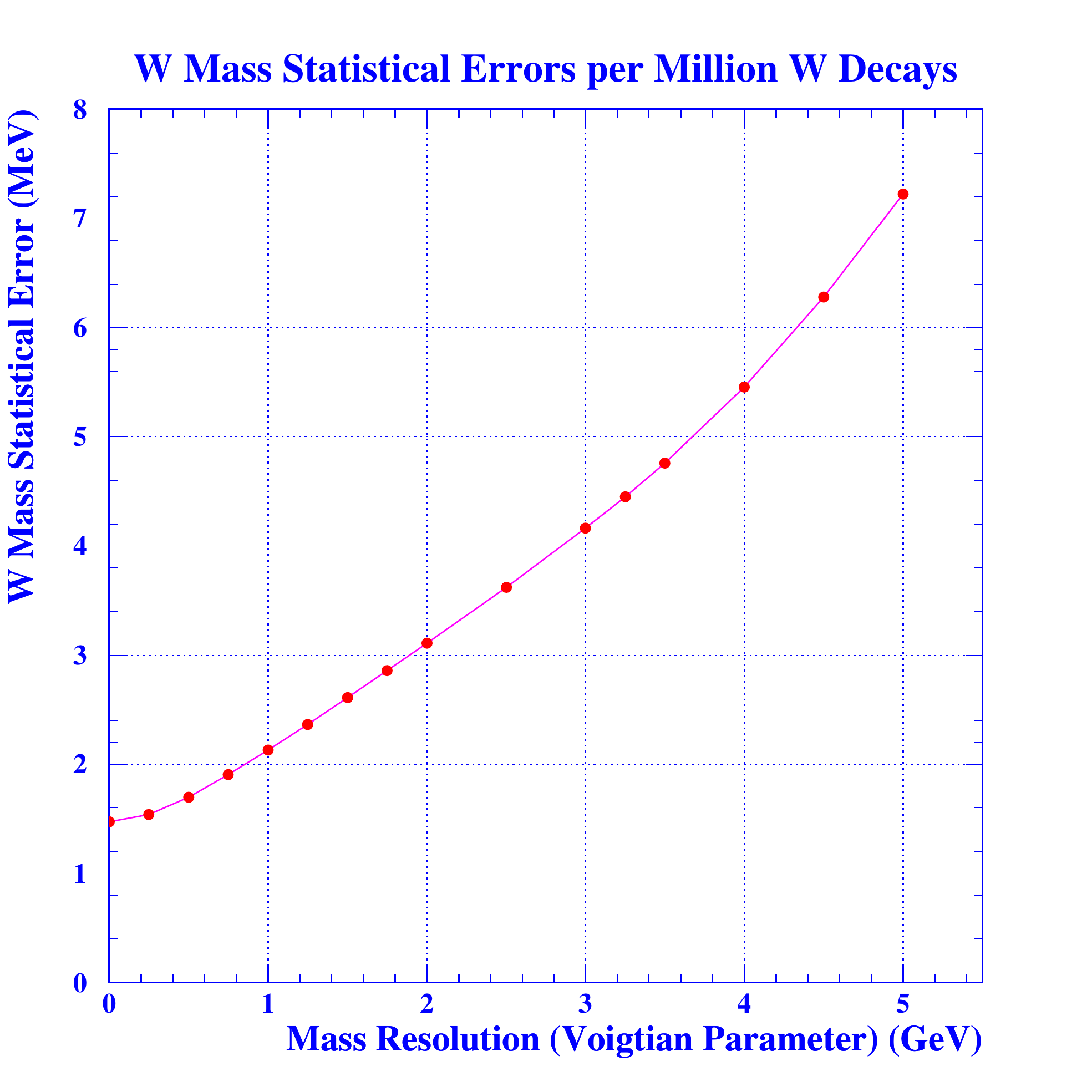}
\caption{Statistical precision on $\MW$ from the Voigtian fit (see text).
} 
\label{fig:mwstat}
\end{figure}

\begin{sloppypar}
Statistical errors from a single cross-section measurement near 
threshold ($\sqrt{s} \approx 2  \MW + 0.5 \gev$) 
are discussed in~\cite{LEP2-YR}. The statistical sensitivity 
factor on $\MW$ for an optimized single cross-section measurement
assuming unpolarized beams, 100\% efficiency and no backgrounds 
is $0.91 \mev/\sqrt{{\cal L}_{\rm int} [\mathrm{ab}^{-1}]}$.
For an integrated luminosity of ${\cal L}_{\rm int} = 100~\mathrm{fb}^{-1}$ this 
translates to $2.9 \mev$. However experimental systematic errors on such a 
single cross-section measurement of $\sim 0.25\%$ enter directly and 
would give a corresponding $4.2 \mev$ experimental systematic uncertainty.
At the ILC, the statistical sensitivity factor can be further improved using
polarized beams colliding with the appropriate helicities corresponding
effectively for practical  
polarization values (80-90\%, 40-60\%) to a factor of up to three 
$WW$-production luminosity upgrade.
\end{sloppypar}

\bigskip 

The method of a
polarized threshold scan is discussed in some detail in ~\cite{Wilson:2001aw} based
on conservative extrapolations from the measurements using the LEP detectors.
The idea is to use the measurement of the threshold dependence of 
the cross-section to determine $\mW$.
The study is based on 100 $\mathrm{fb}^{-1}$ allocated to 5 scan points 
near threshold and 1 scan point at $170 \gev$. Data are collected mostly 
with $e^{-}_{L} e^{+}_{R}$ but other combinations 
of two-beam, single-beam and no beam polarization are used to control 
the backgrounds and polarization systematics. 
The $170 \gev$ point has little sensitivity to $\mW$ but helps to 
constrain the efficiency systematics. 
The overall experimental error on the $W$~mass (excluding beam energy
systematic and eventual theoretical errors) is estimated to be $5.2 \mev$.

A critical external input needed to interpret the threshold 
dependence of the cross-section in terms of $\mW$ is knowledge 
of the center-of-mass energies. Various measurements sensitive to the
center-of-mass energy can  
be made using $e^+e^- \to \ell \ell \gamma$ ($\ell = e, \mu$) events.
From knowledge of the polar angles of the leptons, under the assumption of a
3-body final  
state, one can measure statistically the luminosity-weighted center-of-mass energy 
with an error of 31 ppm for the proposed scan. This translates into a $\mW$ error 
of $2.5 \mev$ per 100$\mathrm{fb}^{-1}$ polarized scan. A related method using 
the momenta of the two leptons (particularly the muons) can determine the center-of-mass energy 
with much better statistical precision. The tracker momentum scale needs 
to be controlled - this is feasible using $Z$'s - and potentially with 
other particles with well measured masses.

In summary, it is estimated that $\mW$ can be measured to $6 \mev$ 
experimental accuracy using this method which uses dedicated running near
threshold. This number includes also the anticipated uncertainties from the
beam energy ($\sim 1.9 \mev$) and from theory ($\sim 2.5 \mev$), where
the corresponding theoretical issues will be discussed in the next
subsection.


\bigskip 

Much of the ILC program is likely to take place at energies 
significantly above the $WW$ threshold in a regime where both $WW$ 
production and single-$W$ production are prevalent. 
Consequently, a direct reconstruction of the hadronic mass can be very
important. 
One can use $WW$ events with one $W$ decaying leptonically ($e, \mu, \tau$)
and the other decaying hadronically, and also single-$W$ events 
with the $W$ decaying hadronically to measure $\mW$ 
from the measured hadronic mass. Beam polarization can be used 
to enhance the cross-sections.
The critical issue is being able to control the jet energy scale. 
A number of approaches are plausible and should be pursued. 
One approach consists of using $Z$($\gamma$) radiative return events where the
$Z$ decays hadronically  
and the photon is unmeasured within or close to the beam-pipe.
Another approach attempts to do a jet-energy 
calibration from first principles using the
individual components that make up the measured jet energy, namely using 
the calibration of the
tracker momentum scale and the calorimeter energy scales at the 
individual particle level determined from for
example calibration samples of well-known particles ($J/\psi$, $K^0_S$,
$\Lambda$, $\pi^0$ etc.).
The latter has the advantage that it 
does not rely directly on the $Z$ mass. Other calibration possibilities 
are using $ZZ$, $Zee$ and Z$\nu\nu$ events. Assuming a sample
of $5\,10^6$ hadronic $Z$'s for calibration one should be able to 
approach a jet-energy scale related statistical error of around $2.0 \mev$ for
$\mW$.  
Systematic limitations in the $Z$ based methods is the knowledge 
of the $Z$ mass (currently $2.1 \mev$) 
- and any residual quark-flavor related systematics that 
make the detector response of hadronic $W$'s different from hadronic $Z$'s.
It seems plausible to strive for an overall error of $5 \mev$ from these
methods. 

\bigskip 

A kinematically-constrained reconstruction of $WW$ pairs 
was the work-horse of LEP2 - but has received little 
attention to date for ILC studies related to $W$~mass measurement.
By imposing kinematic constraints, the LEP2 experiments were able 
to compensate for modest jet energy resolution. At ILC, the constraints are no 
longer as valid (beamstrahlung) the detector resolution is much better 
(of the same order as $\Gamma_W$), and until recently, it seemed that the beam energy could not 
be determined with adequate precision at high energy.
Lastly, at the order of precision that is being targeted, it seems unwise to bank on  
the fully hadronic channel where it is quite possible that final-state interactions such as 
color reconnection may cause the mass information to be corrupted.
So it seems that the kinematically-constrained reconstruction method is 
most pertinent to the $q \bar q e \nu_e$ and $q \bar q \mu \nu_\mu$ channels.

\begin{sloppypar}
Recent work exploring the reconstruction of the center-of-mass energy using 
the measured muon momenta in $e^+e^- \to \mu^+ \mu^- (\gamma)$ events 
indicates that it is very feasible to measure the luminosity-weighted center-of-mass energy 
with high precision, and that this approach is promising 
also at relatively high center-of-mass energies.
\end{sloppypar}

\begin{sloppypar}
In addition, given the impetus for potentially running the ILC at a center-of-mass energy of 
around $250 \gev$, not far above LEP2, 
there seems a clear potential to improve the $\mW$ measurement by including information 
from the leptons in the mass estimate.
This lower energy regime should be the most favorable for 
beamstrahlung and beam-energy determination outlook.
Probably by performing kinematically constrained fits 
that build on the existing methods one would be able to get complementary 
information which would be significantly uncorrelated in several of 
the main systematics with the direct reconstruction method.
This deserves more study - but errors at the $5 \mev$ level or less may be
achievable. 
\end{sloppypar}

\bigskip 
To summarize, 
the ILC facility has three principal ways of measuring $\mW$. 
Each method can plausibly measure $\mW$ to a precision in the $5 \mev$ range. 
The three methods are largely uncorrelated. If all three methods do live 
up to their promise, one can target an overall uncertainty 
on $\mW$ in the $3-4$~MeV range.


%

\subsubsection{Theory aspects concerning the $WW$ threshold scan
\protect\footnotemark}
\footnotetext{Ansgar Denner, Stefan Dittmaier}
\providecommand{\epem}{e^+e^-}

\noindent While in the previous subsection the experimental precision for the $W$~boson
mass measurement at the LC was discussed, this subsection deals with the
correspondingly required theory calculations and precisions, in particular for
the $WW$ threshold scan.

The theoretical uncertainty (TU) for the direct mass reconstruction at
LEP2 has been estimated to be of the order of $\sim {5-10}\mev$
\cite{Jadach:2001cz,Cossutti}, based on results of 
{\sc YFSWW}~\cite{Jadach:2001uu} and
{\sc RacoonWW}~\cite{Denner:2000bj}, 
which used the double-pole approximation (DPA) for the
calculation of the NLO corrections. This is barely sufficient for the
accuracies aimed at a LC.
These shortcomings of the theoretical predictions have been cured by
dedicated calculations.  

\begin{sloppypar}
In \cite{Denner:2005es,Denner:2005fg} the total cross section for
the charged-current four-fermion production processes
$\epem\to\nu_\tau\tau^+\mu^-\bar\nu_\mu$, $u\bar{d}\mu^-\bar\nu_\mu$,
${u}\bar{d}s\bar{c}$ was presented including the complete electroweak
NLO corrections and all finite-width effects. This calculation was
made possible by using the complex-mass scheme for the description of
the W-boson resonances and by novel techniques for the evaluation of
the tensor integrals appearing in the calculation of the one-loop
diagrams. The full ${\cal O}(\alpha)$ calculation, improved by
higher-order effects from ISR, reduced the remaining TU due to unknown
electroweak higher-order effects to a few $0.1\%$ for scattering
energies from the threshold region up to $\sim500\gev$; above this
energy leading high-energy logarithms, such as Sudakov logarithms,
beyond one loop have to be taken into account to match this accuracy
\cite{Kuhn:2007ca}.  At this level of accuracy, also improvements in
the treatment of QCD corrections to semileptonic and hadronic
$\epem\to4f$ processes are necessary. The corrections beyond DPA, 
were assessed by comparing predictions in DPA from the generator {\sc RacoonWW}
to results from the full four-fermion calculation~\cite{Denner:2005es,Denner:2005fg},
as coded in the follow-up program {\sc Racoon4f} (which is not yet public).
This comparison revealed
effects on the total cross section without cuts of 
$\sim 0.3\%\;(0.6\%)$ for CM energies ranging from $\sqrt{s}\sim200\gev$
($170\gev$) to $500\gev$.  The difference to the DPA increases to
$0.7{-}1.6\%$ for $\sqrt{s}\sim 1{-}2\tev$.  At threshold, the full
${\cal O}(\alpha)$ calculation corrects the IBA by about $2\%$. While
the NLO corrections beyond DPA have been calculated only for the
processes $\epem\to\nu_\tau\tau^+\mu^-\bar\nu_\mu$,
$u\bar{d}\mu^-\bar\nu_\mu$, ${u}\bar{d}s\bar{c}$ so far, the effect
for the other four-fermion processes, which interfere with ZZ production,
should be similar. Once the corrections to those channels are needed,
they can be calculated with the available methods.
\end{sloppypar}

Using methods from effective field theory, the total cross section for
4-fermion production was calculated near the $W$ pair production
threshold \cite{Beneke:2007zg,Actis:2008rb}. These calculations used
unstable-particle effective field theory to perform an expansion in
the coupling constants, $\Gamma_W/\MW$, and in the non-relativistic
velocity $v$ of the $W$ boson up to NLO in
$\Gamma_W/\MW\sim\alpha_{\rm ew}\sim v^2$.  In \cite{Beneke:2007zg} the
theoretical error of an $\MW$ determination from the threshold scan
has been analysed. As a result, the resummation of next-to-leading
collinear logarithms from initial-state radiation is mandatory to
reduce the error on the $W$ mass from the threshold scan below $30\mev$.
It was found that the remaining uncertainty of the pure NLO EFT
calculation is $\delta\MW\approx10{-}15\mev$ and is reduced to about
$5\mev$ with additional input from the NLO four-fermion calculation in
the full theory.  In order to reduce this error further, in
\cite{Actis:2008rb} the (parametrically) dominant
next-to-next-to-leading order (NNLO) corrections (all associated with
the electromagentic Coulomb attraction of the intermediate W~bosons)
in the EFT have been calculated leading to a shift of
$\delta\MW\sim3\gev$ and to corrections to the cross section at the
level of $0.3\%$. The effect of typical angular cuts on these
corrections was shown to be completely negligible. Thus, one may
conclude that the inclusive partonic four-fermion cross section near
the W-pair production threshold is known with sufficient precision.

In summary, all building blocks for a sufficiently precise prediction
of the W-pair production cross section in the threshold region are
available. They require the combination of the NLO calculation of the
full four fermion cross section with the (parametrically) dominant
NNLO corrections, which are calculated within the EFT. 
For the precise determination of the
cross section at energies above $500\gev$ the leading
two-loop (Sudakov) corrections should be included in addition
to the full NLO corrections.
Combining the theoretical uncertainties with the anticipated precision from a
threshold scan (see the previous subsection) a total uncertainty of $7 \mev$
can be estimated~\cite{Baur:2001yp}.

\subsubsection{Theory predictions for $\MW$ in the SM and MSSM
\label{sec:quantum-MWtheo}
\protect\footnotemark}
\footnotetext{Sven Heinemeyer, Georg Weiglein, Lisa Zeune}

\begin{sloppypar}
\noindent The precise measurement of the $W$~boson mass can be used to test new physics
models via their contribution to quantum corrections to $\MW$. However, this
requires a precise prediction of $\MW$ in the respective models. Here we will
concentrate on the prediction of $\MW$ in the Minimal Supersymmetric Standard
Model (MSSM).
\end{sloppypar}

The prediction of $\MW$ in the MSSM depends on the masses, mixing angles and
couplings of all MSSM particles.
Sfermions, charginos, neutralinos and the MSSM Higgs bosons enter 
already at one-loop level and can give substantial contributions to $\MW$. 
Consequently, it is expected to obtain restrictions on the MSSM parameter
space in the comparison of the $\MW$ prediction and the experimental
value of \refeq{mwexp}. 

\begin{sloppypar}
The results for the general MSSM can be obtained in an extensive parameter
scan~\cite{Heinemeyer:2013dia}. The ranges of the various SUSY parameters are given
in \refta{tab:scanparam}. $\mu$ is the Higgsino mixing parameter, 
$M_{\tilde {F}_i}$ denotes the soft SUSY-breaking parameter for sfermions of
the $i$th family for left-handed squarks ($F = Q$), right-handed up- and
down-type squarks ($F = U, D$), left-handed sleptons ($F = L$) and
right-handed sleptons ($F = E$). $A_f$ denotes the trilinear sfermion-Higgs
couplings, $M_3$ the gluino mass parameter and $M_2$ the SU(2) gaugino mass
parameter, where the U(1) parameter is fixed as $M_1 = 5/3 \sw^2/\cw^2
M_2$. $M_A$ is the $\cp$-odd Higgs boson mass and $\tan\beta$ the ratio of the two
Higgs vacuum expectation values. 
\end{sloppypar}

\begin{table}[htb!]
{\footnotesize
\begin{tabular}{llll}
\hline
Parameter &  Minimum &  Maximum \\
\hline
$\mu$ & -2000       & 2000 \\
$M_{\tilde{E}_{1,2,3}}=M_{\tilde{L}_{1,2,3}}$ & 100       & 2000 \\
$M_{\tilde{Q}_{1,2}}=M_{\tilde{U}_{1,2}}=M_{\tilde{D}_{1,2}}$ & 500       & 2000 \\
$M_{\tilde{Q}_{3}}$     & 100       & 2000 \\
$M_{\tilde{U}_{3}}$     & 100       & 2000 \\
$M_{\tilde{D}_{3}}$     & 100       & 2000 \\
$A_e=A_{\mu}=A_{\tau}$   & -3$\,M_{\tilde{E}}$      & 3$\,M_{\tilde{E}}$     \\
$A_{u}=A_{d}=A_{c}=A_{s}$& -3$\,M_{\tilde{Q}_{12}}$  & 3$\,M_{\tilde{Q}_{12}}$ \\
$A_b$ & -3$\,$max($M_{\tilde{Q}_{3}},M_{\tilde{D}_{3}}$)  &
3$\,$max($M_{\tilde{Q}_{3}},M_{\tilde{D}_{3}}$) \\ 
$A_t$ & -3$\,$max($M_{\tilde{Q}_{3}},M_{\tilde{U}_{3}}$)  &
3$\,$max($M_{\tilde{Q}_{3}},M_{\tilde{U}_{3}}$) \\ 
$\tab$     & 1       & 60 \\
$M_3$     & 500     & 2000 \\
$M_A$     & 90      & 1000\\
$M_2$     &100      &1000\\
\hline
\end{tabular}
}
\caption{Parameter ranges. All parameters with mass dimension are given
  in GeV.
}
\label{tab:scanparam}
\end{table}

\begin{sloppypar}
All MSSM points included in the results have the neutralino as LSP
and the sparticle masses pass the lower mass limits from direct searches
at LEP.  
The Higgs and SUSY masses are calculated using {\tt FeynHiggs}
(version
2.9.4)~\cite{Frank:2006yh,Degrassi:2002fi,Heinemeyer:1998np,Heinemeyer:1998yj,Hahn:2009zz}.   
For every point it was tested whether it is allowed by direct Higgs searches
using the code {\tt HiggsBounds}  (version 4.0.0)
\cite{Bechtle:2008jh,Bechtle:2011sb}. 
This code tests the MSSM points against
the limits from LEP, Tevatron and the LHC.
\footnote{An updated version of {\tt HiggsBounds} became available at 
{\tt http://higgsbounds.hepforge.org} after this study was completed.
}
\end{sloppypar}

\begin{figure}
\centering
\includegraphics[width=0.8\columnwidth]{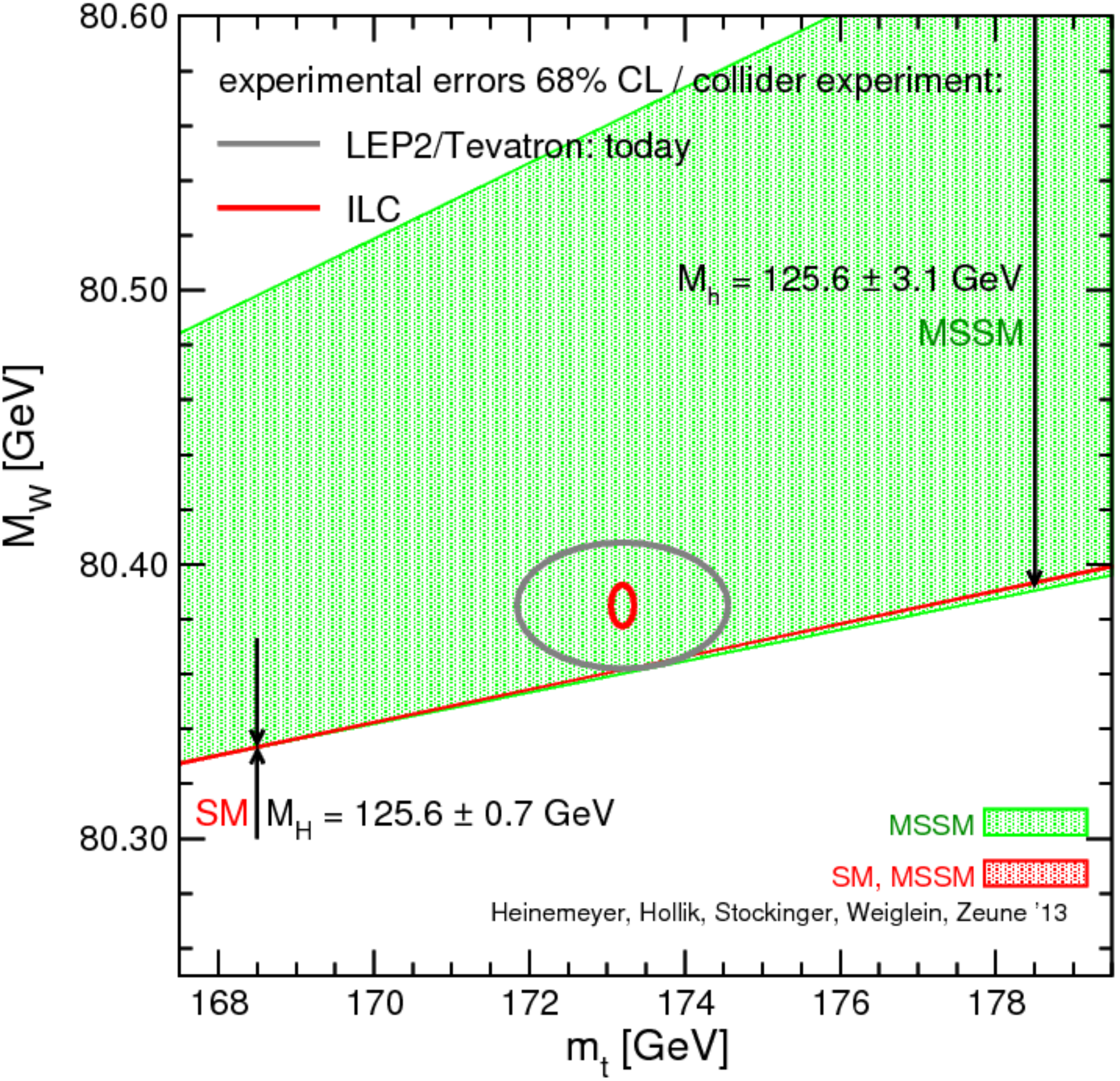}
\caption{Prediction for $\MW$ as a function of $\mt$.
The plot shows the $\MW$ prediction assuming the light
$\cp$-even Higgs $h$ in the region $125.6 \pm 3.1 \gev$. 
The red band indicates the overlap region of the SM and the MSSM with 
$M_H^{\rm SM}=125.6\pm 3.1$~GeV.
All points are allowed by {\tt HiggsBounds}.
The gray ellipse indicates the current experimental uncertainty, whereas the
red ellipse shows the anticipated future ILC/GigaZ precision.}
\label{fig:mtmwmh125}
\end{figure}

The evaluation of $\MW$ includes the full one-loop result and all
known higher order corrections of SM- and SUSY-type, for details
see~\cite{Heinemeyer:2013dia,Heinemeyer:2006px} and references therein.  The
results for $\MW$ are shown in \reffi{fig:mtmwmh125} as a function of
$\mt$. In the plot the green region indicated the MSSM $\MW$
prediction assuming the light $\cp$-even Higgs $h$ in the region
$125.6 \pm 3.1 \gev$.  The red band indicates the overlap region of
the SM and the MSSM.  The leading one-loop SUSY contributions arise
from the stop sbottom doublet. However requiring $M_h$ in the region
$125.6 \pm 3.1 \gev$, restricts the parameters in the stop
sector~\cite{Heinemeyer:2011aa} and with it the possible $\MW$
contribution.  Large $\MW$ contributions from the other MSSM sectors
are possible, if either charginos, neutralinos or sleptons are light.

\begin{sloppypar}
The gray ellipse indicates the current experimental uncertainty,
see \refeqs{mwexp}, (\ref{mtexp}), whereas the
red ellipse shows the anticipated future ILC/GigaZ precision.
While at the current level of precision SUSY might be considered as slightly
favored over the SM by the $\MW$-$\mt$ measurement, no clear conclusion can be
drawn. The small red ellipse, on the other hand, indicates the discrimination
power of the future ILC/GigaZ measurements. With the improved precision a
small part of the MSSM parameter space could be singled out. The
comparison of the SM and MSSM predictions with the ILC/GigaZ precision could
rule out either of models.
\end{sloppypar}


\subsection{\boldmath{$Z$} pole observables}
\label{sec:quantum-sw2eff}

Other important EWPOs are the various observables related to the $Z$
boson, measured in four-fermion processes, $e^+ e^- \to \gamma,Z \to f
\bar f$, at the $Z$~boson pole.  We review the theoretical precision
of SM predictions for various $Z$~boson pole observables and the
anticipated experimental precision at GigaZ. As for $\MW$, we also
review the potential of a precise measurement and prediction of
$\sweff$ to obtain information about the MSSM parameter space.

\newpage
\subsubsection{Theoretical prospects
\label{sec:quantum-sw2efftheo}
\protect\footnotemark}
\footnotetext{Ayres Freitas}

Near the $Z$-peak the differential cross section for $e^+e^- \to f\bar{f}$ can
be written as\footnote{For a review, see \cite{ALEPH:2005ab}.}
\begin{eqnarray}
&\frac{d\sigma}{d\cos\theta}
 = {\cal R}_{\rm ini} \biggl [
  \frac{9}{2}\pi\, \nonumber\\
&\frac{\Gamma_{ee}\Gamma_{ff}(1-{\cal P}_e{\cal A}_e)
   (1+\cos^2\theta) + 2({\cal A}_e-{\cal P}_e){\cal A}_f\cos\theta}%
  {(s-\MZ^2)^2-\MZ^2\Gamma_Z^2} + \sigma_{\rm non-res} \biggr ], 
  \label{xsecz} \nonumber \\
&\mbox{\rm where}\nonumber \\
&\Gamma_{ff} = {\cal R}^f_V g_{Vf}^2 + {\cal R}^f_A g_{Af}^2, \qquad
\Gamma_Z = \sum_f \Gamma_{ff}, \\
&{\cal A}_f = 2\frac{g_{Vf}/g_{Af}}{1+(g_{Vf}/g_{Af})^2}
 = \frac{1-4|Q_f|\sin^2\theta^f_{\rm eff}}{1-4\sin^2\theta^f_{\rm eff} +
     8(\sin^2\theta^f_{\rm eff})^2}.
\end{eqnarray}
Here $\Gamma_Z$ is the total $Z$ decay width,
$\Gamma_{ff}$ is the partial width for the decay $Z\to f\bar{f}$,
and $g_{Vf}$/$g_{Af}$ are the
effective vector/axial-vector couplings that mediate this decay. These effective
couplings include higher-order loop corrections to the vertex, except for
QED and QCD corrections to the external $f\bar{f}$ system, which are
captured by the radiator functions ${\cal R}^f_V$ and ${\cal R}^f_A$. The factor
$\cal R_{\rm ini}$, on the other hand, accounts for QED radiation in the
initial-state. (Specifically, as written in Eq.~\eqref{xsecz}, it describes
these effects \emph{relative} to the final-state radiation contribution for
$e^+e^-$.)

Eq.~\eqref{xsecz} explicitly spells out the leading $Z$-pole contribution, while
additional effects from photon exchange and box corrections are included in the
remainder $\sigma_{\rm non-res}$.

\begin{sloppypar}
The ratio of 
$g_{Vf}$ and $g_{Af}$ is commonly parametrized through the effective weak
mixing angle $\sin^2\theta^f_{\rm eff}$. It can be determined from the angular
distribution with respect to $\cos\theta$ or from the dependence on the initial
electron polarization ${\cal P}_e$. On the other hand, the partial and total
widths are determined from the total cross section $\sigma(s)$ for different
values of $s$ and from branching ratios (see the previous subsection).
\end{sloppypar}

For leptonic final states, the effective weak mixing angle
$\sin^2\theta^\ell_{\rm eff}$ has been calculated in the SM to the complete
two-loop order 
\cite{sineffl2,Awramik:2004ge,sineffl2a}, 
and 3- and 4-loop corrections of order ${\cal
O}(\alpha\als^2)$ \cite{sineffl3} and ${\cal O}(\alpha\alpha_{\rm
s}^3)$ \cite{sineffl4} are also known. Furthermore, the leading ${\cal
O}(\alpha^3)$ and ${\cal O}(\alpha^2\als)$  contributions for large
values of $m_t$ \cite{sineffmt} or $m_H$ \cite{sineffmh} have been computed.

\begin{sloppypar}
The current uncertainty from unknown higher orders is estimated to amount to
about $4.5\times 10^{-5}$ \cite{sineffl2a}, which mainly stems from missing ${\cal
O}(\alpha^2\als)$ and ${\cal O}(N_f^2\alpha^3,\,N_f^3\alpha^3)$ contributions beyond the
leading $m_t^4$ and $m_t^6$ terms, respectively. (Here $N_f^n$ denotes diagrams
with $n$ closed fermion loops. Based on experience from lower
orders, the ${\cal O}(\alpha^3)$ diagrams with several closed fermion loops are
expected to be dominant.) The calculation of these corrections requires three-loop
vertex integrals with self-energy sub-loops and general three-loop self-energy
integrals, which realisitically can be expected to be worked out in the
forseeable future. The remaining ${\cal O}(\alpha^3)$ and four-loop terms should
amount to $\sim 10^{-5}$.%
 \footnote{This estimate can be made more precise only after
aforementioned calculations have been completed.}
\end{sloppypar}

For quark final states, most 2-loop corrections to $\sin^2\theta^q_{\rm eff}$
have been computed \cite{sineffl2a,sineffb,Zqqaas}, but only the ${\cal
O}(N_f\alpha^2)$ and ${\cal O}(N_f^2\alpha^2)$ contributions are known for the
electroweak 2-loop corrections, while the diagrams without closed fermion loops
are still missing. However, based on experience from the leptonic weak mixing
angle, they are expected to amount to
$\,\raisebox{-.1ex}{$_{\textstyle<}\atop^{\textstyle\sim}$}\, 10^{-5}$. However,
the ${\cal O}(\alpha\als^2)$ also not known in this case, leading
to an additional theory error of $\sim 2\times 10^{-5}$. The calculation of
the missing ${\cal O}(\alpha\als^2)$ corrections, as well as the ${\cal
O}(\alpha^2\als)$ corrections, involves general 3-loop vertex
corrections to $Z \to q\bar{q}$, which will only be possible with serious progress in
calculational techniques.

When extracting $\sin^2\theta^\ell_{\rm eff}$ from realistic observables
(left-right (LR) and forward-backward (FB) asymmetries, see the next
subsection), the initial- and 
final-state QED radiator functions ${\cal R}_i$ must be taken into account. In general, the QED
corrections are known to ${\cal O}(\alpha)$ for the differential cross section
and to ${\cal O}(\alpha^2)$ for the integrated cross section (see
Ref.~\cite{zfitter} for a summary). However, for
the LR asymmetry they complete cancel up to NNLO \cite{QEDLR}, while for the FB
asymmetry they cancel if hard photon contributions are excluded, i.e.\ they
cancel up to terms of order $E_\gamma/\sqrt{s}$ \cite{QEDLR, QEDFB, hollikee}.
Therefore, a sufficiently precise result for the soft-photon contribution with
$E_\gamma < E_\gamma^{\rm cut}$ can be obtained using existing calcations for small enough $E_\gamma^{\rm
cut}$, while the hard-photon contribution ($E_\gamma > E_\gamma^{\rm cut}$) can
be evaluated with numerical Monte-Carlo methods.
A similar procedure can be carried out for final-state QCD effects for $\sin^2\theta^q_{\rm eff}$
although the corrections beyond NLO are not fully implemented in exisiting programs (see below).

\begin{sloppypar}
For the branching fraction $R_b = \Gamma_b/\Gamma_{\rm had}$ and the total width
$\Gamma_Z$, two-loop
corrections of ${\cal O}(\alpha\als)$, ${\cal O}(N_f\alpha^2)$, and
${\cal O}(N_f^2\alpha^2)$ are known \cite{Zqqaas,rb,gz}. Assuming geometric
progression of the perturbative series, the remaining higher-order contributions
are estimated to contribute at the level of $\sim 1.5\times 10^{-4}$ and $0.5$~MeV,
respectively. As before,
the contribution from electroweak two-loop diagrams without closed fermion loops
is expected to be small. The dominant missing contributions are the same as for
$\sin^2\theta^q_{\rm eff}$.
\end{sloppypar}

\begin{sloppypar}
The current status of the theoretical calculations and prospects for the near
future are summarized in Tab.~\ref{sumz}.
Note that $\sigma_{\rm non-res}$ is suppressed by $\Gamma_Z/\MZ$ compared to the
leading pole term, so that the known one-loop corrections are sufficient to
reach NNLO precision at the $Z$ pole.
\end{sloppypar}
\begin{table}[t]
{\scriptsize
\begin{tabular}{lrlc}
\hline
Quantity & Cur. theo. error & Lead. missing terms & Est.\ improvem.
\\
\hline\hline
$\sin^2\theta^\ell_{\rm eff}$ & $4.5\times 10^{-5}\qquad$ & 
 ${\cal O}(\alpha^2\als)$, ${\cal O}(N_f^{\ge 2}\alpha^3)$ & factor 3--5 \\ 
\hline
$\sin^2\theta^q_{\rm eff}$ & $5\times 10^{-5}\qquad$ & ${\cal O}(\alpha^2)$, 
 ${\cal O}(N_f^{\ge 2}\alpha^3)$ & factor 1--1.5 \\[-.5ex]
&& \sl [${\cal O}(\alpha\als^2)$, ${\cal O}(\alpha^2\als)$]
& \sl [factor 3--5] \\ 
\hline
$R_b$ & $\sim 1.5\times 10^{-4}\qquad$ & ${\cal O}(\alpha^2)$, 
 ${\cal O}(N_f^{\ge 2}\alpha^3)$ & factor 1--2 \\[-.5ex]
&& \sl [${\cal O}(\alpha\als^2)$, ${\cal O}(\alpha^2\als)$]
& \sl [factor 3--5] \\ 
\hline
$\Gamma_Z$ & $0.5\text{ MeV}\qquad$ & ${\cal O}(\alpha^2)$, 
 ${\cal O}(N_f^{\ge 2}\alpha^3)$ & factor 1--2 \\[-.5ex]
&& \sl [${\cal O}(\alpha\als^2)$, ${\cal O}(\alpha^2\als)$]
& \sl [factor 3--5] \\ 
\hline
\end{tabular}
}
\caption{Some of the most important precision observables for $Z$-boson
production and decay (first column), their present-day estimated theory error
(second column), the dominant missing higher-order corrections (third column),
and the estimated improvement when these corrections are available (fourth
column). In many cases, the leading parts in a large-mass expansion are already
known, in which case the third column refers to the remaining pieces at the
given order. The numbers in the last column are rough order-of-magnitude
guesses. Entries in {\sl [italics]} indicate contributions that probably will
require very significant improvements in calculational techniques to be
completed. }
\label{sumz}
\end{table}

The known corrections to the effective weak mixing angles and the leading
corrections to the partial widths
are implemented in programs such as {\tt Zfitter} \cite{zfitter,zfitter2} and
{\tt Gfitter}~\cite{gfitter} (see also \refse{sec:quantum-fit}), while the
incorporation of the recent full fermionic two-loop corretions is in progress. 
However, these programs are based on a
framework designed for NLO but not NNLO corrections. In particular,
there are mismatches between the electroweak NNLO corrections to the
$Zf\bar{f}$ vertices and QED/QCD corrections to the external legs due
to approximations and factorization assumptions. Another problem is
the separation of leading and sub-leading pole terms in
Eq.~\eqref{xsecz} \cite{sineffl2a}. While these discrepancies may be
numerically small, it would be desireable to construct a new framework
that treats the radiative corrections to $Z$-pole physics
systematically and consistenty at the NNLO level and beyond. Such a
framework can be established based on the pole scheme \cite{pole},
where the amplitude is expanded about the complex pole $s=\MZ^2-i
\MZ\Gamma_Z$, with the power counting $\Gamma_Z/\MZ \sim \alpha$.

\subsubsection{Experimental prospects
\label{sec:quantum-sw2effexp}
\protect\footnotemark}
\footnotetext{Klaus Moenig}
\newcommand{\ALR}    {A_{\mathrm{LR}}}
\newcommand{\ppl}  {{\cal P}_{e^+}}
\newcommand{\pmi}  {{\cal P}_{e^-}}
\newcommand{\ppm}  {{\cal P}_{e^\pm}}
\newcommand{\peff}  {{\cal P}_{\rm{eff}}}
\newcommand {\cAe} {\mbox{${\cal A}_{e}$}}
\newcommand {\cAb} {\mbox{${\cal A}_{b}$}}
\newcommand {\Rbz} {R_{b}^0}
\newcommand{\vef}{g_{V_e}}
\newcommand{\aef}{g_{A_e}}
\newcommand{\vf}{g_{V_f}}
\newcommand{\af}{g_{A_f}}
\newcommand{\sweffl}{\sin^2\theta_{\mathrm{eff}}^{\ell}}
\newcommand{\sweffb}{\sin^2\theta_{\mathrm{eff}}^{b}}
\newcommand{\swefff}{\sin^2\theta_{\mathrm{eff}}^{f}}


The effective weak mixing angle $\sweffl$ can be measured at a linear
collider running at the $Z$-mass using the left-right asymmetry
\cite{Hawkings:1999ac}.  
With at least the electron beam polarised with a polarisation of 
${\cal P}$,
$\sweffl$ can be obtained via

\begin{eqnarray}
\label{eq:alrdef}
\ALR & = & \frac{1}{{\cal P}}\frac{\sigma_L-\sigma_R}{\sigma_L+\sigma_R}\\
     & = &   \cAe \nonumber \\
     & = & \frac{2 \vef \aef}{\vef^2 +\aef^2} \nonumber \\
  {\vef}/{\aef} & = & 1 - 4 \sweffl \nonumber
\end{eqnarray}
independent of the final state.
With $10^9$ $Z$'s, an electron polarisation of 80\% and no positron
polarisation the statistical error is $\Delta \ALR = 4 \cdot 10^{-5}$.
The error from the polarisation measurement is
$\Delta \ALR/\ALR = \Delta {\cal P}/{\cal P}$.
With electron polarisation only and 
$\Delta {\cal P}/{\cal P} = 0.5\%$  one has
$\Delta \ALR = 8 \cdot 10^{-4}$, much larger than the statistical precision. 
If also positron polarisation is available ${\cal P}$ in equation
(\ref{eq:alrdef}) has to be replaced by 
$\peff \, = \, \frac{\ppl+\pmi}{1+\ppl\pmi}$. 
For $\pmi(\ppl) = 80\%(60\%)$, due to error propagation, the
error in $\peff$ is a factor of three to four smaller than the error on
$\ppl,\, \pmi$ depending on the correlation between the two
measurements. If one takes, however, data on all four polarisation
combinations the left-right asymmetry can be extracted without absolute
polarimetry \cite{Blondel:1987wr} and basically without increasing the error if
the positron polarisation is larger than 50\%. Polarimetry, however, is
still needed for relative measurements like the difference of absolute
values of the positive and the negative helicity states.
Assuming conservatively $\Delta \ALR = 10^{-4}$ leads to $\Delta \sweffl
= 0.000013$, more than a factor 10 better than the LEP/SLD result. 

\begin{sloppypar}
The largest possible uncertainty comes from the know\-ledge of the beam
energy. $\sqrt{s}$ must be known with $1\,{\rm MeV}$ relative to the
$Z$-mass. The absolute precision can be calibrated in a $Z$-scan, however a
spectrometer with a relative precision of $10^{-5}$ is needed not to be
dominated by this uncertainty. Similarly the beamstrahlung must be known to a
few percent relative between the calibration scans and the pole
running. However both requirements seem to be possible.
\end{sloppypar}

Apart from  $\sweffl$ also some other $Z$-pole observables can be measured at
a LC. Running at the $Z$ peak gives access to the polarised
forward backward asymmetry for $b$-quarks which measures $\sweffb$ and the
ratio of the $b$- to the hadronic partial width of the $Z$-boson
$\Rbz=\Gamma_{b \overline{b}}/\Gamma_{had}$. Both quantities profit from
the large statistics and the much improved $b$-tagging capabilities of an ILC
detector compared to LEP. 

$\Rbz$ can be measured 
using the same methods as at LEP. The statistical error will be almost
negligible and the systematic errors shrink due to the better $b$-tagging. In
total $\Delta \Rbz =  0.00014$ can be reached which is an improvement of a
factor five compared to the present value \cite{Hawkings:1999ac}.

$\sweffb$ can be measured from the left-right-forward-backward asymmetry
for $b$-quarks, $A_{\rm FB,LR}^b = 3/4 {\cal P} \cAb$. $\cAb$ depends on
$\sweffb$ as shown in eq.~(\ref{eq:alrdef}), however in general one has 
${\vf}/{\af} =  1 - 4 q_f \swefff$ and due to the small $b$-charge the
dependence is very weak. At present $\sweffb$ is known with a precision of
0.016 from $A_{\rm FB,LR}^b$ measured at the SLC and the
forward-backward asymmetries for $b$-quarks at LEP combined with $\sweffl$
measurements at LEP and SLC \cite{zpole}. Using the left-right-forward-backward
asymmetry only at the ILC an improvement by more than a factor 10
seems realistic \cite{Hawkings:1999ac}.

The total $Z$-width $\Gamma_Z$ can be obtained from a scan of the resonance
curve. The statistical error at GigaZ will be negligible and the systematic
uncertainty will be dominated by the precision of the beam energy and the
knowledge of beamstrahlung. If a spectrometer with a precision of $10^{-5}$
can be built, $\Gamma_Z$ can be measured with $1\,{\rm MeV}$
accuracy \cite{Hawkings:1999ac}. However no detailed study on the uncertainty due to
beamstrahlung exists.

\subsubsection{Constraints to the MSSM from 
\boldmath{$\sweff$} 
\label{sec:sw2eff-constraints}
\protect\footnotemark}
\footnotetext{Sven Heinemeyer, Georg Weiglein}

As for $\MW$ we review examples showing how the MSSM parameter space
could be constrained by a precise measurement of $\sweff$. We also
discuss the relevance of this measurement in a combined $\MW$-$\sweff$
analysis. 

In the first example it is investigated
whether the high accuracy achievable at
the GigaZ option of the LC would provide sensitivity to indirect effects of
SUSY particles even in a scenario where the (strongly interacting) 
superpartners are so heavy that they escape detection at the
LHC~\cite{Heinemeyer:2007bw}. 

\begin{sloppypar}
We consider in this context a scenario with very heavy squarks and a 
very heavy gluino. It is based on the values of the SPS~1a$'$ benchmark
scenario~\cite{Allanach:2002nj}, but the squark and gluino
mass parameters
are fixed to 6~times their SPS~1a$'$ values. The other masses are 
scaled with a common scale factor 
except $\MA$ which we keep fixed at its SPS~1a$'$ value.
In this scenario 
the strongly interacting particles are too heavy to be detected at the
LHC, while, depending on the scale-factor, some colour-neutral particles
may be in the LC reach. In \reffi{fig:sw2eff-theo} we show the prediction for
$\sweff$ in
this SPS~1a$'$ inspired scenario as a function of the lighter chargino
mass, $\mcha{1}$. The prediction includes the parametric
uncertainty, $\si^{\rm para-LC}$, induced by the LC measurement of $\mt$, 
$\de\mt = 100 \mev$ (see \refse{Top-QCD}), and the numerically more
relevant prospective future uncertainty on $\De\al^{(5)}_{\textup{had}}$,
$\de(\De\al^{(5)}_{\textup{had}})=5\times10^{-5}$. 
The MSSM prediction for $\sweff$
is compared with the experimental resolution with GigaZ precision,
$\si^{\rm LC} = 0.000013$, using for simplicity the current
experimental central value. The SM prediction (with 
$M_H^{\rm SM}=\Mh^{\rm MSSM}$) is also shown, applying again the parametric 
uncertainty $\si^{\rm para-LC}$.
\end{sloppypar}

Despite the fact that no coloured SUSY 
particles would be observed at the LHC in this scenario, the LC with
its high-precision 
measurement of $\sweff$ in the GigaZ mode could resolve indirect effects
of SUSY up to $m_{\tilde\chi^\pm_1} \lsim 500 \gev$. This means that the
high-precision measurements at the LC with GigaZ option could be
sensitive to indirect effects of SUSY even in a scenario where SUSY
particles have {\em neither \/} been directly detected at the LHC nor the
first phase of the LC with a centre of mass energy of up to $500 \gev$.


\begin{figure}[htb!]
\begin{center}
\includegraphics[width=6.7cm,height=5.7cm]{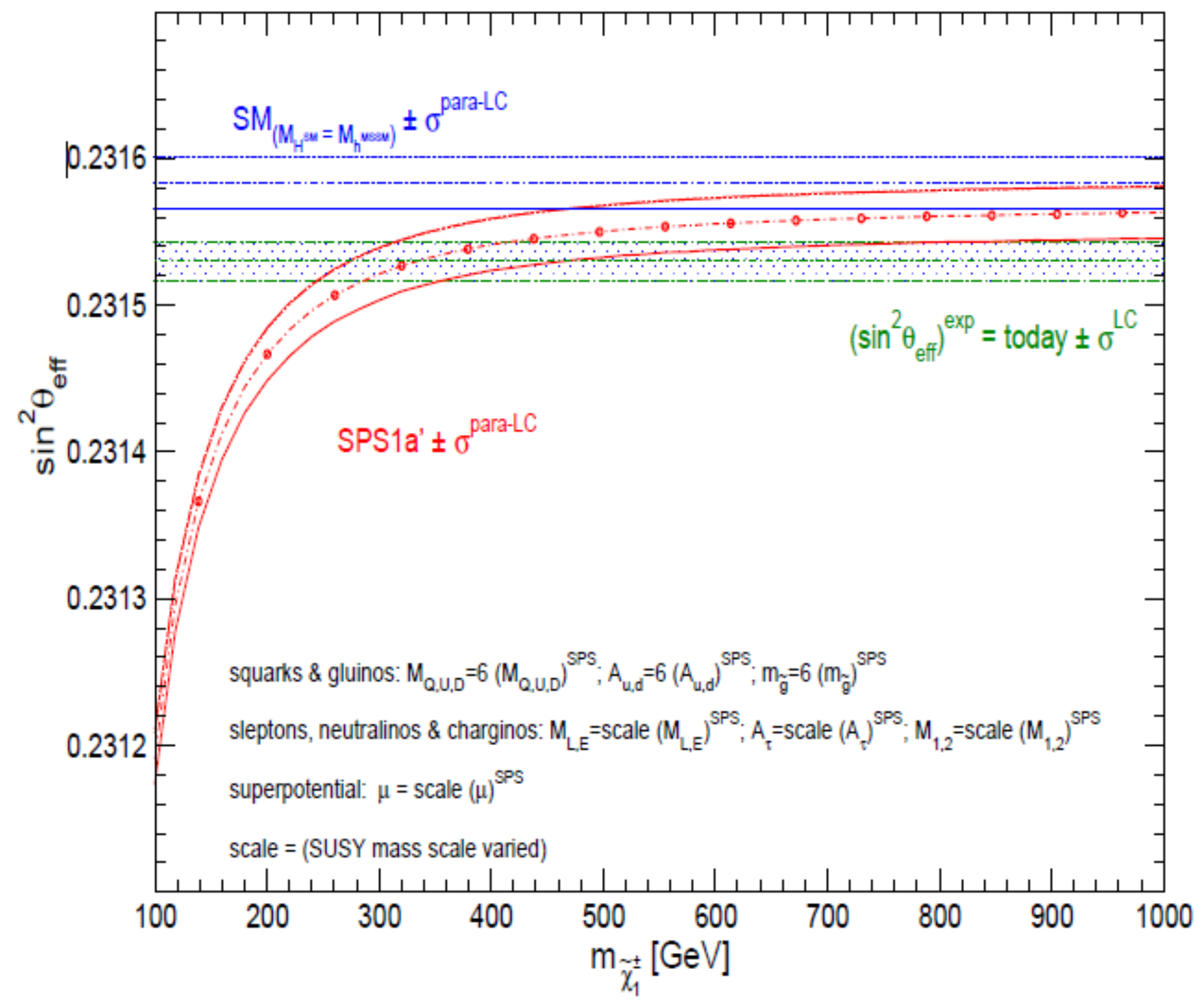}
\hspace{.3em}
\caption{
Theoretical prediction for $\sweff$ in the SM and the MSSM (including
prospective parametric theoretical uncertainties) compared to
the experimental precision at the LC with GigaZ option.  
An SPS~1a$'$ inspired scenario is used, where the squark and gluino
mass parameters
are fixed to 6~times their SPS~1a$'$ values. The other mass 
parameters are varied with a common scalefactor.}
\label{fig:sw2eff-theo} 
\end{center}
\end{figure}

\bigskip
We now analyse the sensitivity of $\sweff$ together with $\MW$ 
to higher-order effects in the MSSM by
scanning over a broad range of the SUSY parameter space. The following SUSY
parameters are varied independently of each other in a random parameter scan
within the given range:
\begin{eqnarray}
 {\rm sleptons} &:& M_{{\tilde L_{1,2,3}},{\tilde E_{1,2,3}}}
                   = 100\dots 2000\gev, \non \\
 {\rm light~squarks} &:& M_{{\tilde Q_{1,2}},{\tilde U_{1,2}},{\tilde D_{1,2}}}
                   = 100\dots 2000\gev, \non \\
 \Stop/\Sbot {\rm ~doublet} &:& M_{{\tilde Q_3},{\tilde U_3}, {\tilde D_3}}
                   = 100\dots 2000\gev, \non\\
 && 
 \quad A_{\tau,t,b} = -2000\dots 2000\gev, \non \\
 {\rm gauginos} &:& M_{1,2}=100\dots2000\gev, 
\\
 && 
 \quad \mgl=195\dots1500\gev, \non \\
 && \mu = -2000\dots2000\gev,\non \\ 
 {\rm Higgs} &:& \MA=90\dots1000\gev, \non \\
 && 
 \quad\tab = 1.1\dots60.
\label{scaninput}
\end{eqnarray}
Only the constraints on the MSSM parameter space
from the LEP Higgs searches~\cite{Barate:2003sz,Schael:2006cr} and the lower
bounds on the SUSY particle masses previous to the LHC SUSY searches were
taken into account. However, the SUSY particles strongly affected by the LHC
searches are the squarks of the first and second generation and the
gluino. Exactly these particles, however, have a very small effect on
the prediction of $\MW$ and $\sweff$ and thus a negligible effect on
this analysis.

\begin{sloppypar}
In \reffi{fig:sw2efftheo-Scans2} we
compare the SM and the MSSM predictions for $\MW$ and $\sweff$
as obtained from the scatter data. 
The predictions within the two models 
give rise to two bands in the $\MW$--$\sweff$ plane with only a relatively 
small overlap region (indicated by a dark-shaded (blue) area).
The parameter region shown in the SM (the medium-shaded (red)
and dark-shaded (blue) bands) arises from varying the mass of the SM
Higgs boson, from $M_H^{\rm SM} = 114\gev$, the old LEP exclusion
bound~\cite{Barate:2003sz} 
(lower edge of the dark-shaded (blue) area), to $400$~GeV (upper edge of the
medium-shaded (red) area), and from varying $\mt$ in the range of 
$\mt = 165 \ldots 175$~GeV. The value of $M_H^{\rm SM} \sim 125.5$~GeV
corresponds roughly to the dark-shaded (blue) strip.
The light shaded (green) and the
dark-shaded (blue) areas indicate allowed regions for the unconstrained
MSSM, where no restriction on the light $\cp$-even Higgs mass has been
applied. The decoupling limit with SUSY masses, in particular of scalar
tops and bottoms, of \order{2 \tev}
yields the upper edge of the dark-shaded (blue) area. 
Including a Higgs mass measurement into the MSSM scan would cut out a
small part at the lower edge of the light shaded (green) area.
\end{sloppypar}
\begin{figure}
\vspace{-4cm}
\begin{center}
\includegraphics[width=0.48\textwidth]
                {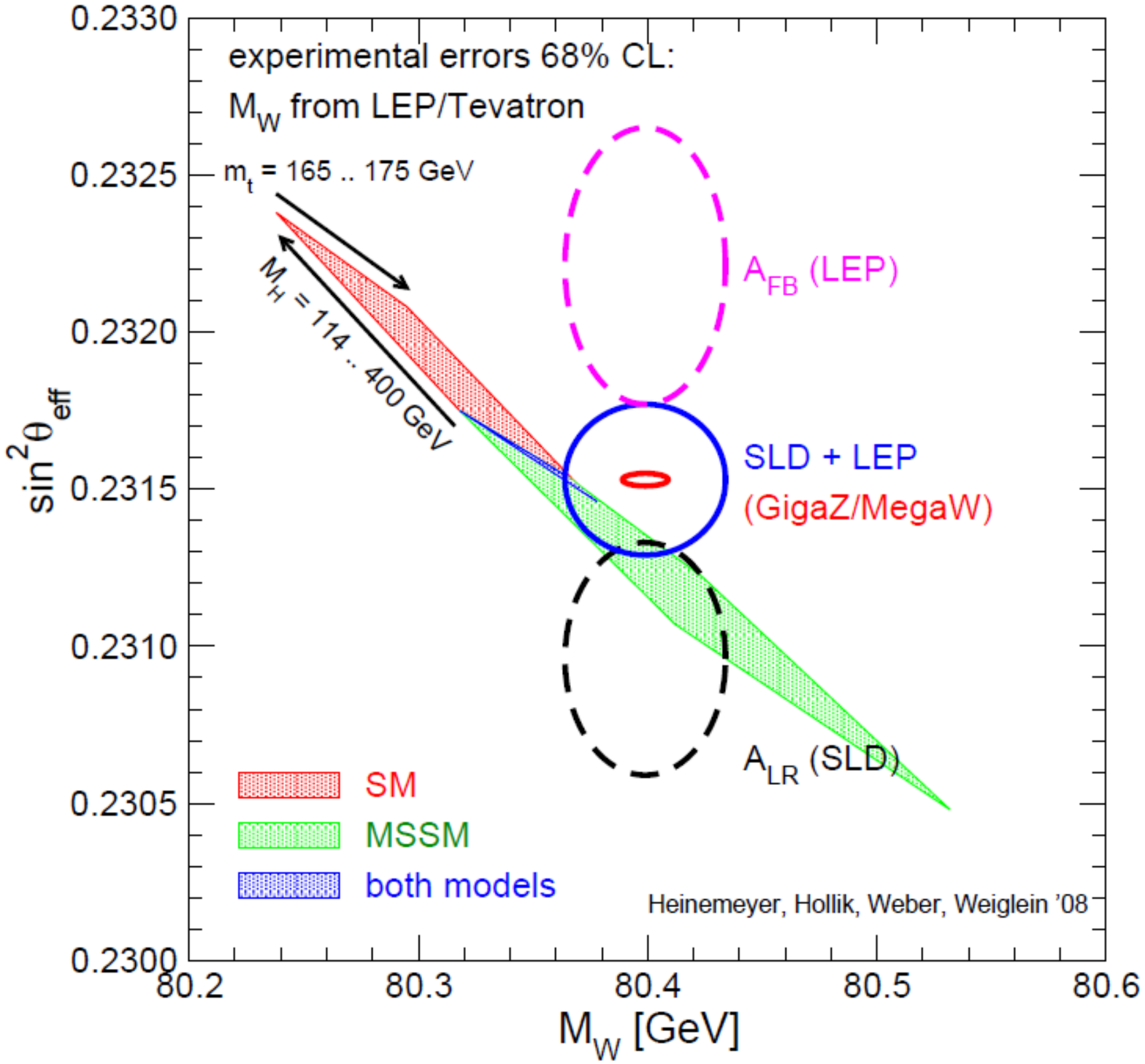}
\end{center}
\vspace{-1em}
\caption{MSSM parameter scan for $\MW$ and $\sweff$ over the
  ranges given in \refeq{scaninput} with 
  $\mt = 165 \ldots 175$~GeV. Todays 68\%~C.L.\ ellipses (from 
  $A_{\rm FB}^b({\rm LEP})$, $A_{\rm LR}^e({\rm SLD})$ and the world
  average) are shown 
  as well as the anticipated GigaZ/MegaW precisions, drawn around todays
  central value}  
\label{fig:sw2efftheo-Scans2} 
\end{figure}
%
\begin{sloppypar}
The 68\%~C.L.\ experimental results
for $\MW$ and $\sweff$ are indicated in the plot. 
The center ellipse corresponds to the current world average given
in \refeq{sweff-exp}. 
Also shown are the error ellipses corresponding to the two individual most
precise measurements of $\sweff$ , based on 
$A_{\rm LR}^e$ by SLD and $A_{\rm FB}^b$ by LEP, corresponding to
\end{sloppypar}
\begin{align}
\label{afb}
A_{\rm FB}^b({\rm LEP}) &: \sweff^{\rm exp,LEP} = 0.23221 \pm 0.00029~, \\[.1em]
\label{alr}
A_{\rm LR}^e({\rm SLD}) &: \sweff^{\rm exp,SLD} = 0.23098 \pm 0.00026~, \\[.1em]
\label{sweff-exp}
                       &~~ \sweff^{\rm exp,aver.} = 0.23153 \pm 0.00016~,
\end{align}
\begin{sloppypar}
where the latter one represents the average~\cite{LEPEWWG}.
The first (second) value prefers a 
value of $M_H^{\rm SM} \sim 32 (437) \gev$~\cite{gruenewaldpriv}. 
The two measurements differ by more than $3\,\si$.
The averaged value of $\sweff$ , as given in \refeq{sweff-exp},
prefers $M_H^{\rm SM} \sim 110$~GeV~\cite{gruenewaldpriv}. 
The anticipated improvement with the GigaZ/MegaW options (the latter one
denoting the $WW$~threshold scan, see \refse{sec:quantum-MW}), indicated as
small ellipse, is shown around the current experimental central data.
One can see that the current averaged value is compatible with the SM
with $M_H^{\rm SM} \sim 125.5$~GeV 
and with the MSSM. The value of $\sweff$ obtained from $A_{\rm LR}^e$(SLD)
clearly favors the MSSM over the SM.
On the other hand, the value of $\sweff$ obtained from $A^b_{\rm FB}$(LEP) 
together with the $\MW$ data from LEP and the Tevatron would correspond to an
experimentally preferred region that deviates from the predictions of both
models. 
This unsatisfactory solution can only be resolved by new measurements, where
the a $Z$~factory, i.e.\ the GigaZ option would be an ideal solution. 
Thus, the unclear experimental situation regarding the two single
most precise measurements entering the combined value for $\sweff$
has a significant impact on the constraints that can be obtained from this
precision observable on possible New Physics scenarios. Measurements at a new 
$e^+e^-$ $Z$~factory, which could be realized in particular with the GigaZ
option of the ILC, would be needed to resolve this issue.
As indicated by the solid light shaded (red) ellipse, the anticipated
GigaZ/MegaW precision of the combined $\MW$--$\sweff$ measurement could
put severe constraints on each of the models and resolve the discrepancy
between the $A_{\rm FB}^b$(LEP) and $A_{\rm LR}^e$(SLD) measurements. 
If the central value of an improved measurement with higher precision should
turn out to be close to the central value favored by the current measurement
of $A_{\rm FB}^b({\rm LEP})$, this would mean that the electroweak
precision observables $\MW$ and $\sweff$ could rule out both the SM and the
most general version of the MSSM. 
\end{sloppypar}


\subsection{The relevance of the top quark mass\protect\footnotemark}
\footnotetext{Sven Heinemeyer and Georg Weiglein}
\label{sec:quantum-mt}



The mass of the top quark, $\mt$, is a fundamental parameter of the 
electroweak theory. It is by far the heaviest of all quark masses and
it is also larger than the masses of all other known fundamental
particles. For details on the experimental determination of $\mt$, see
\refse{sec:thresholdmeasurements}. 
The top quark is deeply connected to many other issues of high-energy
physics: 
\begin{itemize}
\item
The top quark could 
play a special role in/for electroweak symmetry breaking.
\item
The experimental uncertainty of $\mt$ induces the largest parametric
uncertainty in the prediction for electroweak precision
observables~\cite{Heinemeyer:2003ud,Heinemeyer:2004gx} and can thus
obscure new physics effects. 
\item
In SUSY models the top quark mass is an important
input parameter and is crucial for radiative electroweak symmetry
breaking and unification. 
\item
Little Higgs models contain ``heavier tops''.
\end{itemize}

The large value of $\mt$ gives rise to a large coupling between the top 
quark and the Higgs boson and is furthermore important for flavor
physics. It could therefore provide a window to new physics. (The correct
prediction of $\mt$ will be a crucial test for any fundamental theory.)
The top-quark mass also plays an important role in electroweak precision
physics, as a consequence in particular of non-decoupling effects being
proportional to powers of $\mt$. A precise knowledge of $\mt$ is
therefore indispensable in order to have sensitivity to possible effects
of new physics in electroweak precision tests.

The current world average for the top-quark mass from the measurement
at the Tevatron and the LHC is~\cite{ATLAS:2014wva},
\begin{align}
\mt &= 173.34 \pm 0.76 \gev~.
\label{mtexp}
\end{align}
The prospective accuracy at the LHC is 
$\de\mt^{\rm exp} \approx 500$~MeV \cite{Agashe:2013hma},
while at the ILC a very precise determination of $\mt$ with an accuracy
of $\de\mt^{\rm exp} \lsim 100$~MeV  will be possible,
see \refse{sec:thresholdmeasurements}. 
This uncertainty contains both the experimental error of the mass parameter
extracted from the $t \bar t$ threshold measurements at the ILC and 
the envisaged theoretical uncertainty from its transition into a suitable
short-distance mass (like the \msbar\ mass).


The relevance of the $\mt$ precision as parametric uncertainty has been
discussed for the $W$~boson mass, $\MW$, in \refse{sec:quantum-MW}, and
for the effective leptonic weak mixing angle, $\sweff$, in
\refse{sec:quantum-sw2eff}. 

Because of its large mass, the top quark is expected to have a large
Yukawa coupling to Higgs bosons, being proportional to $\mt$.
In each model where the Higgs boson mass is not a free
parameter but predicted in terms of the the other model parameters
(as e.g.\ in the MSSM), the diagram in \reffi{fig:mhiggs} contributes
to the Higgs mass. This diagram gives rise to a leading $\mt$
contribution of the form
\BE
\De\MH^2 \sim \gf \; N_C \; C \; \mt^4~,
\end{equation}
where $\gf$ is the Fermi constant, $N_C$ is the color factor, and the
coefficient $C$ depends on the specific model. Thus the experimental
error of $\mt$ necessarily leads to a parametric error in the Higgs
boson mass evaluation. 

\begin{figure}[htb]
\vspace{-10em}
\BC
\includegraphics[width=0.40\textwidth]{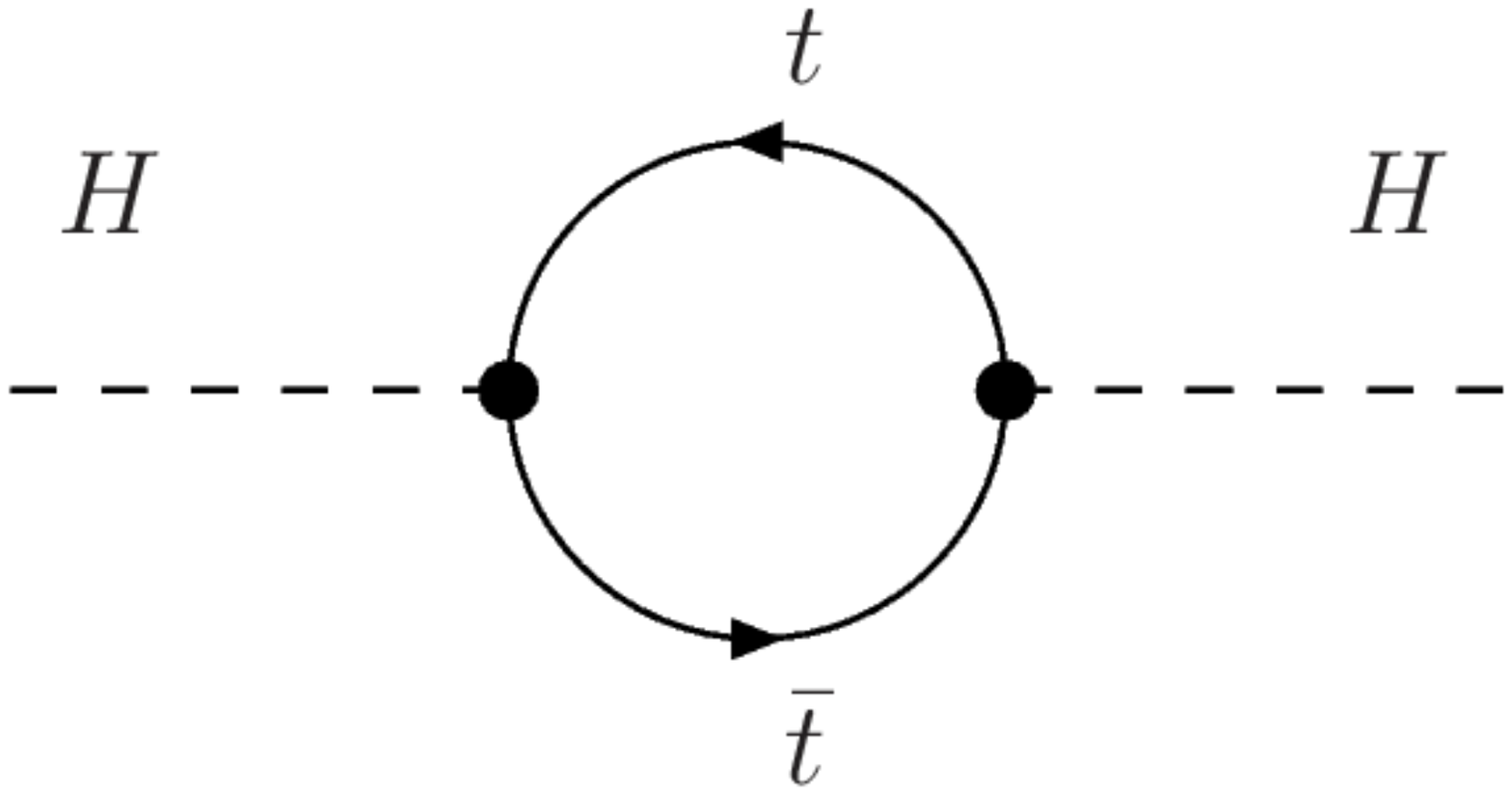}
\EC
\vspace{-10em}
\caption{Loop contribution of the top quark to the Higgs boson mass.}
\label{fig:mhiggs}
\end{figure}


Taking the MSSM as a specific example 
(including also the scalar top contributions and the appropriate
renormalization) $N_C \, C$ is given for the light $\cp$-even Higgs
boson mass in leading logarithmic approximation by  
\BE
N_C \, C = \frac{3}{\sqrt{2}\,\pi^2\,\SQb} \; 
\log \left( \frac{\mste\mstz}{\mt^2} \right)~.
\end{equation}
Here $m_{\tilde t_{1,2}}$ denote the two masses of the scalar tops.
The current
precision of $\de\mt \sim 1$~GeV  leads to an uncertainty of 
$\sim 2.5\%$ in the prediction of $\MH$, while the ILC will yield a
precision of $\sim 0.2\%$.  
These uncertainties have to be compared with the anticipated precision of
the future Higgs boson mass measurements. With a precision of 
$\de\MH^{\rm exp,LHC} \approx 0.2$~GeV  the relative
precision is at the level of $\sim 0.2\%$. It is apparent that only the
LC precision of $\mt$ will yield a parametric error small enough to
allow a precise comparison of the Higgs boson mass prediction and its
experimental value. 

Another issue that has to be kept in mind here (in SUSY as in any other
model predicting $\MH$) is the intrinsic theoretical uncertainty due to
missing higher-order corrections.
Within the MSSM currently the uncertainty for the lightest $\cp$-even Higgs is
estimated to  
$\de\Mh^{\rm intr,today} 
\approx 2-3$~GeV~\cite{Heinemeyer:2004gx,Degrassi:2002fi}%
\footnote{We are not aware of any such estimate in other New Physics
models.}%
. In the future one can hope for an improvement down to 
$\lsim 0.5$~GeV  or better~\cite{Heinemeyer:2004gx}, 
i.e.\ with sufficient effort on higher-order corrections it should be possible
to reduce the intrinsic theoretical uncertainty to the level of 
$\de\MH^{\rm exp, LHC}$. 

Confronting the theoretical prediction of $\MH$ with a precise
measurement of the Higgs boson mass constitutes a very sensitive test
of the MSSM (or any other model that predicts $\MH$), which allows one to
obtain constraints on the model 
parameters. However, the sensitivity of the $\MH$ measurement
cannot directly be translated into a prospective indirect determination
of a single model parameter. In a realistic
situation the anticipated experimental errors of {\em all} relevant SUSY
parameters have to be taken into account.
For examples including these parametric errors see
\citeres{Heinemeyer:2003ud,Weiglein:2004hn}.


\subsection{Prospects for the electroweak fit to the SM Higgs mass
\protect\footnotemark}
\footnotetext{Andreas Hoecker, Roman Kogler, Martin Gr\"unewald}
\label{sec:quantum-fit}



\begin{sloppypar}
The global fit to electroweak precision data allows among other constraints 
to extract information on the Higgs mass from Higgs loops modifying the values 
of $Z$ boson asymmetry observables and the $W$ 
mass~\cite{ALEPH:2010aa,LEPEWWG,Alcaraz:2006mx,ErlerNakamura:2010zzi,Baak:2011ze}.
Assuming the new boson discovered by the ATLAS~\cite{Aad:2012gk} 
and CMS~\cite{Chatrchyan:2012ufa} experiments at the LHC to be the SM Higgs boson, the 
elextroweak fit is overconstrained and 
can be used to quantify the compatibility of the mass (and couplings) of the 
discovered boson with the electroweak precision data in an overall
goodness-of-fit measure. Similarly, it allows to confront indirect
determinations of the $W$ boson mass, the effective weak mixing angle
predicting the $Z$ asymmetries, and the top quark mass with the
measurements. The LHC and a next generation electron--positron collider
have the potential to significantly increase the precision of most of
the observables that are relevant to the  fit. This section reports on a
prospective study of the electroweak fit following the approach
published in earlier works by the Gfitter
group~\cite{Flacher:2008zq,Baak:2011ze,Baak:2012kk} (and compares
briefly to a corresponding fit from the LEPEWWG).
\end{sloppypar}

\begin{sloppypar}
For the study aiming at a comparison of the accuracies of the measured
and predicted electroweak observables, the central values of the input
observables are chosen to agree with the SM prediction for a Higgs
mass of 125.8\:GeV.  Total experimental uncertainties of 6~MeV for
$M_W$, $1.3 \cdot10^{-5}$ for $\sinleff$, $4\cdot10^{-3}$ for
$R^{0}_{\ell}$, and 100~MeV for $m_t$ (interpreted as pole mass) are
used. The exact achieved precision on the Higgs mass is irrelevant for
this study.  For the hadronic contribution to the running of the QED
fine structure constant at the $Z$ pole, $\dalphaHadMZ$, an
uncertainty of $4.7\cdot10^{-5}$ is assumed (compared to the currently
used uncertainty of
$10\cdot10^{-5}$~\cite{Davier:2010nc,Baak:2012kk}), which benefits
below the charm threshold from the completion of BABAR analyses and
the ongoing program at VEPP-2000, and at higher energies from improved
charmonium resonance data from BES-3, and a better knowledge of
$\alpha_s$ from the $R^{0}_{\ell}$ measurement and reliable lattice QCD
predictions.  The other input observables to the electroweak fit are
taken to be unchanged from the current settings~\cite{Baak:2012kk}.
\end{sloppypar}

For the theoretical predictions, the calculations detailed
in~\cite{Baak:2011ze} and references therein are used. They feature among
others the complete $\mathcal{O}(\alpha_s^4)$ calculation of the QCD Adler
function~\cite{Baikov:2008jh, Baikov:2012er} and the full two-loop and leading
beyond-two-loop prediction of the $W$ mass and the effective weak
mixing angle~\cite{Awramik:2003rn,Awramik:2006uz,Awramik:2004ge}. 
An improved prediction of $R^0_b$ is invoked that includes the
calculation of the complete fermionic electroweak two-loop (NNLO) corrections 
based on numerical Mellin-Barnes integrals~\cite{Freitas:2012sy}.
The calculation of the vector and axial-vector couplings in Gfitter 
relies on accurate 
parametrisations~\cite{Hagiwara:1994pw,Hagiwara:1998yc,Cho:1999km,Cho:2011rk}.

\begin{sloppypar}
The most important theoretical uncertainties in the fit are those affecting the 
$M_W$ and $\sinleff$ predictions. They arise from three 
dominant sources of unknown higher-order corrections:
${\cal O}(\alpha^2\alpha_s)$ terms beyond the known 
contribution of ${\cal O}(G_{\scriptscriptstyle F}^2 \alpha_s \mt^4)$, 
${\cal O}(\alpha^3)$ electroweak three-loop corrections, and  
${\cal O}(\alpha_s^3)$ QCD terms, see \refse{sec:quantum-sw2efftheo}. The
quadratic sums of the above corrections amount to 
$\deltatheo M_W=4\:\mev$ and $\deltatheo\sinleff=4.7 \cdot 10^{-5}$,
which are the theoretical ranges used in present electroweak fits. We assume 
in the following that theoretical developments have let to improved uncertainties 
of $\deltatheo M_W=2\:\mev$ and $\deltatheo\sinleff=1.5 \cdot 10^{-5}$,
see \refta{tab:results_best}.
Within the \Rfit scheme employed here~\cite{Hocker:2001xe,Charles:2004jd},
theoretical uncertainties are treated as uniform likelihoods in the fit, 
corresponding to an allowed offset from the predicted value within the defined 
range (we discuss the difference with respect to standard Gaussian theoretical 
uncertainties below). 
\end{sloppypar}

\begin{table}[thb!] 
{\scriptsize
\begin{tabular*}{0.5\textwidth}{lccc} 
\hline\noalign{\smallskip}
Parameter                   
& Input value          &  Free in fit   & Predicted fit result      \\
\noalign{\smallskip}\hline\noalign{\smallskip}
$M_{H}$ {\tiny [GeV]}                &  $125.8 \pm 0.1$        & yes      &   $125.0^{\,+12}_{\,-10}$ \\
\noalign{\smallskip}\hline
$M_{W}$ {\tiny [GeV]}               &  $80.378\pm 0.006$       & --       &   $80.361\pm 0.005$             \\
$\Gamma_{W}$ {\tiny [GeV]}          &  --                       & --      &   $2.0910\pm 0.0004$           \\
\noalign{\smallskip}\hline\noalign{\smallskip}
$M_{Z}$ {\tiny [GeV]}               &  $91.1875\pm 0.0021$     & yes      &   $91.1878 \pm 0.0046$          \\
$\Gamma_{Z}$ {\tiny [GeV]}          &  --                       & --       &   $2.4953\pm 0.0003$           \\
$\sigma_{\rm had}^{0}$ {\tiny [nb]}  &  --                       & --        &  $41.479\pm 0.003$             \\
$R^{0}_{\ell}$                      &  $20.742\pm 0.003$         & --       &  --                             \\
$A_{\rm FB}^{0,\ell}$                 &  --                       & --       &  $0.01622 \pm 0.00002 $        \\
$A_\ell$                           &  --                      & --        &  $0.14706 \pm 0.00010 $       \\
$\sinleff$                        &  $0.231385\pm 0.000013$   & --        & $0.23152\pm 0.00004 $       \\
$A_{c}$                           & --                        & --        &  $0.66791\pm 0.00005 $        \\
$A_{b}$                           & --                        & --        &  $0.93462\pm 0.00002 $        \\
$A_{\rm FB}^{0,c}$                 &  --                       & --        &  $0.07367\pm 0.00006 $         \\
$A_{\rm FB}^{0,b}$                 &  --                       & --        &  $0.10308\pm 0.00007 $         \\
$R^{0}_{c}$                       &  --                       & --        &  $0.17223\pm 0.00001 $        \\
$R^{0}_{b}$                       &  --                       & --        &  $0.214746\pm 0.000004 $      \\
\noalign{\smallskip}\hline\noalign{\smallskip}
$\mc$ {\tiny [GeV]}                &  $1.27^{\,+0.07}_{\,-0.11}$  & yes         & --                          \\
$\mb$ {\tiny [GeV]}                &  $4.20^{\,+0.17}_{\,-0.07}$  & yes         & --                          \\
$m_{t}$ {\tiny [GeV]}              &  $173.18\pm0.10$            & yes         & $173.3\pm 1.2    $         \\
$\dalphaHadMZ$ $^{(\bigtriangleup)}$ &  $2757.0\pm  4.7$ & yes    & $2757 \pm 10$                \\
$\alpha_s(M_{Z}^{2})$                & --                         & yes         & $0.1190\pm 0.0005$           \\
\noalign{\smallskip}\hline\noalign{\smallskip}
$\deltatheo M_W$ {\tiny [MeV]}    & $[-2.0,2.0]_{\rm theo}$          & yes         &   --                        \\
$\deltatheo \sinleff$ $^{(\bigtriangledown)}$  & $[-1.5,1.5]_{\rm theo}$ & yes    &  --                        \\
\noalign{\smallskip}\hline
\noalign{\smallskip}
\end{tabular*} \\
{\ft
$^{(\bigtriangleup)}$In units of $10^{-5}$. $^{(\bigtriangledown)}$Rescaled due to $\alpha_s$ dependency.
}}
\caption{Input values and fit results for the observables and
  parameters of the global electroweak fit in a hypothetical future scenario. 
  The first and second columns list respectively the observables/parameters used in the
  fit, and their experimental values or phenomenological estimates
  (see text for references). The subscript ``theo'' labels theoretical
  error ranges.  The third column indicates whether a parameter is
  floating in the fit and in the fourth column the fit results are given
  without using the corresponding experimental or phenomenological
  estimate in the given row.\label{tab:results_best} }
\end{table} 

\begin{figure}[!t]
\centering
\epsfig{file=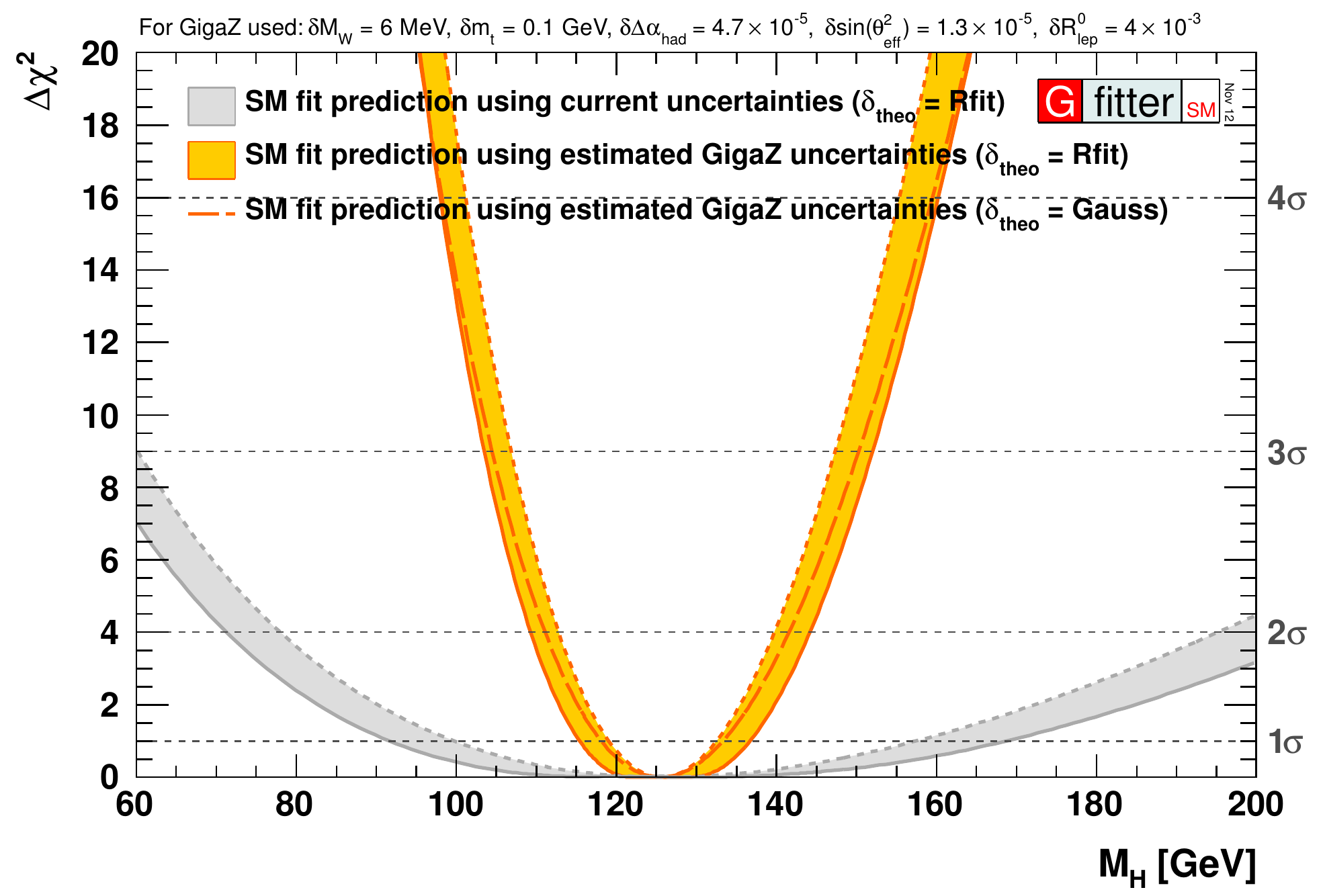, scale=0.41}
\epsfig{file=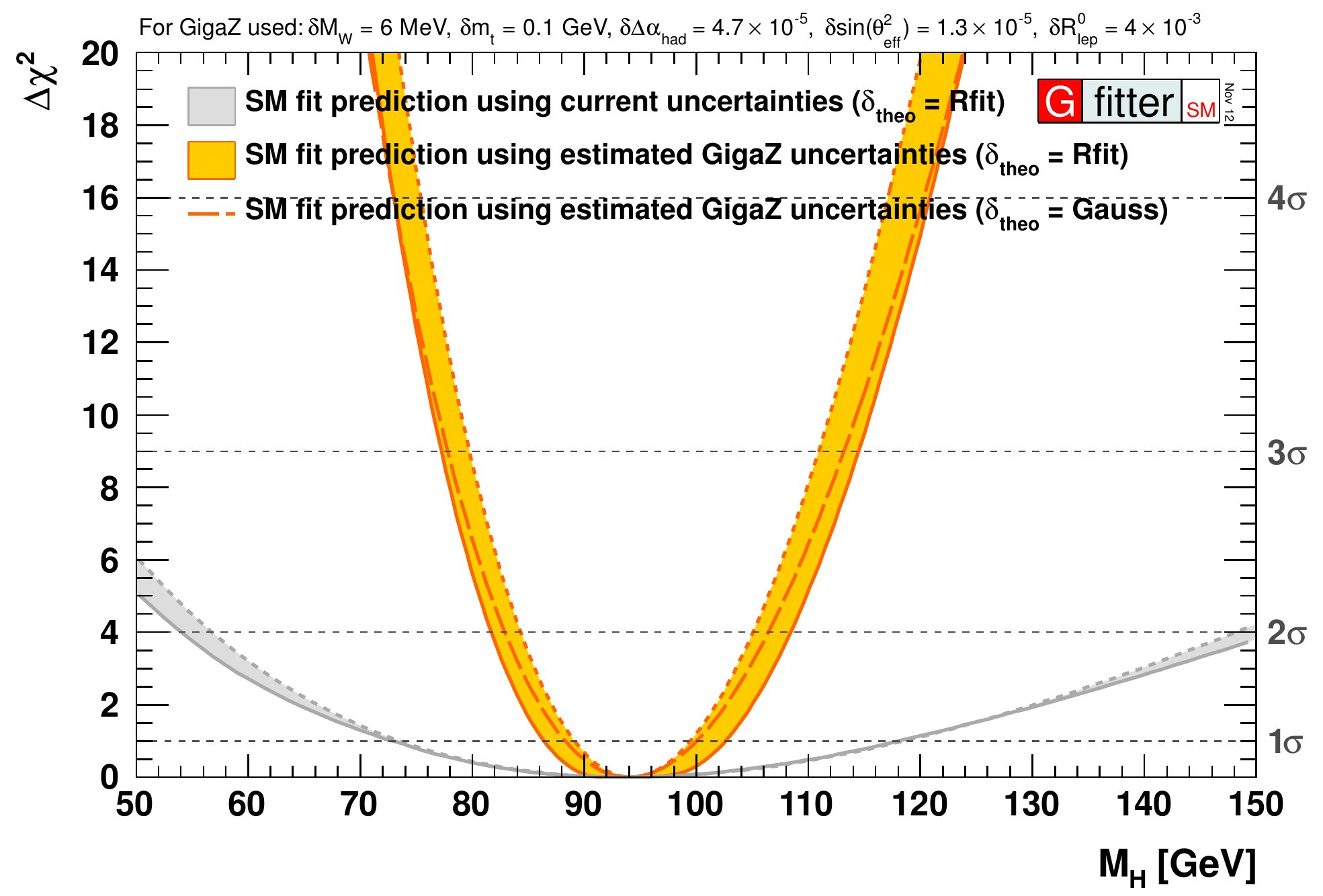, scale=0.41}
  \caption[]{$\Delta\chi^2$ profiles as a function of the Higgs mass for 
    electroweak fits compatible with an SM Higgs boson of mass 125.8\:\gev (left)
    and 94\:\gev (right), respectively. The measured Higgs boson mass 
    is not used as input in the fit. The grey bands show the results obtained
    using present uncertainties~\cite{Baak:2012kk}, and the yellow bands 
    indicate the results for the hypothetical 
    future scenario given in Table~\ref{tab:results_best} (left plot)
    and corresponding input data shifted to accommodate a 94\:\gev Higgs 
    boson but unchanged uncertainties (right plot). The right axes depict
    the corresponding Gaussian 'sigma' lines. The thickness of the bands 
    indicates the effect from the theoretical uncertainties treated according
    to the \Rfit prescription. The long-dashed line in each plot shows the 
    curves one would obtain when treating the theoretical uncertainties in a 
    Gaussians manner just like any other uncertainty in the fit. }
\label{fig:scans}
\end{figure}

\begin{figure}[!t]
\centerline{
\epsfig{file=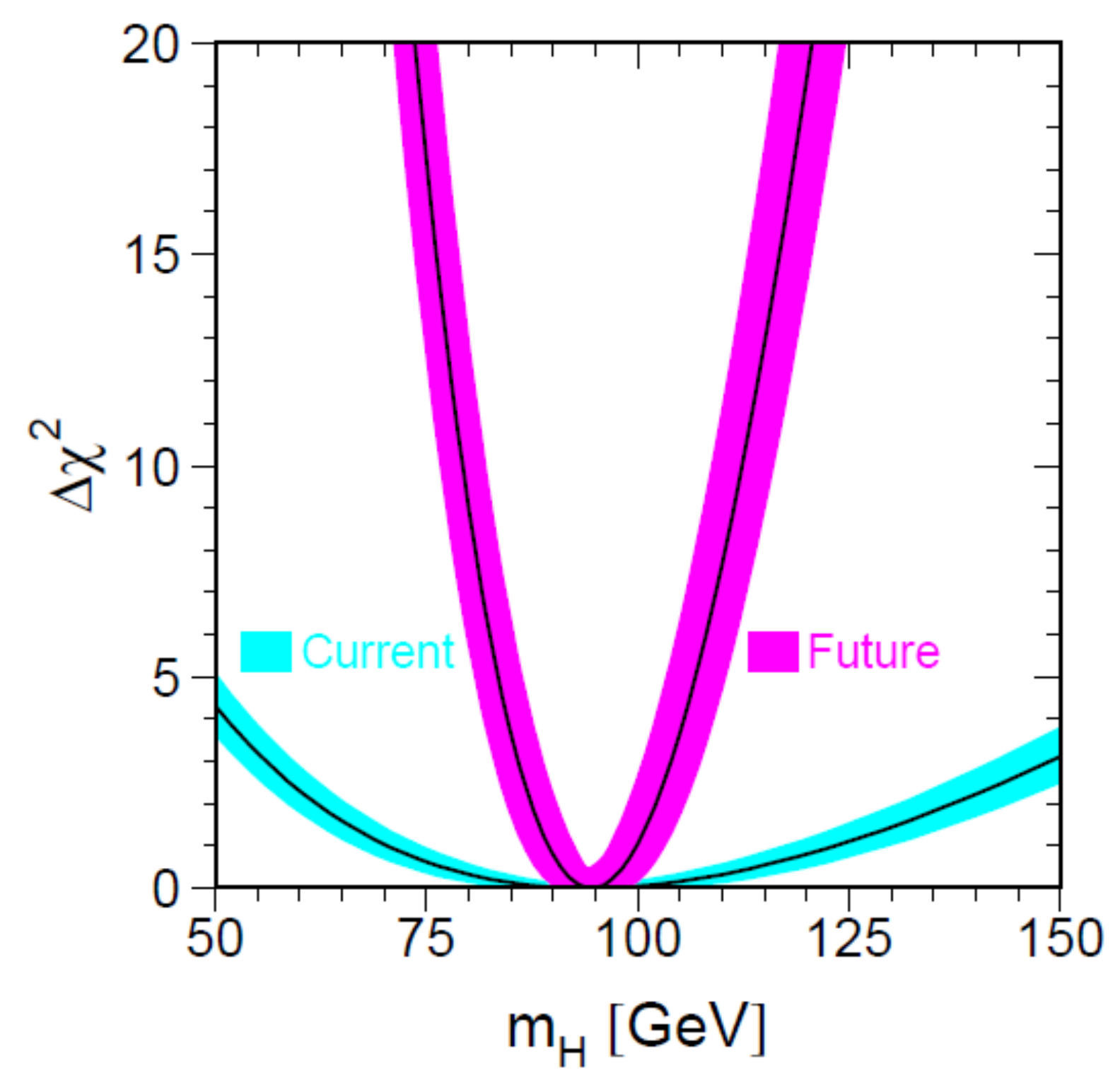, scale=0.55}}
  \caption[]{$\Delta\chi^2$ profiles as a function of the Higgs mass for 
    electroweak fits compatible with an SM Higgs boson with mass 
    94\:\gev using the LEPEWWG approach~\cite{LEPEWWG}.
The blue (pink) parabola shows the current (future) fit (see text).}
\label{fig:lepewwg}
\end{figure}

Table~\ref{tab:results_best} gives the input observables and values used (first 
and second columns) and 
the predictions obtained from the fit to all input data except for the one 
that is predicted in a given row (last column). It allows to compare the 
accuracy of direct and indirect determinations. To simplify the numerical 
exercise the $Z$-pole asymmetry observables are combined into a single input 
$\sinleff$, while for the reader's convenience the fit predictions are 
provided for all observables.

\begin{sloppypar}
The indirect prediction of the Higgs mass at 125\:\gev achieves an uncertainty of 
$^{\,+12}_{\,-10}\:\gev$. For $M_W$ the prediction with an estimated uncertainty of 
5\:\mev is similarly accurate as the (assumed) measurement, while the 
prediction of $\sinleff$ with an uncertainty of $4\cdot10^{-5}$ is three times 
less accurate than the experimental precision. The fit would therefore
particularly benefit from additional  
experimental improvement in $M_W$. It is interesting to notice that the accuracy of 
the indirect determination of the top mass (1.2\:\gev) becomes similar to that of 
the present experimental determination. An improvement beyond, say, 200\:\mev 
uncertainty cannot be exploited by the fit. The input values of $M_Z$ and $\dalphaHadMZ$ 
are twice more accurate than the fit predictions, which is sufficient to not
limit the fit but further improvement would certainly be useful.
\end{sloppypar}

Keeping the present theoretical uncertainties in the prediction of 
$M_W$ and $\sinleff$ would worsen the accuracy of the $M_H$ prediction 
to $^{\,+20}_{\,-17}\:\gev$, whereas neglecting theoretical uncertainties altogether 
would improve it to $\pm7\:\gev$. 
This emphasises the importance of the required theoretical work. 

\begin{sloppypar}
Profiles of $\Delta\chi^2$ as a function of the Higgs mass for present and 
future electroweak fits compatible with an SM Higgs boson of mass 125.8\:\gev 
and 94\:\gev, respectively, are shown in Fig.~\ref{fig:scans} (see caption for 
a detailed description). The measured Higgs boson mass is not used as input 
in these fits. If the experimental input data, currently predicting 
$M_H=94^{\,+25}_{\,-22}\:\gev$~\cite{Baak:2012kk}, were left unchanged with respect
to the present values, but had uncertainties as in Table~\ref{tab:results_best},
a deviation of the measured $M_H$ exceeding $4\sigma$ could be established 
with the fit (see right-hand plot in Fig.~\ref{fig:scans}). Such a conclusion 
does not strongly depend on the treatment of the theoretical uncertainties 
(\Rfit versus Gaussian) as can be seen by comparison of the solid yellow and 
the long-dashed yellow  $\Delta\chi^2$ profiles. 
\end{sloppypar}

A similar result has also been obtained by the LEPEWWG, as can be seen in
\reffi{fig:lepewwg}~\cite{LEPEWWG}. 
The $\Delta\chi^2$ profile of their fit is shown as a function of the Higgs
mass. The blue band shows the current result with a best fit point at 
$\sim 94 \gev$ with an uncertainty of $\sim \pm 30 \gev$. The pink parabola
shows the expected improvement under similar assumptions as in
\reffi{fig:scans}. This confirms that a strong improvement of the fit
can be expected taking into 
account the anticipated future LC accuracy for the electroweak precision data.


\subsection{The muon magnetic moment and new physics\protect\footnotemark}
\footnotetext{Dominik St\"ockinger}
\label{sec:quantum-amu}



One of the prime examples of precision observables sensitive to
quantum effects are the magnetic moments $(g-2)$ of the electron and
muon. In particular after the measurements at Brookhaven
\cite{Bennett:2006}, the muon magnetic moment $a_\mu=(g_\mu-2)/2$ has
reached a sensitivity to all sectors of the SM and to many New Physics
Models (NPM).  
The currently observed deviation between the experimental value and
the SM prediction is particularly well compatible with NPM which can
also be tested at a LC. Before the startup of a future LC, new $a_\mu$
measurements are planned at Fermilab \cite{FNALProposal} and J-PARC
\cite{Mibe}. For these reasons it is of interest to briefly discuss
the conclusions that can be drawn from current and future $a_\mu$
results on LC physics.

Like many LC precision observables, $a_\mu$ is a flavour- and
CP-conserving quantity; unlike the former it is
chi\-ra\-li\-ty-flipping and therefore particularly sensitive to
modifications of the muon Yukawa coupling or more generally the muon
mass generation mechanism. A simple consideration, however,
demonstrates that like a LC, $a_\mu$ is generically sensitive to NPM
with new weakly interacting particles at the weak scale \cite{czmar}.

\begin{sloppypar}
Because of the similar quantum field theory operators relevant for
$m_\mu$ and $a_\mu$, contributions of a NPM at some scale $\Lambda$ to
both quantities, $a_\mu(\mbox{N.P.})$ and $\delta m_\mu(\mbox{N.P.})$,
are linked as
\begin{equation}
\label{CzMbound} a_\mu(\mbox{N.P.})={\cal O}(1)\times
\left(\frac{m_\mu}{\Lambda}\right)^2 \times \left(\frac{\delta
m_\mu(\mbox{N.P.})}{m_\mu}\right). 
\end{equation}
All coupling constants and loop factors are contained in the constant
$C := \delta m_\mu(\mbox{N.P.})/m_\mu$, which is 
highly model-dependent.  A first consequence of
this relation is that new physics can explain the currently observed
deviation of~\cite{Gnendiger:2013pva} (based on~\cite{Davier:2010nc}),
\begin{align}
a_\mu^{\rm exp} - a_\mu^{\rm SM} &= (28.7 \pm 8.0)\times10^{-10}~,
\label{amuexp}
\end{align}
only if $\Lambda$ is at the TeV
scale or smaller (assuming no fine-tuning in the muon mass, $|C|<1$). 
\end{sloppypar}

Equation (\ref{CzMbound}) also illustrates how widely different
contributions to $a_\mu$ are possible. 
\begin{itemize}
\item For models with new weakly interacting particles (e.g.\ $Z'$,
 $W'$, see \refse{sec:quantum-agc}, little Higgs or universal extra dimension
models) one typically obtains perturbative contributions to the muon mass
$C={\cal O}(\alpha/4\pi)$. Hence, for weak-scale masses these models predict
very small contributions to $a_\mu$ and might be challenged by the
future more precise $a_\mu$ measurement, see e.g.\
\cite{Blanke:2007db,AppelqDob}. Models of this kind can only 
explain a significant contribution to $a_\mu$ if the new particles
interact with muons but are otherwise hidden from searches. An example
is the model with a new gauge boson associated to a gauged lepton
number $L_\mu-L_\tau$ \cite{LmuLtau}, where a gauge boson mass of
${\cal O}(100\mbox{ GeV})$ is viable, If this model is the origin of
the observed $a_\mu$ deviation it would be highly desirable to search
for the new $Z'$, corresponding to the $L_\mu-L_\tau$-symmetry. This
would be possible at the LHC in part of the parameter space but also
at the LC in the process $e^+e^-\to\mu^+\mu^-Z'$ \cite{LmuLtau}. 

\item 
For SUSY models one obtains an additional factor
$\tan\beta$, the ratio of the two Higgs vacuum expectation
values, see e.g.\  \cite{dsreview} and references therein. A numerical
approximation for the SUSY contributions is given by
\begin{equation}
a_\mu^{\rm SUSY} \approx  13\times10^{-10}\left(\frac{100\,\rm
    GeV}{M_{\rm SUSY}}\right) ^2\, \tan\beta\ \mbox{sign}(\mu),
\label{amususy}
\end{equation}
\begin{sloppypar}
where $M_{\rm SUSY}$ denotes the common superpartner mass scale and
$\mu$ the Higgsino mass parameter. It agrees with the generic result
\refeq{CzMbound} for $C={\cal O}(\tan\beta\times\alpha/4\pi)$ and is
exactly valid if all SUSY 
masses are equal to $M_{\rm SUSY}$. The
formula shows that the observed deviation could be explained e.g.\ 
for relevant SUSY masses (smuon, chargino and neutralino masses) of
roughly 200 GeV and $\tan\beta\sim10$ or 
SUSY masses of 500 GeV and $\tan\beta\sim50$. This is well in
agreement with current bounds on weakly interacting SUSY particles and
in a very interesting range for a LC.
This promising situation has motivated
high-precision two-loop calculations of $a_\mu^{\rm SUSY}$
\cite{Fargnoli:2013zda,Fargnoli:2013zia}, which depend on all sfermion,
chargino and neutralino masses and will benefit particularly from precise
SUSY mass measurements at a LC.

\end{sloppypar}

\item Models with large $C\simeq1$ are of interest since there the
  muon mass is essentially given by new physics loop effects. Some
  examples of such 
  radiative muon mass generation models are given in
  \cite{czmar}. For examples within SUSY see e.g.\ 
  \cite{Borzumati:1999sp,Crivellin:2010ty}. In such models $a_\mu$ can
  be large even for particle masses at the TeV scale, potentially
  beyond the direct reach of a LC. The possibility to test such
  models using precision observables at the LC has not yet been
  explored in the literature. 
\end{itemize}

Fig.\ \ref{fig:susyplotsamu} illustrates the complementarity of
$a_\mu$ and LC measurements in investigating SUSY.
\begin{figure}
\setlength{\unitlength}{1cm}
\centering
\begin{picture}(7.2,5.5)
\epsfysize=4.8cm
\put(0,0.7){\epsfbox{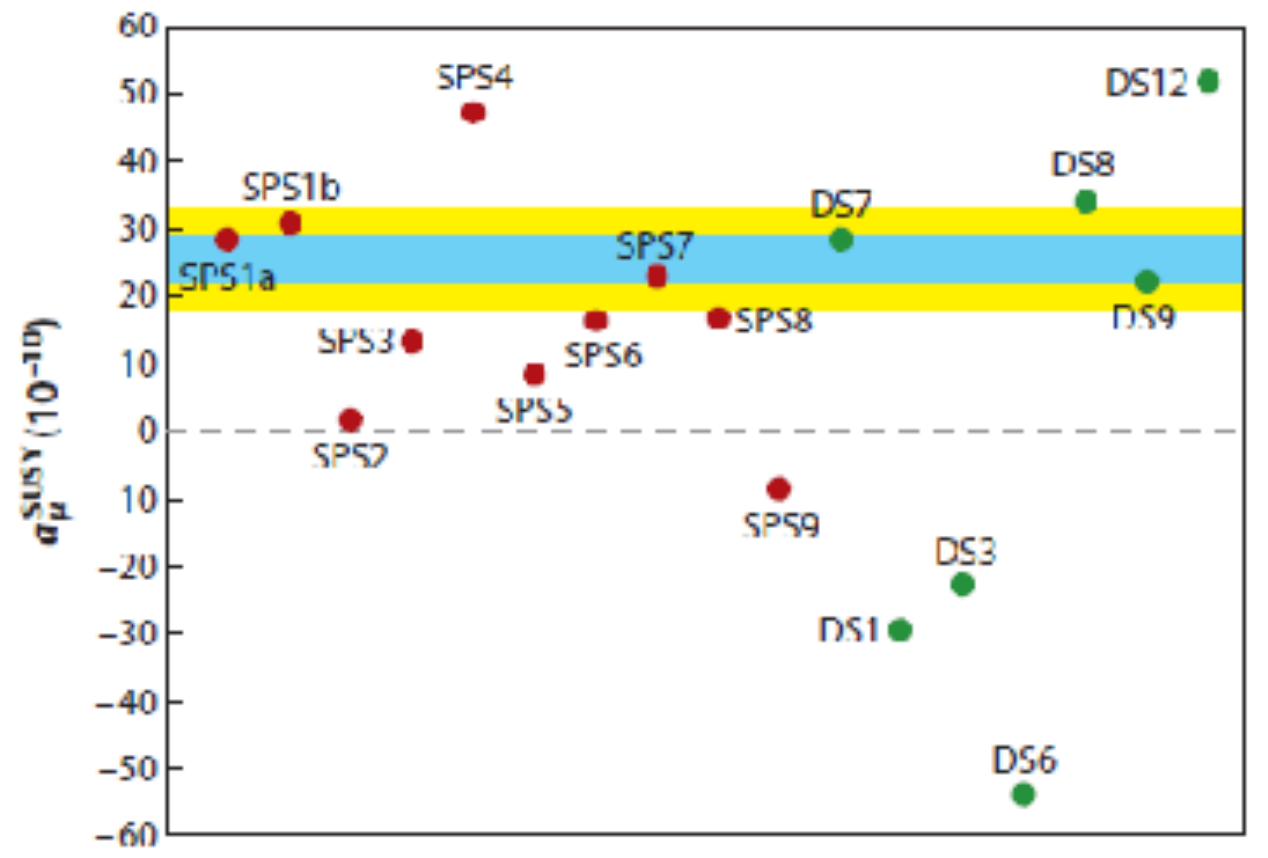}}
\end{picture}
\begin{picture}(8,5.2)
\epsfysize=5.5cm
\epsfxsize=6.5cm
\put(0.5,-,1){
\epsfbox{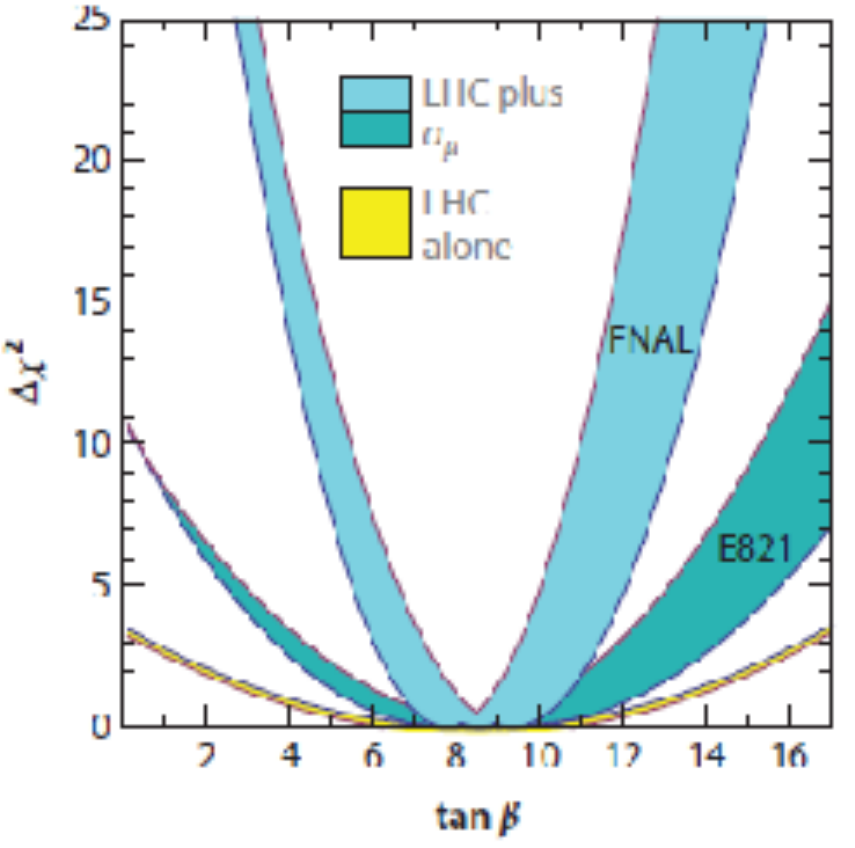}
}
\end{picture}
\\
\caption{\label{fig:susyplotsamu}(a) SUSY contributions to $a_\mu$ for
the SPS benchmark points (red), and for the ``degenerate solutions''
from Ref.\ \cite{Adam:2010uz}. 
The yellow and blue band indicate the current and an improved experimental
result, respectively. 
(b) Possible future $\tan\beta$
determination  assuming that a slightly modified MSSM point SPS1a (see
text) is realized. The bands show the $\Delta\chi^2$ parabolas from
LHC-data alone (yellow)~\cite{PlehnRauchNew}, including the $a_\mu$
with   current precision (dark blue) and with prospective precision
(light blue).  The width of the blue curves results from the expected
LHC-uncertainty of the parameters (mainly smuon and chargino
masses)~\cite{PlehnRauchNew}. Taken from \cite{MdRRS}.  } 
\end{figure} 
%
\begin{sloppypar}
The upper plot shows the $a_\mu(\rm SUSY)$-values for the SPS benchmark
points~\cite{Allanach:2002nj}, of which only the weakly interacting sector is relevant. The contributions span a wide range and can be positive or negative.%
\footnote{
Most of the points are ruled out by LHC searches for colored
particles. However, for our purposes only the weakly interacting
particles are relevant, and these are not excluded.
}%
The discriminating power of the
current (yellow band) and an improved (blue band) measurement is
evident from Fig.~\ref{fig:susyplotsamu}(a). The green points illustrate
that the LHC alone is not sufficient to discover SUSY and measure all
its parameters.
They correspond to ``degenerate solutions'' as defined in 
\citere{Adam:2010uz} --- different SUSY parameter points which cannot be
distinguished at the LHC alone.  They have very different $a_\mu$ predictions, 
in particular different signs for $\mu$, and hence $a_\mu$ can
resolve such LHC degeneracies. However, the LC can go much further and
rule out the wrong parameter choices with far higher significance
\cite{Adam:2010uz}.
\end{sloppypar}

The lower plot of Fig.\ \ref{fig:susyplotsamu} illustrates that the SUSY
parameter $\tan\beta$ can be measured more precisely by combining
LHC-data with $a_\mu$. It is based on the assumption that SUSY is
realized, found at the LHC and the origin of the observed $a_\mu$
deviation in \refeq{amuexp}. To fix an example, we use a slightly
modified SPS1a benchmark point with $\tan\beta$ scaled down to
$\tan\beta=8.5$ such that $a_\mu^{\rm SUSY}$ is equal to an assumed
deviation   $\Delta a_\mu=255\times
10^{-11}$.\footnote{
The following conclusions are
neither very sensitive to the actual $\tan\beta$ value nor to the
actual value of the deviation $\Delta a_\mu$.}  Ref.\
\cite{PlehnRauchNew} has shown that then mass measurements at the LHC
alone are sufficient to determine $\tan\beta$ to a precision of
$\pm4.5$ only.  The corresponding $\Delta\chi^2$ parabola is shown in
yellow in the plot.  In such a situation one can study the SUSY
prediction for $a_\mu$ as a function of $\tan\beta$ (all other
parameters are known from the global fit to LHC data) and compare it
to the measured value, in particular after an improved
measurement. The plot compares the LHC $\Delta\chi^2$ parabola with
the ones obtained from including $a_\mu$, $\Delta\chi^2=[(a_\mu^{\rm
SUSY}(\tan\beta)-\Delta a_\mu)/\delta a_\mu]^2$ with the errors
$\delta a_\mu=80\times 10^{-11}$ (dark blue) and $34\times 10^{-11}$
(light blue).  
Here the widths of the parabolas mainly originate in the
experimental uncertainties of the relevant electroweak particles, such
as smuons and charginos. It can be seen that on the one hand future
measurements of $a_\mu$ would drastically improve the $\tab$
determination. On the other hand, an LC measurement of the electroweak
masses would also be important to obtain a very good fit to $\tab$.

Ref.\ \cite{Adam:2010uz} has also studied the impact of a LC on the
$\tan\beta$-determination in a similar context, and a similar
improvement was found as in the case of $a_\mu$. Here it is noteworthy
that in the MSSM, $\tan\beta$ is a universal quantity entering all
sectors, like $\sin\theta_W$ in the SM, but that $a_\mu$ and LC
measurements are sensitive to $\tan\beta$ in different sectors, the
muon Yukawa coupling and sparticle masses, respectively. These
examples show how the LC will complement information from $a_\mu$ and
test NPM compatible with $a_\mu$.

The situation would be quite different if the $a_\mu$ deviation is
real but not due to weak-scale new particles but to  very light,
sub-GeV new particles, as suggested e.g.\ in \cite{Pospelov:2008zw}. In such a
case, such new light dark-force particles could be probed by dedicated
low-energy precision experiments such as the next generation $a_\mu$
measurements but the full understanding of whatever physics at the
electroweak scale will have been found at the LHC would be the task of
a future LC.


\subsection{Anomalous gauge boson couplings}
\label{sec:quantum-agc}

\subsubsection{Electroweak Gauge Boson Interactions: Effective Field Theory 
and Anomalous Couplings\protect\footnotemark}
\footnotetext{Nicolas Greiner}
\label{sec:quantum-agc-theo}
%

One possibility to search for new physics in the electroweak sector is the
precision investigation of the couplings of the electroweak gauge bosons. 
At the LC at tree-level, the incoming leptons
interact via an exchange of an electroweak gauge boson. This allows
for precise studies of trilinear gauge couplings in $e^{+} e^{-} \to
W^{+} W^{-}$ as well as quartic gauge couplings occurring in a variety
of final states like $e^{+} e^{-} \to V V V $ with $V V V$ being
$W W Z$ or $W W \gamma$.  In contrast to a hadron collider the
advantages are the absence of parton distribution functions so that
the center of mass energy at which the hard scattering takes place is
exactly known.  This also allows to tune the beam energy according to
the occurring resonances similar to what has already be done at LEP.
The second advantage is the clean environment. At a hadron collider
the most likely processes involve QCD radiation and therefore jets in
the final state. Triple or quartic gauge boson scatterings are
typically detected via VBF processes which however have to be
discriminated from irreducible background processes.

One approach to parametrize new physics in a model independent way is to write down
an effective Lagrangian with all possible vertices and general coupling
constants. For the trilinear electroweak gauge couplings (TGC) this has been suggested in
\cite{Hagiwara:1986vm} for instance, resulting in the following effective Lagrangian including anomalous TGCs: 
\begin{eqnarray}
{\cal L}_{\rm TGC}&=&ig_{WWV}\left(g_1^V(W_{\mu\nu}^+W^{-\mu}-W^{+\mu}W_{\mu\nu}^-)V^\nu
\right.\nonumber\\
&&\left. +\kappa^V W_\mu^+W_\nu^-V^{\mu\nu}
+\frac{\lambda^V}{M_W^2}W_\mu^{\nu+}W_\nu^{-\rho}V_\rho^{\mu}
\right.\nonumber\\&&\left.
+ig_4^VW_\mu^+W^-_\nu(\partial^\mu V^\nu+\partial^\nu V^\mu)
\right. \nonumber\\
&&\left. -ig_5^V\epsilon^{\mu\nu\rho\sigma}(W_\mu^+\partial_\rho W^-_\nu-\partial_\rho W_\mu^+W^-_\nu)V_\sigma
\right.\nonumber\\&&\left.
+\tilde{\kappa}^V W_\mu^+W_\nu^-\tilde{V}^{\mu\nu}
+\frac{\tilde{\lambda}^V}{m_W^2}W_\mu^{\nu+}W_\nu^{-\rho}\tilde{V}_\rho^{\mu}
\right)\;,
\label{eq:Ltgc}
\end{eqnarray}
with
 $V=\gamma,Z$;  $W_{\mu\nu}^\pm = \partial_\mu W_\nu^\pm - \partial_\nu W_\mu^\pm$, 
 $V_{\mu\nu} = \partial_\mu V_\nu - \partial_\nu V_\mu$ and 
 $\tilde{V}_{\mu,\nu}=\epsilon_{\mu \nu \rho \sigma} V_{\rho \sigma}/2$.
The overall coupling constants are given by $g_{WW\gamma}=-e$ and
 $g_{WWZ}=-e\cot\theta_W$ (with $\cos\theta_W = \MW/\MZ$).
In the same spirit, one can write down an effective Lagrangian describing quartic gauge boson couplings (QGC) as follows~\cite{Beyer:2006hx}:
\begin{eqnarray}
 &&{\cal L}_{\rm QGC}=e^2\left(g_1^{\gamma \gamma}A^{\mu}A^{\nu}W^{-}_{\mu}W^{+}_{\nu} 
 -g_2^{\gamma \gamma}A^{\mu}A_{\mu}W^{-\nu}W^{+}_{\nu}\right)
 \nonumber\\&&
 +e^2\frac{c_w}{s_w}\left(g_1^{\gamma Z} A^{\mu} Z^{\nu}(W^{-}_{\mu}W^{+}_{\nu} +W^{+}_{\mu}W^{-}_{\nu}) 
\right.
\nonumber\\&&
\left.
\mbox{\hspace{1cm}}
 -2g_2^{\gamma Z}A^{\mu} Z_{\mu}W^{-\nu}W^{+}_{\nu}\right) \nonumber \\&&
 +e^2\frac{c^2_w}{s^2_w}\left(g_1^{Z Z} Z^{\mu}Z^{\nu}W^{-}_{\mu}W^{+}_{\nu} 
 -g_2^{Z Z} Z^{\mu}Z^{\mu}W^{-\nu}W^{+}_{\nu}\right) \nonumber \\&&
 +\frac{e^2}{2s^2_w}\left(g_1^{WW}W^{-\mu}W^{+\nu}W^{-}_{\mu}W^{+}_{\nu}-g_2^{WW}(W^{-\mu}W^{+}_{\mu})^2\right)
 \nonumber \\&&
 +\frac{e^2}{4s^2_wc^4_w}h^{ZZ}(Z^{\mu}Z_{\mu})^2\;.
 \label{eq:Lqgc}
\end{eqnarray}
In the SM the couplings in \refeq{eq:Ltgc} are given by
\begin{equation}
 g_1^{\gamma,Z} =\kappa^{\gamma, Z} =1, \quad g_{4,5}^{\gamma,Z} =\tilde{\kappa}^{\gamma, Z} =1, 
 \quad \lambda^{\gamma,Z}=\tilde{\lambda}^{\gamma,Z}=0 \;,
\end{equation}
whereas the SM values of the QGCs are
\begin{equation}
 g_1^{VV'}=g_2^{VV'}=1 (VV' = \gamma\gamma, \gamma Z, ZZ, WW),\quad h^{ZZ}=0\;.
\end{equation}
In the context of the recent discovery of a particle compatible 
with a SM Higgs boson \cite{lhchiggs} it will be interesting to study the couplings
of the Higgs boson to the electroweak gauge bosons. A parametrization of trilinear couplings can be found in
\cite{Eboli:1998vg,GonzalezGarcia:1999fq}, for instance, and reads
\begin{eqnarray}
{\cal L}_{\rm TGC}^{H} &=& 
g_{H \gamma \gamma} H A_{\mu \nu} A^{\mu \nu} + 
g^{(1)}_{H Z \gamma} A_{\mu \nu} Z^{\mu} \partial^{\nu} H \nonumber \\ 
&+& g^{(2)}_{H Z \gamma} H A_{\mu \nu} Z^{\mu \nu}
+ g^{(1)}_{H Z Z} Z_{\mu \nu} Z^{\mu} \partial^{\nu} H \nonumber \\
&+& g^{(2)}_{H Z Z} H Z_{\mu \nu} Z^{\mu \nu} +
g^{(2)}_{H W W} H W^+_{\mu \nu} W_{-}^{\mu \nu} \; \nonumber \\
&+&g^{(1)}_{H W W} \left (W^+_{\mu \nu} W_{-}^{\mu} \partial^{\nu} H 
+h.c.\right)\,.
\label{eq:Htgc} 
\end{eqnarray}
Note that none of the terms in \refeq{eq:Htgc} has a SM contribution
as the $HVV$ vertex in the SM is given by
\begin{equation}
 {\cal L}_{\rm SM}^{H}= \frac{1}{2} \frac{g}{\cos \theta_W} \MZ H Z_{\mu} Z^{\mu} + g \MW W^{+}_{\mu} W^{-\mu}\;.
 \label{eq:Hsm}
\end{equation}
In \refeqs{eq:Ltgc}, (\ref{eq:Lqgc}), (\ref{eq:Htgc}) the number of possible
additional interaction terms in the Lagrangian is restricted by the
requirement of electroweak gauge- and Lorentz invariance.  
If one loosens this requirement, there would be many more possibilities as discussed for instance in~\cite{Gaemers:1978hg}.

A slightly different approach to a model independent parametrisation of new physics is based on the idea 
of an effective field theory (EFT) \cite{eft}, where additional, higher dimensional operators are added to the SM Lagrangian, 
\begin{equation}
{\cal L}_{\rm eff}= {\cal L}_{\rm SM} +\sum_{n=1}^{\infty}\sum_{i}\frac{f_{i}^{(n)}}{\Lambda^n}{\cal O}_i^{(n+4)}.
\label{eq:Left}
\end{equation}
As the Lagrangian is required to have dimension four, this means that higher dimensional operators
are accompanied by dimensionful coupling constants. It is not possible to construct operators of
dimension five that are Lorentz- and gauge invariant, so the first additional operators are of
dimension six. A general analysis of dimension six operators has been presented
in \cite{operators}. The choice of the basis of these operators is however not unique, and especially for
operators involving electroweak gauge bosons a number of different choices have been discussed in the literature; a common
representation can be found in \cite{Hagiwara:1993ck}. 
In the effective field theory approach one first specifies the particle content of the theory
and derives the corresponding vertices and coupling constants from there.
At a first glance the two approaches, i.~e. the EFT and the effective Lagrangian approach,  
may lead to the same results, as one can express the coupling constants
of \refeqs{eq:Ltgc}, (\ref{eq:Lqgc}), (\ref{eq:Htgc}) as functions of the coefficients $f_{i}^{(n)}/\Lambda^n$ of
\refeq{eq:Left}~\cite{Hagiwara:1993ck}, as follows:
\begin{eqnarray}
 g_1^Z &=& 1 + f_W\frac{m_Z^2}{2\Lambda^2}, \nonumber\\
 \kappa_Z &=& 1 + \left[ f_W -\sin^2 \theta_W(f_B +f_W)\right]\frac{m_Z^2}{2\Lambda^2},\nonumber \\
 \kappa_{\gamma} &=& 1 +(f_B + f_W)\frac{m_W^2}{2\Lambda^2},\nonumber \\
 \lambda_{\gamma} &=& \lambda_{Z} = \frac{3 m_W^2 g^2}{2\Lambda ^2} f_{WWW} \; . 
 \label{eq:translation}
\end{eqnarray}
The corresponding Lagrangian using the EFT approach of \refeq{eq:Left} leading to \refeq{eq:translation} is given by
~\cite{Hagiwara:1993ck}
\begin{eqnarray}
 {\cal L}_{eff}&=& {\cal L}_{SM} +\frac{f_B}{\Lambda^2}(D_{\mu}\phi)^{\dagger}\hat B^{\mu \nu}(D_{\nu} \phi) \nonumber\\
&&+
 \frac{f_W}{\Lambda^2}(D_{\mu}\phi)^{\dagger}\hat W^{\mu \nu}(D_{\nu} \phi) \nonumber \\ && 
+\frac{f_{WWW}}{\Lambda^2}\textmd{Tr}\left[ \hat W_{\mu \nu} \hat W ^{\nu \rho} \hat W_{\rho}^{\mu}\right]\;,
\end{eqnarray}
with $\hat B^{\mu \nu}=i\frac{g'}{2}B^{\mu \nu}$ and $\hat W^{\mu \nu}=ig\frac{\sigma^a}{2}W^{a,\mu \nu}$.
However, the EFT approach offers a better interpretation of the origin
of these additional couplings as we will describe in more detail next.

The scale $\Lambda$ denotes the energy scale at which the structure of
the full theory is resolved. At lower energies, the heavy degrees of
freedom of this full theory are considered to be integrated out,
appearing as higher-dimensional operators in the EFT that describes
the low energy physics. One example for such an effective field theory
is Fermi's theory of weak interactions. At an energy scale well below
the $W$ boson mass the weak interaction of leptons and neutrinos can
be described by a four fermion operator of dimension six. The
corresponding scale $\Lambda$ in an EFT description of weak
interaction would then be the $W$ boson mass.  For energies well below
the (usually unknown) scale $\Lambda$, the higher dimensional
operators are suppressed by powers of $\Lambda$.  This ensures that
the higher dimensional operators are more suppressed than lower
dimensional operators, i.e. dimension eight operators can usually be
neglected compared to dimension six operators. In the limit
$\Lambda \to \infty$ one recovers the SM.  The effective field theory
is only valid at energies well below $\Lambda$. As soon as one
approaches this scale the operators of dimension greater than six are
no longer suppressed. They contribute equally and can no longer be
neglected.  At this point the effective field theory breaks down and
has to be replaced by the UV completion of the underlying full
theory. Therefore the effective field theory provides a handle on the
energy range in which it is valid, which cannot be deduced from the
effective Lagrangians of \refeqs{eq:Ltgc}, (\ref{eq:Lqgc}), (\ref{eq:Htgc}).

One very important feature of higher dimensional operators is their high energy behavior. Due to their
higher dimension, the effects of these operators increase with energy and would eventually violate
unitarity. The energy at which (tree level) unitarity is violated depends on the operator and in general
also depends on the helicity \cite{Gounaris:1994cm}.
Typically this problem is solved by introducing form factors which suppress the effects of the operators
hence rendering the cross section unitary. These form factors are however completely arbitrary as long
as they preserve unitarity and from the viewpoint of an effective field theory they are not needed
because at this energy the effective theory is no longer valid \cite{Degrande:2012wf}.

\begin{sloppypar}
The effects of anomalous couplings in electroweak gauge boson
interactions in the production of multiple gauge bosons have been
calculated both for $e^{+} e^{-}$ colliders
\cite{anom:lep} as well as for hadron colliders 
\cite{anom:lhc,Eboli:1998vg} and many available results also include
next-to-leading order QCD and/or electroweak corrections. For the
extraction of limits on anomalous TGCs and QGCs it is essential that
precise predictions of the relevant processes are provided in form of
Monte Carlo programs including the effects of anomalous couplings. The
implementation of anomalous couplings in publicly available Monte
Carlo programs ranges from specific processes to a general
implementation at the level of the Lagrangian. For $e^+e^-$ colliders
anomalous couplings for the production of four fermions (and a photon)
are contained in \texttt{RacoonWW} \cite{racoonww,Denner:2001vr,Denner:2001bd},
including NLO EW corrections to four-fermion production in
double-pole approximation. A broader implementation of anomalous
couplings for $e^+e^-$ colliders is provided in \texttt{WHIZARD}
\cite{whizard}, which can also be used for hadron
colliders. \texttt{VBFNLO} \cite{vbfnlo} provides NLO QCD predictions
for processes at hadron colliders including trilinear and quartic
couplings as well as anomalous couplings of electroweak gauge bosons
to the Higgs boson.
\texttt{CalcHEP} and \texttt{CompHEP} \cite{calchep,comphep} can
import anomalous couplings from \texttt{LanHEP} \cite{lanhep} which generates
them at the level of the Lagrangian.  \texttt{FeynRules} also
can generate anomalous couplings at the Lagrangian level and the corresponding
Feynman rules can be implemented via the UFO format \cite{ufo} to any Monte
Carlo program that supports this format, as for instance \texttt{MadGraph} \cite{mg5}. 
\end{sloppypar}


\subsubsection{Anomalous gauge couplings: experimental prospects
\protect\footnotemark}
\footnotetext{ Nicolas Greiner, Sven Heinemeyer, Doreen Wackeroth}
\label{sec:quantum-agc-exp}

\providecommand{\Cdgz}{\ensuremath{\Delta g^\mathrm{Z}_1}}
\providecommand{\Cdgg}{\ensuremath{\Delta g^\mathrm{\gamma}_1}}
\providecommand{\Cdkz}{\ensuremath{\Delta \kappa_\mathrm{Z}}}
\providecommand{\Cdkg}{\ensuremath{\Delta \kappa_{\gamma}}}
\providecommand{\Ckg}{\ensuremath{\kappa_{\gamma}}}
\providecommand{\Ckz}{\ensuremath{\kappa_{\mathrm{Z}}}}
\providecommand{\Clg}{\ensuremath{\lambda_{\gamma}}}
\providecommand{\Clz}{\ensuremath{\lambda_{\mathrm{Z}}}}
\providecommand{\Cgv}[1]{\ensuremath{g^V_{#1}}}
\providecommand{\Cgz}[1]{\ensuremath{g^Z_{#1}}}
\providecommand{\Cgg}[1]{\ensuremath{g^{\gamma}_{#1}}}
\providecommand{\Ckzt}{\ensuremath{\tilde{\kappa}_\mathrm{Z}}}
\providecommand{\Clzt}{\ensuremath{\tilde{\lambda}_\mathrm{Z}}}
\providecommand{\Ckgt}{\ensuremath{\tilde{\kappa}_{\gamma}}}
\providecommand{\Clgt}{\ensuremath{\tilde{\lambda}_{\gamma}}}

We briefly review the capabilities of an LC to measure triple and
quartic gauge couplings (based on \citere{Baer:2013cma} and references
therein). 
As mentioned earlier, the effects of higher dimensional operators are
suppressed at low energies and their impact increases with increasing
center of mass energy. Therefore a general pattern is the deviation
from the SM best visible in the high energy tails of distributions 
like $p_T$, $H_T$ or invariant masses. 

The couplings among the electroweak gauge bosons are directly given by
the structure of the gauge group, see the previous section. 
This structure can thus directly be
determined by a measurement of the gauge boson interactions.  
Particularly sensitive is the process $e^+e^- \to W^+W^-$, since any
``naive'' change in the gauge couplings would lead to a violation of
unitarity, and small changes lead to relatively large variations.

To date, electroweak precision observables together with the LEP data 
yielded the strongest constraints on anomalous couplings
\cite{Burgess:1993vc,Aihara:1995iq}. For the triple gauge couplings
the bounds are \cite{Aihara:1995iq}
\begin{eqnarray}
 \Delta g_1^Z &= -0.033 \pm 0.031,\nonumber \\
 \Delta \kappa_{\gamma} &= 0.056 \pm 0.056,\nonumber \\
 \Delta \kappa_Z &= -0.0019 \pm 0.044,\\
 \lambda_{\gamma} &= -0.036 \pm 0.034,\nonumber \\
 \lambda_Z &= 0.049 \pm 0.045 .\nonumber
\end{eqnarray}
\begin{sloppypar}
The bounds currently available from LHC data are weaker but approach the
precision of the LEP results \cite{ATLAS:tgc}. 
\end{sloppypar}

Turning to the ILC, the different types of couplings can be disentangled
experimentally 
by analyzing the production angle distribution of the $W$ boson
and the $W$ polarization structure, which can be obtained from the
decay angle distributions.  Anomalous couplings for $WW\gamma$ and $WWZ$ 
result in similar final state distributions. However, using beam
polarization, they can be disentangled, where a large beam polarization,
in particular for the left-handed $e^-$ is required. Also positron
polarization is required for an optimal
resolution~\cite{MoortgatPick:2005cw}. 

A fast detector simulation analysis was performed for 
$\sqrt{s} = 500 \gev$ and $800 \gev$~\cite{Menges}.
The results for single parameter fits are shown in 
\refta{tab:tgc}. Correlations in the multi-parameter fits were taken
into account where possible. For $\sqrt{s} = 800 \gev$ 
they are relatively small, not increasing the uncertainties by more than
$\sim 20\%$. At $\sqrt{s} = 500 \gev$ the effect is larger, and
uncertainties can increase by up to a factor of two, see
also \citere{AguilarSaavedra:2001rg}. 

\begin{table}
\centering
\renewcommand{\arraystretch}{1.2}
\begin{tabular}[c]{|c|c|c|}
\hline
coupling & \multicolumn{2}{|c|}{error $\times 10^{-4}$} \\
\cline{2-3}
         & $\sqrt{s}=500\gev$ & $\sqrt{s}=800\gev$ \\
\hline
\hline
  \Cdgz  &$ 15.5 \phantom{0} $&$ 12.6 \phantom{0} $\\
  \Cdkg  &$  3.3 $&$  1.9 $\\
  \Clg   &$  5.9 $&$  3.3 $\\
  \Cdkz  &$  3.2 $&$  1.9 $\\
  \Clz   &$  6.7 $&$  3.0 $\\
\hline
\hline         
  \Cgz{5}&$ 16.5 \phantom{0} $&$ 14.4 \phantom{0} $\\
  \Cgz{4}&$ 45.9 \phantom{0} $&$ 18.3 \phantom{0} $\\
  \Ckzt  &$ 39.0 \phantom{0} $&$ 14.3 \phantom{0} $\\
  \Clzt  &$  7.5 $&$  3.0 $\\
  \hline
\end{tabular}
\caption{
  Results of the single parameter fits ($1 \sigma$) to the different 
    triple gauge couplings at the ILC for $\sqrt{s}=500$ GeV with
 ${\cal L}=
    500$ fb$^{-1}$ and $\sqrt{s}=800$~GeV with ${\cal L}=1000$~fb$^{-1}$;
    ${\cal P}_{e^-} = 80\%$ and ${\cal P}_{e^+} = 60\%$ has been used.
  Taken from~\cite{Menges}.}
\label{tab:tgc} 
\end{table}

\begin{figure}[htb!]
  \centering
  \includegraphics[width=0.4\linewidth,bb=33 17 492 468]{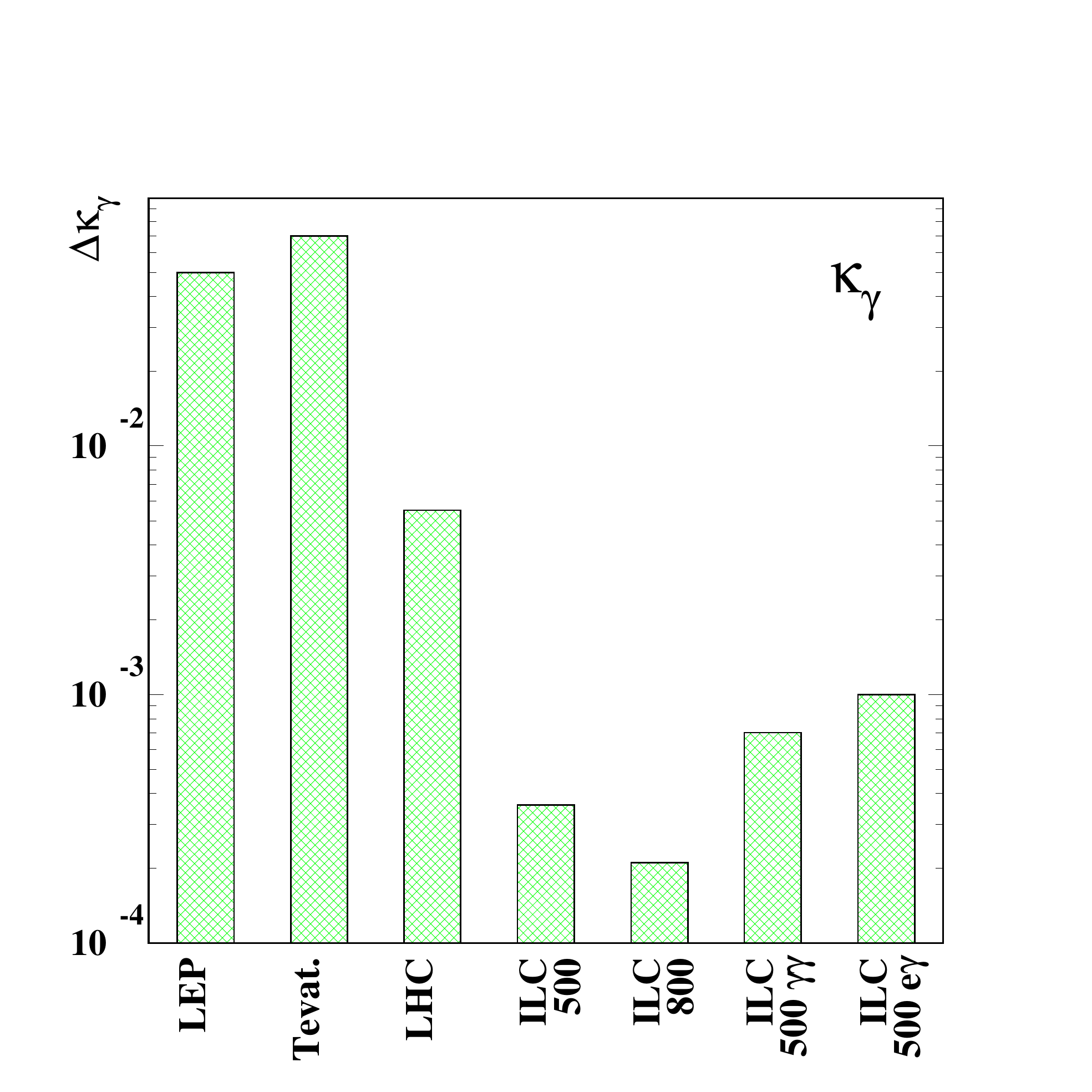}
  \hspace*{5mm}
  \includegraphics[width=0.4\linewidth,bb=33 17 492 468]{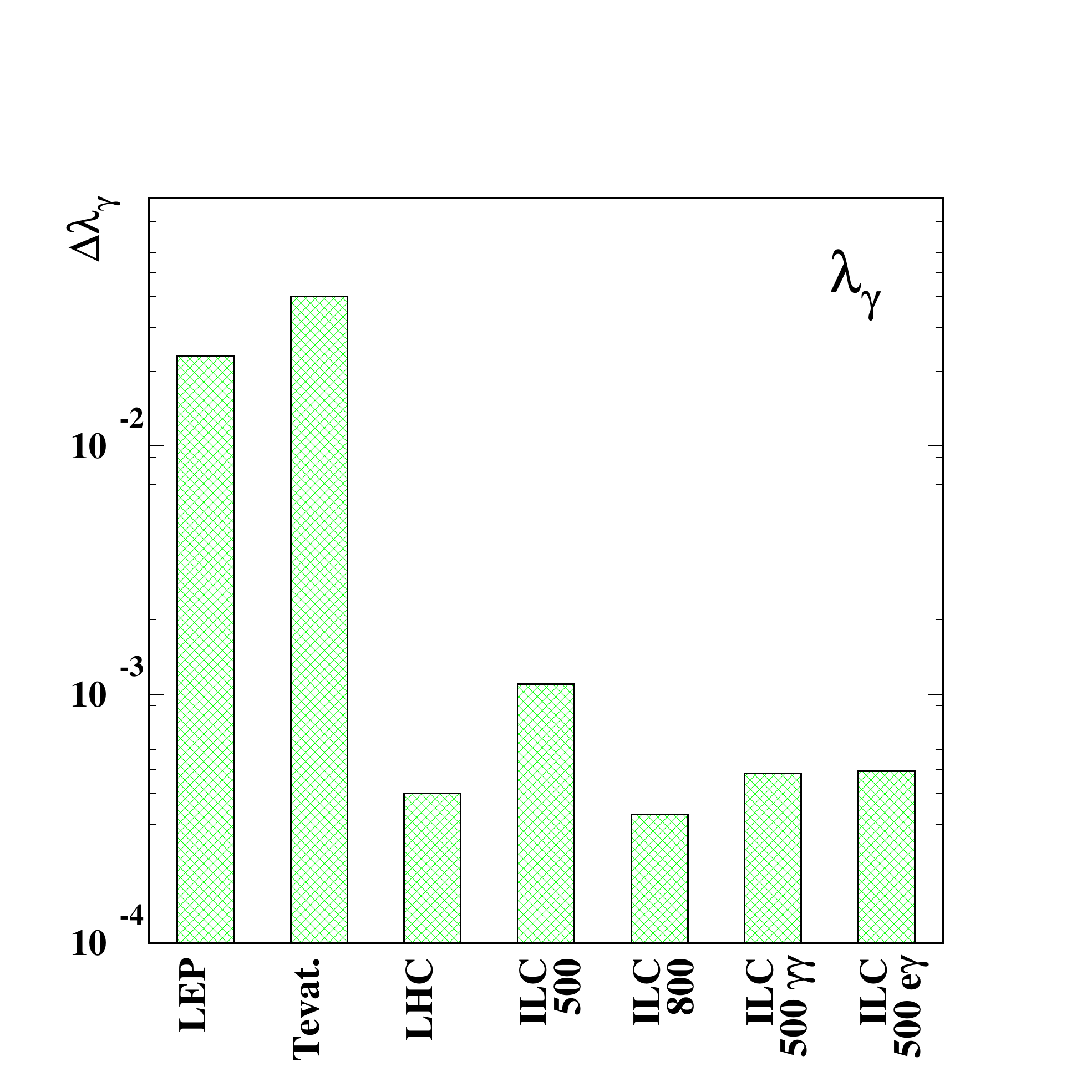}
  \caption{
    Comparison of $\Delta \kappa_\gamma$ and $\Delta
      \lambda_\gamma$ at different machines. For LHC and ILC three
      years of running are assumed (LHC: $300$~fb$^{-1}$, ILC
      $\sqrt{s}=500$ GeV: $500$ fb$^{-1}$, ILC $\sqrt{s}=800$ GeV:
 $1000$~fb$^{-1}$).
 If available the results from multi-parameter fits have
      been used. Taken from \cite{Baer:2013cma}.}
 \label{fig:tgc}
\end{figure}
 
Additional information on the triple gauge couplings can be obtained
when going to the $e\gamma$ and $\gamma \gamma$ options at the ILC. 
In this environment the $WW\ga$ couplings can be measured without
the $WWZ$ couplings entering the analysis. It was
shown~\cite{Monig:2004rg,Monig:2005ge} that $\kappa_{\gamma}$ can be measured
better in $e^+e^-$ collisions, while for $\lambda_\gamma$ the $e\gamma$ and
$\gamma\gamma$ modes can add relevant information. 
Figure~\ref{fig:tgc} shows the results for $\kappa_\gamma$ and 
$\lambda_\gamma$
obtained at different machines. The measurement of $\kappa_\gamma$ can be
improved substantially at the ILC. The other coupling, $\lambda_\gamma$, on the
other hand can be measured with similar accuracy at the LHC and the
various ILC options.


\begin{sloppypar}
Apart from the triple electroweak gauge boson couplings, the ILC
is also sensitive to the quartic couplings. Two processes are
important in this context: 
$e^+e^- \to VVV$ (triple gauge boson production, $V = W^\pm, Z$) and 
$e^+e^- \to VV' l_1 l_2$ ($l_{1,2} = e, \nu$, $V = W^\pm, Z$),
see \citere{Beyer:2006hx} and references therein.
This study uses complete six-fermion
matrix elements in unweighted event samples, fast simulation of the
ILC detector and a multidimensional parameter fit of the set of
anomalous couplings.  It also includes a study of triple weak boson
production which is sensitive to the same set of anomalous couplings.
It was shown that, under the assumption of custodial symmetry,
sensitivities for $h^{ZZ}$ and $g_2^{WW}$ at and below the level of 
$\sim 5\%$ can be found~\cite{Beyer:2006hx} for $\sqrt{s} = 1$ TeV and 
$1$~ab$^{-1}$ (see also \cite{Baer:2013cma}). 

As mentioned earlier, apart from the investigation of di-boson and
triple gauge boson production processes, constraints on the
coefficients of higher dimensional operators that lead to new
trilinear gauge couplings can also be obtained from their
contributions to EWPOs.  For instance, modifications of gauge boson
self energies induced by these higher dimensional operators can be
described with the help of $S$,$T$ and $U$ parameters
\cite{stu,Peskin:1991sw} and their extensions \cite{Barbieri:2004qk}, 
and by precisely measuring these oblique parameters the effects of these
operators can be severely constrained~\cite{Hagiwara:1993ck,Mebane:2013cra,Mebane:2013zga}.
Typically, bounds from EWPOs mainly affect those operators that contribute
already at tree level to the observables. Effects of operators contributing only at the
one loop level are suppressed and therefore their bounds are weaker compared to
the bounds that can be derived from direct measurements~\cite{Mebane:2013cra,Mebane:2013zga}.

Recently, constraints on anomalous quartic gauge couplings have been
obtained from studies of $WW\gamma$ and $WZ\gamma$ production
\cite{Chatrchyan:2014bza} and like-sign $WWjj$ production \cite{Aad:2014zda}
at the 8 TeV LHC.
\end{sloppypar}


\subsection{New gauge bosons\protect\footnotemark}
\footnotetext{Stephen Godfrey}
\label{sec:quantum-bsm-gb}



Extra gauge bosons, $Z'$'s and $W'$'s,  
are a feature of many models of physics beyond the SM
\cite{Cvetic:1995zs,Rizzo:2006nw,Leike:1998wr,Langacker:2008yv,Hewett:1988xc}. 
Examples of such models are Grand Unified theories based on groups
such as $SO(10)$ or $E_6$ \cite{Hewett:1988xc},  
Left-Right symmetric models \cite{lrmodels}, Little Higgs models 
\cite{Perelstein:2005ka,ArkaniHamed:2002qy,Schmaltz:2004de,Han:2003wu},
and Technicolour models \cite{Chivukula:1994mn,Simmons:1996ws,Hill:1994hp,Lane:1995gw}
 to name a few.  In addition, resonances that arise as Kaluza-Klein excitations in theories
of finite size extra dimensions \cite{Hewett:2002hv}
would also appear as new gauge bosons in high energy experiments.
It is therefore quite possible that the discovery of a new gauge boson could be one of the
first pieces of evidence for physics beyond the SM.  Depending on the model, the dominant
$Z'$ decay may be either into leptons or jets, leading to a resonance in the reconstructed dilepton or 
dijet invariant mass distribution respectively.

Currently, the highest mass bounds on most extra neutral gauge bosons are obtained by searches at
the Large Hadron Collider by the ATLAS and CMS experiments.  The most recent results 
based on dilepton resonance searches in $\mu^+\mu^-$ and $e^+e^-$ final states
use data from the  
7 TeV proton collisions collected in 2011 and more recent 8 TeV data collected in 2012.  
ATLAS \cite{ATLAS-ZP-2012} obtains the  exclusion limits at 95\% C.L.
$M(Z^\prime_{\rm SSM})>2.49$~TeV, $M(Z^\prime_\eta)>2.15$~TeV, 
$M(Z^\prime_\chi)>2.24$~TeV and $M(Z^\prime_\psi)> 2.09$~TeV 
using only the 8 TeV (6 fb$^{-1}$) data set 
and CMS \cite{CMS-ZP-2012} obtains 95\% C.L. exclusion
limits of $M(Z^\prime_{\rm SSM})>2.59$~TeV and $M(Z^\prime_\psi)> 2.26$~TeV
using the 7 TeV (5 fb$^{-1}$) and 8 TeV (4 fb$^{-1}$) data sets.
It is expected that 
the LHC should be able to see evidence for $Z'$'s up to $\sim$5~TeV 
once the LHC reaches its design energy and luminosity 
\cite{Godfrey:1994qk,Rizzo:1996ce,hep-ph/0201093,Diener:2010sy,Erler:2011ud} and to 
distinguish between models up to $M_{Z'}\simeq 2.1$~TeV (95\% C.L.) \cite{Osland:2009tn}.

It is expected that the LHC will be able to discover $W'$'s up to masses of $\sim 5.9$~TeV 
in leptonic final states assuming SM couplings \cite{Rizzo:1996ce}. 
Based on searches for a new $W$ boson decaying to a 
charged lepton and a neutrino using the transverse mass variable
CMS \cite{CMS:2012bza}
excludes the existence of a SSM $W'$ boson with a mass below 2.85~TeV at 95\% C.L. 
using the $\sqrt{s}=8$~TeV, ${\cal L}_{\rm int}=3.7$~fb$^{-1}$ dataset
while ATLAS excludes the
existence of a $W^*$ with a mass below 2.55~TeV at 95\% C.L. 
using the 7 TeV dataset with ${\cal L}_{\rm int}=4.7$~fb$^{-1}$ \cite{ATLAS:2012loa}.

For models that predict $Z'$ or $W'$ bosons that decay to two quarks, searches have been performed
that require two well-separated jets with high transverse momentum.  The CMS collaboration
excludes the existence of a SSM $Z'$ boson with mass below 1.6 TeV at 95\% C.L. 
and a SSM $W'$ with mass below 2.12~TeV  using the $\sqrt{s}=8$~TeV, ${\cal L}_{\rm int}=4.0$~fb$^{-1}$ 
dataset \cite{CMS:2012eza}.  The CMS 
collaboration also developed a dedicated search for $b\bar{b}$ resonances and excluded 
existence of a SSM $Z'$ boson with mass below 1.5 TeV at 95\% C.L. in the $b\bar{b}$ 
channel \cite{CMS:2012dxa}. For 
models with larger branching fractions to $b$-quarks the limit improves considerably, excluding 
a larger mass range.


If a narrow resonance were discovered, the crucial next step
would be to measure its properties and
determine the underlying theory.  While LHC measurements \cite{Cvetic:1995zs,Li:2009xh}
and low energy precision measurements \cite{Diener:2011jt}
can to some extent constrain new gauge boson couplings, precise measurements will
need a LC.

\subsubsection{New Gauge Boson Studies at High Energy \boldmath{$e^+ e^-$}
  Colliders}

Although the LHC will have explored the energy regime accessable to on-shell $Z'$ production by the 
time a LC is built, a high energy $e^+e^-$ collider will be sensitive to new gauge bosons with
$M_{Z' , W'} \gg \sqrt{s}$.
In $e^+e^-$ collisions below the on-shell production threshold, 
extra gauge bosons manifest themselves as deviations
from SM predictions due to interference between the new physics and the SM $\gamma /Z^0$ contributions.  
$e^+e^-\to f\bar{f}$ reactions are characterized by relatively clean, simple final states
where $f$ could be leptons ($e$, $\mu$, $\tau$) or  quarks ($u$, $d$, $s$, $c$, $b$, $t$), for both 
polarized and unpolarized $e^\pm$.  The baseline ILC configuration envisages 
electron beam polarization greater than 80\% and positron beam polarization of $\sim$30\% might
be initially achieved, eventually increasing to $\sim$60\%.
The basic $e^+e^-\to f\bar{f}$ processes can be parametrized in terms of four 
helicity amplitudes which 
can be determined by measuring various observables: the leptonic cross section,
$\sigma(e^+e^-\to \mu^+\mu^-)$,
the ratio of the hadronic to the QED point cross section $R^{\rm had}=\sigma^{\rm had}/\sigma_0$,
the leptonic forward-backward asymmetry, $A^\ell_{\rm FB}$, the leptonic  longitudinal asymmetry, 
$A^\ell_{\rm LR}$, the hadronic longitudinal asymmetry, $A^{\rm had}_{\rm LR}$, 
the forward-backward asymmetry for specific quark or lepton 
flavours, $A^f_{\rm FB}$, the $\tau$ polarization asymmetry, 
$A_{pol}^\tau$, and the polarized forward-backward asymmetry for 
specific fermion flavours, $A^f_{\rm FB}({\rm pol})$ \cite{Godfrey:1996uq}
(see also \refse{sec:quantum-sw2eff}).
The indices $f=\ell, \; q$, $\ell =(e,\mu,\tau)$, $q=(c, \; b)$, 
and $had=$`sum over all hadrons' indicate the final state fermions.
Precision measurements of these observables 
for various final states ($\mu^+\mu^-$, $b\bar{b}$, $t\bar{t}$) can 
be sensitive to extra gauge boson masses that by far exceed the
direct search limits that are expected at the LHC
\cite{Godfrey:1994qk,hep-ph/0201093,Godfrey:1996uq,Osland:2009dp}.  
Further,  precision measurements of cross sections
to different final state fermions using polarized beams can be used to constrain the gauge boson
couplings and help distinguish the underlying theory
\cite{Osland:2009dp,Leike:1996qq,Riemann:2001,Godfrey:2005pm,riemann,Linssen:2012hp,Battaglia:2012ez}.  
A deviation for one observable is always possible as 
a statistical fluctuation.  In addition, different observables have different sensitivities to
different models (or more accurately to different couplings).  As a consequence, 
a more robust strategy is to combine many observables to obtain a $\chi^2$ figure of merit.

The ILC sensitivity to $Z'$'s is based on high statistics precision cross section measurements 
so that the reach will depend on the integrated luminosity.  
For many models a 500~GeV $e^+e^-$ 
collider with as little as 50~fb$^{-1}$ integrated luminosity would see the effects of a $Z'$ with 
masses as high as $\sim 5$~TeV \cite{Godfrey:1994qk}.  
The results of a recent study \cite{Osland:2009dp}
is shown in Fig.~\ref{fig:discovery}. That study
finds that a 500~GeV ILC with 500~fb$^{-1}$ and a 1~TeV ILC with 1~ab$^{-1}$
can see evidence or rule out a $Z'$ with masses that can exceed  $\sim 7$ and $\sim 12$~TeV
for many models, for the two respective energies \cite{Osland:2009dp}.  
These recent results also
consider various polarizations for the $e^-$ and $e^+$ beams and show that beam polarization 
will increase the potential reach of the ILC, see also
\citere{MoortgatPick:2005cw}. 


\begin{figure}[t]
\begin{center}
\includegraphics[width=0.45\textwidth]{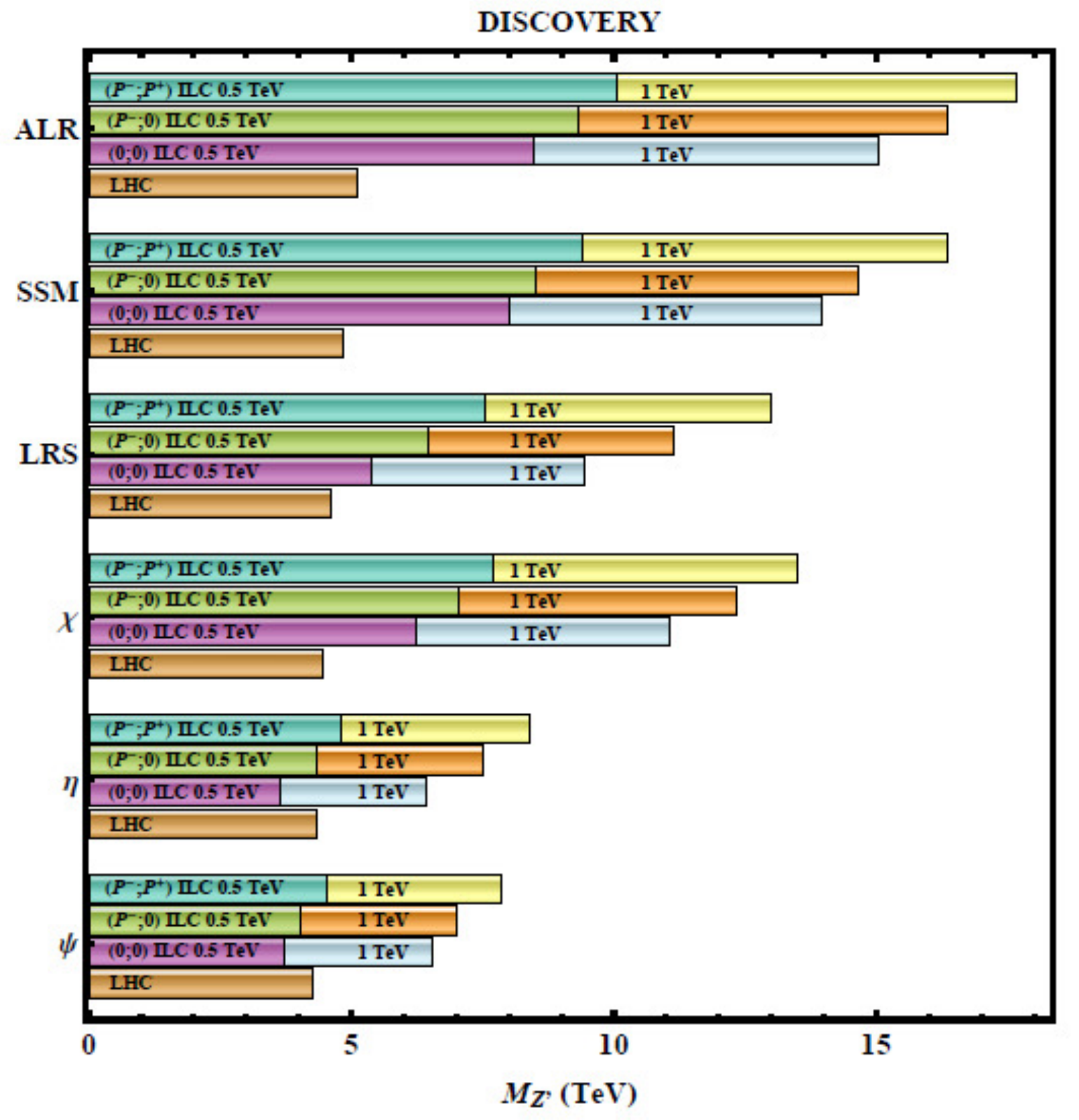}
\caption{
Discovery reach of the ILC with $\sqrt{s}=0.5$ (1.0) TeV and 
${\cal L}_{{\rm int}}=500$ (1000)~fb$^{-1}$. The discovery reach of the LHC 
for $\sqrt{s}=14$~TeV and 100~fb$^{-1}$ via the Drell-Yan process 
$pp\to \ell^+\ell^- +X$ are shown for comparison. From Ref.~\cite{Osland:2009dp} with kind 
permission of The European Physical Journal (EPJ).
}
\label{fig:discovery}
\end{center}
\end{figure}

\subsubsection{Measurement of \boldmath{$Z'$} Couplings at High Energy 
\boldmath{$e^+ e^-$} Colliders}


If a $Z'$ were discovered at the LHC, 
measurements of 2-fermion processes at the ILC could provide valuable 
constraints on its couplings and discriminate between models. 
Fig.~\ref{fig:id} (top panel) shows the expected resulting precision on $Z'$ couplings to leptons 
 for $\sqrt{s}=500$~GeV and ${\cal L}_{\rm int}=1$~ab$^{-1}$
for 3 values of $M_{Z'}$ for several representative models \cite{Godfrey:2005pm}.
In this figure, the KK case should
not be taken too literally as the couplings do not in fact correspond to the KK $Z'$ couplings 
but are an effective coupling, reflecting that in this model there are both photon and $Z$ KK 
excitations roughly degenerate in mass.  The point is simply that the KK model can be 
distinguished from other models. One notes that there is a two-fold ambiguity in the signs of 
the lepton couplings since all lepton observables are bilinear products of the couplings.  Hadronic
observables can be used to resolve this ambiguity since for this case the quark and lepton couplings
enter the interference terms linearly.  
Studies \cite{Godfrey:2005pm,Osland:2009dp} have demonstrated that 
beam polarization plays an important role in the measurement of the
$Z'$-fermion couplings and therefore in the discrimination between models.

Rather than measure 
the $Z'$-fermion couplings one could pose the question; if measurements resulted from 
a true BSM model, could one rule out other possibilities?  A recent analysis given in
Ref.~\cite{Osland:2009dp} showed that the ILC could discriminate models for $Z'$ masses up to
4-8~TeV for a 500~GeV ILC and up to 6-11~TeV for a 1~TeV ILC, depending on the true model. 
This exceeds the corresponding discovery reach at the LHC and is only
slightly lower than the discovery reach at the ILC due to the relatively large differences between
angular distributions for $e^+e^- \to f\bar{f}$ for the different models.  
More crucially, the ILC 
is significantly more powerful for measuring $Z'$ couplings than is possible at the LHC.
These results 
are based on purely leptonic processes.  Measurements of $c$- and $b$-quark pair production 
cross sections would contribute important complementary information for identifying the underlying 
theory.

If deviations from the SM were observed but there was no direct evidence for a $Z'$ from the LHC
one could still exclude a "tested" model for any value of $M_{Z'}$ below some value 
for a given set of ILC measurements.  To see how one can extract such limits consider 
normalized couplings defined by $C_{v,a}^{fN}={C_{v,a}^{f'}}\sqrt{s/(M_{Z'}^2-s)}$.  
Fig.~\ref{fig:id} (bottom panel) 
shows contraints on ``normalized'' couplings for a 10~TeV $Z'$ and  $\sqrt{s}=3$~TeV and
${\cal L}_{\rm int}=1$~ab$^{-1}$ \cite{riemann,Linssen:2012hp}.  One can see how, if a model with a
10~TeV $Z'$ were the true model, other models could be excluded.  
Ref.~\cite{Osland:2009dp} finds that for the models they considered 
one might be able to distinguish between $Z'$ models, at 95\% C.L., up to 
$M_{Z'}\simeq 3.1$~TeV (4.0 TeV) for unpolarized (polarized) beams at the 0.5~TeV ILC
and 5.3 TeV (7.0 TeV) at the 1~TeV ILC.  
Presented another way, they find that if one of the 6 models they studied
is true, the other 5 candidates can be ruled out by a 500~GeV ILC for $Z'$ masses up to 4-8 TeV, 
depending on the true model.  This discrimination reach is only slightly below the
discovery reach due to order-one differences among the angular distributions in $e^+e^- \to f\bar{f}$ 
predicted by the different models and in all cases is significantly higher than that of the LHC.\\

\begin{figure}[htb!]
\begin{center}
%
\includegraphics[width=0.36\textwidth]{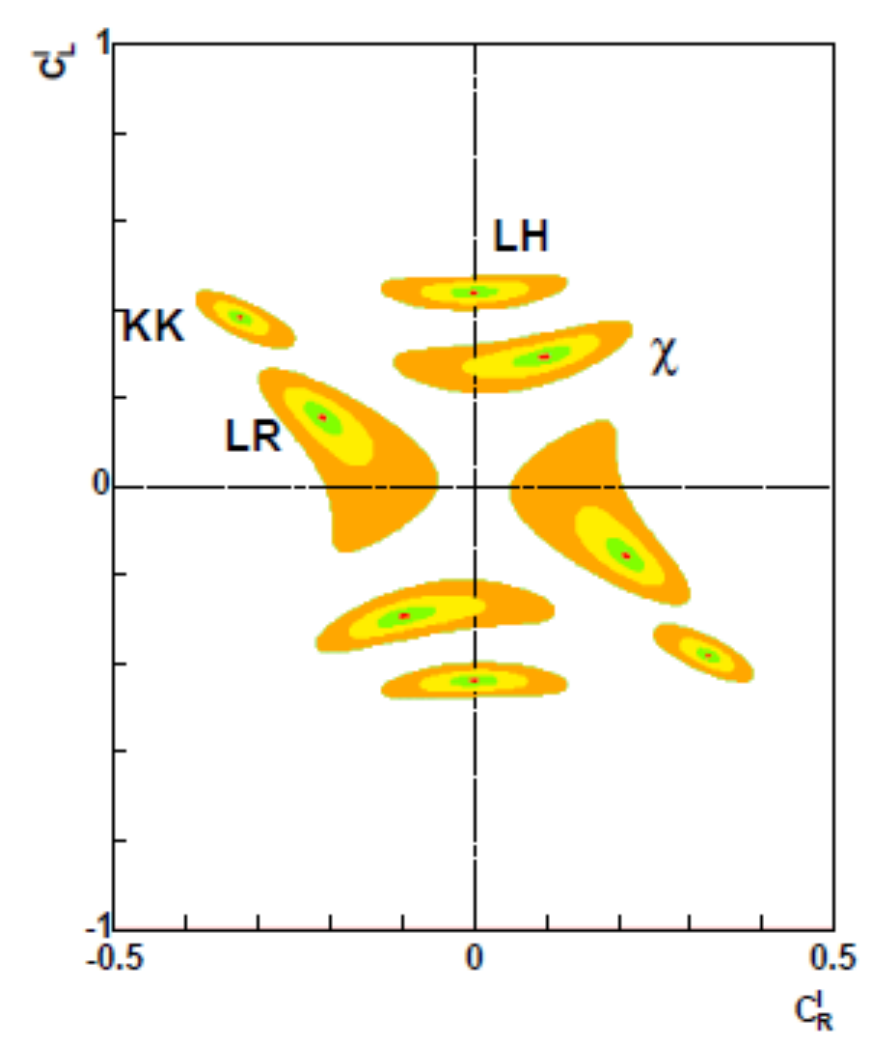}  
\vspace{1cm}
\includegraphics[width=0.35\textwidth]{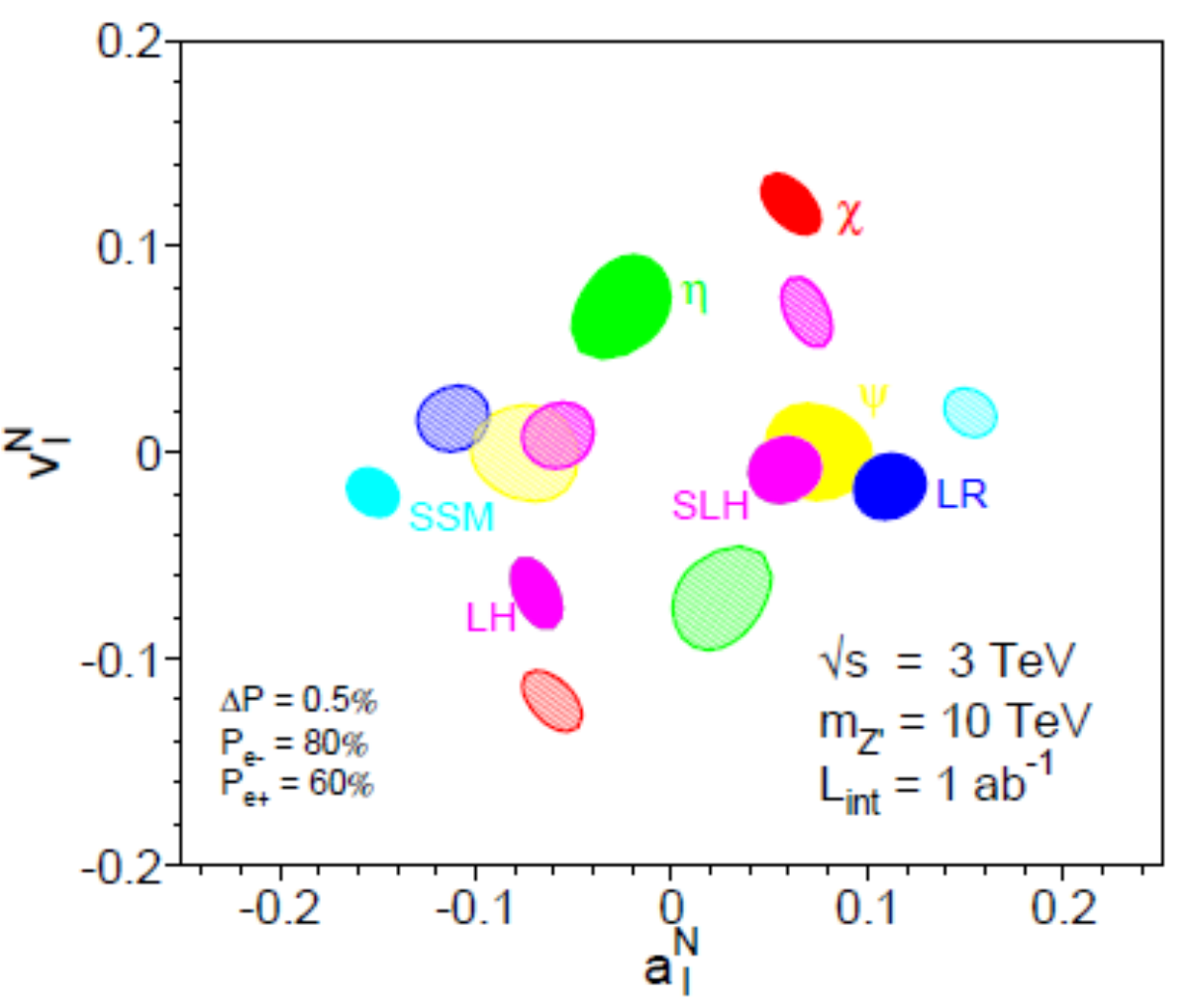} 
\caption{
Top: Resolving power (95\% CL) for $M_{Z'}=1$, 1.5, and 2~TeV and
$\sqrt{s}=500$~GeV, ${\cal L}_{\rm int}=1$ab$^{-1}$, $|P_{e^-}|=$80\%,
$|P_{e^+}|=$60\%, for leptonic couplings based on the leptonic
observables $\sigma$, $A_{\rm LR}$, $A_{\rm FB}$. The couplings
correspond to the $E_6$ $\chi$, LR, LH, and KK models. From
Ref.~\cite{Godfrey:2005pm}.  Bottom: Expected resolution at CLIC with
$\sqrt{s}=3$~TeV and ${\cal L}= 1$~ab$^{-1}$ on the "normalised"
leptonic couplings of a 10~TeV $Z'$ in various models, assuming lepton
universality.  The mass of the $Z'$ is assumed to be unknown. The
couplings correspond to the $E_6$ $\chi$, $\eta$, and $\psi$, the SSM,
LR, LH and SLH models.  The couplings can only be determined up to a
twofold ambiguity.  The degeneracy between the $\psi$ and SLH models
might be lifted by including other channels in the analysis
($t\bar{t}$, $b\bar{b}$, ...).  From \citeres{riemann,Linssen:2012hp}.
}
\label{fig:id}
\end{center}
\end{figure}

\subsubsection{Discovery and Identification of \boldmath{$W'$} Bosons 
in \boldmath{$e^+e^-$}}

While there is a broad literature on $Z'$ properties, $W'$ studies for high energy $e^+e^-$ colliders
are rather limited.  One study showed that the process $e^+ e^-\to \nu \bar{\nu} \gamma$ would be sensitive
to $W'$ masses up to several TeV depending on the model, the centre of mass energy, and the
assumed luminosity \cite{Godfrey:2000hc}.   For example, evidence for a SSM $W'$ could be seen 
up to $M_{W'}=$4.3, 5.3, and 6.0~TeV for $\sqrt{s}=0.5$, 1.0, and 1.5~TeV respectively with 
${\cal L}_{\rm int}=500$~fb$^{-1}$ while a LR $W'$ could only be detected up to
$M_{W'}=$1.2, 1.6, and 1.9~TeV for the same collider parameters.   Another process that has
been considered is $e\gamma \to \nu q +X$ where the photon is 
produced by a backscattered laser or is a Weizs\"acker-Williams photon \cite{Godfrey:2000pw}.
These processes yield discovery limits for $W'_{\rm SSM}$ 
of 4.1 (2.5), 5.8 (3.6) and 7.2 (4.5)~TeV for the 
backscattered laser (Weizs\"acker-Williams) cases and for the three values 
for $\sqrt{s}$ and ${\cal L}_{\rm int}$ 
given above. Limits for the LR model are substantially lower.  

In general we do not expect an $e^+e^-$ collider
to be sensitive to $W'$'s with masses larger than could be discovered at the LHC. 
If new gauge bosons were discovered 
first in other processes, the ILC could  measure $W'$ (and $Z' \nu \bar{\nu}$) couplings 
which would complement measurements made at the LHC.





\section[Supersymmetry]{Supersymmetry\protect\footnotemark}
\footnotetext{Editors: H.~Baer, M.~Battaglia, J.~Kalinowski\\
Contributors: A.~Arbey, P.~Bechtle, A.~Bharucha, F.~Br\"ummer, S.Y.~Choi,
A.~Freitas, J.~Heisig,	J.~List, F.~Mahmoudi,
G.~Moortgat-Pick, W.~Porod, S.~Porto,
K.~Rolbiecki\\
H.B.  would like to thank D. Mickelson and A. Mustafayev for providing 
several figures.}
\label{susy}

\subsection{Introduction and overview}
\label{sec:susy1}

The recent discovery of a Higgs-like resonance at $M_h=(125.15\pm
0.24)$~GeV by the Atlas and CMS experiments at the CERN LHC seemingly
completes the identification of all matter states predicted to exist
by the Standard Model (SM) of particle physics. In spite of this
extraordinary achievement, the SM remains beset by an array of
shortcomings which strongly suggest that new physics exists at, or
around, the TeV energy scale. Chief among these is the gauge hierarchy
problem, which arises if fundamental scalar fields (such as the Higgs
field) do exist. In this case, the scalar field mass term diverges
quadratically, and we would expect the Higgs field to have mass far
beyond the 125~GeV level unless an exquisite degree of fine-tuning
between bare and loop corrections is invoked at each order in
perturbation theory.

Along with the gauge hierarchy problem, the SM is lacking in that it
provides no particle to explain cold dark matter (CDM) in the
universe, it does not allow for baryogenesis in the early universe, it
does not allow for the suggested unification of SM forces, it contains 
no solution to the strong CP problem and it
provides no avenue for a sensible inclusion of quantum gravity into
its structure.

\begin{sloppypar}
While a variety of solutions to the gauge hierarchy problem have been
proposed, weak scale supersymmetry
\cite{Wess:1973kz,Salam:1974jj,Salam:1974ig,Witten:1981nf,Kaul:1981wp},
or SUSY, is the most theoretically engaging and one which also appears
to be, at least indirectly, supported by experimental data.
Supersymmetry is a quantum space-time symmetry that predicts a
correspondence between bosonic and fermionic degrees of freedom. In
SUSY theories, scalar fields inherit the protective chiral symmetry
enjoyed by fermions, reducing their quadratic divergence to merely
logarithmic. Since the log of a large number can be small, the
required tuning between bare mass and loop mass is greatly reduced,
allowing disparate mass scales to co-exist within the same theoretical
structure.
\end{sloppypar}

To be phenomenologically viable, supersymmetrised versions of the SM
must include {\it soft} SUSY breaking\cite{Girardello:1981wz}, 
{\it i.e.} only those SUSY breaking terms which maintain 
the cancellation of quadratic divergences. 
In the Minimal Supersymmetric Standard Model (MSSM), a variety of new matter
states-- spin 0 squarks and sleptons along with
additional Higgs bosons and spin $1\over 2$ charginos, neutralinos and
gluinos--  are expected to exist at or around the weak scale. 

The MSSM has received some indirect experimental support from the  
measured values of the strong and electro-weak forces: these unify to a
single value at energy scales $M_{GUT}\sim 2\times 10^{16}$~GeV under
renormalisation group (RG) evolution.  Also, the measured value of the
top quark ($m_t\simeq 173.2$ GeV) turns out to be sufficiently large
as to induce a radiatively-driven breaking of electroweak symmetry.
In addition, while the SM allows for a Higgs mass within a wide range,
$100<M_H<1000$~GeV, the MSSM restricts the lightest SUSY Higgs
boson $100<M_h<135$~GeV. The fact that the newly discovered
Higgs-like state falls within the narrow mass range predicted by SUSY
may also be regarded as an indirect support of this picture.  Simple
arguments based on electro-weak naturalness would suggest that
super-partners should exist at or below the $\sim 1$ TeV scale,
motivating a significant effort for their search at the LHC and
inspiring the physics program of a future $e^+e^-$ linear collider.
Finally, SUSY provides us with at least three viable candidates for
dark matter: the lightest neutralino $\tilde{\chi}^0_1$ (a WIMP
candidate) the gravitino $\tilde{G}$ and the axino $\tilde{a}$ (the
spin-1/2 super-partner of the axion)\cite{Steffen:2008qp}.

SUSY theories also offer at least three mechanisms for baryogenesis,
including weak scale baryogenesis (now nearly excluded in the MSSM),
thermal and non-thermal leptogenesis and Affleck-Dine baryo- and
leptogenesis\cite{Dine:2003ax}.  Local SUSY (supergravity) theories
necessarily include spin-2 gravitons and spin-3/2 gravitinos, and
reduce to Einstein's general relativity in the classical limit.

This chapter provides an overview of the capabilities of a linear
$e^+e^-$ collider in the search for supersymmetry, in view of the
constraints and indications derived from present experimental data, in
particular the LHC results from the 7 and 8~TeV data
for the SUSY direct searches and the
Higgs properties. The limits derived in these searches seem to
require SUSY particles beyond the TeV scale, seemingly in
contradiction to the aforementioned arguments based on electro-weak
naturalness. However, it is important to observe that the strongly-interacting
SUSY particles-- which LHC is most sensitive to-- are also those with
less direct connection to the electro-weak naturalness. 
Taken in this context, there remains a huge role for LHC operation at 13-14~TeV 
and for subsequent operation of a linear $e^+e^-$ collider of sufficient centre of mass energy,
$\sqrt{s}$, to play a decisive role in the search for, and proof of, SUSY. 
Indeed, even if no SUSY particles are seen at the LHC at 13-14~TeV, 
then a 0.5-1 TeV linear $e^+e^-$-collider 
may still retain its role as {\it discovery} machine 
for SUSY\cite{Baer:2013faa,Baer:2013vqa} 
in that {\it the most natural SUSY models require light higgsinos} 
with mass $\sim 100-200$ GeV which can easily elude LHC searches
(due to the small energy release from their compressed spectra),
but which can easily be detected in $e^+e^-$ collisions of sufficient 
energy $\sqrt{s}>2m(higgsino)$.

If supersymmetric matter is indeed found at LHC or the $e^+e^-$-LC, 
then a program of precision measurements, which can be made in high
energy $e^+e^-$ collisions, will be crucial for pinning down SUSY
particle masses, mixings and other properties. 
From such measurements, it may be possible to clarify the role of SUSY in cosmic
dark matter production and possibly also in baryogenesis, thus
establishing even more closely the link between particle physics and
cosmology.  If indeed a desert exists between the weak scale and some
high scale such as $M_{GUT}$ or $M_{string}$, then it may be possible
to extrapolate SUSY parameters to these ultra-high scales, thus
testing ideas about unification, SUSY breaking, and string theory. We
will conclude that a linear $e^+e^-$ collider of sufficient energy and
luminosity is absolutely needed for providing a detailed
experimental exploration of the intriguing concept
of weak-scale supersymmetry, if it is realised in nature.

\subsection{Models of Supersymmetry}
\label{sect:susymodels}

The superfield formalism provides an algorithm for the direct
supersymmetrization of the SM\cite{Baer:2006rs,Drees:2004jm}.  In this
case, each SM matter fermion of a given chirality is elevated to a
chiral superfield which also contains a spin-0 superpartner. The SM
gauge fields are elevated to gauge superfields which also contain
spin-$1\over 2$ gauginos. The SM Higgs doublet is embedded in a chiral
superfield necessitating introduction of spin-$1\over 2$ higgsinos.
The addition of extra higgsinos carrying gauge quantum numbers
destroys the elegant anomaly cancellation mechanism in the SM, unless
one introduces as well a second Higgs/higgsino doublet superfield
carrying opposite weak hypercharge.

\begin{sloppypar}
The resulting supersymmetrized SM enjoys exact, rigid supersymmetry--
but this is known not to be true since it would imply {\it e.g.} the
existence of spin-0 partners of the electron (selectrons) with the
same mass as the electron: such matter states would easily have been
detected long ago. Hence, SUSY must be a broken symmetry.  SUSY can be
broken explicitly by adding by hand {\it soft SUSY breaking} (SSB)
terms to the Lagrangian. These terms include mass terms for spin-0
superpartners, mass terms for each gaugino, and bilinear and trilinear
scalar interactions (so-called $B$ and $A$ terms).
\end{sloppypar}

\begin{sloppypar}
In addition, a plethora of terms are allowed in the superpotential
which violate baryon and lepton number conservation, and lead to rapid
proton decay. Such terms are suppressed by invoking an $R$-parity
(which naturally arises in SUSY GUT theories based on $SO(10)$). If
$R$-parity is conserved, then SUSY particles can only be produced in
pairs at colliders, SUSY particles must decay to other SUSY particles,
and the lightest SUSY particle must be absolutely stable, perhaps
offering a good dark matter candidate.
\end{sloppypar}

The resulting theory, called the Minimal Supersymmetric Standard
Model, or MSSM, is the direct supersymmetrization of the SM that is
consistent with all known constraints.  It includes more than 100 adjustable
parameters\cite{Baer:2006rs}, most of these consisting of flavor 
or $CP$ violating
terms.  Under the assumption of minimal flavor violation (MFV) and
minimal $CP$-violation (MCPV), these are set to zero, so that FV and
CPV arise solely from the Yukawa sector.  The pMSSM model with 19
adjustable weak scale parameters is a popular model for this approach.

\subsubsection{Gravity mediation}

An appealing approach to SUSY breaking 
comes from invoking {\it local} SUSY, or supergravity (SUGRA). 
If SUSY is local, then one must necessarily
include a graviton-gravitino supermultiplet. One may include a 
so-called \emph{hidden sector} of fields whose sole purpose is to 
allow for spontaneous breaking of SUSY via the superHiggs mechanism\cite{Cremmer:1982en}. 
Under the superHiggs mechanism, hidden sector fields acquire a 
SUSY breaking vev $\langle F\rangle\sim m^2$ so that the gravitino 
gains a mass $m_{3/2}\sim m^2/M_P$ while the graviton remains 
massless: if $m_{3/2}\sim M_{weak}$, then $m\sim 10^{11}$ GeV. 

\begin{sloppypar}
The above-mentioned soft SUSY breaking terms arise via tree level 
gravitational interactions with magnitude $\sim m_{3/2}$. More
generally, ``gravity-mediated supersymmetry breaking'' denotes any
theory in which supersymmetry breaking is communicated to the visible sector
by $M_P$-suppressed interactions at the tree level, not necessarily
just involving the gravitational multiplet, and therefore gives
soft parameters of the order $m_{3/2}$. If $m_{3/2}\sim M_{weak}$, 
then in the limit $M_P\rightarrow \infty$ while keeping $m_{3/2}$ 
constant we obtain a theory with weak-scale rigid supersymmetry plus
soft SUSY breaking terms.
\end{sloppypar}

The minimal supergravity model (mSUGRA\cite{Arnowitt:2012gc} or
CMSSM\cite{Kane:1993td}) assumes all matter scalars and both 
Higgs fields receive a common soft mass $m_0$ at some high scale, 
usually taken to be $M_{GUT}\simeq 2\times 10^{16}$ GeV, the scale 
where gauge couplings unify in the MSSM.
Likewise, all gauginos receive a common mass $m_{1/2}$, and all $A$ 
terms are set to a common value $A_0$.  While this ansatz is
simple, and receives some experimental motivation in that such choices
suppress flavor and $CP$-violating terms, one must remember that it is
at best merely a simplifying assumption that is not likely to remain
true for realistic models\cite{Nath:1997qm}.  

\begin{sloppypar}
One of the virtues of SUSY models defined
at a high scale such as $Q=M_{GUT}$ is that the large top quark Yukawa
coupling drives exactly the right scalar Higgs field $m_{H_u}^2$ to negative
squared values, resulting in a radiatively driven breakdown of
electroweak symmetry (REWSB)\cite{Ibanez:1982fr}. Upon EWSB, the
$B\mu$ parameter may be traded for a parameter $\tan\beta =v_u/v_d$, 
the ratio of Higgs field vevs, and the magnitude of the Higgsino mass 
parameter $\mu$ is fixed to yield the measured $Z$-boson mass.
Then all sparticle masses and mixings, and hence production and decay
rates, are determined by the well-known parameter set:
$m_0,\ m_{1/2},\ A_0,\ \tan\beta,\ \ {\rm and}\ sign(\mu ).$
However, many more parameters are allowed if one deviates from the 
simplistic assumption listed above, resulting
in models with {\it non-universal} soft SUSY breaking terms.
\end{sloppypar}

\subsubsection{GMSB and AMSB} 

\begin{sloppypar}
In addition to models of gravity-mediated SUSY breaking, other
possibilities exist. One of these is \emph{gauge-mediated SUSY breaking}, 
or GMSB\cite{Dine:1995ag, Meade:2008wd}. 
In this class of theories, the hidden sector couples to a messenger
sector (which carries SM gauge quantum numbers) which acts as an intermediary
between the visible and hidden sectors. 
In GMSB, loop diagrams containing messenger states induce visible sector soft 
SUSY breaking terms. 
\end{sloppypar}

The gravitino again gets a mass $m_{3/2}\sim \langle F\rangle/M_P$, 
while the sparticles gain soft masses of the order 
$\frac{g^2}{16\pi^2}\frac{F}{M}$, where $M$ is the messenger mass and $g$ is
any MSSM gauge coupling. For $M\ll M_P$, the
SUSY particles may still be at the TeV scale while gravitinos can be
much lighter, so that the gravitino may play the role of the LSP. In 
the simplest GMSB models, the trilinear SSB terms are suppressed, so 
there is little mixing in the top squark sector. Thus, these models 
have trouble generating a light Higgs scalar of mass $\sim 125$ GeV 
as is now required by data\cite{Arbey:2011ab,Baer:2012uy}.  More 
general gauge mediation models \cite{Craig:2013wga} are now required 
for phenomenological viability.

\begin{sloppypar}
A third possibility is \emph{anomaly-mediated SUSY
breaking} \cite{Randall:1998uk,Giudice:1998xp}. In any model of SUSY
breaking mediation, there are contributions to SSB terms arising from
the super-Weyl anomaly. These are however suppressed by a loop factor
with respect to $m_{3/2}$ and therefore subdominant in gravity
mediation or GMSB. They become relevant in \emph{sequestered models}
where the gravity- and gauge-mediated soft masses are negligible,
e.g.~because the hidden sector is spatially separated from the visible
sector in extra dimensions.
\end{sloppypar}

In AMSB, the SSB terms are governed by the RG beta functions and
anomalous dimensions divided by loop factors.  In this case, the
wino-like neutralino turns out to be LSP, while $m_{3/2}\sim 25-50$
TeV, thus solving the cosmological gravitino problem.  Since minimal
versions of these models fail to generate a large $A$-term, they
also seem disfavored by the recently measured Higgs boson mass.
Moreover, the minimal anomaly mediated model predicts tachyonic 
sleptons, which is an even more serious shortcoming. However, 
various string-inspired modifications of the minimal
framework do lead to viable
phenomenology\cite{Choi:2005uz,Baer:2010uy,Dudas:2012wi,Acharya:2012tw}.

\subsubsection{Hybrid mediation schemes}

Embedding the MSSM into a more fundamental model at high scales, for
instance into the effective field theory of some superstring
compactification, can naturally lead to hybrid mediation scenarios.
These are attractive also from the phenomenological point of view.

\begin{sloppypar}
An example, motivated from both heterotic and type IIB string models, is
\emph{mirage mediation} \cite{Choi:2004sx,Choi:2005ge,Endo:2005uy}: If
gravity-mediated contributions to the gaugino masses are only mildly
suppressed, they may be of similar magnitude as the anomaly-mediated
contributions. A combination of gravity and anomaly mediation allows to
interpolate between unified gaugino masses at the GUT scale (as
predicted by the simplest gravity-mediated GUT models) and unified
gaugino masses at some arbitrary lower \emph{mirage scale} (after adding
the anomaly-mediated contributions, since these are given by the very
same beta function coefficients that govern the gaugino mass RGEs). An
immediate consequence is a compressed low-scale gaugino mass spectrum if
the mirage scale is low \cite{Baer:2006id,Asano:2012sv,Badziak:2012yg,
Krippendorf:2013dqa}. This allows for a lower gluino mass without
conflicting with the LHC search bounds, thus possibly reducing the fine-tuning. 
Depending on the underlying model, a ``natural SUSY''
pattern for the squark masses, with sub-TeV stops but multi-TeV first-
and second-generation squarks, may also be realized
\cite{Krippendorf:2012ir,Asano:2012sv}. Sub-TeV charginos and
neutralinos are common in these models.
Such models, realized within the MSSM, do have problems generating a light
Higgs scalar with $M_h\simeq 125$ GeV\cite{Baer:2007eh} while
maintaining naturalness\cite{Baer:2014ica}.
\end{sloppypar}

A more extreme example is the case where the gravity-mediated
contributions to the gaugino masses vanish altogether, e.g.~because they
are forbidden by some symmetry under which the goldstino superfield is
charged \cite{Giudice:1998xp}. In this case (which suffers from extreme
fine-tuning with regards to electroweak symmetry breaking) the squarks
and sleptons have gravity-mediated masses up to around $100$ TeV, while
the gaugino masses follow the anomaly mediation pattern and are lighter
by a loop factor
\cite{Wells:2004di,ArkaniHamed:2006mb,Hall:2011jd,Ibe:2011aa,Ibe:2012hu}. The
LSP is a wino-like neutralino which is nearly degenerate with a
wino-like chargino.

Alternatively, for a high messenger scale just below the scale of grand
unification (which is well motivated within certain F-theory and
heterotic models \cite{Marsano:2008jq,Brummer:2011yd}), gauge mediation
can coexist with gravity mediation. This is because the GUT scale is
about a loop factor below the Planck scale. Generic models of high-scale
gauge mediation tend to have problems with flavour constraints
\cite{Feng:2007ke,Hiller:2008sv}, which should be solved similarly as in
ordinary gravity mediation. Such hybrid gauge-gravity mediation models
naturally allow to obtain near-degenerate higgsino-like charginos and
neutralinos with masses around the electroweak scale, while the rest of
the spectrum can be in the multi-TeV range
\cite{Brummer:2011yd,Brummer:2012zc}.
Models with mixed gauge, gravity and anomaly mediation are also a 
possibility\cite{Altunkaynak:2010xe}.

All the above hybrid mediation scenarios have in common that the
coloured superpartners may be difficult to see at the LHC, either
because they are heavy or because the spectrum is compressed. In
particular, large parameter space regions survive the constraints from
LHC8. At the same time, at least some of the charginos and neutralinos
are often light enough to be produced, detected, and studied at 
a linear $e^+e^-$ collider.

\subsection{Naturalness and Fine-tuning}
\label{sec:susy1-1}

The main reason we expect supersymmetric matter states to arise with
masses around the electroweak scale derives from the notion of
electroweak naturalness. A model is considered to be natural in the
electroweak sector if there are no large, unnatural cancellations
(fine-tunings) required in deriving the measured values of both $M_Z$
and $M_h$.

A quantitative measure of fine-tuning of a supersymmetric model was introduced over twenty five years ago, while SUSY was being searched for at 
LEP~\cite{Ellis:1986yg,Barbieri:1987fn,Dimopoulos:1995mi}). 
The so-called {\it Barbieri-Giudice} measure,
$\Delta_{BG}$, is defined as
\be
\Delta_{BG}\equiv max_i\left[ c_i\right]\ \ {\rm where}\ \ c_i=\left|\frac{\partial\ln M_Z^2}{\partial\ln a_i}\right|
=\left|\frac{a_i}{M_Z^2}\frac{\partial M_Z^2}{\partial a_i}\right|
\label{eq:DBG}
\ee
where the set $a_i$ constitute the fundamental parameters of the model.
Thus, $\Delta_{BG}$ measures the fractional change in $M_Z^2$ due to fractional variation in 
model parameters $a_i$. 
The $c_i$ are known as {\it sensitivity co-efficients}\cite{Feng:2013pwa}.

For models
with parameters defined at very high scales ({\it e.g.} at $\Lambda=M_{GUT}$), 
as those discussed above,
the evaluation of $\Delta_{BG}$ requires to
express $M_Z^2$ in terms of high-scale parameters 
using semi-analytic solutions of the 
 renormalization group
equations for the corresponding
soft term and $\mu$~\cite{Abe:2007kf,Martin:2007gf,Feng:2013pwa}.  

The $\Delta_{BG}$ measure picks off the co-efficients of the various terms and
recales by the soft term squared over the $Z$-mass squared: {\it e.g.}
$c_{M_{Q_3}^2}=0.73\cdot (M_{Q_3}^2/M_Z^2)$.  For example, if one allows
$M_{Q_3}\sim 3$ TeV (in accord with requirements from the measured
value of $M_h$) the result is $c_{M_{Q_3}^2}\sim 800$ and so
$\Delta_{BG}\ge 800$.  In this case, one expects SUSY would be
electro-weak fine-tuned to about 0.1\%.  However, in constrained SUSY
models where the high scale parameters are related, then cancellations
between positive and negative contributions can
occur. For instance, in models with universal scalar masses, then
third generation fine-tuning is greatly reduced in the focus point
region.  More generally, in models of gravity-mediated SUSY breaking,
then for any hypothesized hidden sector, the SUSY soft breaking terms
are all calculated as numerical co-efficients times the gravitino mass
$m_{3/2}$\cite{Soni:1983rm}.  

These shortcomings can be cured by modifying the definition of the 
fine-tuning measure.
In the calculation of the SUSY mass spectrum, the actual
fine-tuning occurs when enforcing the {\it electroweak minimization condition}
which is written as
\be 
\frac{M_Z^2}{2} = \frac{M_{H_d}^2+\Sigma_d^d - (M_{H_u}^2+\Sigma_u^u) \tan^2\beta}{\tan^2\beta -1} 
-\mu^2 .
\label{eq:mZsSig}
\ee 
In the above expression, $M_{H_u}^2$ and $M_{H_d}^2$ are weak scale
soft SUSY breaking masses while the terms $\Sigma_d^d$ and
$\Sigma_u^u$ incorporate a variety of radiative corrections (a
complete list of one-loop corrections is provided in Ref.
\cite{Baer:2012cf}.)  

For typical SUSY models with parameters defined
at some high scale $\Lambda$ (where $\Lambda$ is frequently taken as
high as $M_{GUT}\simeq 2\times 10^{16}$ GeV), the positive value of
$M_{H_u}^2(\Lambda)$ is driven radiatively to negative values at the
weak scale (owing to the large top quark Yukawa coupling) so that
electroweak symmety is radiatively broken.  In models where large
TeV-scale values of $-M_{H_u}^2$ are generated at the weak scale, then
a compensating value of $\mu^2$ must be dialed/tuned to enforce the
measured value of $M_Z\simeq 91.2$ GeV.

The amount 
of fine-tuning required in Eq. \ref{eq:mZsSig} can be quantified by defining 
the
{\it  electro-weak fine-tuning measure}\cite{Baer:2012up,Baer:2012cf,Baer:2013gva} 
\be 
\Delta_{EW} \equiv max_i \left|C_i\right|/(M_Z^2/2)\;, 
\ee 
%
  where $C_{H_d}=M_{H_d}^2/(\tan^2\beta -1)$,
  $C_{H_u}=-M_{H_u}^2\tan^2\beta /(\tan^2\beta -1)$ and $C_\mu
  =-\mu^2$.  Also, $C_{\Sigma_u^u(k)} =-\Sigma_u^u(k)\tan^2\beta
  /(\tan^2\beta -1)$ and $C_{\Sigma_d^d(k)}=\Sigma_d^d(k)/(\tan^2\beta
  -1)$, where $k$ labels the various loop contributions included in
  Eq. \ref{eq:mZsSig}.  

Since $\Delta_{EW}$ depends only upon the weak scale SUSY spectrum, it
is {\it model independent} (within the MSSM) in that different models
giving rise to exactly the same spectrum will have the same values of
$\Delta_{EW}$.
For models with parameters defined at the weak scale, such as the pMSSM, then 
$\Delta_{BG}\approx \Delta_{EW}$ since the sensitivity co-efficients  
$c_{\mu}=C_{\mu}$ and $c_{H_u}=C_{H_u}$.

For $\tan\beta\agt 5$ and neglecting radiative corrections, the
condition Eq. \ref{eq:mZsSig} reduces to $M_Z^2/2\simeq
-M_{H_u}^2-\mu^2$, so that models with weak scale naturalness require
that $-M_{H_u}^2\sim M_Z^2$ and also $\mu^2\sim M_Z^2$.  The first of
these conditions obtains crisis when $M_{H_u}^2$ is driven to small rather
than large negative values during the process of radiative electroweak
symmetry breaking.  The second condition implies a spectrum of light
higgsino-like ``electroweakinos'' (i.e.\ charginos and neutralinos) 
with mass the closer to $M_Z$ the
better:
\begin{itemize}
\item[] $m_{\tilde{\chi}_1^\pm},\ m_{\tilde{\chi}_{1,2}^0}\sim |\mu |\sim 100-250$~GeV.
\end{itemize}
Such light higgsinos would be accessible at an $e^+e^-$ linear
collider of centre-of-mass energy, $\sqrt{s}=250$-500~GeV, i.e.\
exceeding twice their mass.  
In such a case,
then a high energy $e^+e^-$ collider would function as a {\it higgsino
  factory}\cite{Baer:2014yta} in addition to a Higgs factory!  
While such light higgsinos might be
produced at some sizeable rates at the LHC, the kinematics of their
visible decay products may make it difficult if not impossible to
observe them in hadronic collisions. The compressed spectra reduce the
transverse momentum of the produced jets and leptons bringing them
below the cuts applied by the triggers and the subsequent offline
event selection criteria.

\subsection{Indirect Constraints}
\label{sec:susy2}

In spite of the many attractive features of SUSY models, no sign of
supersymmetric matter has yet emerged and dark matter is still to be
observed at ground-based direct detection experiments.  Here, we
review the constraints on SUSY particle masses and parameters derived
from precision measurements of low energy processes and the dark
matter relic density.  Constraints from the direct search for SUSY
particles at the LHC will be addressed in the following Section.

\subsubsection{Flavour Physics}
\label{sec:susy2-1}

Flavour physics provides indirect information about supersymmetry
which can play an important and complementary role compared to direct
searches at colliders.  Several decays of $b$ hadrons which are
suppressed in the SM may offer sensitivity to SUSY through additional
contributions mediated by supersymmetric particles, which do not
suffer the same suppression and may substantially modify the decay
rate.  The main processes of interest are the $\bar B \to X_s \gamma$,
$B_s \to \mu^+\mu^-$ and $B_u\to\tau\nu_\tau$ decays.

The decay $\bar B \to X_s \gamma$ is a loop-induced flavour changing
neutral current (FCNC) process that offers high sensitivity to
supersymmetry due to the fact that additional contributions to the
decay rate-- in which SM particles are replaced by SUSY particles such
as charged Higgs, charginos or top squarks-- are not suppressed by a loop
factor relative to the SM contribution.  Within a global effort, a
perturbative QCD calculation to the NNLL level has been
performed~\cite{Misiak:2006zs}, leading to~\cite{Mahmoudi:2008tp}:
\begin{equation}\label{final1}
\mathrm{BR}(\bar B \to X_s \gamma)_{\rm NNLL} =  (3.08 \pm 0.23) \times
10^{-4},
\end{equation}
for a photon energy cut at $E_\gamma = 1.6$ GeV, and using the updated
input parameters of PDG~\cite{Beringer:1900zz}.
The non-perturbative corrections to this decay mode are
sub-leading~\cite{Benzke:2010js} and their error is included in the above
prediction.
The averaged experimental value by the HFAG group~\cite{Amhis:2012bh} gives
\begin{equation}
  {\mathrm{BR}}(\bar B \rightarrow X_s \gamma)_{\rm exp} = (3.43 \pm
0.21 \pm 0.07) \times 10^{-4},
\end{equation}
where the first error is the combined statistical and systematic
uncertainties and the second represents the photon energy
extrapolation.  The SM prediction and the experimental average are
hence consistent at the $1.2 \sigma$ level, and therefore this decay
has a restrictive power on the SUSY parameter space. Recently, the
first practically complete NLL calculation of the decay rate in the
MSSM has been finalised~\cite{Greub:2011ji}. The dominant SUSY
contributions are provided by diagrams with top-squarks and charginos,
which grow linearly with $\tan \beta$\cite{Baer:1996kv}. This decay is therefore
particularly constraining in the regions with large $\tan \beta$ or spectra
with both light top-squarks and charginos. The charged Higgs contributions on the
other hand are not $\tan \beta$ enhanced.

\begin{sloppypar}
Recently, the purely leptonic decay of $B_s \to \mu^+\mu^-$ has
received special attention due to the progress on both experimental
results and theory calculations. This rare decay is very sensitive to
supersymmetric contributions which are free from the helicity
suppression of the SM diagrams. 
The recent observation of this decay by the LHCb~\cite{Aaij:2012ct} and CMS\cite{cms_Bs} 
experiments allows for a combined determination of its branching fraction to be
\begin{equation}
\mathrm{BR}(B_s\to\mu^+\mu^- )=(2.9\pm 0.7)\times\  10^{-9} .
\end{equation}
While this is in accord with the SM prediction of $(3.53 \pm 0.38) \times 10^{-9}$~\cite{Mahmoudi:2012un},
it also provides a stringent limit on the viable parameter space of many supersymmetric models. 
The SUSY contributions to the decay {\it amplitudes} are dominated by Higgs-mediated
penguin diagrams~\cite{Babu:1999hn,Mizukoshi:2002gs,Haisch:2012re} and
are proportional to
\begin{equation} 
-\mu A_t \, \frac{\tan^3\beta}{(1+\epsilon_b \, \tan\beta)^2} \;
\frac{M_t^2}{M_{\tilde t}^2} \,
\frac{M_b  M_\mu}{4\sin^2\theta_W M_W^2 M_A^2}.
\end{equation}
The sensitivity of $B_s \to \mu^+ \mu^-$ to
SUSY contributions is significant in regions at large $\tan \beta$ and 
small to
moderate $M_A$ values, regions which are also probed by direct SUSY
particle searches at ATLAS and CMS, in particular $H/A \rightarrow
\tau^+ \tau^-$.
As a result, while the constraints derived from the current LHCb
result remove a large fraction of points at large $\tan \beta$ and low
$M_A$, nonetheless for
intermediate $\tan \beta$ values and/or large masses of the
pseudo-scalar Higgs boson $A$, the branching fraction in the MSSM does
not deviate much from its SM prediction, leaving a sizeable fraction
of SUSY parameter regions totally unconstrained \cite{Arbey:2012ax}.
\end{sloppypar}

The decay $B\to K^* \mu^+ \mu^-$ gives also access to angular
distributions, in addition to the differential branching fraction, and
offers a variety of complementary observables. However, these
observables suffer from large uncertainties, in particular due to form
factors. A set of optimised observables has been defined from ratios
of angular coefficients to minimise hadronic uncertainties, while
preserving the sensitivity to new physics 
effects~\cite{Matias:2012xw,Descotes-Genon:2013vna}. 
They have
been recently measured by the LHCb collaboration\cite{Aaij:2013qta} 
highlighting a
tension in several binned observables. While these tensions remain
even when including the SUSY contributions, the overall agreement with
the MSSM predictions is within 1$\sigma$-level for an appropriate
choice of the model parameters\cite{Mahmoudi:2014mja}.

Finally, the purely leptonic decay of $B_u\to\tau\nu_\tau$ is
sensitive to supersymmetry through the exchange of a charged Higgs boson
already at tree level, which does not suffer from the helicity
suppression of the SM contribution with the exchange of a $W$ boson.
The branching ratio of $B_u \to \tau \nu_\tau$ in supersymmetry relative
to the SM is given by
\begin{equation}
\frac{\mathrm{BR}(B_u\to\tau\nu_\tau)_{\mathrm{MSSM}}
}{\mathrm{BR}(B_u\to\tau\nu_\tau)_{\mathrm{SM}}}=\left[1-\frac{m_B^2}{M_{H^+}^2}\, \frac{\tan^2\beta}{1+\epsilon_0\tan\beta}\right]^2,
\end{equation}
where $\epsilon_0$ is an effective coupling parametrising the
non-holomorphic correction to the down-type Yukawa coupling induced by
gluino exchange.  This decay is therefore also very sensitive to the
MSSM parameter region at large $\tan \beta$ and small $M_{H^+}$
values, and much less sensitive to other SUSY parameters.  The
branching fraction for the decay is calculated in the SM to be
$(1.10\pm 0.29)\times 10^{-4}$~\cite{Eriksson:2008cx}, which exhibits
a slight tension with the experimental averaged value of $(1.14\pm
0.22)\times 10^{-4}$~\cite{Amhis:2012bh}.

\subsubsection{Muon Magnetic Moment}
\label{sec:susy2-2}

The SUSY contribution to the muon magnetic moment is given by \cite{Moroi:1995yh}
\begin{equation}
\Delta a_\mu^{SUSY}\propto \frac{M_\mu^2\mu M_i\tan\beta}{M_{SUSY}^4}
\end{equation} 
where $i=1,2$ stands for electroweak gaugino masses and $M_{SUSY}$ is
the characteristic sparticle mass circulating in the muon-muon-photon
vertex correction: $M_{\tilde{\mu}_{L,R}}$,
$M_{\tilde{\nu}_\mu}$ and $M_{\tilde{\chi}_i}$.

The anomalous magnetic moment of the muon $a_\mu\equiv\frac{(g-2)_\mu}{2}$ was
measured by the Muon $g$-2 Collaboration~\cite{Bennett:2006fi} which
gives a $3.6\sigma$ discrepancy when compared to the SM calculations
based on $e^+e^-$ data~\cite{Davier:2010nc}, $\Delta a_\mu
=a_\mu^{meas}-a_\mu^{SM}[e^+e^-]=(28.7\pm 8.0)\times 10^{-10}$.  As
discussed in more detail in Chapter~\ref{sec:quantum}, 
the SM prediction depends on
the estimate of the hadronic vacuum polarisation contribution. Using
$\tau$-decay data rather than low energy $e^+e^-$ annihilation data
reduces the discrepancy to $2.4\sigma$ giving $\Delta a_\mu
=a_\mu^{meas}-a_\mu^{SM}[\tau]=(19.5\pm 8.3)\times 10^{-10}$.

Attempts to explain the muon $g$-2 anomaly using supersymmetry usually
invoke sparticle mass spectra with relatively light smuons and/or large
$\tan\beta$ (see {\it e.g.} Ref.~\cite{Feng:2001tr}).  Some SUSY models
where $M_{\tilde{\mu}_{L,R}}$ is correlated with squark masses (such as
mSUGRA) are now highly stressed to explain the $(g-2)_\mu$
anomaly, given the bounds from the LHC direct searches. In addition,
since naturalness favours a low value of $|\mu |$, tension again arises
between a large contribution to $\Delta a_\mu^{SUSY}$ and naturalness
conditions. The current $3\sigma$-deviation is clearly not
sufficient to prove the existence of new physics, but in the future,
progress can be expected both on the experimental side (due to a new
measurement at Fermilab with fourfold improved precision
\cite{FNALProposal}) as well as on the theoretical 
side~\cite{Benayoun:2014tra,Blum:2013xva}.

\subsubsection{Dark Matter and Cosmological Constraints}
\label{sec:susy2-3}

During the past several decades, a very compelling and simple scenario
has emerged to explain the presence of dark matter in the universe
with an abundance roughly five times that of ordinary baryonic matter.
The WIMP miracle scenario posits that weakly interacting massive
particles would be in thermal equilibrium with the cosmic plasma at
very high temperatures $T \ge M_{\mathrm WIMP}$. As the universe
expands and cools, the WIMP particles would freeze out of thermal
equilibrium, locking in a relic abundance that depends inversely on
the thermally-averaged WIMP (co)-annihilation cross
section~\cite{Lee:1977ua,Gondolo:1990dk}:
\be
\Omega_{\chi} h^2\simeq \frac{s_0}{\rho_c/h^2}\left(\frac{45}{8 \pi^2 g_*}\right)^{1/2}
\frac{x_f}{M_P}\frac{1}{\langle\sigma v\rangle}
\ee
where $s_0$ is the present entropy density, $\rho_c$ is the critical closure density, 
$g_*$ measures the degrees of freedom, $x_f=m/T_f$ is the inverse freeze-out 
temperature rescaled by the WIMP mass, 
$M_P$ is the reduced Planck mass and $\langle\sigma v\rangle$ is the thermally
averaged WIMP annihilation cross section with $v$ being the WIMP relative velocity.
 The WIMP ``miracle'' occurs in that a weak
strength annihilation cross section gives roughly the measured relic
abundance provided the WIMP mass is also of order the weak 
scale~\cite{Baltz:2006fm}.

The lightest neutralino of SUSY models has been touted as a
prototypical WIMP
candidate~\cite{Goldberg:1983nd,Ellis:1983ew,Jungman:1995df}. The
precise determination of the dark matter relic density,
$\Omega_{\mathrm{CDM}} h^2$, obtained from the cosmic microwave
background (CMB) by the WMAP satellite experiment
first~\cite{Komatsu:2010fb} and the Planck mission~\cite{Ade:2013zuv},
now stands as a reference constraint for SUSY models. While the
comparison of the measured abundance of CDM with the neutralino dark
matter relic density, $\Omega_{\chi} h^2$, computed in an assumed SUSY
scenario, is affected by cosmological uncertainties which may be
large\cite{Gelmini:2006pw}, it is certainly appropriate to require at
least that SUSY models do not violate the upper bound on the CDM
abundance, after accounting for these uncertainties.  An predicted
overabundance of thermally-produced WIMPs may in fact be allowed in
some specific models with either R-parity violating WIMP decays, late
WIMP decays to an even lighter LSP (e.g.\ axino or gravitino) or by
late time entropy injection from moduli or saxion decays.

Despite the WIMP ``miracle'', SUSY theories where the lightest
neutralino plays the role of a thermally produced WIMP, have a relic
abundance $\Omega_{\chi}h^2$ spanning over a broad range of values
from several orders of magnitude larger than the value derived from
the CMB spectrum in the case of a bino-like neutralino, and 
up to two-to-three orders of magnitude lower in the case of
wino- or higgsino-like neutralinos \cite{Baer:2010wm}
 with a mass of order 100~GeV, see Fig.~\ref{fig:Oh2MN1}.
A wino- or higgsino-like neutralino LSP in the generic MSSM gives a relic 
density compatible with the CMB data for masses in the range 0.9--3~TeV, while
 bino-like or mixed neutralinos may match the CMB data for lighter masses.
A deficit is, in principle, acceptable since the neutralino may
not be the only source of dark matter and its relic density should not
necessarily saturate the measured value. As an example, in the case of
the axion solution to the strong CP problem within the SUSY context,
dark matter is due to a mixture of axions and neutralinos\cite{Baer:2011hx}.
For the case of bino-like LSPs where the abundance might be expected to 
exceed the WMAP/Planck value, then an efficient annihilation mechanism -- 
such as co-annihilation, resonance annihilation or mixed bino-higgsino or 
mixed wino-bino annihilation -- is needed. Such enhanced annihilation mechanisms
define specific patterns of the masses of one or more SUSY particles
compared to the lightest neutralino, which are important for searches at colliders.
\begin{figure}[hb!]
\begin{center}
\includegraphics[width=7.0cm]{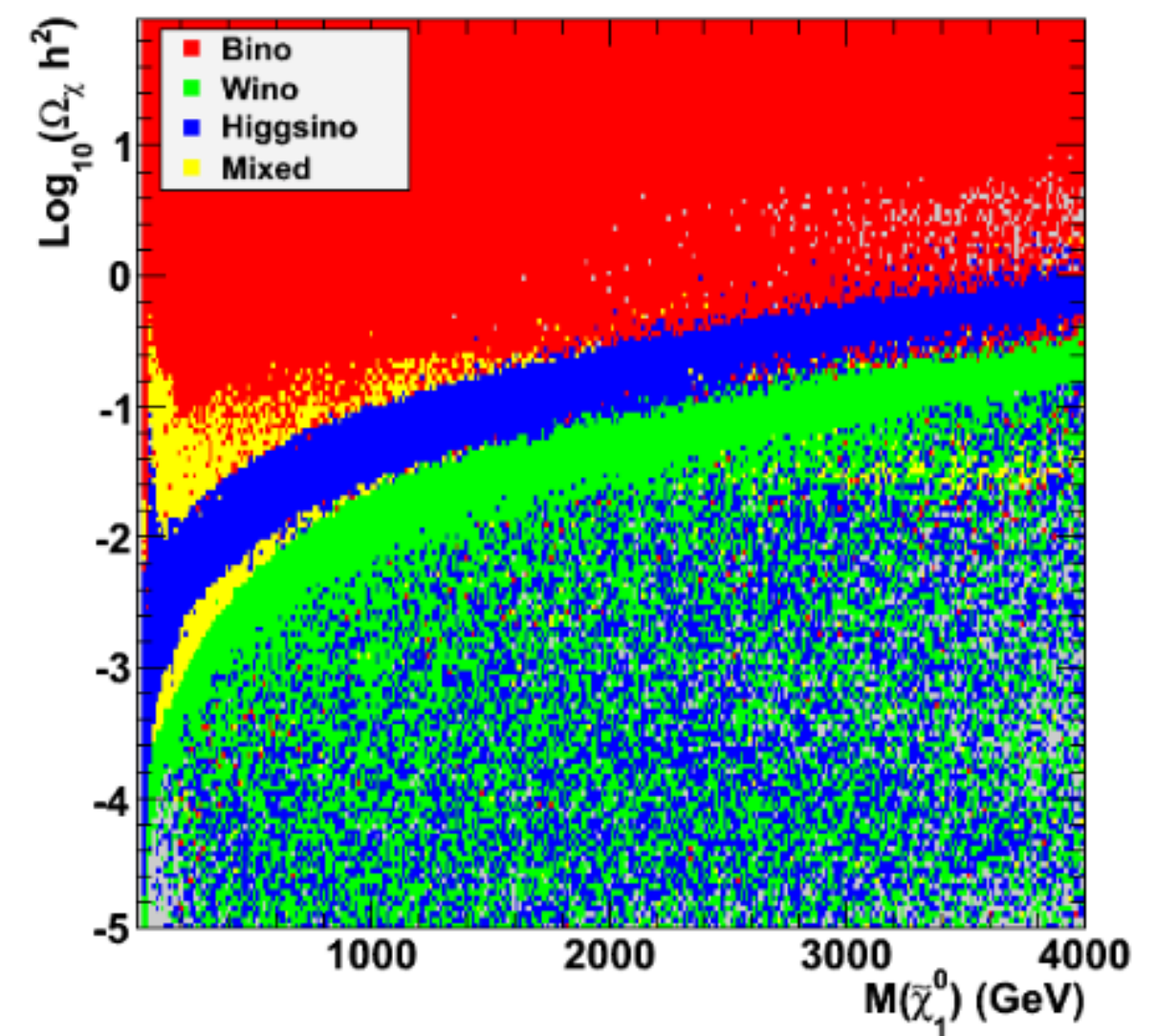} 
\end{center}
\caption{Neutralino relic density as a function of the neutralino LSP mass from a scan of the pMSSM parameter space. 
The colours indicate the nature of the neutralino LSP with the largest occurrence in each bin.}
\label{fig:Oh2MN1}
\end{figure}
\begin{figure}[hb!]
\begin{center}
\includegraphics[width=7.0cm]{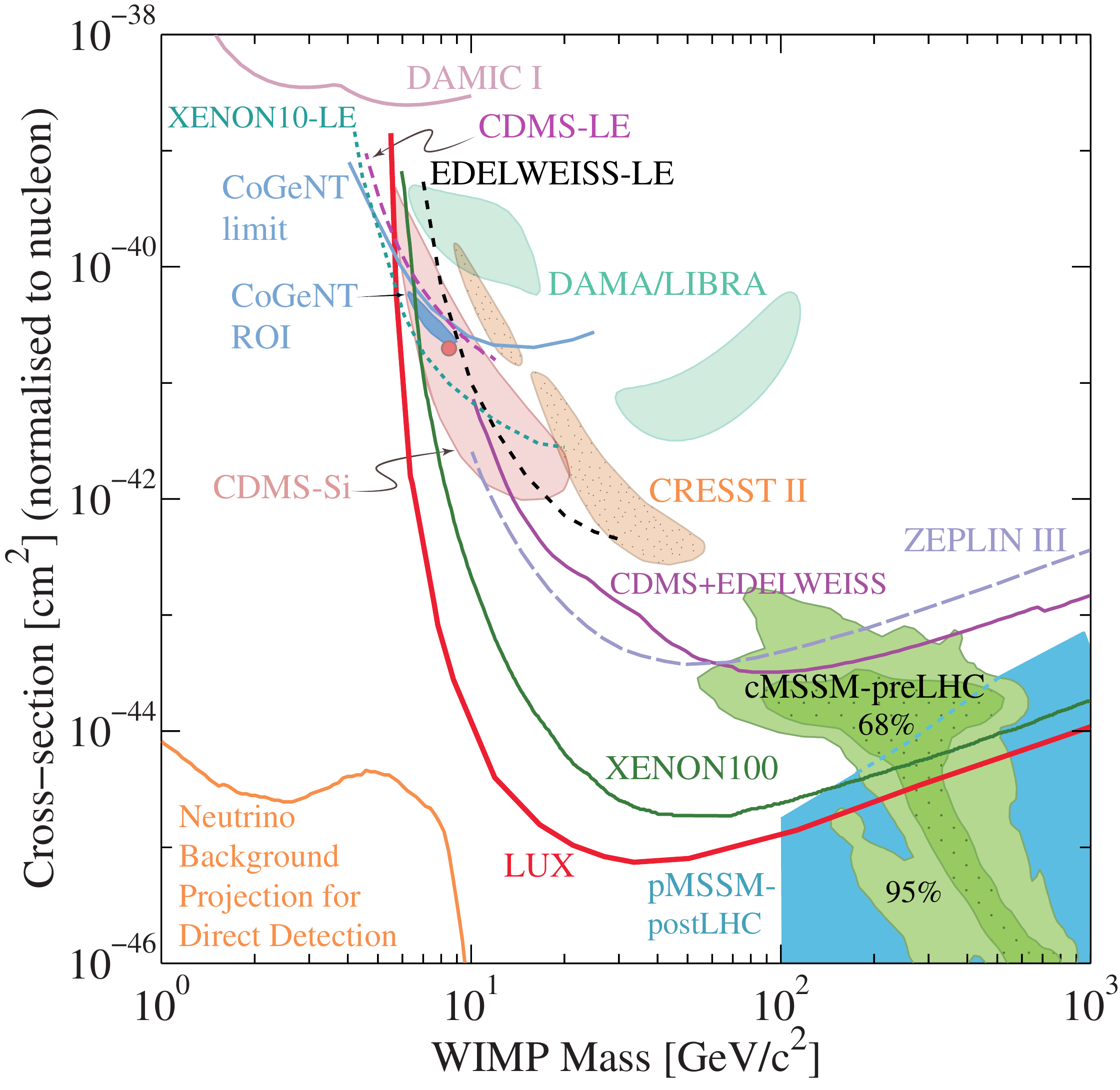} 
\end{center}
\caption{Limits on the $\chi$-$p$ spin-independent scattering cross section 
vs.the $\chi^0_1$ mass. 
The shaded regions include MSSM points compatible with recent LHC SUSY searches 
and Higgs mass results\cite{Drees:2012ji}. Also indicated is the most stringent recent 
limit from the LUX experiment\cite{Akerib:2013tjd}.
}
\label{fig:DD}
\end{figure}

%
\begin{figure}[hb!]
\begin{center}
\includegraphics[width=7.0cm]{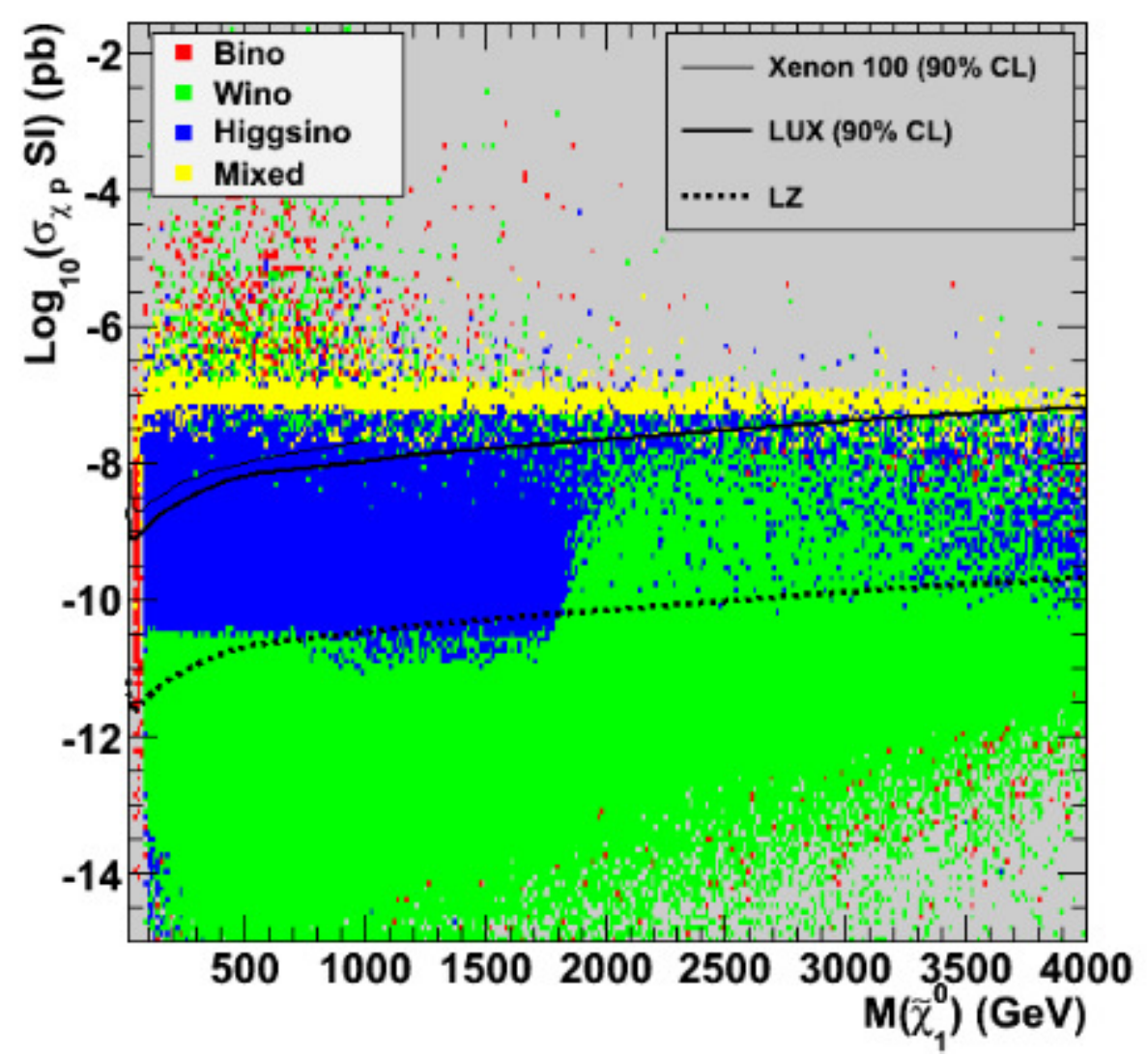} 
\end{center}
\caption{Neutralino-nucleon spin-independent scattering cross section 
vs.the $\chi^0_1$ mass. The colours indicate the nature of the neutralino LSP with the largest occurrence in each bin}
\label{fig:DD2}
\end{figure}

\begin{sloppypar}
The relic abundance constraint is now complemented by upper limits on
WIMP-nucleon scattering cross sections from underground dark
matter direct detection experiments. The $\tilde\chi p$
spin-independent scattering process receives SUSY contributions from
scalar quark exchange and t-channel Higgs exchange
\cite{Jungman:1995df}. The latter dominates over a vast region of the
parameter space. The scattering cross section retains a strong
sensitivity on the scalar Higgs boson mass and $\tan \beta$
\cite{Arbey:2011aa}.  
Limits on spin-independent 
$\chi$-nucleon scattering from the
 initial run of the LUX experiment\cite{Akerib:2013tjd} are shown in
Figure~\ref{fig:DD} along with some expected SUSY parameter space.
\end{sloppypar}

\begin{sloppypar}
There is a large number of recent results reported by experiments
using crystals \cite{Bernabei:2010mq,Angloher:2011uu}, 
semiconductors \cite{Ahmed:2010wy,Agnese:2013rvf} and noble 
gases \cite{Aprile:2012nq,Akerib:2013tjd}
as sensitive material. The excess of events reported by some of these
experiments~\cite{Bernabei:2010mq,Angloher:2011uu,Aalseth:2012if,Agnese:2013rvf}, which would appear to point to a very light WIMP,
are confronted by the stringent limits set by negative results in the
searches by the Xenon-based detectors, Xenon-100\cite{Aprile:2013doa} and
LUX\cite{Akerib:2013tjd}.  
These limits are cutting into the region of scattering
cross sections typical of the MSSM (see Fig.~\ref{fig:DD2}) and therefore
provide some meaningful bounds, even if systematics and model
dependencies due to the assumed dark matter profile in the galaxy are
known to be sizeable\cite{Bovy:2012tw}. In particular, the Xenon-100 and LUX
bounds-- if taken at face value-- exclude a sizeable fraction of the
viable SUSY points with neutralino dark matter at small values of the
$\mu$ and $M_2$ parameters, which would give chargino and neutralino
pair production observables at a linear collider with $\sqrt{s}$ below
1 TeV and small fine tuning, as discussed above. In the case where WIMPs make up only
a portion of the total dark matter abundance (perhaps the bulk is composed of axions),
then these direct detection predictions would have to be rescaled by a factor
$\xi=\Omega_\chi^{TP} h^2/0.12$ in which case the search limits are much less constraining.
\end{sloppypar}

In gravity-mediation, the gravitino mass sets the scale for the soft breaking terms 
so that one expects gravitinos to have mass comparable to the SUSY partners. 
While gravitinos may decouple from collider physics, they can be produced at 
large rates proportional to $T_R$ in the early universe. The gravitino decay
rate to SUSY particles is suppressed by $1/M_P^2$ so that they may decay well after 
BBN has started, thus upsetting the successful prediction of light element production
from Big Bang Nucleosynthesis\cite{Kawasaki:2008qe}. 
To avoid this so-called ``gravitino problem''\cite{Weinberg:1982zq},
one typically requires $T_R\alst 10^5$ GeV for $m_{3/2}<5$ TeV.  
Alternatively, if the gravitino is very heavy-- $m_{3/2}\agt 5$ TeV-- then
gravitinos typically decay before the onset of BBN.
In addition, overproduction of gravitinos may lead to overproduction 
of LSPs from gravitino decay. 
To avoid over-production of WIMPs arising from thermally-produced gravitinos, 
one must typically obey the less restrictive bound $T_R\alst 10^9$ GeV.

Besides the case of neutralino dark matter, it is possible that
gravitinos are the lightest SUSY particles, and
could be responsible for dark matter.
The case of gravitino LSPs with a weak scale value of $m_{3/2}$ is called the
super-WIMP scenario and is again highly restricted by BBN bounds on late decaying
WIMP to gravitino decays.
Also, superWIMP gravitino LSPs can be thermally overproduced as dark matter
unless constraints are again imposed 
on the reheating temperature\cite{Moroi:1993mb,Bolz:2000fu,Pradler:2006hh}. 
For weak-scale gravitino dark matter, a reheating temperature above $10^9$~GeV 
can only be achieved in small corners of model parameter space which impose
strict bounds on the superparticle mass spectrum\cite{Heisig:2013sva}. 

Alternatively, the gravitino mass might be far below the weak scale; 
this scenario is a viable option and occurs naturally in GMSB scenarios.
For such a small gravitino mass, the goldstino couplings are enhanced 
which helps to evade the BBN constraints on NLSP decay to gravitinos.
In addition, expectations for thermal overproduction of gravitino dark matter 
in GMSB are modified and can depend as well on the messenger 
mass scale\cite{Choi:1999xm,Fukushima:2013vxa}.



\subsection{Constraints from LHC}
\label{sec:susy2-4}

The searches performed by ATLAS and CMS on the 7 and 8~TeV LHC data in
channels with jets, leptons and missing transverse energy (MET) have
already significantly re-shaped our views of the high energy frontier
in relation to SUSY. Searches for signatures of production and decay
of supersymmetric particles with large MET have failed to reveal any
significant excess of events compared to SM expectations.

A variety of final states have been probed in LHC searches which are
sensitive to the production and decay modes of both strongly- and
weakly-interacting SUSY particles. The results of searches for gluinos
and squarks of the first two generations are easy to interpret in
generic models. The analyses of the almost 25~fb$^{-1}$ of combined 7~TeV and 8~TeV
have led to mass limits in the range of $m_{\tg}\agt 1-1.3$~TeV 
and $m_{\tq}\agt 0.4-1.8$~TeV for scalar quarks of the first two
generations. There is an important exception to these limits which
originates from scenarios with compressed spectra giving rise to
highly degenerate masses and correspondingly low transverse energies
from the produced jets and leptons: the visible energy from such
compressed spectra often falls below analysis cuts or even the trigger
thresholds, which causes generic LHC limits to collapse.

\begin{figure}[h!]
\begin{center}
\includegraphics[width=8.0cm]{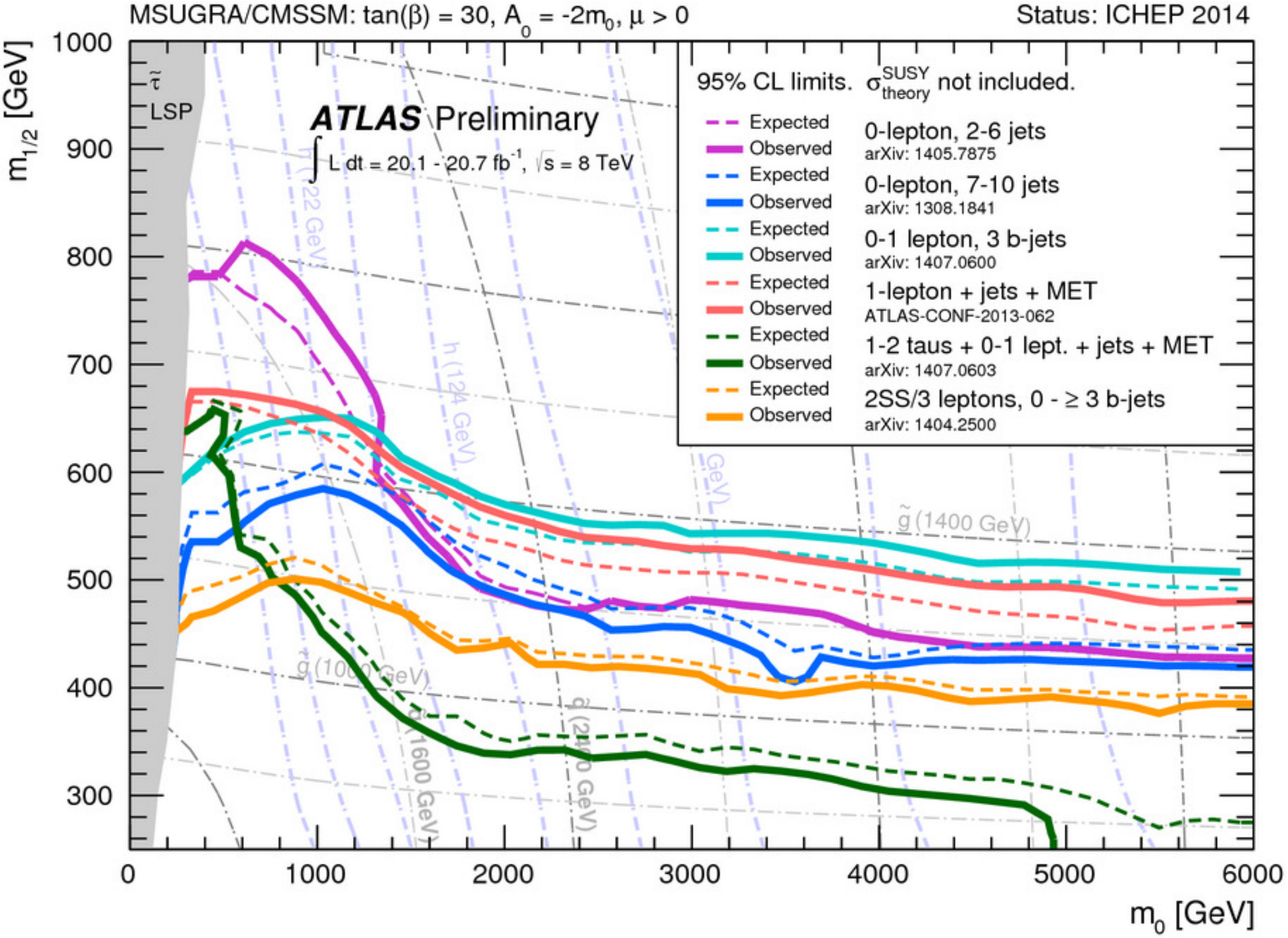}%
\end{center}
\caption{95\% CL exclusion limits for MSUGRA/CMSSM models with $\tan \beta=30$, 
$A_{0}=-2m_0$ and $\mu>0$ presented in the 
$[M_{0}, M_{1/2}]$ plane obtained by the ATLAS experiment with 20~fb$^{-1}$ of data at 8~TeV (from \cite{ATLAS-CMSSM}).}
\label{fig:ATLAS-CMSSM}
\end{figure}

These results have rapidly excluded most of the benchmark points
adopted in the last two decades of SUSY studies and have put
significant pressure on highly constrained SUSY models
such as the CMSSM/mSUGRA model (discussed above) where SUSY soft terms 
are unified at a high scale.
In fact, the LHC searches have excluded regions of parameter space which had
been clearly preferred by fits performed on the pre-LHC data, pushing
the masses of squarks and gluinos beyond 1-2~TeV
(see Figure~\ref{fig:ATLAS-CMSSM}). 
To further aggravate the crisis of such highly constrained models, 
it has also become difficult to accommodate a lightest Higgs boson with mass
$\sim$125~GeV in the CMSSM, 
except for very specific parameter values~\cite{Arbey:2011ab,ehowp}.  
In view of this, adopting more generic MSSM models without implicit correlations
between the masses of the various SUSY particles, such as the
so-called phenomenological MSSM (pMSSM), has become presently more
common for studying SUSY theories at the LHC and at linear $e^+e^-$ colliders. 

Still, the benchmark studies carried out for linear
colliders keep much of their validity with respect to the sensitivity
and accuracy of the measurements, even if the underlying models used
in those studied have already been excluded by the LHC data.

\begin{sloppypar}
Contrary to the case of constrained models, the mass limits for
strongly interacting sparticles (in particular the gluino $\tilde g$
and the scalar quarks of the first two generations $\tilde q$) have
little impact on the mass scale of their weakly-interacting
counterparts (charginos, neutralinos and scalar leptons) in generic
models of Supersymmetry, such as the pMSSM
\cite{Conley:2010du,Conley:2011nn,Sekmen:2011cz,Arbey:2011un}.
Searches for weakly-interacting SUSY particle partners at LHC, of
which the first results have recently been reported, are more
model-dependent than the case of gluino and squark searches, since
they depend not only on the mass splitting with respect to the
lightest neutralino, but also on the mass hierarchy of the neutralinos
and sleptons, as well as on the neutralino mixing matrix: {\it e.g.}
the neutralino decay channels which yield multiple lepton final states
used as experimental signatures include $\tilde{\chi}_2^0 \to
\tilde{\ell} \ell$, $Z\tilde{\chi}_1^0$ or $\ell^+
\ell^-\tilde{\chi}_1^0$.
\begin{figure}[h!]
\begin{center}
\begin{tabular}{c}
\includegraphics[width=6.0cm]{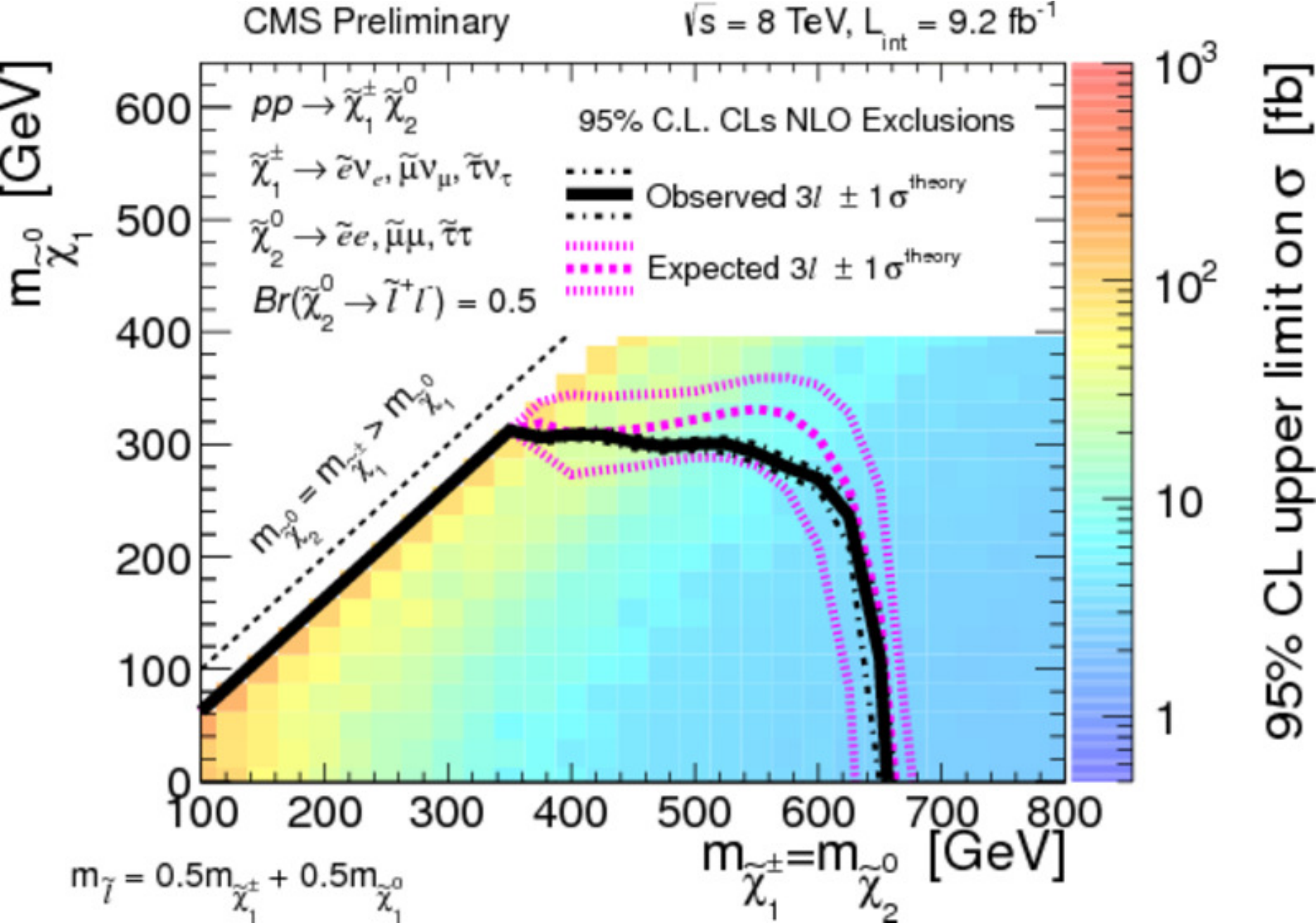} \\
\includegraphics[width=6.0cm]{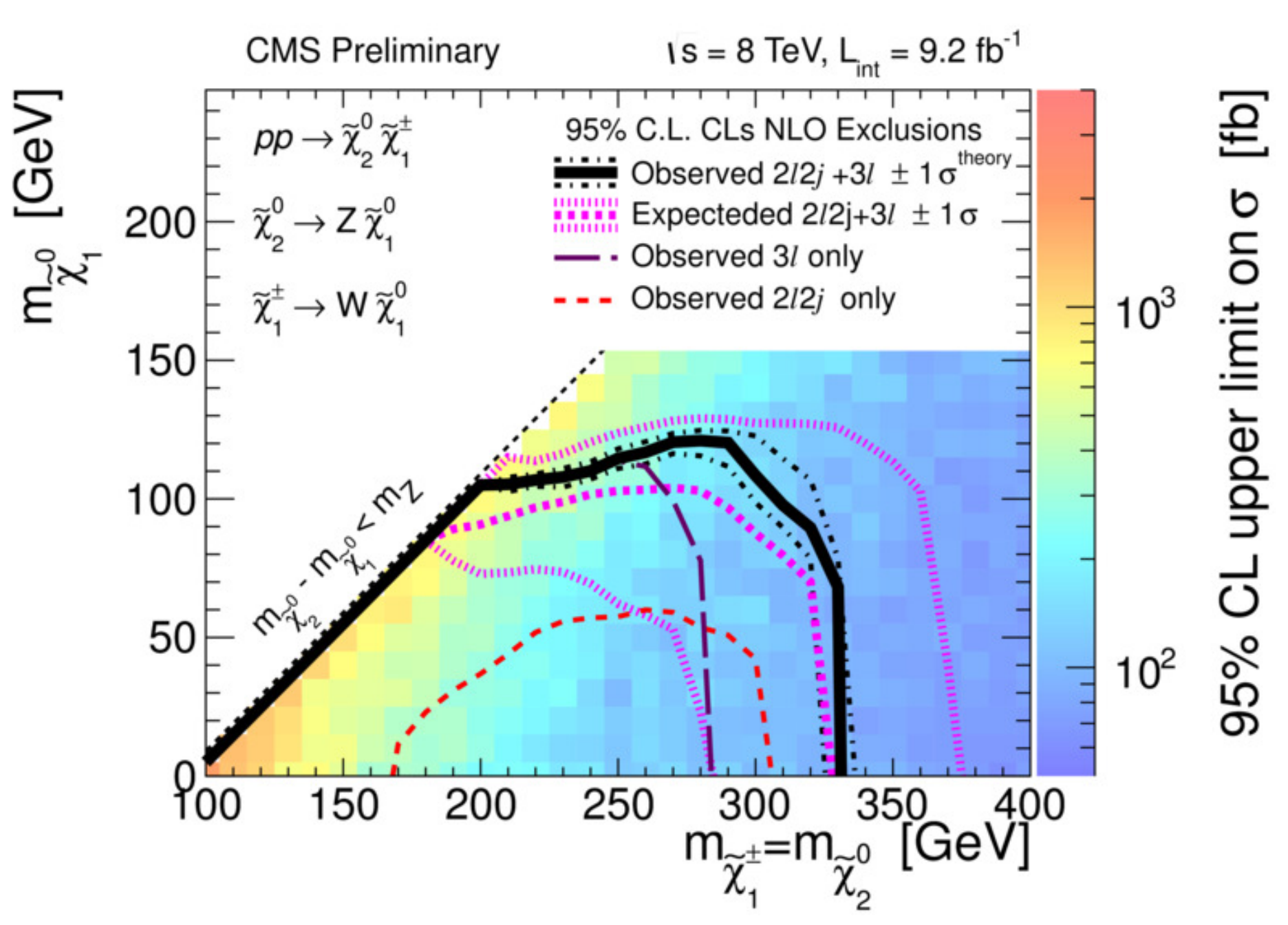} \\
\end{tabular}
\end{center}
\caption{95\% CL exclusion limits on the chargino-neutralino production NLO cross section times branching fraction in the 
flavor-democratic scenario, for the three-lepton (upper panel), 
di-lepton $WZ$+MET and tri-lepton (lower panel) CMS searches with 
9.2~fb$^{-1}$ of data at 8~TeV (from \cite{CMS-chi}).}
\label{fig:CMS-chi}
\end{figure}
These searches are probing charginos and neutralinos of mass up to
$\sim$300--650~GeV, under these specific conditions (see
Figure~\ref{fig:CMS-chi}).  Extensive scans of the pMSSM have shown
that significant regions of parameters giving rise to relatively light
weakly interacting SUSY particles still remain unexplored and will not
be probed even after the first operation of the LHC at its design
energy of 14~TeV~\cite{Conley:2010du,Conley:2011nn,Arbey:2011un}.
\end{sloppypar} 

There are regions in SUSY parameter space that are not well covered by 
the searches for missing energy and require more exotic search strategies.
One example are scenarios where an electrically or colour-charged NLSP 
becomes long-lived on collider time-scales. This situation occurs either through
strongly suppressed couplings of the LSP or through kinematic suppression.
The former case naturally occurs in GMSB models where the lighter stau 
often is the NLSP\@. The clean signature of the resulting highly ionising
charged tracks at the LHC typically lead to stronger limits on sparticle masses
in such a model \cite{Heisig:2012zq,CMS1305.0491}. The latter case occurs, 
e.g., in scenarios with a wino- or higgsino-like neutralino LSP being almost 
mass degenerate to the lightest chargino.
Another example of exotic SUSY signatures are models with $R$-parity
violating couplings.

\begin{sloppypar}
The recent observation of a Higgs-like particle with mass $\simeq$
125~GeV at the LHC is opening new perspectives for SUSY searches at
colliders. The mass of the newly discovered particle sets some
non-trivial constraints on the SUSY parameters.  In particular, the
relatively large mass value observed implies strong restrictions on
the scalar top mass and the mixing in the top
sector~\cite{Baer:2011ab,Arbey:2011ab}.  Heavy scalar top quarks
and/or large mixing are required to bring the $h$ boson mass around
125~GeV.  The first measurement of the yields (or signal strengths $\mu$) 
in the decay channels studied so far-- including $\gamma
\gamma$, $ZZ^*$ and $WW^*$-- (although limited in accuracy and only at
the level of upper limits in the important $bb$ and $\tau \tau$
channels) will add further constraints.  In particular, if interpreted
within the SUSY framework, the data point towards a decoupling
scenario, with a relatively heavy $A$ boson. A possible enhancement in
the $\gamma \gamma$ channel, observed by ATLAS 
and recently confirmed by the updated ATLAS study with 13~fb$^{-1}$ of 8~TeV data,
may be a first hint of deviation from the SM expectations and could be
explained through a reduction of the $b \bar b$ width as an effect of
SUSY particle loops with intermediate, positive values of $\mu \tan
\beta$~\cite{Arbey:2012dq,Arbey:2012bp}, or the contribution of light
staus~\cite{Carena:2012gp,Carena:2012xa,Giudice:2012pf} or
charginos~\cite{Arbey:2012dq}.  Several of the preferred scenarios
complying with $M_h \simeq$ 125~GeV and low values of the fine tuning
parameter have sbottom particles lighter than the stops with multiple
decay modes with comparable rates~\cite{CahillRowley:2012kx}.  This
allows them to evade in part the constraints from direct LHC searches
which assume a single dominant decay channel.
\end{sloppypar}

One of the indirect probes on the scale of SUSY particles is fine
tuning.  The gradual exclusion of SUSY particles at lower masses as a
consequence of LHC searches naively affects the value of the fine tuning
parameter, $\Delta$, for the surviving SUSY models. It has been noted
that in generic MSSM models, fine tuning is mostly determined by the
$\mu$ parameter and an acceptably low fine tuning corresponds to small
to moderate value of $|\mu|$.  If fine tuning is taken as a criterion
to select MSSM scenarios compatible with the 125~GeV Higgs mass,
(setting $\Delta < 100$ as has been proposed~\cite{CahillRowley:2012rv}
\footnote{In this study, the authors implement the BG fine-tuning measure
applied to 19 uncorrelated parameters in the pMSSM which is assumed 
valid up to a scale $\Lambda \simeq 20$ TeV. The $\Lambda =20$ TeV scale 
induces an additional factor of 3 in the fine tuning evaluation.}) 
a constraint on the mass scale of
weakly-interacting sparticles is implicitly derived with values of
$m_{\chi^{\pm}_1}\le$ 270~GeV.  This would match particularly well
with the reach of a linear $e^+e^-$ collider with $\sqrt{s}$ energy in
the range 0.5-1.0~TeV.

In summary, despite the far reaching constraints derived by the direct
searches for SUSY production at the LHC, specific classes of models
exist in the general MSSM and in constrained models such as NUHM2, 
which are consistent with the current bounds and have SUSY particles 
within reach of an $e^+ e^-$ collider operating at $\sqrt{s}\sim 0.25-0.5$~TeV and above.  
A recent study showed that over 20\% of the
viable pMSSM models, not yet excluded by the combined LHC searches at
7 and 8~TeV, have the lightest chargino, $\chi^{\pm}_1$, accessible at
$\sqrt{s}$ = 0.5~TeV increasing to 58\% for $\sqrt{s}$ = 1~TeV and
94\% for 2~TeV~\cite{CahillRowley:2012kx}. 
In addition, a study of natural SUSY NUHM2 parameter space in the $\mu$ vs. $m_{1/2}$
parameter plane shows the LHC8 and LHC14 reach (assuming 300 fb$^{-1}$) which will
cover only a portion of the $\Delta_{EW}<30$ favored parameter space. 
However, a $\sqrt{s}=0.5-0.6$ TeV $e^+e^-$ collider would access the entire
low $\Delta_{EW}$ parameter space, thus either discovering light higgsinos or ruling out
natural SUSY: see Fig. \ref{fig:muvsmhf}.
\begin{figure}[h!]
\begin{center}
\begin{tabular}{c}
\includegraphics[width=7.0cm]{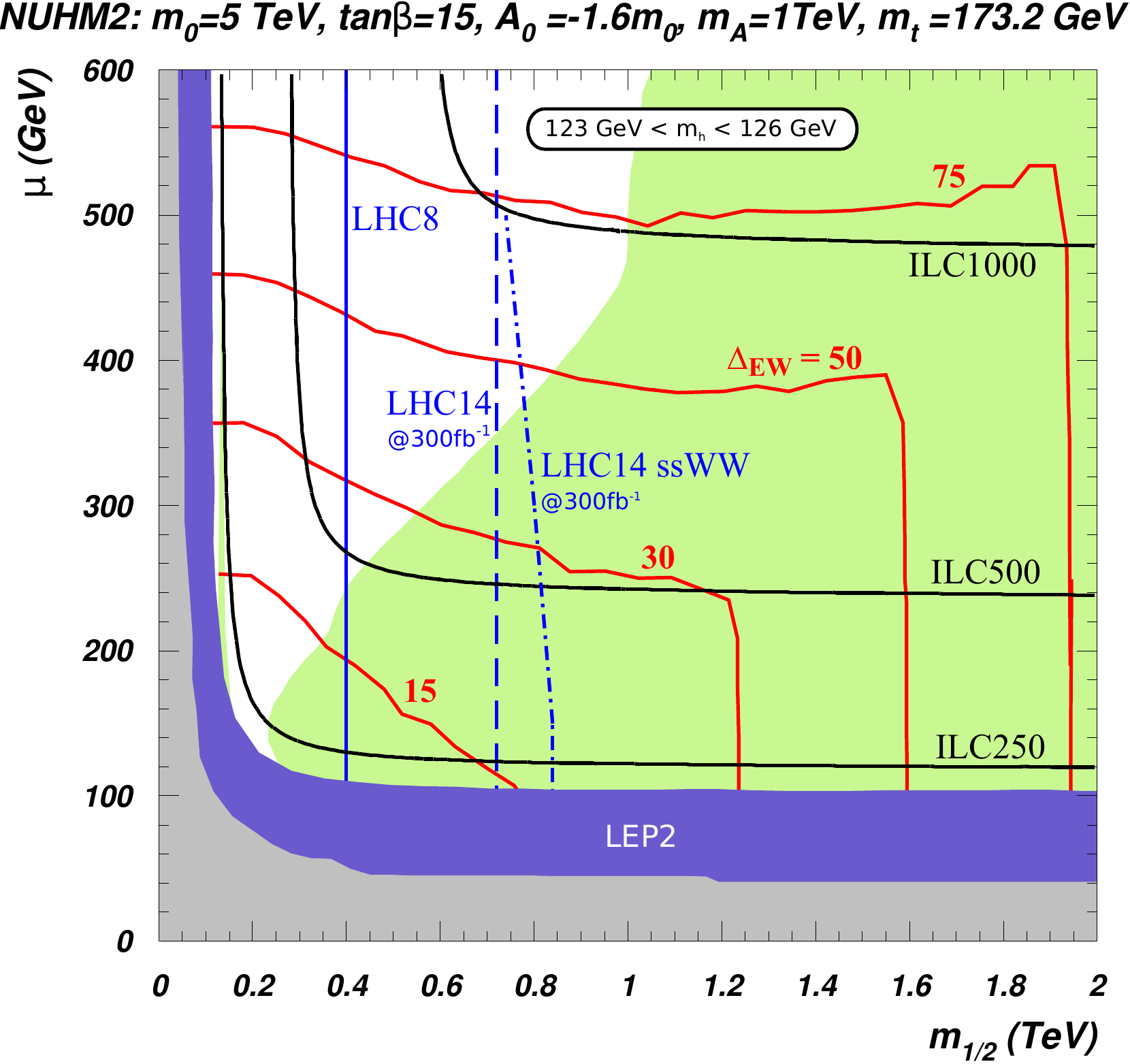}
\end{tabular}
\end{center}
\caption{
Plot of $\Delta_{EW}$ contours in the
$m_{1/2}\ vs.\ \mu$ plane of NUHM2 model for $A_0=-1.6 m_0$ and $m_0=5$ TeV
and $\tan\beta =15$.
We also show the region accesses by LHC8 gluino pair searches, and the region
accessible to LHC14 searches with 300 fb$^{-1}$ of integrated luminosity.
We also show the reach of various ILC machines for higgsino pair production.
The green-shaded region has $\Omega_{\tz_1}^{std}h^2<0.12$.
Figure from \cite{Baer:2013xua}.
}
\label{fig:muvsmhf}
\end{figure}
These considerations highlight the role of a high-energy $e^+ e^-$ collider as a
complementary discovery machine compared to the LHC.

\subsection{Linear Collider Capabilities}
\label{sec:susy4}

As mentioned earlier, a linear $e^+e^-$ collider operating with 
$\sqrt{s}\agt 2m(sparticle)$ can serve as a discovery machine, not only in 
models like natural SUSY, but also in dark matter motivated cases such as
the stau co-annihilation region or in $R$-parity violating models where the
LSP decays hadronically so that the SUSY signal is buried beneath QCD multijet backgrounds
at the LHC.

\begin{sloppypar}
Since SUSY is expected to (more than) double the number of physical
particles over a possibly wide mass spectrum, an $e^+e^-$ collider
with (1) a broad energy range, (2) the capability to precisely tune
its $\sqrt{s}$ energy at well-defined values corresponding to new
particle production thresholds, (3) the added analysing power afforded
by beam polarisation and (4) possibly different beam species ($\gamma
\gamma$, $e^- e^-$) appears ideally suited for a program of detailed,
high precision studies.  The cross sections for pair production of
SUSY particles are in the range 0.1-30 fb for masses of 200, 400 and
1200~GeV at $\sqrt{s}$ = 0.5, 1 and 3~TeV, respectively. For
comparison, those for the two SM processes $e^+e^- \rightarrow W^+ W^-
\nu \bar \nu$ and $e^+e^- \rightarrow \mu^+ \mu^- \nu \bar \nu$--
which are the irreducible backgrounds to chargino and smuon pairs
production-- are 2, 10 and 25 fb and 25, 35 and 45 fb, respectively,
at the same collision energies. These cross sections ensure a
favourable signal-to-background ratio after appropriate selection cuts
and make the study of SUSY particle pair production at a linear
collider extremely promising!  
\end{sloppypar}

Typical values of sparticle production cross sections are shown as a
function of the collider energy, $\sqrt{s}$ in
Fig. \ref{fig:sigma_rns}.  If the fine tuning and naturalness
arguments summarized in the previous section are taken as guidance, it
is possible to identify scenarios where LHC searches may cover only a
part of the parameter space, while a $\sqrt{s}=0.5-0.8$~TeV $e^+e^-$ collider 
would access the entire parameter space corresponding to low
$\Delta_{EW}$ values.
 These considerations highlight the possible role of 
a linear $e^+e^-$ collider as a SUSY discovery machine, complementary to the LHC.

\begin{figure}[h!]
\begin{center}
\begin{tabular}{c}
\includegraphics[width=8.0cm]{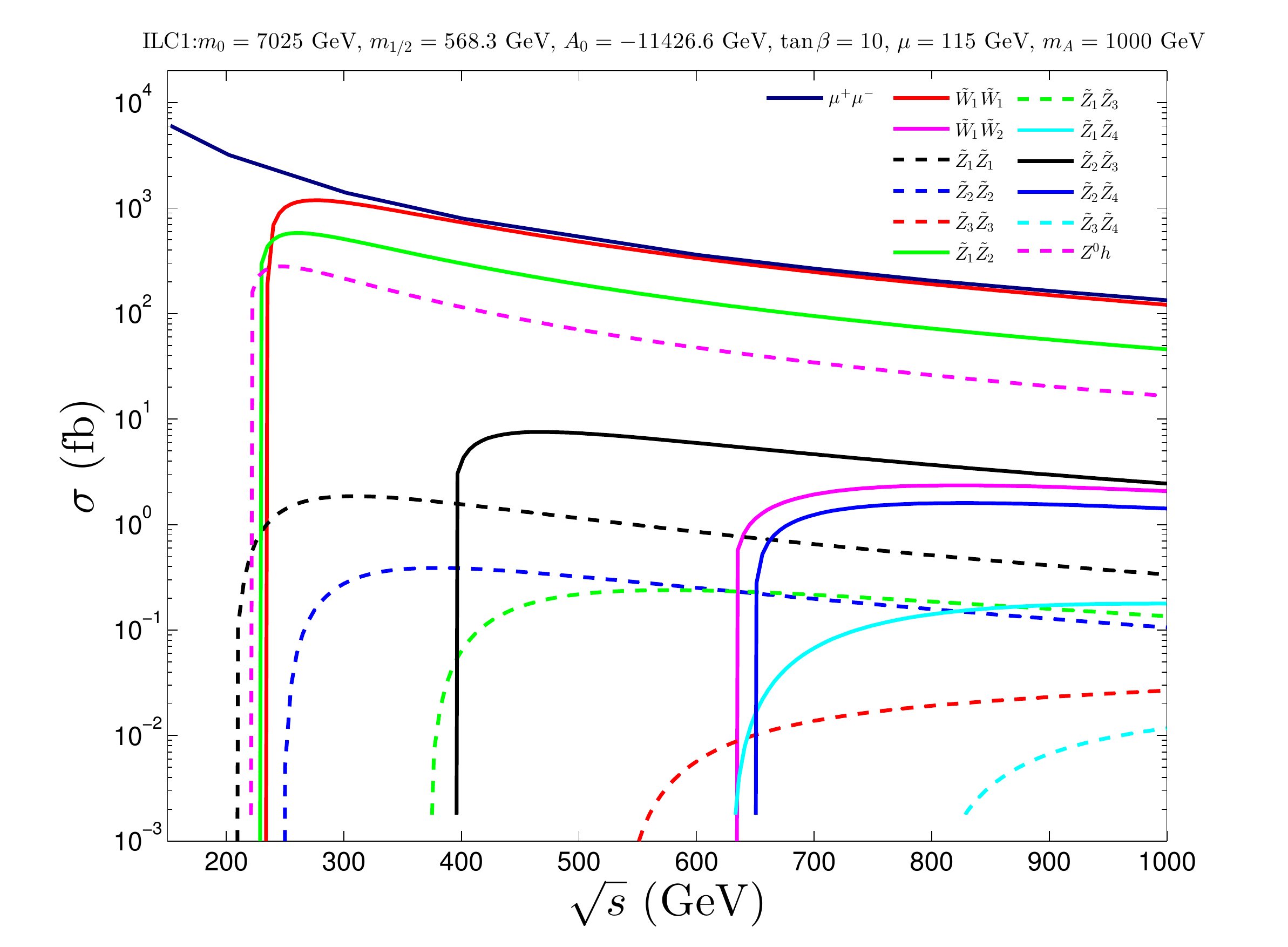}
\end{tabular}
\end{center}
\caption{
Sparticle production cross sections vs. $\sqrt{s}$
at a Higgsino factory for a radiatively-driven natural SUSY 
benchmark point\cite{Baer:2014yta}.
}
\label{fig:sigma_rns}
\end{figure}

If SUSY exists, one of the major undertakings of collider physics is
the precise determination of the quantum numbers and decay properties
of the SUSY particle partners. At a linear collider, the masses of
SUSY particles can be determined either by the end points of the
energy distribution of the visible SM particle emitted in two-body
decays (or even 3-body decays) or-- more precisely but more demanding
for the accelerator design and tuning-- by dedicated energy scans at
the onset of the pair production process.  For typical SUSY spectra--
having particles spaced from tens to hundreds of GeV-- threshold scans
set specific requirements on the accelerator design implying the
flexibility to deliver collisions at several $\sqrt{s}$ energies with
comparable luminosity and within the operating plan.

The capability of a linear collider in the study of SUSY has been
studied for the last twenty years with increasing realism from the
adoption of detailed simulation and reconstruction. New techniques for
the optimal reconstruction of physics observables, such as the parton
energy or the jet flavour, have been developed and new detector concepts
and sensor technologies, tailored to the requirements of the linear
collider physics program have been introduced and demonstrated under
realistic operating conditions. Supersymmetry has played an important
role in setting these requirements and shaping the detector
concepts. The recent studies for the ILC Letters of Intent
(LoI)~\cite{Abe:2010aa,Aihara:2009ad} and also the CLIC Conceptual
Design report (CDR)~\cite{clic-cdr} have adopted full Geant-4~\cite{g4}
based simulation and detailed reconstruction, accounting for machine
induced backgrounds. In most cases, the SUSY signatures can be clearly
discriminated from the SM processes.  Inclusive SUSY production often
appears to be the major source of background for specific processes.  In
fact, different SUSY cascade decay chains\cite{Baer:1988kx} may lead to
the same final states.  The ability to fully reconstruct the events with
excellent energy resolution and to suppress some processes by changing
the beam energy and, possibly, the beam polarisation offer excellent
tools for ensuring an efficient study of each individual channel of
interest.  For example, the interference of the contribution of
$\tilde{\chi}^+_1 \tilde{\chi}^-_1$ decays with $\tilde{e}^+_L
\tilde{e}^-_L$ in $WW$ + missing energy and $\tilde{\chi}^0_2
\tilde{\chi}^0_2$ decays with $\tilde{\nu} \tilde{\nu}$ in $hh$ +
missing energy is studied in detail with full simulation in
\cite{Alster:2011he} and the separation of neutral and charged sleptons
of the first/second generation in \cite{Freitas:2004re}.  Another
important source of background is due to two-photon events, which may
obscure the production of sfermion pairs, in particular in scenarios
with small mass splitting. This background source can be controlled by
ensuring electron tagging capability in the detector down to very small
angles~\cite{Bambade:2004tq}.

\subsubsection{Particle Property Measurements}

{\it Mass Measurements}\\[.2em]
a) { In the continuum}\\[-.8em]
\label{sec:susy4-1}

\begin{sloppypar}
\noindent The precise and unambiguous determination of SUSY particle masses is
essential for the reconstruction of the theory fundamental parameters
and for determining that the nature of the new physics is indeed
supersymmetric.  Mass reconstruction can be performed at an $e^+e^-$
linear collider by the reconstruction of the kinematics in SUSY
particle pair production and by threshold energy scans.  Threshold
scans also provide us with access to the particle width, which is
important since the narrow width approximation largely used in the
context of the SM fails in general theories of new
physics~\cite{Berdine:2007uv}.
\end{sloppypar}

\begin{sloppypar}
In the two-body decay process $\tilde{A} \to B \tilde{C}$ of a SUSY
particle $\tilde{A}$ into a lighter sparticle $\tilde{C}$ and a SM
particle $B$, the masses of the parent and daughter sparticle can be
extracted from the position of the kinematic edges of the energy
spectrum of $B$ since $\tilde{A}$ is produced with fixed, known energy
in the pair production $e^+e^- \rightarrow \tilde{A} \tilde{A}$.  The
technique was first proposed in \cite{Tsukamoto:1993gt} for two-body
decays of sleptons and charginos, for squarks in \cite{Feng:1993sd}
and three-body and cascade decays in \cite{Baer:1996vd} and later
extended to other two-body decays~\cite{Martyn:1999tc}.  In the case
of neutralino and chargino decays into bosons, where the daughter mass
$M_B$ cannot be neglected (as in the case of squark and slepton
decays), the relation between the energy endpoints and the masses of
the particle involved in the decay process are given by:
\begin{eqnarray}
\mathrm {E_{BH,BL}}= \gamma \left( E_B^{*} \pm \beta E_B^{*} \right)
\label{formula:eleh}
\end{eqnarray}
where
\begin{eqnarray}
E_B^{*} = \frac{M_A^2 + M_B^2 - M_C^2}{2 M_A}, \\
\mbox{\rm with}\quad \gamma = \frac{\sqrt{s}}{2 M_A},\quad\mbox{\rm and}\quad
\beta = \sqrt{\frac{1 - 4 M_A^2}{s}} .
\label{formula:eleh1}
\end{eqnarray}
These formulae can be extended in a straightforward way to the case in
which the particle $\tilde A$ is not directly produced in the $e^+e^-$
collisions but originates from the decay of a heavier particle,
$\tilde{A}^{\prime}$, by replacing $s$ with $E_A^2$, where $E_A$ is its
energy. In the case of cascading decays $\tilde{A}^{\prime} \to
\tilde{A} B^{\prime} \to B \tilde{C}$, $E_A$ is obtained as
$\sqrt{s}-E_{B^{\prime}H} < E_A < \sqrt{s}-E_{B^{\prime}L}$.
\end{sloppypar}

The determination of the lower and upper endpoints of the energy
spectrum constrains the ratio of the mass of $\tilde A$ to that of
$\tilde C$. If the mass of $\tilde C$-- in most cases the lightest
neutralino-- is independently known, then $M_{\tilde A}$ can be
extracted. The accuracy in the extraction of the masses by the endpoint
technique depends on the resolution in determining $E_B$, which may be
the resolution in measuring the momentum of a lepton in the case of
sleptons or the energy of a jet (di-jet) in the case of a scalar quark
(chargino or neutralino decaying into a boson).  Excellent energy and
momentum resolution are therefore essential. The energy of the beams at
collision must also be known accurately because this enters in the
determination of $\beta$. Beam-beam effects which induce radiation off
the beam particles before collision are responsible for distortions of
the luminosity spectrum, which must be precisely measured from collision
data.

Detailed analyses, based on full Geant-4 detector simulation,
digitisation and reconstruction and including the inclusive SM
backgrounds, have validated earlier results on the expected accuracy on
the mass determination for sleptons, gaugino and squarks at $\sqrt{s}$ =
0.5 and 3~TeV. Studies for the ILD and SiD LoIs performed for the ILC
parameters at 0.5~TeV~\cite{Abe:2010aa,Aihara:2009ad}, have shown that
the kinematic endpoints of the energy spectrum of $W$ and $Z$ bosons
produced in decays of chargino and neutralinos (see
Figure~\ref{fig:chi-mass}), respectively, can be determined with an
accuracy of better than 1~GeV, thanks to the excellent performance of
energy flow with highly segmented calorimeters in the reconstruction of
parton energy~\cite{Thomson:2009rp}. Kinematic fitting imposing equal
masses of pair produced particle can be applied to improve the energy
resolution. This translates into relative statistical accuracies in the
determination of the $\tilde \chi^{\pm}_1$, $\tilde \chi^0_2$ and
$\tilde \chi^0_1$ masses of 1\%, 0.5\% and 0.7\%, respectively.  These
results confirm, with the realism of full simulation and reconstruction
and full SM backgrounds, the findings of earlier studies indicating that
the masses of gaugino could be measure to a relative statistical
accuracy of $\sim$1\%.
\begin{figure}[htb]
\begin{center}
\includegraphics[width=4.0cm]{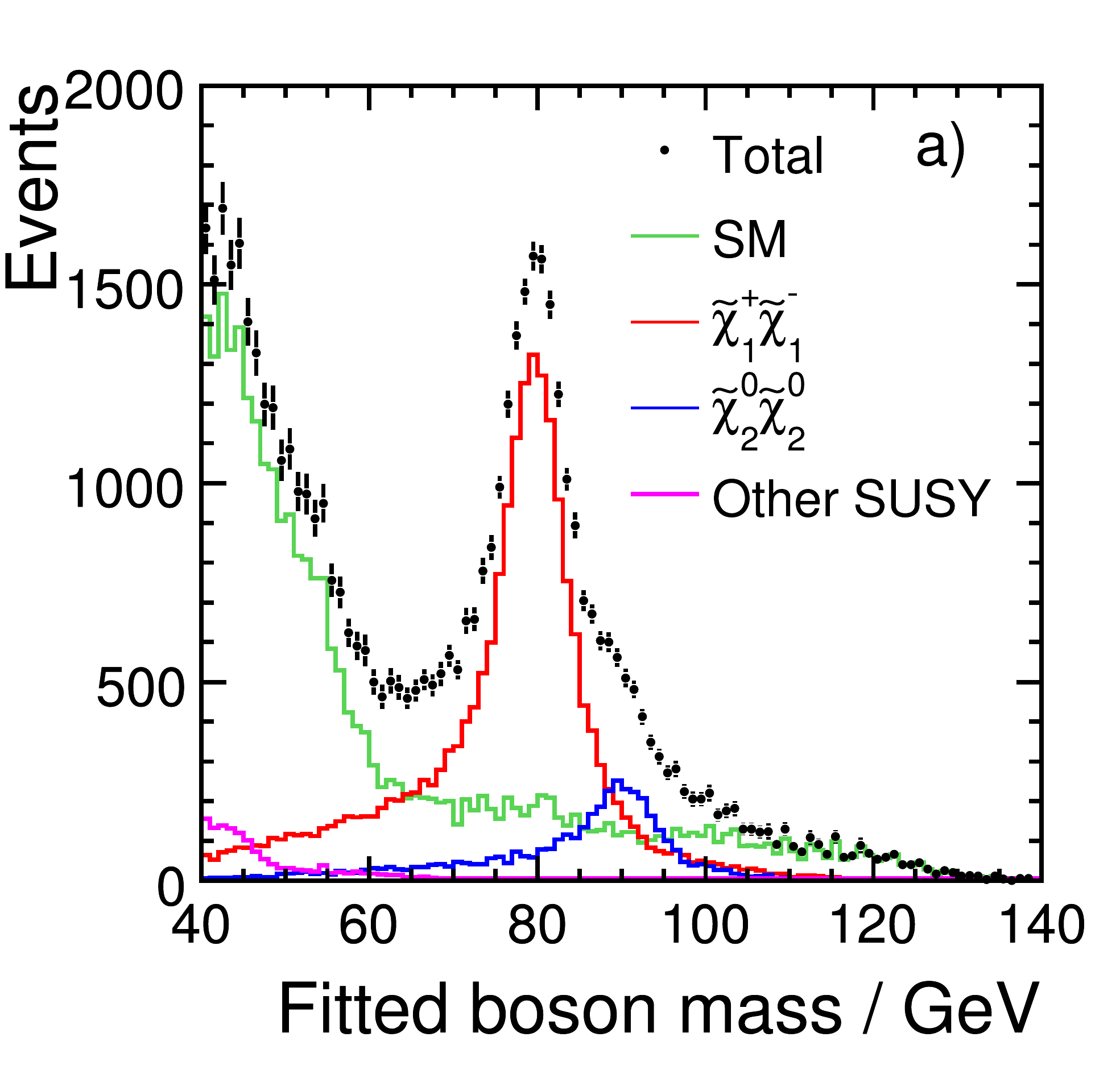}
\includegraphics[width=4.0cm]{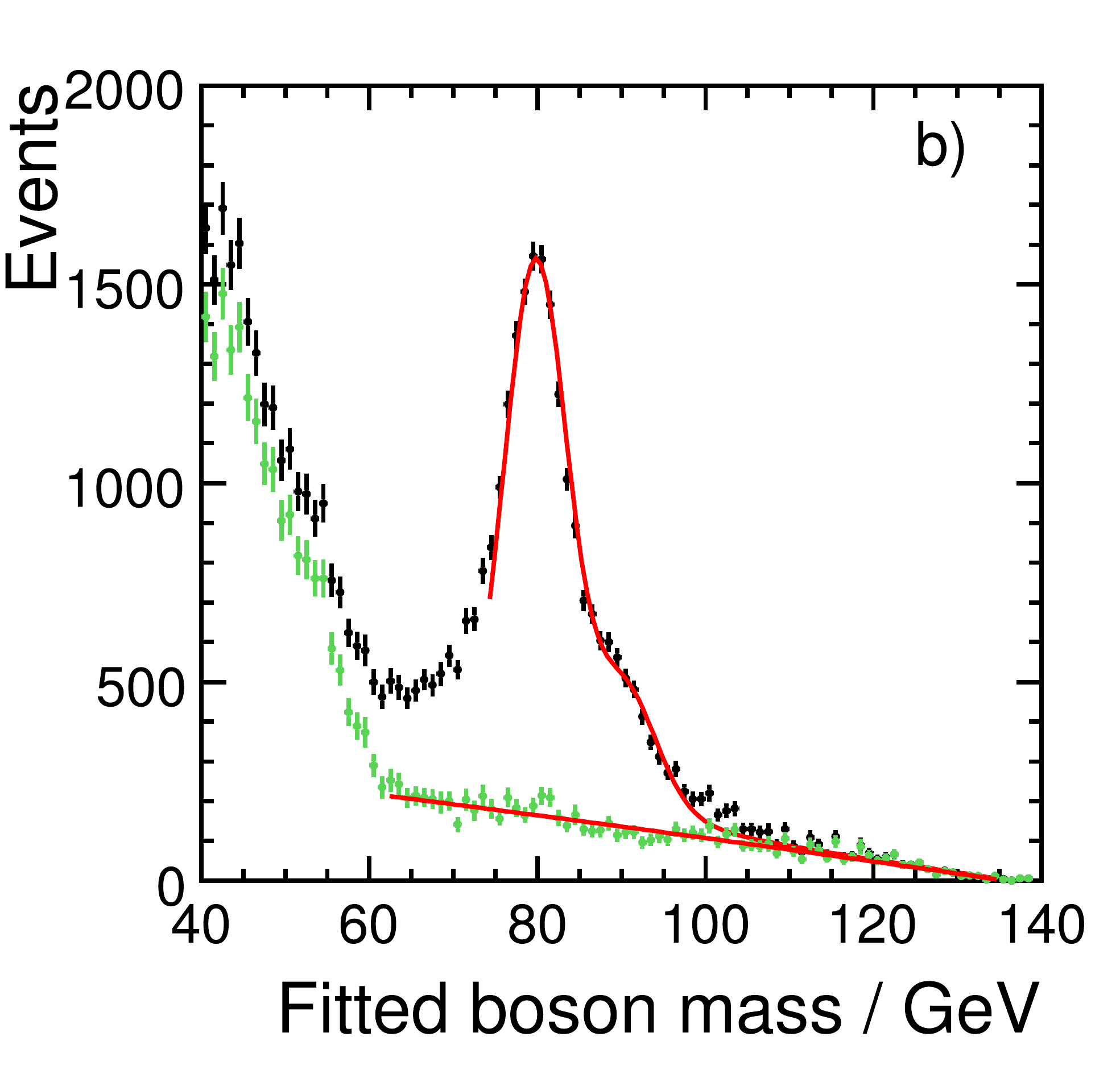}
\includegraphics[width=4.0cm]{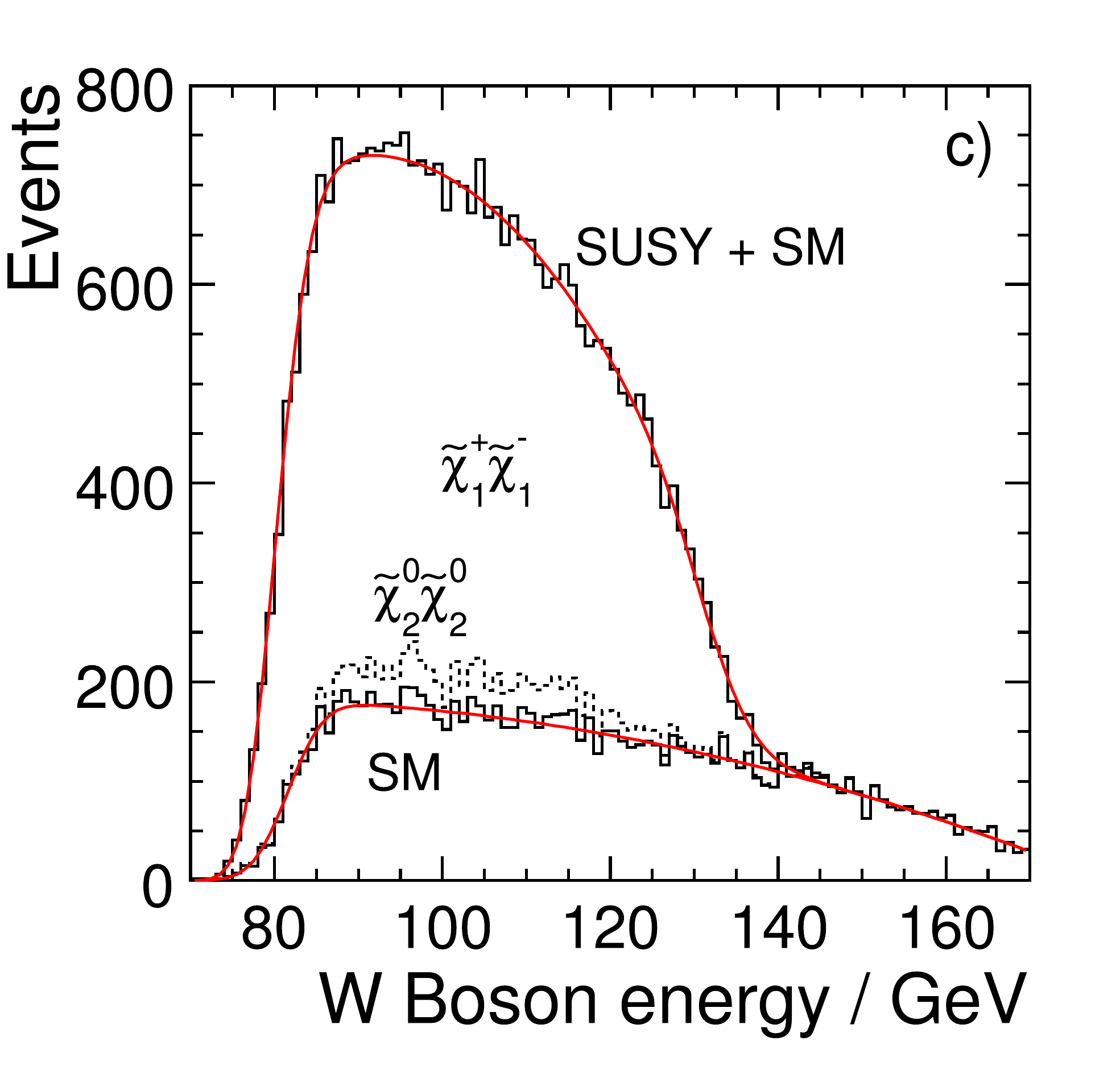}
\includegraphics[width=4.0cm]{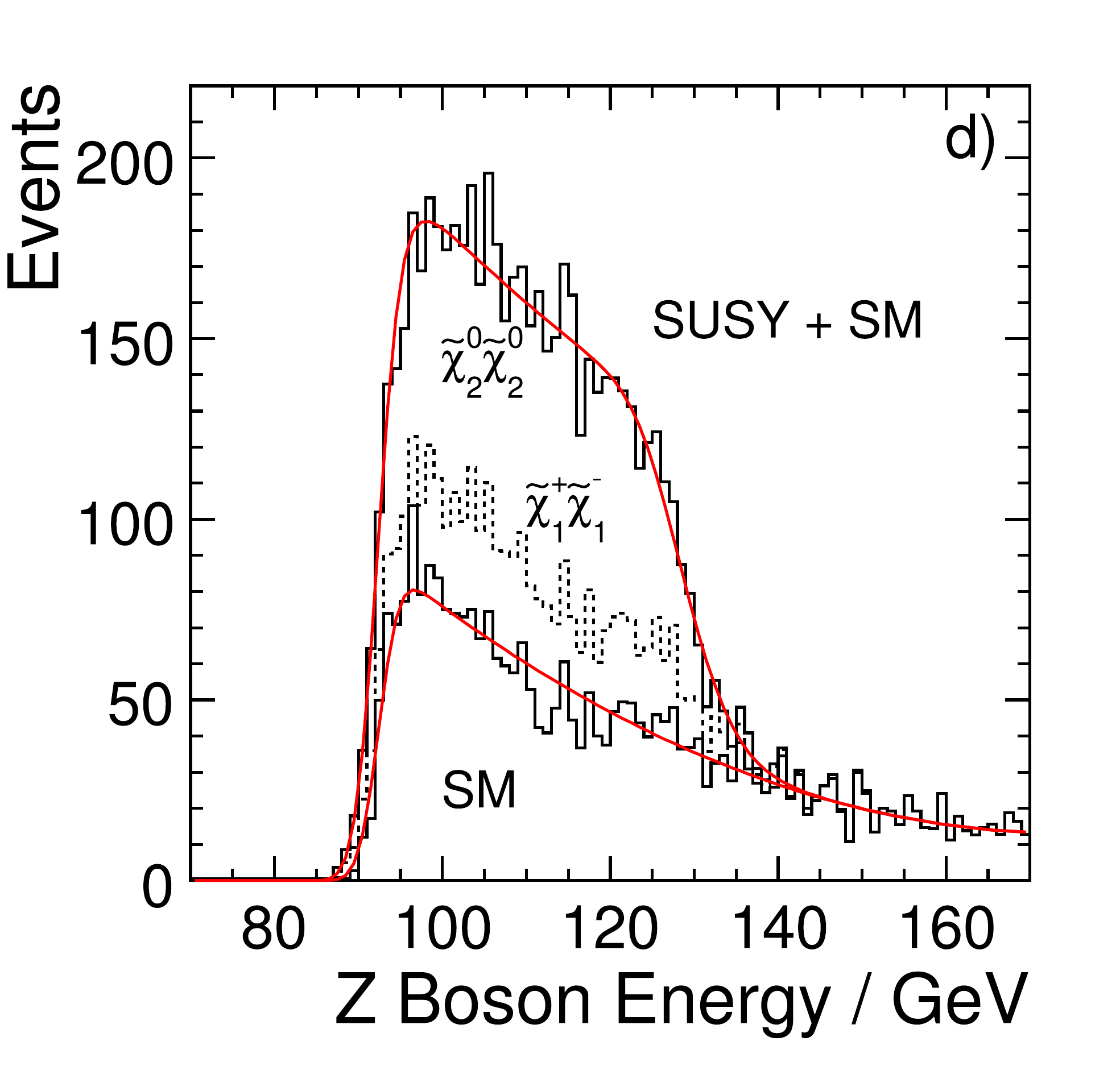}
\caption{Di-jet mass (upper plots) and energy spectra (lower plots) for chargino and neutralino production at 0.5~TeV 
(from~\cite{Abe:2010aa}).}
\label{fig:chi-mass}
\end{center}
\end{figure}

The excellent momentum resolution, required by the study of the
Higgs-strahlung process, implies that the accuracy on the mass
determination is dominated by the beamstrahlung effects. Not only the
dominant modes, such as $\tilde \mu^+_R \tilde \mu^-_R \rightarrow \mu^+
\mu^- \tilde \chi^0_1 \tilde \chi^0_1$, but also the sub-dominant
process $\tilde \mu^+_L \tilde \mu^-_L \rightarrow \mu^+ \mu^- \tilde
\chi^0_1 \tilde \chi^0_1$ can be studied in the 2-lepton + missing
energy final state.  Scalar $\tilde t$ and $\tilde b$ quarks can be
observed almost up to the kinematical threshold for the pair production
process even in the case of small mass splitting with the
$\tilde{\chi}^0_1$ with the signal cross section measured with a
statistical accuracy of $\sim$15\% for the case of the
$\tilde{b}$~\cite{Aihara:2009ad}. These scenarios at small mass
splitting are of special relevance in relation to the dark matter relic
density since stop or sbottom co-annihilation may be responsible for
reducing $\Omega_{\chi} h^2$ to values compatible with the WMAP results
and are very difficult for LHC searches. In addition, an $e^+e^-$
collider of sufficient energy to produce scalar top pairs can determine
the stop mixing angles through a study of the $e_L^+e_R^- \rightarrow
\tilde{t}_1 \tilde{t}_1$ and $e_R^+e_L^- \rightarrow \tilde{t}_1
\tilde{t}_1$ production with polarised beams along with study of the decays into
multiple channels with comparable rate: such cases are difficult, if not
impossible, at the LHC.

\begin{sloppypar}
Much of the accuracy demonstrated by the detailed ILC studies at
0.5~TeV is preserved at multi-TeV energies, as confirmed by some of
the studies carried out for the CLIC CDR~\cite{clic-cdr}, which
focused on 3~TeV $e^+e^-$ collisions.  Chargino and neutralino masses
in the range 600 - 1000~GeV can be determined with a relative
statistical accuracy of 1 - 2\% with unpolarised beams and 2~ab$^{-1}$
of data~\cite{Alster:2011he,clic-cdr}. The mass of $\tilde \mu_R$ of
1.1~TeV is again determined to $\sim$2\% with unpolarised beams and
1\% with polarised electrons and positrons, accounting for
backgrounds~\cite{Blaising:2012vd,clic-cdr}.  In addition to the
weakly interacting SUSY particles, multi-TeV collisions may access
scalar quark pair production, providing unique accuracy on their
masses. In the case of a 1.1~TeV right-handed squark of the first
generation a detailed study performed for 2~ab$^{-1}$ of integrated
luminosity at $\sqrt{s}$ = 3~TeV demonstrated a relative statistical
accuracy on the mass of 0.5\%~\cite{Simon:2012fj}. The linear collider
opportunities for precision study of SUSY particles extend to
three-body decays \cite{Chen:2011cya} of gauginos
\cite{Baer:2003ru,Baer:2004zk}, sleptons \cite{Kraml:2007sx} and
scalar quarks \cite{Jimbo:2012qw}, which are more difficult for the
LHC. In the study of these processes, SUSY becomes a possible
background to the searches where different production and decay
channels lead to the same final state or topology. In these cases,
special attention must be paid to the use of tight cuts on
discriminants based on neural networks or multivariate techniques
which may induce strong biases on the kinematics and configuration of
the selected events.\\[.2em]
\end{sloppypar}

\noindent {b) At the threshold}\\[-.8em]

\begin{sloppypar}
\noindent An $e^+e^-$ linear collider with tunable beam energy
can determine the sparticle masses by performing energy 
scans of their pair production cross section near
threshold. In principle, this method often provides a better
mass accuracy compared to the kinematic end-point
method discussed above, and also, in most cases, a
constraint on the particle width. Threshold energy
scans put significant requirements on the machine performance 
and versatility. Not only the beam energy
needs to be varied over a broad range, but since the
cross section at threshold is small a large amount of
integrated luminosity must be dedicated to each scan. Effects from 
beamstrahlung, 
finite sparticle widths, and Sommerfeld rescattering \cite{sp0,sp1} 
are important at threshold, while SUSY
backgrounds are reduced, at least for the lighter states.
It turns out to be preferable to concentrate the luminosity in a small number of
scan points \cite{Blair:2001cz}. Measurements at energies very close to the
kinematic threshold are most sensitive to the width
while those on the cross section rise above threshold are most sensitive 
to the mass. In general, on can achieve few per-mille precision 
for the mass determination from a threshold scan. In absolute numbers, the
uncertainty for the width measurement is comparable, but since 
electroweak sparticle widths are typically a factor 1000 smaller than their
mass, only an upper bound on the width 
can be established in most cases. With an $e^-e^-$ running option for the ILC,
on the other hand,
the selectron masses and widths can be measured with up to tenfold better precision
than in $e^+e^-$ collisions \cite{sp1}, which is due to the fact that
$\te_R^-\te_R^-$ and $\te_L^-\te_L^-$ pairs are produced in a s-wave rather than
a p-wave, leading to a steep $\propto \beta$ rise near threshold.
\end{sloppypar}

A comparison of ILC mass measurements
for various sparticles via continuum and via threshold
measurements is shown in Table~\ref{tab:mass_meas} (from
Ref.~\cite{sp1,Martyn:2004ew,Weiglein:2004hn}).
Note that the threshold scans require some rough a priori knowledge of the 
sparticle masses and take significant amount of the running time at various energy
points, which will reduce
the statistics available at the highest energy. There
have been a few detailed studies of run plan scenarios 
including threshold scans for SUSY particles which
show the feasibility to acquire data at the thresholds
of a few important processes while accumulating a sizeable data 
set at the highest operational energy \cite{Battaglia:2002ey}.
The scenarios adopted in those studies are now made
obsolete by the recent LHC bounds, but the findings
are still applicable in a general sense.\\

\begin{table}[h]
\centering
\begin{tabular}{|c|c|c|c|c|}
\hline
\boldmath $e^+e^-$ & $m$ & $\delta m_{\rm c}$ & $\delta m_{\rm th}$ & $\Gamma_{\rm th}$ \\
\hline
$\tmu_R$ & 143.0 & 0.2 & 0.2 & $<$0.5 \\
$\tmu_L$ & 202.1 &     & 0.5 & --- \\
$\te_R$  & 143.0 & 0.1 & 0.15& $<$0.4 \\
$\te_L$  & 202.1 & 0.8 & 0.3 & $<$0.4\\
$\tnu_{e}$ & 186.0 & 1.2 & 0.8 & $<$0.7 \\
$\ttau_1$ & 133.2 & 0.3 & & \\
$\tilde{\chi}^\pm_1$ & 176.4 & 1.5 & 0.55 & \\
$\tilde{\chi}^\pm_2$ & 378.2 & 3   & & \\
$\tilde{\chi}^0_1$   & 96.1  & 0.1 & & \\
$\tilde{\chi}^0_2$   & 176.8 & 2   & 1.2 & \\
$\tilde{\chi}^0_3$   & 358.8 & 3--5 & & \\
$\tilde{\chi}^0_4$   & 377.8 & 3--5 & & \\
\hline
\end{tabular}\\
\begin{tabular}{|c|c|c|c|}
\hline
\boldmath $e^-e^-$ & $m$ & $\delta m_{\rm th}$ & $\Gamma_{\rm th}$ \\
\hline
$\te_R$  & 143.0 & 0.05 & $0.21\pm0.05$ \\
$\te_L$  & 202.1 & 0.25 & $0.25\pm0.04$ \\
\hline
\end{tabular}
\vskip 0.2cm
\caption{\it Expected precision on sparticle masses (in GeV) for the SPS1a
  scenario \cite{Allanach:2002nj} using polarized $e^\pm$ beams ($P_L(e^-) = 0.8,
  P_L(e^+) = 0.6$). $\Delta m_{\rm c}$ is from decay kinematics measured in the
  continuum (${\cal L} = 200/500/1000 \text{fb}^{-1}$ at $\sqrt{s} =
  400/500/750$~GeV), and $\delta m_{\rm th}$ and $\delta \Gamma_{\rm th}$ 
are from
  threshold scans (${\cal L}=100$~fb$^{-1}$ for $e^+e^-$ and  ${\cal
    L}=5$~fb$^{-1}$ for $e^-e^-$). From
  Ref.~\cite{sp1,Martyn:2004ew,Weiglein:2004hn}.} 
\label{tab:mass_meas}
\end{table}

\noindent{\it Cross Sections, Width and Branching fractions}

\noindent Decays of charginos and neutralinos into bosons, such as $\tilde
\chi^{\pm}_1 \rightarrow W^{\pm} \tilde \chi^0_1$ and $\tilde \chi^{0}_2
\rightarrow Z \tilde \chi^0_1$ or $\tilde{\chi}_1^0 h$, are well suited to $e^+e^-$
collider capabilities. 
The four-jet + missing energy final states can be studied with
good accuracy thanks to the small background and the excellent di-jet
mass resolution ensuring separation of $W$ from $Z$ or $h$ masses. Production
cross sections of pairs of chargino and neutralino with mass of 216~GeV
have been studied at 0.5~TeV and the statistical uncertainty on the
cross section has been estimated at 0.6\% and 2\% respectively.  It is
interesting to observe that decays of SUSY particles, in particular
neutralinos into the lightest Higgs boson, $h$, are common and even
enhanced in specific models and combinations of MSSM
parameters~\cite{Djouadi:2001fa,Gori:2011hj,Baer:2012ts,Arbey:2012fa}.
This opens up an interesting perspective of studying SUSY processes
through the reconstruction of $h$ pairs + missing energy in four jet
events, where Higgs boson production is selected from that of other bosons by
di-jet mass (see Figure~\ref{fig:n2h}) and also $b$-tagging.
A further possibility is the study of single Higgs boson
plus missing $E$ production
via $e^+e^-\to\tilde{\chi}_1^0\tilde{\chi}_2^0$ with the decay 
$\tilde{\chi}_2^0\to\tilde{\chi}_1^0h$.
\begin{figure}[h!]
\begin{center}
\includegraphics[width=7.0cm]{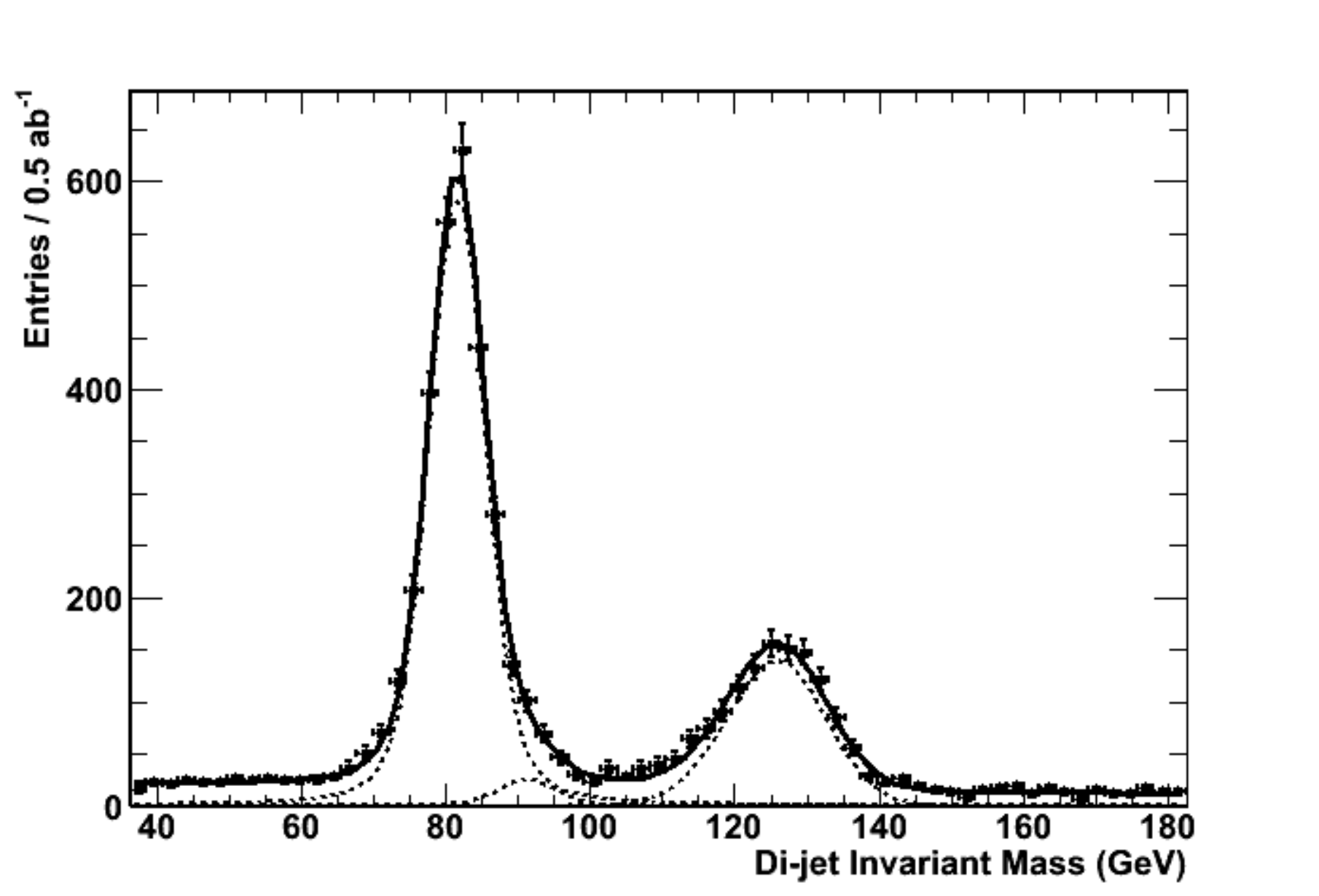}%
\end{center}
\caption{Di-jet invariant mass distribution
in inclusive 4-jet + missing energy SUSY events produced in $\sqrt{s}$=3~TeV $e^+e^-$ collisions for 
0.5~ab$^{-1}$ of fully simulated events. The result of the fit to extract the boson content is shown by the continuous line with the 
individual $W$, $Z$ and $h$ components represented by the dotted lines (from~\cite{Arbey:2012fa}).}
\label{fig:n2h}
\end{figure}

\noindent In addition, the determination of the dependence of the cross section
for production of gaugino pairs, including $\tilde{\chi}^0_2
\tilde{\chi}^0_2$ and $\tilde{\chi}^+_1 \tilde{\chi}^-_1$, with the beam
polarisation and energy is important to establish the nature of the
$\tilde{\chi}^0_2$ and measure the chargino mixing angles and the $\mu$
parameter~\cite{Desch:2003vw}.\\

\noindent {\it $\tau$--polarization}

\noindent The measurement of $\tau$ polarisation,
$P_{\tau}$, in $\tilde \tau_1$ decays offers sensitivity to the mixing
of interaction and mass eigenstates in the stau
sector~\cite{Godbole:2004mq}.  $P_{\tau}$ is extracted from the energy
spectrum of the pion emitted in the 1-prong decay $\tau \to \pi
\nu$. Again, the $\pi$ energy spectrum depends on the collision energy
and thus on beamstrahlung.  Nonetheless, using realistic parameters for
the ILC, the $\tau$ polarisation can be determined to a 15\% accuracy
(see Figure~\ref{fig:tau-pol}).\\
\begin{figure}
\begin{center}
\includegraphics[width=3.5cm]{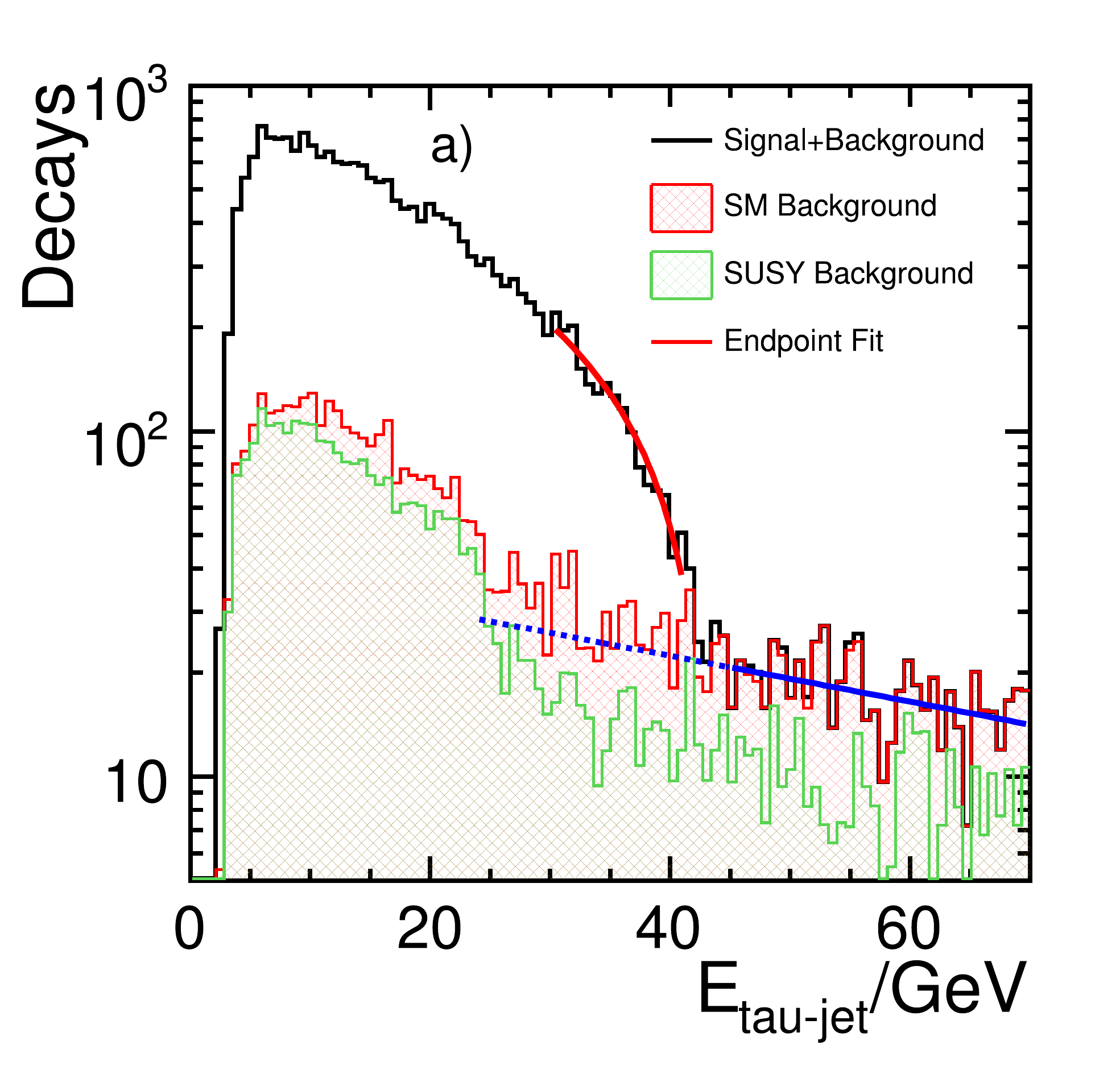}
\includegraphics[width=3.5cm]{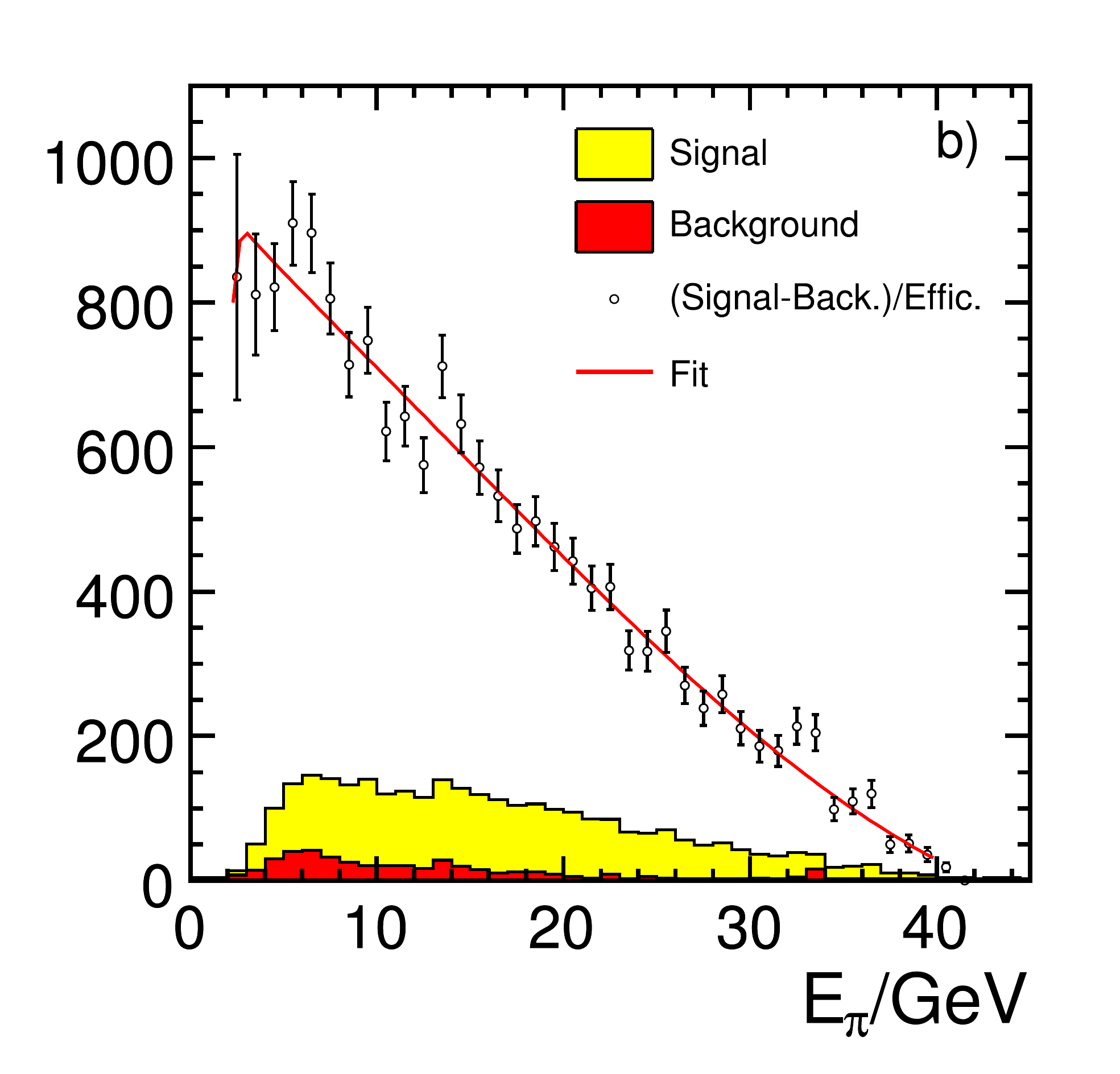}
\caption{Energy spectrum of reconstructed $\tau$ leptons from $\tilde \tau_1$ decays (left) and energy distribution of the 
pions from 1-prong decays with the fit for the determination of the polarisation for fully simulated $e^+e^-$ events at 
0.5~TeV (from~\cite{Abe:2010aa}).}
\label{fig:tau-pol}
\end{center}
\end{figure}

\noindent {\it CP--violating asymmetries}

\noindent The sub-leading, two-body decay
$\tilde \chi^0_i \to \tilde \ell_{R} \ell \to \ell \ell \tilde \chi^0_1$
is sensitive to CP asymmetries in the triple product of the final
particle momenta.  This measurement, which would open the way to the
detection of SUSY CP phases, is discussed below in more detail.  While
the measurement may be possible also at the LHC, the sensitivity of a
linear collider is expected to be far superior.  A detailed analysis,
based on full simulation and reconstruction and which makes use of event
kinematics, obtained values of $|M_1|$, $\mu$ and $M_2$ to a relative
accuracy of 1\% or better and the CP phases to $\sim$10\% resolving the
sign ambiguity, for states accessible at $\sqrt{s}$ = 0.5~TeV using
polarised electron and positron beams~\cite{Kittel:2011rk}.

\subsubsection{Testing the SUSY Character}

\begin{sloppypar}
One of the most important aspects of new physics searches is to really
identify the new physics model. Concerning SUSY theories, such an
identification requires measurements beyond just determining the mass and 
spin of the new particle. 
In order to prove that the new physics candidate is indeed
the SUSY partner of the corresponding SM particle, one also has to
measure precisely their couplings\cite{Feng:1995zd} and their quantum numbers. 
In this context also the special feature of carrying a Majorana character has 
to be proven for the neutral gauginos.
\end{sloppypar}

\vspace{.2cm}
\noindent {\it Spin determination}\\
\vspace{-.2cm}

\noindent The spin is one of the fundamental characteristics of all particles and it
must be determined experimentally for any new particles so as to clarify
the nature of the particles and the underlying theory. In particular, this
determination is crucial to distinguish the supersymmetric interpretation
of new particles from other models.

In supersymmetric theories, spin-1 gluons and electroweak gauge bosons, and
spin-0 Higgs bosons are paired with spin-1/2 gluinos, electroweak gauginos
and higgsinos, which mix to form charginos and neutralinos in the
non-colored sector. This calls for a wide spectrum of necessary attempts
to determine the nature of the new particles experimentally.

The measurement of the spins in particle cascades at LHC is quite
involved \cite{Barr:2004ze,Datta:2005zs,Athanasiou:2006ef}. While the invariant mass distributions
of the particles in decay cascades are characteristic for the spins of the
intermediate particles involved, detector effects strongly reduce the
signal in practice.

In contrast, the spin measurement at $e^+e^-$ colliders is
straightforward \cite{Battaglia:2005zf,Choi:2006mr}. A sequence of techniques-- increasing in
complexity-- can be exploited to determine the spin of supersymmetric
particles in pair production of sleptons, charginos and neutralinos
in $e^+e^-$ collisions:
\begin{itemize}
\item[(a)] rise of the excitation curve near the threshold,
\item[(b)] angular distribution in the production process,
\item[(c)] angular distribution in decays of the polarized particles and
\item[(d)] angular correlations between decay products of two particles.
\end{itemize}
\begin{sloppypar}
Within the general theoretical framework it can be proven that the second
step (b) is already sufficient in the slepton sector, although in general
the final state analysis is required to determine the spin unambiguously
in the chargino and neutralino sectors.
\end{sloppypar}
\begin{figure}[h]
\centering
\includegraphics[width=0.22\textwidth,height=0.22
                 \textwidth,angle=0]{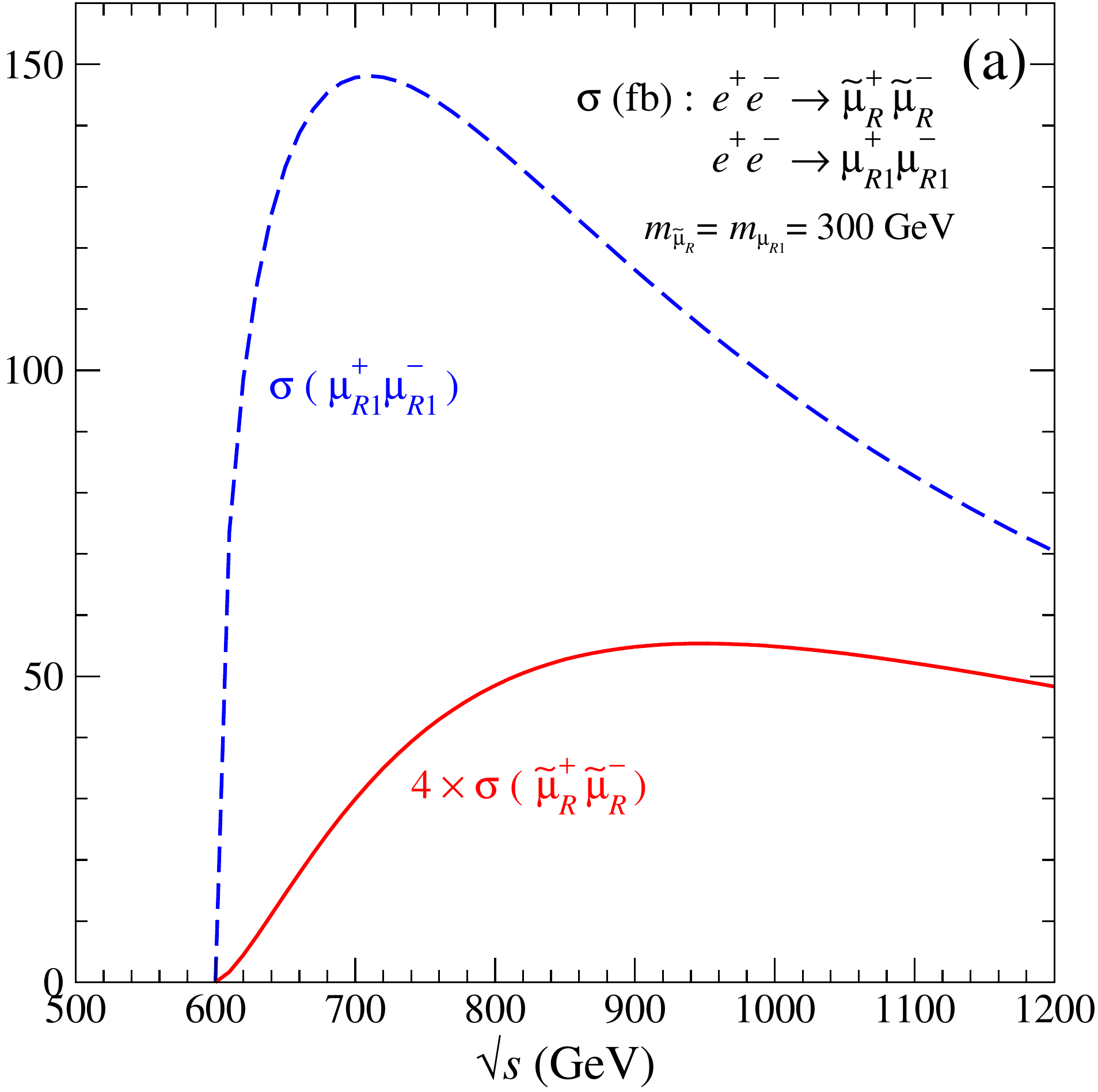}
\hskip 0.5cm
\includegraphics[width=0.22\textwidth,height=0.22
                 \textwidth,angle=0]{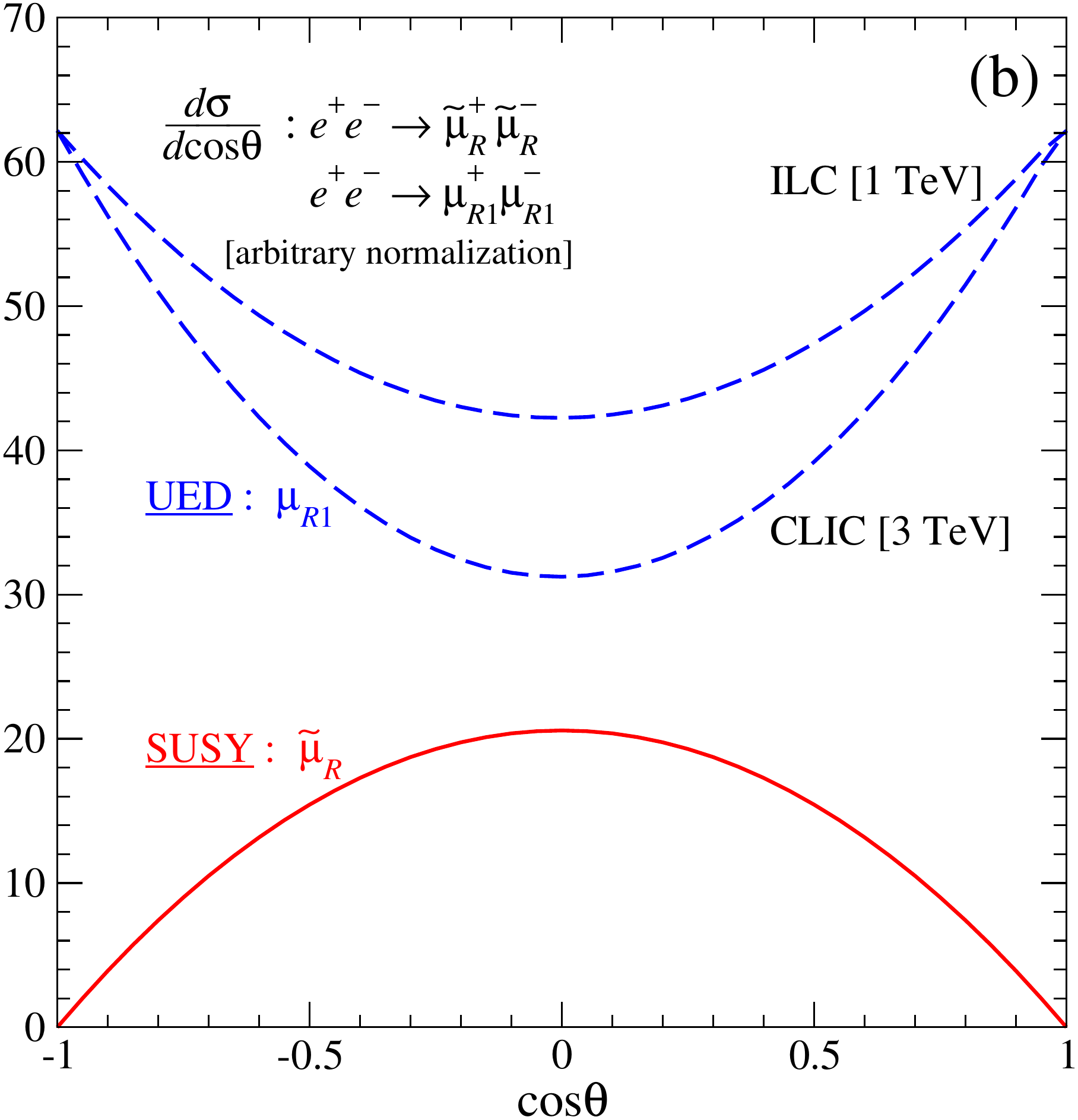}
\caption{\it The threshold excitation (a) and the angular distribution (b)
         in pair production of smuons in the MSSM, compared with the first
         spin-1/2 Kaluza-Klein muons in a model of universal extra
         dimensions; for details,
         see Ref.$\,$\cite{Choi:2006mr}.}
\label{fig_spin_1}
\end{figure}

As shown clearly in Fig.$\,$\ref{fig_spin_1}, the threshold excitation
curve and the production angle distribution for smuons in the MSSM are
characteristically different from those for the first Kaluza-Klein muons
in a model of universal extra dimension. Even though the $p$-wave onset of
the excitation curve is generally a necessary but not sufficient condition,
the $\sin^2\theta$ law for the angular distribution in the production of
sleptons (for selectrons close to threshold) is a unique signature of the
fundamental spin-zero character.
\begin{figure}[htb]
\begin{center}
\includegraphics[width=8.4cm,height=8.4cm,angle=0]{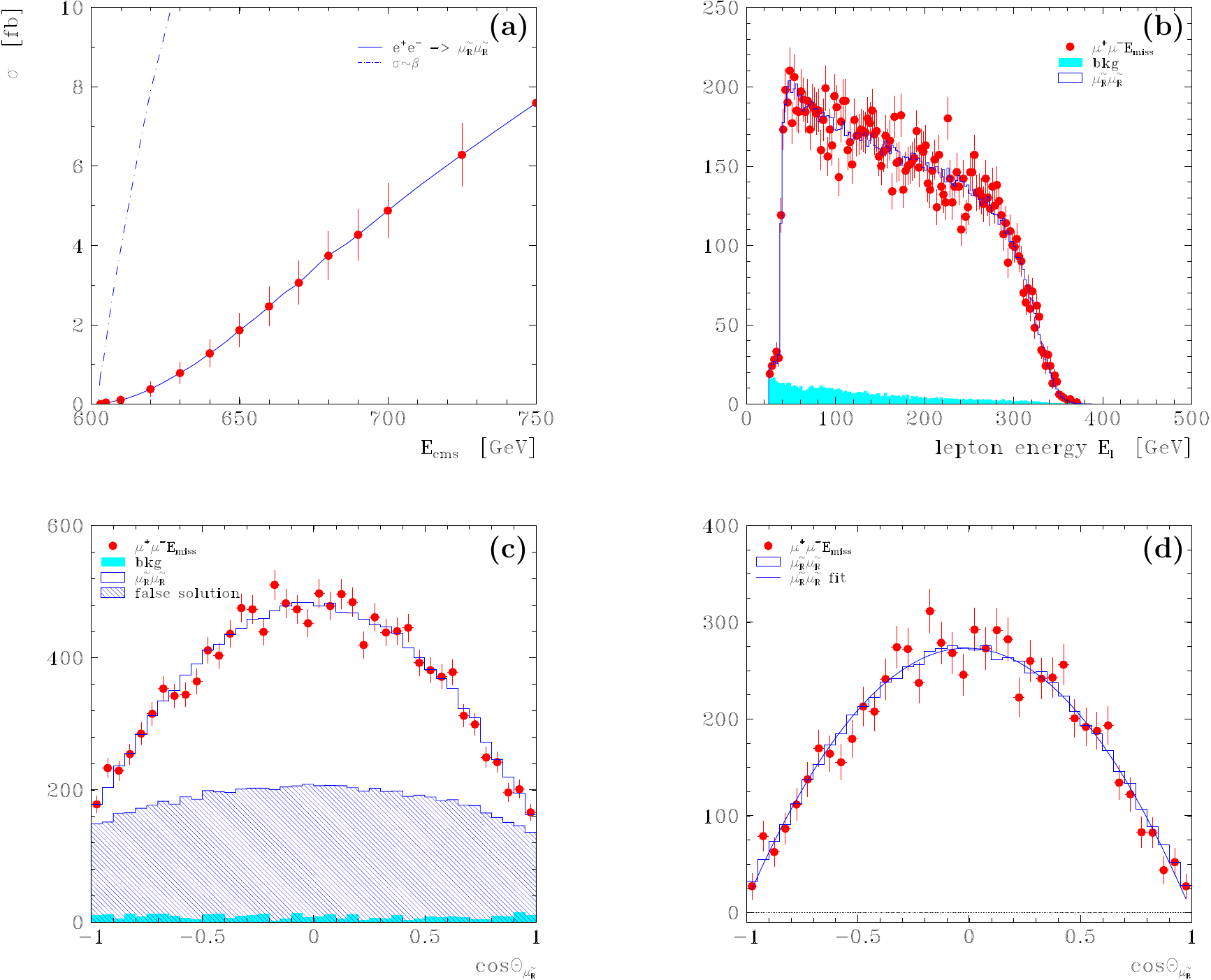}
\end{center}
\vskip 0.0cm
\caption{\it
  (a) The unpolarized cross section of $e^+e^-\to\tilde{\mu}_R^+
  \tilde{\mu}^-_R$ production close to threshold,
  including QED radiation, beamstrahlung and width effects;
  the statistical errors correspond to ${\cal L}=10\,{\rm fb}^{-1}$
  per point,
  (b) energy spectrum $E_\mu$ from $\tilde{\mu}_R^-\to \mu^-\tilde{\chi}^0_1$
  decays; polar angle distribution $\cos\theta_{\tilde{\mu}_R}$
  (c) with and (d) without contribution of false solution.
  The simulation for the energy and polar angle distribution.
  The simulation for the energy and polar angle distribution
  is based on polarized beams with $(P_{e^-},P_{e^+})
  = (+0.8, -0.6)$ at $\sqrt{s}=1\,{\rm TeV}$ and ${\cal L}=500\,
  {\rm fb}^{-1}$.
  For details, see Ref.$\,$\cite{Choi:2006mr}.
}
\label{fig:smuon_all}
\end{figure}

The
measurement of the cross section for smuon pair production
$\tilde{\mu}^+_R\tilde{\mu}^-_R$ can be carried out by identifying acoplanar
$\mu^+\mu^-$ pairs [with respect to the $e^\pm$ beam axis] accompanied
by large missing energy carried by the invisible lightest neutralino
$\tilde{\chi}^0_1$ in the decays $\tilde{\mu}^\pm_R\to\mu^\pm\tilde{\chi}^0_1$.
In addition, initial and final state QED radiations, beamstrahlung and detector
effects, etc. needs to be taken into account for reconstructing the
theoretically predicted distributions. As shown in Fig.$\,$\ref{fig:smuon_all}
through a detailed simulation, the characteristic $p$-wave threshold excitation and
the production, as well as the flat decay distribution for
the process $e^+e^-\to\tilde{\mu}^+_R\tilde{\mu}^-_R$ followed by
the decays $\tilde{\mu}^\pm_R\to\mu^\pm\tilde{\chi}^0_1$, 
can be reconstructed experimentally.

\begin{sloppypar}
Unlike the slepton sector, the chargino and neutralino sectors in general have
much more involved patterns. Neither the onset of the excitation curves near
threshold nor the angular distribution in the production processes
provides unique signals of the spin of charginos and neutralinos. However,
decay angular distributions of polarized charginos and neutralinos,
as generated naturally in $e^+e^-$ collisions, can provide an
unambiguous determination of the spin-1/2 character of the particles
albeit at the expense of more involved experimental analyses \cite{Choi:2006mr}.
Using polarized electron and/or positron beams will in general
assure that the decaying spin-1/2 particle is polarized; reasonable
polarization analysis power is guaranteed in many decay processes.
\end{sloppypar}

\begin{sloppypar}
Generally, quantum interference among helicity amplitudes-- reflected typically
in azimuthal angle distributions and correlations-- may provide another method for
determining spins \cite{Buckley:2007th}, although this method depends strongly on the
masses of the decay products and the $\sqrt{s}$ energy, as the quantum interference
disappears with increasing energy.
\end{sloppypar}

To summarize, the spin of sleptons, charginos and neutralinos can be determined
in a model-independent way at $e^+e^-$ colliders. Methods similar to 
those applied to slepton pair production  can be applied in the squark sector.
For gluinos,  a quite different methodology is required since these are not
produced at tree level in $e^+e^-$ collisions.

\vspace{.2cm}
\noindent{\it Yukawa couplings}\\
\vspace{-.2cm} 

\noindent The SM/SUSY coupling relations are not affected by SUSY
breaking and therefore the couplings of the SM particle are the same
as those of their SUSY partners. That means, for instance, that the
SU(3), SU(2) and U(1) gauge couplings $g_S$, $g$ and $g'$ have to be
identical to the corresponding SUSY Yukawa couplings $g_{\tilde{g}}$,
  $g_{\tilde{W}}$ and $g_{\tilde{B}}$.  These tests are of fundamental
    importance. Concerning the test of the SUSY-QCD Yukawa couplings,
    first examinations could be performed at the LHC via determining
    the couplings in $\tilde{q}\tilde{g}$, $\tilde{g}\tilde{g}$ and
    $\tilde{q}\tilde{q}$ productions\cite{Freitas:2007fd}.
These SUSY-QCD Yukawa studies have been accomplished by the analysis at a LC 
in\cite{Brandenburg:2008gd}, 
so that one expects in total an uncertainty of about 5\%-10\% in the 
determination of the SUSY-QCD Yukawa couplings.

The SUSY-EW Yukawa coupling, however, is one of the final targets of LC
experiments which should provide a complete picture of the
electroweak gaugino sector with a resolution at the level 
of at least 1\%~\cite{Choi:2001ww}. Under the assumption that the $SU(2)$
and $U(1)$ parameters have been determined in the chargino/ higgsino
sector (see Sect.\ref{sec:susy5}), we test precisely the equality of
the Yukawa and gauge couplings via measuring polarized cross sections:
varying the left-handed and right-handed Yukawa couplings has
consequences on the measured cross sections. Depending on the electron 
(and positron) beam polarization and on the luminosity, a per cent 
level precision can be achieved.
%

\begin{figure}
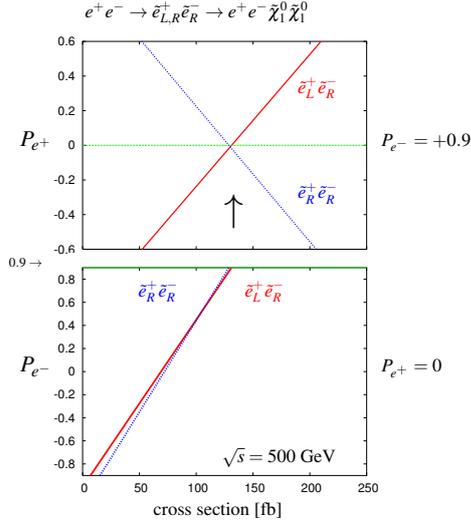

\vspace{3cm}
\hspace{2cm}
\begin{picture}(6,10)
\setlength{\unitlength}{1cm}
\put(-.1,0){\mbox{\includegraphics[height=.14\textheight]{Intro/Figures/sel-elp-povar}}}
\put(0.4,.0){\mbox{\includegraphics[height=.015\textheight,width=.22\textheight]{Intro/Figures/box}}}
\put(-.1,-3){\mbox{\includegraphics[height=.14\textheight]{Intro/Figures/sel-elvar}}}
\put(-.25,1.7){\footnotesize $P_{e^+}$}
\put(-.25,-1.3){\footnotesize $P_{e^-}$}
\put(-.4,0.1){\tiny $0.9\to$}
\put(4.5,1.7){\scriptsize $P_{e^-}=+0.9$}
\put(3.4,2.4){\scriptsize\color{red} $\tilde{e}^+_L\tilde{e}^-_R$}
\put(3.4,1){\scriptsize\color{blue} $\tilde{e}^+_R\tilde{e}^-_R$}
\put(2.7,-.3){\scriptsize\color{red}$\tilde{e}^+_L\tilde{e}^-_R$}
\put(1.3,-.3){\scriptsize\color{blue} $\tilde{e}^+_R\tilde{e}^-_R$}
\put(2.43,.7){\Large$\uparrow$}
\put(2.4,-2.5){\scriptsize $\sqrt{s}=500$~GeV}
\put(4.5,-1.3){\scriptsize $P_{e^+}=0$}
\put(0.6,3.4){\scriptsize $e^+e^-\to\tilde{e}^+_{L,R}\tilde{e}^-_{R}\to e^+e^- \tilde{\chi}^0_1\tilde{\chi}^0_1$}
\put(1.5,-3.2){\scriptsize cross section [fb]}
\end{picture}
\vspace{3.5cm}
\caption{Polarized cross section versus $P(e^-)$ (left panel) or
$P(e^+)$ (right panel) for $e^+e^-\tilde{e}\tilde{e}$-production with direct
decay in $\tilde{\chi}^0_1e$ in a scenario where the non-coloured spectrum is
similar to a SPS1a-modified scenario but with  $m_{\tilde{e}_L}=200$~GeV, 
$m_{\tilde{e}_R}=195$~GeV~\cite{MoortgatPick:2005cw}. }
\label{fig:susy_selectrons-susy}
\end{figure}

\vspace{.2cm}
\noindent{\it Quantum numbers}\\
\vspace{-.2cm}

\noindent One of the important tasks at future experiments is to determine
model-independently the underlying quantum numbers of any new
particles and check whether they correspond to their Standard Model
counterparts. For instance, a particularly challenging measurement is
the determination of the chiral quantum numbers of the sfermions.
Although these are scalar particles, they have to carry the chiral
quantum numbers of their Standard Model partners.  Since chirality can
be identified in the high energy limit via helicity and its
conservation, it will be part of the charge of a linear collider to
prove such an association.  Since the limits from LHC for the
electroweak SUSY spectrum are not very strong, it is still the case
that a rather light spectrum selectrons, smuons, staus continues to be
viable.

In $e^+e^-$ collisions, the associated production reactions
$e^+e^-\rightarrow \tilde{e}_L^+\tilde{e}_R^-,\ \tilde{e}_R^+\tilde{e}_L^-$ occur
only via $t$-channel exchange, whereas the pair production 
reactions $\tilde{e}_L\tilde{e}_L$, $\tilde{e}_R\tilde{e}_R$ occur 
also via $s$-channel $\gamma$ and $Z$ exchange.
Since $m_{\te_L}$ is in general not equal to $m_{\te_R}$, then 
the electron energy distribution endpoints will be different for 
each of the four possible reactions as will the positron energy distributions.
Furthermore, the total cross sections for each reaction depend strongly 
on beam polarization so that by dialing the polarization, one can 
move between distinct spectra possibilities which allows one to disentangle
the individual $\te_L$ and $\te_R$ masses, and to distinguish which one
is which: {\it e.g.} measure their chiral quantum numbers, 
see~\ref{fig:susy_selectrons-susy}. The masses of $m_{\tilde{e}_L}=200$~GeV, 
$m_{\tilde{e}_R}=195$~GeV are close, 
both particles decay directly to $\tilde{\chi}^0_1 e$.

The polarization of $P(e^+)$ is mandatory in such cases.  
An example from Ref. \cite{snowmass} using an Isajet simulation
is shown in Fig. \ref{fig:eLeR}.
\begin{figure}[hb!]
\begin{center}
\includegraphics[width=7.0cm,angle=0]{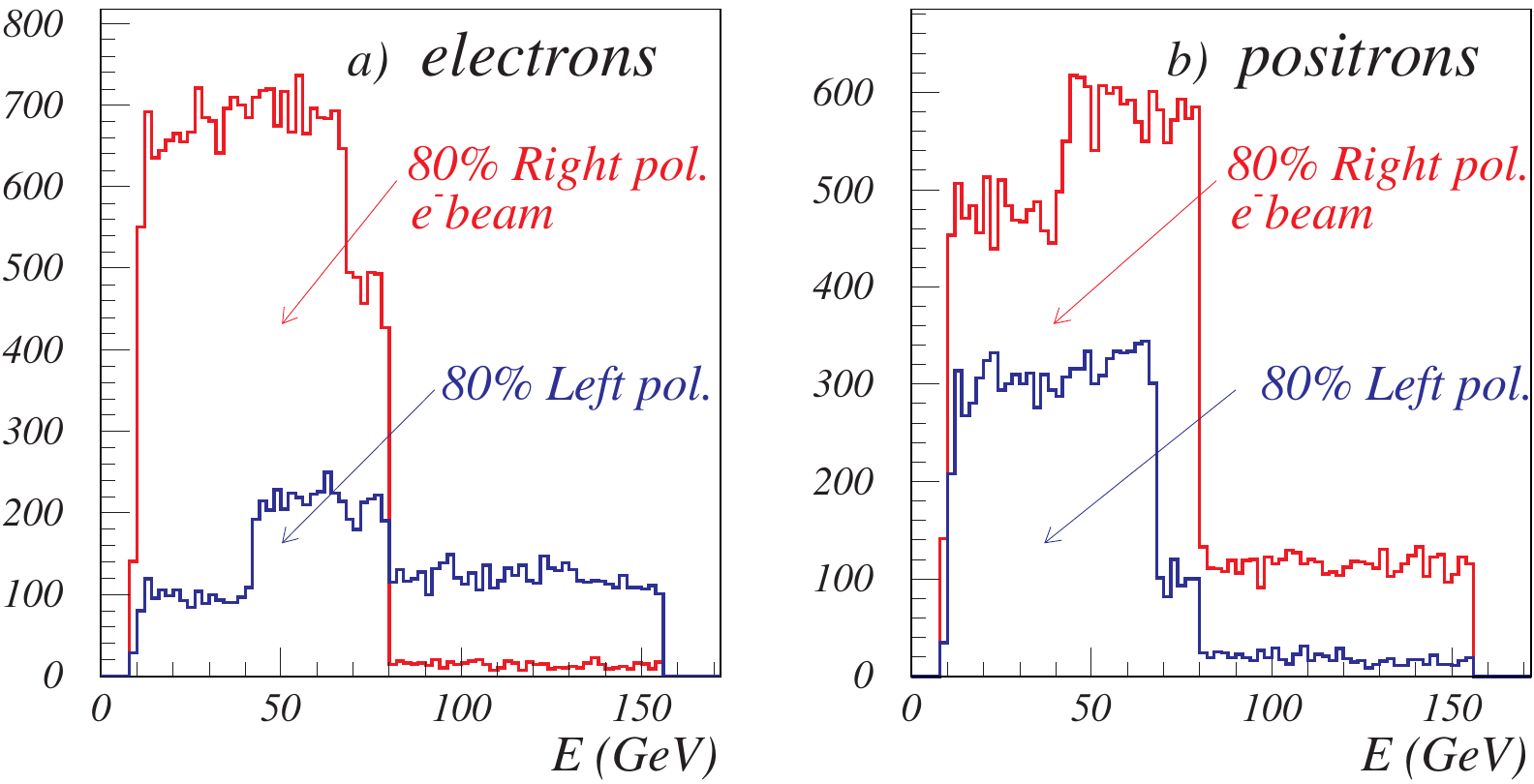}
\end{center}
\caption{Electron and positron energy
 distributions for selectron pair production
with the indicated beam polarizations and an integrated luminosity 
of 50 fb$^{-1}$ at $\sqrt{s}=500$~GeV
(E.~Goodman, U.~Nauenberg {\it et al.} in Ref. \cite{snowmass}).
}
\label{fig:eLeR}
\end{figure}

\vspace{.2cm}
\noindent{\it Majorana character}\\
\vspace{-.2cm}

Experimental tests of the Majorana character of gluinos and
neutralinos will provide non-trivial insight into the realization
of SUSY in nature.  
There are several powerful methods for probing
the nature of neutralinos in $e^\pm e^-$ collisions with polarized
beams.

The parallelism between self-conjugate neutral vector gauge bosons and their
fermionic supersymmetric partners induces the Majorana nature of these
particles in the minimal $N=1$ supersymmetric extension of the standard
model (MSSM). Therefore, experimental tests of the Majorana character of
colored gluinos and non-colored electroweak neutralinos would provide non-trivial
insight into the realization of SUSY in nature, since extended supersymmetric
models can include Dirac gauginos and/or higgsinos\cite{Benakli,Kribs:2007ac,Choi:2010an}.

A theoretical basis for formulating a solid testing ground for Dirac gauginos
is provided by a model with a continuous global U(1) $R$
symmetry \cite{Kribs:2007ac,Choi:2010an} under which the Grassmann coordinates
transform as $\theta \to e^{i\xi}\theta$,
{\it i.e.} $R(\theta)=1$. It implies that the component fields of a supersymmetric
superfield differ by the $R$-charge. Since the gauge superfields $\hat{G}$ are
real, they must have a zero $R$-charge, $R(\hat{G})=0$, implying that $R=0$ for
the gauge vector fields $G^\mu$ and $R=1$ for the spin-1/2 gauginos
$\tilde{G}^\alpha$.
Every term in the superpotential must have $R=2$ to provide a $R$-symmetric
potential while any soft-SUSY breaking terms must have $R=0$.

When the $R$-charges of the MSSM matter, $H$-Higgs and gauge-vector
superfields are assigned as in Table~\ref{tab:Rcharges}, not only the
supersymmetric $\mu$ term and the baryon- and lepton-number breaking
terms but also the soft-SUSY breaking Majorana mass terms and
trilinear $A$ terms are forbidden. As a result, the sfermion
left-right mixing and the proton decay through dimension-five
operators are absent (while Majorana neutrino masses can be
generated).

\begin{table}[h]
\centering
\begin{tabular}{|c|c|c|}
\hline
 Field & Superfield & $R$ charge      \\
\hline \hline
 Matter       & $\hat{L}, \hat{E}^c$                     &  $1$  \\
              & $\hat{Q}, \hat{D}^c, \hat{U}^c$          &  $1$  \\
 $H$-Higgs    & ${\hat{H}}_{d,u}$                        &  $0$   \\
 $R$-Higgs    & ${\hat{R}}_{d,u}$                        &  $2$  \\
 Gauge Vector & $\hat{G} = \{\, G^\mu, \tilde{G}^\alpha\}$  &  $0$   \\
 Gauge Chiral & $\hat{\Sigma} = \{ \sigma, \tilde{G}'^\alpha \}$  & $0$ \\
\hline
\end{tabular}
\vskip 0.2cm
\caption{\it The $R$-charges of the matter, Higgs and gauge superfields in the minimal
             $R$-symmetric supersymmetric standard model \cite{Kribs:2007ac}.
        }
\label{tab:Rcharges}
\end{table}

\begin{sloppypar}
Since the gaugino Majorana-type mass terms and the conventional higgsino $\mu$
term are forbidden in the $R$-symmetric theory, the superfield content of the
minimal theory needs to be extended so as to give non-zero masses to gluinos,
electroweak gauginos and higgsinos. The simplest extension, called the minimal
$R$-symmetric supersymmetric standard model (MRSSM) \cite{Kribs:2007ac}, is
to introduce new chiral superfields
$\hat{\Sigma}=\{\,\sigma, \tilde{G}'^\alpha\}$ in the adjoint representation
of the SM gauge group in addition to the standard vector superfields
as well as two iso-doublet chiral superfields $\hat{R}_d$ and $\hat{R}_u$
($R$-Higgs) to complement the standard $H$-Higgs superfields $\tilde{H}_d$ and
$\hat{H}_u$. (For a simpler formulation, see Ref.$\,$\cite{Davies:2011mp}.)
\end{sloppypar}

In the color sector the original MSSM $R=1$ gluino $\tilde{g}^a$ and the new
$R=-1$ gluino $\tilde{g}'^a$ ($a=1-8$) are coupled by the SUSY-breaking but
$R$-symmetric Dirac mass term so that they can be combined into a single
Dirac fermion field $\tilde{g}^a_D = \tilde{g}^a_L + \tilde{g}'^a_R$ with
$R=1$. Note that $\tilde{g}_D$ is not self-conjugate any more, i.e.
$\tilde{g}^C_D\neq \tilde{g}_D$ as the anti-gluino carries $R=-1$. In the
similar manner the original electroweak gauginos, $R=1$
$\tilde{B}$ and $\tilde{W}^i$ ($i=1-3$) and $R=-1$ $H$-higgsinos, $\tilde{H}_u$
and $\tilde{H}_d$ are coupled with the new electroweak gauginos, $R=-1$
$\tilde{B}'$ and $\tilde{W}'^i$ ($i=1-3$) and $R=1$ $R$-higgsinos,
$\tilde{R}_u$ and $\tilde{R}_d$, giving rise to four Dirac neutralinos
$\tilde{\chi}^0_{D1,\ldots,D4}$ with $R=1$ and four Dirac anti-neutralinos
with $R=-1$.

The extension from the minimal model MSSM with Majorana gluinos and neutralinos
to the $R$-symmetric MRSSM with Dirac gluinos and neutralinos as well as new
$R$-Higgs bosons and adjoint scalar fields $\sigma$ leads to a lot of
distinct phenomenological consequences on sparticle productions at the
LHC and $e^\pm e^-$ colliders \cite{Choi:2010an,Collider_phenomenology},
flavor and CP problems \cite{Kribs:2007ac,Fok:2012fb} and cold dark matter
issues \cite{Dark_matter}.

There are several methods to investigate the nature of gluinos
at the LHC. In the original form, decays to heavy stop/top quarks are exploited
\cite{gluino_Majorana} to test whether the final state in the fermion decay $
\tilde{g} \to \tilde{t} \bar{t} + {\tilde{t}}^\ast t$ is self-conjugate.
The standard production processes for investigating the nature of
gluinos \cite{Choi:2008pi} are the production of a pair of equal-chirality
squarks, $q_Lq'_L\to \tilde{q}_L\tilde{q}'_L$ and
$q_Rq'_R\to\tilde{q}_R\tilde{q}'_R$.
While the cross section for the scattering processes with
equal-chirality quarks is non-zero in the Majorana theory, it vanishes in the Dirac
theory. Owing to the dominance of $u$-quarks over $d$-quarks in the proton,
the Majorana theory predicts large rates of like-sign dilepton final states from
squark pair production with an excess of positively charged leptons\cite{Baer:1995va} while they
are absent, apart from a small number of remnant channels, in the Dirac theory.
(In a realistic analysis one has to include gluino production processes which can
also feed the like-sign dilepton signal but can be discriminated by extra jet
emission from the gluino decays.) In addition, the nature of neutralinos could
be checked at the LHC if cascade squark-decay chains involving
intermediate sleptons and neutralinos are identified, as the final-state
$q\ell^\pm$ invariant mass distributions are distinct \cite{Choi:2010gc}.

An $e^\pm e^-$ collider with polarized beams is an ideally clean and powerful
instrument for testing the nature of neutralinos. In parallel to the squark pair
production through quark-quark collisions, the processes $e^-e^-\to
\tilde{e}^-_R\tilde{e}^-_R$ or $\tilde{e}^-_L\tilde{e}^-_L$ with equal chirality
indices and $e^+e^-\to\tilde{e}^+_L\tilde{e}^-_R$ or $\tilde{e}^+_R\tilde{e}^-_L$
are forbidden due to the conserved $R$ charge in the Dirac case while the
processes occur in general in the Majorana case as in the
MSSM \cite{Choi:2008pi,Choi:2010gc,Keung:1983nq,AguilarSaavedra:2003hw}.

\begin{figure}[t]
\centering
\begin{center}
\hskip -1cm
\includegraphics[width=0.18\textwidth,height=0.24
                 \textwidth,angle=0]{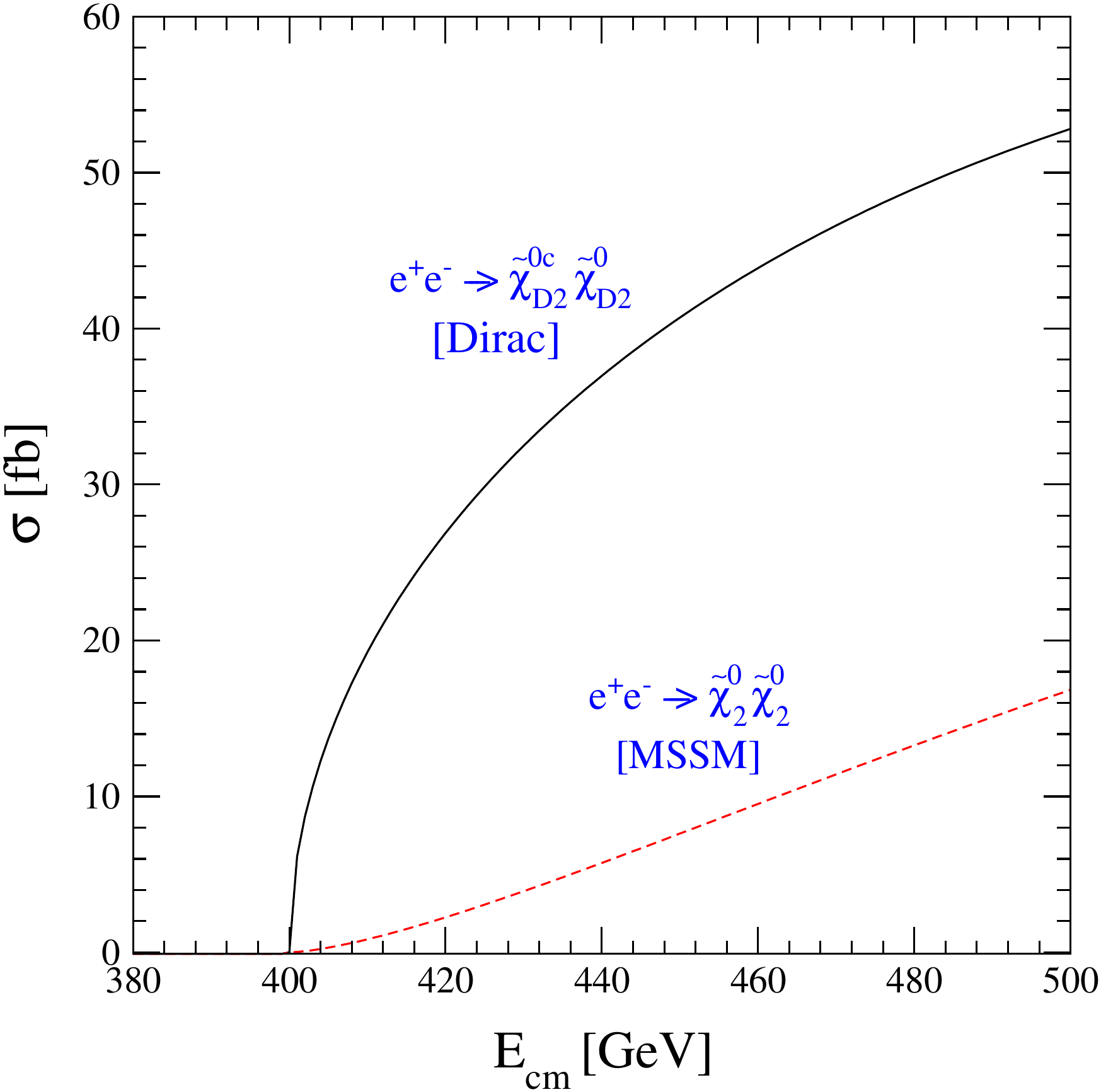}
\hskip 0.5cm
\includegraphics[width=0.18\textwidth,height=0.24
                 \textwidth,angle=0]{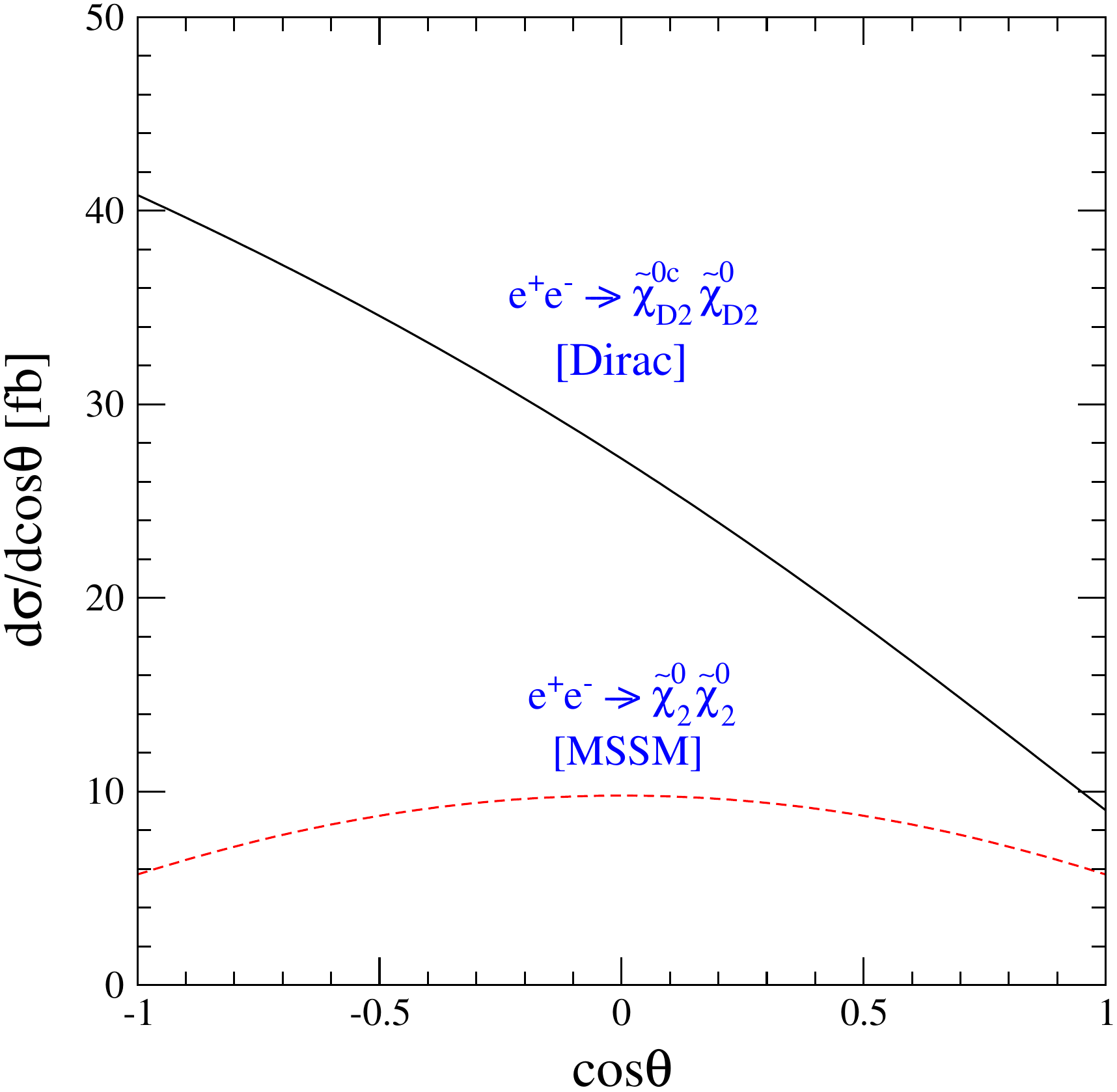}
\end{center}
\vspace{.9em}
\caption{\it Left: the total cross sections for pair
         production of wino-like neutralinos near threshold in the MSSM and
         the Dirac theory. Right: dependence of the cross sections
         on the production angle $\theta$ for $\sqrt{s}=E_{\rm cm}=500$~GeV.
         The sparticle masses in both plots are $m_{\tilde{\chi}_2^0}
         = m_{\tilde{\chi}_{D2}^0} = 200$~GeV and $m_{\tilde{e}_L}
         = 400$~GeV (For the details, see Ref.$\,$\cite{Choi:2010gc}).}
\label{fig:n2n2}
\end{figure}

Another powerful experimental test for characterising the nature of
neutralinos is based on the threshold behavior of the neutralino
diagonal-pair production and its polar-angle distribution (Fig.~\ref{fig:n2n2}). 
In the case
with Dirac neutralinos $\tilde{\chi}^0_D$, the cross section for the
process $e^+e^-\to \tilde{\chi}^0_{Di}\tilde{\chi}^0_{Di}$ ($i=1-4$)
exhibits a typical sharp $s$-wave excitation and a forward-backward
asymmetric angular distribution, while in the case with Majorana
neutralinos the cross section for neutralino diagonal pair production
in $e^+e^-$ collisions is excited in the characteristic slow $p$-wave
and the angular distribution is forward-backward symmetric
\cite{Choi:2010gc}.

\begin{sloppypar}
  To summarize, the gluinos, the electroweak gauginos and the
  electroweak higgsinos are either Majorana or Dirac fermions in
  extended supersymmetric models.  The $e^\pm e^-$ colliders and the
  LHC provide us with various complementary and powerful tests for
  probing the nature of new fermionic states from which we can get
  non-trivial insight into the realization of SUSY in nature and find
  new directions for collider phenomenology as well as many related
  fields.
\end{sloppypar}

\subsection{From SUSY Measurements to Parameter Determination}
\label{sec:susy5}

The measurements which can be performed from operating a linear
collider with a large enough energy $\sqrt{s}\ge$0.5~TeV and
luminosity, to collect of order of 0.5-2~ab$^{-1}$ of data, can be
turned into precise predictions on the fundamental MSSM parameters of
the theory Lagrangian, on their evolution to the unification scale,
and on the relic density of light neutralinos in the universe inferred
from collider data.  These quantities are crucial to understand the 
underlying structure and to identify the SUSY model
and its connections to cosmology. In this section, we discuss the
extraction of these parameters based on the anticipated accuracy of
measurements of SUSY particle properties at a linear collider.

\subsubsection{General strategy}

\begin{sloppypar}
Since the general MSSM depends already on over 100
new parameters, it is a true challenge to measure all parameters in as
model-independent fashion as is possible. Therefore often model
assumptions-- in particular on the SUSY breaking mechanism and mass
unifications-- are made (see Section~\ref{sect:susymodels}) resulting in a 
reduction to just 4--6 SUSY parameters. 
Then for unravelling the underlying SUSY model one needs a
model-independent strategy for measuring the parameters. Since the
current results from LHC point towards the TeV-scale for the coloured SUSY
partners, it is clear that one would need a combined approach from LHC and
the LC to resolve the SUSY puzzle. 
The determination of the fundamental SUSY parameters at low energy would 
allow a critical test of the theory:
extrapolating the mass parameters to the GUT scale points to
which SUSY breaking scheme might be realized in nature. \
Such extrapolations would be an important achievement that
illustrates well the complementarity of data from the LHC and a linear
collider~\cite{Blair:2000gy,AguilarSaavedra:2005pw,Desch:2003vw} 
(see also Section \ref{sec:extrap}).
\end{sloppypar}

The fundamental parameters of the gaugino/higgsino sector are the $U(1)$
and $SU(2)$ gaugino masses $M_1$ and $M_2$, and the higgsino mass
parameter $\mu$, where  also $M_i$ and $\mu$ can contain CP-violating
phases.  In addition, also $\tan\beta$ enters the mixing of this
electroweak particle SUSY sector.
These parameters can-- very accurately and independently of the
underlying SUSY breaking scheme-- be determined at a LC. 
This has been shown in many detailed studies\cite{Tsukamoto:1993gt,CKMZ,Baer:2003ru}.

In case the full spectrum, 
$\tilde{\chi}^0_i$, $\tilde{\chi}^{\pm}_j$, $i=1,\ldots,4$, $j=1,2$, is 
accessible, the determination of the fundamental parameters via
measurements of masses and cross sections seems to be trivial and is
therefore not discussed here in detail. In this case, however, stringent
tests of the closure of the system can be designed. Models with
additional chiral and vector superfields extend the gaugino/higgsino
sector. Since unitary matrices diagonalize the system, powerful sum rules
can be set up for the couplings and a unique test whether the observed 
4--system is closed or not might be possible. These sum rules for
couplings can be directly converted into high energy
sum rules for production cross sections of neutralinos\cite{CKMZ}:
\begin{eqnarray}
{\rm lim}_{s\to \infty}s \Sigma^{4}_{i\le j}\sigma_{ij} &=&
\frac{\pi\alpha^2}{48\cos^4\theta_W\sin^4\theta_W}\times \nonumber \\ 
& & [64 \sin^4\theta_W -8 \sin^2\theta_W +5 ]
\label{equ_sumrule}
\end{eqnarray}
In this case, one also has to provide a 
measurement for the production $\tilde{\chi}^0_1\tilde{\chi}^0_1$. 
This final state is invisible in R-parity invariant
theories where $\tilde{\chi}_1^0$ is the LSP. 
Nevertheless, it can be studied indirectly by photon tagging
in the final state $\gamma\tilde{\chi}^0_1\tilde{\chi}^0_1$, which can
be observed with a rather high accuracy at a LC. 
More details on photon tagging are included in the 'light higgsino' section.
   
The powerful test via sum rules stresses the importance of upgrading
the collider to achieve high $\sqrt{s}$ energies, if physics dictates
it, in addition to combining LC and LHC results.  In order to
reconstruct the complete MSSM Lagrangian and evolve the parameters to
the GUT scale~\cite{Blair:2002pg}, it is generally needed to combine
the linear collider measurements with those of squarks and gluinos
(and possibly heavier gauginos) observed probably first at the
LHC. Results at 0.5~TeV and 3~TeV are discussed in
\cite{AguilarSaavedra:2005pw} and \cite{clic-cdr}.

\subsubsection{Parameter determination with $\tilde{\chi}^{\pm}_1$, $\tilde{\chi}^{0}_{1,2}$ only}

Even if only $\tilde{\chi}^0_{1,2}$
and $\tilde{\chi}^{\pm}_1$ were accessible, the precise measurements of 
the masses as well as polarized cross section for
$\tilde{\chi}^+_1\tilde{\chi}^-_1$, $\tilde{\chi}^0_1\tilde{\chi}^0_2$
in different beam polarization configurations
is sufficient to determine the fundamental SUSY
parameters and allow mass predictions of the heavier particles, yet unseen 
SUSY states. 

The diagonalization of the two chargino system can be parametrized by
two mixing angles $\phi_L$, $\phi_R$.
Defining the mixing angles in the unitary matrices diagonalizing the 
chargino mass matrix ${\cal M}_C$ by $\phi_L$ and $\phi_R$ for the left--
and right--chiral fields, the fundamental SUSY parameters $M_2$, $|\mu|$,
$\cos \Phi_\mu$ and $\tan\beta$ can be derived from the chargino
masses and the cosines $c_{2L,R}=\cos 2\phi_{L,R}$ of the mixing 
angles~\cite{6A,SONG}.

If only the light charginos $\tilde{\chi}^\pm_1$ can be produced, the mass
$m_{\tilde{\chi}^\pm_1}$ as well as 
both mixing parameters $\cos 2\phi_{L,R}$ can be measured. 
The quantities $\cos 2\phi_{L,R}$ can be determined
uniquely if the polarized cross sections are measured at one energy 
including transverse beam polarization, or else if the longitudinally
polarized cross sections are measured at two different energies.

The heavy chargino mass is 
bounded from above after $m_{\tilde{\chi}^\pm_1}$ and $\cos 2\phi_{L,R}$
are measured experimentally. At the same time, it is bounded from below
by not observing the heavy chargino in mixed light$-$heavy pair production:
\begin{eqnarray}
     {\textstyle{\frac{1}{2}}} \sqrt{s} - m_{\tilde{\chi}^\pm_1} 
\,\leq\, m_{\tilde{\chi}^\pm_2} 
\,\leq\, \sqrt{\, m^2_{\tilde{\chi}^\pm_1}
           +\, 4 m^2_W/| \cos 2\phi_L-\cos 2\phi_R|}.\nonumber\\
\end{eqnarray}

\begin{sloppypar}
If both the light chargino mass $m_{\tilde{\chi}^\pm_1}$ and the heavy
chargino mass $m_{\tilde{\chi}^\pm_2}$ can be measured, the fundamental
parameters $M_2$, $\mu$, $\tan\beta$ can be extracted unambiguously. 
However, if $\tilde{\chi}^\pm_2$ is not accessible, 
their determination depends on the CP properties 
of the higgsino sector. 
\end{sloppypar}

\noindent
{\bf (A)}\, If the higgsino sector is 
     \underline{CP invariant}\footnote{Analyses of electric dipole
     moments strongly suggest that CP violation in the higgsino sector
     will be very small in the MSSM if this sector is 
     non--invariant at all \cite{EDM,WYSONG}.}, one can 
     determine 
     $m^2_{\tilde{\chi}^\pm_2}$ from the condition 
$\cos\Phi_\mu =\pm 1$, up to at most 
     a two--fold ambiguity: see Ref's.~\cite{SONG,6A}.
     This ambiguity can be resolved  as well as
the gaugino parameter $M_1$ be determined if observables 
from the neutralino sectror, in particular,  the mixed--pair 
$\tilde{\chi}^0_1   \tilde{\chi}^0_2$ 
production cross sections and $m_{\tilde{\chi}^0_{1,2}}$
are included, see Fig.~\ref{fig:mixingangles}.

\noindent
{\bf (B)}\, If $\tilde{\chi}^\pm_2$ is not accessible, 
the parameters $M_2$, $\mu$, $\tan\beta$, $\cos\Phi_{\mu}$ 
cannot be determined in a 
     \underline{CP non--invariant} theory in the chargino sector alone. 
     They remain dependent on the unknown heavy chargino
     mass $m_{\tilde{\chi}^\pm_2}$. However, two trajectories can be generated
     in $\{M_2, \mu; \tan\beta\}$ space, parametrized by 
     $m_{\tilde{\chi}^\pm_2}$ and classified by the two possible
values $\Phi_\mu$ and
     $(2\pi -\Phi_\mu)$ for the phase of the higgsino mass parameter.
Including information from the neutralino sector, namely the measured
masses and the polarized cross sections of the two
     light neutralino states $\tilde{\chi}^0_1$ and $\tilde{\chi}^0_2$, 
     the heavy chargino mass $m_{\tilde{\chi}^\pm_2}$ can be predicted
in the MSSM and subsequently the entire set of fundamental gaugino and
     higgsino parameters can be determined uniquely~\cite{CKMZ,WYSONG}: 
the symmetric neutralino mass matrix ${\cal M}_N$ is diagonalized by
a unitary matrix, defined such that the mass eigenvalues 
$m_{\tilde{\chi}^0_i}$ of the four Majorana fields $\tilde{\chi}^0_i$
are positive.

The squared mass eigenvalues of ${\cal M}_N {\cal M}^\dagger_N$ are  
solutions of the characteristic equations \cite{CKMZ}
\begin{equation}
m_{\tilde{\chi}^0_i}^8-a\, m_{\tilde{\chi}^0_i}^6
+b\, m_{\tilde{\chi}^0_i}^4-c\, m_{\tilde{\chi}^0_i}^2+d=0 
\label{eq:characteristic}
\end{equation}
for $i=1,2,3,4$
with the invariants $a$, $b$, $c$ and $d$ given by the fundamental
$SU(2)$ and $U(1)$ gaugino mass parameters $M_2$ and $M_1$, and the higgsino
mass parameter $\mu$, {\it i.e.} the moduli $M_2$, $|M_1|$, $|\mu|$ and the
phases $\Phi_1$, $\Phi_\mu$. Each of the four invariants 
$a$, $b$, $c$ and $d$  is a binomial of $Re(M_1)=|M_1|\,\cos\Phi_1$ 
and $Im(M_1)=|M_1|\,\sin\Phi_1$. Therefore, each of the characteristic 
equations in the set (\ref{eq:characteristic}) for the neutralino mass 
squared $m^2_{\tilde{\chi}^0_i}$ can be rewritten in the form
\begin{eqnarray}
Re(M_1)^2&+&Im(M_1)^2+ u_i\, Re(M_1)+ v_i\, Im(M_1)=w_i\nonumber\\
&&  
\label{eq:Mphase}
\end{eqnarray}
for $i=1-4$.
The coefficients $u_i$, $v_i$ and $w_i$ are functions of the 
parameters $M_2$, $|\mu|$, $\Phi_\mu$, $\tan\beta$ and the mass
eigenvalue $m^2_{\tilde{\chi}^0_i}$ for fixed $i$. The coefficient $v_i$ 
is necessarily proportional to $\sin\Phi_\mu$ because physical neutralino
masses are CP--even; the sign ambiguity for $\sin\Phi_\mu$, a result of
the two--fold cos solution $\Phi_\mu\leftrightarrow (2\pi-\Phi_\mu)$,
transfers to the associated sign ambiguity in the CP--odd quantity
$Im(M_1)$, {\it i.e.} in $\sin\Phi_1$.

\begin{sloppypar}
The characteristic equation (\ref{eq:Mphase}) defines a circle
in the $Re(M_1), Im(M_1)$ plane for each neutralino mass 
$m_{\tilde{\chi}^0_i}$. With only two light neutralino masses 
$m_{\tilde{\chi}^0_1}$ and $m_{\tilde{\chi}^0_2}$ measured, we are left with
a two--fold ambiguity.
The intersection points of the two 
     crossing points depend on the  unknown heavy chargino mass 
     $m_{\tilde{\chi}^\pm_2}$. 
By measuring the pair--production cross sections 
     $\sigma_L\{\tilde{\chi}^0_1\tilde{\chi}^0_2\}$ and
     $\sigma_R\{\tilde{\chi}^0_1\tilde{\chi}^0_2\}$, a unique solution, 
     for both the parameters $m_{\tilde{\chi}^\pm_2}$ and $Re(M_1), 
     Im(M_1)$ can be found at the same time~\cite{CKMZ}. 
     As a result, the additional measurement of the cross sections leads to a 
     unique solution for $m_{\tilde{\chi}^\pm_2}$ and subsequently to a unique 
     solution for $\{M_1, M_2; \mu; \tan\beta\}$ (assuming that the discrete 
     CP ambiguity in the associated signs of $\sin\Phi_\mu$ and 
     $\sin\Phi_1$ has been resolved by measuring the normal $\tilde{\chi}^0_2$ 
     polarization).
\end{sloppypar}
\begin{figure}
\vspace{-5cm}
\hspace*{-6.cm}
\includegraphics[width=20cm,height=15cm]{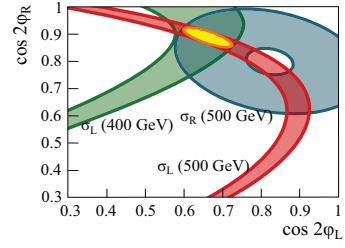}
\vspace{-6cm}
\caption{Determination of the chargino mixing angles $\cos 2
  \Phi_{L,R}$ from LC measurements in $e^+e^-\to
  \tilde{\chi}^+_1\tilde{\chi}^-_1$ with polarized beams at different
  cms energies. The electroweak part of the spectrum in this scenario
  is a modified benchmark scenario SPS1a.
\label{fig:mixingangles} }
\end{figure}


\subsubsection{Sensitivity to heavy virtual particles via spin correlations}

Detection of charginos and neutralinos provides not only a way to
measure electroweakino sector parameters (discussed in the previous
sections) but is also sensitive to heavy virtual particles exchanged
in chargino or neutralino production. Chargino production in the MSSM
proceeds by exchange of photon and $Z$ boson in $s$-channel or
sneutrino exchange in $t$-channel.
 

In a study Ref.~\cite{Desch:2006xp}, it was shown that the mass of a
multi-TeV sneutrino can be measured up to precision of $10\%$ at the
ILC.  Forward-backward asymmetries of the final state leptons and
quarks from chargino decays.  These asymmetries are spin-dependent
observables: therefore, a correct evaluation of such asymmetries
requires inclusion of spin correlations between production and decay
of charginos.  The asymmetry is in turn a highly sensitive probe of a
particle exchanged in the $t$-channel, in this case mediated by a
heavy sneutrino.  This dependence, showing also the importance of
including spin correlations, can be seen in Fig.~\ref{fig:afb2}.

\begin{figure}
 \begin{center}
  \includegraphics[width=7.0cm]{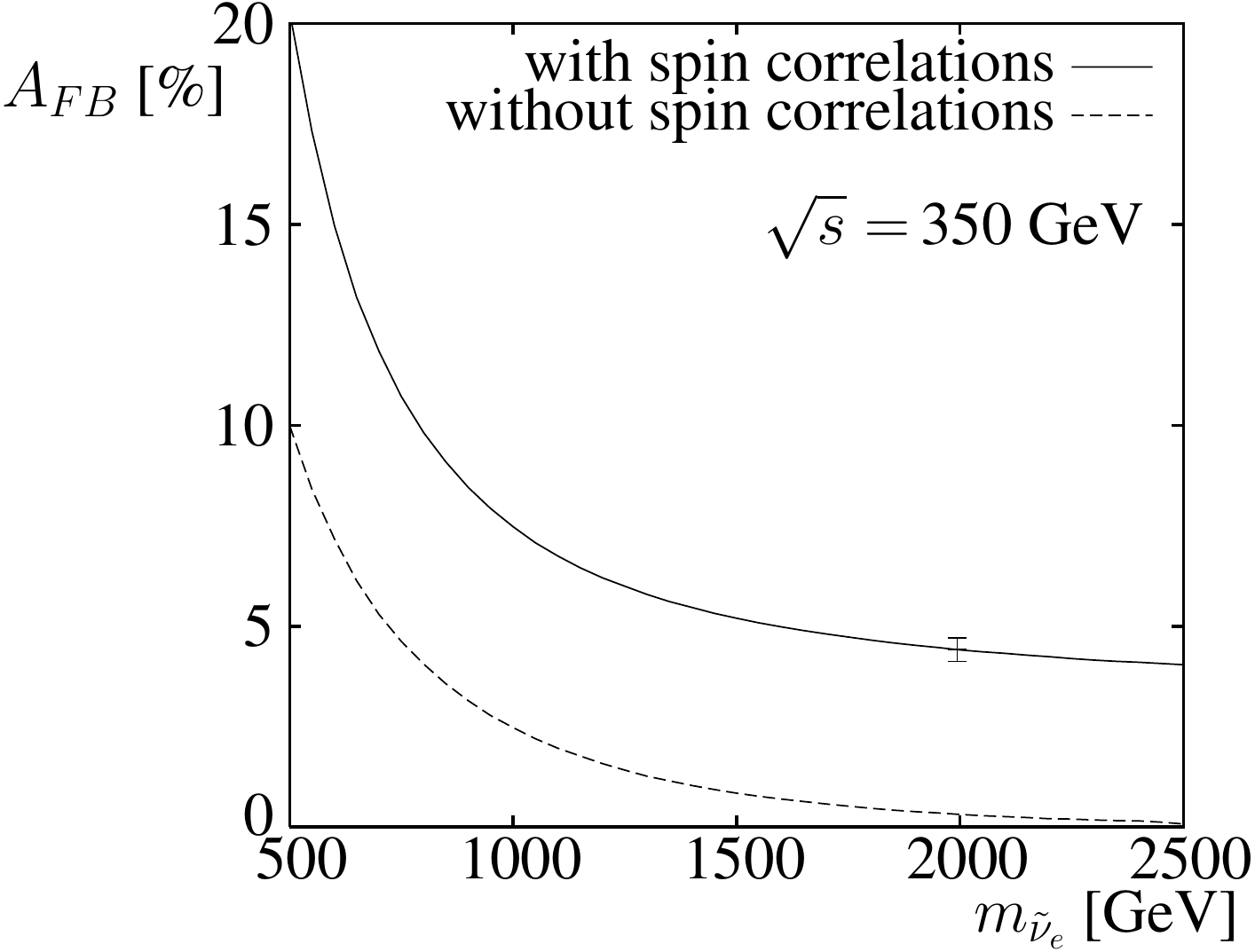}
 \end{center}
\caption{Forward-backward asymmetry of $e^-$ in $e^+e^-\to
  \tilde{\chi}^+_1\tilde{\chi}^-_1$,
  $\tilde{\chi}^-_1\to\tilde{\chi}^0_1\ell^-\bar{\nu}$ as a function
  of $m_{\tilde{\nu}}$ at $\sqrt{s}=350$~GeV and with $P(e^-)=-90\%$,
  $P(e^+)=+60\%$. For a nominal value of $m_{\tilde{\nu}}=1994$~GeV
  the statistical error in the asymmetry is shown\cite{Desch:2006xp}. 
\label{fig:afb2}}
\end{figure}

In a scenario studied in Ref.~\cite{Desch:2006xp}, the following set
of parameters has been assumed:
\begin{eqnarray}
&& M_1 = 60\ \mathrm{GeV},\; M_2 = 121\ \mathrm{GeV},\; \mu=540\ \mathrm{GeV} \nonumber \\
&& \tan\beta = 20,\; m_{\tilde{\nu}} = 2\ \mathrm{TeV}. 
\end{eqnarray}
Using the light chargino production cross sections and mass, 
together with forward-backward asymmetries of decay products, a $\chi^2$ fit has been performed to 
obtain the relevant MSSM parameters. 
The mass of the otherwise kinematically inaccessible sneutrino could be determined with a precision of
\begin{equation}
 m_{\tilde{\nu}} = 2000 \pm 100 \ \mathrm{GeV}
\end{equation}
when forward-backward asymmetries for both leptonic and hadronic decays of chargino are used. 

\subsubsection{Sensitivity to heavy virtual particles via loop effects}

\begin{table*}[tb!]
\renewcommand{\arraystretch}{1.1}
\begin{center}
 \begin{tabular}{lclclclc}
\hline
\multicolumn{4}{c}{\footnotesize{S1}}& \multicolumn{4}{c}{\footnotesize{S2}}\\ 
\multicolumn{4}{c}{\downbracefill}&\multicolumn{4}{c}{\downbracefill}\\
Parameter & Value & Parameter & Value& Parameter & Value & Parameter & Value\\
\hline
$M_1$ & 125& $M_2$ & 250&$M_1$ & 106& $M_2$ & 212\\
$\mu$ & 180 & $M_{A^0}$ & 1000&$\mu$ & 180 & $M_{A^0}$ & 500\\
$M_3$ & 700 & $\tan\beta$ & 10&$M_3$ & 1500 & $\tan\beta$ & 12\\
$M_{e_{1,2}}$ & 1500 & $M_{e_{3}}$ & 1500&$M_{e_{1,2}}$ & 125 & $M_{e_{3}}$ & 106\\
$M_{l_{i}}$ & 1500 & $M_{q_{1,2}}$ & 1500 &$M_{l_{i}}$ & 180 & $M_{q_{i}}$ & 1500 \\
$M_{{q/u}_{3}}$ & 400 & $A_f$ & 650&$M_{{u}_{3}}$ & 450 & $A_f$ & -1850\\
\hline
\end{tabular}
\caption{Table of parameters (with the exception of $\tan\beta$ in GeV), for scenarios 1 (S1) and 2 (S2). Here $M_{(l/q)_{i}}$ and
$M_{(e/u)_{i}}$ represent the left and right handed mass parameters for of a
slepton/squark of generation $i$ respectively, and $A_f$ is the trilinear
coupling for a sfermion $\tilde f$.\label{tab:s1}}
\end{center}
\end{table*}

With the accuracy achievable at a linear collider, one requires loop
corrections in order to draw meaningful conclusions about the
underlying new physics parameters.  For the electroweakino sector, a
study was carried out in Ref.~\cite{Bharucha:2012ya} where one-loop
predictions of the cross section and forward-backward asymmetry for
chargino pair production and of the accessible chargino and neutralino
masses were fitted to expected measurements. A number of one-loop
calculations in the gaugino-higgsino sector can be found in the
literature~\cite{Lahanas:1993ib,Pierce:1993gj,Pierce:1994ew,Eberl:2001eu,
  Fritzsche:2002bi,Oller:2003ge,Oller:2005xg,Drees:2006um,Schofbeck:2006gs,Schofbeck:2007ib,
  Fowler:2009ay,AlisonsThesis,bfmw,Bharucha:2012re}.  Although complex
parameters were not considered in Ref.~\cite{Bharucha:2012ya}, the
renormalization was performed following
Ref's.~\cite{Fowler:2009ay,AlisonsThesis, bfmw,Bharucha:2012re}, where
a dedicated renormalisation scheme in the complex MSSM was defined, in
order that the analysis could easily be extended to the complex case.
At tree level, there are four real parameters to be used in the fit:
$M_1$, $M_2$, $\mu$ and $\tan\beta$, as well as the sneutrino mass,
provided it is beyond the direct reach of the LC.  The study aimed to
provide information about the sensitivity to the remaining MSSM
parameters which contribute to the masses and production amplitude via
virtual effects.  In the fit, the polarised cross-sections and forward
backward asymmetry for chargino production as well as the
$\tilde{\chi}_1^{\pm},\tilde{\chi}_2^{\pm}$ and $\tilde{\chi}^0_{1},
\tilde{\chi}^0_{2}, \tilde{\chi}^0_{3}$ masses-- calculated at NLO in
an on-shell scheme as described in Ref.~\cite{Bharucha:2012ya}-- were
used.  Note that the masses are assumed to have been measured at the
LC using the threshold scan method: however the change in fit
precision if the masses were obtained from the continuum was also
investigated~\cite{AguilarSaavedra:2001rg}.  Further details of the
fit method and errors are given in Ref.~\cite{Bharucha:2012ya}.  The
fit was performed for two scenarios, S1 and S2, shown in
table~\ref{tab:s1}.\footnote{Note that S2 corresponds to S3 in
  Ref.~\cite{Bharucha:2012ya}.}  The scenarios were chosen such as to
be compatible with the current status of direct LHC
searches~\cite{ATLAS:2013lla,Chatrchyan:2012lia}, indirect limits,
checked using {\texttt{micrOmegas
    2.4.1}}~\cite{Belanger:2006is,Belanger:2010gh}, and flavour
physics constraints i.e.\ the branching ratio $\mathcal{B}(b\to
s\gamma)$ and $\Delta(g_\mu-2)/2$.  Note that although in S1, $M_h$ is
not compatible with the recent Higgs results from the
LHC~\cite{ATLAS:2013mma,Chatrchyan:2013lba}, this could easily be
rectified by changing $A_t$ which would have minimal effects on the
results.  The one-loop corrections to the polarised cross-section and
forward backward asymmetry for $e^+e^-\to
\tilde{\chi}^+_1\tilde{\chi}^-_1$ are calculated in full within the
MSSM, following~\cite{bfmw,Bharucha:2012re}, including soft and hard
radiation.

\begin{sloppypar}
For S1, the inputs for the fit included: the masses of the charginos
($\tilde{\chi}_1^{\pm},\,\tilde{\chi}_2^{\pm}$) and three lightest
neutralinos
($\tilde{\chi}^0_{1},\,\tilde{\chi}^0_{2},\,\tilde{\chi}^0_{3}$), the
production cross section $\sigma(\tilde{\chi}^+_1\tilde{\chi}^-_1)$
with polarised beams at $\sqrt{s} = 350$ and $500$ GeV, the
forward-backward asymmetry $A_{FB}$ at $\sqrt{s} = 350$ and $500 \gev$
and the branching ratio $\mathcal{B}(b\to s\gamma)$ calculated using
{\texttt{micrOmegas}}~\cite{Belanger:2006is,Belanger:2010gh}.
\end{sloppypar}

For S2, the inputs for the fit
were the same as in S1, 
with $\sqrt{s}=400$ GeV instead of $350$ GeV and supplemented by the Higgs boson mass $M_h$. 
The sneutrino mass would have been measured. 
The results for S1, given in Table~\ref{tab:ressc1}, 
show the fit to the 8 MSSM parameters: $M_1$, $M_2$, $\mu$, $\tan\beta$,
$m_{\tilde{\nu}}$, $\cos\theta_{\tilde{t}}$, $m_{\tilde{t}_1}$, and
$m_{\tilde{t}_2}$.
We find that the gaugino and higgsino mass parameters are determined with an 
accuracy better than 1\%, while $\tan\beta$ is determined with an accuracy of 
$5\%$, and  2-3\% for the sneutrino mass.
The limited access to the stop sector (Table~\ref{tab:ressc1}) could nevertheless lead to
hints allowing a well-targeted search at the LHC. 
In Table~\ref{tab:ressc1}, we also compare the fit results obtained using masses
of the charginos and neutralinos from threshold scans to those obtained using masses from the
continuum. For the latter, the fit
quality deteriorates,
clearly indicating the need to measure these masses via
threshold scans.
The results for S2 in table~\ref{tab:ressc1} show that the fit is further 
sensitive to $m_{\tilde{t}_2}$, with an accuracy better than $20\%$. In
addition, an upper limit on the mass of the heavy Higgs boson can be placed at 1000~GeV,
at the 2$\sigma$ level. 

\begin{table*}[tb!]
\renewcommand{\arraystretch}{1.1}
\begin{center}
\begin{tabular}{lr@{}l@{ }l r@{}l@{ }l r@{}l@{ }l }\hline
&\multicolumn{6}{c}{\footnotesize S1}&\multicolumn{3}{c}{\footnotesize S2}\\
&\multicolumn{6}{c}{\downbracefill}&\multicolumn{3}{c}{\downbracefill}\\
Parameter & \multicolumn{3}{c}{Threshold fit} & \multicolumn{3}{c}{Continuum fit} & \multicolumn{3}{c}{Threshold fit}\\
\hline
$M_1$ &  $125 $ & $\pm 0.3 $ & $ (\pm 0.7) $ & $125 $ & $\pm 0.6 $ & $ (\pm 1.2)$ &   $106$ &$\pm 0.3$ & $(\pm 0.5) $  \\  
$M_2$ & $250 $ & $\pm 0.6 $ & $ (\pm 1.3) $  & $250 $ & $\pm 1.6 $ & $ (\pm 3)$ & $212$ &$\pm 0.5 $ & $ (\pm 1.0) $   \\  
$\mu$ &  $180 $ & $\pm 0.4 $ & $ (\pm 0.8) $ & $180 $ & $\pm 0.7 $ & $ (\pm 1.3)$ &  $180$ &$\pm 0.4 $ & $ (\pm 0.9) $  \\  
$\tan\beta$ & $10 $ & $ \pm 0.5 $ & $ (\pm 1) $  & $10 $ & $\pm 1.3 $ & $ (\pm 2.6) $ &  $12$ &$\pm 0.3 $ & $ (\pm 0.7) $ \\  
$m_{\tilde{\nu}}$ & $1500 $ & $\pm 24 $ & $ (^{+60}_{-40}) $  & $1500 $ & $\pm 20 $ & $ (\pm 40) $&\multicolumn{3}{c}{$-$}\\
$\cos\theta_{\tilde{t}}$ & $0.15$&$^{+0.08}_{-0.06} $ & $ (^{+0.16}_{-0.09}) $ & $0 $ & $\pm 0.15 $ & $ (^{+0.4}_{-0.3}) $ & \multicolumn{3}{c}{$-$} \\ 
$m_{\tilde{t}_1}$ & $400 $ & $^{+180}_{-120} $ & $ (^{\textrm{at limit}}_{\textrm{at
limit}}) $  & \multicolumn{3}{c}{$-$} &  $430$&$^{+200}_{-130} $ & $ (^{+300}_{-400}) $  \\  
$m_{\tilde{t}_2}$ & $800$ & $^{+300}_{-170} $ & $ (^{+1000}_{-290}) $  &
$800$ & $^{+350}_{-220} $ & $  (^{\textrm{at limit}}_{\textrm{at
limit}}$) & $1520$&$^{+200}_{-300} $ & $ (^{+300}_{-400}) $  \\
$m_{A^0}$ &\multicolumn{3}{c}{$-$}&\multicolumn{3}{c}{$-$}& \multicolumn{2}{r}{$<650$}  & $ (<1000) $ \\ \hline
\end{tabular}\hspace{.8cm}
\end{center}
\caption{Fit results (masses in GeV) for S1 (left) and S2 (right), for masses obtained from threshold scans
(threshold fit) 
and from the continuum (continuum fit). Numbers in brackets denote $2\sigma$
errors.\label{tab:ressc1}}
\end{table*}

Therefore, incorporating NLO corrections was shown to be required for the
precise determination of the fundamental electroweakino 
parameters at the LC, and to provide sensitivity to the parameters
describing particles contributing via loops.
This work will soon be extended to consider both the sensitivity to complex parameters and
the neutralino production cross-section.

\subsubsection{Challenging scenarios: light higgsinos with sub-GeV mass gaps}


In the MSSM, higgsino-like charginos and neutralinos are preferred to
have masses of the order of the electroweak scale by naturalness
arguments, as discussed in Section~\ref{sec:susy1-1} of this review.
If gauginos are heavy, such light $\tilde{\chi}^0_1$,
$\tilde{\chi}^0_2$ and $\tilde{\chi}^\pm_1$ states will be almost mass
degenerate and very challenging to be studied at the LHC.  On the
other hand, the clean experimental environment afforded by the ILC may
allow to perform a measurement of their
properties.  An analysis to assess the
prospects of light higgsino measurements at the ILC, based on detailed
simulations, is presented in~\cite{Berggren:2013vfa}.  
Two scenarios with light charginos and
neutralinos and mass splitting between them in the range of 0.8-2.7~GeV, but all
the other SUSY particle masses in the multi-TeV range were chosen 
(i.e.\ $\mu\sim 170$~GeV, 
$M_1\sim 5$~TeV, $M_2\sim 10$~TeV, $\tan\beta\sim 48$).

For such small mass differences, the decay products of chargino are 
soft pions and leptons, 
while the largest decay mode  of $\tilde{\chi}^0_2$ is to photon and LSP. 
Despite the fact  that these 
final states will suffer from large SM backgrounds, 
a suitable set of cuts provides separation of the
the signal~\cite{Berggren:2013vfa}. 
The effective tool
for background rejection here is the tag 
of ISR photons recoiling against the chargino or neutralino system.

The masses of chargino and neutralino $\tilde{\chi}^0_2$ are then
reconstructed from the distribution of the reduced center-of-mass
energy of the system recoiling against the hard ISR photon.  The
expected mass resolution ranges from $1.5$ to $3.3$~GeV depending on
the scenario.  The mass difference between $\tilde{\chi}^\pm_1$ and
the LSP is measured by fitting energy distribution of soft pions in
the respective decays.  The accuracy up to $40$~MeV can be obtained
for $m_{\tilde{\chi}^\pm_1} - m_{\tilde{\chi}^0_1} = 770$~MeV.
Finally, the polarised cross sections for chargino pair production and
$\tilde{\chi}^0_1 \tilde{\chi}^0_2$ can be measured with order of per
cent statistical accuracy.  These results are greatly encouraging for
the potential of a linear collider to tackle even such difficult scenarios.
Still, detailed studies with full detector simulation and
reconstruction and the incorporation of machine--induced backgrounds
will be necessary to fully quantify this potential.

The fundamental MSSM parameters $M_1$, $M_2$, $\mu$ and $\tan\beta$
can be extracted from these types of observables.  For the specific
benchmarks chosen, the $\mu$ parameter can be determined to $\pm 4\%$.
For the gaugino mass parameters, $M_1$ and $M_2$, the lower bounds can
be set in the multi-TeV range, depending on the value of $\tan\beta$,
which cannot be fixed from the above measurements alone, see
Fig.~\ref{fig:higgsinofit}.  If the uncertainties could be reduced by
a factor of 2 by including additional observables or increasing the
integrated luminosity, the constraints on gaugino mass parameters
would be significantly more restrictie and less dependent on
$\tan\beta$.

\begin{figure}
 \begin{center}
  \includegraphics[width=7.0cm]{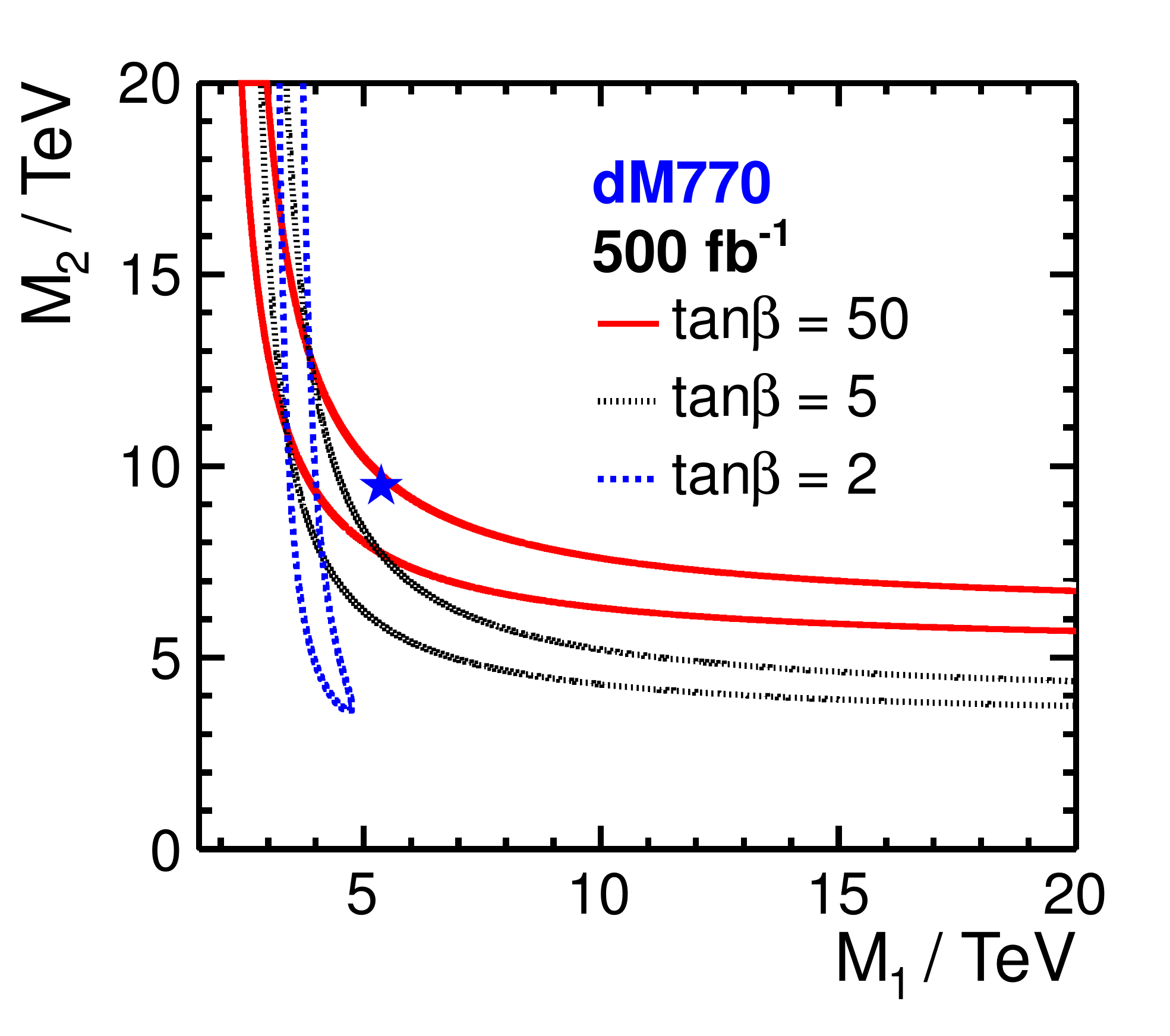}
 \end{center}
\caption{The contours for determination of $M_1$ and $M_2$ in scenario with $m_{\tilde{\chi}^\pm_1} - m_{\tilde{\chi}^0_1} = 770$~MeV. The star denotes input values. See Ref.~\cite{Berggren:2013vfa} for more details. \label{fig:higgsinofit}}
\end{figure}

\subsubsection{Parameter fits}


The determinaton of SUSY parameters in global fits using hypothetical
measurements at the ILC has been studied in
detail~\cite{Bechtle:2009ty} using the Fittino~\cite{Bechtle:2004pc}
package for model points such as
SPS1a'~\cite{Allanach:2002nj}. However, this point is now excluded
from generic searches for SUSY at the LHC (see e.g.~\cite{Aad:2012fqa}
for early exclusions). Since then, no new complete analysis have been
preformed for parameter determinations in global fits using data from
low energy precision experiments, cosmological measurements, Higgs
mass and rate measurements, up-to-date LHC constraints on SUSY
production, and hypothetical ILC measurements. Therefore, in this
section we revert to the existing SPS1a' results, keeping in mind that
measurements of SUSY production properties at a currently realistic
SUSY point would be less favourable both for the LHC and for the
ILC. The reason is that the higher mass scale of first- and
second-generation squarks and gluinos very strongly reduces the
statistics in potential SUSY cascade decay signatures at the LHC.  At
the same time, given the current LHC bounds,
the resolution of the small mass splittings between particles in the
cascade decays typically required to allow light gauginos and sleptons
at $m_{\tilde{\ell},\tilde{\chi}}\leq 250$ or $500$\,GeV 
is more challenging, however, yet possible at the ILC.

\begin{table*}
  \caption{Result of the fit of the CMSSM model to the precision
    measurements and to the hypothetical results from LHC with ${\cal
      L}^{\mathrm{int}}=300\,\mathrm{fb}^{-1}$ and ILC.}
  \label{tab:susy:fits:LELHCILC:mSUGRA}
   \begin{center}
      \begin{tabular}{lcccc}
        \hline
        Parameter & Nominal value & Fit & LHC Uncertainty & ILC Uncertainty \\
        \hline
        $\tan\beta$       &  10   &  9.999           & $\pm$   0.36  & $\pm$   0.050   \\
        $M_{1/2}$ (GeV)          & 250   &  249.999   & $\pm$   0.33  & $\pm$   0.076  \\
        $M_0$ (GeV)            & 100   &  100.003    & $\pm$   0.39  & $\pm$   0.064   \\
        $A_0$ (GeV)            & $-$100  & $-$100.0   & $\pm$  12.0  & $\pm$   2.4   \\
      \hline
      \end{tabular}
    \end{center}
\end{table*}

As a relative comparison between the possible LHC and LHC+ILC
performance, either SUSY models constrained at the GUT scale (such as
the CMSSM) or models defined at the TeV scale can be used. 
The CMSSM results from~\cite{Bechtle:2009ty} are shown in
Table~\ref{tab:susy:fits:LELHCILC:mSUGRA}. The LHC result is based on
actual precision measurements from B-factories and on $(g-2)_{\mu}$,
on the neutralino relic abundance $\Omega_{CDM}h^2$, 
on LEP1 SM precision measurements, and on
hypothetical LHC measurements of the Higgs mass and of kinematical
quantities measured in SUSY cascade decays. For a detailed list
see~\cite{Bechtle:2009ty}. For the ILC, realistically modelled studies
of Higgs mass, cross-section and branching fraction measurements,
hypothetical measurements of kinematical edges in SUSY decays, and a
large amount of measurements of cross-section times branching
fractions for every kinematically accessible SUSY decay chain at
sufficient rate is assumed. A time-consuming running scenario with
measurements at $\sqrt{s}=400,500$ and $1000$\,GeV at different
combinations of beam polarizations is employed to disentangle the
mixing of the gauginos and heavy sleptons. 

\begin{sloppypar}
The results in Table~\ref{tab:susy:fits:LELHCILC:mSUGRA} clearly show a
significant improvement by a factor of about 5 between the LHC results
and the same fit but now including additional ILC information. 
However, an even stronger
improvement is observed when moving towards a SUSY model with
significantly more freedom in the parameter choice. One possiblity is
the pMSSM. Here, a minimal flavour-violating MSSM with unification in
the first two generations is constructed at the TeV scale, here called
the MSSM18. The value $m_{t}$ is kept fixed due to the high expected accuracy at
the ILC. This is a very favourable asumption for the LHC, because for
a fit without information on $m_{t}$ from the ILC, the parametric
uncertainties-- especially on the Higgs mass-- would be expected to
degrade the precision of the fit result from the LHC. For details on
the model, see~\cite{Bechtle:2009ty} again.
\end{sloppypar}

\begin{figure}
\vspace{-5cm}
  \includegraphics[width=0.49\textwidth]{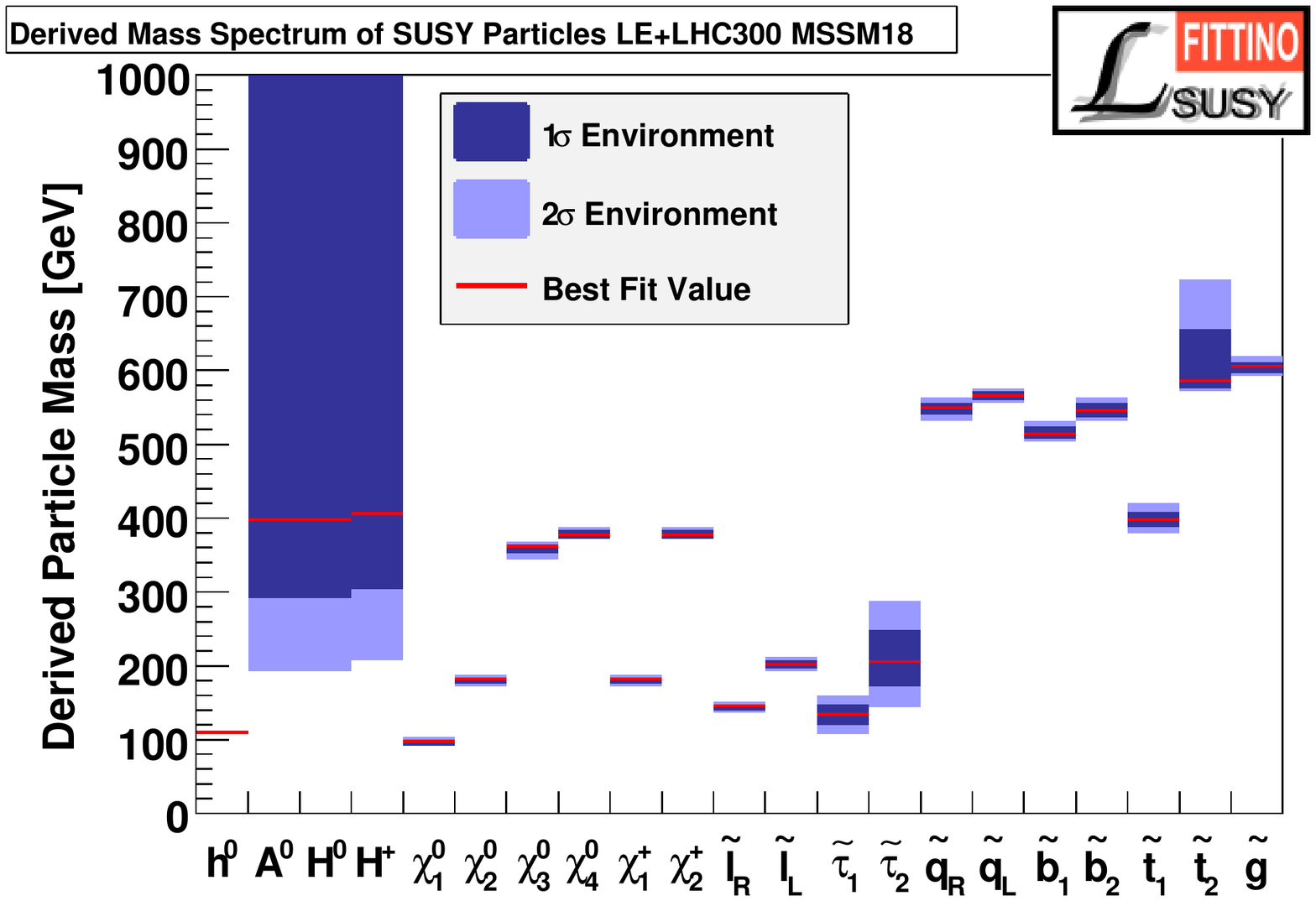}
  \caption{SUSY mass spectrum consistent with the existing low-energy
    measurements and the hypothetical LHC measurements at ${\cal
      L}^{\mathrm{int}}=300\,\mathrm{fb}^{-1}$ for the MSSM18 model.
    The uncertainty ranges represent model dependent uncertainties of
    the sparticle masses and not direct mass measurements.}
  \label{fig:susy:fits:LELHC:MSSM18:massDist300fb}
\end{figure}

\begin{figure}
\vspace{-5cm}
  \includegraphics[width=0.49\textwidth]{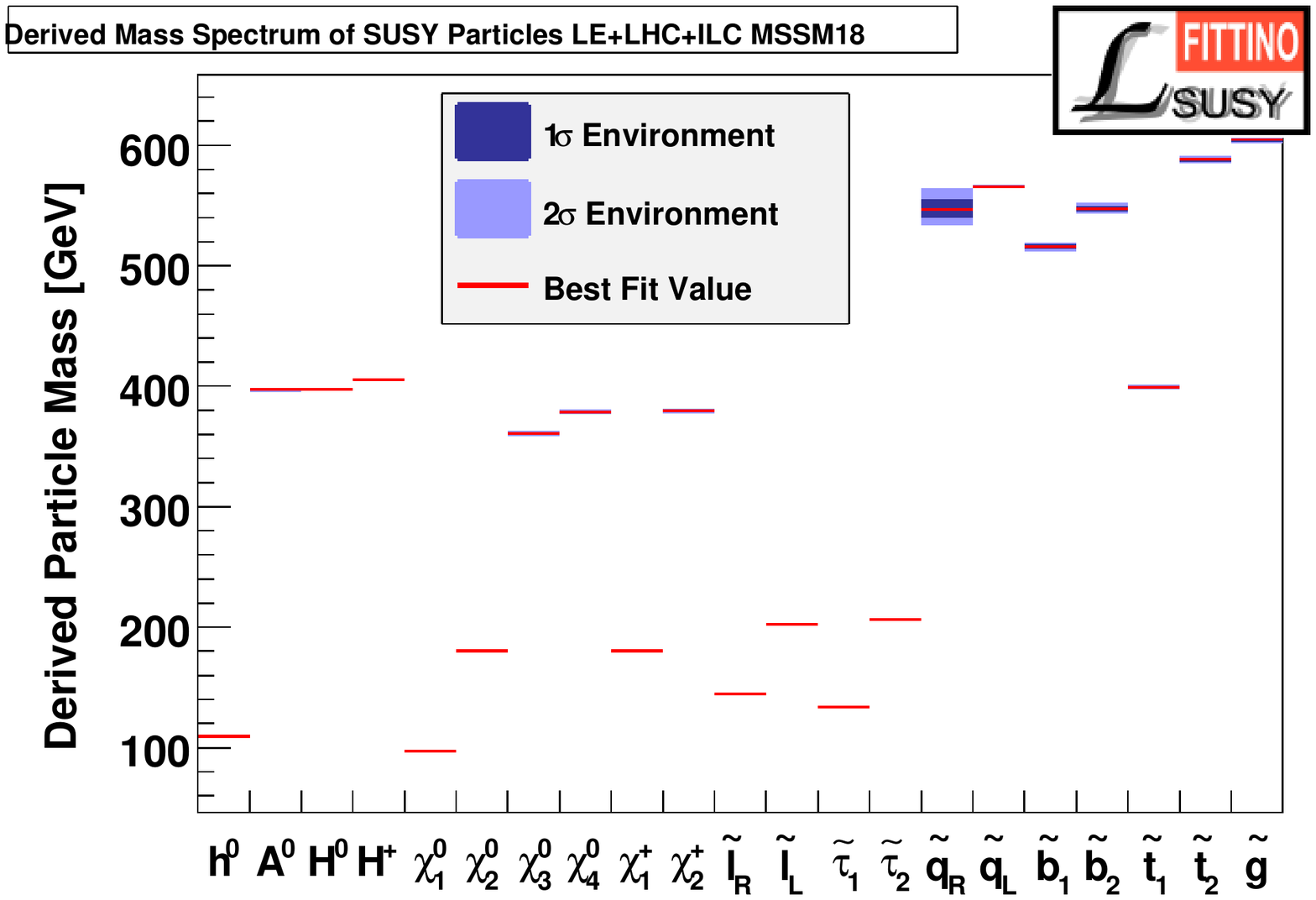}
  \caption{Derived mass distributions of the SUSY particles using low
    energy measurements, hypothetical results from LHC with ${\cal
      L}^{\mathrm{int}}=300\,\mathrm{fb}^{-1}$ and hypothetical
    results from ILC. When comparing to
    Fig.~\ref{fig:susy:fits:LELHC:MSSM18:massDist300fb}, please note
    the difference in the scale.}
  \label{fig:susy:fits:LELHCILC:MSSM18:massDist}
\end{figure}

For a graphical comparison of the power of the ILC at a very
favourable, albeit now excluded model point, see the difference
between the LHC precision of a model-dependent determination of a SUSY
mass spectrum in Fig.~\ref{fig:susy:fits:LELHC:MSSM18:massDist300fb}
and the corresponding spectrum for the added ILC information in
Fig.~\ref{fig:susy:fits:LELHCILC:MSSM18:massDist}. An enormous
improvement is observed in the heavy Higgs sector, stemming from the
hypothetical direct measurements of the heavy Higgs bosons at the ILC,
while they would have remained inaccessible at the LHC. Also for the
other masses, improvements of a factor of 10 to 100 are
possible~\cite{Bechtle:2009ty}. For the SPS1a' like MSSM18, the
added benefit of the ILC over the LHC is much more apparent due to the
larger freedom in the model. For a model with only four free
parameters, such as the CMSSM, a few measurements with relatively good
precision are enough to constrain the parameters in a reasonable
range, such as for the LHC in the hypothetical SPS1a' CMSSM. However,
once the less accessible states decouple from the more accessible
ones, such as in the MSSM18, the direct information on states like the
light CP-even Higgs boson $h$ and the squark mass scales does not
suffice to constrain less accessible states anymore (like the heavy
Higgses) since they are controlled by additional parameters like $m_A$
and $X_{f}$ in the MSSM18: these cannot be easily accessed otherwise. 
At the $e^+e^-$ LC, however, the high precision measurements of the
full Higgs sector (as for SPS1a') and the very high precision
measurements of sparticle masses and couplings,  would have allowed to
disentangle the mixings and mass parameters in the gaugino, the heavy
slepton and the stop sector individually. Such determinations reduce the
model-dependence dramatically and improve the fit precision
accordingly, by providing independent precise probes of all degrees of
freedom of the model.

\subsubsection{Extrapolation to GUT scale}
\label{sec:extrap}

As discussed in Section~\ref{sect:susymodels}, many of the commonly
used SUSY models impose strong assumptions at the high scale inspired
by suppositions on the SUSY breaking mechanism.  In the CMSSM-- with
the input parameters $m_0$, $m_{1/2}$, $\tan\beta$, $A_0$, sign$\mu$
at the GUT scale $M_{\rm GUT}\approx 2\cdot 10^{16}$~GeV-- all gauge
couplings $\alpha_{1,2,3}$ and also all gaugino masses $M_{1,2,3}$ and
scalar masses unify at $M_{\rm GUT}$.

\begin{sloppypar}
Generally, in order to test such model-dependent assumptions, one can
start from a precisely measured particle spectrum at lower energies
and extrapolate the underlying parameter to higher energies, up to
$M_{\rm GUT}$, as described in \cite{Blair:2002pg}. The evolution of
the parameters happens via applying the renormalization group
equations (RGE).  In practically all studies, it is assumed to combine
measurements of the non-colored spectrum at the LC with measurements
of the coloured spectrum at the LHC.
\end{sloppypar}

As one example, we choose benchmark `Model I' from Ref.~\cite{Linssen:2012hp} 
with the GUT scale parameters $m_0=966$~GeV,
$m_{1/2}=800$~GeV, $A_0=0$, $\tan\beta=51$, sign$(\mu )=+1$ which
determine the particle spectrum at low energy.  In \cite{Linssen:2012hp}, 
it has been shown that the masses of neutralinos
and the sleptons of the first two generation can be measured with a precision
of 1-2\% at a 3-TeV collider. 
In addition, one assumes to measure the gluino mass $m_g=1812$~GeV with 5\% precision at the LHC
and at a 3\% level for all other sfermion masses at the LC. Based on the mass
and cross section measurements of the neutralino/chargino sector, one
can reconstruct the quantities at tree-level: $M_1$, $M_2$, $\mu$ and
$\tan\beta$. 

Since we measure on-shell masses, but use  $\overline{DR}$ parameters for
the evolution of parameters, the corresponding shifts must be calculated.
This intertwines the different sectors: naively one would expect
that the relative precision of the masses transfers
one to one to the precision on the gaugino mass parameters. However,
in case of the gluino mass parameters, the uncertainty due to the squark
mass measurements can increases the uncertainty on $M_3$ by up to a factor 2, 
e.g.\ instead of a five per-cent uncertainty one obtains roughly a ten
per-cent uncertainty. At the level of one-loop RGEs, the relative uncertainties
are approximately scale invariant as at this level $M_i/\alpha_i$ is an
RGE invariant. However, at the two-loop level, also the trilinear $A$-parameters
of the third generation enter and, thus, one should know them 
to a precision of at least 40\% as otherwise the uncertainties
at the high-scale can be significantly worse compared to the one at the electroweak
scale. The trilinear couplings can be determined via cross section measurements
and sfermion decays involving Higgs bosons 
(or decays of heavy Higgs bosons into sfermions)\cite{Blair:2002pg,Adam:2010uz}.
Under the above assumption, we find a unification of the
gaugino mass parameters to about 10\%: see Figure~\ref{fig:gut-susy} (top panel).

In the evaluation of the sfermion mass parameters, also the gaugino mass
parameters enter where in particular $M_3$ is important for the
evolution of the squark mass parameters. In case of third generation
sfermions and the Higgs mass parameters, also large Yukawa couplings
as well as the  $A$-parameters enter the RGEs and intertwine them
in a non-trivial way. Taking the same assumptions as above,  we find
a clear overlap between all scalar mass parameters when running up to
the GUT scale, see Fig.~\ref{fig:gut-susy} (middle and bottom panel),
pointing clearly to the $1000$~GeV region for $m_0$. 

\begin{figure}[h]
\begin{center}
\includegraphics[width=5.0cm]{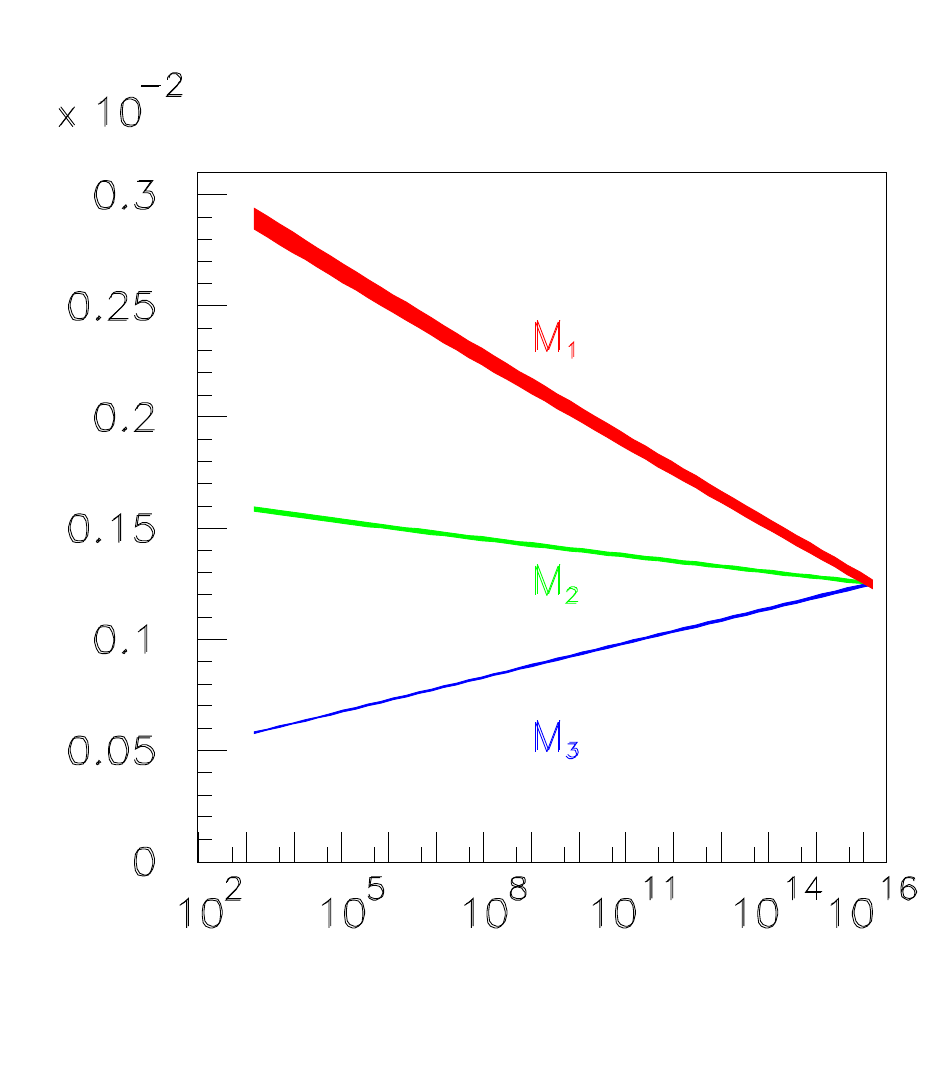}
\includegraphics[width=5.0cm]{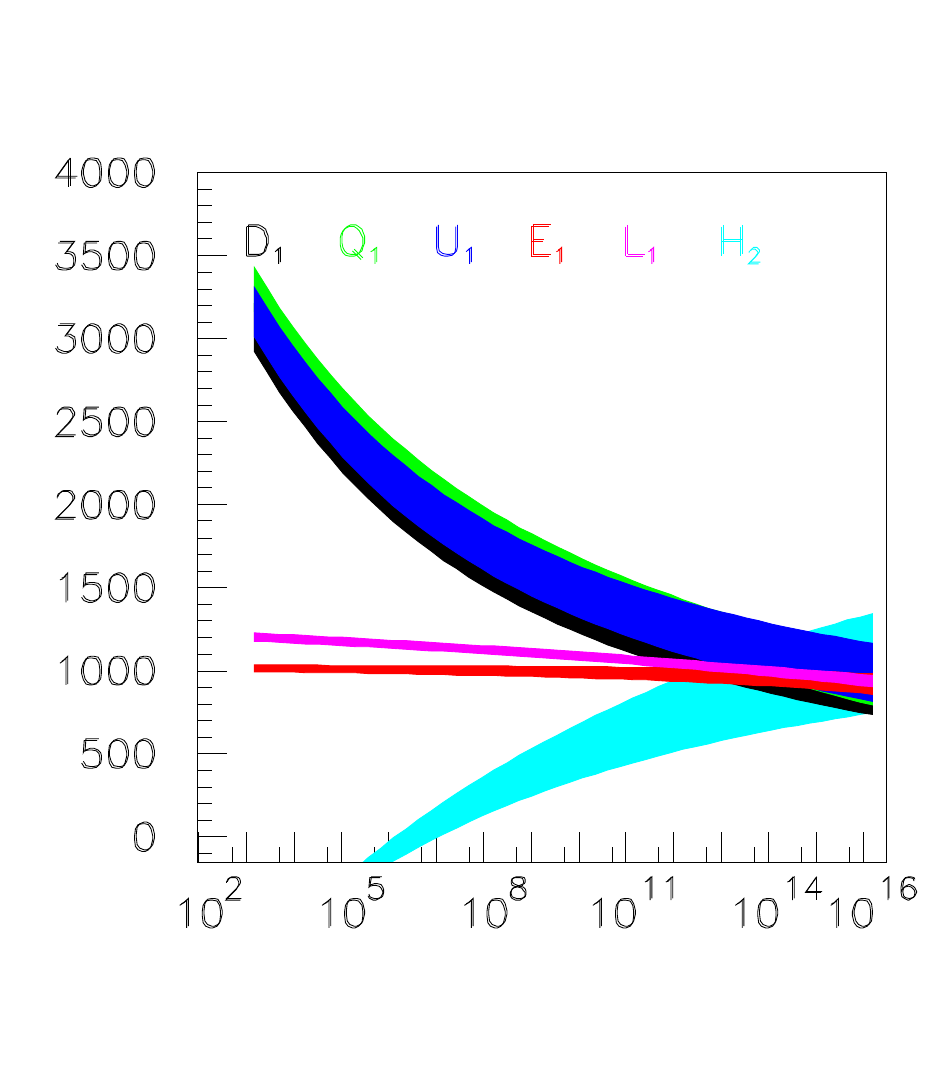}
\includegraphics[width=5.0cm]{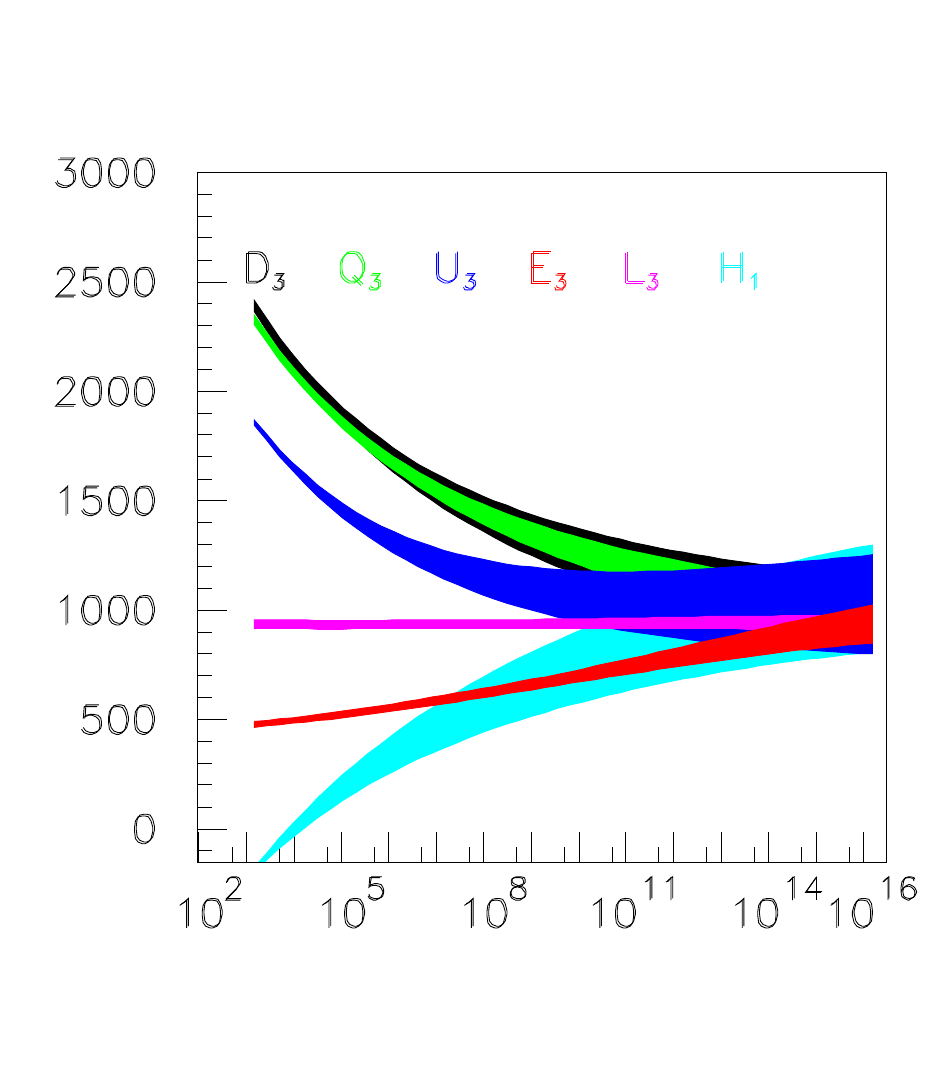} 
\end{center}
\caption{\label{fig:gut-susy}
Evolution of gaugino and sfermion (1. and 3. generation)
  parameters in the CMSSM for $m_0=966$~GeV, $m_{1/2}=800$~GeV,
  $A_0=0$, $\tan\beta=51$, sign$\mu=+1$\cite{Linssen:2012hp} to the GUT scale.}
\end{figure}

\subsection{Lepton Flavour and CP Violation}
\label{sec:susy6}

The general structure of supersymmetry admits several possible
extensions to the MSSM, either by switching on new couplings or
introducing new parameters, such as CP-violating phases
or adding new fields, each resulting in new, specific phenomenology.
Because of its versatility and the limited SM backgrounds, a linear
collider is best suited to investigate these scenarios.  In this
section, we review the sources of lepton flavour
and CP violation in extended SUSY models and their phenomenology in
$e^+e^-$ collisions.

\subsubsection{Lepton Flavour Violation}

A significant body of data from atmospheric, solar, reactor and
accelerator neutrino experiments \cite{neutrinodata} have revealed the
non-zero value of neutrino masses and oscillations with near-maximal
$\nu_\mu$-$\nu_\tau$ and large $\nu_e$-$\nu_\mu$ mixing.  A very
attractive explanation for the smallness of neutrino masses and their
mixings is a seesaw mechanism embedded within the framework of SUSY
models.  In this case\cite{seesaw}, masses and mixings in the neutrino
system are caused by very heavy right-handed Majorana neutrinos with
masses close to the GUT scale. Even if the sfermion mass matrices are
diagonal at the GUT scale, flavour violating mixings are induced
radiatively\cite{Borzumati:1986qx}. A substantial $\nu_\mu$-$\nu_\tau$
mixing leads to large $\tilde{\mu}_L$-$\tilde{\tau}_L$ and
$\tilde{\nu}_\mu$-$\tilde{\nu}_\tau$ mixings.  It is natural to expect
that charged lepton flavour violation (cLFV) should occur at some
level thus raising the interesting possibility of observing these
processes in low-energy rare decays $\mu\to e \gamma$, or
$\tau\to\mu\gamma$, or $\mu$-$e$ conversion \cite{muegamma,muth} and
at a high energy $e^+e^-$ collider.

In the Standard Model, cLFV processes are strongly suppressed due to
the GIM mechanism.  However, in SUSY, virtual super-partner loops may
provide an enhancement \cite{muth} making them observable.  Moreover,
if sleptons are directly produced, cLFV can also be directly tested in
their production and decay processes.  For nearly degenerate sleptons,
supersymmetric LFV contributions to low-energy rare decay processes
are suppressed as $\Delta m_{\tilde \ell}/m_{\tilde \ell}$ through the
superGIM mechanism and constraints from the yet unobserved radiative
decays $\ell_i\to \ell_j\gamma$ are not very stringent.  On the other
hand, in direct decays of sleptons, this kind of supersymmetric lepton
flavour violation is suppressed only as $\Delta m_{\tilde
  \ell}/\Gamma_{\tilde \ell}$\cite{ArkaniHamed:1996au}.  Since
$m_{\tilde \ell}/\Gamma_{\tilde \ell}$ can be large, spectacular
signals may be expected leading to possible discoveries at the LHC and
in particular at future lepton collider experiments.  Among the
possibilities considered so far, there is slepton pair production at a
linear collider as well as signals from electro-weak gaugino
production and their subsequent cascade decays $\tilde{\chi}^0_2\to
\tilde{\chi}^0_1 + e^\pm\mu^\mp$, $\tilde{\chi}^0_2\to
\tilde{\chi}^0_1 + \mu^\pm\tau^\mp$, at both a linear collider and the
LHC \cite{ArkaniHamed:1996au,Hisano:1998wn,atLC,atLHC}.

%
At a LC, the cLFV signals can be looked for directly in slepton pair production, for example
\begin{eqnarray}
e^+e^- & \rightarrow &
\tilde{\ell}^-_i\tilde{\ell}^+_j \rightarrow  \tau^+\mu^- \tilde{\chi}^0_1
\tilde{\chi}^0_1, \nonumber \\
e^+e^- & \rightarrow &
\tilde{\nu}_i\tilde{\nu}^c_j  \rightarrow  \tau^+\mu^- \tilde{\chi}^+_1
\tilde{\chi}^-_1 \label{eq:pairlfv}
\end{eqnarray}
or indirectly  via sleptons produced singly in  chain decays  of heavier charginos and/or neutralinos 
$\tilde{\chi}_2\to \ell_i\tilde{\ell}_j$, $\tilde{\ell}_j\to\ell_k\tilde{\chi}_1$:
\begin{eqnarray}
e^+e^- & \rightarrow &
\tilde\chi^+_2\tilde{\chi}^-_1   \rightarrow  \tau^+\mu^- 
 \tilde\chi^+_1\tilde\chi^-_1  \nonumber\label{eq:pairprod}\\
e^+e^- & \rightarrow &
\tilde\chi^0_2  \tilde{\chi}^0_1 \rightarrow  \tau^+\mu^-
\tilde\chi^0_1\tilde\chi^0_1. \label{eq:singleprod}
\end{eqnarray}
With  $\tilde{\chi}^\pm_1 \rightarrow \tilde{\chi}^0_1 f\bar{f}'$, and
$\tilde{\chi}^0_1$ escaping detection, the signature therefore would
be $\tau^{\pm}\mu^{\mp}+jets+ {E\!\!\!/}_T$,  $\tau^{\pm}\mu^{\mp}+
\ell + {E\!\!\!/}_T$, or $\tau^{\pm}\mu^{\mp}+ {E\!\!\!/}_T$, 
depending on hadronic or leptonic
$\tilde{\chi}^\pm_1$ decay mode. 

In the case of narrow widths and small mass differences between the sleptons of different generations, 
$\Delta \tilde{m}_{ij} \ll \tilde{m} = \frac{1}{2}(m_2+m_3)$ 
and
$\tilde{m}\overline{\Gamma}_{ij} 
\simeq (\tilde{m}_i\Gamma_i+\tilde{m}_j\Gamma_j)/2\ll
\tilde{m}^2$,  
and assuming a pure 2-3 inter-generation mixing between $\tilde\nu_\mu$ and $\tilde\nu_\tau$, generated by a
 near-maximal mixing angle $\tilde{\theta}_{23}$, and ignoring any mixings
 with $\tilde\nu_e$\footnote{Complete expressions are usually used for phenomenological investigations.}, 
the cross sections for $\tau^+\mu^-$ in the final state simplify considerably~ \cite{ArkaniHamed:1996au,Kalinowski:2001tq}. 
For $\tau^+\mu^-$ produced in the decays of a pair of sleptons, Eq.~(\ref{eq:pairprod}), the cross section  
can be approximated as:     
\begin{eqnarray}
\sigma^{\rm pair}_{23} = \chi_{23}(3-4 \chi_{23})  
\sin^2 2\tilde{\theta}_{23} \;
\times \sigma_0\times 
Br, 
\label{eq:pair}
\end{eqnarray}
whereas for $\tau^+\mu^-$ produced from  the gaugino decay, Eq.~(\ref{eq:singleprod}), it takes the form
\begin{eqnarray}
\sigma^{\rm casc}_{23}=\chi_{23} \sin^2 2\tilde{\theta}_{23}\;
\times \sigma_0\times 
Br .
\label{eq:casc}
\end{eqnarray}
\begin{sloppypar}
Here the cLFV effect is 
taken into account by the factors $\sin^2 2\tilde{\theta}_{23}$ and 
$\chi_{23} \equiv x_{23}^2/2(1+x_{23}^2)$
where $x_{23} \equiv \Delta \tilde{m}_{23}/\overline{\Gamma}_{23}$. The difference between Eq.~(\ref{eq:pair}) and
Eq.~(\ref{eq:casc}) is due to the correlated slepton pair production
in the processes Eq.~(\ref{eq:pairprod}).  
In the above expressions, $\sigma_0$ is the corresponding sparticle
pair-production cross section in $e^+e^-$ collision and
${Br}$ is the product of relevant branching ratios for the
corresponding decay chains without cLFV contributions. 
\end{sloppypar}

\begin{sloppypar}
The potential of exploring the cLFV at a LC has recently been revisited in final states with 
$\tau\mu$ \cite{Carquin:2011rg}  and $e\mu$ \cite{Abada:2012re}. 
Both analyses adopted the cMSSM framework with benchmark points chosen to be consistent with the
limits from the LHC searches and cosmological relic LSP density. The benchmarks feature relatively 
low values of $m_0$ (compared to $m_{1/2}$) 
to provide a relatively light slepton spectrum accessible at a LC while avoiding the LHC bounds on the strongly 
interacting sector.  To assess the sensitivity of the cross section measurements to the LFV terms  
$(\delta_{LL,RR})_{ij}$, where  the 
 flavour-mixing entries encode  the inter-generation elements of the slepton mass matrix
$(\delta_{XX})_{ij}=(M^2_{XX})^{ij}/(M^2_{XX})^{ii}$, $(X=L,R)$, 
Fig.~\ref{fig:susy-lfv1} shows current constraints and possible LFV effects for reference points. 
Despite the SM and SUSY charged current backgrounds, the expected number of 
signal events should allow us to probe cLFV in extensive regions of the SUSY seesaw parameter space.   
Both direct slepton pair production and sleptons produced in cascade decays may provide interesting signals in the 
cosmologically-favoured region of the supersymmetric parametric space. In comparison to the LHC, the LC could provide
additional insights by virtue of its greater kinematic range for slepton production and its sensitivity to both RR and 
LL mixing.
\end{sloppypar}
\begin{figure}[h]
\begin{center}
\includegraphics[width=4.0cm]{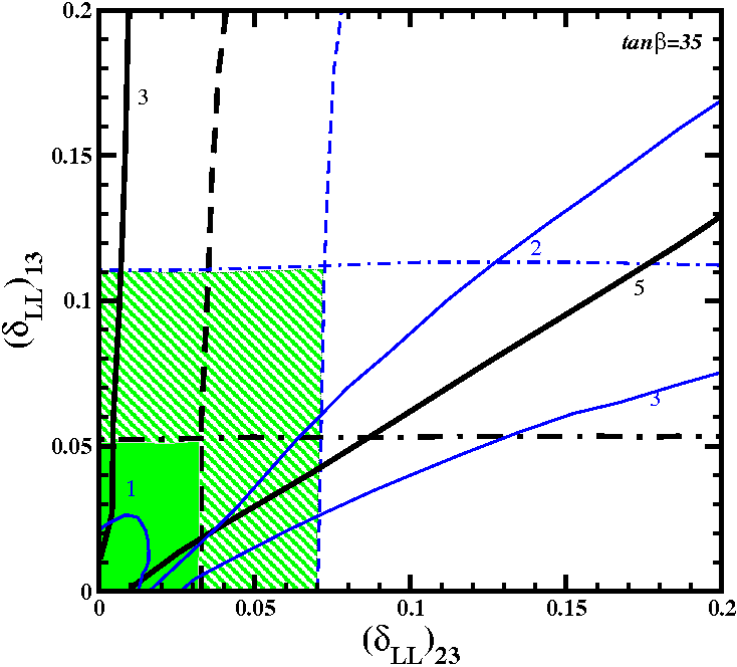}
\includegraphics[width=4.0cm]{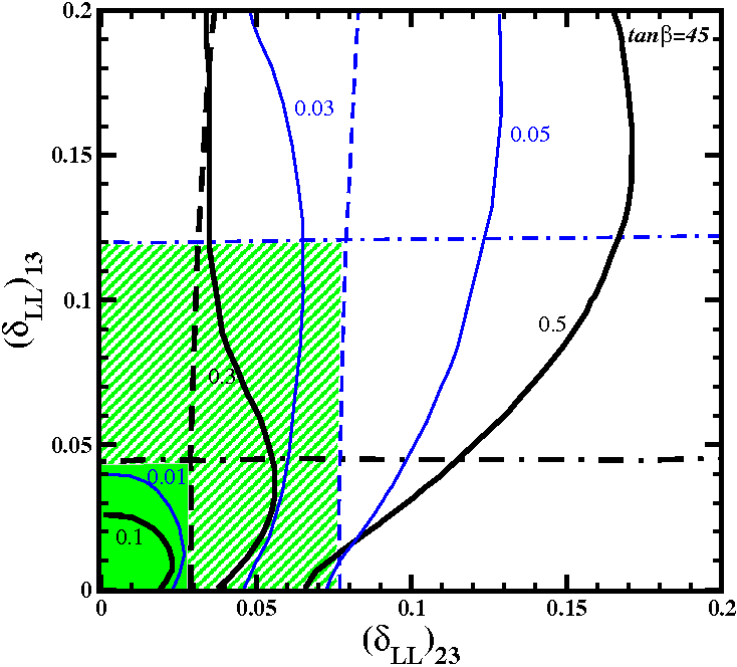} 
\includegraphics[width=4.0cm]{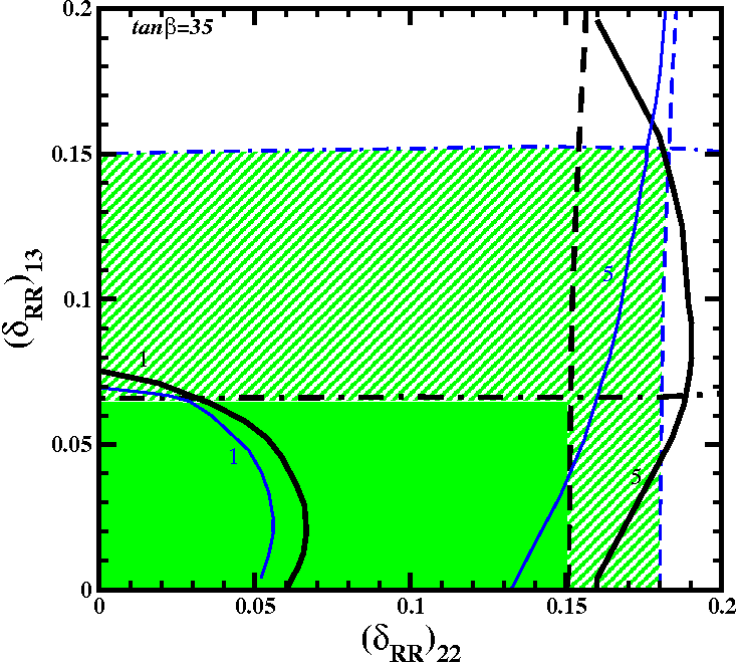} 
\includegraphics[width=4.0cm]{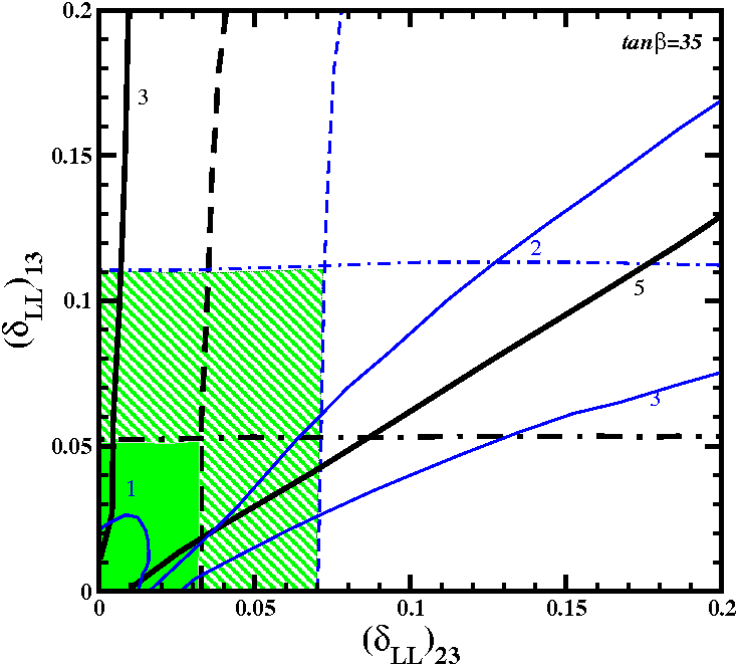}
\end{center}
\caption{Constraints on the magnitudes of the
mixing parameters and possible LFV effects for reference points from \cite{Carquin:2011rg}.  The shaded areas are those
allowed by current limits on 
BR($\tau\rightarrow e \gamma$) (dot-dash line) and 
BR($\tau\rightarrow \mu \gamma$) (dash line) using four different reference points  
(shown by the thick lines bounding the solid shaded areas and the thin blue lines bounding the ruled shaded areas). 
The solid lines are contours of $\sigma(e^+e^-\rightarrow \tau ^\pm\mu^\mp +2 \chi^0)$ in fb for
$\sqrt{s}=2000 {\rm ~GeV}$.}
\label{fig:susy-lfv1}
\end{figure}

Lepton flavour violation can also reveal itself in other processes such as $e^+e^-\to\tilde{\chi}^+_i\tilde{\chi}^-_j$.  
This process proceeds through $s$-channel $\gamma/Z$ and also $t$-channel $\tilde{\nu}_e$ exchange.  In the LFV scenario,
the $\tilde{\nu}_e$ is a mixture of three mass eigenstates. The production cross section for chargino pair production may 
change by a factor of two or more in the presence of $\tilde{\nu}_e-\tilde{\nu}_\tau$ mixing even if current bounds on LFV 
rare lepton decays are significantly improved (see Fig.\ref{fig:susy-lfv2}) \cite{HohenwarterSodek:2007az}. The effect of 
$\tilde{\nu}_e-\tilde{\nu}_\mu$ mixing, due to stronger experimental bounds, is less dramatic, as seen in the right panel 
of Fig.\ref{fig:susy-lfv2}.
\begin{figure}[h]
\begin{center}
\hspace*{-0.5cm}
\includegraphics[width=0.23\textwidth]{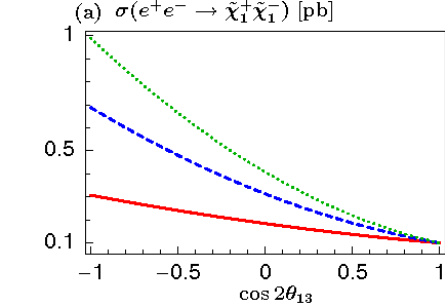}
\includegraphics[width=0.23\textwidth]{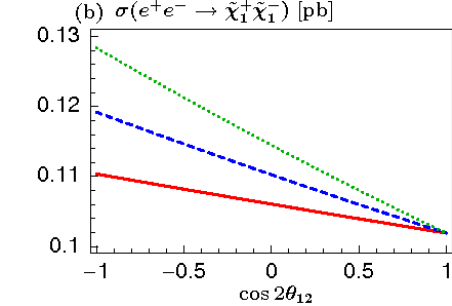} 
\end{center}
\caption{Cross section $\sigma(e^+e^-\to\tilde{\chi}^+_1\tilde{\chi}^-_1)$ as a function of the mixing parameter $\cos2\theta_{13}$ (a) and $\cos2\theta_{12}$ (b) at a LC with cm energy of 500 GeV and polarised beams: $P_L$=-0.9 for electrons and $P_L$=0.6 for positrons.  Details of assumed scenarios (a) and (b) are in \cite{HohenwarterSodek:2007az}.}
\label{fig:susy-lfv2}
\end{figure}

\subsubsection{CP Violation}

Since the first observation of CP violation almost fifty years ago, 
the ``cryptic message from Nature'' it conveys still needs to be 
deciphered in full. An attractive feature of SUSY is that it allows for new sources of CP violation which are needed if the 
baryon-antibaryon asymmetry observed in the universe is to be explained by particle physics.  Compared to the case of CP-conserving SUSY, 
new CP phases appearing in supersymmetry may change masses, cross sections, decay branching ratios, etc.\ providing many  possible ways 
to detect and measure them at colliders.  
Since such observables are CP-even, CP-violating effects may be distinguished from fortuitous combinations of parameters not invoking CP-violating 
phases only by the joint analyses of several CP-even observables. For example, an observation of $s$-wave excitation above respective thresholds 
of three non-diagonal pairs of neutralinos  \cite{Choi:2001ww}, or the observation of simultaneous sharp $s$-wave excitations of the production 
cross section $\sigma(e^+e^- \to \tilde{\chi}^0_i\tilde{\chi}^0_j)$  ($i \ne j$) near threshold and the $f\bar f$ invariant mass distribution near 
the end point  of the decay  $\tilde{\chi}^0_i\to \tilde{\chi}^0_j f\bar f$ \cite{Choi:2003hm} is a qualitative, unambiguous evidence for CP 
violation in the neutralino system. A linear collider of sufficient energy can perform all these measurements.

The most direct way to detect CP-violation is to construct CP-odd observables which cannot be mimicked by other parameters of the theory. 
Such quantities typically involve asymmetries constructed as triple products of momenta and/or spin vectors.  
Due to spin correlations, such asymmetries show unique hints for CP phases already at tree level.  Triple product asymmetries have been 
proposed in many theoretical papers in which neutralino production with two- and three-body decays, charginos with two- and three-body decays, 
also with transversely polarised beams, have been studied in the past \cite{Choi:2005gt}.   At tree level, 
the neutralino and chargino sector has two independent CP phases: 
for instance of $M_1$ and $\mu$ when rotating away the phase of $M_2$.  Assuming the phase of $\mu$-- 
strongly constrained by EDM bounds-- to be small, the phase of $M_1$ could lead to CP sensitive triple product asymmetries of up to 20\%: see Fig.~\ref{fig:susy-CPEW}.  
As mentioned above, a recent analysis performed with full event simulation and reconstruction \cite{Kittel:2011rk} shows that these 
asymmetries constructed from $(\vec{p}_{e^-}\times \vec{p}_{\ell^+_N})\cdot\vec{p}_{\ell^-_F}$  can be measured to $\pm 1$\% from neutralino 
two-body decays into slepton and lepton followed by slepton decay: $\tilde{\chi}^0_j\to\tilde{\ell}^-\ell^+_N\to\tilde{\chi}^0_1\ell^-_F\ell^+_N$. 
 From a fit to the measured neutralino cross-sections, masses and CP-asymmetries, $|M_1|$ and $|\mu|$ can be determined to a few per mil, 
$M_2$ to a few percent, $\Phi_1$ to 10\% as well as $\tan\beta$ and $\Phi_\mu$ to 16\% and 20\%, respectively. 

The sfermion sector brings in the CP phase of the trilinear scalar coupling $\Phi_{A}$. The sensitivity of the linear collider to the CP phase in 
the stop sector has been looked at recently \cite{salimkhani} by analysing a chain decay 
$\tilde{t}_1\to\tilde{\chi}^0_2 (\to \tilde{\chi}^0_1\ell^{\mp}_N \ell_F^\pm)\, + t (\to W^+b)$.  
Such decays allow one to construct two triple 
products originating from the covariant product in the spin-spin-dependent part of the amplitude, namely 
$A_{\ell_1}\sim \vec{p}_{\ell_1^\mp} \cdot ( \vec{p}_W\times \vec{p}_t)$  calculated in the reconstructed $\tilde{\chi}^0_2$ rest frame, and 
$A_{\ell\ell}\sim  \vec{p}_b \cdot(\vec{p}_{\ell^+} \times \vec{p}_{\ell^-})$ calculated in the reconstructed $W$ rest frame.   
The right panel of Fig.~\ref{fig:susy-CPEW} shows that CP sensitive asymmetries can reach 10-15\%.   Under the assumption of accurate momentum 
reconstruction, this asymmetry could be measured for 2 $ab^{-1}$ (1 $ab^{-1}$) of data collected at $\sqrt{s}=1$ TeV in the region of a maximal 
CP violating angle, $1.10 \pi< \Phi_{A_t} < 1.5\pi$ ($1.18\pi < \Phi_{A_t} < 1.33\pi$).

\begin{figure}[t] %
\includegraphics[width=0.41\textwidth]{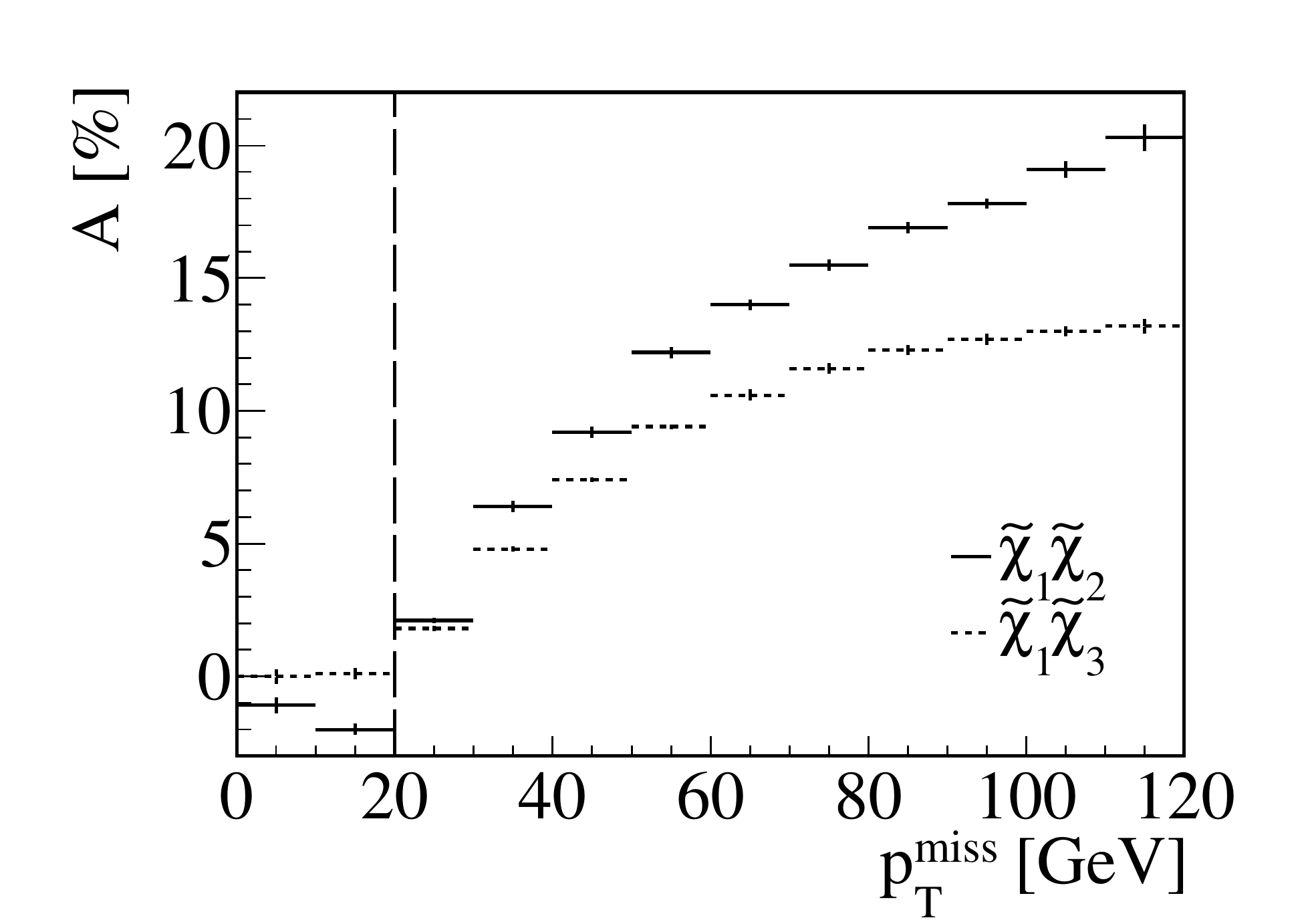} 
\includegraphics[width=0.4\textwidth]{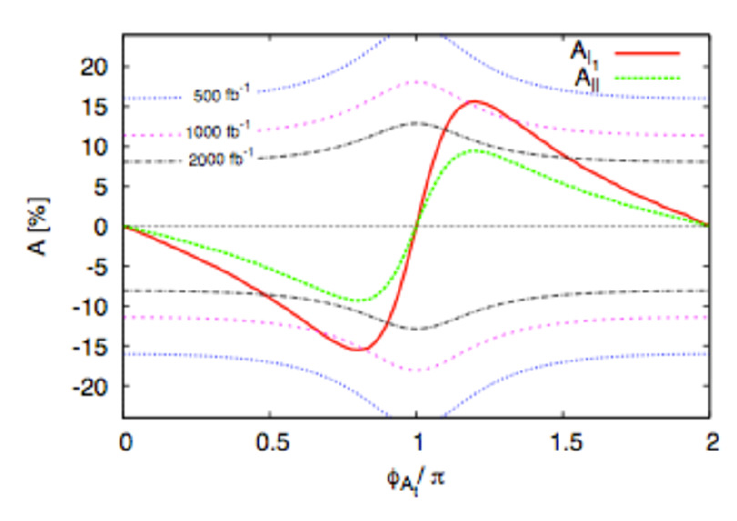}
\caption{\label{fig:susy-CPEW}  Top panel: $p^{\rm miss}_T$ dependence of CP asymmetries in neutralino-pair production and 
decay processes (from \cite{Kittel:2011rk}). Bottom panel: Asymmetries, $A_{\ell_1}$ and $A_{\ell\ell}$ as functions of $\Phi_{A_t}$ 
(from \cite{salimkhani}).}
\end{figure}

\begin{sloppypar}
Finally, it is worthwhile to recall that the CP-odd observables can
also be constructed in the non-diagonal chargino pair production
process $e^+e^-\to\tilde{\chi}^\pm_1 \tilde{\chi}^\mp_2$ from
unpolarised cross sections at one-loop
\cite{Osland:2007xw}. Obviously, at tree level the CP-asymmetry
$A^{12}\sim \int [d\sigma(\tilde{\chi}^-_1
  \tilde{\chi}^+_2)-d\sigma(\tilde{\chi}^-_2
  \tilde{\chi}^+_1)]d\cos\theta$ (with the polar angle $\theta$ of
$\tilde{\chi}^-_j$ with respect to the $e^-$ momentum direction)
vanishes even in CP-noninvariant theories.  In order to obtain a
non-zero asymmetry in the chargino production it requires another
source of non-trivial imaginary contribution to the amplitude. Such a
term can be generated by the absorptive part of a loop diagram when
some of the intermediate state particles in loop diagrams go
on-shell. The CP-odd asymmetry is generated due to interference
between the imaginary part of loop integrals and imaginary parts of
couplings. Numerical analyses show that the asymmetries can be of the
order of a few percent and in principle measurable, allowing for
discovery of the CP-violating phases via simple event counting
experiments.
\end{sloppypar}

\subsection{Beyond the MSSM}
\label{sec:susy3}

\subsubsection{The NMSSM}

The supersymmetric $\mu$ problem arises because the higgsino mass
$\mu$ term in the MSSM superpotential is not a SUSY breaking term, but
instead preserves SUSY. Thus, naively one would expect $\mu\sim M_P$
instead of $M_{weak}$; this possibility seems phenomenologically
disallowed.  One solution, endemic to gravity mediation, is for the
$\mu$ term to be forbidden by some symmetry, such as a Peccei-Quinn
(PQ) symmetry, but then to re-generate it via interactions with either
the PQ sector\cite{Kim:1983dt} or the hidden
sector\cite{Giudice:1988yz}.  An alternative possibility occurs by
extending the MSSM with an additional gauge singlet superfield $N$,
where the $\mu$ term then arises from its coupling to the Higgs fields
in the superpotential, $ \lambda N H_uH_d$. This extension is known
as the Next-to-Minimal SUSY extension of the SM, or NMSSM.  In the
NMSSM, an effective $\mu=\lambda x$ term is expected to be generated
around the electro-weak scale when the scalar component of the singlet
$N$ acquires a vacuum expectation value $x=\langle N\rangle$.
Moreover, the NMSSM is additionally motivated in that it provides
additional quartic contributions to the light Higgs scalar mass $M_h$,
thus perhaps more easily accommodating the rather large value $M_h\sim
125$ GeV, which otherwise requires TeV scale top-squarks which some
authors consider to have a conflict with naturalness.  Further
reduction in the fine-tuning of the NMSSM can be achieved by
introducing extra matter terms \cite{Hall:2012mx}.  Independently, a
bottom-up approach for addressing the fine-tuning problem, via
``natural SUSY'', calls for the third generation sfermions and the
higgsino to be light, while the rest of the superpartners can be
heavy.  However, the higgsino cannot then be the sole dark matter
candidate since higgsinos annihilate too rapidly into $WW$ and $ZZ$.

Within the extended Higgs sector of the NMSSM, the new singlino state,
with mass below that of the higgsino, might serve as a dark matter
particle, or the LSP might have a significant singlino component.  The
phenomenology of different scenarios for the mixing character of the
lightest neutralino --singlino, higgsino, gaugino-like-- 
has been systematically analyzed in the plane of the NMSSM-specific 
Yukawa couplings
$\lambda-\kappa$-plane, cf.\ also Fig.~\ref{fig_stefano}. 
\begin{figure}[tbh!] %
\centerline{
\includegraphics[width=0.45\textwidth]{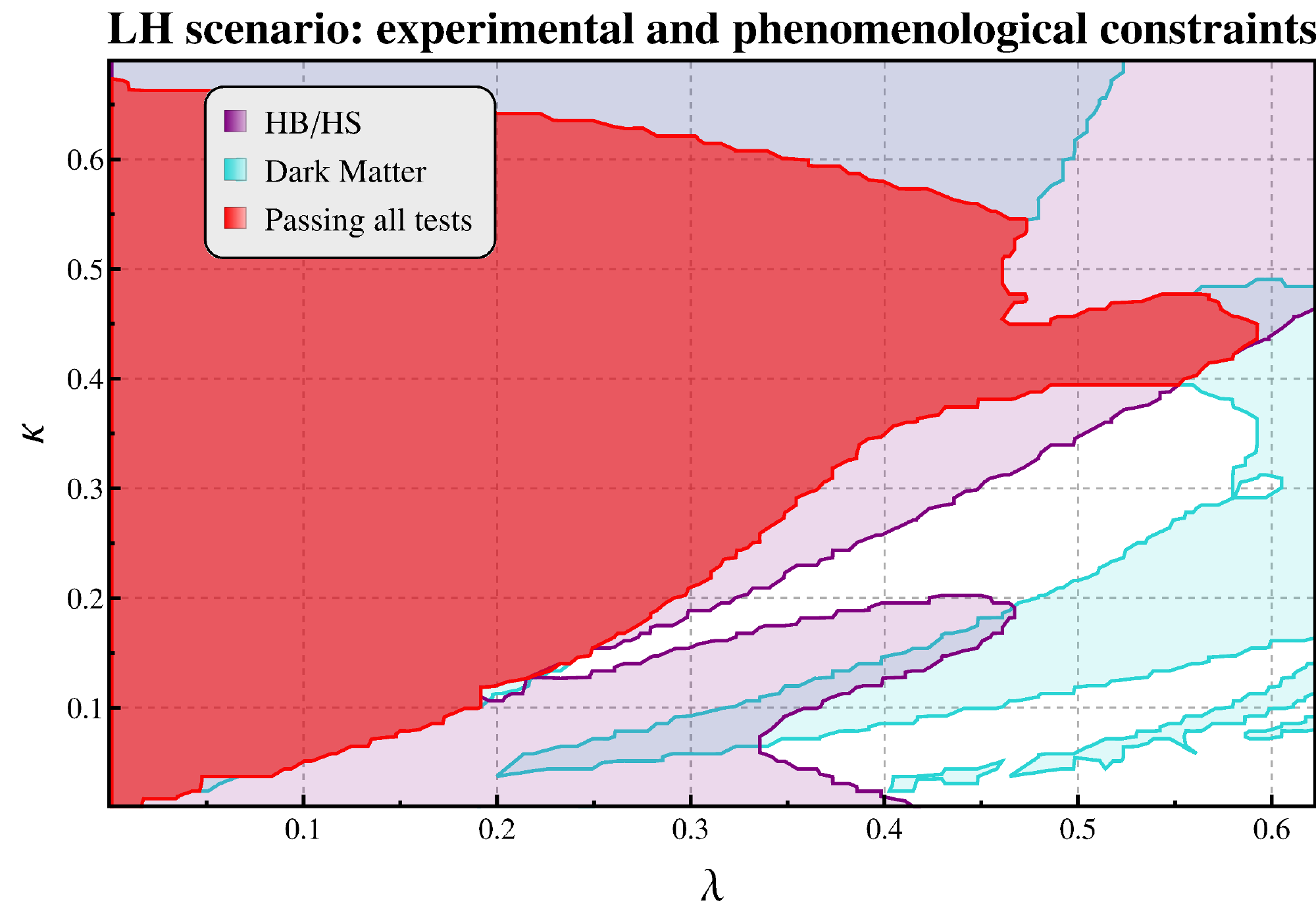}
}
\caption{\label{fig_stefano} 
Lightest neutralino $\tilde{\chi}^0_1$ is mainly higgsino-like: 
regions in the ($\lambda$--$\kappa$)-plane 
allowed by experimental and phenomenological constraints. 
The light-blue-shaded regions delimited by the light blue boundary pass 
dark matter constraints. The coloured regions delimited by the 
purple boundary pass checks within \texttt{HiggsBounds}\cite{Bechtle:2013wla} 
and 
\texttt{HiggsSignals}\cite{Bechtle:2013xfa}. 
The red area is allowed by all the constraints~\cite{Moortgat-Pick:2014uwa}.}
\end{figure}
In the first case, the decay width of the higgsino to the singlino is of
order 100 MeV.  The pattern of decays can be 
rich (see Fig.~\ref{fig:susy-nmssm}), providing us with clear signatures which
can be studied at a LC of sufficient energy. The precision measurement
of these decay branching ratios will illuminate the structure of the
extended model \cite{Das:2012rr}.  These decay products are quite
soft, however, and are expected to be virtually invisible under the
standard LHC trigger conditions. Whether or not these particles can be
seen at the LHC, the linear collider would again be needed for a
complete study, which requires the determination of their branching
fractions. The singlino-higgsino mixing angle, which determines the
annihilation cross section of the LSP and the thermal dark matter
density, could be measured at the LC through a determination of the
higgsino width using a threshold scan, as discussed above, or by
precision measurements of the NMSSM mass eigenvalues.
\begin{figure}[tbh!] %
\centerline{
\includegraphics[width=0.45\textwidth]{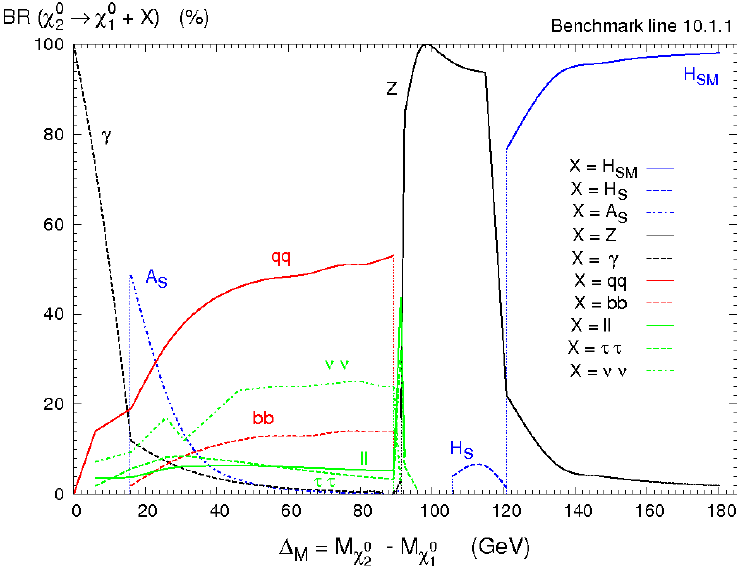}
}
\caption{\label{fig:susy-nmssm}  
Neutralino decay $\tilde{\chi}^0_2\to\tilde{\chi}^0_1 + X$ branching fractions 
as function of the mass splitting $\Delta M = M_{\chi^0_2} - M_{\chi^0_1}$ (from \cite{Das:2012rr}).}
\end{figure}

\begin{sloppypar}
The LC capabilities in distinguishing between the NMSSM and the MSSM,
when the observable particle spectrum and the corresponding decay
chains are very similar in pattern, has been studied in
detail\cite{Choi:2001ww,MoortgatPick:2005vs}.  From data taken in
$e^+e^-$ collisions at three different centre-of-mass energies, the
distinction is possible.  When exploiting the available information by
applying a global fit, just two $\sqrt{s}$ choices can be sufficient,
depending on the mixing character of the lightest neutralino 
states~\cite{MoortgatPick:2005vs,Moortgat-Pick:2014uwa}.
 If the full neutralino/chargino spectrum is 
accessible at the maximum collider energy, sum rules for the production cross sections, yielding a different energy 
behaviour in the two models, may also be exploited. In scenarios with dominant couplings of a mostly-singlino LSP to the 
NLSP particle, as predicted for large values of the $x$ parameter, the existence of displaced vertices leads to a particularly 
interesting signature that can be precisely resolved with the excellent detector resolution envisaged at a linear collider.
\end{sloppypar}

\subsubsection{$R$-Parity Violation}
\label{sec:susyRPV}

The signatures for the SUSY searches discussed so far are based on the
assumption that $R$-parity, the additional quantum number
distinguishing SUSY particles from their SM counterparts, is conserved
leading to final states with significant missing energy, due to the
escaping LSPs.  Introducing $R$-parity violation (RpV) changes
drastically the SUSY phenomenology.  $R$-parity violating couplings
allow for single production of SUSY particles and their decays to SM
particles. The latter aspect makes RpV SUSY much harder to detect at
the LHC due to the absence of missing transverse energy, so that the
currently explored region is significantly smaller than in the
R-parity conserving case, even when assuming mass unification at the
GUT scale\cite{Baer:1996wa}.  Although the LSP is not stable, there
are models with small $R$-parity violation which naturally yield a
consistent cosmology incorporating primordial nucleosynthesis,
leptogenesis and gravitino dark matter\cite{r9}; axion dark matter is also a possibility. 
Since the gravitino decays into SM particles are doubly suppressed by the Planck mass and
the small $R$-parity breaking parameter, its lifetime can exceed the
age of the Universe by many orders of magnitude, and the gravitino
remains a viable dark matter candidate \cite{r10}.  

Bi-linear $R$-parity violation (BRpV) has phenomenological motivations in
neutrino mixing \cite{Porod:2000hv} as well as in leptogenesis \cite{r12}. In
this case, the mixing between neutrinos and neutralinos leads to one
massive neutrino at tree level and the other two via loop effects
\cite{r13}.  Once the parameters are adjusted to satisfy the neutrino
constraints, the lightest neutralino typically decays inside the
detector volume \cite{Porod:2000hv}.  Since the parameters that determine the
decay properties of the LSP are the same parameters that lead to
neutrino masses and oscillations, there are strong correlations
between the neutralino branching ratios and the neutrino mixing
angles, {\it e.g.},
\begin{equation}
BR(\tilde{\chi}_1^0 \to W^\pm\mu^\mp) /BR(\tilde{\chi}_1^0 \to W^\pm\tau^\mp)	
\sim  \tan^2\theta_{23}.
\end{equation}
\begin{sloppypar}
By measuring the ratio of the branching fractions for
$\tilde{\chi}^0_1 \to W^\pm\mu^\mp$ and $W^\pm\tau^\mp$, the neutrino
mixing angle $\sin^2 \theta_{23}$ could be determined to percent-level
precision, as illustrated in Fig.~\ref{fig:susy-BRpV2}.  The
characteristic decay $\tilde{\chi}^0_1\to W^\pm l^\mp$ gives
background-free signatures at an $e^+e^-$ linear collider, possibly
with a detectable lifetime of the $\tilde{\chi}^0_1$ depending on the
strength of the BRpV couplings. In the hadronic decay mode of the
$W^\pm$, these events can be fully reconstructed and the
$\tilde{\chi}^0_1$ mass can be measured to $\cal{O}$(100)~MeV
depending on the assumed cross-section \cite{List:2013dga}.  The LC
results could then be checked against the measurements from neutrino
oscillation experiments to prove
that BRpV SUSY is indeed the origin of the structure of the neutrino sector.
\begin{figure}[tbh!]
\begin{center}
\hspace{2cm}
\includegraphics[width=8.0cm]{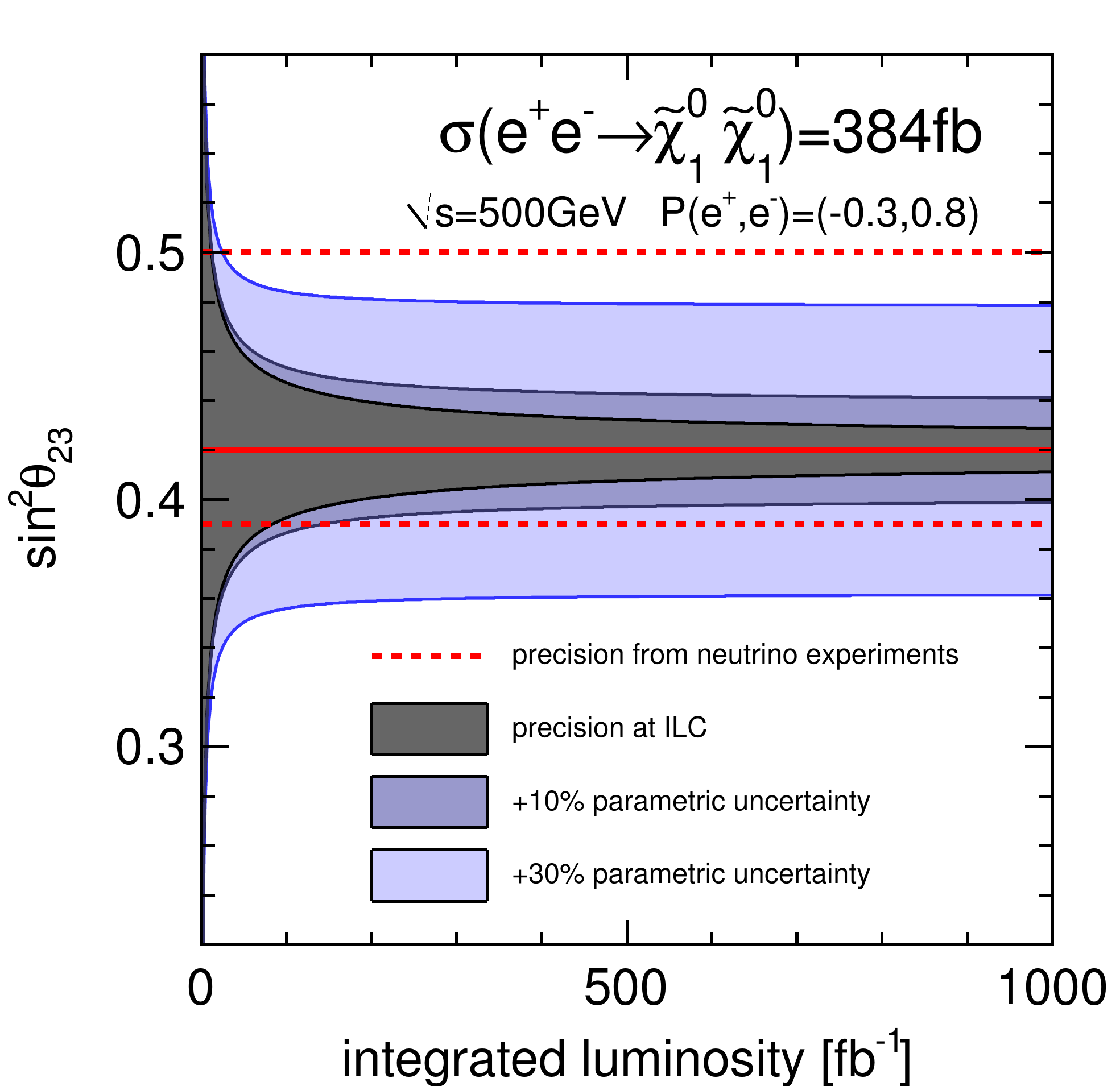}
\end{center}
\caption{Achievable precision on $\sin^2\theta_{23}$ from BRpV decays
  of the $\tilde{\chi}^0_1$ as a function of the produced number of
  neutralino pairs compared to the current precision from neutrino
  oscillation measurements. Over a large part of the $m_{1/2}$ vs
  $m_0$ plane, the neutralino pair production cross-section of the
  order of 100 fb~\cite{List:2013dga}.}
\label{fig:susy-BRpV2}
\end{figure}
\end{sloppypar}

Finally, in the case of trilinear R-parity violation (TRpV), the
exchange of sparticles can contribute significantly to SM processes
and may even produce peak or bump distortions to the distribution of
cross sections \cite{Kalinowski:1997bc}.  Below threshold, these new
spin-0 exchanges may manifest themselves via indirect effects on
observables such as cross sections and asymmetries which can be
precisely measured in $e^+e^-$ collisions, including spectacular
decays \cite{Bomark:2011ye}. It has been shown recently that the
observed enhancement of the semi-leptonic and leptonic decay rates of
$B \rightarrow \tau\nu$ modes can be explained in the framework of
TRpV \cite{Deshpande:2012rr}.  However, in such cases it would be
important to identify the actual source among the possible
non-standard interactions as many different new physics scenarios may
lead to very similar experimental signatures. At the LC, a technique
based on a double polarisation asymmetry formed by polarising both
beams in the initial state has been proposed \cite{Tsytrinov:2012ma}.
This is particularly suitable to directly test for $s$-channel
$\tilde{\nu}$ exchange.  Again, the availability of both $e^-$ and
$e^+$ polarisation plays a crucial r{\^o}le in identifying the new
physics scenario (see Fig.\ref{fig:susy-ident}). In contrast, the
left-right asymmetry, $A_{LR}$, obtained with only electron
polarisation, does not appear to be useful for this purpose.
\begin{figure}[tbh!] %
\newcommand{\Lumint}{{\cal L}_{\rm int}}
\centerline{
\includegraphics[width=0.24\textwidth]{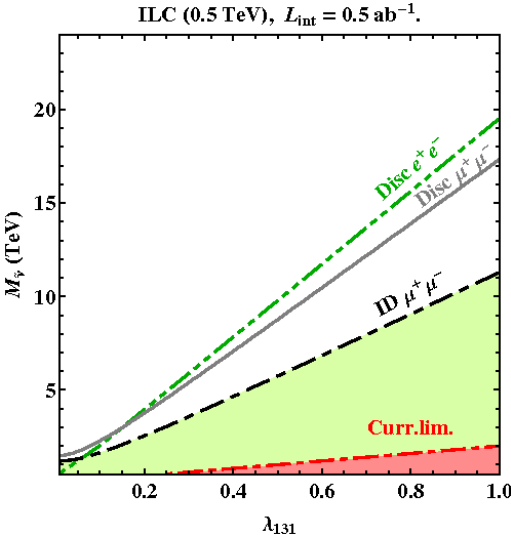}
\includegraphics[width=0.24\textwidth]{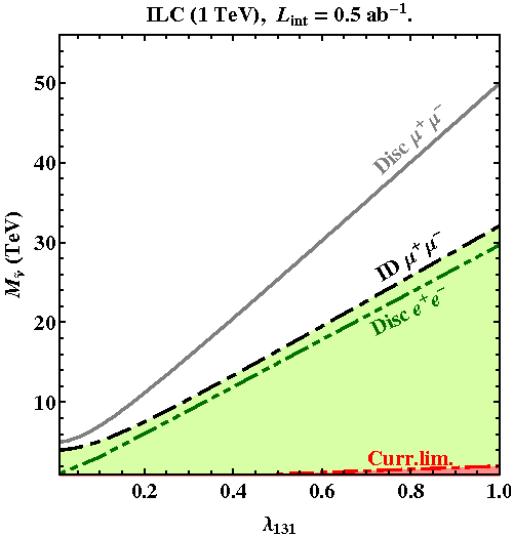}
}
\caption{\label{fig:susy-ident}  Discovery reach at 95\% C.L. in Bhabha scattering for the sneutrino mass as a function of 
$\lambda_{131}$ at $\sqrt{s}=0.5~{\rm TeV}$ (left panel) and 1~TeV (right panel), for $\Lumint=0.5~{\rm ab}^{-1}$. For
comparison, the discovery reach on $M_{\tilde\nu}$ in muon pair production for $\lambda_{232}=0.5\times M_{\tilde\nu}/{\rm TeV}$ 
is also shown (from \cite{Tsytrinov:2012ma}).}
\end{figure}

\subsubsection{$R$ symmetry}

\begin{sloppypar}
In the $R$-parity conserving MSSM, the gravitino, gluino, and other
gaugino mass terms can be introduced once supersymmetry is broken.
However, it has recently been realised that by requiring an
additional $R$-symmetry~\cite{Fayet1,Fayet2} 
beyond $R$-parity, which can be continuous or
discrete, exact or approximate, is not only phenomenologically viable,
but may allow sizable flavor-violating operators without generating
large FCNC or CP violation.  A continuous $U(1)_R$ symmetry on the
MSSM, where gauginos and squarks have $R$-charges $R= +1$, and the
Higgs scalars have $R = 0$, not only forbids baryon- and lepton-number
changing terms in the superpotential, but also dimension-five
operators mediating proton decay~\cite{operators}.  
\end{sloppypar}

$R$ symmetry also removes some of
the potentially unwanted parameters of the theory, such as trilinear
$A$-terms for the scalars, the $\mu$-term and Majorana gaugino masses,
while Majorana neutrino masses are allowed.  The absence of $\mu$ and
$A$ terms helps to solve the flavor problem without flavor-blind
mediation.  However, since gauginos must get masses, adjoint chiral
super-fields for each gauge factor are introduced to generate
$R$-symmetry preserving Dirac gaugino masses.  Similarly, the Higgs
sector is extended by adding multiplets $R_u$ and $R_d$ with the
appropriate charges to allow $R$-symmetric $\mu$-terms with $H_u$ and
$H_d$ respectively.  The scalar components of the Higgs (and not the
$R$-fields) acquire VEVs that break electro-weak symmetry, thereby
preserving the $R$-symmetry. This general class of models goes under
the name of the Minimal $R$-symmetric Supersymmetric Standard Model
(MRSSM) \cite{Kribs:2007ac,Choi:2010an}.

The phenomenology of MRSSM is quite different from that of the MSSM.
Since the mixing with additional scalars reduces the tree-level Higgs
mass, loop corrections must play even more significant role than in
the MSSM. Recently it has been shown
\cite{Bertuzzo:2014bwa,Diessner:2014ksa} that additional contributions
from TeV-scale chiral adjoint superfields and $R$-Higgses allow to
accommodate a light Higgs boson of mass $\sim 125$~GeV more comfortably
than in models such as the cMSSM even for stop masses of order 1 TeV
and absence of stop mixing. Moreover important constraints from
electroweak precision observables are imposed on parameters entering
the Higgs mass calculation, in particular the $W$ boson mass, because
R-symmetry necessarily introduces an SU(2) scalar triplet that
develops a vev.  A full one-loop calculation 
\cite{Diessner:2014ksa,Benakli:2012cy}
shows that regions of parameters can be found consistent with the
measured Higgs and $W$ boson masses.

\begin{sloppypar}
Because gauginos are Dirac, scalars can naturally be lighter than
gauginos.  The scalar component of the adjoint SU(3) super-field, a
sgluon, can be relatively light and accessible at the LHC
\cite{Choi:2009jc,Kotlarski:2011zz,Kotlarski:2013lja}.  
The Dirac neutralinos can easily be tested at a
LC by investigating the threshold production behaviour of the diagonal
pair production (Fig.\ref{fig:susy-mrssm}) or by angular
distributions.  In contrast to standard Higgs, the $R$-Higgs bosons do
not couple singly to SM fields, and all standard-type channels are
shut for the single production. Nevertheless, if they are not too
heavy, the $R$-Higgs bosons can be produced in pairs at the LHC, via
the Drell-Yan mechanism, and at prospective $e^+e^-$ colliders (see
Fig.\ref{fig:susy-mrssm}).
\end{sloppypar}

\begin{figure}[tbh!] %
\newcommand{\Lumint}{{\cal L}_{\rm int}}
\centerline{
\includegraphics[width=0.22\textwidth]{SUSY/Figures/nn_prod_hybrid}\hspace{3mm}
\includegraphics[width=0.22\textwidth]{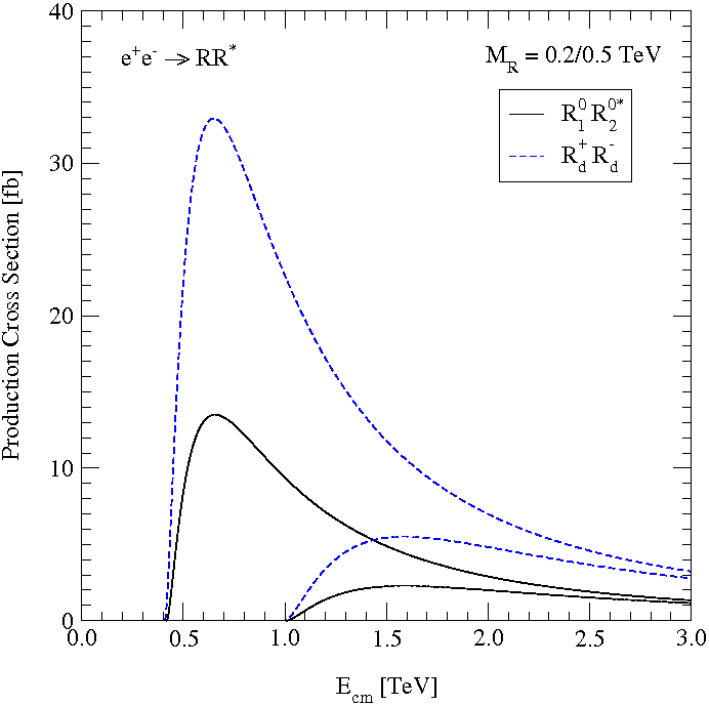}}
\caption{\label{fig:susy-mrssm}  Left panel: pair production of wino-like neutralinos near threshold in the MSSM and the Dirac theory 
(from \cite{Choi:2010gc}. Right panel: production of the neutral and charged $R$-Higgs boson pairs at TeV $e^+e^-$  colliders 
(from \cite{Choi:2010an}).}
\end{figure}

$R$-symmetry allows {\it either} Yukawa or $A$-terms, but not 
both. With  the  neutrino Yukawas zero, large $A$-terms for sneutrinos are thus natural in the MRSSM.   With three singlet superfields $N_i$, a $6\times 6$ sneutrino mass matrix can feature 
large off-diagonal $A$-terms mixing the left and right handed 
sneutrinos.  In such a framework,  a mixed 
sneutrino can serve as a successful candidate for dark matter, an appropriate Majorana neutrino 
masses can be generated and striking lepton-flavor violation signals can be expected at both 
LHC and linear colliders \cite{Kumar:2009sf}.


\subsection{Relevance of $e^-e^-$, $e\gamma$ and $\gamma\gamma$ options for SUSY searches}
\label{sec:egamma}

Linear colliders offer an impressive capacity to discover and untangle
new physics such as supersymmetry in $e^+e^-$ collisions.
Their ability to adapt to the specific needs of various scenarios of new 
physics is augmented by the possibility to run such machines 
in $e^-e^-$, $e\gamma$ or $\gamma\gamma$ modes.
In the latter two cases, the $\gamma$s are generated via laser backscattering 
off of the incoming electron beams. Each of these options offers 
new avenues for understanding supersymmetry.


By operating in $e^-e^-$ mode, a vast array of SM background processes that 
could be problematic at $e^+e^-$ colliders are automatically turned off. 
One might counter that most SUSY production reactions are also turned off 
in $e^-e^-$ mode as well. Reactions like 
$e^-e^-\rightarrow \te_R^-\te_R^-$,\ $\te_L^-\te_L^-$,\ $\te_L^-\te_R^-$ and 
$\te_R^-\te_L^-$ provide distinctive SUSY signals\cite{Feng:1998ud,Feng:2001ce,Freitas:2001zh,Blochinger:2002zw,Freitas:2003yp,Wagner:2008zz}.
These take place via $t$-channel neutralino exchange. 
An advantage of $e^-e^-$ collisions is obtained in threshold scans: 
whereas $e^+e^-\to \te_R^+\te_R^-,\ \te_L^+\te_L^-$ suffer the usual $\beta^3$
suppression 
factor typical of scalar pair production, the $e^-e^-\to\te_R^-\te_R^-$
and $\te_L^-\te_L^-$ reactions are only suppressed by $\beta^1$. 
This offers
better  accuracy in the selectron mass measurement via the unsuppressed 
threshold production of selectron pairs. This is especially important in that 
threshold scans for $\beta^3$ suppressed processes will require 
very high integrated
luminosity while similar or better measurements can be made on $\beta^1$
suppressed processes at much lower integrated luminosity. 

Since $\te_R^-\te_R^-$ takes place via pure bino exchange, 
the total rate for this reaction will be highly sensitive to the bino mass
(assuming a nearly pure binno-like gaugino) 
all by itself even if $m_{\tilde B}$ is far beyond direct production (in such 
a case, perhaps the LSP would be the lightest Higgsino with $m_{\tilde B}$ 
much heavier).
Furthermore, using beam polarization, 
one might dial up individual reactions $\te_L^-\te_L^-$, $\te_R^-\te_R^-$ or 
$\te_L^-\te_R^-$. 
It has also been emphasized that $\te_{L,R}^-\te_{L,R}^-$ production would be
an excellent environment for testing possible rare, perhaps flavor violating, slepton decay
modes to the low background environment\cite{Feng:1998ud}.


The possibility of $e\gamma$ colisions is important in several cases
relevant for SUSY
searches\cite{Choudhury:1994vi,Kiers:1996np,Barger:1997qu}.  The first
scenario is offered by the reaction $e\gamma\rightarrow \te_{L,R}\tz_i$, or
single production of selectrons. In this case, even if
$\te_{L,R}^+\te_{L,R}^-$ is beyond the maximal $\sqrt{s}$ of an
$e^+e^-$ collider, then if $m_{\tz_1}$ is light, single production of
sleptons may take place for $\sqrt{S}>m_{\te}+m_{\tz_1}$.  The utility
of an $e\gamma$ collider has also been considered for GMSB SUSY models
where one may produce $\te_{L,R}\tG$ where $m_{\tG}$ may be very
light\cite{Ghosal:1997dv}, and in models with $R$-parity
violation\cite{Ghosh:1997kz}.


\begin{sloppypar}
A linear collider 
running in $\gamma\gamma$ mode (two backscattered laser beams)
has been considered in~\cite{Mayer:2000ui,Mayer:2003sw} 
for chargino pair production and in~\cite{Berge:2000cb} for sfermion
production. 
For $\gamma\gamma$ collisions, the couplings are pure QED so that the
production cross sections depend only on the mass of the charged sparticles
which are being produced.
For both these cases, an advantage can be gained by scattering polarized laser
light on polarized beam to gain polarized photon collisions.
A variety of helicity studies can then be made on the various sparticle 
pair production
processes. 
\end{sloppypar}

\subsection{Summary and Conclusions}
\label{sec:susy7}

It is timely to re-assess the physics opportunities related to SUSY
models for an $e^+e^-$ linear collider before the start of LHC operation
at 13-14 TeV. The run at 7 and 8 TeV has been marked by the discovery
of a Higgs-like boson with a mass $M_h\sim 125$~GeV and has
provided us with important bounds on the mass of new particles from
dedicated SUSY searches.
 These LHC results are complemented by
important data on dark matter, from the precision determinations of
its relic density from the CMB spectrum to much improved bounds on its
scattering cross section from underground search experiments.  

The
combination of the relatively light mass of the newly-discovered
Higgs-like particle, easily interpretable within SUSY, and the
compelling evidence for dark matter, which can be explained as due to
relic neutralinos or gravitinos, have reinforced the interest for
supersymmetric models.  The combined 7+8~TeV LHC data have already set
significant bounds on the masses of strongly-interacting SUSY
particles in the jets$+MET$ channel and have started addressing the detection of
weakly-interacting particles in $\ell$s+MET and $h$+MET channels and
more model-independent searches for neutralino LSPs and
nearly-degenerate squark-neutralino scenarios with monojets.

All these searches will have a powerful impact on supersymmetric models with
the Run-2 data taking at 13-14~TeV and higher luminosity. However,
despite the broad range and the ingenuity of the LHC searches,
scenarios with nearly degenerate sparticle-neutralino LSP masses,
compressed spectra, multiple decay modes with comparable rates and
some of the 'natural' SUSY spectra may prove difficult for the LHC to
probe in full.  In fact, if we take guidance from the concept of
naturalness and the fine tuning of supersymmetric models, we are
brought to consider natural SUSY models which contain a spectrum of
light higgsino particles. In these models, gluinos and scalar quarks
may be as heavy as several-TeV, with TeV-values stops required to be
highly mixed in order to lift $M_h$ up into the 125~GeV range. 
Such `natural' SUSY spectra would be characterised by electro-weak fine-tuning at
the level of $\sim 10\%$ and their concomitant light higgsinos could be
readily detected and studied at an $e^+e^-$ linear collider of sufficient 
energy. When the higgsino mass $\mu$ sets the scale for fine-tuning, then
we expect a center-of-mass energy $E_{CM}$ to probe 
electroweak fine tuning of $E_{CM}>2\mu\sim\sqrt{2\Delta_{EW}}M_Z$.

In these scenarios, the combination of clean environment, well-known
beam energy, adjustable centre-of-mass energy and availability of
polarized beams at the $e^+e^-$ linear collider will provide us with
the tools required for precision measurements of masses, spins and
other quantum numbers of these new states. Precision mass and spin
measurements can be performed either by kinematic measurements in the
continuum or via threshold scans. An $e^+e^-$ collider should be able
to extract precision values of scattering cross sections, branching
fractions, angular distributions of final state particles and decay
widths.

These precision measurements will lead to the  extraction of the fundamental
SUSY Lagrangian parameters and test
the unification at very high energy scales.  All together
these measurements will
provide us with a unique 
window onto 
the energy scales associated
with grand unification.

\begin{sloppypar}
Production of SUSY particles at an $e^+e^-$ linear collider 
may allow for tests of the
Majorana nature of gauginos, flavor-violating decays, CP-violating
processes, R-parity violating reactions (which can also elude LHC
searches), $R$-symmetry effects and the presence of additional matter
states such as the added singlets in extended models.
In the event that just a few SUSY particles are produced at some
energy scale, then the linear collider can still determine the 
fundamental SUSY parameters in a model-independent way 
and can
still test higher mass scales through virtual particle exchange, such as
  sneutrino exchange
effects in chargino pair production, and additional
SUSY parameters via loop effects, for instance, to Higgs branching fractions.
\end{sloppypar}

The knowledge obtained from combining the data of the LHC, an $e^+e^-$
linear collider and dark matter experiments will be crucial for
understanding the nature of dark matter and, possibly, test models of
baryogenesis.

From all these facets, it is clear that a linear $e^+e^-$ collider
operating in the $\sim 0.25-1$ TeV range can play a major role in
the study of supersymmetry-- ranging from discovery to precision
measurements-- and will provide a new and more refined view as to
the next level in the laws of physics as we know them.





\section[Connection to astroparticle physics and
cosmology]{Connection to astroparticle physics and 
cosmology\protect\footnotemark }
\footnotetext{Editors: G.~Belanger, K.~Olive\\
Contributors: Y.~Mambrini, P.~Serpico}
\label{astro}

\subsection{Introduction}
\label{sec:astro1}

While an enormous amount of energy is spent on the search for physics
beyond the standard model, perhaps the most compelling reason for
expecting new physics is dark matter (DM). The evidence
for DM is overwhelming.  On galactic scales, one observes
relatively flat rotation curves \cite{rot} which can not be accounted
for by the observed luminous component of the galaxy.  The simplest
interpretation of these observations are that nearly all spiral galaxies
are embedded in a large galactic halo of DM which lead to
rather constant rotational velocities at large distances from the center
of the galaxy.  X-ray emission from hot gas surrounding large elliptical
galaxies and clusters of galaxies also require a large potential well
(to gravitationally bind the hot gas) which can not be accounted for by
the galaxy or gas itself \cite{xray}.  Gravitational lensing also
implies large gravitational potentials from unseen matter on the scale
of clusters of galaxies \cite{lense}.  In addition, there are
observations of both x-ray emitting hot gas and gravitational lensing in
the same systems \cite{both} which all point to the presence of dark
matter.

On larger, scales, baryon acoustic oscillations \cite{Percival:2007yw} indicate a
matter component $\Omega_m = \rho_m/\rho_c \simeq 0.25$, where $\rho_c =
1.88 \times 10^{-29} h^2 $~g cm$^{-3}$ is the critical energy density for
spatial flatness. However, the baryon density of the universe from big
bang nucleosynthesis (BBN) \cite{Yang:1983gn} is restricted to $\Omega_B h^2
\lesssim 0.03$ where $h = 0.71$ is the Hubble parameter in units of 100
km/s/Mpc. Furthermore, both the estimate from baryon acoustic
oscillations and nucleosynthesis are in complete agreement with the
determination of both the total matter density and baryon density from
the cosmic microwave background anisotropy spectrum \cite{Hinshaw:2012aka}
 which
yields a dark matter density of \be \Omega h^2 = 0.1196 \pm 0.0031 .
\label{wmapden} \ee

As we will see, there are no candidates for the dark matter of the
Universe found in the Standard Model (SM). Thus the body of evidence for dark
matter clearly points to physics beyond the SM.  Below, we
will briefly describe some of the well-studied candidates for dark
matter with an emphasis on their relevance for a future LC.

\subsection{Candidates}
\label{s:cand}

With the discovery of the Higgs boson~\cite{lhch}, 
the SM field content is complete.
As DM must be stable or long lived, 
a priori there are only two 
possible candidates for DM in the SM.
While baryonic dark matter may account for some of the dark matter
in galactic halos, it can not make up the bulk of the dark matter in the Universe.
As noted above, BBN limits the baryon density to less than 25\% of the total
amount of non-relativistic matter in the universe, which is consistent with the
determination of the baryon density from microwave background anisotropies.
However,  a dominant component of 
baryonic dark matter  even on the galactic scale is problematic \cite{Hegyi:1985yu}.
Put simply, baryons tend to clump and form stellar-like objects.
While massive objects such as white dwarfs or neutron stars or black holes may be dark,
they are typically associated with heavy element production and a significant number
of these objects would produce excessive metallicity. 
Smaller jupiter-like objects would require a very special
mass distribution to avoid constraints from luminosity density in the red and infrared.
More concrete constraints are obtained from microlensing observations \cite{pac}
where the contribution of such objects (collectively known as MACHOs) is 
limited to less than 25\% of the halo for masses   
$2 \times 10^{-7} M_\odot < M < 1 M_\odot$.

Another potential possibility for a DM candidate in the SM
is a neutrino. Indeed, neutrino oscillation experiments
indicate that at least one neutrino has a mass in excess of 0.05 eV. 
This would correspond to a cosmological contribution, $\Omega_\nu h^2 > 5\times 10^{-4}$.
However there are upper limits to the sum of neutrino masses from
large scale structure considerations.
In particular, using CMB data (notably PLANCK, WMAP 9-years, ACT and SPT) and including observations from BAO and HST, one finds that the sum of neutrino
masses is constrained to be $\sum m_{\nu} < 0.22$~eV corresponding to
$\Omega_{\nu} h^2 < 2.4 \times 10^{-3}$~\cite{Giusarma:2014zza}.

the 7-year WMAP data and including observations from
SDSS and HST, one finds that the sum of neutrino masses is constrained to be
$\sum m_\nu < 0.39$ eV corresponding to $\Omega_\nu h^2 < 4 \times 10^{-3}$
\cite{Giusarma:2013pmn}.

At this time, if there is any firm indication of physics beyond the SM,
it comes from our understanding of dark matter in the Universe.
While not all dark matter candidates can be probed by a future
linear collider, we are will restrict our attention to those that can.
Thus we will not discuss possibilities such as sterile neutrinos or axions below
and we concentrate on those candidates with potential signatures at the LC.

\subsubsection{Supersymmetric Candidates}

The supersymmetric extension of the SM
is one of the most studied example of physics beyond the SM
and is currently being tested at the LHC. Its motivations (which we will not review here)
include the stabilization of the weak scale hierarchy, gauge coupling unification,
radiative electroweak symmetry breaking, and the prediction of a light Higgs boson
($m_h \lesssim 130$ GeV) which has been borne out by experiment \cite{lhch}.
In models with R-parity conservation, another prediction of supersymmetric
models, is the existence of one stable particle, which if neutral, may be candidate 
for the DM. This is the lightest supersymmetric particle of LSP.  
Below, we review some of the most studied 
realizations of the low energy supersymmetry.

\begin{sloppypar}
For the most part, we will restrict our attention here to the minimal
supersymmetric standard model (MSSM) (though see below for a
discussion of the next to minimal model or NMSSM). The minimal model
is defined by the superpotential
\begin{equation}
W =  \bigl( y_e H_1 L e^c + y_d H_1 Q d^c + y_u H_2 Q u^c \bigr) +  \mu H_1 H_2 ,
\label{WMSSM}
\end{equation}
Beyond the parameters associated with the SM, the
superpotential introduces a mixing term between the two Higgs doublets in the MSSM.
The bulk of the new parameters are associated with supersymmetry breaking
and are associated with soft scalar masses, gaugino masses, and so-called 
bi- and tri-linear terms, $B$ and $A$. There are well over 100 new parameters
in the minimal theory and we are thus forced to make some (well motivated)
simplifications as we discuss below. 
\end{sloppypar}

\paragraph{The CMSSM}

As is clear, supersymmetry must be broken, and one way of transmitting 
the breaking of supersymmetry to the low energy sector of the theory is through
gravity.  Indeed the extension of global supersymmetry to supergravity
is in some sense necessary to ensure the (near) vanishing of the
cosmological constant in models with weak scale supersymmetry breaking.
Gravity mediated supersymmetry breaking imposes a number of boundary conditions
on the supersymmetry breaking masses at some high energy renormalization scale,
which is usually taken to be the same scale at which gauge coupling unification occurs, $M_{GUT}$.
In gravity mediated models, one often finds that all scalar masses are equal at $M_{GUT}$
defining a universal scalar mass $m_0$. Similarly, all gaugino mass and trilinear terms
are also universal at $M_{GUT}$, with values $m_{1/2}$ and $A_0$ respectively.

In these gravity mediated supersymmetry breaking models,
supersymmetry breaking masses and gauge and Yukawa couplings are
are run down from the universality scale and often trigger electraoweak symmetry
breaking as one or both of the soft Higgs masses, $m_{1,2}^2$ run negative.
In true minimal supergravity models or mSUGRA, the scalar mass is 
equal to the gravitino mass, $m_0 = m_{3/2}$, and the $B$-term is given
by $B_0 = A_0 - m_0$.  One consequence of the latter relation
is the determination of the two Higgs vacuum expectation values
as the soft masses are run down to the weak scale. Since
one combination of the two vevs determines the $Z$ gauge boson mass, it is common
to choose the two vev's as input parameters (the other combination is the ratio of 
vevs and defined as $\tan \beta = v_2/v_1$) and discard the relation between $B_0$
and $A_0$. Instead both $B$ and $\mu$ can be calculated at the weak scale from $M_Z$ and $\tan \beta$.
If the relation between the gravitino mass and $m_0$, is also dropped, we have 
the constrained version of the MSSM known as the CMSSM.

\begin{sloppypar}
The CMSSM is therefore a four parameter theory (the sign of $\mu$ must
also be specified).  For given values of $\tan \beta$, $A_0$, and
$sgn(\mu)$, the regions of the CMSSM parameter space that yield an
acceptable relic density and satisfy other phenomenological
constraints may be displayed in the $(m_{1/2}, m_0)$ plane.  In
Fig. \ref{fig:tb40}~\cite{Buchmueller:2013psa}, the dark (blue) shaded
region corresponds to that portion of the CMSSM plane with $\tan \beta
= 40$, $A_0 = 2.5 m_0$, and $\mu > 0$ such that the computed relic
density yields the PLANCK value given in Eq. (\ref{wmapden}).  For
this choice of $\tan \beta$ and $A_0$, the relic density strip is
v-shaped. Inside the `v', the annihilation cross sections are too
small to maintain an acceptable relic density and $\Omega_\chi h^2$ is
too large.  The upper side of the `v', at large $m_0$, is produced by
co-annihilation processes between the LSP and the next lightest
sparticle, in this case the $\tilde t$~\cite{stop}. These enhance the
annihilation cross section and reduce the relic density.  This occurs
when the LSP and NLSP are nearly degenerate in mass.  The lower side
of the `v', at lower $m_0$, is produced by coannihilations between the
LSP and the $\tilde \tau$~\cite{stau-co}.  The dark (brown) shaded
regions outside of the `v' have either $m_{\tilde t}< m_\chi$ or
$m_{\tilde \tau}< m_\chi$ and are excluded.  Also shown in the figure
is the constraint from $b \to s \gamma$ \cite{bsgex} (shaded green)
which excludes the stop-coannihilation strip in the portion of the
plane shown.  Contours of constant Higgs mass are shown by the black
curves. Higgs masses are computed using FeynHiggs \cite{fh} and carry
a roughly 1.5 GeV uncertainty. The thick purple line corresponds to
the ATLAS limit on supersymmetry searches \cite{lhc}.  The area to
left of the line is excluded.  Finally, the solid green contour
corresponds to the 95\% CL upper limit to ratio of the branching
fraction of $B_s \to \mu^+ \mu^-$ relative to the SM~\cite{bmm}.
\end{sloppypar} 

\begin{figure}
  \includegraphics[width=.4\textwidth]{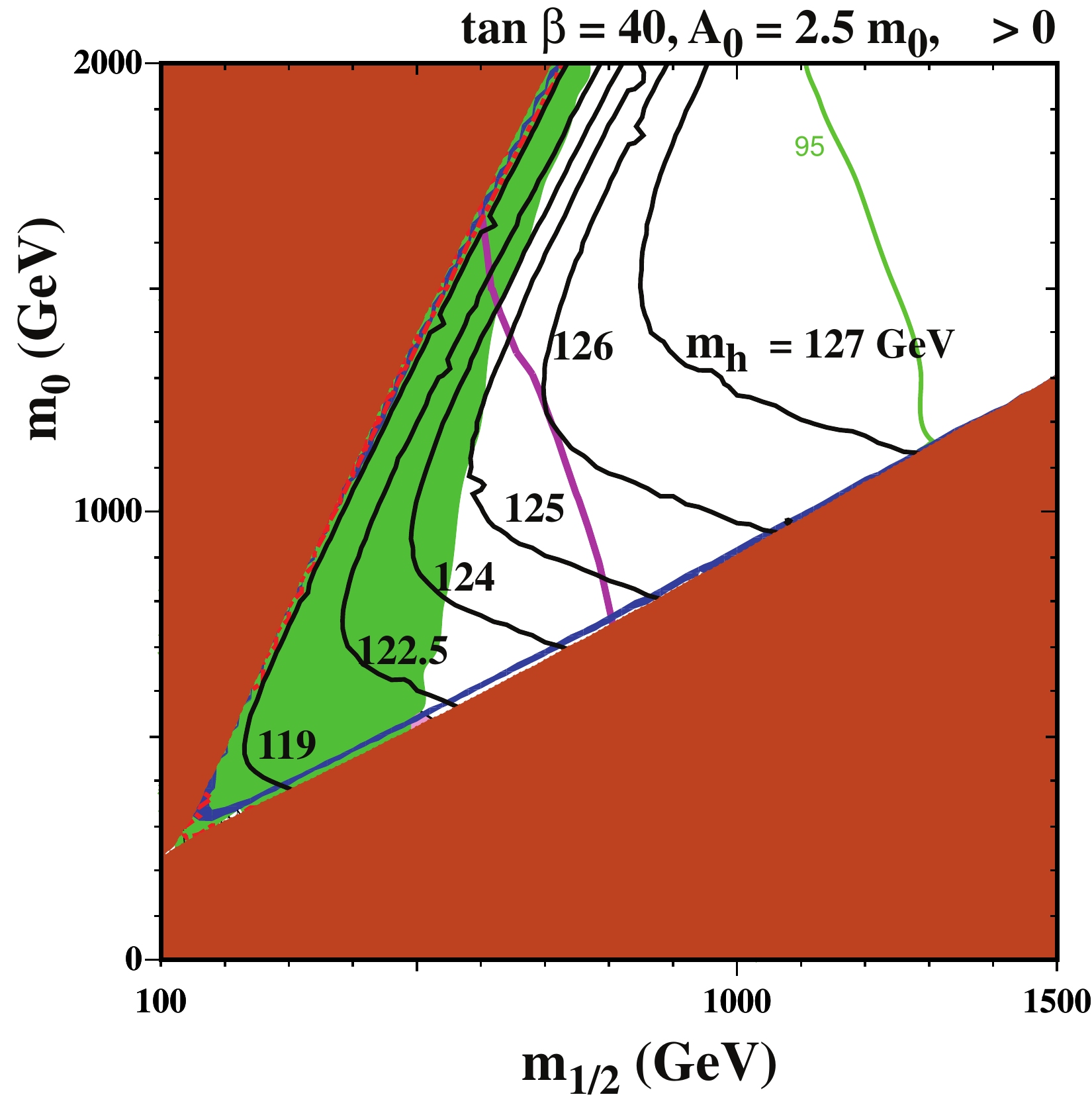}
   \caption{\it The $(m_{1/2}, m_0)$ plane for $\tan \beta = 40$ and
   $\mu > 0$, assuming $A_0 = 2.5 m_0$, $m_t = 173.2$~GeV and
   $m_b(m_b)^{\overline {MS}}_{SM} = 4.25$~GeV. Contours and shaded
   regions are described in the text.}  \label{fig:tb40}
\end{figure}

Note that the choice $A \ne 0$ is made to ensure a sufficiently large Higgs mass.
For $A_0 = 0$, the maximum Higgs mass along the stau-coannihilation strip is
only slight greater than 120 GeV, far short of the value reported in the recent
LHC results \cite{lhch}. Therefore, only the upper end of the strip is compatible with
a Higgs mass around 125 -- 126 GeV and a branching fraction for $B_s \to \mu^+ \mu^-$
sufficiently close to the SM value.

\paragraph{NUHM}

One possible generalization of the CMSSM is the so-called NUHM
in which the Higgs soft masses are not constrained to be equal to 
$m_0$. Indeed, as the Higgses are typically found in separate multiplets in 
a grand unified theory, one or both of the Higgs soft masses may be independent.
In the NUHM1, we may set $m_1 = m_2 \ne m_0$, where $m_{1,2}$ are the 
soft masses associated with $H_{1,2}$. Instead of $m_{1,2}$, one may choose
{\em either} $\mu$ {\em or} the Higgs pseudo-scalar mass, $m_A$ (which is 
a surrogate for $B$) as a free parameter in addition to $m_0$.
In the NUHM2, both $m_1$ and $m_2$ are free and one can equivalently
choose {\em both} $\mu$ {\em and} $m_A$ as free parameters. 

In Fig. \ref{fig:muma}~\cite{Buchmueller:2013psa}, we show one example of a $\mu, m_A$ plane
with $\tan \beta = 10$, $m_{1/2} = m_0 = 1200$ GeV,
and $A_0 = 2.5 m_0$. The strips of acceptable relic density now form a cross-like 
shape. Outside the cross, the relic density is too large. 
The horizontal part of the crosses are due to an enhanced cross section through
rapid s-channel annihilation through the heavy Higgses. For $m_{1/2} = 1200$ GeV,
the neutralino mass, is roughly 520 GeV and the funnel-like region occurs
when $m_A \approx 2 m_\chi$. In contrast, the vertical part of the cross occurs
when $\mu$ becomes sufficiently small that the the LSP picks up a 
significant Higgsino component (at large $|\mu|$, it is almost pure bino) 
which enhances certain final state annihilation channels such as $W^+ W^-$.

\begin{figure}
  \includegraphics[width=.4\textwidth]{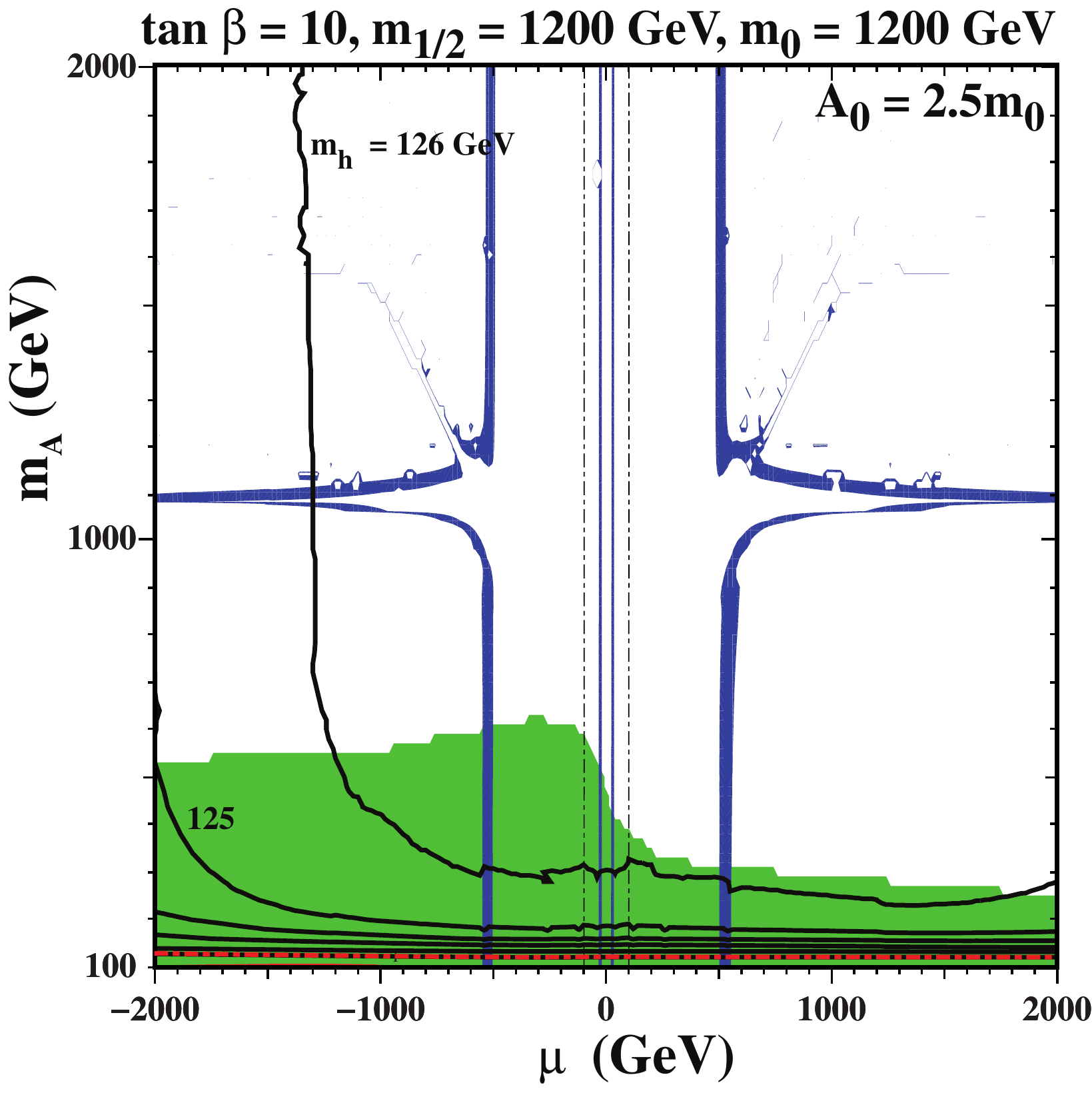}
   \caption{\it The $(\mu, m_A)$ plane for  
$\tan \beta = 10$, $m_{1/2} = m_0 = 1200$ GeV,
assuming $A_0 = 2.5 m_0$, $m_t = 173.2$~GeV and
$m_b(m_b)^{\overline {MS}}_{SM} = 4.25$~GeV. Contours and shaded regions are
described in the text. }
  \label{fig:muma}
\end{figure}

The region in Fig. \ref{fig:muma} with low $m_A$  is excluded by $b \to s \gamma$
and is slightly more pronounced when $\mu < 0$.
At $\tan \beta  = 10$,  the
branching fraction for $B_s \to \mu^+ \mu^-$ is sufficiently small. On the other hand, the Higgs mass is 
$\approx 126$ GeV across much of the plane.
The vertical dashed black lines at small $|\mu|$ correspond to a chargino mass
at the lower limit of 104 GeV.

\paragraph{The pMSSM}

As noted earlier, the most general MSSM contains more than 100 free parameters and is therefore not a convenient framework for phenomenological studies. However with a few well-motivated assumptions (R-parity conservation, no new CP phases, the sfermion mass matrices and trilinear couplings are flavour diagonal, the first two generations are degenerate and their trilinear coupling is negligible) the number of free parameters can be reduced to a more manageable number. This is the so-called 
phenomenological MSSM (pMSSM) with  19 free parameters in addition to the SM parameters:  
the gaugino mass parameters, $M_1,M_2,M_3$, the ratio of the Higgs VeVs, $\tan\beta=v_1/v_2$, the higgsino mass parameter, $\mu$, and the pseudoscalar mass, $m_A$, ten sfermion mass parameters, $m_{\tilde Q_i},m_{\tilde U_i},m_{\tilde D_i},m_{\tilde L_i},m_{\tilde E_i}$ $i=2,3$ and
three trilinear couplings $A_t,A_b,A_\tau$ . This model which is not tied to a specific symmetry breaking mechanism, leads to a much broader set of predictions  for experimental observables at the LHC or in the DM sector. 

\begin{sloppypar}
Relaxing the relation between the parameters of the electroweak-ino
sector, which are most relevant for DM observables and those of the
coloured sector, most relevant for LHC, not only relaxes some of the
limits from SUSY searches at LHC but also influences the expectations
for DM observables~\cite{pMSSM}.  In the pMSSM, the neutralino LSP can
have any composition, making it much more likely than in the CMSSM to
have a very small value for the relic density. Indeed, a significant
higgsino(or wino) component both lead to enhance annihilation in W
pair final states as well as to enhance gaugino/higgsino
coannihilations. On the other hand a higgsino LSP faces severe
constraints from direct detection, see next section.  Enhanced
annihilation through a Higgs funnel can occur for any value of
$\tan\beta$ and for any DM mass provided $m_{LSP}\approx
m_H/2$. Finally coannihilations can occur with any supersymmetric
partners that is sufficiently degenerate in mass with the LSP.
\end{sloppypar}

\begin{sloppypar}
\paragraph{NMSSM}
The Next-to-Minimal Supersymmetric Standard Model
(NMSSM) is a simple extension of the MSSM that contains
an additional gauge singlet superfield. The vev of this singlet
induces an effective $\mu$ term that is naturally of the order
of the electroweak scale, thus providing a solution to
the naturalness problem~\cite{Ellwanger:2009dp}. The model contains one additional
neutralino state, the singlino, as well as three scalar
($h_1,h_2,h_3$) and two pseudoscalar ($a_1,a_2$) Higgs bosons.
 An
important feature of the model is that the singlet fields can
be very light and escape the LEP bounds. This is because
these fields mostly decouple from the SM fields. 
 Furthermore large mixing with the singlet can modify the properties of the SM-like Higgs, allowing quite naturally for $m_h=126$~GeV as well as possibly an enhanced rate for its decay into two-photons.
With regard to DM, the NMSSM shares many of the characteristics of the MSSM. The main differences occur  when the LSP has some singlino component and/or  when the Higgs sector contains new light states that play a role in dark matter interactions. For example new Higgs states can greatly enhance  DM annihilation when their mass is twice that of the LSP or can provide  new annihilation channels when they can be produced in the final state. As a consequence, the NMSSM allows for the possibility of light neutralinos (much below $M_Z/2$), that annihilate efficiently through the exchange of light
Higgs singlets or  into light Higgs singlets ~\cite{Vasquez:2010ru}. The model also accommodates the possibility of a gamma-ray line at 130 GeV, without violating any other constraints from cosmic rays. This requires fine-tuning of the parameters such that 1) the mass of a pseudoscalar is precisely twice the neutralino mass  and 2) the annihilation of the pseudoscalar is dominantly into two-photons rather than into quarks ~\cite{Das:2012ys}.
\end{sloppypar}

\subsubsection{Universal Extra Dimensions}

Extra dimension models also propose a WIMP DM candidate. The UED
scenario~\cite{Appelquist:2000nn} where all SM particles are allowed
to propagate freely in the bulk is of particular interest. In this
model momentum conservation in the extra dimensions entails
conservation of a KK number. Orbifolding is required to obtain chiral
zero modes from bulk fermions, and breaks extra dimensional momentum
conservation. However, there remains a discrete subgroup, KK parity,
thus the lightest KK-odd particle is stable. In the minimal universal
extra dimension model (MUED) the DM candidate is in general a vector
particle, $B_1$, the Kaluza-Klein (KK) level 1 partner of the U(1)
gauge boson.  In the MUED model all KK states of a given level have
nearly the same mass at tree-level, $n/R$, where $R$ is the size of the
compact dimension. The mass degeneracy is lifted only by SM masses and
by radiative corrections. These mass splittings are however small for
all weakly interacting particles. This means that co-annihilation
channels naturally play an important role in the computation of the
relic abundance of dark matter.  Furthermore since the level 2
particles are close to twice the mass of those of level 1,
annihilation or co-annihilation processes can easily be enhanced by
resonance effects. When including level 2 particles in the
computation, the preferred scale for DM was found to be around 1.35
TeV, see line c1 in Fig.~\ref{fig:ued}~\cite{Belanger:2010yx}.  Going
beyond the MUED framework one can treat mass splittings as free
parameters, shifting significantly the preferred DM mass, for example
in the limit where the coannihilation processes are negligible the DM
mass is around 800~GeV, see line a1 in Fig.~\ref{fig:ued}.  The
measurement of the Higgs mass and of its couplings at the LHC can be
used to put a lower limit on the scale $R$. Indeed light KK particles,
in particular the KK top, lead to an increase in the $hgg$ coupling
and to a decrease in the $h\gamma\gamma$ coupling, and to a lower
bound on $R>500$~GeV~\cite{Belanger:2012mc}.  One characteristics of
MUED DM is that annihilation in the galaxy has a large fraction into
fermions leading to strong signal into positrons, however the large
mass scale makes the signature unlikely to be
observable~\cite{Hooper:2007qk}.

\begin{figure}[tb]
\begin{center}
  \includegraphics[width=.4\textwidth]{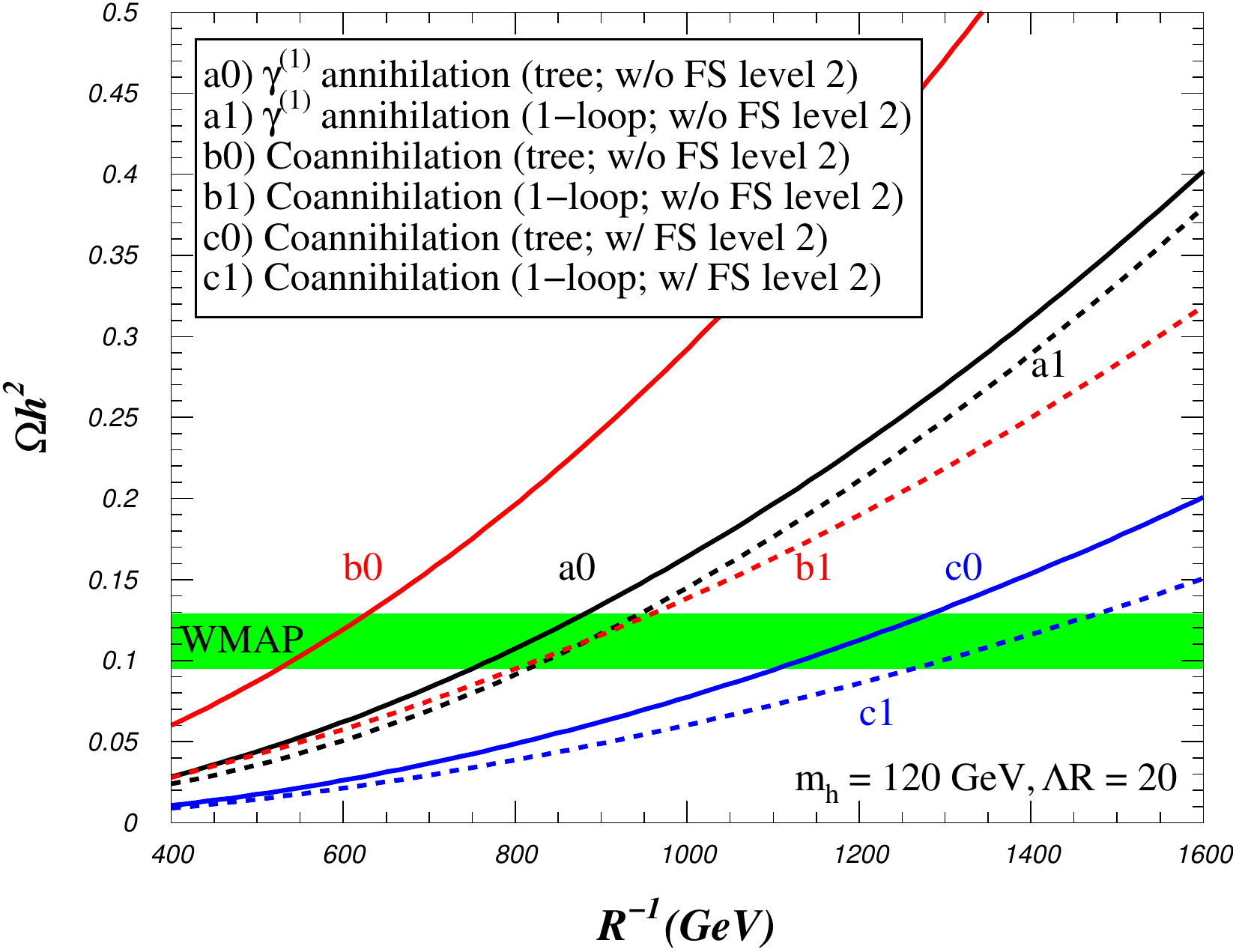}
\caption{$\Omega h^2$ as function of $R^{-1}$ for $m_h=120$~GeV and $\Lambda R=20$
including different processes as specified on the figure. Here 1-loop stands for one-loop couplings between level 2 and SM particles~\cite{Belanger:2010yx}. }
\label{fig:ued}
\end{center}
\end{figure}

\subsubsection{Higgs Portal Models}

The Higgs portal refers to a class of models where the Higgs  connects the dark matter (hidden) sector to the SM. Several possibilities have been considered with either a scalar, a vector or a fermion as DM.
The simplest extension of the SM is the addition of a real singlet scalar field, $S$, which can be made stable by imposing a $Z_2$ symmetry. If the true vacuum of the theory satisfies $\langle S \rangle=0$, thereby precluding
mixing of $S$ and the SM Higgs boson and the existence of cosmologically problematic domain walls.
The terms to be added to the SM Lagrangian are
\begin{equation} \! \Delta {\cal L}_S = -{1\over 2} m_S^2 S^2 - {1\over 4}
\lambda_S S^4 -  {1\over 4} \lambda_{hSS}  H^\dagger H  S^2 .
\end{equation}
A second possibility is to couple 
the Higgs doublet to a massive    vector field $X_\mu$ from the hidden sector. $X_\mu$ can be associated with a hidden U(1) and  becomes massive due to the Higgs or St\"uckelberg mechanism in the hidden sector. A third possibility is the one where  dark matter  can consist of Majorana fermions $\chi$   which interact with  the SM fields
only through the Higgs-portal. In both cases the stability of the DM particle is ensured by a $Z_2$
parity, whose origin is  model--dependent. For example, in the  vector case it
stems  from a natural parity symmetry of abelian gauge sectors with minimal
field content \cite{Lebedev:2011iq}.   The relevant  terms in the 
Lagrangians  are 
\begin{eqnarray} 
\!&&\Delta {\cal L}_V = {1\over 2} m_V^2 V_\mu V^\mu\! +\! {1\over 4} \lambda_{V} 
(V_\mu V^\mu)^2\! +\! {1\over 4} \lambda_{hVV}  H^\dagger H V_\mu V^\mu ,
\nonumber \\ \! &&\Delta {\cal L}_f = - {1\over 2} m_f \bar \chi \chi -  {1\over 4}
{\lambda_{hff}\over \Lambda} H^\dagger H \bar \chi \chi \;.   \end{eqnarray}
Related ideas and analyses can be found in  \cite{Lebedev:2011iq,singlet,all1,Peter,Andreas:2010dz,all2} and more  recent studies of Higgs-portal scenarios have appeared in
\cite{all3,Mambrini:2011ri,Chu:2011be}. 

\begin{sloppypar}
In these models, the Higgs is responsible for both DM annihilation and elastic scattering of DM with nuclei.
Thus, cosmological measurements made by the WMAP
 and PLANCK satellites~\cite{Hinshaw:2012aka}
 basically determine the couplings of the Higgs to DM and thus the spin-independent DM-nucleon cross section for a given DM mass.  
 The same coupling will also determine the Higgs partial decay  widths into invisible DM particles if
$m_{\rm DM} \leq \frac12 m_h$. The discovery of a Higgs boson  with  a mass $m_h =125$ GeV with a small invisible decay branching ratio
is incompatible with dark matter with $m_{\rm DM} \le 55 $ GeV.   
This applies
in particular to the case of  scalar DM with a mass of 5--10 GeV considered, 
for instance, in  Ref.~\cite{Andreas:2010dz}.
Fig.~\ref{Fig:SigmaSIbis}, displays the 
predictions for the spin--independent  DM--nucleon cross section  $\sigma_{\rm
SI}$  after imposing the WMAP and  ${\rm BR}^{\rm inv} < 10\%$ constraints (allowing the invisible width to be 20\% does not change the result significantly). 
The upper band  corresponds to the fermion Higgs-portal DM
and is excluded by XENON100 while scalar and vector DM are both
allowed for a wide range of masses. The
typical value for the scalar  $\sigma_{\rm SI }$ is a few times $10^{-9}$ pb,
whereas $\sigma_{\rm SI }$ for vectors is larger by a factor  of 3 which
accounts for the number of  degrees of freedom.
We note that a large fraction of the 
parameter space will be probed by  XENON1T except for a small region where
$m_{\rm DM}\approx m_h/2$ and  the Higgs--DM coupling is extremely  small.
\end{sloppypar}

\begin{figure}
\centering
    \includegraphics[width=.5\textwidth]{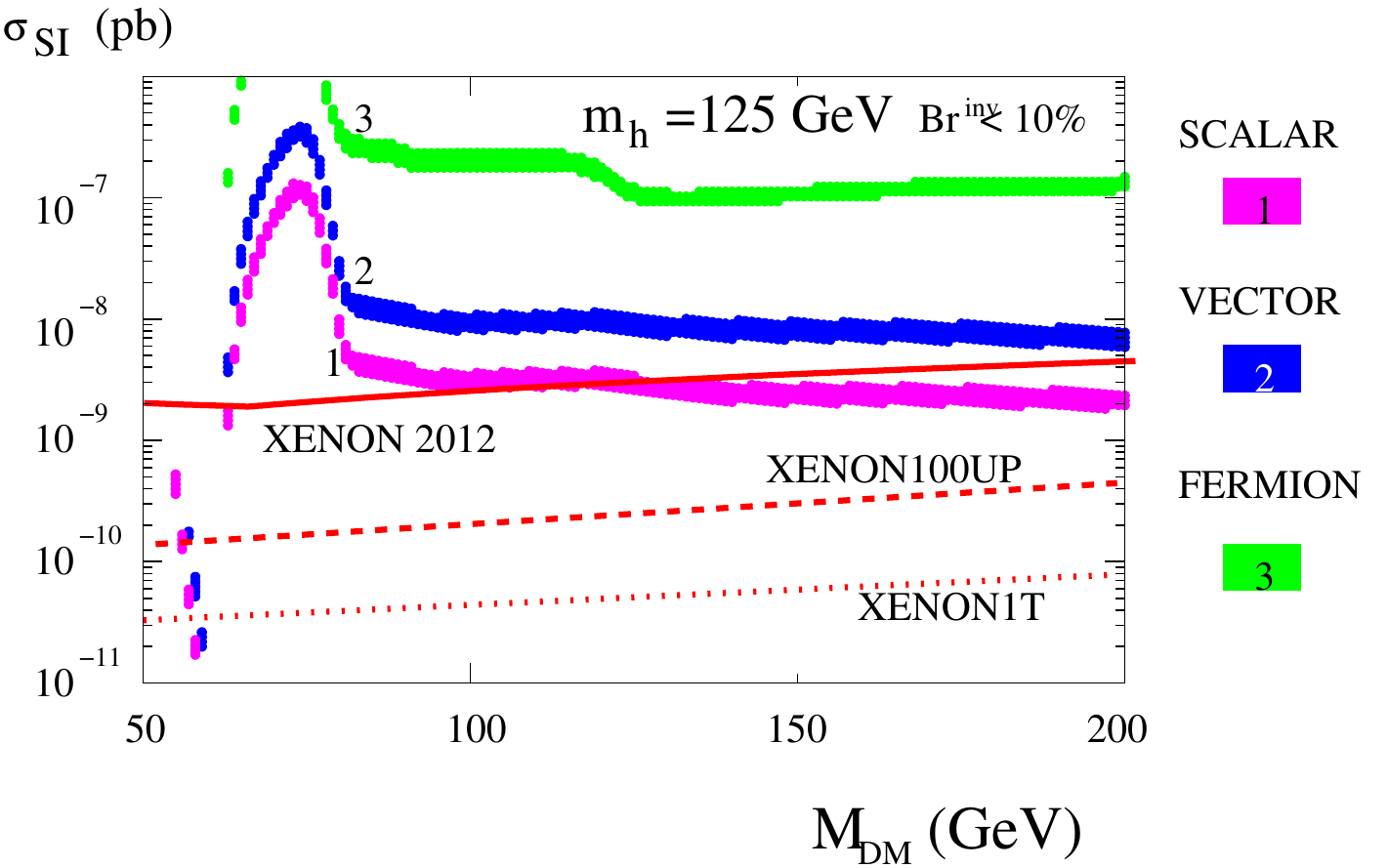}
          \caption{
Spin independent DM--nucleon cross section 
versus DM mass.
The upper band (3)  corresponds to fermion DM, the middle one (2)
to vector DM and the lower one (1)   to scalar DM. The solid, 
dashed and dotted lines represent XENON100 (2012 data \cite{Aprile:2012nq}), XENON100 upgrade
and XENON1T sensitivities, respectively.
 }
\label{Fig:SigmaSIbis}
\end{figure}

\subsubsection{Extended scalar sector}

The Higgs discovery has revived the interest in models with an
extended scalar sector. In such models an unbroken discrete symmetry
which could be leftover from a broken gauge group at a higher scale
guarantees the stability of the lightest scalar, the DM candidate.
One of the nice feature of these models is that the quartic couplings
between the SM-like doublet and other scalars helps stabilize the
scalar potential by giving a contribution that counteracts the effect
of the top Yukawa that drives the SM potential to the metastability
region~\cite{Sher:1988mj,Gonderinger:2009jp}.  The archetype of scalar
dark matter models is the inert doublet model~\cite{Barbieri:2006dq}
where the second doublet has no vev, and no coupling to quarks and
leptons.  Models with only additional singlets~\cite{singlet}, with a
doublet and singlet ~\cite{Belanger:2012vp,Belanger:2012zr} or with
higher multiplets~\cite{multiplet} have also been proposed and
different discrete symmetries to stabilize the dark matter were
considered~\cite{Belanger:2012vp,Belanger:2012zr}.

\begin{sloppypar}
In the inert doublet model, the DM can be either a scalar or
pseudoscalar. After imposing constraints on the model from
perturbativity, stability, direct searches for charged Higgs and
electroweak precision tests, several studies have found that a value
of the relic density in accordance with PLANCK can be reproduced in
the low mass $m_{\rm DM}<60$~GeV, intermediate $60<m_{\rm DM}<110$~GeV 
and high mass range ($m_{\rm DM}>500$~GeV)~\cite{Barbieri:2006dq,inert}.  
The low and intermediate mass
ranges are severely constrained by Higgs measurements and direct
detection. In the low mass region, DM annihilation proceeds through
Higgs exchange and as in the portal models is constrained by the upper
limit on the Higgs invisible width.  In the intermediate region
annihilation into $W$ pairs (including virtual $W$'s) start to
dominate. However, the Xenon and LUX upper limits forces the DM mass
to be near $m_h/2$ and $m_W$.  For DM masses above $m_W$ the
annihilation into $W$ pairs becomes very efficient thus leading to too
low a value for the relic density unless the DM mass is larger than
500 GeV, These allowed mass ranges can be extended in models with more
particles in the inert sector and/or in models which also involve
semi-annihilation~\cite{Belanger:2012zr}.  The collider signatures in
the Higgs sector involve invisible decays (already severely
constrained) and a modification of the two-photon decay width due to
the charged Higgs contribution~\cite{Krawczyk:2013jta}.  At the LC,
the inert Higgses can be directly produced and their decays into real
or virtual gauge bosons exploited to determine the masses of all inert
scalars ~\cite{Aoki:2013lhm}.
\end{sloppypar}

\subsection{Dark Matter at the LHC}
Direct searches for supersymmetry at the LHC have had a significant
impact on the allowable regions of the supersymmetric parameter space
particularly in the context of the CMSSM. An example of this is shown
by the purple curve in Fig. \ref{fig:tb40}. For relatively low $m_0$,
the most recent results from ATLAS place a lower bound on $m_{1/2}$ of
roughly 840 GeV.  Perhaps of greater significance is the discovery of
the Higgs boson at 125-126 GeV. While consistent with general
predictions in supersymmetric models that $m_h \lesssim 128 - 130$ GeV, a
125 GeV Higgs lies at the edge of what can be obtained and pushes the
model to require large contributions from stop mixing (hence a large
value of $A_0$ in the CMSSM) and relatively large SUSY masses. Of
course large SUSY masses are consistent with the lack of discovery of
supersymmetric particles at the LHC, and are consistent with little or
no departures from the SM in rare B decays. Of course this can not be
viewed as a ringing endorsement for supersymmetry. Indeed the past
prospect of resolving the discrepancy between theory and experiment
for the anomalous magnetic moment of the muon, has now essentially
evaporated.

\begin{sloppypar}
To account for the recent LHC results along with other low energy observables,
it is better to perform a global likelihood analysis which can identify regions of the 
parameter space which best fit the data. 
It is well established that Markov-Chain Monte-Carlo (MCMC) algorithms 
offer an efficient technique for sampling a large parameter space such as 
the CMSSM or its variants.  MCMC has been utilized in the Mastercode \cite{mcweb}
framework to perform a frequentist analysis of the CMSSM and other variants of the model.
The MCMC technique is used to sample the SUSY parameter space,
and thereby construct the $\chi^2$ probability function,
$P(\chi^2,N_{\rm dof})$. This accounts for the number of degrees of
freedom, $N_{\rm dof}$, and thus provides a quantitative measure for the
quality-of-fit such that $P(\chi^2,N_{\rm dof})$ can be used to estimate the
absolute probability with which the CMSSM describes the experimental data. 
\end{sloppypar}

\begin{sloppypar}
The results of the mastercode analysis include the parameters of the
best-fit points as well as the 68 and
95\%~C.L.\ regions  
found with default implementations of the phenomenological, experimental
and cosmological constraints. These include precision electroweak data,
the anomalous magnetic moment of the muon,  
$B$-physics observables, the Higgs
  boson mass, $m_h$, and the cold DM density.
In addition it includes
the constraint imposed by the experimental
upper limit on the spin-independent DM scattering
cross section from LUX \cite{Akerib:2013tjd}. 
The results described here are taken from \cite{Buchmueller:2008qe,Buchmueller:2011sw,Buchmueller:2012hv,Buchmueller:2013rsa}.
\end{sloppypar}

In Fig. \ref{fig:mc}, we show the resulting 68 \% (shown in red)
and 95 \% (shown in blue) CL limits from the mastercode analysis \cite{Buchmueller:2013rsa} 
in the $m_0,m_{1/2}$
plane corresponding to $\Delta \chi^2 = 2.3$ and 5.99
relative to the best fit point  (note the axes are reversed compared to Fig. \ref{fig:tb40}).
Results which include the ATLAS constraints at 20/fb are shown by solid curves. The best
fit point is at ($m_0, m_{1/2}$) = (5650,2100) GeV and is shown by the filled star. 
At the best fit point, we also have $A_0 \simeq -780$ GeV, and $\tan \beta = 51$
We see in Fig.~\ref{fig:mc} that the 95\% CL region in the CMSSM extends 
to $m_0 \agt 6000$ GeV and $m_{1/2} \agt 3000$ GeV. 
Note that the CMSSM fit features two
disconnected 68\%~CL `islands', the one at lower $m_0$ and $m_{1/2}$
corresponding to the stau coannihilation region, and that at larger $m_0$
and $m_{1/2}$ corresponding to the s -channel rapid-annihilation funnel region
(the best fit point in the lower island has $\tan \beta = 21$. 
The low mass island is only disfavoured at the level of 
$\Delta \chi^2 \sim 0.7$, reflecting the relative flatness of
the global $\chi^2$ function. 

\begin{figure}
\hskip -.1in
  \includegraphics[width=.5\textwidth]{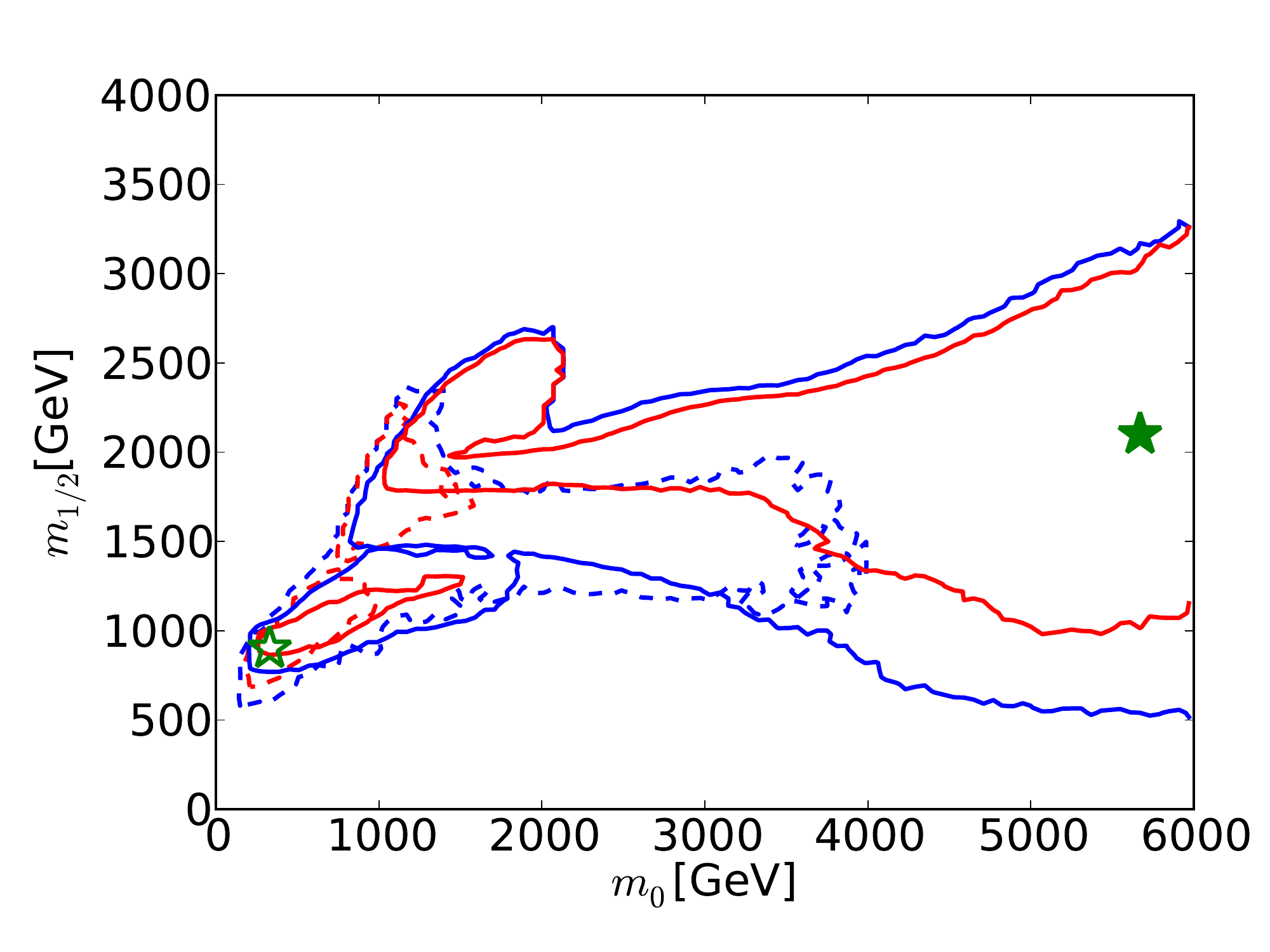}
   \caption{\it The $(m_0, m_{1/2})$ planes in the CMSSM including the ATLAS 20/fb  jets + ${/ \hspace{-.7em} E_T}$,
   BR$(B_{s, d} \to \mu^+\mu^-)$, $m_h$, $\Omega_\chi h^2$, LUX, and other contraints.
The most recent results are indicated by solid lines
  and filled stars, and previous fit based on $\sim 5$/fb of LHC data
  is indicated by dashed lines and open stars.    The blue lines denote
  68\%~CL  contours, and the red lines denote 95\%~CL contours.
  }
  \label{fig:mc}
\end{figure}

The impact of the recent LHC results can be seen by comparing the solid curves to the
dashed in Fig. \ref{fig:mc}. 
The pre-LHC expectations \cite{Buchmueller:2008qe,Buchmueller:2011sw} were
driven by a large extent by $g_\mu -2$. The initial best fit result 
was found at quite low susy masses with $(m_0, m_{1/2}) \sim
(90, 360)$ GeV and had a p-value of 37\%. The entire pre-LHC 68\% CL region is now
excluded at 95\% CL, though much of the initial 95\% CL region is still valid.  
The dashed curves in Fig. \ref{fig:mc} represent the status of 
the CMSSM after 5/fb data were collected though assuming a 125 GeV Higgs boson mass.
The best fit point in this case is at low $m_0, m_{1/2}$ shown by an open star. 
The p-value in this case is only 8.8\%. Thus already at 5/fb, the LHC results had greatly diminished
the probability that the CMSSM improves the fit relative to the SM.
The current results have a p-value of 5.1\% which is close to the SM
value.  Of course the SM p-value does not include the dark matter
constraint as there is no candidate for dark matter within the SM.

\subsection{Other searches}

\subsubsection{Direct Detection}

Direct searches of DM particles through their scattering off nuclei in
a large detector can establish that the DM matter is indeed made of a
new stable particle.  The elastic scattering of weakly interacting
massive particles (WIMPs) off nuclei taking place at low momentum
transfer can be conveniently described in terms of an effective
Lagrangian interaction of DM with quarks and gluons giving rise to
either spin independent or spin dependent interactions.  

The spin
independent (SI) cross section for WIMPs on nuclei adds coherently and
is proportional to the square of the number of nucleons, it therefore
usually dominates for heavy nuclei.  The spin independent cross
section receives a contribution from Higgs exchange, $Z$ exchange
(except for Majorana fermions) and from interactions with new colored
fermions/scalars (for example new quarks in extra dimension models or
squarks in supersymmetry). The latter contribution is however
constrained by the non-observation of new colored particles at the
LHC. 

The spin dependent (SD) cross section depends solely on the
nucleon that contributes to the nucleus spin, and is dominant only for
light nuclei.  The SD cross section receives contributions from Z
exchange and/or from interactions with new colored fermions/scalars.
In order to easily compare results obtained using different nuclei,
limits are normally expressed in terms of the SI or SD interaction
with protons and neutrons.  

At the microscopic level a positive signal
in several DM direct searches could altogether lead to information on
up to four independent quantities that depend on the details of the DM
model, the SD/SI interactions on protons and neutrons. Note however
that when scalar interactions are dominated by Higgs exchange the
cross-section on protons and neutrons are almost equal.  Furthermore,
if the DM has a mass comparable or below that of the nucleus, the
shape of the nucleus recoil energy distribution can also be used to
extract some rough information of the DM mass.

\begin{figure}[t]
\begin{center}
\hskip -0.9in
  \includegraphics[width=.5\textwidth]{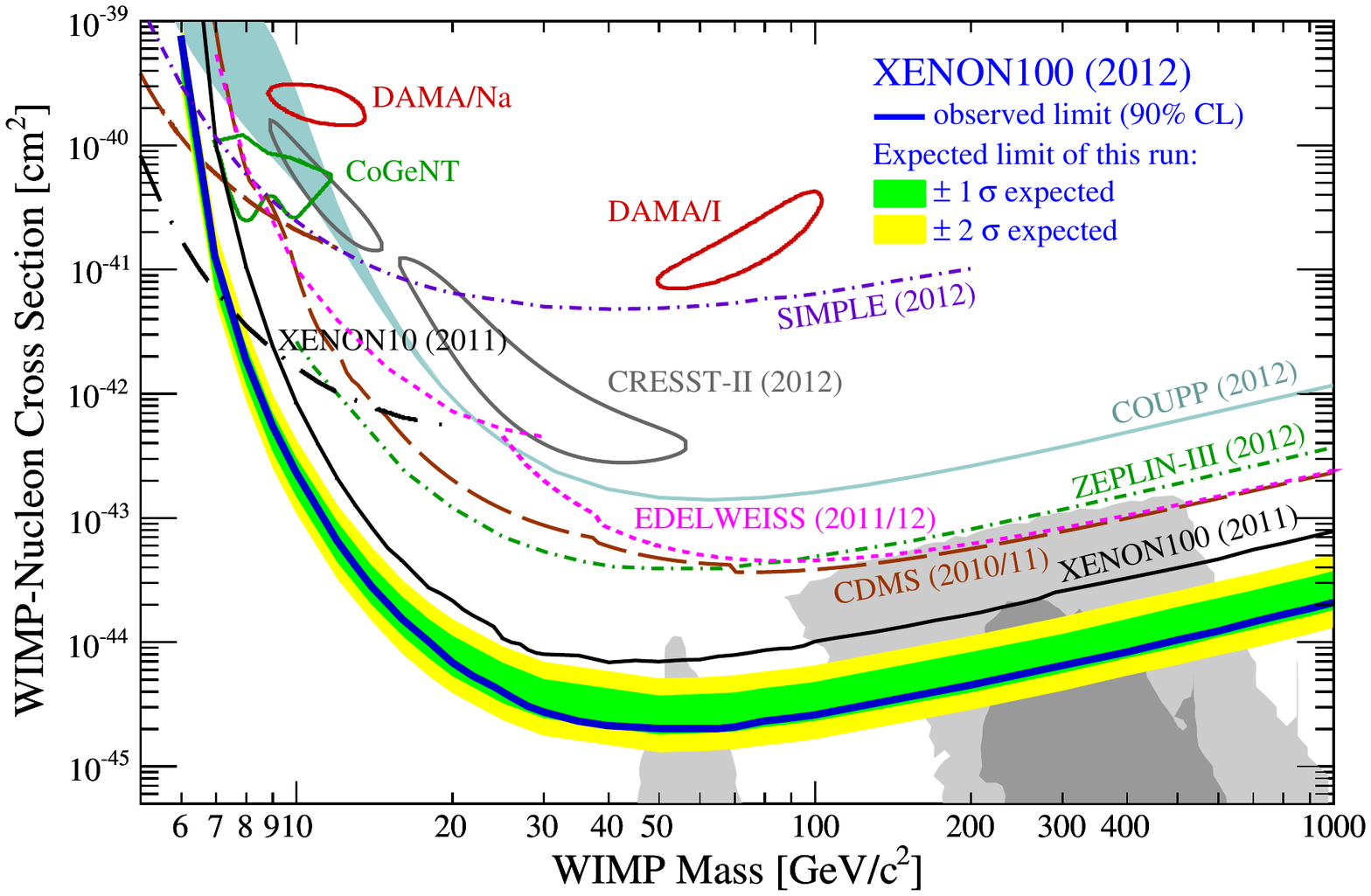}
  \vskip -.65in
\caption{Limits on spin-independent direct detection cross section $\sigma_{\rm{SI}}$  on protons \textit{vs.}  dark matter mass $m_{DM}$. In grey
the preferred region in the CMSSM, from a combination of~\cite{Strege:2011pk,Fowlie:2012im,Buchmueller:2011ab}. }
\label{fig:si}
\end{center}
\end{figure}

\begin{sloppypar}
Several experiments have been taking data, some claiming potential
signals compatible with the detection of a weakly interacting massive
particle. This includes DAMA~\cite{Bernabei:2010mq} which observes an
annual modulation, CoGeNT~\cite{Aalseth:2011wp},
CRESST~\cite{Angloher:2011uu} and CDMS-Si~\cite{Agnese:2013rvf} which
also have found signals that would be compatible with dark matter in
the range 5-30 GeV. These observations are however in conflict with
the negative search results by other collaborations notably CDMS,
Edelweiss~\cite{Armengaud:2012kd}, XENON~\cite{Aprile:2012nq} or
LUX~\cite{Akerib:2013tjd}.  The large ton scale detectors that are
planned, such as XENON should improve by more than one order of
magnitude the current sensitivity thus resolving the apparent conflict
in SI results at low masses and probing a large number of DM models,
see fig.~\ref{fig:si} for a comparison of current limits with
expectations in the CMSSM. In particular, the case where the
neutralino is a mixed gaugino/higgsino state is challenged by current
limits as illustrated in fig.~\ref{fig:higgsino} where
$P=min(f_h,1-f_h)$ and $f_h$ is the higgsino fraction.  Finally
COUPP~\cite{Behnke:2008zza}, KIMS~\cite{Kim:2012rza},
Picasso~\cite{Archambault:2009sm} (Xenon10~\cite{Angle:2008we}) have
set limits on spin dependent interactions on protons (neutrons).
\end{sloppypar}

\begin{figure}[tb]
\begin{center}
\vspace{-9cm}
\hspace*{1.5cm}
  \includegraphics[width=.5\textwidth]{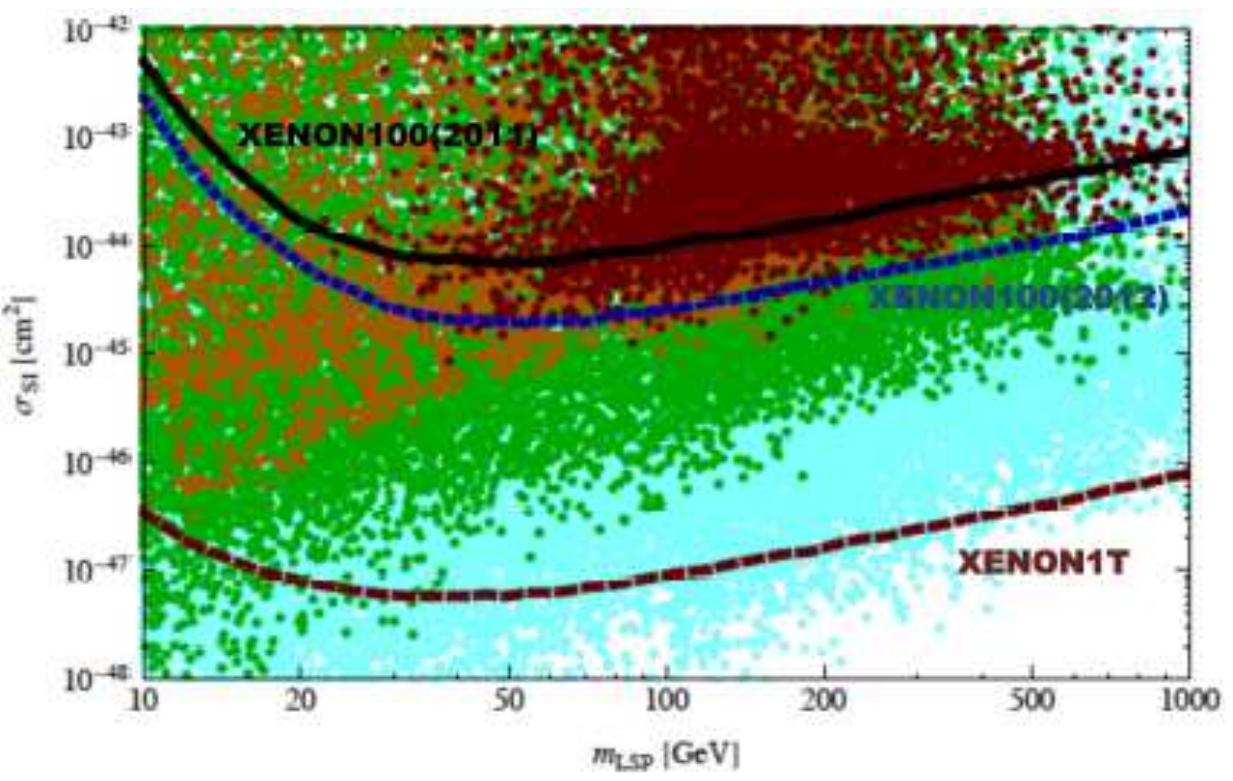}
\caption{Spin-independent direct detection cross section $\sigma_{\rm{SI}}$  on protons \textit{vs.}  dark matter mass $m_{LSP}$, from ~\cite{Perelstein:2012qg}. The  black (blue) line are the 90\% CL limits from the XENON100(2011) \cite{Aprile:2011hi} and (2012) results \cite{Aprile:2012nq}. The dashed brown line is the projected sensitivity of the XENON1T experiment \cite{Aprile:2012zx}. The color code shows the  with $P>0.2$ (red), $0.1<P<0.2$ (orange)
$0.01<P<0.1$ (green) and $0.001<P<0.01$ (blue). Note however that the relic density constraint is not imposed here.}
\label{fig:higgsino}
\end{center}
\end{figure}

\subsubsection{Indirect Detection}

In general, the goal of the ongoing generation of indirect detection
experiments sensitive to dark matter is:
\begin{itemize}
\item[I.] to probe the vanilla ``WIMP paradigm'', at least for particles
of masses at the electroweak scale and characterized by s-wave annihlation cross sections $\langle \sigma v\rangle$.
\item[II.] to clarify some of the ``anomalies'' presently claimed.
\item[III.] in the case of independent detection (at colliders or direct detection), provide one or several cross-checks taking advantage of   the multi-messenger characteristics of this detection strategy.
\end{itemize}

\begin{sloppypar}
Concerning the first task, it is worth pointing out that for some
channels Fermi-LAT has already reached the sensitivity to test this
paradigm up to a few tens of GeV (dwarf
spheroidals~\cite{Ackermann:2011wa,GeringerSameth:2011iw}, diffuse
gamma-ray halo signal~\cite{Ackermann:2012rg}) or even more for the
Galactic Center~\cite{Hooper:2012sr}.
\end{sloppypar}

In general, we expect that probing the $\sim$100 GeV mass scale will
be within reach with a decade worth of data, see for example the
forecasts in~\cite{Baltz:2008wd}.  
Preliminary results from Fermi-LAT also comfort these expectations, 
see~\cite{Anderson:fermi}.
Especially for candidates
annihilating into leptons, such a goal seems also within reach of
Planck, which probes dark matter energy deposition at early times via
its impact on the reionization (see e.g.~\cite{Galli:2009zc}).

Needless to say, if new states are below the TeV scale, these WIMP
candidates are also in the right ball-park to be probed directly or
indirectly by a future ILC, hence the complementarity of the two
approaches.

The current generation of ground-based imaging atmospheric cherenkov
telescopes (IACTs) is less sensitive to theoretically preferred values
of $\langle \sigma v\rangle$.  Nonetheless, they are already more
sensitive than Fermi-LAT to TeV scale dark matter, see
e.g.~\cite{Abramowski:2011hc}, and with the future CTA they may probe
not-yet-excluded regions of parameter space for viable particle
physics models, see e.g.~\cite{Bergstrom:2010gh}.  Typically, the
Galactic Center is among the most promising targets, provided that the
DM distribution is comparable to expectations based on pure cold dark
matter simulation or even enhanced as a consequence of
baryons~\cite{Doro:2012xx}. Dwarf spheroidal have also been studied by
IACTs see e.g.~\cite{Aleksic:2013xea,Abramowski:2014tra} and show some
potential for interesting complementary constraints, since they are
affected by very different systematics.

\begin{sloppypar}
Performances similar to Fermi-LAT (but more dependent on astrophysical
modeling of cosmic ray transport) are expected by high-precision
measurements of cosmic ray antimatter, most notably antiprotons and,
possibly, anti-deuterons ~\cite{Cirelli:2013hv}. Positrons are
significantly sensitive to astrophysical backgrounds (see
e.g.~\cite{Serpico:2011wg}) and both their primary and secondary
fluxes show a larger dependence from source distribution (in space and
time) as well as from the medium properties (e.g. their E-losses
crucially depend on B-field and interstellar radiation fields). While
they remain challenging for a {\it robust detection} of dark matter,
they may be useful for cross-checks of tentative signals. AMS-02 and,
concerning anti-deuterons, GAPS, are expected to achieve the needed
precision and sensitivity for such competitive results.
\end{sloppypar}

A caveat is mandatory: ultimately, if sufficient statistics is accumulated, the main limitations will come from the degree of understanding of the astrophysical foregrounds, so that
most of these projections must be taken with a grain of salt. 

\begin{sloppypar}
An example of the second type of goal has been provided in the recent
past by the multi-messenger constraints on the dark matter
interpretation of the PAMELA positron fraction ``rise'' (where the
relevance of the point just made clearly manifested) or, at present,
by the tentative hint for a $\sim 130\,$GeV ``gamma-ray
line''~\cite{Weniger:2012tx}.  For this kind of task, statistics helps
a lot but it is clearly not enough. Cross-checks and tests with
different techniques and possibly improved resolutions are needed.
Fortunately, current (HESS) or planned (CTA) IACTs may provide such a
tool. This is also an arena where the proposed satellite experiment
Gamma-400~\cite{Galper:2012fp} might contribute, thanks to its
superior resolution (see e.g.~\cite{Bergstrom:2012vd}).
\end{sloppypar}

The third possibility has been heavily discussed in recent years in the context of direct detection ``anomalies''  ~\cite{Bernabei:2010mq,Aalseth:2011wp,Angloher:2011uu,Agnese:2013rvf}. If interpreted in terms of ``light'' dark matter,
a wealth of indirect detection cross-checks can be thought of, see e.g.~\cite{Hooper:2012ft,Hooper:2012gq}.
We conclude by pointing out that  especially in this context (cross-checking direct detection potential signals), neutrino signals from the center of the Sun
(and possibly the Earth) are of particular relevance. In fact, they probe a similar combination (albeit not equal!) of DM-baryon cross-section and local density of dark matter as direct detection
experiments.  
Significant advances are expected by the IceCube in its current configuration, including the Deep-core (see e.g.~\cite{Silverwood:2012tp}).
Further progress may be also possible if the R\&D PINGU low-energy extension will be realized~\cite{Koskinen:2011zz} (the same would apply to
comparable programs in the Mediterranean sea such as those pursed within Km3Net, of course). Finally, it is worth pointing out that this
is also one of the few ways to potentially detect indirectly  p-wave annihilating WIMPs, since the equilibrium flux is only dependent on the DM
scattering cross section.

\subsection{Dark matter at the ILC}

The goal of colliders with regard to the DM issue is first 
to search for a new particle, stable at the collider scale,  
and as a second step to  determine the microscopic properties of this particle. These can then be used to reconstruct DM observables such as the relic abundance (within a cosmological model), the DM annihilation cross section in the galaxy and of the DM scattering cross sections on nucleons thus  checking the  self-consistency of DM interpretation of different signals and  the compatibility of specific dark matter models with  observations. 

The issues that will be most relevant  at the ILC  will  be influenced by the forthcoming results of new physics searches at the LHC and of DM searches in direct and indirect detection. 
At the LHC the generic DM signature consists of jets (and leptons) plus large missing transverse energy.   With this signature, it is highly non trivial to then resolve the underlying theory as well as the nature of the DM candidate. For this one needs
a precise determination of their properties such as  masses, spins and couplings, as was shown in many specific models~\cite{Baltz:2006fm,Mitsou:2013rwa,Berggren:2013bua}. This is where the  ILC has an important role to play.  
Failing discoveries of new particles, the role of the ILC will be to search for the DM candidate as well as for other weakly interacting  particles that might have escaped the LHC searches. Indeed
the direct production of electroweak particles not only suffer from small rates at the LHC, but often feature a  compressed spectra that  can make their identification challenging. At the ILC  new electroweak particles can be easily produced  provided the center of mass  energy is sufficient to cross the mass threshold.

\begin{sloppypar}
It might well be that the only kinematically accessible new particle
at the first stage of the ILC is the DM particle itself. In this case
DM radiative production can be used. The signal is a single high
energy photon, emitted from the incoming beam or from the exchanged
particle, and missing energy.  Effective operators that describe the
interaction of electrons with DM particles can be used to parametrise
the effect of new physics.  In this model independent approach, it has
been shown that for DM annihilation cross section compatible with the
relic abundance of dark matter, the cross section for radiative DM
production at the ILC can be large enough to observe this process
above the irreducible background from radiative neutrino production
~\cite{Bartels:2012ex}.  The electron and positron beam polarizations
can be used to significantly enhance the signal and suppress the
background from radiative neutrino production simultaneously
~\cite{Chae:2012bq}.  Furthermore the energy spectrum of the ISR
photon can be exploited to extract information on the WIMP mass and
cross-section, at the percent level~\cite{Bartels:2012ex}.  Similar
conclusions were reached for radiative neutralino production in the
MSSM~\cite{Bartels:2012rg},
distinguishing between models through a shape discrimination analysis of
the photon energy spectrum which is affected by the particle exchange
in t-channel~\cite{Konar:2009ae}.
\end{sloppypar}

A measurement of the invisible width of the Higgs also provides a
unique opportunity to determine the Higgs coupling to DM particles
directly when $m_{DM}< m_h/2$. This is an essential ingredient in
determining the spin independent direct detection cross section
($\sigma^{SI}$) in models dominated by Higgs
exchange~\cite{Djouadi:2012zc}. A refined upper limit on the invisible
width will constrain the maximal allowed value for $\sigma^{SI}$ for
light DM~\cite{Kanemura:2010sh}.

\begin{sloppypar}
Parameter determination in order to reconstruct DM observables and in
particular the relic density amounts to determining the DM mass and
its couplings, the mass of the particles exchanged in either t-channel
or s-channel and the mass splittings between the DM and the new
particles that can participate in coannihilation processes.  Many
studies have examined within the context of specific DM scenarios
whether a high enough precision can be achieved so that a meaningful
comparison with observables can be made~
\cite{Baltz:2006fm,Allanach:2004xn,Matsumoto:2009ki}.  To illustrate
what could be achieved we will consider the most studied model, a
supersymmetric model with a neutralino LSP and assume that some of the
supersymmetric spectrum is kinematically accessible.
\end{sloppypar}

The measurements of the masses of the chargino  and of the heavier neutralinos  (e.g. through a threshold scan) together with the determination of their mass splitting with the LSP using  the endpoints of the energy spectrum of the SM particle produced together with the LSP in the decay of the heavier SUSY particle, allow a 
 reconstruction of the four elements of the neutralino mass matrix. Moreover since
 the  $e^+e^-$ production cross sections of charginos and heavier neutralinos are  sensitive to the gaugino/higgsino mixing  they can provide crucial  information on the nature of the LSP. 
In a scenario where only electroweakinos are accessible at the LHC and the ILC, it was shown that with the ILC measurements at the percent level or better, the  value of $\Omega h^2$ could be inferred with an uncertainty around $10\%$ ~\cite{Baltz:2006fm}.  Of particular importance in this scenario is  the need to get a lower bound on the mass of the heavy pseudoscalar to ensure that its contribution to DM annihilation is negligible~\cite{Battaglia:2009zzb}. In other scenarios, where neutralino annihilation is strongly enhanced because the pseudoscalar exchange in the s-channel is nearly on resonance, a  determination of the pseudoscalar mass to about 3\% and its width to 20\% -  is required to infer the DM relic density at the $10\%$ level~\cite{Allanach:2004xn}. For these measurements it might be necessary to run the ILC at energies above 1 TeV.
When  co-annihilation processes play an important role,  the mass splitting of the coannihilating particle with the LSP - for example the stau NLSP, must be measured at the percent level - which  requires the measurement of masses at the few permil level~\cite{Allanach:2004xn}.
An issue that comes up is the impact of radiative corrections, which  introduce  more degrees of freedom from particles appearing only in higher order loops in the reconstruction of the neutralino mass matrix. Nevertheless it was shown in~\cite{Bharucha:2012ya}  that parameters of the electroweakino sector could still be determined at better than the percent level and that indirect information on the mass of e.g. the pseudoscalar could be extracted. 

In conclusion, despite intensive ongoing efforts to search for dark matter at colliders and in astrophysics, the nature of the DM, even whether it is a new weakly interacting particle is far from being solved. While
near future results from the LHC are expected to provide crucial clues - even to discover new particles -  it is clear that  a high precision machine such at the ILC, designed with a high enough energy to probe most of the BSM spectrum,  is needed for a  verification of the DM paradigm.

%
%




\section{Summary}
\label{sum}

Exciting times in high energy physics are just ahead. Discovering a
Higgs boson at the LHC in exactly the range predicted by electroweak precision
measurements confirms the successful strategy in particle physics of 
confronting direct discoveries with theoretical predictions of
 virtual effects in indirect searches.
Within the current theoretical and experimental uncertainties the
properties of the Higgs boson are in agreement with the predictions of the
SM.
Higher precision measurements are required to reveal whether nature
can be described via the SM only or whether physics beyond the SM
is required at some higher scale. The direct 
measurement of the total width of the 
Higgs within a few percent accuracy as well as the measurement of all 
Higgs couplings to fermions and bosons at the percent level are 
crucial to pin down the correct model of electroweak symmetry breaking. 
In this context also high precision for the Higgs 
mass is required.
With such an accuracy one gets 
high sensitivity to virtual effects and even small 
traces of BSM physics become
measurable. 
In order to really establish the Brout-Englert-Higgs mechanism
also the Higgs self-couplings would be required. 
An accuracy of 10-20\% would constitute a first test
whether the 
Higgs potential provides indeed the required structure for the vacuum
to generate the BEH mechanism.
As we have discussed in this report, the full physics programme of the
Linear Collider could perfectly well fulfill all these requirements.

Further footprints of new physics can be detected in the measurement
of the electroweak couplings of the top quark with a unique precision
at the Linear Collider. Exploiting asymmetries with polarized beams
allows to determine the electroweak top quark form factors at the per
cent level, that's up to one order of magnitude more precise than
expectation from corresponding analyses at the LHC with
$\sqrt{s}=14$~TeV and 300~fb$^{-1}$. Polarized beams are required to
fit all factors simultaneously and to measure the asymmetry.

Highest precision in measuring the top quark mass is mandatory 
to match the precision of the theoretical predictions with the 
expected experimental precision of the electroweak precision observables
that are strongly sensitive to the effects of virtual
particles far beyond the kinematic limit.  
In order to uniquely relate the measured quantity to a
well defined mass scheme the top quark measurement via a threshold
scan is required and one can determine the
mass of the top quark 
with an uncertainty of $\Delta m_{top}^{\overline{MS}}=100$~MeV.

The LC has also an overwhelming potential for the discovery of further
electroweak interacting particles and, in particular, of a cold dark
matter candidate. The potential of the LC to resolve even challenging
scenarios, for instance, via applying the ISR method and to determine
precisely the interaction character of dark matter candidate via
applying polarized beams.

As it is shown in many
reports~\cite{AguilarSaavedra:2001rg,DBD,Linssen:2012hp,Diberder,MoortgatPick:2005cw,Moortgat-Pick:2013awa}
as well as discussed here in detail, a Linear Collider with precisely
tunable energy in the range of $\sqrt{s}=91$~GeV up to $\ge 1$TeV,
high luminosity and polarized beams provides the required flexibility
and precision to tackle these physics questions left by the LHC and is
well prepared for even the `Unexpected'. With the currently promising 
activities towards the realization of the International Linear Collider 
in Japan one could even discuss the optimization of the 
physics potential in HEP via a time of 
concurrent running\cite{Weiglein:2004hn} 
of the LHC and the LC. The 
described physics goals as well as not-yet-thought physics questions
could be addressed by this option. 

The physics world has changed on July, 4th, 
2012 with the discovery of the Higgs boson at the LHC. 
Cucial milestones in particle physics
are expected to be achieved in the near future 
with data in pairs from the upgraded LHC and from a future Linear Collider.
In combination with astroparticle physics, 
a new era for pinning down the structure of our micro-- as well as 
macrocosm has just started.


\begin{acknowledgements}
\begin{sloppypar}
Several authors acknowledges the support of the DFG through the grant 
SFB676 `Particles, Strings and the Early Universe'.\\
This work was supported by European Commission through the 
contract PITN-GA-2012-316704
(HIGGSTOOLS).\\
This work is supported in part by the Creative
  Scientific Research Grant No.\,18GS0202 of the Japan Society for
  Promotions of Science (JSPS), the JSPS Grant-in-Aid for Science
  Research No.\,22244031, and the JSPS Specially Promoted Research
  No.\,23000002.\\
This work is part of the D-ITP consortium, a program of the Netherlands
Organisation for Scientific Research (NWO) that is funded by the Dutch
Ministry of Education, Culture and Science (OCW).\\
G.~Moortgat-Pick would like to thank A.A. Mikhailichenko for useful discussions
and valuable comments on collider aspects.\\
C.~Grojean is supported by the Spanish Ministry MICNN under contract
FPA2010-17747 and by the European Commission under the ERC Advanced
Grant 226371 MassTeV and M.M.~M\"uhlleitner is supported by the
DFG/SFB-TR9 {\it Computational Particle Physics}.  \\
M.~Asano
acknowledges support from the German Research Foundation (DFG) through
grant BR 3954/1-1 and DFG TRR33 "The Dark Universe".\\ 
S.~Matsumoto
acknowledges supports from the MEXT, Japan through grants
Nos. 22244031 \& 26287039, and also from the WPI Initiative, MEXT,
Japan. \\ 
K.~Rolbiecki has been supported by the MICINN, Spain, under contract
FPA2013-44773-P, Consolider-Ingenio CPAN CSD2007-00042 and the Spanish
MINECO {\em Centro de excelencia Severo Ochoa Program} under
grant SEV-2012-0249.\\
S.~Godfrey was supported in part by the Natural Sciences and Engineering
Research Council of Canada under grant number 121209-2009 SAPIN.\\
The work of S.Y.~Choi was supported by Basic Science Research Program through
the National Research Foundation (NRF) funded by the Ministry of 
Education, Science and Technology (2012-0002746).\\
The work was partly supported by Polish National Center for
Science, grant NCN OPUS 2012/05/B/ST2/03306 (2012- 2016)
and the grant NCN DEC-2012/05/B/ST2/02597, and by BMBF,
DAAD PPP Poland Project 56269947, "Dark Matter at Colliders"
(M.~Krawczyk), grants RFBR 11-02-00242, NSh-3802.2012.2 (I.Ginzburg).
\\
A.S.\ Kronfeld is supported in part by the German Excellence
Initiative and the European Union Seventh Framework Programme under
grant agreement No.~291763 as well as the European Union's Marie Curie
COFUND program.  Fermilab is operated by Fermi Research Alliance, LLC,
under Contract No.\ DE-AC02-07CH11359 with the United States
Department of Energy. \\ The work of K. A. Olive was supported in part
by DOE grant DE-SC0011842 at the University of Minnesota.
\end{sloppypar}
\end{acknowledgements}

\end{document}